# Predictions of charge density distributions for nuclei with Z ≥ 8*

Yun Dong Wang,[1] Tian Shuai Shang,[1] Hui Hui Xie,[1] Peng Xiang Du,[1] Jian Li,[1,†] and Haozhao Liang[2,3,4]

[1] *College of Physics, Jilin University, Changchun 130012, China*

[2] *Department of Physics, Graduate School of Science, The University of Tokyo, Tokyo 113-0033, Japan*

[3] *Quark Nuclear Science Institute, The University of Tokyo, Tokyo 113-0033, Japan*

[4] *RIKEN Interdisciplinary Theoretical and Mathematical Sciences Program, Wako 351-0198, Japan*

A deep neural network (DNN) has been developed to accurately predict nuclear charge density distributions for nuclei with proton numbers $Z \geq 8$. By incorporating essential nuclear structure features, the model achieves a significant improvement in predictive accuracy over conventional methods. The charge density distributions are analyzed using a Fourier-Bessel (FB) series expansion, and the DNN is trained on a comprehensive dataset derived from relativistic continuum Hartree-Bogoliubov (RCHB) theory calculations. The model demonstrates exceptional performance, with root-mean-square deviations of 0.0123 fm and 0.0198 fm for charge radii on the training and validation sets, respectively—remarkably surpassing the precision of the original RCHB calculations. Beyond advancing nuclear physics research, this high-precision model provides critical data for applications in atomic physics, nuclear astrophysics, and related fields.

## I. INTRODUCTION

Nuclear charge density is one of an essential properties of nucleus it reflects a large amount of nuclear structure information, including shell structure and pairing effect [1, 2]. Nuclear charge density distribution can also be extended to other applications, such as nuclear astrophysics, atomic physics, quantum electrodynamics, and related fields [3–5].

High-energy electron elastic scattering [6–8] and muonic atom spectroscopy [9–11] represent the most precise experimental techniques for determining nuclear charge density distributions. Among these, electron-nucleus scattering has emerged as a particularly powerful tool for investigating charge density distributions of atomic nuclei. However, current experimental approaches are predominantly limited to stable nuclei [12–14] and a select number of long-lived radioactive isotopes [6], leaving significant gaps in our understanding of exotic nuclear systems. This scarcity of experimental data underscores the critical importance of developing comprehensive theoretical datasets to advance nuclear structure research.

Given the inherent limitations of experimental data and the increasing scientific demand for precise characterization and prediction of nuclear charge density distributions, microscopic theoretical approaches have become indispensable in nuclear structure research. Among various microscopic frameworks, density functional theory (DFT) [15–19] has emerged as the most comprehensive approach for characterizing nuclear charge properties, encompassing both nonrelativistic and relativistic formulations. As a fully microscopic approach, DFT provides a self-consistent description of nuclear systems without requiring additional phenomenological parameters. However, comparative studies with methods incorporating Garvey-Kelson local relations [20] reveal that the predictive accuracy of DFT for charge properties still requires substantial improvement. The present dataset offers a robust benchmark for refining these theoretical approaches across diverse nuclear regions.

To overcome the limitations of DFT, we employ machine learning (ML) techniques in this study, leveraging our comprehensive dataset that integrates both theoretical calculations and experimental constraints. ML methods have become increasingly important in nuclear physics research [21–26], demonstrating particular success in predicting nuclear charge radii through various approaches. These include kernel ridge regression [27, 28], Bayesian neural networks [29, 30], and convolutional neural networks [31, 32], all of which have achieved notable accuracy. However, applications to charge density distributions remain relatively unexplored, primarily due to two key challenges: (1) the lack of a uniffed framework that simultaneously describes both charge density distributions and charge radii, and (2) the absence of globally applicable charge density tables.

In one of our previous works [33], a DNN model (DNN(i2)) was developed using proton number ($Z$) and neutron number ($N$) as inputs, achieving significant improvements over RCHB theory [34, 35] with PC-PK1 [36] in predicting charge density. The current study builds upon this foundation by incorporating three key nuclear structure features into input features: the distances from $Z$ and $N$ to magic number $V_Z$ and $V_N$ for shell effects together with a pairing parameter $\Delta$ ($\Delta = \frac{1}{2}[(-1)^Z + (-1)^N]$). The refined model is trained on dataset of RCHB-derived FB coefficients while being rigorously constrained by experimental charge radii

* This work was supported by the National Natural Science of China (No.12475119).

† jianli@jlu.edu.cn

data, establishing a new benchmark for nuclear charge density predictions.

## II. METHODS

### A. Fourier-Bessel analysis

The nuclear charge density distribution was analyzed using a model-independent approach based on the FB series expansion method, originally introduced by Dreher *et al.* [37].

#### 1. Mathematical formalism

For a spherically symmetric charge density distribution $\rho(r)$ with finite spatial extent, a physical cutoff radius $R$ was defined such that

$$\rho(r) = 0 \quad \text{for} \quad r > R. \tag{1}$$

The FB series expansion of $\rho(r)$ within the nuclear volume ($r \leq R$) takes the form of

$$\rho(r) = \sum_{k=1}^{N} a_k j_0\left(\frac{k\pi r}{R}\right), \tag{2}$$

where $j_0(x) = \sin x/x$ is the zeroth-order spherical Bessel function, and the first $N(= Rq_{\max}/\pi)$ coefficients $a_k(k = 1, 2, ..., N)$ of this series expansion are determined from the experimental data directly. In this study, $N$ is taken as 17. That is because the 67 nuclei are given by FB analysis experimentally, with a maximum of 17 coefficients [6]. The cutoff radius of a nucleus is correlated with its mass number $A$. Specifically,

$$R = \begin{cases} 8 \text{ fm}, & \text{for } A < 50, \\ 9 \text{ fm}, & \text{for } 50 \leq A < 100, \\ 10 \text{ fm}, & \text{for } 100 \leq A < 150, \\ 12 \text{ fm}, & \text{for } A \geq 150. \end{cases} \tag{3}$$

#### 2. Momentum-space representation

The charge form factor $F(q)$, which represents the momentum-space distribution of the nuclear charge density distribution, is obtained through the FB transform of the charge density, i.e.,

$$F(q) = \frac{4\pi}{Z}\int_0^\infty \rho(r) j_0(qr) r^2 dr. \tag{4}$$

The fact that an inverse relationship exists enables direct determination of the FB coefficients $a_k$ from the charge form factor as

$$a_k = \frac{q_k^2}{2\pi R} F(q_k). \tag{5}$$

Here, $q_k = k\pi/R$ denotes the discrete momentum transfer values corresponding to the $k$-th component of the expansion. This formulation establishes a crucial connection between the form factor and the underlying charge density parameters.

#### 3. Normalization and constraints

The charge number normalization condition ensures the conservation of total charge in the FB construction, which is expressed as

$$4\pi \int_0^\infty \rho(r) r^2 dr = Z, \tag{6}$$

where $Z$ represents the total nuclear charge. The integral on the left-hand side of Eq. (6) can be analytically calculated, yielding

$$\int_0^\infty \rho(r) r^2 dr = \sum_{k=1}^{N} \frac{(-1)^{k+1} a_k R^3}{(k\pi)^2}. \tag{7}$$

This formulation guarantees proper charge conservation in the constructed charge density.

#### 4. Moment analysis

The $n$-th moment of the charge distribution is given by

$$\langle r^n \rangle = \frac{4\pi}{Z}\int_0^R \rho(r) r^{n+2} dr. \tag{8}$$

In particular, the charge radius $R_c$ used in this paper is defined as root-mean-square (RMS) radii, which is the square root of the second moment

$$R_c = \sqrt{\langle r^2 \rangle}. \tag{9}$$

#### 5. The data set

This study utilizes a comprehensive data set of nuclear charge density distributions systematically computed across the entire nuclide chart. The calculations were performed within the RCHB theory framework [34, 35, 38, 39] using the PC-PK1 relativistic energy density functional [36]. This approach provides a consistent and theoretically sound basis for investigating nuclear charge distributions throughout the nuclear landscape.

In this study, the RCHB theory was employed to systematically compute nuclear charge density distributions and determine the corresponding FB coefficients using Eqs. (2) and (6). The resulting dataset serves as the foundation for DNN training. To ensure physical reliability, we incorporate both calculated RMS radii and experimental charge radius data into the loss function optimization. This dual approach enhances the accuracy of both global charge density distributions and fine-scale charge properties across the nuclear chart.

### B. Deep neural network approach

#### 1. *Network architecture specification*

The present work employs a six-layer fully connected neural network with the following architecture:

$$\mathcal{N} : \mathbb{R}^5 \to \mathbb{R}^{17}. \tag{10}$$

The layer configuration consists of an input layer, four hidden layers, and an output layer. The information flow through the network follows the transformation rule

$$a^{(n)} = f\left(W^{(n)} a^{(n-1)} + b^{(n)}\right), \tag{11}$$

where $a^{(n)}$ denotes the output of the $n$th layer, $W^{(n)}$ is the weight matrix, $b^{(n)}$ is the bias vector, and $f(x) = \tanh(x)$ serves as the activation function.

#### 2. *Input feature engineering*

The input vector $x \in \mathbb{R}^5$ encodes nuclear structure information:

$$x = [Z, N, V_Z, V_N, \Delta]^T, \tag{12}$$

where $Z$ is proton number, $N$ is neutron number, $V_Z$ and $V_N$ are the distances from proton and neutron number to the nearest magic number, and $\Delta = \frac{1}{2}\left[(-1)^Z + (-1)^N\right]$ related to paring effect.

#### 3. *Output representation*

The network predicts FB coefficients for charge density distributions:

$$y = [a_1, a_2, ..., a_{17}]^T \in \mathbb{R}^{17}. \tag{13}$$

#### 4. *Optimization framework*

Maximum likelihood estimation(MLE) [40] was used to find the suitable weight and bias, i.e.,

$$\theta^{\mathrm{MLE}} = \arg\min_{\theta} \mathrm{Loss}(\theta), \tag{14}$$

where $\theta = \{W, b\}$, $\mathrm{Loss}(\theta)$ is the loss function of the DNN.

The composite loss function implements a physics-informed training strategy. The training of the neural network is divided into two steps in this work. The first step is training DNN to predict the FB coefficients derived from the RCHB theory. The corresponding loss function reads

$$\mathrm{Loss}_1(y^t, y^p) = \frac{1}{N_s}\sum_{k=1}^{N_s}(y_k^t - y_k^p)^2\,, \tag{15}$$

where $y^t$ denotes the result of RCHB, $y^p$ represents the network output value, and $N_s$ is the batch-size hyperparameter of the network.

The second step is training DNN with the experimental charge radii and the corresponding loss function reads

$$\begin{aligned}\mathrm{Loss}_2(y^t, y^p) = (1-\lambda)\frac{1}{N_s}\sum_{k=1}^{N_s}(y_k^t - y_k^p)^2 \\ + \lambda\frac{1}{N_s}\sum_{k=1}^{N_s}(R_{c,k}^t - R_{c,k}^p)^2,\end{aligned} \tag{16}$$

where $\lambda \in [0, 1]$ is the weight hyperparameter. $R_c^t$ is the experimental values, and $R_c^p$ is the charge radii from DNN.

All hyperparameters used for DNN in this work are listed in Table 1.

## III. DATA RECORDS

As detailed in Sec. II, the DNN training employs a two-stage optimization process. In the initial phase, the RCHB-calculated FB coefficients for all 1,014 nuclei [41, 42] were treated as the data set, with Eq. (15) serving as the loss function. The subsequent refinement phase incorporates experimental values through the composite loss function defined in Eq. (16), which optimizes the preliminary DNN model from the first stage. This hierarchical training strategy ultimately yields an optimized neural network model for nuclear charge density distribution predictions.

The DNN outputs the FB coefficients, which are subsequently transformed into charge radii, the second, forth, and sixth moments using Eqs. (8) and (9). The optimized DNN was used to predict the nuclei listed in AME2020 with $Z \geq 8$. All numerical outputs from the DNN, including both the FB coefficients, derived charge radii and high-order moments, are systematically presented in Tables 3 and 4 of the appendix. And the coefficients are rounded to eight decimal places, while the radii and higher-order moments are rounded to four decimal places.

TABLE 1. The hyperparameters of DNN.

| $i$th layer | Name | Number of neurons | Activation function |
|---|---|---|---|
| 0 | Input Layer | 5 | |
| 1-4 | Dense Layer | 50 | Tanh |
| 5 | Output Layer | 17 | |
| **Other hyperparameters** | | **Values and properties** | |
| Normalization factor | | 14.85 | |
| Batch size | | 10 | |
| Objective function | | $\text{Loss}_1, \text{Loss}_2$ | |
| Optimizer | | Adam | |
| $\lambda$ in Eq.(16) | | 0.7 | |
| Learning rate | | 0.0001 for $\text{Loss}_1$ and $\text{Loss}_2$ | |

The complete data set output by DNN, have been uploaded to the Science Data Bank. A direct link to the dataset is available at (https://doi.org/10.57760/sciencedb.28479).

## IV. TECHNICAL VALIDATION

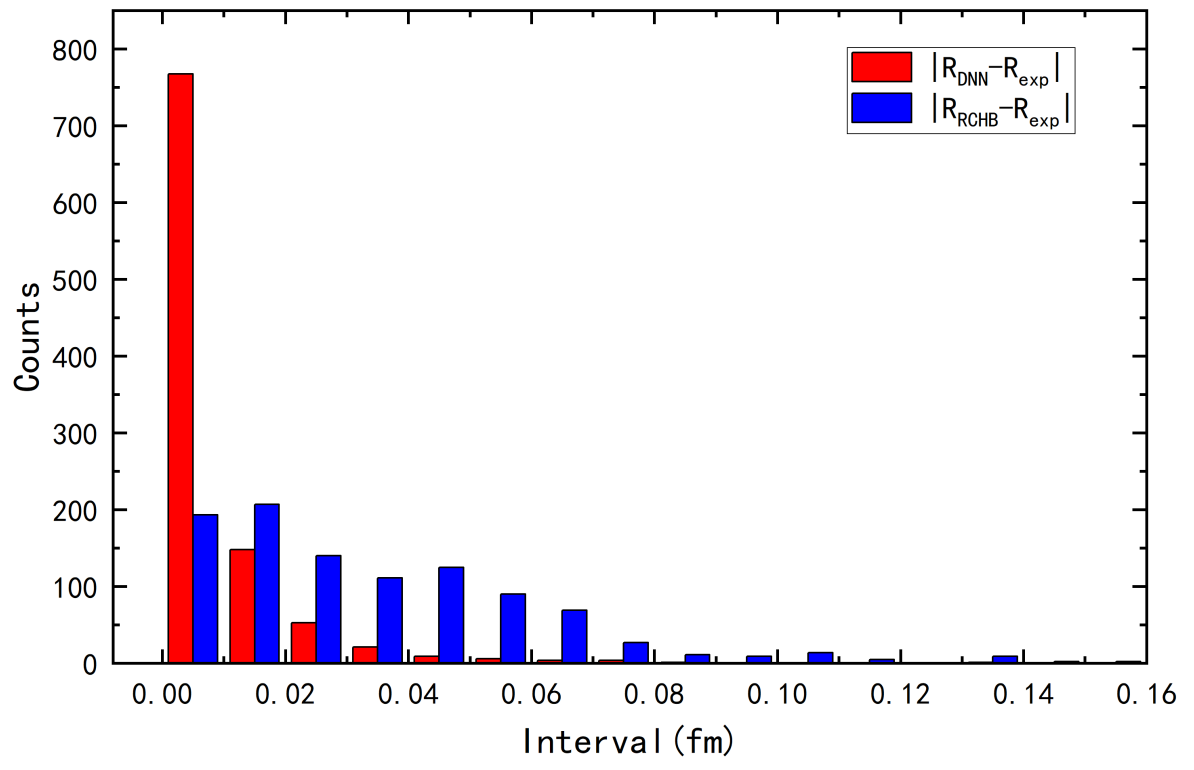

Fig. 1. Comparative distributions of absolute charge radius deviations between DNN predictions and RCHB calculations relative to experimental values. The red indicates deviations for DNN, and blue for RCHB.

To validate the DNN predictions, we computed charge radii from the network's FB coefficients and compared them with experimental values. A systematic comparison was performed between these deviations and those obtained from RCHB theoretical calculations across the complete dataset of 1,014 nuclei. As shown in Fig. 1, the DNN-derived charge radii demonstrate superior precision, with deviations from experimental values predominantly confined within the 0 — 0.01 fm range. In contrast, the RCHB predictions exhibit significantly broader dispersion in their deviations. RCHB has proven to be a successful DFT for investigating nuclear structure. However, due to inherent limitations such as its reliance on an approximation of spherical nuclei, there remains improvement in achieving a globally accurate description of nuclei. While enhancing the theoretical accuracy of RCHB poses significant challenges, the application of DNN offers a relatively straightforward pathway toward this goal. In future studies, ML is expected to play a crucial role in optimizing theoretical models of nuclear structure, thereby providing more accurate nuclear structure data.

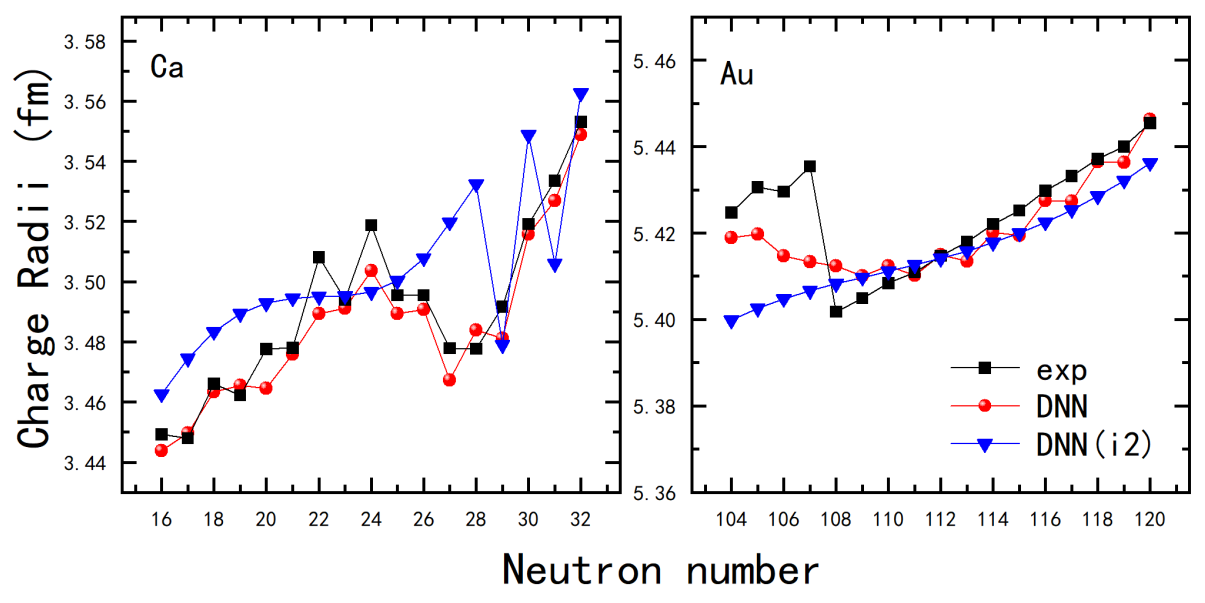

Fig. 2. A comparison of charge radii for Ca and Au isotopes: DNN predictions versus DNN(i2) predictions and experimental values.

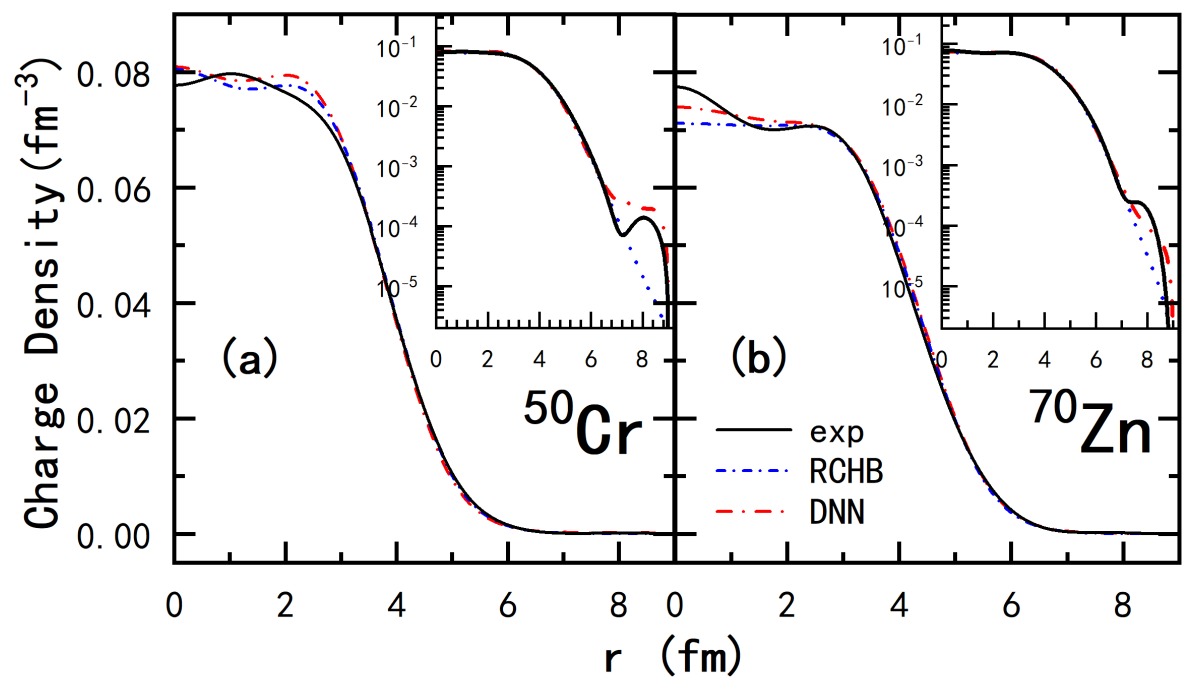

Fig. 3. Charge density distributions of $^{50}$Cr, and $^{70}$Zn. The results from the RCHB theory and the corresponding experimental data are also included for comparison.

TABLE 2. Comparison of charge radii for the test set (Ni, Pd, Hg, and Bi isotopes) among predictions from the DNN model, the earlier DNN(i2) model, RCHB calculations, and experimental values. The experimental data comprise new measurements reported since 2021. The corresponding RMSE for the DNN, DNN(i2), and RCHB models are denoted $\sigma$.

| | $A$ | $R_c^{\mathrm{DNN}}$ | $R_c^{\mathrm{DNN(i2)}}$ | $R_c^{\mathrm{RCHB}}$ | $R_c^{\mathrm{exp}}$ | Ref. |
|---|---|---|---|---|---|---|
| Ni | 54 | 3.7487 | 3.7851 | 3.7220 | 3.7366 | [43] |
| | 55 | 3.7424 | 3.7808 | 3.7074 | 3.7252 | [44] |
| | 56 | 3.7307 | 3.7785 | 3.6974 | 3.7226 | [44] |
| | 59 | 3.7965 | 3.7967 | 3.7423 | 3.7820 | [45] |
| | 63 | 3.8468 | 3.8618 | 3.8058 | 3.8420 | [45] |
| | 65 | 3.8657 | 3.8828 | 3.8365 | 3.8560 | [45] |
| | 66 | 3.8816 | 3.8895 | 3.8513 | 3.8700 | [45] |
| | 67 | 3.8851 | 3.8949 | 3.8703 | 3.8730 | [45] |
| | 70 | 3.9239 | 3.9099 | 3.8914 | 3.9100 | [45] |
| Pd | 98 | 4.4154 | 4.4303 | 4.3950 | 4.4192 | [46] |
| | 99 | 4.4361 | 4.3940 | 4.4025 | 4.4316 | [46] |
| | 100 | 4.4480 | 4.4834 | 4.4192 | 4.4532 | [46] |
| | 101 | 4.4733 | 4.4528 | 4.4280 | 4.4646 | [46] |
| | 112 | 4.5879 | 4.5881 | 4.5422 | 4.5957 | [46] |
| | 114 | 4.5997 | 4.6033 | 4.5586 | 4.6094 | [46] |
| | 116 | 4.6134 | 4.6166 | 4.5132 | 4.6189 | [46] |
| | 118 | 4.6274 | 4.6279 | 4.5872 | 4.6268 | [46] |
| Hg | 207 | 5.4976 | 5.4834 | 5.4952 | 5.4923 | [47] |
| | 208 | 5.5199 | 5.4917 | 5.5100 | 5.5033 | [47] |
| Bi | 187 | 5.4621 | 5.4253 | 5.4186 | 5.4345 | [48] |
| | 188 | 5.4501 | 5.4330 | 5.4250 | 5.4907 | [48] |
| | 189 | 5.4632 | 5.4400 | 5.4279 | 5.4428 | [48] |
| | 191 | 5.4653 | 5.4519 | 5.4381 | 5.4473 | [48] |
| $\sigma$ | | 0.0149 | 0.0273 | 0.0374 | | |

A comparison of charge radii derived from the FB coefficients of the DNN and the previous DNN model (DNN(i2)) is also presented. As shown in Fig. 2, the Ca and Au isotopic chains—previously identified in [33] as poorly described by DNN(i2) are now more accurately reproduced in the present work, particularly in capturing the complex evolution of their charge radii.

Furthermore, we quantitatively evaluated the prediction accuracy by computing the root-mean-square errors (RMSE) between theoretical and experimental charge radii. The DNN achieves RMSE values of 0.0123 fm for training set and 0.0198 fm for validation set, demonstrating remarkable consistency with experimental values. Notably, these results represent 71.4% and 54.0% improvement, respectively, over the RCHB theoretical predictions with RMSE = 0.0430 fm.

In addition, the DNN exhibits excellent predictive capability for charge density distributions. Due to limited experimental data, chromium (Cr) and zinc (Zn) are selected as representative cases in Fig. 3. The DNN results show significant improvements over the RCHB method, not only optimizing the central density to better match experimental trends but also successfully reproducing the characteristic tail structure—a key feature absent in RCHB calculations.

## V. USAGE NOTES

This study presents a global dataset of nuclear charge density distribution predictions generated through DNN model, and the charge radii, high-order moments calculated by charge density distribution. The primary objectives of this work are to systematically document the complete dataset, and provide accessible charge density distributions for the nuclear physics community and interdisciplinary researchers. The dataset offers significant utility for fundamental and applied nuclear physics research, with particular relevance to:

1. The nuclear charge density distribution is a fundamental property of the nucleus. It serves as a critical benchmark for validating, refining, and advancing nuclear structure models. Theoretical frameworks like DFT provide a microscopic description of charge densities, yet face challenges in accurately treating nuclei with complex correlations. To bridge this gap between theory and experiment, a DNN model that refines the predictions of RCHB calculations was developed. This approach significantly improves the theoretical accuracy, providing more reliable charge density distributions for a broad range of nuclei.

2. In atomic physics, the spectroscopic properties of

atoms are profoundly influenced by electron-nucleus hyperfine interactions, which are highly sensitive to the nuclear charge density distribution. Since the electromagnetic interaction between electrons and the nucleus is governed by the Coulomb force, the precise spatial distribution of the nuclear charge directly determines the Coulomb potential experienced by the electrons. An accurate description of the nuclear charge density is therefore essential for modeling the effective Coulomb field, ultimately affecting atomic energy levels, transition rates, and hyperfine structures. A reliable nuclear charge density model allows for rigorous corrections to the electron wavefunctions near the nucleus, where the relativistic and quantum electrodynamic (QED) effects dominate. Consequently, improving the accuracy of nuclear charge densities not only advances nuclear structure studies but also enables more precise predictions of atomic spectra, supporting tests of fundamental physics, including QED and searches for new physics beyond the Standard Model.

3. The nuclear charge density distribution, as a fundamental observable in nuclear structure studies, not only constrains key parameters in the nuclear matter equation of state but also provides essential inputs for nuclear astrophysics research. The nuclear charge density is closely related to the symmetry energy of nuclear matter and its density dependence—a relationship that directly influences our understanding of neutron star structure and heavy-element nucleosynthesis processes. Furthermore, the nuclear charge density serves as a critical input for nuclear reaction network calculations in extreme astrophysical environments. Therefore, the precise nuclear charge density distributions are of paramount importance for advancing nuclear astrophysics research.

## CODE AVAILABILITY

A total of 1014 measured nuclei [41, 42] were utilized in this study. The dataset was randomly partitioned into training and validation sets with an 8:2 ratio, and this division remained consistent throughout all subsequent training procedures. A separate test set consisting of 23 nuclei with charge radii measured since 2021 was employed for performance evaluation. The results summarized in Table 2 demonstrate that the DNN model reduces the RMSE in predictions of charge radius by approximately 45% and 60% compared to the the previous DNN model (DNN(i2)) and RCHB approaches, respectively.

The computational implementation leverages the PYTORCH deep learning framework . The robust convergence behavior was demonstrated with the model typically reaching optimal performance within 10,000 training epochs. This training process requires approximately 30 minutes of computation time when executed on an NVIDIA GeForce RTX 4060 GPU. The trained network exhibits remarkable inference efficiency, generating predictions in mere milliseconds - a feature that facilitates rapid deployment and real-time applications.

The hyperparameters of DNN are all in Table 1 and the dataset is attached at the end of the article in the form of an appendix.

## ACKNOWLEDGEMENTS

This work is supported by the National Natural Science Foundation of China (Grant No.12475119 ), the Key Laboratory of Nuclear Data Foundation(JCKY2025201C154) , and the JSPS Grant-in-Aid for Scientific Research (S) under Grant No.20H05648.

## APPENDIX

TABLE 3. Nuclear charge density distribution FB coefficients.

| $Z$ | $N$ | $a_1$ | $a_2$ | $a_3$ | $a_4$ | $a_5$ | $a_6$ | $a_7$ | $a_8$ | $a_9$ | $a_{10}$ | $a_{11}$ | $a_{12}$ | $a_{13}$ | $a_{14}$ | $a_{15}$ | $a_{16}$ | $a_{17}$ |
|---|---|---|---|---|---|---|---|---|---|---|---|---|---|---|---|---|---|---|
| 8 | 3 | 0.01839419 | 0.04039983 | 0.03131848 | 0.00400708 | -0.00826989 | -0.00640241 | -0.00293219 | -0.00122300 | 0.00080751 | -0.00027369 | -0.00007487 | 0.00022823 | -0.00021985 | -0.00015593 | -0.00003307 | 0.00008313 | -0.00010921 |
| 8 | 4 | 0.02043784 | 0.04267001 | 0.02959763 | 0.00117381 | -0.01131378 | -0.00760564 | -0.00156642 | -0.00093862 | 0.00009871 | -0.00035186 | 0.00003623 | 0.00020064 | -0.00021395 | -0.00006704 | 0.00001610 | -0.00015532 | -0.00016963 |
| 8 | 5 | 0.01947830 | 0.04216031 | 0.03181686 | 0.00287531 | -0.00998659 | -0.00693113 | -0.00215171 | -0.00096519 | 0.00055109 | -0.00019558 | 0.00001299 | 0.00026405 | -0.00022948 | -0.00021116 | -0.00010945 | 0.00016657 | -0.00009240 |
| 8 | 6 | 0.02005430 | 0.04276412 | 0.03050799 | 0.00199837 | -0.01107007 | -0.00742206 | -0.00095994 | -0.00058103 | 0.00004341 | -0.00039412 | 0.00005332 | 0.00015545 | -0.00014650 | -0.00008637 | 0.00001179 | -0.00011645 | -0.00009829 |
| 8 | 7 | 0.01765818 | 0.04043640 | 0.03258796 | 0.00505576 | -0.00840960 | -0.00667995 | -0.00209365 | -0.00057660 | 0.00067655 | -0.00032554 | -0.00004484 | 0.00014251 | -0.00008441 | -0.00018436 | -0.00004347 | 0.00015605 | 0.00002509 |
| 8 | 8 | 0.01828531 | 0.04140825 | 0.03167865 | 0.00399986 | -0.00954859 | -0.00679584 | -0.00070303 | -0.00006617 | 0.00043219 | -0.00045997 | -0.00003500 | 0.00008043 | -0.00005678 | -0.00005708 | 0.00005224 | -0.00007562 | 0.00001770 |
| 8 | 9 | 0.01741945 | 0.04048078 | 0.03279244 | 0.00494578 | -0.00876371 | -0.00678205 | -0.00176772 | -0.00035552 | 0.00063699 | -0.00030875 | -0.00003316 | 0.00011527 | -0.00004036 | -0.00019206 | -0.00005519 | 0.00019790 | 0.00007869 |
| 8 | 10 | 0.01947708 | 0.04309213 | 0.03183601 | 0.00277005 | -0.01079923 | -0.00669858 | 0.00025449 | -0.00002673 | 0.00000526 | -0.00038822 | 0.00008297 | 0.00007860 | -0.00003270 | -0.00012325 | -0.00001005 | -0.00001467 | 0.00003977 |
| 8 | 11 | 0.01873473 | 0.04259625 | 0.03318942 | 0.00329297 | -0.01042109 | -0.00677722 | -0.00098098 | -0.00041723 | 0.00021652 | -0.00020032 | 0.00006132 | 0.00014607 | -0.00003865 | -0.00025068 | -0.00013168 | 0.00027163 | 0.00008735 |
| 8 | 12 | 0.02059121 | 0.04526189 | 0.03239714 | 0.00100972 | -0.01257837 | -0.00671909 | 0.00085801 | -0.00004447 | -0.00031492 | -0.00028647 | 0.00014125 | 0.00007022 | -0.00000136 | -0.00016222 | -0.00005532 | 0.00006179 | 0.00007676 |
| 8 | 13 | 0.01988129 | 0.04496871 | 0.03386958 | 0.00122456 | -0.01245722 | -0.00689462 | -0.00060456 | -0.00050895 | -0.00005424 | -0.00008439 | 0.00009156 | 0.00017413 | -0.00003832 | -0.00028334 | -0.00018709 | 0.00035652 | 0.00010642 |
| 8 | 14 | 0.02144911 | 0.04742369 | 0.03319174 | -0.00082625 | -0.01446790 | -0.00694684 | 0.00107347 | -0.00007290 | -0.00047117 | -0.00022580 | 0.00015957 | 0.00005716 | 0.00003449 | -0.00020019 | -0.00007532 | 0.00013681 | 0.00012725 |
| 8 | 15 | 0.01983203 | 0.04503911 | 0.03355198 | 0.00048714 | -0.01252217 | -0.00621752 | -0.00022965 | -0.00059699 | -0.00010227 | -0.00000844 | 0.00010585 | 0.00015821 | -0.00002004 | -0.00029878 | -0.00020922 | 0.00041792 | 0.00015256 |
| 8 | 16 | 0.02045488 | 0.04559306 | 0.03222414 | -0.00002781 | -0.01216965 | -0.00469413 | 0.00209353 | -0.00008396 | -0.00023158 | -0.00010974 | 0.00017023 | 0.00002917 | 0.00004120 | -0.00021506 | -0.00011100 | 0.00023562 | 0.00021036 |
| 8 | 17 | 0.01897605 | 0.04286242 | 0.03136450 | 0.00030631 | -0.01085821 | -0.00460446 | 0.00049875 | -0.00039132 | 0.00032956 | 0.00001978 | 0.00012080 | 0.00014651 | -0.00006923 | -0.00031022 | -0.00021977 | 0.00048131 | 0.00020579 |
| 8 | 18 | 0.01959293 | 0.04365858 | 0.03027068 | -0.00035457 | -0.01061734 | -0.00307091 | 0.00257649 | 0.00000918 | 0.00037296 | -0.00007230 | 0.00015378 | 0.00008010 | -0.00005906 | -0.00023968 | -0.00014622 | 0.00033826 | 0.00025193 |
| 8 | 19 | 0.01836286 | 0.04063444 | 0.02826908 | -0.00044676 | -0.00933943 | -0.00330585 | 0.00081696 | 0.00003630 | 0.00107212 | -0.00013049 | 0.00011286 | 0.00023215 | -0.00023696 | -0.00033587 | -0.00022766 | 0.00050906 | 0.00017578 |
| 8 | 20 | 0.01901339 | 0.04179177 | 0.02747439 | -0.00128733 | -0.00947068 | -0.00231351 | 0.00214861 | 0.00009234 | 0.00104701 | -0.00021792 | 0.00011467 | 0.00020657 | -0.00022340 | -0.00026742 | -0.00015751 | 0.00035126 | 0.00019126 |
| 9 | 4 | 0.02220550 | 0.04449932 | 0.02871186 | -0.00028067 | -0.00984507 | -0.00565211 | -0.00161099 | -0.00051204 | 0.00031448 | -0.00032003 | -0.00000547 | 0.00022825 | -0.00025702 | -0.00007460 | -0.00003526 | -0.00003798 | -0.00020007 |
| 9 | 5 | 0.02227521 | 0.04676825 | 0.03159508 | -0.00002212 | -0.01151306 | -0.00650080 | -0.00129776 | -0.00006346 | 0.00018592 | -0.00026809 | 0.00000821 | 0.00006158 | -0.00002994 | -0.00015107 | -0.00004980 | 0.00023911 | 0.00007340 |
| 9 | 6 | 0.02192436 | 0.04466777 | 0.02936813 | 0.00017493 | -0.00988151 | -0.00574609 | -0.00124018 | -0.00030965 | 0.00015423 | -0.00035236 | 0.00001563 | 0.00017047 | -0.00016318 | -0.00009352 | -0.00003073 | -0.00002022 | -0.00013267 |
| 9 | 7 | 0.02068936 | 0.04469457 | 0.03134959 | 0.00177351 | -0.00985267 | -0.00638007 | -0.00161064 | 0.00002761 | 0.00012917 | -0.00040468 | -0.00003384 | -0.00007315 | 0.00013134 | -0.00012434 | 0.00003715 | 0.00016210 | 0.00014286 |
| 9 | 8 | 0.02037672 | 0.04268801 | 0.02890156 | 0.00155396 | -0.00837602 | -0.00558786 | -0.00148985 | -0.00009559 | 0.00026190 | -0.00047041 | -0.00004680 | 0.00005068 | -0.00002693 | -0.00005580 | 0.00004862 | -0.00006487 | -0.00003948 |
| 9 | 9 | 0.02072415 | 0.04487222 | 0.03113042 | 0.00118481 | -0.01035563 | -0.00647263 | -0.00146405 | 0.00007561 | 0.00000416 | -0.00035794 | -0.00002177 | -0.00009045 | 0.00016552 | -0.00012972 | 0.00001911 | 0.00019337 | 0.00016978 |
| 9 | 10 | 0.02155602 | 0.04493611 | 0.02973874 | -0.00021607 | -0.01055267 | -0.00589197 | -0.00078402 | -0.00018490 | -0.00012137 | -0.00031481 | 0.00004885 | 0.00008872 | -0.00003027 | -0.00011602 | -0.00004101 | 0.00004120 | -0.00001993 |
| 9 | 11 | 0.02212398 | 0.04732012 | 0.03174813 | -0.00082852 | -0.01225419 | -0.00644966 | -0.00089254 | -0.00008398 | -0.00031852 | -0.00016809 | 0.00004184 | -0.00001797 | 0.00013019 | -0.00018047 | -0.00008796 | 0.00031576 | 0.00017060 |
| 9 | 12 | 0.02249151 | 0.04724041 | 0.03081289 | -0.00221026 | -0.01294707 | -0.00629666 | -0.00048696 | -0.00029336 | -0.00034854 | -0.00016097 | 0.00007952 | 0.00012642 | -0.00003930 | -0.00015030 | -0.00010994 | 0.00015467 | 0.00000490 |
| 9 | 13 | 0.02309947 | 0.04929139 | 0.03234924 | -0.00269269 | -0.01392169 | -0.00643161 | -0.00074927 | -0.00025832 | -0.00044663 | -0.00000702 | 0.00004647 | 0.00006407 | 0.00007346 | -0.00021256 | -0.00017643 | 0.00044288 | 0.00016790 |
| 9 | 14 | 0.02303094 | 0.04905811 | 0.03187526 | -0.00389829 | -0.01501697 | -0.00680732 | -0.00057594 | -0.00039379 | -0.00039049 | -0.00008606 | 0.00007490 | 0.00016590 | -0.00005756 | -0.00018972 | -0.00015379 | 0.00026018 | 0.00003436 |
| 9 | 15 | 0.02311297 | 0.04909456 | 0.03161379 | -0.00344877 | -0.01364353 | -0.00555877 | -0.00051778 | -0.00045448 | -0.00048294 | 0.00008214 | 0.00005840 | 0.00007470 | 0.00005777 | -0.00022866 | -0.00021010 | 0.00050212 | 0.00017958 |
| 9 | 16 | 0.02241711 | 0.04706289 | 0.02959988 | -0.00393740 | -0.01288011 | -0.00483528 | 0.00010069 | -0.00064509 | -0.00038639 | 0.00001745 | 0.00011527 | 0.00010544 | -0.00002433 | -0.00019163 | -0.00016038 | 0.00029436 | 0.00009410 |
| 9 | 17 | 0.02257043 | 0.04694524 | 0.02879342 | -0.00404973 | -0.01202402 | -0.00374110 | 0.00037253 | -0.00025149 | -0.00023527 | 0.00006567 | 0.00010200 | 0.00005992 | -0.00001137 | -0.00023678 | -0.00022137 | 0.00053305 | 0.00019602 |
| 9 | 18 | 0.02189666 | 0.04466565 | 0.02619380 | -0.00471163 | -0.01088768 | -0.00289792 | 0.00086552 | -0.00042698 | 0.00009815 | -0.00002686 | 0.00014412 | 0.00013039 | -0.00013656 | -0.00021667 | -0.00018304 | 0.00036109 | 0.00012069 |
| 9 | 19 | 0.02215916 | 0.04443480 | 0.02478977 | -0.00538861 | -0.01071061 | -0.00229804 | 0.00107197 | 0.00034461 | 0.00033657 | -0.00014364 | 0.00012626 | 0.00014256 | -0.00021276 | -0.00025341 | -0.00024106 | 0.00054961 | 0.00014402 |
| 9 | 20 | 0.02161197 | 0.04214433 | 0.02176215 | -0.00606140 | -0.00908053 | -0.00157890 | 0.00104306 | 0.00005290 | 0.00080287 | -0.00029088 | 0.00014343 | 0.00024420 | -0.00033870 | -0.00025002 | -0.00019023 | 0.00035673 | 0.00004866 |
| 9 | 21 | 0.02254359 | 0.04384280 | 0.02217502 | -0.00754812 | -0.01062652 | -0.00119124 | 0.00153436 | 0.00034929 | 0.00043077 | -0.00023024 | 0.00017547 | 0.00024678 | -0.00037132 | -0.00030405 | -0.00031220 | 0.00063269 | 0.00010923 |
| 9 | 22 | 0.02218973 | 0.04361345 | 0.02216233 | -0.00848108 | -0.01096325 | -0.00087297 | 0.00170910 | -0.00052063 | 0.00031296 | -0.00013643 | 0.00022687 | 0.00028074 | -0.00038914 | -0.00031285 | -0.00029868 | 0.00055756 | 0.00012073 |
| 10 | 5 | 0.02553638 | 0.04910239 | 0.02748641 | -0.00488023 | -0.01255846 | -0.00550984 | -0.00085271 | 0.00002638 | 0.00004004 | -0.00022946 | 0.00000234 | 0.00023780 | -0.00029720 | 0.00001451 | -0.00005641 | -0.00004224 | -0.00022905 |
| 10 | 6 | 0.02598625 | 0.04920258 | 0.02600461 | -0.00523624 | -0.01244156 | -0.00559289 | -0.00003869 | 0.00015960 | -0.00043110 | -0.00041749 | 0.00004568 | 0.00012509 | -0.00019109 | 0.00011746 | 0.00008117 | -0.00034354 | -0.00025523 |
| 10 | 7 | 0.02422918 | 0.04711046 | 0.02689289 | -0.00286307 | -0.01017459 | -0.00500236 | -0.00089799 | -0.00001332 | -0.00017903 | -0.00041695 | 0.00000230 | 0.00008710 | -0.00009558 | 0.00002092 | 0.00004509 | -0.00016717 | -0.00014743 |
| 10 | 8 | 0.02474921 | 0.04744313 | 0.02572074 | -0.00308203 | -0.00962813 | -0.00434426 | 0.00035373 | 0.00031331 | -0.00041965 | -0.00056364 | 0.00002899 | 0.00000799 | -0.00003857 | 0.00011490 | 0.00014841 | -0.00040715 | -0.00016685 |
| 10 | 9 | 0.02412604 | 0.04722311 | 0.02679246 | -0.00336215 | -0.01065148 | -0.00515232 | -0.00081180 | -0.00005777 | -0.00032947 | -0.00037874 | 0.00001888 | 0.00004755 | -0.00003114 | 0.00001016 | 0.00004058 | -0.00013959 | -0.00009769 |
| 10 | 10 | 0.02549404 | 0.04955692 | 0.02702734 | -0.00483627 | -0.01215085 | -0.00485675 | 0.00079595 | 0.00028596 | -0.00062888 | -0.00038219 | 0.00008147 | 0.00002336 | -0.00003194 | 0.00007169 | 0.00007414 | -0.00026797 | -0.00012115 |
| 10 | 11 | 0.02490216 | 0.04942982 | 0.02816151 | -0.00529573 | -0.01335176 | -0.00575237 | -0.00053461 | -0.00016837 | -0.00053874 | -0.00019515 | 0.00005486 | 0.00009650 | -0.00005428 | -0.00002877 | -0.00004759 | 0.00000485 | -0.00007144 |
| 10 | 12 | 0.02592196 | 0.05133527 | 0.02837946 | -0.00636067 | -0.01448557 | -0.00549554 | 0.00092267 | 0.00024799 | -0.00071239 | -0.00026991 | 0.00010458 | 0.00003355 | -0.00002417 | 0.00002505 | 0.00002418 | -0.00014284 | -0.00006953 |
| 10 | 13 | 0.02526631 | 0.05102294 | 0.02943473 | -0.00677167 | -0.01555455 | -0.00639612 | -0.00060494 | -0.00027238 | -0.00058290 | -0.00009971 | 0.00005903 | 0.00014409 | -0.00008396 | -0.00007308 | -0.00010889 | 0.00013248 | -0.00004344 |
| 10 | 14 | 0.02601177 | 0.05243888 | 0.02951881 | -0.00726463 | -0.01614551 | -0.00615222 | 0.00083129 | 0.00019994 | -0.00068696 | -0.00026950 | 0.00012564 | 0.00003797 | -0.00001394 | -0.00004621 | -0.00000118 | -0.00004465 | -0.00001336 |
| 10 | 15 | 0.02510184 | 0.05070683 | 0.02885660 | -0.00733875 | -0.01525885 | -0.00571631 | -0.00053374 | -0.00060632 | -0.00064024 | -0.00001475 | 0.00007622 | 0.00013002 | -0.00006134 | -0.00009764 | -0.00012586 | 0.00018656 | -0.00000511 |
| 10 | 16 | 0.02555977 | 0.05085896 | 0.02751326 | -0.00730476 | -0.01356936 | -0.00353571 | 0.00155648 | -0.00027221 | -0.00067772 | -0.00007194 | 0.00013896 | -0.00000159 | 0.00001944 | -0.00004184 | -0.00002888 | 0.00002492 | 0.00004872 |
| 10 | 17 | 0.02483463 | 0.04864469 | 0.02551217 | -0.00805127 | -0.01285300 | -0.00335474 | 0.00026524 | -0.00081819 | -0.00050523 | 0.00006182 | 0.00012470 | 0.00008820 | -0.00007348 | -0.00011355 | -0.00013709 | 0.00023409 | 0.00005060 |
| 10 | 18 | 0.02518572 | 0.04873251 | 0.02406809 | -0.00841846 | -0.01157422 | -0.00134114 | 0.00216446 | -0.00046599 | -0.00037826 | -0.00000318 | 0.00015017 | 0.00003100 | -0.00007436 | -0.00005621 | -0.00008009 | 0.00013250 | 0.00009091 |
| 10 | 19 | 0.02463784 | 0.04608893 | 0.02092648 | -0.00945421 | -0.01073189 | -0.00139721 | 0.00098162 | -0.00050616 | -0.00000046 | -0.00009164 | 0.00016016 | 0.00014534 | -0.00023048 | -0.00014797 | -0.00016064 | 0.00027747 | 0.00004362 |
| 10 | 20 | 0.02492946 | 0.04645948 | 0.01974665 | -0.01023601 | -0.01020892 | -0.00014007 | 0.00217194 | -0.00032243 | 0.00011303 | -0.00015504 | 0.00013976 | 0.00013925 | -0.00024940 | -0.00007542 | -0.00011063 | 0.00017013 | 0.00004883 |
| 10 | 21 | 0.02476559 | 0.04537736 | 0.01862213 | -0.01144017 | -0.01075401 | -0.00050807 | 0.00136362 | -0.00050548 | 0.00012204 | -0.00017083 | 0.00020442 | 0.00022018 | -0.00035353 | -0.00020206 | -0.00021514 | 0.00036817 | 0.00004093 |
| 10 | 22 | 0.02495991 | 0.04722563 | 0.02027480 | -0.01218998 | -0.01225543 | -0.00014296 | 0.00242014 | -0.00075110 | -0.00020799 | -0.00005116 | 0.00020047 | 0.00015954 | -0.00028718 | -0.00013213 | -0.00018705 | 0.00034458 | 0.00011610 |
| 10 | 23 | 0.02462941 | 0.04595125 | 0.01937517 | -0.01299128 | -0.01223339 | -0.00022250 | 0.00150132 | -0.00096924 | -0.00008841 | -0.00007349 | 0.00024071 | 0.00028022 | -0.00042465 | -0.00026392 | -0.00030390 | 0.00055690 | 0.00009078 |
| 10 | 24 | 0.02460900 | 0.04724571 | 0.02064580 | -0.01343758 | -0.01353499 | -0.00035631 | 0.00223139 | -0.00104233 | -0.00029017 | -0.00005378 | 0.00022631 | 0.00019826 | -0.00033942 | -0.00019361 | -0.00023092 | 0.00046963 | 0.00014864 |
| 11 | 6 | 0.02737707 | 0.05152097 | 0.02628033 | -0.00693602 | -0.01281508 | -0.00502424 | -0.00056439 | 0.00026791 | -0.00020373 | -0.00024665 | -0.00001311 | 0.00014982 | -0.00019931 | 0.00010222 | 0.00001432 | -0.00014819 | -0.00020295 |
| 11 | 7 | 0.02668391 | 0.05150448 | 0.02704509 | -0.00579518 | -0.01290537 | -0.00579791 | -0.00070276 | 0.00054111 | -0.00043673 | -0.00033785 | -0.00000902 | -0.00009891 | 0.00010446 | 0.00006205 | 0.00009454 | -0.00005393 | 0.00003670 |
| 11 | 8 | 0.02639232 | 0.04954413 | 0.02467099 | -0.00578452 | -0.01065584 | -0.00466535 | -0.00090360 | 0.00006382 | -0.00038597 | -0.00039868 | -0.00001985 | -0.00000513 | 0.00000215 | 0.00011591 | 0.00013402 | -0.00030126 | -0.00013182 |
| 11 | 9 | 0.02675806 | 0.05157220 | 0.02657631 | -0.00652249 | -0.01332379 | -0.00586272 | -0.00078895 | 0.00035151 | -0.00062164 | -0.00028134 | 0.00000444 | -0.00012175 | 0.00014879 | 0.00005235 | 0.00008321 | -0.00003574 | 0.00005544 |
| 11 | 10 | 0.02707729 | 0.05162806 | 0.02618510 | -0.00770699 | -0.01357749 | -0.00533818 | -0.00061217 | -0.00004315 | -0.00060091 | -0.00019415 | 0.00002223 | 0.00005826 | -0.00004258 | 0.00007283 | 0.00002166 | -0.00012384 | -0.00011013 |
| 11 | 11 | 0.02748366 | 0.05319370 | 0.02750632 | -0.00814556 | -0.01533332 | -0.00606078 | -0.00058512 | 0.00016520 | -0.00078301 | -0.00009695 | 0.00003289 | -0.00001875 | 0.00005838 | 0.00001115 | -0.00004559 | 0.00014266 | 0.00004824 |
| 11 | 12 | 0.02734283 | 0.05300416 | 0.02753307 | -0.00907232 | -0.01589646 | -0.00601011 | -0.00062609 | -0.00015356 | -0.00068141 | -0.00008431 | 0.00003784 | 0.00011771 | -0.00009213 | 0.00002472 | -0.00006373 | 0.00003023 | -0.00008904 |
| 11 | 13 | 0.02773478 | 0.05398745 | 0.02823154 | -0.00908854 | -0.01653737 | -0.00618228 | -0.00072105 | -0.00003610 | -0.00077054 | 0.00000420 | 0.00002994 | 0.00008554 | -0.00003960 | -0.00003538 | -0.00014687 | 0.00029945 | 0.00003782 |
| 11 | 14 | 0.02721015 | 0.05353718 | 0.02859760 | -0.00961806 | -0.01728827 | -0.00661038 | -0.00086206 | -0.00027476 | -0.00063090 | -0.00009703 | 0.00005148 | 0.00017254 | -0.00014140 | -0.00005150 | -0.00012266 | 0.00015615 | -0.00006353 |
| 11 | 15 | 0.02764892 | 0.05345182 | 0.02733671 | -0.00955718 | -0.01580325 | -0.00523689 | -0.00073217 | -0.00043005 | -0.00079676 | 0.00009588 | 0.00004102 | 0.00009826 | -0.00005020 | -0.00005923 | -0.00017441 | 0.00034854 | 0.00004004 |
| 11 | 16 | 0.02709760 | 0.05207586 | 0.02581831 | -0.01029940 | -0.01496845 | -0.00441033 | -0.00045917 | -0.00087809 | -0.00072612 | 0.00008557 | 0.00008345 | 0.00009907 | -0.00006947 | -0.00004029 | -0.00010555 | 0.00015040 | -0.00001688 |
| 11 | 17 | 0.02758548 | 0.05162144 | 0.02386176 | -0.01059768 | -0.01370558 | -0.00299404 | 0.00008315 | -0.00057737 | -0.00073707 | 0.00012856 | 0.00009797 | 0.00005430 | -0.00006882 | -0.00007054 | -0.00017181 | 0.00035009 | 0.00005960 |
| 11 | 18 | 0.02710528 | 0.04986959 | 0.02141453 | -0.01163749 | -0.01249236 | -0.00196564 | 0.00041520 | -0.00093844 | -0.00047934 | 0.00007500 | 0.00013590 | 0.00008931 | -0.00013589 | -0.00006773 | -0.00012087 | 0.00019391 | 0.00002032 |
| 11 | 19 | 0.02762337 | 0.04918219 | 0.01878771 | -0.01262058 | -0.01200053 | -0.00112540 | 0.00096820 | -0.00017566 | -0.00036306 | -0.00005680 | 0.00014982 | 0.00009094 | -0.00021522 | -0.00009981 | -0.00018470 | 0.00035556 | 0.00003679 |
| 11 | 20 | 0.02727735 | 0.04729513 | 0.01565656 | -0.01382036 | -0.01040821 | -0.00022297 | 0.00098688 | -0.00049436 | 0.00003382 | -0.00017717 | 0.00016701 | 0.00015714 | -0.00029611 | -0.00010657 | -0.00012582 | 0.00018375 | -0.00001901 |
| 11 | 21 | 0.02780328 | 0.04806523 | 0.01576369 | -0.01482361 | -0.01173679 | -0.00009233 | 0.00137106 | -0.00013210 | -0.00020838 | -0.00014691 | 0.00019267 | 0.00018327 | -0.00036228 | -0.00015374 | -0.00025128 | 0.00044408 | 0.00001347 |
| 11 | 22 | 0.02699773 | 0.04771171 | 0.01663437 | -0.01516017 | -0.01206556 | -0.00012340 | 0.00119366 | -0.00094174 | -0.00019179 | -0.00007935 | 0.00022173 | 0.00022065 | -0.00036940 | -0.00017962 | -0.00022632 | 0.00038643 | 0.00003658 |
| 11 | 23 | 0.02749607 | 0.04787750 | 0.01599904 | -0.01586623 | -0.01222979 | 0.00046711 | 0.00128772 | -0.00061577 | -0.00027349 | -0.00006208 | 0.00020928 | 0.00030458 | -0.00049171 | -0.00021762 | -0.00036856 | 0.00064873 | 0.00002247 |
| 11 | 24 | 0.02631583 | 0.04734479 | 0.01745134 | -0.01575430 | -0.01288643 | -0.00015972 | 0.00094121 | -0.00127913 | -0.00018253 | -0.00007913 | 0.00023500 | 0.00030540 | -0.00046218 | -0.00025507 | -0.00029887 | 0.00055318 | 0.00006095 |
| 11 | 25 | 0.02761934 | 0.04643872 | 0.01241656 | -0.01836287 | -0.01178911 | 0.00140310 | 0.00154062 | -0.00053640 | -0.00008715 | -0.00019889 | 0.00025576 | 0.00037630 | -0.00061017 | -0.00028545 | -0.00041166 | 0.00071376 | 0.00000759 |
| 11 | 26 | 0.02721082 | 0.04524240 | 0.01007445 | -0.02006721 | -0.01153427 | 0.00142041 | 0.00168103 | -0.00068590 | 0.00018894 | -0.00036622 | 0.00031926 | 0.00032289 | -0.00055018 | -0.00032366 | -0.00027459 | 0.00048591 | 0.00002678 |
| 11 | 27 | 0.02865223 | 0.04433327 | 0.00479059 | -0.02275517 | -0.01070873 | 0.00250942 | 0.00210727 | 0.00008408 | 0.00027965 | -0.00056590 | 0.00035008 | 0.00037786 | -0.00066937 | -0.00036836 | -0.00036110 | 0.00058795 | -0.00004333 |
| 11 | 28 | 0.02839292 | 0.04325914 | 0.00236781 | -0.02419995 | -0.00977287 | 0.00265099 | 0.00194940 | -0.00004832 | 0.00067983 | -0.00073223 | 0.00038394 | 0.00033764 | -0.00059688 | -0.00040685 | -0.00020598 | 0.00032858 | -0.00005646 |
| 12 | 7 | 0.02906787 | 0.05388375 | 0.02533329 | -0.00867916 | -0.01317041 | -0.00499341 | -0.00053873 | 0.00035116 | -0.00034574 | -0.00020554 | -0.00002997 | 0.00003756 | -0.00004968 | 0.00016630 | 0.00011031 | -0.00026697 | -0.00014493 |
| 12 | 8 | 0.02974365 | 0.05404047 | 0.02389988 | -0.00858025 | -0.01140187 | -0.00364292 | 0.00052760 | 0.00052526 | -0.00054538 | -0.00033935 | -0.00000502 | -0.00003897 | 0.00001484 | 0.00023642 | 0.00021295 | -0.00049776 | -0.00018234 |
| 12 | 9 | 0.02900629 | 0.05388177 | 0.02500489 | -0.00925639 | -0.01353989 | -0.00509973 | -0.00064202 | 0.00012772 | -0.00053476 | -0.00016643 | -0.00001329 | -0.00000369 | 0.00002010 | 0.00015050 | 0.00011155 | -0.00025438 | -0.00010720 |
| 12 | 10 | 0.02999278 | 0.05546502 | 0.02539781 | -0.00995208 | -0.01390927 | -0.00430082 | 0.00070788 | 0.00050915 | -0.00063501 | -0.00019447 | 0.00002519 | -0.00000726 | -0.00001345 | 0.00018646 | 0.00012566 | -0.00032582 | -0.00014250 |
| 12 | 11 | 0.02924670 | 0.05505313 | 0.02628393 | -0.01062520 | -0.01594560 | -0.00569317 | -0.00058431 | 0.00000905 | -0.00067170 | -0.00004721 | 0.00001255 | 0.00007182 | -0.00005827 | 0.00010167 | -0.00000269 | -0.00007004 | -0.00009800 |
| 12 | 12 | 0.02991024 | 0.05615335 | 0.02659451 | -0.01066210 | -0.01569896 | -0.00493924 | 0.00071714 | 0.00044642 | -0.00067139 | -0.00016944 | 0.00006255 | 0.00001412 | -0.00003613 | 0.00011227 | 0.00006281 | -0.00018763 | -0.00009834 |
| 12 | 13 | 0.02909589 | 0.05540460 | 0.02728766 | -0.01117872 | -0.01744039 | -0.00622072 | -0.00074226 | -0.00013491 | -0.00069693 | -0.00004780 | 0.00003766 | 0.00013919 | -0.00013194 | 0.00002692 | -0.00008859 | 0.00008176 | -0.00008412 |
| 12 | 14 | 0.02957034 | 0.05620401 | 0.02751226 | -0.01064488 | -0.01666299 | -0.00549044 | 0.00061038 | 0.00033178 | -0.00065197 | -0.00025398 | 0.00011411 | 0.00002615 | -0.00004795 | 0.00000541 | 0.00002134 | -0.00007877 | -0.00004593 |
| 12 | 15 | 0.02889064 | 0.05480524 | 0.02650855 | -0.01152622 | -0.01677181 | -0.00547681 | -0.00085518 | -0.00061604 | -0.00073410 | 0.00003253 | 0.00005677 | 0.00013155 | -0.00011082 | -0.00000781 | -0.00010195 | 0.00012503 | -0.00005803 |
| 12 | 16 | 0.02944004 | 0.05499906 | 0.02515215 | -0.01125321 | -0.01409968 | -0.00291905 | 0.00098151 | -0.00035818 | -0.00064164 | 0.00000331 | 0.00010354 | -0.00001102 | 0.00000579 | 0.00002939 | 0.00001900 | -0.00005323 | -0.00000924 |
| 12 | 17 | 0.02903960 | 0.05317701 | 0.02272597 | -0.01273051 | -0.01420446 | -0.00299811 | -0.00025780 | -0.00107814 | -0.00066887 | 0.00016394 | 0.00009513 | 0.00007571 | -0.00008314 | -0.00001377 | -0.00008837 | 0.00012939 | -0.00001070 |
| 12 | 18 | 0.02945583 | 0.05306886 | 0.02092345 | -0.01303953 | -0.01209509 | -0.00070127 | 0.00143402 | -0.00074790 | -0.00049282 | 0.00010460 | 0.00010870 | 0.00000961 | -0.00005906 | 0.00003248 | -0.00001791 | 0.00001387 | 0.00001762 |
| 12 | 19 | 0.02932592 | 0.05085102 | 0.01718944 | -0.01483221 | -0.01191240 | -0.00083805 | 0.00054180 | -0.00094102 | -0.00034998 | 0.00004196 | 0.00014281 | 0.00009250 | -0.00018186 | -0.00004642 | -0.00009427 | 0.00013628 | -0.00000263 |
| 12 | 20 | 0.02962339 | 0.05086732 | 0.01547273 | -0.01574608 | -0.01075336 | 0.00065889 | 0.00162057 | -0.00066865 | -0.00018637 | -0.00001513 | 0.00010784 | 0.00008216 | -0.00019233 | 0.00002419 | -0.00004188 | 0.00003017 | -0.00001116 |
| 12 | 21 | 0.02933670 | 0.04963457 | 0.01438488 | -0.01680244 | -0.01171756 | -0.00004303 | 0.00089037 | -0.00090041 | -0.00020813 | -0.00006086 | 0.00018693 | 0.00015638 | -0.00029232 | -0.00010589 | -0.00014185 | 0.00021680 | -0.00000069 |
| 12 | 22 | 0.02900157 | 0.05058612 | 0.01618039 | -0.01672966 | -0.01242073 | 0.00016799 | 0.00155874 | -0.00098737 | -0.00032200 | -0.00000609 | 0.00016158 | 0.00012346 | -0.00024953 | -0.00005009 | -0.00010556 | 0.00018487 | 0.00003915 |
| 12 | 23 | 0.02858056 | 0.04914314 | 0.01528809 | -0.01735314 | -0.01276566 | -0.00015907 | 0.00075944 | -0.00124769 | -0.00026542 | -0.00005730 | 0.00021988 | 0.00024422 | -0.00039335 | -0.00018681 | -0.00023241 | 0.00040495 | 0.00003012 |
| 12 | 24 | 0.02816644 | 0.04978816 | 0.01670473 | -0.01710529 | -0.01335183 | -0.00041167 | 0.00120030 | -0.00121448 | -0.00030449 | -0.00009470 | 0.00019496 | 0.00017609 | -0.00031196 | -0.00013609 | -0.00014545 | 0.00030191 | 0.00006715 |
| 12 | 25 | 0.02857972 | 0.04762283 | 0.01183239 | -0.01969890 | -0.01231490 | 0.00061281 | 0.00101334 | -0.00110165 | -0.00005275 | -0.00020393 | 0.00026657 | 0.00029054 | -0.00047900 | -0.00026094 | -0.00025457 | 0.00045612 | 0.00003304 |
| 12 | 26 | 0.02892493 | 0.04783026 | 0.00974015 | -0.02155462 | -0.01229662 | 0.00104758 | 0.00166608 | -0.00081983 | 0.00001765 | -0.00025406 | 0.00023953 | 0.00020424 | -0.00039515 | -0.00017490 | -0.00013757 | 0.00026928 | 0.00004036 |
| 12 | 27 | 0.02989775 | 0.04557448 | 0.00335447 | -0.02461002 | -0.01062149 | 0.00218520 | 0.00162756 | -0.00046448 | 0.00041601 | -0.00050359 | 0.00034433 | 0.00028037 | -0.00051152 | -0.00034433 | -0.00019346 | 0.00032413 | -0.00001032 |
| 12 | 28 | 0.03006665 | 0.04605552 | 0.00231774 | -0.02586016 | -0.01069554 | 0.00242410 | 0.00193343 | -0.00033107 | 0.00046474 | -0.00051024 | 0.00029401 | 0.00022229 | -0.00043979 | -0.00024951 | -0.00009200 | 0.00016096 | -0.00001730 |
| 12 | 29 | 0.03009872 | 0.04414430 | -0.00036613 | -0.02702612 | -0.01003823 | 0.00284421 | 0.00186133 | -0.00029199 | 0.00064656 | -0.00066656 | 0.00041786 | 0.00029206 | -0.00054508 | -0.00044291 | -0.00018893 | 0.00033568 | 0.00000132 |
| 13 | 8 | 0.03068931 | 0.05618596 | 0.02461444 | -0.00990274 | -0.01320276 | -0.00506042 | -0.00060146 | 0.00034680 | -0.00035870 | -0.00011922 | -0.00004663 | -0.00009334 | 0.00014022 | 0.00020397 | 0.00022648 | -0.00039861 | -0.00006616 |
| 13 | 9 | 0.03073769 | 0.05686749 | 0.02524276 | -0.01065852 | -0.01568870 | -0.00627084 | -0.00057676 | 0.00062390 | -0.00059233 | -0.00005226 | -0.00001084 | -0.00015667 | 0.00018514 | 0.00016489 | 0.00014295 | -0.00013199 | 0.00006742 |
| 13 | 10 | 0.03096452 | 0.05717126 | 0.02570468 | -0.01136068 | -0.01564405 | -0.00548413 | -0.00046904 | 0.00022557 | -0.00055515 | 0.00000540 | -0.00001438 | 0.00000008 | 0.00002804 | 0.00015796 | 0.00008252 | -0.00018330 | -0.00007146 |
| 13 | 11 | 0.03105143 | 0.05737283 | 0.02576466 | -0.01173696 | -0.01715527 | -0.00616302 | -0.00046831 | 0.00039397 | -0.00076293 | 0.00004860 | 0.00001964 | -0.00002401 | 0.00002810 | 0.00011553 | -0.00001990 | 0.00008541 | 0.00003465 |
| 13 | 12 | 0.03085967 | 0.05737000 | 0.02653402 | -0.01205401 | -0.01725112 | -0.00585716 | -0.00054207 | 0.00006088 | -0.00066631 | 0.00001525 | 0.00001966 | 0.00008454 | -0.00007900 | 0.00008845 | -0.00003464 | -0.00000077 | -0.00007306 |
| 13 | 13 | 0.03098933 | 0.05724454 | 0.02618343 | -0.01206216 | -0.01776622 | -0.00599746 | -0.00063660 | 0.00012278 | -0.00078798 | 0.00006817 | 0.00003480 | 0.00009965 | -0.00011471 | 0.00004651 | -0.00015019 | 0.00027196 | 0.00000783 |
| 13 | 14 | 0.03044889 | 0.05697191 | 0.02721914 | -0.01194378 | -0.01798048 | -0.00616782 | -0.00078235 | -0.00014876 | -0.00066927 | -0.00007811 | 0.00006192 | 0.00015637 | -0.00016715 | -0.00001793 | -0.00012312 | 0.00015178 | -0.00006078 |

TABLE 3. Nuclear charge density distribution FB coefficients.

| $Z$ | $N$ | $a_1$ | $a_2$ | $a_3$ | $a_4$ | $a_5$ | $a_6$ | $a_7$ | $a_8$ | $a_9$ | $a_{10}$ | $a_{11}$ | $a_{12}$ | $a_{13}$ | $a_{14}$ | $a_{15}$ | $a_{16}$ | $a_{17}$ |
|---|---|---|---|---|---|---|---|---|---|---|---|---|---|---|---|---|---|---|
| 13 | 15 | 0.03091748 | 0.05653584 | 0.02512835 | -0.01238946 | -0.01673902 | -0.00497132 | -0.00080451 | -0.00039686 | -0.00080253 | 0.00015442 | 0.00004388 | 0.00012073 | -0.00013066 | 0.00001620 | -0.00017792 | 0.00031609 | -0.00000191 |
| 13 | 16 | 0.03061507 | 0.05592995 | 0.02431302 | -0.01299116 | -0.01565493 | -0.00405405 | -0.00067751 | -0.00091701 | -0.00069526 | 0.00016332 | 0.00006593 | 0.00008618 | -0.00007431 | 0.00000847 | -0.00007862 | 0.00010536 | -0.00003024 |
| 13 | 17 | 0.03116588 | 0.05499223 | 0.02128165 | -0.01377779 | -0.01450181 | -0.00280203 | -0.00024717 | -0.00073518 | -0.00076041 | 0.00022195 | 0.00008447 | 0.00006833 | -0.00011325 | 0.00001446 | -0.00014778 | 0.00026987 | 0.00000752 |
| 13 | 18 | 0.03098275 | 0.05402969 | 0.01936297 | -0.01487439 | -0.01319788 | -0.00172959 | -0.00001213 | -0.00116121 | -0.00052085 | 0.00018938 | 0.00010460 | 0.00006135 | -0.00009820 | -0.00000606 | -0.00006662 | 0.00009633 | -0.00000092 |
| 13 | 19 | 0.03162187 | 0.05276513 | 0.01538805 | -0.01639982 | -0.01258615 | -0.00091778 | 0.00053991 | -0.00050862 | -0.00048022 | 0.00006611 | 0.00013430 | 0.00006642 | -0.00019012 | -0.00001576 | -0.00012891 | 0.00022231 | -0.00000436 |
| 13 | 20 | 0.03161464 | 0.05168663 | 0.01261878 | -0.01789626 | -0.01112754 | 0.00001945 | 0.00059105 | -0.00084365 | -0.00013869 | -0.00002065 | 0.00013779 | 0.00008536 | -0.00019063 | -0.00004078 | -0.00004659 | 0.00004180 | -0.00002240 |
| 13 | 21 | 0.03181682 | 0.05128862 | 0.01177830 | -0.01874006 | -0.01211882 | 0.00006656 | 0.00087983 | -0.00047262 | -0.00033076 | -0.00004155 | 0.00017390 | 0.00014604 | -0.00031829 | -0.00007426 | -0.00018795 | 0.00029698 | -0.00002556 |
| 13 | 22 | 0.03085224 | 0.05109878 | 0.01348569 | -0.01852152 | -0.01241214 | -0.00011970 | 0.00057819 | -0.00120101 | -0.00027229 | -0.00001420 | 0.00018783 | 0.00017848 | -0.00030562 | -0.00012363 | -0.00015787 | 0.00025324 | 0.00001076 |
| 13 | 23 | 0.03116195 | 0.05043210 | 0.01207229 | -0.01920539 | -0.01244154 | 0.00038755 | 0.00065750 | -0.00087345 | -0.00036028 | -0.00003035 | 0.00018874 | 0.00028816 | -0.00047531 | -0.00014779 | -0.00031815 | 0.00051435 | -0.00003213 |
| 13 | 24 | 0.02981528 | 0.05004730 | 0.01436253 | -0.01847010 | -0.01298995 | -0.00037578 | 0.00019555 | -0.00149339 | -0.00023627 | -0.00009284 | 0.00020958 | 0.00028409 | -0.00042484 | -0.00021877 | -0.00024574 | 0.00043811 | 0.00002730 |
| 13 | 25 | 0.03133713 | 0.04873379 | 0.00793956 | -0.02184960 | -0.01178176 | 0.00129029 | 0.00089897 | -0.00077964 | -0.00015299 | -0.00017115 | 0.00023456 | 0.00034273 | -0.00056571 | -0.00022442 | -0.00035089 | 0.00056477 | -0.00003804 |
| 13 | 26 | 0.03113444 | 0.04788196 | 0.00552886 | -0.02375774 | -0.01134435 | 0.00138641 | 0.00104142 | -0.00091143 | 0.00017456 | -0.00031675 | 0.00028593 | 0.00025477 | -0.00044421 | -0.00028547 | -0.00018921 | 0.00032447 | 0.00001522 |
| 13 | 27 | 0.03269524 | 0.04657142 | -0.00092824 | -0.02713878 | -0.01043641 | 0.00266220 | 0.00163425 | -0.00016133 | 0.00027457 | -0.00047540 | 0.00032962 | 0.00029580 | -0.00056163 | -0.00031524 | -0.00027280 | 0.00040974 | -0.00005624 |
| 13 | 28 | 0.03264851 | 0.04590044 | -0.00337996 | -0.02881063 | -0.00939960 | 0.00291272 | 0.00154473 | -0.00024116 | 0.00072052 | -0.00061840 | 0.00035187 | 0.00022459 | -0.00043566 | -0.00037277 | -0.00009744 | 0.00014147 | -0.00003740 |
| 13 | 29 | 0.03277949 | 0.04503345 | -0.00450588 | -0.02933808 | -0.00979881 | 0.00332090 | 0.00188263 | -0.00001285 | 0.00048189 | -0.00063790 | 0.00040175 | 0.00030834 | -0.00059406 | -0.00041009 | -0.00027198 | 0.00042332 | -0.00004063 |
| 13 | 30 | 0.03152880 | 0.04458321 | -0.00266063 | -0.02873065 | -0.00977418 | 0.00280723 | 0.00151995 | -0.00055465 | 0.00070418 | -0.00064373 | 0.00043257 | 0.00027018 | -0.00049899 | -0.00049470 | -0.00017591 | 0.00034696 | 0.00005377 |
| 14 | 8 | 0.03344064 | 0.05977257 | 0.02467440 | -0.01082947 | -0.01255864 | -0.00381662 | 0.00070443 | 0.00088548 | -0.00023580 | -0.00004938 | -0.00004427 | -0.00012700 | 0.00016648 | 0.00026351 | 0.00029203 | -0.00052720 | -0.00008610 |
| 14 | 9 | 0.03248032 | 0.05932285 | 0.02575289 | -0.01130017 | -0.01499822 | -0.00541971 | -0.00029242 | 0.00048312 | -0.00034617 | 0.00006881 | -0.00003918 | -0.00009693 | 0.00016609 | 0.00019485 | 0.00019330 | -0.00031605 | -0.00001445 |
| 14 | 10 | 0.03342692 | 0.06018149 | 0.02558230 | -0.01164408 | -0.01428758 | -0.00407255 | 0.00086186 | 0.00084289 | -0.00035369 | -0.00000979 | -0.00000314 | -0.00008196 | 0.00009269 | 0.00020115 | 0.00018638 | -0.00034427 | -0.00006496 |
| 14 | 11 | 0.03247229 | 0.05940152 | 0.02632334 | -0.01221477 | -0.01668829 | -0.00556852 | -0.00027733 | 0.00030900 | -0.00053761 | 0.00008789 | -0.00000027 | 0.00000585 | 0.00002217 | 0.00013362 | 0.00004477 | -0.00010180 | -0.00003254 |
| 14 | 12 | 0.03315963 | 0.06003054 | 0.02625890 | -0.01184993 | -0.01537074 | -0.00430992 | 0.00088774 | 0.00071336 | -0.00044820 | -0.00007372 | 0.00005548 | -0.00004875 | 0.00003267 | 0.00010787 | 0.00010173 | -0.00019107 | -0.00003470 |
| 14 | 13 | 0.03215771 | 0.05892979 | 0.02679601 | -0.01238407 | -0.01760293 | -0.00570377 | -0.00044328 | 0.00007391 | -0.00063748 | 0.00000888 | 0.00004763 | 0.00009462 | -0.00010096 | 0.00003765 | -0.00007479 | 0.00008126 | -0.00003749 |
| 14 | 14 | 0.03270550 | 0.05952348 | 0.02683610 | -0.01146863 | -0.01583449 | -0.00451689 | 0.00079954 | 0.00050123 | -0.00049780 | -0.00021795 | 0.00012840 | -0.00002743 | -0.00000537 | -0.00001984 | 0.00003753 | -0.00006092 | 0.00001055 |
| 14 | 15 | 0.03200152 | 0.05819904 | 0.02582260 | -0.01267472 | -0.01668413 | -0.00482714 | -0.00067973 | -0.00051693 | -0.00065682 | 0.00008553 | 0.00006436 | 0.00009941 | -0.00009310 | -0.00000313 | -0.00009276 | 0.00012467 | -0.00002511 |
| 14 | 16 | 0.03273467 | 0.05852325 | 0.02437723 | -0.01228175 | -0.01323351 | -0.00206404 | 0.00094087 | -0.00026054 | -0.00041145 | 0.00006567 | 0.00009325 | -0.00005600 | 0.00005711 | 0.00001756 | 0.00005797 | -0.00007723 | 0.00002321 |
| 14 | 17 | 0.03238174 | 0.05682604 | 0.02176596 | -0.01425169 | -0.01421589 | -0.00260797 | -0.00039766 | -0.00112247 | -0.00056100 | 0.00025288 | 0.00007459 | 0.00005027 | -0.00004830 | 0.00000928 | -0.00005425 | 0.00008196 | 0.00000259 |
| 14 | 18 | 0.03296813 | 0.05671393 | 0.01964223 | -0.01462129 | -0.01148771 | -0.00024302 | 0.00107315 | -0.00076460 | -0.00030317 | 0.00018896 | 0.00007285 | -0.00003268 | 0.00001437 | 0.00004496 | 0.00004587 | -0.00006932 | 0.00002194 |
| 14 | 19 | 0.03301476 | 0.05470765 | 0.01550379 | -0.01703374 | -0.01207446 | -0.00074614 | 0.00015892 | -0.00114242 | -0.00031333 | 0.00016762 | 0.00010372 | 0.00004728 | -0.00009609 | -0.00000731 | -0.00002903 | 0.00002874 | 0.00000505 |
| 14 | 20 | 0.03347670 | 0.05458787 | 0.01322466 | -0.01819841 | -0.01038119 | 0.00084941 | 0.00108374 | -0.00080645 | -0.00009029 | 0.00011339 | 0.00005536 | 0.00002093 | -0.00007465 | 0.00005569 | 0.00004279 | -0.00010341 | -0.00001541 |
| 14 | 21 | 0.03308569 | 0.05316097 | 0.01203448 | -0.01923112 | -0.01176405 | -0.00005624 | 0.00038644 | -0.00113986 | -0.00019856 | 0.00004744 | 0.00014150 | 0.00010623 | -0.00019613 | -0.00006516 | -0.00007111 | 0.00009217 | 0.00000209 |
| 14 | 22 | 0.03259843 | 0.05365351 | 0.01369132 | -0.01881482 | -0.01179313 | 0.00021531 | 0.00085189 | -0.00111550 | -0.00020542 | 0.00004598 | 0.00010853 | 0.00007657 | -0.00015403 | -0.00002575 | -0.00002474 | 0.00005113 | 0.00002406 |
| 14 | 23 | 0.03208338 | 0.05207099 | 0.01283125 | -0.01938326 | -0.01264748 | -0.00028511 | 0.00014015 | -0.00145965 | -0.00026527 | -0.00002347 | 0.00017945 | 0.00021589 | -0.00033112 | -0.00015794 | -0.00018175 | 0.00030336 | 0.00002136 |
| 14 | 24 | 0.03160122 | 0.05248974 | 0.01414441 | -0.01892936 | -0.01264413 | -0.00048940 | 0.00038970 | -0.00136988 | -0.00021735 | -0.00010037 | 0.00014822 | 0.00013796 | -0.00022860 | -0.00012482 | -0.00007532 | 0.00018305 | 0.00005409 |
| 14 | 25 | 0.03219378 | 0.05030350 | 0.00868785 | -0.02198632 | -0.01201380 | 0.00043319 | 0.00032088 | -0.00134269 | -0.00004911 | -0.00017284 | 0.00022110 | 0.00025301 | -0.00039659 | -0.00023584 | -0.00019670 | 0.00033930 | 0.00002711 |
| 14 | 26 | 0.03274618 | 0.05031148 | 0.00563837 | -0.02430679 | -0.01131213 | 0.00100075 | 0.00081130 | -0.00101689 | 0.00015081 | -0.00021012 | 0.00017572 | 0.00014136 | -0.00026284 | -0.00015555 | -0.00004080 | 0.00009821 | 0.00002766 |
| 14 | 27 | 0.03390961 | 0.04813189 | -0.00126972 | -0.02783433 | -0.00999072 | 0.00218881 | 0.00099436 | -0.00074282 | 0.00047991 | -0.00041142 | 0.00028831 | 0.00020564 | -0.00037087 | -0.00031782 | -0.00010609 | 0.00016221 | 0.00000064 |
| 14 | 28 | 0.03418233 | 0.04835215 | -0.00309172 | -0.02943220 | -0.00946903 | 0.00257150 | 0.00120811 | -0.00052061 | 0.00065156 | -0.00041630 | 0.00022632 | 0.00013164 | -0.00026676 | -0.00023080 | 0.00002234 | -0.00003940 | -0.00001530 |
| 14 | 29 | 0.03412054 | 0.04642883 | -0.00551266 | -0.03037067 | -0.00915522 | 0.00289584 | 0.00123341 | -0.00056125 | 0.00074878 | -0.00057852 | 0.00036182 | 0.00020686 | -0.00038695 | -0.00042425 | -0.00009294 | 0.00016248 | 0.00002025 |
| 14 | 30 | 0.03316541 | 0.04689514 | -0.00285432 | -0.02936519 | -0.00982269 | 0.00213260 | 0.00109149 | -0.00076303 | 0.00064164 | -0.00051039 | 0.00031319 | 0.00013984 | -0.00028903 | -0.00035365 | -0.00001540 | 0.00010504 | 0.00007881 |
| 14 | 31 | 0.03282235 | 0.04487865 | -0.00449640 | -0.02960761 | -0.00915531 | 0.00265878 | 0.00097921 | -0.00086325 | 0.00078423 | -0.00064729 | 0.00043055 | 0.00026721 | -0.00045896 | -0.00055606 | -0.00017417 | 0.00037679 | 0.00010630 |
| 15 | 9 | 0.03413588 | 0.06166316 | 0.02586119 | -0.01046432 | -0.01335313 | -0.00364869 | 0.00164677 | 0.00120942 | -0.00048637 | 0.00001156 | 0.00004795 | -0.00032938 | 0.00038792 | 0.00017444 | 0.00027015 | -0.00028516 | 0.00017768 |
| 15 | 10 | 0.03454481 | 0.06189625 | 0.02589443 | -0.01114193 | -0.01297910 | -0.00275796 | 0.00151292 | 0.00080732 | -0.00038976 | 0.00009733 | 0.00000656 | -0.00019047 | 0.00024050 | 0.00017870 | 0.00021280 | -0.00031194 | 0.00005000 |
| 15 | 11 | 0.03434948 | 0.06156637 | 0.02578809 | -0.01132927 | -0.01414896 | -0.00305833 | 0.00176004 | 0.00100451 | -0.00059082 | 0.00010505 | 0.00006137 | -0.00020402 | 0.00021363 | 0.00013629 | 0.00010616 | -0.00006201 | 0.00013828 |
| 15 | 12 | 0.03438653 | 0.06159142 | 0.02619372 | -0.01156010 | -0.01379777 | -0.00256769 | 0.00155329 | 0.00068546 | -0.00044089 | 0.00008843 | 0.00003812 | -0.00012264 | 0.00012934 | 0.00011039 | 0.00010000 | -0.00013176 | 0.00004912 |
| 15 | 13 | 0.03424739 | 0.06107987 | 0.02573721 | -0.01160012 | -0.01429244 | -0.00253610 | 0.00157100 | 0.00072629 | -0.00056539 | 0.00012805 | 0.00006409 | -0.00009513 | 0.00007092 | 0.00007312 | -0.00001928 | 0.00012385 | 0.00011154 |
| 15 | 14 | 0.03396890 | 0.06095656 | 0.02650972 | -0.01135774 | -0.01399256 | -0.00244730 | 0.00138935 | 0.00048429 | -0.00039875 | -0.00000702 | 0.00007804 | -0.00007574 | 0.00005558 | 0.00000327 | 0.00002261 | 0.00001178 | 0.00006821 |
| 15 | 15 | 0.03421288 | 0.06029336 | 0.02437267 | -0.01214863 | -0.01317798 | -0.00144294 | 0.00122955 | 0.00011583 | -0.00054777 | 0.00023990 | 0.00004911 | -0.00006177 | 0.00004707 | 0.00005600 | -0.00004869 | 0.00015695 | 0.00008007 |
| 15 | 16 | 0.03420189 | 0.05998591 | 0.02358115 | -0.01251783 | -0.01157164 | -0.00021530 | 0.00137712 | -0.00033860 | -0.00037257 | 0.00024966 | 0.00004596 | -0.00009785 | 0.00011369 | 0.00004346 | 0.00004651 | -0.00004026 | 0.00005186 |
| 15 | 17 | 0.03460609 | 0.05867511 | 0.01966706 | -0.01465096 | -0.01179672 | -0.00004440 | 0.00112283 | -0.00048247 | -0.00056768 | 0.00034267 | 0.00004643 | -0.00005996 | 0.00003237 | 0.00008757 | -0.00003238 | 0.00008498 | 0.00003400 |
| 15 | 18 | 0.03469951 | 0.05805447 | 0.01776703 | -0.01570483 | -0.01030714 | 0.00107959 | 0.00125328 | -0.00084127 | -0.00029826 | 0.00030462 | 0.00003210 | -0.00006427 | 0.00005873 | 0.00006824 | 0.00004311 | -0.00007955 | 0.00001938 |
| 15 | 19 | 0.03537604 | 0.05645789 | 0.01251077 | -0.01876523 | -0.01082511 | 0.00094408 | 0.00116262 | -0.00054870 | -0.00038352 | 0.00022642 | 0.00005882 | -0.00003132 | -0.00004173 | 0.00008117 | -0.00001618 | 0.00000507 | -0.00001437 |
| 15 | 20 | 0.03562207 | 0.05581407 | 0.00986385 | -0.02033019 | -0.00938380 | 0.00184311 | 0.00103626 | -0.00083193 | -0.00002863 | 0.00015286 | 0.00002015 | -0.00001104 | -0.00002574 | 0.00006173 | 0.00005921 | -0.00016423 | -0.00003949 |
| 15 | 21 | 0.03570854 | 0.05482547 | 0.00803433 | -0.02188124 | -0.01067307 | 0.00155377 | 0.00111091 | -0.00062179 | -0.00026447 | 0.00013121 | 0.00006925 | 0.00004792 | -0.00015579 | 0.00004191 | -0.00006472 | 0.00005345 | -0.00005065 |
| 15 | 22 | 0.03478510 | 0.05479148 | 0.01006431 | -0.02128086 | -0.01063102 | 0.00158502 | 0.00079948 | -0.00118312 | -0.00015108 | 0.00014337 | 0.00004061 | 0.00007080 | -0.00013443 | 0.00000962 | -0.00002844 | 0.00000692 | -0.00002676 |
| 15 | 23 | 0.03497370 | 0.05372209 | 0.00799473 | -0.02252886 | -0.01105736 | 0.00173311 | 0.00071993 | -0.00100193 | -0.00030001 | 0.00012968 | 0.00005669 | 0.00017329 | -0.00029810 | -0.00000556 | -0.00017344 | 0.00024081 | -0.00006665 |
| 15 | 24 | 0.03371136 | 0.05354962 | 0.01051396 | -0.02157461 | -0.01132444 | 0.00117831 | 0.00025191 | -0.00149219 | -0.00014171 | 0.00005433 | 0.00004162 | 0.00015489 | -0.00023187 | -0.00006006 | -0.00008987 | 0.00015139 | -0.00001936 |
| 15 | 25 | 0.03521859 | 0.05198006 | 0.00348490 | -0.02552802 | -0.01061396 | 0.00237381 | 0.00074640 | -0.00098482 | -0.00013329 | -0.00000375 | 0.00008517 | 0.00022154 | -0.00037071 | -0.00006761 | -0.00019591 | 0.00026993 | -0.00007708 |
| 15 | 26 | 0.03527303 | 0.05134149 | 0.00083388 | -0.02764175 | -0.01004650 | 0.00255055 | 0.00071458 | -0.00110394 | 0.00022027 | -0.00013596 | 0.00009686 | 0.00014623 | -0.00024923 | -0.00011699 | -0.00004268 | 0.00002939 | -0.00004625 |
| 15 | 27 | 0.03654379 | 0.04995972 | -0.00512563 | -0.03077245 | -0.00953123 | 0.00346697 | 0.00137377 | -0.00043275 | 0.00024652 | -0.00029849 | 0.00017809 | 0.00017440 | -0.00035280 | -0.00015638 | -0.00011672 | 0.00011402 | -0.00008415 |
| 15 | 28 | 0.03667004 | 0.04952969 | -0.00755379 | -0.03247412 | -0.00841943 | 0.00378632 | 0.00115186 | -0.00047399 | 0.00071339 | -0.00043177 | 0.00016248 | 0.00011896 | -0.00023269 | -0.00020155 | 0.00004658 | -0.00014623 | -0.00008449 |
| 15 | 29 | 0.03645480 | 0.04844591 | -0.00824785 | -0.03266510 | -0.00902635 | 0.00395921 | 0.00159869 | -0.00027792 | 0.00041750 | -0.00046791 | 0.00024369 | 0.00017600 | -0.00037098 | -0.00023874 | -0.00010388 | 0.00011575 | -0.00006083 |
| 15 | 30 | 0.03540109 | 0.04805121 | -0.00674878 | -0.03224470 | -0.00884212 | 0.00356432 | 0.00109237 | -0.00079345 | 0.00064524 | -0.00047148 | 0.00022886 | 0.00014473 | -0.00027698 | -0.00029536 | -0.00000765 | 0.00002223 | 0.00000220 |
| 15 | 31 | 0.03523751 | 0.04692264 | -0.00723661 | -0.03164155 | -0.00851857 | 0.00410010 | 0.00115172 | -0.00072938 | 0.00039514 | -0.00043445 | 0.00026552 | 0.00026965 | -0.00046944 | -0.00031352 | -0.00020237 | 0.00032743 | -0.00002053 |
| 15 | 32 | 0.03408648 | 0.04636788 | -0.00566494 | -0.03092965 | -0.00840771 | 0.00326560 | 0.00050874 | -0.00117329 | 0.00067825 | -0.00049047 | 0.00025916 | 0.00020970 | -0.00034392 | -0.00038962 | -0.00007643 | 0.00021142 | 0.00006224 |
| 16 | 10 | 0.03764994 | 0.06346305 | 0.02184399 | -0.01305533 | -0.01057208 | -0.00079140 | 0.00168895 | 0.00043579 | -0.00026309 | 0.00029056 | -0.00000004 | -0.00022121 | 0.00022539 | 0.00022517 | 0.00026291 | -0.00039368 | 0.00000558 |
| 16 | 11 | 0.03651445 | 0.06342361 | 0.02463255 | -0.01145432 | -0.01087056 | -0.00039753 | 0.00251244 | 0.00078936 | -0.00033136 | 0.00019031 | 0.00004040 | -0.00024162 | 0.00025694 | 0.00014939 | 0.00020466 | -0.00026158 | 0.00009074 |
| 16 | 12 | 0.03741517 | 0.06311081 | 0.02218994 | -0.01279064 | -0.01051890 | -0.00032220 | 0.00208024 | 0.00056405 | -0.00015846 | 0.00023910 | 0.00005655 | -0.00023189 | 0.00020016 | 0.00013087 | 0.00020871 | -0.00026238 | 0.00005771 |
| 16 | 13 | 0.03623499 | 0.06287087 | 0.02474387 | -0.01140301 | -0.01086124 | 0.00005303 | 0.00256800 | 0.00074179 | -0.00022269 | 0.00015108 | 0.00006975 | -0.00021100 | 0.00018867 | 0.00006291 | 0.00013004 | -0.00011441 | 0.00011408 |
| 16 | 14 | 0.03699036 | 0.06258768 | 0.02257976 | -0.01210890 | -0.01009676 | 0.00003208 | 0.00230816 | 0.00060596 | -0.00002588 | 0.00010576 | 0.00012139 | -0.00026186 | 0.00021224 | 0.00000556 | 0.00018446 | -0.00016755 | 0.00012595 |
| 16 | 15 | 0.03606832 | 0.06225180 | 0.02414701 | -0.01101132 | -0.00882413 | 0.00190455 | 0.00288809 | 0.00045910 | -0.00008885 | 0.00019762 | 0.00006549 | -0.00021425 | 0.00020669 | 0.00001882 | 0.00011338 | -0.00007945 | 0.00011967 |
| 16 | 16 | 0.03693647 | 0.06184129 | 0.02141480 | -0.01141575 | -0.00611426 | 0.00383939 | 0.00362428 | 0.00054698 | 0.00015181 | 0.00019331 | 0.00007384 | -0.00025187 | 0.00022741 | 0.00001832 | 0.00016050 | -0.00015133 | 0.00011405 |
| 16 | 17 | 0.03638440 | 0.06081241 | 0.01992189 | -0.01318436 | -0.00705502 | 0.00375197 | 0.00296314 | -0.00007112 | -0.00006519 | 0.00030842 | 0.00002058 | -0.00017379 | 0.00017554 | 0.00005971 | 0.00009442 | -0.00011137 | 0.00006892 |
| 16 | 18 | 0.03709456 | 0.06030490 | 0.01712864 | -0.01394946 | -0.00489075 | 0.00539025 | 0.00376770 | 0.00032815 | 0.00021083 | 0.00021703 | -0.00000114 | -0.00017200 | 0.00014916 | 0.00006595 | 0.00011322 | -0.00014236 | 0.00006132 |
| 16 | 19 | 0.03712992 | 0.05873702 | 0.01251191 | -0.01816159 | -0.00704516 | 0.00400175 | 0.00224614 | -0.00047161 | 0.00001575 | 0.00028119 | -0.00003250 | -0.00008890 | 0.00008065 | 0.00009641 | 0.00007695 | -0.00017116 | -0.00001137 |
| 16 | 20 | 0.03768980 | 0.05845329 | 0.01015296 | -0.01911735 | -0.00531835 | 0.00538697 | 0.00280768 | 0.00000846 | 0.00032137 | 0.00019516 | -0.00009492 | -0.00005915 | 0.00004422 | 0.00011657 | 0.00008204 | -0.00018330 | -0.00003435 |
| 16 | 21 | 0.03735256 | 0.05711259 | 0.00783595 | -0.02191752 | -0.00774564 | 0.00382195 | 0.00160632 | -0.00073155 | 0.00008057 | 0.00023082 | -0.00005701 | -0.00001236 | -0.00001088 | 0.00008769 | 0.00004546 | -0.00015031 | -0.00005746 |
| 16 | 22 | 0.03694581 | 0.05720392 | 0.00939838 | -0.02060610 | -0.00680558 | 0.00451216 | 0.00225367 | -0.00034217 | 0.00021341 | 0.00017032 | -0.00007729 | -0.00002938 | -0.00000222 | 0.00008303 | 0.00005421 | -0.00010132 | -0.00002177 |
| 16 | 23 | 0.03644519 | 0.05584949 | 0.00761126 | -0.02295791 | -0.00876409 | 0.00335571 | 0.00103587 | -0.00109270 | 0.00003432 | 0.00020615 | -0.00006743 | 0.00005196 | -0.00009090 | 0.00005275 | -0.00000299 | -0.00003261 | -0.00005617 |
| 16 | 24 | 0.03607946 | 0.05588454 | 0.00882355 | -0.02161600 | -0.00789955 | 0.00350594 | 0.00149648 | -0.00067003 | 0.00017316 | 0.00006728 | -0.00006777 | -0.00000441 | -0.00002939 | 0.00003189 | 0.00005071 | -0.00005051 | -0.00001081 |
| 16 | 25 | 0.03663486 | 0.05411856 | 0.00292266 | -0.02645710 | -0.00895495 | 0.00343859 | 0.00065819 | -0.00121137 | 0.00015305 | 0.00010359 | -0.00006529 | 0.00009878 | -0.00014894 | 0.00001733 | -0.00001081 | -0.00003014 | -0.00007823 |
| 16 | 26 | 0.03715780 | 0.05391333 | 0.00048475 | -0.02771126 | -0.00809515 | 0.00393478 | 0.00107151 | -0.00077658 | 0.00034259 | 0.00003158 | -0.00008627 | 0.00006378 | -0.00010287 | 0.00004157 | 0.00003940 | -0.00011010 | -0.00007591 |
| 16 | 27 | 0.03796509 | 0.05235874 | -0.00537846 | -0.03171550 | -0.00813411 | 0.00429578 | 0.00089260 | -0.00081204 | 0.00049320 | -0.00013170 | -0.00001257 | 0.00010115 | -0.00016179 | -0.00004282 | 0.00004161 | -0.00016329 | -0.00011349 |
| 16 | 28 | 0.03807896 | 0.05244768 | -0.00618179 | -0.03201120 | -0.00751177 | 0.00456799 | 0.00103974 | -0.00049827 | 0.00065314 | -0.00014560 | -0.00006093 | 0.00009250 | -0.00013161 | 0.00000342 | 0.00007647 | -0.00021602 | -0.00012232 |
| 16 | 29 | 0.03780589 | 0.05092969 | -0.00825351 | -0.03356295 | -0.00791164 | 0.00453701 | 0.00105682 | -0.00065472 | 0.00063897 | -0.00029792 | 0.00004579 | 0.00009561 | -0.00017011 | -0.00010919 | 0.00006796 | -0.00017898 | -0.00009032 |
| 16 | 30 | 0.03705977 | 0.05084792 | -0.00615135 | -0.03192643 | -0.00792690 | 0.00402440 | 0.00084435 | -0.00081348 | 0.00052981 | -0.00017250 | 0.00000079 | 0.00007867 | -0.00013296 | -0.00005572 | 0.00006339 | -0.00012627 | -0.00005205 |
| 16 | 31 | 0.03650532 | 0.04931283 | -0.00738813 | -0.03282619 | -0.00793631 | 0.00424955 | 0.00069931 | -0.00103823 | 0.00057641 | -0.00029282 | 0.00007717 | 0.00012422 | -0.00020918 | -0.00016847 | 0.00002655 | -0.00003651 | -0.00002850 |
| 16 | 32 | 0.03618012 | 0.04810339 | -0.00613961 | -0.02859992 | -0.00549671 | 0.00389787 | -0.00077389 | -0.00161154 | 0.00035560 | 0.00022335 | -0.00002407 | 0.00020634 | -0.00023785 | -0.00003955 | -0.00007233 | 0.00008137 | -0.00011896 |
| 16 | 33 | 0.03481790 | 0.04602595 | -0.00536724 | -0.02734737 | -0.00470481 | 0.00429481 | -0.00149296 | -0.00194301 | 0.00047512 | 0.00021524 | 0.00000014 | 0.00039872 | -0.00046180 | -0.00013347 | -0.00021100 | 0.00031565 | -0.00016798 |
| 17 | 11 | 0.03803008 | 0.06452303 | 0.02341535 | -0.01128259 | -0.00907988 | 0.00157538 | 0.00386536 | 0.00107457 | -0.00049613 | 0.00016722 | 0.00013405 | -0.00036118 | 0.00035010 | 0.00009990 | 0.00020767 | -0.00016237 | 0.00021347 |
| 17 | 12 | 0.03840097 | 0.06407185 | 0.02185199 | -0.01272631 | -0.00923553 | 0.00119288 | 0.00249921 | 0.00032817 | -0.00023530 | 0.00034608 | 0.00006550 | -0.00026102 | 0.00024193 | 0.00010953 | 0.00018417 | -0.00019057 | 0.00011490 |
| 17 | 13 | 0.03790495 | 0.06381670 | 0.02286988 | -0.01142015 | -0.00847681 | 0.00237854 | 0.00371661 | 0.00099377 | -0.00025181 | 0.00021818 | 0.00011361 | -0.00029107 | 0.00024517 | 0.00004460 | 0.00011880 | -0.00000145 | 0.00020832 |
| 17 | 14 | 0.03801523 | 0.06335564 | 0.02178863 | -0.01234595 | -0.00860691 | 0.00170706 | 0.00250184 | 0.00036996 | 0.00001975 | 0.00026827 | 0.00009241 | -0.00026200 | 0.00021651 | 0.00000360 | 0.00014737 | -0.00007757 | 0.00015982 |
| 17 | 15 | 0.03785678 | 0.06306041 | 0.02168896 | -0.01143681 | -0.00647352 | 0.00412971 | 0.00388371 | 0.00079118 | -0.00010631 | 0.00025865 | 0.00007943 | -0.00026143 | 0.00022589 | 0.00002521 | 0.00009150 | 0.00002551 | 0.00017351 |
| 17 | 16 | 0.03818551 | 0.06266455 | 0.02018730 | -0.01191953 | -0.00469955 | 0.00536438 | 0.00381377 | 0.00038434 | 0.00013480 | 0.00029747 | 0.00004740 | -0.00024628 | 0.00022714 | 0.00001974 | 0.00012734 | -0.00009995 | 0.00011675 |
| 17 | 17 | 0.03835082 | 0.06156640 | 0.01677065 | -0.01440225 | -0.00534679 | 0.00547788 | 0.00386624 | 0.00041130 | -0.00016069 | 0.00029794 | 0.00003368 | -0.00021063 | 0.00018111 | 0.00007097 | 0.00006824 | -0.00003340 | 0.00008796 |
| 17 | 18 | 0.03871178 | 0.06109364 | 0.01468752 | -0.01538517 | -0.00378451 | 0.00656282 | 0.00377672 | 0.00018231 | 0.00018032 | 0.00027368 | -0.00002302 | -0.00016698 | 0.00015340 | 0.00006060 | 0.00008833 | -0.00012952 | 0.00004443 |
| 17 | 19 | 0.03935843 | 0.05956783 | 0.00867062 | -0.02022101 | -0.00586012 | 0.00540729 | 0.00304950 | 0.00003988 | -0.00010064 | 0.00025596 | -0.00001427 | -0.00013138 | 0.00009646 | 0.00009955 | 0.00004832 | -0.00010534 | -0.00000130 |
| 17 | 20 | 0.03977778 | 0.05922629 | 0.00625251 | -0.02157300 | -0.00437014 | 0.00628022 | 0.00261833 | -0.00013941 | 0.00033689 | 0.00020533 | -0.00010635 | -0.00006461 | 0.00006757 | 0.00009058 | 0.00007350 | -0.00021046 | -0.00005762 |
| 17 | 21 | 0.03974731 | 0.05798478 | 0.00364129 | -0.02420014 | -0.00652789 | 0.00523138 | 0.00230470 | -0.00026209 | -0.00002940 | 0.00023526 | -0.00005112 | -0.00004694 | 0.00000238 | 0.00009861 | 0.00000703 | -0.00008037 | -0.00005484 |
| 17 | 22 | 0.03906470 | 0.05788718 | 0.00533850 | -0.02316201 | -0.00555091 | 0.00577129 | 0.00201899 | -0.00052476 | 0.00026067 | 0.00024298 | -0.00012412 | -0.00000846 | -0.00000719 | 0.00008044 | 0.00002443 | -0.00010006 | -0.00006227 |
| 17 | 23 | 0.03899417 | 0.05673640 | 0.00310569 | -0.02513021 | -0.00696889 | 0.00507917 | 0.00158917 | -0.00067141 | -0.00002886 | 0.00027954 | -0.00010097 | 0.00004233 | -0.00009324 | 0.00009225 | -0.00005591 | 0.00004741 | -0.00007299 |
| 17 | 24 | 0.03808790 | 0.05652203 | 0.00499564 | -0.02404411 | -0.00637043 | 0.00505877 | 0.00117713 | -0.00088498 | 0.00027266 | 0.00020299 | -0.00015392 | 0.00003869 | -0.00005625 | 0.00005417 | 0.00000854 | -0.00002528 | -0.00006464 |
| 17 | 25 | 0.03905828 | 0.05524287 | -0.00078048 | -0.02811186 | -0.00733972 | 0.00504795 | 0.00120035 | -0.00082223 | 0.00004310 | 0.00020566 | -0.00010764 | 0.00009194 | -0.00015202 | 0.00007308 | -0.00007195 | 0.00005891 | -0.00009262 |
| 17 | 26 | 0.03927773 | 0.05481062 | -0.00317749 | -0.02998175 | -0.00677542 | 0.00528239 | 0.00092670 | -0.00088777 | 0.00041109 | 0.00009874 | -0.00014607 | 0.00008360 | -0.00010372 | 0.00004836 | 0.00001505 | -0.00011650 | -0.00011675 |
| 17 | 27 | 0.03987491 | 0.05386360 | -0.00707416 | -0.03241552 | -0.00750499 | 0.00527375 | 0.00147983 | -0.00044180 | 0.00024042 | -0.00004957 | -0.00004029 | 0.00007748 | -0.00015570 | 0.00002236 | -0.00001824 | -0.00006667 | -0.00010444 |
| 17 | 28 | 0.04006497 | 0.05366350 | -0.00898146 | -0.03371189 | -0.00646777 | 0.00557518 | 0.00101307 | -0.00044135 | 0.00070974 | -0.00016277 | -0.00010428 | 0.00008588 | -0.00010389 | 0.00000279 | 0.00008263 | -0.00026169 | -0.00015133 |
| 17 | 29 | 0.03954790 | 0.05256199 | -0.00920052 | -0.03374704 | -0.00736376 | 0.00545231 | 0.00149914 | -0.00038162 | 0.00032192 | -0.00016810 | -0.00000161 | 0.00008608 | -0.00017684 | -0.00001758 | -0.00000474 | -0.00007342 | -0.00009575 |
| 17 | 30 | 0.03880560 | 0.05204758 | -0.00840161 | -0.03346868 | -0.00685133 | 0.00531283 | 0.00082209 | -0.00082006 | 0.00058244 | -0.00012277 | -0.00007616 | 0.00009663 | -0.00013393 | -0.00002542 | 0.00005199 | -0.00014540 | -0.00009585 |
| 17 | 31 | 0.03807097 | 0.05004956 | -0.00794777 | -0.03099664 | -0.00539286 | 0.00611731 | 0.00024433 | -0.00126234 | 0.00010769 | 0.00023236 | -0.00004377 | 0.00025835 | -0.00035080 | 0.00002550 | -0.00016572 | 0.00014496 | -0.00021389 |
| 17 | 32 | 0.03722939 | 0.04925105 | -0.00677606 | -0.02943249 | -0.00453342 | 0.00541720 | -0.00073491 | -0.00163506 | 0.00037162 | 0.00034145 | -0.00012830 | 0.00024653 | -0.00027284 | 0.00003729 | -0.00008893 | 0.00006089 | -0.00020353 |

TABLE 3. Nuclear charge density distribution FB coefficients.

| $Z$ | $N$ | $a_1$ | $a_2$ | $a_3$ | $a_4$ | $a_5$ | $a_6$ | $a_7$ | $a_8$ | $a_9$ | $a_{10}$ | $a_{11}$ | $a_{12}$ | $a_{13}$ | $a_{14}$ | $a_{15}$ | $a_{16}$ | $a_{17}$ |
|---|---|---|---|---|---|---|---|---|---|---|---|---|---|---|---|---|---|---|
| 17 | 33 | 0.02752955 | 0.04713644 | 0.01481192 | -0.01736872 | -0.01019853 | 0.00407063 | 0.00213311 | -0.00026835 | 0.00008498 | 0.00056068 | 0.00007645 | 0.00054306 | -0.00077669 | -0.00002221 | -0.00046906 | 0.00079287 | -0.00016907 |
| 17 | 34 | 0.02686009 | 0.04641460 | 0.01497029 | -0.01683730 | -0.01016372 | 0.00301851 | 0.00154960 | -0.00052636 | 0.00018310 | 0.00050274 | 0.00000893 | 0.00052170 | -0.00070641 | 0.00000189 | -0.00039668 | 0.00071779 | -0.00011633 |
| 17 | 35 | 0.02889654 | 0.04511452 | 0.00520344 | -0.02610579 | -0.01106723 | 0.00756892 | 0.00270281 | -0.00141280 | 0.00027830 | 0.00028187 | -0.00000085 | 0.00106315 | -0.00143446 | -0.00024137 | -0.00109506 | 0.00156977 | -0.00020598 |
| 18 | 11 | 0.04038393 | 0.06497740 | 0.01864097 | -0.01485317 | -0.00875938 | 0.00142826 | 0.00166276 | -0.00058021 | -0.00042188 | 0.00054972 | 0.00006400 | -0.00027599 | 0.00028051 | 0.00014103 | 0.00021536 | -0.00024945 | 0.00010138 |
| 18 | 12 | 0.04130467 | 0.06440888 | 0.01543534 | -0.01664790 | -0.00845310 | 0.00115398 | 0.00067750 | -0.00094204 | -0.00010381 | 0.00070298 | 0.00005740 | -0.00026356 | 0.00024514 | 0.00010790 | 0.00020301 | -0.00020582 | 0.00008593 |
| 18 | 13 | 0.04012808 | 0.06413624 | 0.01812486 | -0.01466092 | -0.00797526 | 0.00210148 | 0.00178583 | -0.00038641 | -0.00010805 | 0.00051316 | 0.00008717 | -0.00027263 | 0.00023506 | 0.00005219 | 0.00016913 | -0.00012135 | 0.00014052 |
| 18 | 14 | 0.04093415 | 0.06361280 | 0.01512588 | -0.01606870 | -0.00759557 | 0.00150270 | 0.00086820 | -0.00066495 | 0.00017442 | 0.00056259 | 0.00010949 | -0.00030334 | 0.00026060 | -0.00000581 | 0.00020091 | -0.00013528 | 0.00014849 |
| 18 | 15 | 0.04003784 | 0.06350458 | 0.01758222 | -0.01375342 | -0.00516553 | 0.00448211 | 0.00266531 | -0.00013847 | 0.00012077 | 0.00042351 | 0.00007381 | -0.00026512 | 0.00022516 | 0.00000205 | 0.00014683 | -0.00008377 | 0.00013464 |
| 18 | 16 | 0.04109277 | 0.06299171 | 0.01395885 | -0.01539552 | -0.00383227 | 0.00523758 | 0.00254517 | -0.00021516 | 0.00025860 | 0.00042915 | 0.00004335 | -0.00023570 | 0.00018750 | 0.00001314 | 0.00014072 | -0.00009971 | 0.00009237 |
| 18 | 17 | 0.04044418 | 0.06256495 | 0.01453118 | -0.01469082 | -0.00225541 | 0.00742551 | 0.00395540 | 0.00024076 | 0.00021309 | 0.00029233 | 0.00001198 | -0.00020639 | 0.00016966 | 0.00001896 | 0.00009686 | -0.00008268 | 0.00008014 |
| 18 | 18 | 0.04141196 | 0.06205876 | 0.01073050 | -0.01694081 | -0.00168592 | 0.00767078 | 0.00363164 | 0.00031606 | 0.00038566 | 0.00025201 | -0.00004569 | -0.00015016 | 0.00010050 | 0.00003498 | 0.00007413 | -0.00006038 | 0.00004733 |
| 18 | 19 | 0.04129112 | 0.06121347 | 0.00816657 | -0.01904555 | -0.00167654 | 0.00833272 | 0.00382174 | 0.00030280 | 0.00035404 | 0.00018304 | -0.00007557 | -0.00012049 | 0.00010301 | 0.00004090 | 0.00005646 | -0.00011321 | 0.00001078 |
| 18 | 20 | 0.04206838 | 0.06102708 | 0.00530227 | -0.02082271 | -0.00107032 | 0.00866184 | 0.00350298 | 0.00051895 | 0.00059029 | 0.00012240 | -0.00015242 | -0.00005370 | 0.00003466 | 0.00005316 | 0.00002981 | -0.00006926 | -0.00001554 |
| 18 | 21 | 0.04155575 | 0.05979375 | 0.00356547 | -0.02283011 | -0.00242803 | 0.00799206 | 0.00298931 | 0.00004871 | 0.00043171 | 0.00016911 | -0.00013399 | -0.00004810 | 0.00004061 | 0.00004971 | 0.00002532 | -0.00009991 | -0.00003820 |
| 18 | 22 | 0.04147283 | 0.05952191 | 0.00392448 | -0.02210124 | -0.00187075 | 0.00798692 | 0.00305138 | 0.00034838 | 0.00054908 | 0.00010264 | -0.00014325 | -0.00003885 | 0.00000429 | 0.00003430 | 0.00001208 | -0.00000878 | -0.00000999 |
| 18 | 23 | 0.04073581 | 0.05832908 | 0.00270023 | -0.02400457 | -0.00325985 | 0.00733287 | 0.00222026 | -0.00028824 | 0.00042571 | 0.00018639 | -0.00016515 | -0.00000558 | -0.00000965 | 0.00004321 | 0.00000157 | -0.00002114 | -0.00004370 |
| 18 | 24 | 0.04069366 | 0.05808308 | 0.00299445 | -0.02293086 | -0.00256792 | 0.00703177 | 0.00234897 | 0.00014026 | 0.00053957 | 0.00002385 | -0.00014394 | -0.00003363 | 0.00000292 | 0.00000543 | 0.00002425 | 0.00001124 | -0.00000532 |
| 18 | 25 | 0.04070074 | 0.05693047 | -0.00104968 | -0.02721653 | -0.00418011 | 0.00683696 | 0.00150825 | -0.00056464 | 0.00045262 | 0.00016565 | -0.00020148 | 0.00004534 | -0.00005009 | 0.00005140 | -0.00001020 | -0.00002583 | -0.00007393 |
| 18 | 26 | 0.04125788 | 0.05692817 | -0.00268549 | -0.02770973 | -0.00356125 | 0.00706438 | 0.00186580 | -0.00008822 | 0.00056773 | 0.00005929 | -0.00020487 | 0.00004879 | -0.00005304 | 0.00005153 | -0.00001565 | -0.00001415 | -0.00005889 |
| 18 | 27 | 0.04138933 | 0.05594618 | -0.00667610 | -0.03157153 | -0.00494863 | 0.00664813 | 0.00120825 | -0.00048014 | 0.00059305 | 0.00004139 | -0.00020439 | 0.00008029 | -0.00006886 | 0.00005280 | 0.00001646 | -0.00013466 | -0.00011877 |
| 18 | 28 | 0.04162623 | 0.05614182 | -0.00699384 | -0.03110504 | -0.00415925 | 0.00693090 | 0.00135557 | -0.00013505 | 0.00070192 | 0.00001701 | -0.00023606 | 0.00010240 | -0.00007654 | 0.00007494 | -0.00000121 | -0.00010213 | -0.00011656 |
| 18 | 29 | 0.04100774 | 0.05473796 | -0.00864100 | -0.03300260 | -0.00525991 | 0.00647013 | 0.00104383 | -0.00049680 | 0.00063979 | -0.00002453 | -0.00018727 | 0.00008972 | -0.00008207 | 0.00004236 | 0.00003524 | -0.00015740 | -0.00012070 |
| 18 | 30 | 0.04073802 | 0.05367773 | -0.00747682 | -0.02955740 | -0.00292995 | 0.00714044 | 0.00085395 | -0.00072802 | 0.00043309 | 0.00040297 | -0.00021688 | 0.00010083 | -0.00008440 | 0.00012438 | -0.00005336 | -0.00004324 | -0.00016636 |
| 18 | 31 | 0.03956958 | 0.05218819 | -0.00760296 | -0.03024640 | -0.00359670 | 0.00677615 | 0.00008985 | -0.00120669 | 0.00039406 | 0.00042226 | -0.00021681 | 0.00015643 | -0.00015042 | 0.00011718 | -0.00004865 | -0.00004680 | -0.00021182 |
| 18 | 32 | 0.02953631 | 0.05225350 | 0.01669029 | -0.01815796 | -0.01085136 | 0.00432719 | 0.00511340 | 0.00079198 | -0.00005142 | 0.00053398 | 0.00011097 | -0.00012624 | -0.00003742 | 0.00008953 | 0.00001670 | 0.00014700 | 0.00011975 |
| 18 | 33 | 0.02914220 | 0.05045295 | 0.01477749 | -0.01880432 | -0.01011419 | 0.00451829 | 0.00325702 | 0.00008470 | 0.00009880 | 0.00058223 | -0.00005444 | 0.00020219 | -0.00035725 | 0.00012477 | -0.00017246 | 0.00035366 | -0.00005973 |
| 18 | 34 | 0.03079190 | 0.05104889 | 0.00715188 | -0.02875535 | -0.01383382 | 0.00673104 | 0.00555110 | -0.00055737 | -0.00047155 | 0.00017625 | 0.00003159 | 0.00026822 | -0.00059207 | 0.00001189 | -0.00047026 | 0.00069067 | 0.00005276 |
| 18 | 35 | 0.03000623 | 0.04895277 | 0.00598198 | -0.02892267 | -0.01282420 | 0.00716279 | 0.00396157 | -0.00128005 | -0.00003432 | 0.00023599 | -0.00006234 | 0.00053156 | -0.00083831 | -0.00006575 | -0.00062589 | 0.00092700 | -0.00002545 |
| 18 | 36 | 0.02966245 | 0.04949663 | 0.00630236 | -0.03006421 | -0.01450794 | 0.00649016 | 0.00526844 | -0.00114088 | -0.00041133 | 0.00002551 | 0.00012421 | 0.00024360 | -0.00057592 | -0.00014248 | -0.00048583 | 0.00080814 | 0.00017311 |
| 19 | 12 | 0.04195300 | 0.06497018 | 0.01512195 | -0.01687409 | -0.00773192 | 0.00229337 | 0.00097853 | -0.00133221 | -0.00033037 | 0.00072593 | 0.00008946 | -0.00029016 | 0.00029107 | 0.00007080 | 0.00018883 | -0.00016331 | 0.00013891 |
| 19 | 13 | 0.04127937 | 0.06454118 | 0.01626673 | -0.01539153 | -0.00682915 | 0.00357134 | 0.00267046 | -0.00025772 | -0.00036584 | 0.00044766 | 0.00014631 | -0.00033599 | 0.00029065 | 0.00002013 | 0.00017027 | -0.00004987 | 0.00021774 |
| 19 | 14 | 0.04160747 | 0.06395202 | 0.01430282 | -0.01661708 | -0.00665674 | 0.00274475 | 0.00090583 | -0.00104834 | 0.00007866 | 0.00063668 | 0.00010475 | -0.00029847 | 0.00026842 | -0.00002648 | 0.00017320 | -0.00007100 | 0.00017945 |
| 19 | 15 | 0.04140642 | 0.06367938 | 0.01461656 | -0.01538779 | -0.00444437 | 0.00547217 | 0.00315395 | -0.00002892 | -0.00016891 | 0.00036389 | 0.00010732 | -0.00029296 | 0.00023959 | -0.00000284 | 0.00013618 | -0.00001291 | 0.00017678 |
| 19 | 16 | 0.04212805 | 0.06329767 | 0.01216272 | -0.01664017 | -0.00318049 | 0.00615579 | 0.00243663 | -0.00057087 | 0.00007926 | 0.00044959 | 0.00004130 | -0.00022015 | 0.00017839 | -0.00000214 | 0.00011253 | -0.00006127 | 0.00009478 |
| 19 | 17 | 0.04219273 | 0.06264506 | 0.01030056 | -0.01744292 | -0.00229575 | 0.00773969 | 0.00416314 | 0.00033218 | -0.00015831 | 0.00020697 | 0.00006404 | -0.00024157 | 0.00018431 | 0.00000840 | 0.00008715 | -0.00002332 | 0.00011753 |
| 19 | 18 | 0.04290786 | 0.06240024 | 0.00774176 | -0.01896300 | -0.00116885 | 0.00839248 | 0.00345345 | -0.00000048 | 0.00020260 | 0.00023182 | -0.00003442 | -0.00014291 | 0.00009900 | 0.00000504 | 0.00004998 | -0.00004175 | 0.00004717 |
| 19 | 19 | 0.04319742 | 0.06153621 | 0.00412936 | -0.02165776 | -0.00188698 | 0.00854005 | 0.00424286 | 0.00050789 | -0.00002888 | 0.00007517 | 0.00001029 | -0.00018657 | 0.00014381 | 0.00000720 | 0.00005070 | -0.00004880 | 0.00007644 |
| 19 | 20 | 0.04388300 | 0.06158914 | 0.00183838 | -0.02305414 | -0.00055148 | 0.00924241 | 0.00335781 | 0.00028161 | 0.00044419 | 0.00006020 | -0.00012217 | -0.00007131 | 0.00006548 | 0.00000083 | 0.00002319 | -0.00007935 | -0.00000328 |
| 19 | 21 | 0.04339537 | 0.06038673 | 0.00040165 | -0.02459405 | -0.00234038 | 0.00838981 | 0.00364032 | 0.00035869 | 0.00007162 | 0.00007282 | -0.00004523 | -0.00012173 | 0.00008961 | 0.00001404 | 0.00001321 | -0.00001842 | 0.00004450 |
| 19 | 22 | 0.04323064 | 0.06005026 | 0.00066613 | -0.02411199 | -0.00105840 | 0.00884673 | 0.00285428 | 0.00008357 | 0.00045403 | 0.00010939 | -0.00014670 | -0.00003111 | 0.00000771 | 0.00000210 | -0.00001552 | 0.00001659 | -0.00000618 |
| 19 | 23 | 0.04260797 | 0.05904534 | -0.00019993 | -0.02519383 | -0.00255685 | 0.00806876 | 0.00282941 | 0.00001185 | 0.00013015 | 0.00015961 | -0.00010976 | -0.00005167 | 0.00002042 | 0.00002773 | -0.00003191 | 0.00008509 | 0.00002858 |
| 19 | 24 | 0.04229075 | 0.05860300 | 0.00021188 | -0.02456290 | -0.00144884 | 0.00815382 | 0.00207983 | -0.00013624 | 0.00052043 | 0.00010209 | -0.00018515 | -0.00000240 | -0.00001811 | -0.00000497 | -0.00001862 | 0.00006943 | -0.00000952 |
| 19 | 25 | 0.04239167 | 0.05798018 | -0.00269565 | -0.02736304 | -0.00325783 | 0.00772160 | 0.00230052 | -0.00017605 | 0.00017165 | 0.00016196 | -0.00015156 | -0.00000505 | -0.00001426 | 0.00004151 | -0.00005030 | 0.00009750 | 0.00001041 |
| 19 | 26 | 0.04283215 | 0.05789929 | -0.00458355 | -0.02870838 | -0.00253963 | 0.00807732 | 0.00185708 | -0.00018680 | 0.00054870 | 0.00007917 | -0.00022849 | 0.00005018 | -0.00003958 | 0.00002869 | -0.00003634 | 0.00001494 | -0.00004578 |
| 19 | 27 | 0.04274627 | 0.05734498 | -0.00667330 | -0.03072028 | -0.00433793 | 0.00738709 | 0.00215680 | -0.00005540 | 0.00024272 | 0.00004859 | -0.00014445 | 0.00000536 | -0.00001142 | 0.00004512 | -0.00002149 | -0.00000843 | -0.00001980 |
| 19 | 28 | 0.04310824 | 0.05750494 | -0.00805269 | -0.03162834 | -0.00333560 | 0.00771785 | 0.00148267 | -0.00006078 | 0.00069887 | -0.00003528 | -0.00024824 | 0.00007549 | -0.00002945 | 0.00004232 | 0.00000892 | -0.00011461 | -0.00009485 |
| 19 | 29 | 0.04213789 | 0.05573790 | -0.00806749 | -0.03121132 | -0.00382011 | 0.00793447 | 0.00186907 | -0.00036551 | 0.00015212 | 0.00024394 | -0.00013640 | 0.00003076 | -0.00003833 | 0.00007962 | -0.00004144 | -0.00004941 | -0.00012023 |
| 19 | 30 | 0.04182511 | 0.05501863 | -0.00772846 | -0.02997830 | -0.00214468 | 0.00829489 | 0.00111974 | -0.00064453 | 0.00046192 | 0.00040868 | -0.00025451 | 0.00008887 | -0.00005914 | 0.00011807 | -0.00004801 | -0.00005999 | -0.00017662 |
| 19 | 31 | 0.03142219 | 0.05467196 | 0.01554192 | -0.02045196 | -0.01067034 | 0.00615390 | 0.00560015 | 0.00099238 | 0.00005126 | 0.00059411 | 0.00007218 | -0.00015535 | 0.00000847 | 0.00004519 | 0.00001289 | 0.00012260 | 0.00008127 |
| 19 | 32 | 0.03127206 | 0.05401474 | 0.01495332 | -0.02000848 | -0.00961749 | 0.00628447 | 0.00514520 | 0.00080689 | 0.00016913 | 0.00060670 | -0.00005688 | -0.00006080 | -0.00008168 | 0.00013809 | -0.00002176 | 0.00013312 | 0.00002734 |
| 19 | 33 | 0.03322139 | 0.05363913 | 0.00562035 | -0.03016018 | -0.01212243 | 0.00921568 | 0.00534153 | -0.00052192 | 0.00012824 | 0.00054037 | -0.00015912 | 0.00034509 | -0.00062316 | 0.00001503 | -0.00053229 | 0.00076773 | -0.00000167 |
| 19 | 34 | 0.03235370 | 0.05288413 | 0.00587967 | -0.03054526 | -0.01263939 | 0.00852421 | 0.00526754 | -0.00078510 | -0.00001144 | 0.00038892 | -0.00014494 | 0.00026835 | -0.00055707 | 0.00004463 | -0.00043751 | 0.00064478 | 0.00003380 |
| 19 | 35 | 0.03190285 | 0.05185044 | 0.00511304 | -0.03101279 | -0.01232465 | 0.00941762 | 0.00486266 | -0.00129739 | 0.00034128 | 0.00052295 | -0.00011718 | 0.00039610 | -0.00069892 | -0.00009737 | -0.00059048 | 0.00095017 | 0.00009687 |
| 19 | 36 | 0.03109894 | 0.05113606 | 0.00533848 | -0.03106855 | -0.01268384 | 0.00834711 | 0.00466345 | -0.00145772 | 0.00014433 | 0.00031073 | -0.00008995 | 0.00027413 | -0.00056255 | -0.00008292 | -0.00045551 | 0.00077495 | 0.00014054 |
| 19 | 37 | 0.03126866 | 0.04998258 | 0.00416571 | -0.02968031 | -0.01018992 | 0.00961589 | 0.00354140 | -0.00193494 | 0.00061652 | 0.00052773 | -0.00015553 | 0.00052318 | -0.00079172 | -0.00016510 | -0.00066516 | 0.00111285 | 0.00009745 |
| 19 | 38 | 0.03055666 | 0.04939349 | 0.00421151 | -0.02984138 | -0.01065622 | 0.00828341 | 0.00333228 | -0.00202978 | 0.00036812 | 0.00026165 | -0.00011168 | 0.00036012 | -0.00059585 | -0.00017400 | -0.00050217 | 0.00090613 | 0.00016087 |
| 19 | 39 | 0.03091923 | 0.04821591 | 0.00302476 | -0.02787520 | -0.00752349 | 0.00977318 | 0.00205234 | -0.00251892 | 0.00091740 | 0.00050157 | -0.00019382 | 0.00066763 | -0.00088449 | -0.00025610 | -0.00075352 | 0.00129177 | 0.00008921 |
| 19 | 40 | 0.03008395 | 0.04778890 | 0.00315228 | -0.02936321 | -0.01025649 | 0.00778553 | 0.00255051 | -0.00210802 | 0.00030771 | 0.00025030 | -0.00019053 | 0.00054050 | -0.00075091 | -0.00011096 | -0.00058912 | 0.00098121 | 0.00005013 |
| 20 | 13 | 0.04320996 | 0.06488512 | 0.01202089 | -0.01846759 | -0.00630177 | 0.00322818 | 0.00046364 | -0.00185838 | -0.00012699 | 0.00084673 | 0.00010660 | -0.00033562 | 0.00035988 | -0.00002024 | 0.00019267 | -0.00010930 | 0.00019622 |
| 20 | 14 | 0.04393480 | 0.06458597 | 0.00957718 | -0.01935271 | -0.00530193 | 0.00304343 | -0.00016612 | -0.00180210 | 0.00032907 | 0.00091136 | 0.00009463 | -0.00037592 | 0.00041675 | -0.00007859 | 0.00022424 | -0.00011433 | 0.00021010 |
| 20 | 15 | 0.04340787 | 0.06396393 | 0.01011427 | -0.01868144 | -0.00429107 | 0.00478278 | 0.00091150 | -0.00149006 | 0.00003100 | 0.00067636 | 0.00007270 | -0.00028073 | 0.00027709 | -0.00004489 | 0.00015378 | -0.00006522 | 0.00015039 |
| 20 | 16 | 0.04457679 | 0.06388468 | 0.00681082 | -0.02053032 | -0.00336756 | 0.00532059 | 0.00072635 | -0.00141816 | 0.00019556 | 0.00068449 | 0.00000339 | -0.00024294 | 0.00024733 | -0.00003237 | 0.00012847 | -0.00005018 | 0.00011154 |
| 20 | 17 | 0.04430689 | 0.06319225 | 0.00622286 | -0.02053506 | -0.00227752 | 0.00708878 | 0.00200432 | -0.00091033 | -0.00000107 | 0.00041701 | -0.00000437 | -0.00017445 | 0.00014074 | -0.00002004 | 0.00006901 | -0.00001830 | 0.00007786 |
| 20 | 18 | 0.04533260 | 0.06323994 | 0.00304572 | -0.02280273 | -0.00213458 | 0.00707866 | 0.00148596 | -0.00084200 | 0.00021090 | 0.00043068 | -0.00009330 | -0.00012158 | 0.00009911 | -0.00000724 | 0.00003402 | 0.00003214 | 0.00005979 |
| 20 | 19 | 0.04530368 | 0.06266513 | 0.00155404 | -0.02334785 | -0.00095630 | 0.00869585 | 0.00268908 | -0.00028671 | 0.00016237 | 0.00015857 | -0.00008107 | -0.00009774 | 0.00006868 | -0.00002823 | 0.00001198 | 0.00000572 | 0.00004798 |
| 20 | 20 | 0.04605681 | 0.06299496 | -0.00061788 | -0.02520616 | -0.00118465 | 0.00837126 | 0.00186910 | -0.00028509 | 0.00038560 | 0.00018261 | -0.00018570 | -0.00002942 | 0.00002248 | -0.00000932 | -0.00002095 | 0.00006688 | 0.00003116 |
| 20 | 21 | 0.04545038 | 0.06177384 | -0.00118599 | -0.02507129 | -0.00054685 | 0.00912591 | 0.00267712 | -0.00003183 | 0.00032177 | 0.00006496 | -0.00012894 | -0.00005013 | 0.00002837 | -0.00003679 | -0.00002078 | 0.00004432 | 0.00003475 |
| 20 | 22 | 0.04556761 | 0.06133220 | -0.00199228 | -0.02550203 | -0.00097832 | 0.00815067 | 0.00184759 | -0.00011219 | 0.00043144 | 0.00014268 | -0.00017157 | -0.00001504 | -0.00002166 | -0.00002182 | -0.00004798 | 0.00013510 | 0.00003341 |
| 20 | 23 | 0.04467193 | 0.06021976 | -0.00191188 | -0.02529797 | -0.00050131 | 0.00880740 | 0.00225717 | -0.00007174 | 0.00041210 | 0.00008282 | -0.00015814 | -0.00001847 | -0.00001591 | -0.00003689 | -0.00004472 | 0.00012163 | 0.00003208 |
| 20 | 24 | 0.04486050 | 0.05982060 | -0.00274295 | -0.02529392 | -0.00070034 | 0.00765411 | 0.00156732 | 0.00001728 | 0.00050977 | 0.00006245 | -0.00017226 | -0.00001418 | -0.00002712 | -0.00003876 | -0.00004027 | 0.00015505 | 0.00003240 |
| 20 | 25 | 0.04438454 | 0.05938752 | -0.00356781 | -0.02655473 | -0.00067365 | 0.00883362 | 0.00215224 | 0.00000598 | 0.00048298 | 0.00004747 | -0.00020339 | 0.00001563 | -0.00003182 | -0.00002493 | -0.00006328 | 0.00014029 | 0.00002721 |
| 20 | 26 | 0.04506050 | 0.05966280 | -0.00489342 | -0.02697182 | -0.00059424 | 0.00860801 | 0.00205731 | 0.00026633 | 0.00053798 | 0.00004684 | -0.00023354 | 0.00004125 | -0.00005089 | -0.00000006 | -0.00009091 | 0.00017355 | 0.00002359 |
| 20 | 27 | 0.04456911 | 0.05936968 | -0.00600905 | -0.02885064 | -0.00131265 | 0.00892387 | 0.00213549 | 0.00016370 | 0.00057493 | -0.00000305 | -0.00024916 | 0.00004337 | -0.00001060 | 0.00000269 | -0.00005385 | 0.00006164 | -0.00000089 |
| 20 | 28 | 0.04505881 | 0.05910322 | -0.00711304 | -0.02864708 | -0.00036984 | 0.00980549 | 0.00249087 | 0.00016063 | 0.00050371 | 0.00025861 | -0.00024153 | 0.00002860 | 0.00000182 | 0.00004943 | -0.00009086 | 0.00004690 | -0.00006424 |
| 20 | 29 | 0.04396494 | 0.05781906 | -0.00728228 | -0.02917749 | -0.00077399 | 0.00962623 | 0.00219867 | -0.00006917 | 0.00049075 | 0.00029377 | -0.00024414 | 0.00002208 | 0.00002069 | 0.00006214 | -0.00005226 | -0.00003007 | -0.00010074 |
| 20 | 30 | 0.03356175 | 0.05801959 | 0.01565424 | -0.02208555 | -0.01153874 | 0.00694832 | 0.00792574 | 0.00154456 | -0.00001727 | 0.00054422 | 0.00016763 | -0.00045157 | 0.00025290 | 0.00001869 | 0.00014821 | -0.00001328 | 0.00028920 |
| 20 | 31 | 0.03323402 | 0.05707947 | 0.01499431 | -0.02141572 | -0.00969218 | 0.00764411 | 0.00682969 | 0.00136037 | 0.00021533 | 0.00058305 | -0.00000228 | -0.00028339 | 0.00013456 | 0.00007922 | 0.00008642 | 0.00000777 | 0.00014320 |
| 20 | 32 | 0.03528029 | 0.05740680 | 0.00589534 | -0.03245488 | -0.01389498 | 0.00908842 | 0.00722009 | -0.00006217 | -0.00030346 | 0.00048595 | -0.00008356 | -0.00003505 | -0.00029984 | 0.00010539 | -0.00028616 | 0.00044792 | 0.00015393 |
| 20 | 33 | 0.03467558 | 0.05626098 | 0.00543692 | -0.03219295 | -0.01251603 | 0.00974163 | 0.00626655 | -0.00041499 | 0.00006222 | 0.00051650 | -0.00019396 | 0.00008370 | -0.00037023 | 0.00008588 | -0.00032062 | 0.00048391 | 0.00008948 |
| 20 | 34 | 0.03434734 | 0.05589541 | 0.00472430 | -0.03347702 | -0.01409087 | 0.00891902 | 0.00682414 | -0.00058928 | -0.00027011 | 0.00038667 | -0.00004158 | -0.00004910 | -0.00029795 | 0.00003321 | -0.00028975 | 0.00051682 | 0.00023088 |
| 20 | 35 | 0.03347569 | 0.05455866 | 0.00472523 | -0.03283574 | -0.01258629 | 0.00954930 | 0.00562863 | -0.00109448 | 0.00017883 | 0.00046502 | -0.00016228 | 0.00008917 | -0.00038078 | 0.00000433 | -0.00032822 | 0.00058203 | 0.00017641 |
| 20 | 36 | 0.03381496 | 0.05429542 | 0.00344131 | -0.03269307 | -0.01253348 | 0.00870481 | 0.00578532 | -0.00102780 | -0.00018874 | 0.00030570 | -0.00005446 | -0.00001703 | -0.00029042 | -0.00001052 | -0.00029283 | 0.00056031 | 0.00024271 |
| 20 | 37 | 0.03284159 | 0.05279950 | 0.00356999 | -0.03180085 | -0.01078716 | 0.00940236 | 0.00433664 | -0.00168141 | 0.00033739 | 0.00042150 | -0.00018982 | 0.00015748 | -0.00040407 | -0.00005131 | -0.00035408 | 0.00067355 | 0.00019032 |
| 20 | 38 | 0.03348114 | 0.05279505 | 0.00200034 | -0.03164231 | -0.01058189 | 0.00849921 | 0.00451400 | -0.00150321 | -0.00009310 | 0.00022056 | -0.00008282 | 0.00003593 | -0.00028600 | -0.00006047 | -0.00030381 | 0.00061269 | 0.00024083 |
| 20 | 39 | 0.03247155 | 0.05115806 | 0.00221550 | -0.03042237 | -0.00855059 | 0.00925710 | 0.00287770 | -0.00225548 | 0.00050911 | 0.00035728 | -0.00022198 | 0.00024821 | -0.00043332 | -0.00012402 | -0.00039559 | 0.00078330 | 0.00019347 |
| 20 | 40 | 0.03300340 | 0.05138348 | 0.00120972 | -0.03093093 | -0.01012365 | 0.00805373 | 0.00378477 | -0.00159313 | -0.00019528 | 0.00024157 | -0.00017111 | 0.00019488 | -0.00040636 | 0.00001773 | -0.00038343 | 0.00066821 | 0.00013263 |
| 20 | 41 | 0.03187025 | 0.04965599 | 0.00141231 | -0.03078849 | -0.00971040 | 0.00853825 | 0.00274572 | -0.00196343 | 0.00015155 | 0.00040421 | -0.00034036 | 0.00047212 | -0.00064702 | 0.00006954 | -0.00050032 | 0.00077938 | -0.00001732 |
| 21 | 14 | 0.04411440 | 0.06430452 | 0.00989117 | -0.01845063 | -0.00452832 | 0.00388933 | 0.00004105 | -0.00166541 | 0.00027740 | 0.00076159 | 0.00014609 | -0.00022318 | 0.00017006 | -0.00007310 | 0.00011254 | 0.00001077 | 0.00015262 |
| 21 | 15 | 0.04421632 | 0.06376866 | 0.00894863 | -0.01822808 | -0.00301378 | 0.00605194 | 0.00209484 | -0.00070136 | -0.00010516 | 0.00044753 | 0.00014887 | -0.00019396 | 0.00010106 | -0.00002334 | 0.00007119 | 0.00004681 | 0.00012183 |
| 21 | 16 | 0.04512860 | 0.06352540 | 0.00626607 | -0.01968663 | -0.00177886 | 0.00642148 | 0.00108776 | -0.00127561 | 0.00016418 | 0.00052868 | 0.00007217 | -0.00012785 | 0.00006872 | -0.00003702 | 0.00005803 | -0.00000884 | 0.00004400 |
| 21 | 17 | 0.04548286 | 0.06277388 | 0.00371734 | -0.02074082 | -0.00082345 | 0.00809255 | 0.00305381 | -0.00032239 | -0.00015429 | 0.00024014 | 0.00010888 | -0.00014908 | 0.00006284 | -0.00001965 | 0.00003144 | 0.00001491 | 0.00006478 |
| 21 | 18 | 0.04641820 | 0.06266347 | 0.00084956 | -0.02251551 | 0.00022384 | 0.00841094 | 0.00205465 | -0.00069953 | 0.00022777 | 0.00027033 | 0.00000061 | -0.00005768 | 0.00000453 | -0.00003777 | 0.00000165 | -0.00000717 | 0.00000027 |
| 21 | 19 | 0.04671466 | 0.06194510 | -0.00212844 | -0.02430264 | 0.00022529 | 0.00923097 | 0.00359471 | 0.00006806 | -0.00002563 | 0.00004897 | 0.00006825 | -0.00012569 | 0.00006548 | -0.00004221 | 0.00000761 | -0.00001899 | 0.00005533 |
| 21 | 20 | 0.04759676 | 0.06213791 | -0.00463120 | -0.02584988 | 0.00146180 | 0.00964411 | 0.00251710 | -0.00016542 | 0.00047358 | 0.00003980 | -0.00007159 | -0.00001681 | 0.00000655 | -0.00006341 | -0.00001871 | -0.00004253 | -0.00001333 |
| 21 | 21 | 0.04689823 | 0.06102223 | -0.00515754 | -0.02642991 | 0.00025829 | 0.00939570 | 0.00336824 | 0.00012946 | 0.00009854 | 0.00001209 | 0.00001746 | -0.00008450 | 0.00003800 | -0.00004525 | -0.00002374 | 0.00001662 | 0.00004906 |
| 21 | 22 | 0.04696123 | 0.06065724 | -0.00554464 | -0.02641457 | 0.00132874 | 0.00947968 | 0.00215089 | -0.00023114 | 0.00050282 | 0.00006230 | -0.00009747 | 0.00001924 | -0.00005372 | -0.00006212 | -0.00006039 | 0.00006594 | -0.00000964 |
| 21 | 23 | 0.04609939 | 0.05978712 | -0.00542008 | -0.02657571 | 0.00027498 | 0.00918340 | 0.00263913 | -0.00010021 | 0.00017771 | 0.00009196 | -0.00005170 | -0.00002432 | -0.00002243 | -0.00002826 | -0.00006588 | 0.00012215 | 0.00003693 |
| 21 | 24 | 0.04603817 | 0.05934127 | -0.00565893 | -0.02634733 | 0.00130155 | 0.00899768 | 0.00149647 | -0.00032241 | 0.00059334 | 0.00004551 | -0.00014133 | 0.00003979 | -0.00007188 | -0.00006693 | -0.00006297 | 0.00012487 | -0.00000794 |
| 21 | 25 | 0.04592122 | 0.05886672 | -0.00747655 | -0.02804591 | 0.00023236 | 0.00923673 | 0.00227786 | -0.00021010 | 0.00024792 | 0.00015691 | -0.00010294 | -0.00000270 | -0.00001562 | -0.00000811 | -0.00007515 | 0.00011311 | 0.00001225 |
| 21 | 26 | 0.04658765 | 0.05889632 | -0.00959115 | -0.02940105 | 0.00104071 | 0.00953794 | 0.00179401 | -0.00019155 | 0.00064492 | 0.00005238 | -0.00018576 | 0.00006092 | -0.00004783 | -0.00003783 | -0.00007897 | 0.00006973 | -0.00002330 |
| 21 | 27 | 0.04548999 | 0.05809464 | -0.00979481 | -0.03121269 | -0.00108065 | 0.00984616 | 0.00282848 | -0.00024648 | 0.00017524 | 0.00030048 | -0.00004794 | -0.00006552 | 0.00005385 | 0.00000254 | -0.00003965 | -0.00005356 | -0.00006948 |
| 21 | 28 | 0.04636625 | 0.05823126 | -0.01231049 | -0.03279918 | -0.00005921 | 0.01032170 | 0.00213100 | -0.00035695 | 0.00057247 | 0.00020446 | -0.00015831 | 0.00003264 | -0.00001039 | -0.00000605 | -0.00005711 | -0.00008801 | -0.00012971 |
| 21 | 29 | 0.03445257 | 0.05908243 | 0.01400997 | -0.02587614 | -0.01297527 | 0.00742453 | 0.00789140 | 0.00161358 | -0.00001216 | 0.00036871 | 0.00026330 | -0.00046203 | 0.00025707 | -0.00008693 | 0.00017588 | -0.00009724 | 0.00020742 |
| 21 | 30 | 0.03468687 | 0.05851739 | 0.01286055 | -0.02555659 | -0.01141463 | 0.00807059 | 0.00721032 | 0.00131805 | 0.00012381 | 0.00043954 | 0.00007900 | -0.00027803 | 0.00006458 | 0.00003596 | 0.00006886 | -0.00001911 | 0.00009523 |
| 21 | 31 | 0.03650933 | 0.05860123 | 0.00534868 | -0.03258257 | -0.01271806 | 0.00999300 | 0.00701280 | 0.00043948 | 0.00013192 | 0.00051123 | -0.00009625 | -0.00001705 | -0.00024787 | 0.00002022 | -0.00026649 | 0.00038184 | 0.00007665 |
| 21 | 32 | 0.03590901 | 0.05769887 | 0.00525891 | -0.03260503 | -0.01252069 | 0.00977850 | 0.00671983 | 0.00005428 | 0.00003466 | 0.00042185 | -0.00012513 | 0.00000572 | -0.00029367 | 0.00007111 | -0.00025048 | 0.00034933 | 0.00006728 |
| 21 | 33 | 0.03548268 | 0.05691454 | 0.00446680 | -0.03306004 | -0.01244126 | 0.01019461 | 0.00642154 | -0.00033445 | 0.00026062 | 0.00055221 | -0.00010267 | 0.00003054 | -0.00030623 | -0.00002132 | -0.00031064 | 0.00051769 | 0.00015391 |
| 21 | 34 | 0.03483678 | 0.05597974 | 0.00445276 | -0.03275602 | -0.01216206 | 0.00957939 | 0.00606247 | -0.00060285 | 0.00012336 | 0.00038795 | -0.00010857 | 0.00000252 | -0.00027893 | 0.00000495 | -0.00025143 | 0.00043593 | 0.00016261 |
| 21 | 35 | 0.03480991 | 0.05509701 | 0.00338357 | -0.03203477 | -0.01059640 | 0.01032649 | 0.00519650 | -0.00104648 | 0.00043245 | 0.00058376 | -0.00015234 | 0.00012510 | -0.00036898 | -0.00004509 | -0.00035832 | 0.00063473 | 0.00016665 |
| 21 | 36 | 0.03416869 | 0.05420958 | 0.00327162 | -0.03176630 | -0.01043717 | 0.00939042 | 0.00482880 | -0.00123294 | 0.00025362 | 0.00035197 | -0.00013569 | 0.00005091 | -0.00027722 | -0.00004835 | -0.00026655 | 0.00052041 | 0.00020033 |

TABLE 3. Nuclear charge density distribution FB coefficients.

| $Z$ | $N$ | $a_1$ | $a_2$ | $a_3$ | $a_4$ | $a_5$ | $a_6$ | $a_7$ | $a_8$ | $a_9$ | $a_{10}$ | $a_{11}$ | $a_{12}$ | $a_{13}$ | $a_{14}$ | $a_{15}$ | $a_{16}$ | $a_{17}$ |
|---|---|---|---|---|---|---|---|---|---|---|---|---|---|---|---|---|---|---|
| 21 | 37 | 0.03431806 | 0.05332165 | 0.00217082 | -0.03058782 | -0.00830312 | 0.01037747 | 0.00375264 | -0.00174159 | 0.00062981 | 0.00058207 | -0.00020226 | 0.00023575 | -0.00043139 | -0.00009182 | -0.00041547 | 0.00076690 | 0.00017215 |
| 21 | 38 | 0.03373463 | 0.05253856 | 0.00189753 | -0.03046358 | -0.00828999 | 0.00920173 | 0.00339217 | -0.00186375 | 0.00040331 | 0.00029588 | -0.00016976 | 0.00012482 | -0.00028600 | -0.00011723 | -0.00029784 | 0.00062394 | 0.00022380 |
| 21 | 39 | 0.03404938 | 0.05164124 | 0.00082563 | -0.02881616 | -0.00571336 | 0.01033322 | 0.00219860 | -0.00238703 | 0.00084984 | 0.00054686 | -0.00024423 | 0.00036194 | -0.00049866 | -0.00016405 | -0.00048691 | 0.00091906 | 0.00017467 |
| 21 | 40 | 0.03326980 | 0.05100224 | 0.00076659 | -0.03001801 | -0.00781621 | 0.00866415 | 0.00247318 | -0.00199957 | 0.00035426 | 0.00031502 | -0.00026294 | 0.00029468 | -0.00041305 | -0.00004667 | -0.00037340 | 0.00068195 | 0.00011472 |
| 21 | 41 | 0.03348744 | 0.05015905 | -0.00048147 | -0.02983262 | -0.00677596 | 0.00971715 | 0.00204331 | -0.00203882 | 0.00056005 | 0.00051596 | -0.00033795 | 0.00053974 | -0.00066777 | -0.00000985 | -0.00054508 | 0.00086679 | -0.00002061 |
| 21 | 42 | 0.03287070 | 0.04919088 | -0.00134797 | -0.03083110 | -0.00794642 | 0.00838267 | 0.00216263 | -0.00193164 | 0.00001500 | 0.00029880 | -0.00032915 | 0.00046152 | -0.00055587 | 0.00008473 | -0.00043137 | 0.00063323 | -0.00005834 |
| 22 | 15 | 0.04565633 | 0.06425507 | 0.00724626 | -0.01872935 | -0.00209471 | 0.00542511 | 0.00026758 | -0.00157388 | 0.00036181 | 0.00064848 | 0.00014662 | -0.00010402 | 0.00002411 | -0.00007441 | 0.00004189 | 0.00004409 | 0.00007071 |
| 22 | 16 | 0.04701084 | 0.06394323 | 0.00307320 | -0.02097690 | -0.00073753 | 0.00575618 | -0.00024224 | -0.00151680 | 0.00061132 | 0.00064795 | 0.00006149 | -0.00008685 | 0.00004530 | -0.00006160 | 0.00007019 | -0.00001594 | 0.00001641 |
| 22 | 17 | 0.04707743 | 0.06326874 | 0.00184040 | -0.02111298 | 0.00051438 | 0.00755842 | 0.00127450 | -0.00111774 | 0.00033939 | 0.00039928 | 0.00007859 | -0.00004193 | -0.00001691 | -0.00006355 | 0.00000900 | -0.00000771 | -0.00000070 |
| 22 | 18 | 0.04835329 | 0.06296179 | -0.00251659 | -0.02389837 | 0.00118673 | 0.00752422 | 0.00056095 | -0.00102025 | 0.00063022 | 0.00039746 | -0.00002232 | -0.00000709 | -0.00002121 | -0.00005365 | 0.00001390 | -0.00001734 | -0.00003458 |
| 22 | 19 | 0.04865687 | 0.06233921 | -0.00482500 | -0.02469985 | 0.00247453 | 0.00918305 | 0.00205036 | -0.00060246 | 0.00051053 | 0.00016032 | 0.00001411 | -0.00000909 | -0.00000627 | -0.00009003 | -0.00001467 | -0.00005667 | -0.00001712 |
| 22 | 20 | 0.04960207 | 0.06227037 | -0.00809088 | -0.02704065 | 0.00272676 | 0.00898256 | 0.00120757 | -0.00047132 | 0.00081571 | 0.00016444 | -0.00010019 | 0.00004562 | -0.00003539 | -0.00007320 | -0.00002406 | -0.00002872 | -0.00004566 |
| 22 | 21 | 0.04906056 | 0.06132209 | -0.00843924 | -0.02697551 | 0.00295602 | 0.00969176 | 0.00205470 | -0.00041276 | 0.00063948 | 0.00006096 | -0.00003295 | 0.00002145 | -0.00002333 | -0.00010612 | -0.00004282 | -0.00002416 | -0.00001491 |
| 22 | 22 | 0.04916478 | 0.06085411 | -0.00932097 | -0.02755786 | 0.00280252 | 0.00889358 | 0.00102582 | -0.00041325 | 0.00081068 | 0.00009263 | -0.00009669 | 0.00005568 | -0.00007146 | -0.00009254 | -0.00005015 | 0.00005675 | -0.00002813 |
| 22 | 23 | 0.04832628 | 0.06003234 | -0.00891054 | -0.02739250 | 0.00276583 | 0.00947726 | 0.00149022 | -0.00052683 | 0.00065787 | 0.00004584 | -0.00007204 | 0.00005190 | -0.00007105 | -0.00010860 | -0.00007199 | 0.00007794 | -0.00000879 |
| 22 | 24 | 0.04864091 | 0.05955962 | -0.01019349 | -0.02758483 | 0.00321715 | 0.00861625 | 0.00044966 | -0.00046397 | 0.00087387 | 0.00008659 | -0.00012204 | 0.00003849 | -0.00004066 | -0.00010284 | -0.00003190 | 0.00006604 | -0.00003594 |
| 22 | 25 | 0.04834420 | 0.05903744 | -0.01157185 | -0.02924724 | 0.00288764 | 0.00973393 | 0.00127631 | -0.00059165 | 0.00071995 | 0.00008888 | -0.00012278 | 0.00006190 | -0.00005007 | -0.00009351 | -0.00007858 | 0.00006510 | -0.00002672 |
| 22 | 26 | 0.04751658 | 0.05876111 | -0.01113321 | -0.03097792 | 0.00129570 | 0.01002861 | 0.00201171 | -0.00043533 | 0.00079550 | 0.00046043 | -0.00006424 | -0.00007394 | 0.00006398 | -0.00006598 | -0.00000835 | -0.00006361 | -0.00007925 |
| 22 | 27 | 0.04718594 | 0.05839421 | -0.01262474 | -0.03330681 | 0.00028121 | 0.01064332 | 0.00232753 | -0.00069277 | 0.00057542 | 0.00033219 | -0.00005791 | -0.00001274 | -0.00000072 | -0.00005948 | -0.00004960 | -0.00008569 | -0.00011695 |
| 22 | 28 | 0.03604608 | 0.06083186 | 0.01238731 | -0.02996376 | -0.01490300 | 0.00781800 | 0.00819061 | 0.00127332 | -0.00000109 | 0.00035885 | 0.00023555 | -0.00034930 | 0.00003332 | -0.00003496 | 0.00008739 | -0.00001908 | 0.00014401 |
| 22 | 29 | 0.03607167 | 0.06008455 | 0.01130538 | -0.02937485 | -0.01319573 | 0.00839047 | 0.00746798 | 0.00116470 | 0.00002536 | 0.00031575 | 0.00015127 | -0.00025765 | -0.00001539 | -0.00000438 | 0.00005139 | -0.00003749 | 0.00004293 |
| 22 | 30 | 0.03756768 | 0.06014191 | 0.00594607 | -0.03313443 | -0.01355017 | 0.00936436 | 0.00749446 | 0.00060939 | -0.00013789 | 0.00038208 | -0.00003271 | -0.00008664 | -0.00024318 | 0.00009342 | -0.00018312 | 0.00024103 | 0.00005612 |
| 22 | 31 | 0.03721139 | 0.05925145 | 0.00530204 | -0.03292019 | -0.01245192 | 0.00986136 | 0.00695506 | 0.00039226 | 0.00000460 | 0.00035080 | -0.00007480 | -0.00003891 | -0.00025133 | 0.00006468 | -0.00019603 | 0.00023804 | 0.00002515 |
| 22 | 32 | 0.03688905 | 0.05854745 | 0.00465546 | -0.03297590 | -0.01277082 | 0.00921613 | 0.00713148 | 0.00019573 | -0.00016637 | 0.00028862 | -0.00000331 | -0.00011990 | -0.00019078 | 0.00003835 | -0.00017457 | 0.00029105 | 0.00015339 |
| 22 | 33 | 0.03628533 | 0.05750858 | 0.00430416 | -0.03271436 | -0.01172954 | 0.00968894 | 0.00634189 | -0.00022000 | 0.00006187 | 0.00032211 | -0.00006451 | -0.00005559 | -0.00021046 | 0.00000710 | -0.00019110 | 0.00031416 | 0.00013335 |
| 22 | 34 | 0.03634535 | 0.05688597 | 0.00324586 | -0.03218183 | -0.01126528 | 0.00895059 | 0.00629537 | -0.00026468 | -0.00015897 | 0.00019409 | 0.00000077 | -0.00012612 | -0.00013530 | -0.00001381 | -0.00016295 | 0.00032786 | 0.00021485 |
| 22 | 35 | 0.03560371 | 0.05571039 | 0.00303490 | -0.03176897 | -0.01005937 | 0.00946869 | 0.00520818 | -0.00086595 | 0.00016122 | 0.00028550 | -0.00008561 | -0.00003065 | -0.00018053 | -0.00004759 | -0.00019521 | 0.00038990 | 0.00019739 |
| 22 | 36 | 0.03592143 | 0.05528949 | 0.00171390 | -0.03121257 | -0.00946566 | 0.00864609 | 0.00517022 | -0.00078779 | -0.00012346 | 0.00010101 | -0.00001369 | -0.00010886 | -0.00008731 | -0.00006975 | -0.00015829 | 0.00037184 | 0.00025777 |
| 22 | 37 | 0.03510983 | 0.05399182 | 0.00160845 | -0.03052850 | -0.00799095 | 0.00922971 | 0.00382623 | -0.00153194 | 0.00028829 | 0.00023234 | -0.00011711 | 0.00002254 | -0.00016424 | -0.00011547 | -0.00021484 | 0.00048420 | 0.00024389 |
| 22 | 38 | 0.03566821 | 0.05381128 | 0.00006151 | -0.03011804 | -0.00745112 | 0.00833300 | 0.00384744 | -0.00134207 | -0.00005915 | 0.00001311 | -0.00004279 | -0.00006618 | -0.00005481 | -0.00012886 | -0.00016628 | 0.00042938 | 0.00028365 |
| 22 | 39 | 0.03486251 | 0.05241285 | 0.00003896 | -0.02903565 | -0.00563793 | 0.00897788 | 0.00231039 | -0.00217464 | 0.00043956 | 0.00016476 | -0.00015247 | 0.00010509 | -0.00016949 | -0.00019616 | -0.00025600 | 0.00060211 | 0.00027455 |
| 22 | 40 | 0.03527029 | 0.05230542 | -0.00129556 | -0.02968911 | -0.00677814 | 0.00786706 | 0.00288908 | -0.00153746 | -0.00011242 | 0.00005834 | -0.00013630 | 0.00008451 | -0.00015142 | -0.00005729 | -0.00023166 | 0.00047526 | 0.00018691 |
| 22 | 41 | 0.03430113 | 0.05069366 | -0.00163677 | -0.03002954 | -0.00642141 | 0.00836172 | 0.00179281 | -0.00200116 | 0.00021123 | 0.00025362 | -0.00028025 | 0.00032160 | -0.00036166 | -0.00000749 | -0.00033115 | 0.00057000 | 0.00005985 |
| 22 | 42 | 0.03480225 | 0.05038192 | -0.00352831 | -0.03040341 | -0.00656931 | 0.00770330 | 0.00235278 | -0.00166587 | -0.00036830 | 0.00013686 | -0.00021235 | 0.00025069 | -0.00027866 | 0.00007256 | -0.00029426 | 0.00044862 | 0.00003458 |
| 22 | 43 | 0.03405452 | 0.04865486 | -0.00487798 | -0.03131014 | -0.00556973 | 0.00859589 | 0.00148865 | -0.00209981 | -0.00008442 | 0.00015406 | -0.00028137 | 0.00041552 | -0.00042784 | 0.00005447 | -0.00033476 | 0.00047668 | -0.00005449 |
| 23 | 16 | 0.04742067 | 0.06431245 | 0.00393024 | -0.01951473 | 0.00005503 | 0.00641977 | 0.00049739 | -0.00149585 | 0.00048084 | 0.00054250 | 0.00014922 | -0.00003997 | -0.00001936 | -0.00010206 | 0.00001175 | 0.00002789 | 0.00003989 |
| 23 | 17 | 0.04812115 | 0.06323244 | 0.00004444 | -0.02103741 | 0.00142366 | 0.00826729 | 0.00246187 | -0.00067250 | 0.00014629 | 0.00029167 | 0.00018857 | -0.00004747 | -0.00003576 | -0.00007220 | -0.00003444 | 0.00004871 | 0.00004769 |
| 23 | 18 | 0.04925517 | 0.06314196 | -0.00325255 | -0.02296476 | 0.00264159 | 0.00821895 | 0.00138724 | -0.00106463 | 0.00057080 | 0.00030045 | 0.00009097 | -0.00001528 | 0.00000910 | -0.00012298 | 0.00000347 | -0.00006027 | 0.00000688 |
| 23 | 19 | 0.04976995 | 0.06208669 | -0.00731131 | -0.02514243 | 0.00300933 | 0.00926182 | 0.00291568 | -0.00045024 | 0.00028748 | 0.00012344 | 0.00015547 | -0.00006625 | 0.00005166 | -0.00010935 | -0.00001367 | -0.00006662 | 0.00004904 |
| 23 | 20 | 0.05087982 | 0.06221351 | -0.01042384 | -0.02688749 | 0.00438531 | 0.00938060 | 0.00189765 | -0.00062817 | 0.00083278 | 0.00008988 | 0.00002822 | -0.00001011 | 0.00007692 | -0.00016279 | 0.00001196 | -0.00015395 | 0.00000890 |
| 23 | 21 | 0.05018630 | 0.06106289 | -0.01106282 | -0.02778572 | 0.00304995 | 0.00948637 | 0.00266974 | -0.00046070 | 0.00036449 | 0.00008031 | 0.00010608 | -0.00003733 | 0.00003823 | -0.00011566 | -0.00004032 | -0.00004036 | 0.00004697 |
| 23 | 22 | 0.05041744 | 0.06092696 | -0.01144805 | -0.02791462 | 0.00405936 | 0.00953317 | 0.00146832 | -0.00075356 | 0.00076987 | 0.00006805 | -0.00000268 | 0.00003515 | -0.00001116 | -0.00016418 | -0.00005080 | -0.00000235 | 0.00001681 |
| 23 | 23 | 0.04960686 | 0.05994974 | -0.01168218 | -0.02891708 | 0.00273777 | 0.00972270 | 0.00187692 | -0.00080255 | 0.00036419 | 0.00020483 | 0.00003261 | 0.00002311 | -0.00004152 | -0.00008965 | -0.00009612 | 0.00006818 | -0.00000899 |
| 23 | 24 | 0.04978896 | 0.05973326 | -0.01209726 | -0.02871862 | 0.00401301 | 0.00953297 | 0.00059604 | -0.00101794 | 0.00082077 | 0.00012539 | -0.00006323 | 0.00005144 | -0.00003135 | -0.00015928 | -0.00006104 | 0.00005767 | -0.00001641 |
| 23 | 25 | 0.04521420 | 0.05960457 | -0.00469923 | -0.03057170 | -0.00251362 | 0.00949743 | 0.00342365 | -0.00051718 | 0.00037810 | 0.00052319 | 0.00015573 | -0.00019656 | 0.00012150 | -0.00008895 | 0.00006775 | -0.00013666 | -0.00004963 |
| 23 | 26 | 0.04682290 | 0.05936553 | -0.00931885 | -0.03306278 | -0.00123974 | 0.01006870 | 0.00269681 | -0.00085361 | 0.00062326 | 0.00041379 | 0.00006477 | -0.00010713 | 0.00005616 | -0.00012846 | 0.00002093 | -0.00011658 | -0.00006526 |
| 23 | 27 | 0.03733461 | 0.06199566 | 0.01132892 | -0.03138133 | -0.01467466 | 0.00808203 | 0.00789040 | 0.00126997 | 0.00013790 | 0.00038535 | 0.00028127 | -0.00038522 | 0.00011142 | -0.00008558 | 0.00013172 | -0.00013218 | 0.00006803 |
| 23 | 28 | 0.03729663 | 0.06177725 | 0.01055132 | -0.03258115 | -0.01496686 | 0.00846981 | 0.00756978 | 0.00093005 | -0.00003768 | 0.00023296 | 0.00020913 | -0.00023163 | -0.00008559 | -0.00003929 | 0.00004606 | -0.00006273 | -0.00001122 |
| 23 | 29 | 0.03891955 | 0.06164986 | 0.00557220 | -0.03366730 | -0.01310749 | 0.00982287 | 0.00734678 | 0.00110891 | 0.00006799 | 0.00037630 | 0.00000967 | -0.00010449 | -0.00017743 | 0.00003146 | -0.00014389 | 0.00012148 | -0.00005540 |
| 23 | 30 | 0.03850595 | 0.06092282 | 0.00567588 | -0.03312184 | -0.01237810 | 0.00992218 | 0.00695427 | 0.00059134 | -0.00000421 | 0.00031716 | -0.00004667 | -0.00005364 | -0.00023694 | 0.00006820 | -0.00015257 | 0.00014607 | -0.00003483 |
| 23 | 31 | 0.03834355 | 0.06000564 | 0.00421315 | -0.03331182 | -0.01198105 | 0.01013664 | 0.00689193 | 0.00050927 | 0.00009718 | 0.00042193 | -0.00000781 | -0.00009028 | -0.00017797 | 0.00000967 | -0.00017835 | 0.00023125 | 0.00004026 |
| 23 | 32 | 0.03776958 | 0.05914687 | 0.00435856 | -0.03269754 | -0.01131961 | 0.00985108 | 0.00643840 | 0.00004601 | 0.00001171 | 0.00027950 | -0.00003602 | -0.00008408 | -0.00017561 | 0.00001494 | -0.00014552 | 0.00021579 | 0.00008645 |
| 23 | 33 | 0.03778818 | 0.05822702 | 0.00280928 | -0.03242947 | -0.01023317 | 0.01027089 | 0.00594485 | -0.00019556 | 0.00017263 | 0.00044912 | -0.00003992 | -0.00005018 | -0.00017861 | -0.00001862 | -0.00020673 | 0.00033035 | 0.00010726 |
| 23 | 34 | 0.03712339 | 0.05730569 | 0.00289478 | -0.03181656 | -0.00965882 | 0.00964371 | 0.00544347 | -0.00058796 | 0.00007193 | 0.00023097 | -0.00004632 | -0.00008340 | -0.00011854 | -0.00004385 | -0.00014057 | 0.00028304 | 0.00017665 |
| 23 | 35 | 0.03728330 | 0.05645531 | 0.00140700 | -0.03115435 | -0.00809067 | 0.01028517 | 0.00469017 | -0.00094193 | 0.00030276 | 0.00045571 | -0.00008059 | 0.00000950 | -0.00018200 | -0.00006444 | -0.00023966 | 0.00044147 | 0.00016521 |
| 23 | 36 | 0.03659670 | 0.05552509 | 0.00135016 | -0.03063603 | -0.00764001 | 0.00936728 | 0.00415208 | -0.00126541 | 0.00017079 | 0.00017194 | -0.00007031 | -0.00005389 | -0.00007434 | -0.00011366 | -0.00014819 | 0.00036582 | 0.00024791 |
| 23 | 37 | 0.03692510 | 0.05474178 | -0.00005772 | -0.02952218 | -0.00562226 | 0.01018848 | 0.00322118 | -0.00167793 | 0.00047700 | 0.00043804 | -0.00012341 | 0.00009308 | -0.00019841 | -0.00012905 | -0.00028655 | 0.00057332 | 0.00021293 |
| 23 | 38 | 0.03627568 | 0.05386918 | -0.00029004 | -0.02918450 | -0.00534380 | 0.00905162 | 0.00267708 | -0.00193506 | 0.00030176 | 0.00010388 | -0.00010232 | 0.00000822 | -0.00005444 | -0.00019359 | -0.00017729 | 0.00047241 | 0.00029963 |
| 23 | 39 | 0.03678788 | 0.05312355 | -0.00161279 | -0.02755193 | -0.00290577 | 0.00996948 | 0.00163171 | -0.00235123 | 0.00068827 | 0.00039437 | -0.00016191 | 0.00020295 | -0.00023620 | -0.00021167 | -0.00035313 | 0.00072845 | 0.00024834 |
| 23 | 40 | 0.03590270 | 0.05225694 | -0.00197192 | -0.02897351 | -0.00460778 | 0.00850957 | 0.00153767 | -0.00216574 | 0.00030705 | 0.00014401 | -0.00019874 | 0.00016762 | -0.00015705 | -0.00013169 | -0.00023839 | 0.00052137 | 0.00020387 |
| 23 | 41 | 0.03624018 | 0.05150387 | -0.00359210 | -0.02917668 | -0.00367862 | 0.00950103 | 0.00113588 | -0.00212267 | 0.00051727 | 0.00040200 | -0.00026369 | 0.00037401 | -0.00038834 | -0.00006397 | -0.00038592 | 0.00065674 | 0.00005117 |
| 23 | 42 | 0.03553555 | 0.05027995 | -0.00488661 | -0.03046039 | -0.00435434 | 0.00845534 | 0.00099990 | -0.00220592 | 0.00009687 | 0.00015528 | -0.00026309 | 0.00031687 | -0.00027756 | -0.00001108 | -0.00026444 | 0.00044779 | 0.00004245 |
| 23 | 43 | 0.03592758 | 0.04955344 | -0.00690483 | -0.03096468 | -0.00298776 | 0.00977717 | 0.00093497 | -0.00208366 | 0.00028680 | 0.00022191 | -0.00024572 | 0.00042708 | -0.00042297 | -0.00003196 | -0.00035018 | 0.00052469 | -0.00004761 |
| 23 | 44 | 0.03537047 | 0.04816385 | -0.00877463 | -0.03208775 | -0.00285480 | 0.00913950 | 0.00088356 | -0.00235248 | -0.00015095 | -0.00003278 | -0.00020542 | 0.00034779 | -0.00029296 | -0.00000395 | -0.00022531 | 0.00032766 | -0.00001368 |
| 24 | 17 | 0.04946944 | 0.06438749 | -0.00036901 | -0.02090644 | 0.00209652 | 0.00699077 | 0.00065326 | -0.00149420 | 0.00065989 | 0.00045807 | 0.00015537 | -0.00002505 | 0.00003108 | -0.00015880 | 0.00001738 | -0.00003404 | 0.00005351 |
| 24 | 18 | 0.05108487 | 0.06367657 | -0.00624936 | -0.02442241 | 0.00328304 | 0.00688437 | -0.00035423 | -0.00145519 | 0.00104931 | 0.00045276 | 0.00004637 | -0.00000998 | 0.00006805 | -0.00016025 | 0.00006224 | -0.00009566 | 0.00001547 |
| 24 | 19 | 0.05152331 | 0.06312045 | -0.00868656 | -0.02503643 | 0.00458349 | 0.00842977 | 0.00134758 | -0.00112978 | 0.00085343 | 0.00023639 | 0.00010394 | -0.00003109 | 0.00011824 | -0.00020386 | 0.00003368 | -0.00015445 | 0.00005180 |
| 24 | 20 | 0.05278084 | 0.06261270 | -0.01341080 | -0.02812153 | 0.00528426 | 0.00830690 | 0.00034560 | -0.00099992 | 0.00122790 | 0.00023087 | -0.00001538 | 0.00001113 | 0.00010823 | -0.00019477 | 0.00004785 | -0.00015904 | 0.00001362 |
| 24 | 21 | 0.05221716 | 0.06200140 | -0.01312967 | -0.02782521 | 0.00520217 | 0.00911270 | 0.00136291 | -0.00101143 | 0.00094180 | 0.00012139 | 0.00006048 | -0.00000836 | 0.00010814 | -0.00022766 | 0.00000436 | -0.00012118 | 0.00006165 |
| 24 | 22 | 0.05272834 | 0.06127997 | -0.01570106 | -0.03033808 | 0.00494382 | 0.00875834 | 0.00003776 | -0.00118415 | 0.00117490 | 0.00020993 | -0.00001750 | 0.00002199 | 0.00005423 | -0.00021681 | 0.00000525 | -0.00005875 | 0.00000543 |
| 24 | 23 | 0.05184462 | 0.06077797 | -0.01457013 | -0.03001514 | 0.00451596 | 0.00951724 | 0.00072016 | -0.00134066 | 0.00088829 | 0.00016079 | 0.00002071 | 0.00002684 | 0.00002807 | -0.00022841 | -0.00004747 | -0.00000429 | 0.00002445 |
| 24 | 24 | 0.04652055 | 0.06102059 | -0.00474922 | -0.03161620 | -0.00322464 | 0.00690700 | 0.00171847 | -0.00091157 | 0.00114923 | 0.00054790 | 0.00019196 | -0.00037835 | 0.00037322 | -0.00026052 | 0.00029695 | -0.00031064 | 0.00009518 |
| 24 | 25 | 0.04538202 | 0.06095522 | -0.00314854 | -0.03253703 | -0.00465482 | 0.00854830 | 0.00326501 | -0.00088114 | 0.00071308 | 0.00045044 | 0.00022098 | -0.00027027 | 0.00018511 | -0.00022200 | 0.00016295 | -0.00020045 | 0.00003907 |
| 24 | 26 | 0.03913724 | 0.06346926 | 0.00998852 | -0.03259052 | -0.01441780 | 0.00717122 | 0.00680377 | 0.00039796 | 0.00037918 | 0.00051764 | 0.00021716 | -0.00044833 | 0.00020882 | -0.00009637 | 0.00021946 | -0.00020097 | 0.00015563 |
| 24 | 27 | 0.03887373 | 0.06337415 | 0.01016001 | -0.03298613 | -0.01448257 | 0.00839869 | 0.00715085 | 0.00061130 | 0.00015669 | 0.00037175 | 0.00017396 | -0.00025635 | -0.00002642 | -0.00004319 | 0.00007916 | -0.00011642 | -0.00001305 |
| 24 | 28 | 0.03991130 | 0.06341442 | 0.00714608 | -0.03329927 | -0.01326411 | 0.00924257 | 0.00684125 | 0.00076662 | 0.00011591 | 0.00045988 | -0.00006372 | -0.00006376 | -0.00023541 | 0.00012137 | -0.00009487 | 0.00007026 | -0.00009063 |
| 24 | 29 | 0.03970581 | 0.06269796 | 0.00647972 | -0.03314019 | -0.01235102 | 0.00985283 | 0.00671651 | 0.00065294 | 0.00003125 | 0.00033212 | -0.00004076 | -0.00004700 | -0.00023593 | 0.00008011 | -0.00011252 | 0.00006497 | -0.00010647 |
| 24 | 30 | 0.03957443 | 0.06173560 | 0.00531843 | -0.03271130 | -0.01174554 | 0.00944144 | 0.00666547 | 0.00048979 | 0.00002370 | 0.00033127 | -0.00003053 | -0.00011314 | -0.00016307 | 0.00006920 | -0.00008724 | 0.00012372 | 0.00003017 |
| 24 | 31 | 0.03921523 | 0.06088002 | 0.00470901 | -0.03264984 | -0.01096621 | 0.00999484 | 0.00633127 | 0.00019060 | -0.00000611 | 0.00027331 | -0.00002671 | -0.00008584 | -0.00016785 | 0.00003028 | -0.00011051 | 0.00013656 | 0.00002318 |
| 24 | 32 | 0.03917604 | 0.05997010 | 0.00343517 | -0.03201351 | -0.01007234 | 0.00937825 | 0.00612577 | 0.00008391 | -0.00004973 | 0.00018715 | -0.00000097 | -0.00015323 | -0.00007561 | 0.00000325 | -0.00006918 | 0.00016182 | 0.00013767 |
| 24 | 33 | 0.03867779 | 0.05898305 | 0.00290893 | -0.03188865 | -0.00925340 | 0.00988745 | 0.00550429 | -0.00040661 | 0.00000267 | 0.00020088 | -0.00002362 | -0.00010563 | -0.00009192 | -0.00003309 | -0.00010181 | 0.00019940 | 0.00013527 |
| 24 | 34 | 0.03878631 | 0.05823401 | 0.00153884 | -0.03113196 | -0.00821613 | 0.00913553 | 0.00524774 | -0.00041689 | -0.00008929 | 0.00005091 | 0.00001060 | -0.00017435 | 0.00001349 | -0.00006617 | -0.00005173 | 0.00019750 | 0.00022781 |
| 24 | 35 | 0.03817257 | 0.05713000 | 0.00114175 | -0.03079049 | -0.00723564 | 0.00962076 | 0.00433756 | -0.00107163 | 0.00006269 | 0.00012475 | -0.00003603 | -0.00009989 | -0.00002170 | -0.00010668 | -0.00009949 | 0.00027079 | 0.00023035 |
| 24 | 36 | 0.03847385 | 0.05658753 | -0.00036654 | -0.03009058 | -0.00622577 | 0.00876483 | 0.00410058 | -0.00097410 | -0.00009159 | -0.00006949 | 0.00000421 | -0.00017010 | 0.00009071 | -0.00013576 | -0.00004350 | 0.00024060 | 0.00029743 |
| 24 | 37 | 0.03780475 | 0.05538788 | -0.00062571 | -0.02938370 | -0.00497438 | 0.00925134 | 0.00293776 | -0.00174702 | 0.00016580 | 0.00004693 | -0.00005997 | -0.00006136 | 0.00002630 | -0.00018832 | -0.00011567 | 0.00036298 | 0.00030474 |
| 24 | 38 | 0.03829896 | 0.05507914 | -0.00226520 | -0.02889981 | -0.00415691 | 0.00830740 | 0.00277906 | -0.00154068 | -0.00005763 | -0.00017252 | -0.00001592 | -0.00013776 | 0.00014514 | -0.00020404 | -0.00005114 | 0.00029843 | 0.00034593 |
| 24 | 39 | 0.03765042 | 0.05380328 | -0.00239650 | -0.02768959 | -0.00255425 | 0.00880426 | 0.00142259 | -0.00237471 | 0.00030412 | -0.00003338 | -0.00008831 | 0.00001206 | 0.00004053 | -0.00027655 | -0.00015750 | 0.00048157 | 0.00035711 |
| 24 | 40 | 0.03798120 | 0.05348734 | -0.00416975 | -0.02874228 | -0.00326719 | 0.00783964 | 0.00165745 | -0.00182645 | -0.00006574 | -0.00011243 | -0.00010352 | -0.00000415 | 0.00007581 | -0.00014566 | -0.00010095 | 0.00033430 | 0.00026806 |
| 24 | 41 | 0.03716209 | 0.05195877 | -0.00489679 | -0.02933098 | -0.00289343 | 0.00843038 | 0.00061036 | -0.00236526 | 0.00019498 | 0.00009379 | -0.00021530 | 0.00020564 | -0.00011952 | -0.00010450 | -0.00020136 | 0.00043192 | 0.00016034 |
| 24 | 42 | 0.03755686 | 0.05147174 | -0.00707391 | -0.03004824 | -0.00271519 | 0.00791772 | 0.00098854 | -0.00204841 | -0.00020139 | -0.00001322 | -0.00017302 | 0.00013886 | -0.00002766 | -0.00003158 | -0.00013508 | 0.00029153 | 0.00013757 |
| 24 | 43 | 0.03691148 | 0.04986781 | -0.00859091 | -0.03117521 | -0.00185857 | 0.00892389 | 0.00028195 | -0.00249741 | 0.00001055 | 0.00001364 | -0.00021469 | 0.00028742 | -0.00018374 | -0.00004470 | -0.00018403 | 0.00032990 | 0.00006044 |
| 24 | 44 | 0.03728386 | 0.04940122 | -0.01060932 | -0.03148345 | -0.00131711 | 0.00857207 | 0.00074707 | -0.00231638 | -0.00037763 | -0.00008831 | -0.00014565 | 0.00019524 | -0.00006532 | -0.00000096 | -0.00012029 | 0.00021241 | 0.00008476 |
| 24 | 45 | 0.03678241 | 0.04782433 | -0.01260020 | -0.03292500 | -0.00011848 | 0.00988731 | 0.00040192 | -0.00257904 | -0.00020179 | -0.00024768 | -0.00012504 | 0.00028850 | -0.00019251 | -0.00006296 | -0.00012741 | 0.00020815 | 0.00003673 |
| 24 | 46 | 0.03709270 | 0.04748470 | -0.01410025 | -0.03277649 | 0.00038442 | 0.00947150 | 0.00082766 | -0.00246559 | -0.00055595 | -0.00030676 | -0.00006633 | 0.00021087 | -0.00009047 | -0.00001587 | -0.00008560 | 0.00013054 | 0.00007229 |
| 25 | 18 | 0.05223687 | 0.06321134 | -0.00951208 | -0.02598866 | 0.00369403 | 0.00762355 | -0.00006837 | -0.00201531 | 0.00059975 | 0.00040760 | 0.00013076 | 0.00004611 | -0.00001055 | -0.00018074 | -0.00002529 | -0.00005085 | 0.00000203 |
| 25 | 19 | 0.05264188 | 0.06225005 | -0.01273918 | -0.02730883 | 0.00431442 | 0.00895070 | 0.00192951 | -0.00125565 | 0.00022693 | 0.00017000 | 0.00022671 | -0.00001540 | 0.00002079 | -0.00016804 | -0.00004462 | -0.00005136 | 0.00005940 |
| 25 | 20 | 0.05388219 | 0.06250317 | -0.01598984 | -0.02918868 | 0.00552952 | 0.00872967 | 0.00076496 | -0.00146485 | 0.00080226 | 0.00014814 | 0.00009665 | 0.00002419 | 0.00008096 | -0.00023660 | -0.00000531 | -0.00014795 | 0.00003504 |
| 25 | 21 | 0.05306277 | 0.06131641 | -0.01642717 | -0.03076504 | 0.00403287 | 0.00968254 | 0.00212741 | -0.00124049 | 0.00028133 | 0.00014327 | 0.00020948 | -0.00001435 | 0.00001549 | -0.00018828 | -0.00006366 | -0.00005038 | 0.00003527 |
| 25 | 22 | 0.05362971 | 0.06115758 | -0.01781667 | -0.03163079 | 0.00486864 | 0.00944683 | 0.00047113 | -0.00167291 | 0.00073953 | 0.00019065 | 0.00007501 | 0.00006443 | -0.00002746 | -0.00023283 | -0.00007539 | -0.00001043 | -0.00000283 |
| 25 | 23 | 0.04529838 | 0.06175207 | -0.00154890 | -0.03290127 | -0.00667899 | 0.00794555 | 0.00441576 | -0.00062819 | 0.00032063 | 0.00049213 | 0.00040058 | -0.00038256 | 0.00024634 | -0.00019937 | 0.00023371 | -0.00032104 | 0.00002610 |
| 25 | 24 | 0.04476463 | 0.06164854 | -0.00060031 | -0.03218619 | -0.00635146 | 0.00695745 | 0.00310909 | -0.00088193 | 0.00078291 | 0.00052371 | 0.00032602 | -0.00044866 | 0.00037071 | -0.00026649 | 0.00033247 | -0.00037444 | 0.00010229 |
| 25 | 25 | 0.04162287 | 0.06450879 | 0.00631840 | -0.03355536 | -0.01166358 | 0.00891202 | 0.00677935 | 0.00033050 | 0.00039962 | 0.00069400 | 0.00017931 | -0.00050096 | 0.00033425 | -0.00007069 | 0.00023982 | -0.00037694 | 0.00003394 |
| 25 | 26 | 0.04118259 | 0.06455231 | 0.00696096 | -0.03436051 | -0.01310547 | 0.00844852 | 0.00642232 | 0.00020033 | 0.00036390 | 0.00058901 | 0.00011145 | -0.00035417 | 0.00014444 | -0.00004281 | 0.00015903 | -0.00026960 | -0.00001202 |
| 25 | 27 | 0.04162172 | 0.06492658 | 0.00549999 | -0.03459139 | -0.01250003 | 0.00988920 | 0.00729963 | 0.00098873 | 0.00035587 | 0.00059462 | 0.00002305 | -0.00029369 | 0.00006809 | 0.00001429 | 0.00004396 | -0.00015390 | -0.00005179 |
| 25 | 28 | 0.04153679 | 0.06478345 | 0.00554055 | -0.03449712 | -0.01240850 | 0.01001902 | 0.00661482 | 0.00080584 | 0.00023577 | 0.00048312 | -0.00010620 | -0.00006860 | -0.00019059 | 0.00010442 | -0.00007593 | -0.00004099 | -0.00018084 |
| 25 | 29 | 0.04143938 | 0.06390449 | 0.00409809 | -0.03437499 | -0.01154603 | 0.01049992 | 0.00709252 | 0.00107266 | 0.00029588 | 0.00049210 | -0.00003849 | -0.00018440 | -0.00005639 | 0.00004445 | -0.00004600 | -0.00002761 | -0.00007258 |
| 25 | 30 | 0.04122510 | 0.06307861 | 0.00366132 | -0.03401029 | -0.01072389 | 0.01052804 | 0.00640499 | 0.00047312 | 0.00021875 | 0.00042677 | -0.00010272 | -0.00010277 | -0.00014249 | 0.00007298 | -0.00007837 | 0.00002883 | -0.00006561 |
| 25 | 31 | 0.04115852 | 0.06220470 | 0.00220377 | -0.03382256 | -0.00971986 | 0.01099765 | 0.00652150 | 0.00051702 | 0.00029563 | 0.00049787 | -0.00006005 | -0.00016983 | -0.00004882 | 0.00003045 | -0.00007128 | 0.00005661 | 0.00000324 |
| 25 | 32 | 0.04078438 | 0.06126222 | 0.00180758 | -0.03332058 | -0.00884319 | 0.01073664 | 0.00580352 | -0.00000991 | 0.00022740 | 0.00034595 | -0.00010070 | -0.00013005 | -0.00007508 | 0.00002377 | -0.00006628 | 0.00008297 | 0.00004208 |
| 25 | 33 | 0.04076831 | 0.06048790 | 0.00049085 | -0.03287575 | -0.00763893 | 0.01123364 | 0.00558624 | -0.00014277 | 0.00035125 | 0.00049358 | -0.00009151 | -0.00014764 | -0.00002603 | 0.00000215 | -0.00008651 | 0.00013747 | 0.00007881 |

TABLE 3. Nuclear charge density distribution FB coefficients.

| $Z$ | $N$ | $a_1$ | $a_2$ | $a_3$ | $a_4$ | $a_5$ | $a_6$ | $a_7$ | $a_8$ | $a_9$ | $a_{10}$ | $a_{11}$ | $a_{12}$ | $a_{13}$ | $a_{14}$ | $a_{15}$ | $a_{16}$ | $a_{17}$ |
|---|---|---|---|---|---|---|---|---|---|---|---|---|---|---|---|---|---|---|
| 25 | 34 | 0.04029232 | 0.05946645 | 0.00005271 | -0.03230316 | -0.00675153 | 0.01068243 | 0.00482800 | -0.00059547 | 0.00027219 | 0.00026167 | -0.00011231 | -0.00014243 | 0.00000129 | -0.00003475 | -0.00005049 | 0.00013489 | 0.00013948 |
| 25 | 35 | 0.04038221 | 0.05880057 | -0.00113810 | -0.03155610 | -0.00533589 | 0.01121996 | 0.00433864 | -0.00086000 | 0.00045075 | 0.00047753 | -0.00013032 | -0.00010932 | -0.00000223 | -0.00003914 | -0.00010339 | 0.00022637 | 0.00014890 |
| 25 | 36 | 0.03986402 | 0.05775340 | -0.00167419 | -0.03100770 | -0.00451712 | 0.01040593 | 0.00355524 | -0.00123760 | 0.00034365 | 0.00017917 | -0.00013563 | -0.00013010 | 0.00006839 | -0.00009887 | -0.00004436 | 0.00019780 | 0.00022089 |
| 25 | 37 | 0.04011333 | 0.05717819 | -0.00277931 | -0.02990386 | -0.00286769 | 0.01097058 | 0.00286457 | -0.00157730 | 0.00058263 | 0.00044658 | -0.00017092 | -0.00004784 | 0.00000797 | -0.00009233 | -0.00013280 | 0.00033224 | 0.00020860 |
| 25 | 38 | 0.03960052 | 0.05617228 | -0.00341910 | -0.02947746 | -0.00221972 | 0.00994445 | 0.00209644 | -0.00187865 | 0.00043419 | 0.00009934 | -0.00016473 | -0.00008740 | 0.00011057 | -0.00016663 | -0.00005841 | 0.00028107 | 0.00028317 |
| 25 | 39 | 0.04003689 | 0.05564362 | -0.00447480 | -0.02795957 | -0.00030216 | 0.01050265 | 0.00126741 | -0.00223683 | 0.00074203 | 0.00039689 | -0.00020604 | 0.00003938 | -0.00000557 | -0.00015768 | -0.00018066 | 0.00045802 | 0.00025473 |
| 25 | 40 | 0.03924855 | 0.05464256 | -0.00510898 | -0.02921833 | -0.00140009 | 0.00939507 | 0.00093697 | -0.00210916 | 0.00046056 | 0.00015472 | -0.00025869 | 0.00005128 | 0.00003630 | -0.00010944 | -0.00010785 | 0.00032019 | 0.00020136 |
| 25 | 41 | 0.03947845 | 0.05412287 | -0.00638353 | -0.02960365 | -0.00090413 | 0.01016421 | 0.00063486 | -0.00207269 | 0.00065864 | 0.00045474 | -0.00031424 | 0.00019685 | -0.00012712 | -0.00003028 | -0.00020971 | 0.00039565 | 0.00007089 |
| 25 | 42 | 0.03882478 | 0.05278998 | -0.00782118 | -0.03070310 | -0.00110627 | 0.00943284 | 0.00036538 | -0.00217241 | 0.00032969 | 0.00021440 | -0.00032667 | 0.00018364 | -0.00005427 | -0.00000418 | -0.00013079 | 0.00025588 | 0.00005807 |
| 25 | 43 | 0.03910573 | 0.05229225 | -0.00947701 | -0.03143304 | -0.00030170 | 0.01046796 | 0.00036794 | -0.00206689 | 0.00049818 | 0.00032867 | -0.00030485 | 0.00024444 | -0.00014747 | -0.00000601 | -0.00017758 | 0.00028011 | -0.00001636 |
| 25 | 44 | 0.03857148 | 0.05084568 | -0.01131682 | -0.03232065 | 0.00013843 | 0.01007303 | 0.00025710 | -0.00230505 | 0.00013707 | 0.00008087 | -0.00028093 | 0.00021207 | -0.00006274 | 0.00000576 | -0.00009870 | 0.00015293 | 0.00001017 |
| 25 | 45 | 0.03883821 | 0.05043153 | -0.01300622 | -0.03317016 | 0.00101689 | 0.01112083 | 0.00047705 | -0.00204824 | 0.00031310 | 0.00003991 | -0.00020721 | 0.00020117 | -0.00009549 | -0.00005630 | -0.00009627 | 0.00013527 | -0.00001754 |
| 25 | 46 | 0.03841047 | 0.04904511 | -0.01484539 | -0.03378885 | 0.00172939 | 0.01092186 | 0.00046736 | -0.00233704 | -0.00006971 | -0.00020715 | -0.00017298 | 0.00018670 | -0.00004587 | -0.00003578 | -0.00003975 | 0.00004368 | 0.00001590 |
| 25 | 47 | 0.03861338 | 0.04872846 | -0.01635458 | -0.03461622 | 0.00248668 | 0.01175182 | 0.00074229 | -0.00191859 | 0.00013900 | -0.00035369 | -0.00007726 | 0.00012415 | -0.00002930 | -0.00013428 | 0.00000238 | -0.00000775 | 0.00001797 |
| 25 | 48 | 0.03828878 | 0.04745762 | -0.01803957 | -0.03497098 | 0.00327137 | 0.01166266 | 0.00072106 | -0.00222991 | -0.00026343 | -0.00058634 | -0.00005319 | 0.00015613 | -0.00004603 | -0.00008978 | 0.00002000 | -0.00005136 | 0.00002868 |
| 26 | 19 | 0.05441632 | 0.06271046 | -0.01612947 | -0.02985148 | 0.00465881 | 0.00816389 | -0.00043549 | -0.00237761 | 0.00052857 | 0.00036750 | 0.00010996 | 0.00008598 | -0.00002206 | -0.00020355 | -0.00006130 | -0.00005076 | -0.00001086 |
| 26 | 20 | 0.05550952 | 0.06277674 | -0.02004361 | -0.03391054 | 0.00442090 | 0.00842911 | -0.00096309 | -0.00228722 | 0.00097228 | 0.00048307 | -0.00000616 | 0.00008002 | 0.00002413 | -0.00019942 | -0.00002960 | -0.00009425 | -0.00005420 |
| 26 | 21 | 0.05497712 | 0.06179160 | -0.01993993 | -0.03317012 | 0.00453014 | 0.00905393 | 0.00004607 | -0.00219980 | 0.00055928 | 0.00027044 | 0.00011086 | 0.00008259 | -0.00003797 | -0.00023306 | -0.00008457 | -0.00003936 | -0.00002203 |
| 26 | 22 | 0.04861762 | 0.06248373 | -0.00875207 | -0.03826297 | -0.00668660 | 0.00674060 | 0.00214386 | -0.00170083 | 0.00080478 | 0.00069199 | 0.00030234 | -0.00035555 | 0.00031460 | -0.00026324 | 0.00026893 | -0.00038130 | 0.00002025 |
| 26 | 23 | 0.04616801 | 0.06221112 | -0.00349531 | -0.03558385 | -0.00778529 | 0.00673316 | 0.00319842 | -0.00131003 | 0.00054652 | 0.00054843 | 0.00039148 | -0.00043015 | 0.00033997 | -0.00027256 | 0.00031472 | -0.00041836 | 0.00005948 |
| 26 | 24 | 0.04285287 | 0.06499875 | 0.00317044 | -0.03669728 | -0.01145489 | 0.00771659 | 0.00505555 | -0.00109343 | 0.00050820 | 0.00076208 | 0.00019150 | -0.00069519 | 0.00065313 | -0.00017378 | 0.00042707 | -0.00053622 | 0.00021458 |
| 26 | 25 | 0.04306031 | 0.06525151 | 0.00347809 | -0.03633474 | -0.01170535 | 0.00851734 | 0.00543925 | -0.00060220 | 0.00043428 | 0.00075069 | 0.00009010 | -0.00048434 | 0.00037580 | -0.00007429 | 0.00026583 | -0.00044037 | 0.00002856 |
| 26 | 26 | 0.04296829 | 0.06568705 | 0.00310556 | -0.03729952 | -0.01275530 | 0.00888716 | 0.00574109 | -0.00045374 | 0.00039932 | 0.00081902 | -0.00003115 | -0.00038764 | 0.00024482 | 0.00000342 | 0.00014542 | -0.00022307 | 0.00007095 |
| 26 | 27 | 0.04301465 | 0.06589071 | 0.00375004 | -0.03636567 | -0.01248135 | 0.00960473 | 0.00610173 | 0.00019211 | 0.00037947 | 0.00071285 | -0.00009468 | -0.00023197 | 0.00004346 | 0.00005837 | 0.00004131 | -0.00018454 | -0.00009424 |
| 26 | 28 | 0.04296124 | 0.06548171 | 0.00302952 | -0.03628658 | -0.01253363 | 0.00969007 | 0.00594919 | 0.00031961 | 0.00029527 | 0.00072628 | -0.00018502 | -0.00009575 | -0.00012064 | 0.00014149 | -0.00007244 | 0.00001074 | -0.00010406 |
| 26 | 29 | 0.04294690 | 0.06492970 | 0.00249328 | -0.03572249 | -0.01115090 | 0.01049771 | 0.00612493 | 0.00045325 | 0.00032603 | 0.00058372 | -0.00015911 | -0.00011260 | -0.00010498 | 0.00009769 | -0.00005686 | -0.00005111 | -0.00012944 |
| 26 | 30 | 0.04285742 | 0.06378127 | 0.00081585 | -0.03570897 | -0.01044513 | 0.01023355 | 0.00573506 | 0.00011601 | 0.00024513 | 0.00057300 | -0.00016526 | -0.00011501 | -0.00009047 | 0.00010387 | -0.00007253 | 0.00005362 | -0.00002977 |
| 26 | 31 | 0.04266653 | 0.06316642 | 0.00047225 | -0.03513374 | -0.00904989 | 0.01107672 | 0.00571858 | 0.00008713 | 0.00034397 | 0.00049449 | -0.00016216 | -0.00012859 | -0.00006929 | 0.00006377 | -0.00005465 | 0.00000492 | -0.00004111 |
| 26 | 32 | 0.04263119 | 0.06208278 | -0.00122888 | -0.03483961 | -0.00818683 | 0.01054563 | 0.00522225 | -0.00018180 | 0.00021696 | 0.00040998 | -0.00015382 | -0.00013690 | -0.00003313 | 0.00005476 | -0.00005734 | 0.00008187 | 0.00004412 |
| 26 | 33 | 0.04226012 | 0.06141415 | -0.00134342 | -0.03416966 | -0.00676116 | 0.01135755 | 0.00495508 | -0.00038585 | 0.00039329 | 0.00039633 | -0.00017513 | -0.00014172 | -0.00001020 | 0.00001788 | -0.00003921 | 0.00005054 | 0.00004577 |
| 26 | 34 | 0.04234982 | 0.06043720 | -0.00313631 | -0.03371923 | -0.00582348 | 0.01062215 | 0.00441348 | -0.00056285 | 0.00021377 | 0.00025332 | -0.00015490 | -0.00015209 | 0.00003821 | 0.00000100 | -0.00003500 | 0.00010302 | 0.00011082 |
| 26 | 35 | 0.04183775 | 0.05972853 | -0.00304196 | -0.03288775 | -0.00436683 | 0.01134196 | 0.00387386 | -0.00093634 | 0.00046254 | 0.00029836 | -0.00019797 | -0.00014157 | 0.00005553 | -0.00003455 | -0.00002293 | 0.00009730 | 0.00012448 |
| 26 | 36 | 0.04209144 | 0.05888737 | -0.00495904 | -0.03242665 | -0.00345592 | 0.01047186 | 0.00335934 | -0.00100043 | 0.00023315 | 0.00011351 | -0.00016790 | -0.00015301 | 0.00010867 | -0.00005239 | -0.00001466 | 0.00012588 | 0.00016587 |
| 26 | 37 | 0.04151334 | 0.05815463 | -0.00471726 | -0.03137099 | -0.00197312 | 0.01104507 | 0.00255941 | -0.00151891 | 0.00054082 | 0.00020573 | -0.00022653 | -0.00011991 | 0.00011021 | -0.00008950 | -0.00001794 | 0.00015593 | 0.00019017 |
| 26 | 38 | 0.04192913 | 0.05746852 | -0.00674031 | -0.03103437 | -0.00118259 | 0.01010940 | 0.00213668 | -0.00145880 | 0.00027023 | -0.00000547 | -0.00018864 | -0.00013542 | 0.00016537 | -0.00010263 | -0.00000388 | 0.00015721 | 0.00020705 |
| 26 | 39 | 0.04137084 | 0.05672221 | -0.00641941 | -0.02968716 | 0.00032640 | 0.01048881 | 0.00111434 | -0.00208348 | 0.00062231 | 0.00011829 | -0.00025366 | -0.00007347 | 0.00014080 | -0.00014577 | -0.00003173 | 0.00023152 | 0.00024018 |
| 26 | 40 | 0.04164174 | 0.05603493 | -0.00826423 | -0.03044427 | -0.00017709 | 0.00973787 | 0.00123064 | -0.00158035 | 0.00027070 | 0.00003091 | -0.00026640 | -0.00001842 | 0.00010107 | -0.00004867 | -0.00004857 | 0.00019080 | 0.00013503 |
| 26 | 41 | 0.04085199 | 0.05515849 | -0.00808365 | -0.03062622 | 0.00006913 | 0.01024886 | 0.00048695 | -0.00193571 | 0.00057062 | 0.00026891 | -0.00038293 | 0.00010892 | 0.00000467 | -0.00000110 | -0.00009317 | 0.00021172 | 0.00005637 |
| 26 | 42 | 0.04118714 | 0.05430253 | -0.01033089 | -0.03104680 | 0.00031975 | 0.00984116 | 0.00081232 | -0.00162494 | 0.00016085 | 0.00013718 | -0.00032894 | 0.00010772 | 0.00001930 | 0.00004332 | -0.00009633 | 0.00017507 | 0.00002436 |
| 26 | 43 | 0.04048405 | 0.05338038 | -0.01084334 | -0.03191065 | 0.00081066 | 0.01058162 | 0.00017665 | -0.00202740 | 0.00042425 | 0.00026422 | -0.00039580 | 0.00017895 | -0.00002197 | 0.00004442 | -0.00009593 | 0.00014076 | -0.00002785 |
| 26 | 44 | 0.04079789 | 0.05254820 | -0.01296356 | -0.03193810 | 0.00139691 | 0.01029418 | 0.00066183 | -0.00182006 | 0.00000414 | 0.00011915 | -0.00030463 | 0.00014100 | 0.00002554 | 0.00006088 | -0.00009181 | 0.00011690 | -0.00000412 |
| 26 | 45 | 0.04021768 | 0.05165739 | -0.01396707 | -0.03323823 | 0.00210347 | 0.01124817 | 0.00025319 | -0.00212129 | 0.00023903 | 0.00009581 | -0.00031912 | 0.00016237 | 0.00001513 | 0.00001825 | -0.00005246 | 0.00004288 | -0.00003038 |
| 26 | 46 | 0.04046614 | 0.05095391 | -0.01562218 | -0.03285570 | 0.00267174 | 0.01088781 | 0.00074527 | -0.00196688 | -0.00015791 | -0.00002288 | -0.00022984 | 0.00012685 | 0.00005677 | 0.00003720 | -0.00005884 | 0.00004554 | 0.00001190 |
| 26 | 47 | 0.04001494 | 0.05013173 | -0.01693982 | -0.03439823 | 0.00348764 | 0.01192369 | 0.00051090 | -0.00209595 | 0.00005202 | -0.00019763 | -0.00020388 | 0.00011481 | 0.00005621 | -0.00003750 | 0.00000886 | -0.00005482 | -0.00000115 |
| 26 | 48 | 0.04018716 | 0.04956457 | -0.01806026 | -0.03365029 | 0.00390653 | 0.01143559 | 0.00089569 | -0.00199195 | -0.00030425 | -0.00026095 | -0.00013997 | 0.00010646 | 0.00006932 | 0.00000067 | -0.00001900 | -0.00001972 | 0.00003413 |
| 26 | 49 | 0.03984465 | 0.04879772 | -0.01959253 | -0.03531112 | 0.00475335 | 0.01242011 | 0.00072474 | -0.00194861 | -0.00012136 | -0.00056446 | -0.00008743 | 0.00007338 | 0.00007107 | -0.00009483 | 0.00006859 | -0.00013978 | 0.00002288 |
| 26 | 50 | 0.03994701 | 0.04833937 | -0.02024871 | -0.03427538 | 0.00503695 | 0.01184869 | 0.00097089 | -0.00190895 | -0.00042931 | -0.00055876 | -0.00005539 | 0.00009773 | 0.00005203 | -0.00003424 | 0.00001763 | -0.00007384 | 0.00004299 |
| 27 | 20 | 0.05590741 | 0.06280794 | -0.02068529 | -0.03434919 | 0.00389978 | 0.00871986 | -0.00028508 | -0.00277775 | 0.00040619 | 0.00039128 | 0.00012699 | 0.00006330 | 0.00001583 | -0.00023684 | -0.00006757 | -0.00009129 | -0.00001998 |
| 27 | 21 | 0.04788921 | 0.06289952 | -0.00655169 | -0.03837679 | -0.00910056 | 0.00708790 | 0.00463410 | -0.00151399 | -0.00022419 | 0.00041076 | 0.00057964 | -0.00045469 | 0.00033639 | -0.00028521 | 0.00027500 | -0.00045480 | 0.00005935 |
| 27 | 22 | 0.04718256 | 0.06297690 | -0.00523901 | -0.03838009 | -0.00924130 | 0.00660040 | 0.00355922 | -0.00172554 | 0.00032723 | 0.00056708 | 0.00046642 | -0.00045734 | 0.00038205 | -0.00030544 | 0.00032116 | -0.00047250 | 0.00006572 |
| 27 | 23 | 0.04441022 | 0.06633369 | 0.00184588 | -0.03778152 | -0.01116888 | 0.00868310 | 0.00565501 | -0.00083750 | 0.00034647 | 0.00086716 | 0.00017598 | -0.00069708 | 0.00068794 | -0.00011977 | 0.00042204 | -0.00066673 | 0.00009490 |
| 27 | 24 | 0.04449791 | 0.06564756 | 0.00035323 | -0.03813153 | -0.00973128 | 0.00911738 | 0.00466062 | -0.00151017 | 0.00042120 | 0.00081422 | 0.00010856 | -0.00064073 | 0.00064641 | -0.00013728 | 0.00039007 | -0.00061176 | 0.00010578 |
| 27 | 25 | 0.04446166 | 0.06651299 | 0.00097833 | -0.03831890 | -0.01091054 | 0.00996995 | 0.00598877 | -0.00069074 | 0.00042340 | 0.00089628 | 0.00005820 | -0.00062622 | 0.00057123 | -0.00008316 | 0.00031053 | -0.00048834 | 0.00012766 |
| 27 | 26 | 0.04426122 | 0.06657515 | 0.00159316 | -0.03831464 | -0.01183830 | 0.00956534 | 0.00548826 | -0.00062620 | 0.00042048 | 0.00085058 | -0.00005948 | -0.00040581 | 0.00031560 | -0.00001064 | 0.00017013 | -0.00033463 | 0.00001447 |
| 27 | 27 | 0.04424069 | 0.06677769 | 0.00174420 | -0.03777047 | -0.01206967 | 0.00999200 | 0.00646072 | 0.00033027 | 0.00045440 | 0.00078281 | -0.00004507 | -0.00042066 | 0.00029501 | -0.00001221 | 0.00014595 | -0.00027568 | 0.00003150 |
| 27 | 28 | 0.04430736 | 0.06660643 | 0.00165416 | -0.03740834 | -0.01190923 | 0.01019310 | 0.00579998 | 0.00026180 | 0.00034435 | 0.00071794 | -0.00018544 | -0.00015376 | -0.00001018 | 0.00009068 | -0.00002474 | -0.00011210 | -0.00013256 |
| 27 | 29 | 0.04425695 | 0.06587468 | 0.00041006 | -0.03726507 | -0.01084788 | 0.01082186 | 0.00640510 | 0.00056014 | 0.00043583 | 0.00067220 | -0.00010664 | -0.00029722 | 0.00013875 | 0.00002069 | 0.00004695 | -0.00014435 | -0.00001251 |
| 27 | 30 | 0.04422913 | 0.06486598 | -0.00066108 | -0.03700122 | -0.00962629 | 0.01111116 | 0.00551299 | -0.00001935 | 0.00036817 | 0.00062905 | -0.00019955 | -0.00014169 | -0.00002378 | 0.00007771 | -0.00004264 | -0.00004974 | -0.00007865 |
| 27 | 31 | 0.04411468 | 0.06423404 | -0.00160787 | -0.03654838 | -0.00841059 | 0.01173891 | 0.00584480 | 0.00011336 | 0.00049719 | 0.00065203 | -0.00014729 | -0.00025386 | 0.00010322 | 0.00002597 | 0.00001706 | -0.00007238 | 0.00001832 |
| 27 | 32 | 0.04396586 | 0.06313149 | -0.00266760 | -0.03617048 | -0.00715210 | 0.01174313 | 0.00491067 | -0.00039660 | 0.00042272 | 0.00052313 | -0.00021911 | -0.00013760 | -0.00000298 | 0.00004839 | -0.00004007 | -0.00000257 | -0.00001772 |
| 27 | 33 | 0.04383755 | 0.06261780 | -0.00333769 | -0.03541341 | -0.00583983 | 0.01233624 | 0.00498500 | -0.00041349 | 0.00059076 | 0.00061789 | -0.00019198 | -0.00021392 | 0.00009275 | 0.00001710 | 0.00000207 | -0.00000764 | 0.00005682 |
| 27 | 34 | 0.04361289 | 0.06145637 | -0.00443833 | -0.03494404 | -0.00455960 | 0.01205410 | 0.00400440 | -0.00085137 | 0.00049788 | 0.00041093 | -0.00024520 | -0.00013266 | 0.00003991 | 0.00000859 | -0.00002656 | 0.00003786 | 0.00004427 |
| 27 | 35 | 0.04354464 | 0.06105348 | -0.00492826 | -0.03392621 | -0.00320895 | 0.01257715 | 0.00385865 | -0.00098693 | 0.00070050 | 0.00057119 | -0.00023787 | -0.00016963 | 0.00009462 | -0.00000241 | -0.00000808 | 0.00005796 | 0.00009782 |
| 27 | 36 | 0.04328480 | 0.05988066 | -0.00608989 | -0.03341590 | -0.00196820 | 0.01203019 | 0.00285505 | -0.00134833 | 0.00058108 | 0.00030082 | -0.00027496 | -0.00011793 | 0.00008785 | -0.00003608 | -0.00001387 | 0.00008172 | 0.00010197 |
| 27 | 37 | 0.04334469 | 0.05956060 | -0.00651468 | -0.03217994 | -0.00061354 | 0.01244178 | 0.00252700 | -0.00156649 | 0.00081418 | 0.00051353 | -0.00028007 | -0.00011476 | 0.00009578 | -0.00002980 | -0.00002247 | 0.00013073 | 0.00013671 |
| 27 | 38 | 0.04308177 | 0.05843321 | -0.00771968 | -0.03169449 | 0.00049556 | 0.01167113 | 0.00154732 | -0.00184551 | 0.00066265 | 0.00019718 | -0.00030268 | -0.00008757 | 0.00012536 | -0.00008206 | -0.00001143 | 0.00013581 | 0.00015135 |
| 27 | 39 | 0.04330954 | 0.05814572 | -0.00818504 | -0.03027940 | 0.00186091 | 0.01194349 | 0.00106579 | -0.00211517 | 0.00092677 | 0.00044488 | -0.00031264 | -0.00004652 | 0.00008660 | -0.00006460 | -0.00004630 | 0.00021248 | 0.00016971 |
| 27 | 40 | 0.04282342 | 0.05706173 | -0.00915750 | -0.03108715 | 0.00142648 | 0.01122113 | 0.00056464 | -0.00192478 | 0.00069689 | 0.00022830 | -0.00038962 | 0.00003890 | 0.00004811 | -0.00002774 | -0.00005732 | 0.00017349 | 0.00007035 |
| 27 | 41 | 0.04285257 | 0.05683088 | -0.00969289 | -0.03144359 | 0.00139273 | 0.01171346 | 0.00047562 | -0.00189273 | 0.00089380 | 0.00051952 | -0.00042352 | 0.00010541 | -0.00002618 | 0.00003479 | -0.00009117 | 0.00018323 | -0.00000256 |
| 27 | 42 | 0.04244794 | 0.05545918 | -0.01122444 | -0.03195737 | 0.00170550 | 0.01130343 | 0.00013337 | -0.00185450 | 0.00060586 | 0.00030446 | -0.00046026 | 0.00016552 | -0.00003582 | 0.00005594 | -0.00009979 | 0.00014586 | -0.00006055 |
| 27 | 43 | 0.04248768 | 0.05526172 | -0.01215538 | -0.03277048 | 0.00187407 | 0.01194356 | 0.00018879 | -0.00186848 | 0.00077470 | 0.00046025 | -0.00042802 | 0.00014592 | -0.00002310 | 0.00004246 | -0.00007984 | 0.00010463 | -0.00007535 |
| 27 | 44 | 0.04213212 | 0.05382029 | -0.01388949 | -0.03302096 | 0.00267540 | 0.01177147 | 0.00005641 | -0.00193090 | 0.00045314 | 0.00024521 | -0.00042783 | 0.00017953 | -0.00001084 | 0.00005419 | -0.00008409 | 0.00007358 | -0.00008983 |
| 27 | 45 | 0.04216897 | 0.05370267 | -0.01493364 | -0.03402309 | 0.00292471 | 0.01240625 | 0.00025919 | -0.00185206 | 0.00063076 | 0.00026834 | -0.00034835 | 0.00008995 | 0.00006588 | -0.00001678 | -0.00001586 | -0.00000887 | -0.00005819 |
| 27 | 46 | 0.04185370 | 0.05235669 | -0.01655257 | -0.03403812 | 0.00388783 | 0.01235437 | 0.00026217 | -0.00196059 | 0.00028807 | 0.00005802 | -0.00033284 | 0.00012744 | 0.00006115 | 0.00000403 | -0.00003134 | -0.00001740 | -0.00005694 |
| 27 | 47 | 0.04186986 | 0.05232544 | -0.01751853 | -0.03504610 | 0.00408717 | 0.01281562 | 0.00052173 | -0.00172841 | 0.00049853 | -0.00002522 | -0.00023112 | -0.00000870 | 0.00018012 | -0.00010250 | 0.00007510 | -0.00013207 | 0.00000203 |
| 27 | 48 | 0.04160685 | 0.05111412 | -0.01894734 | -0.03488401 | 0.00504278 | 0.01282102 | 0.00053329 | -0.00187607 | 0.00013172 | -0.00022546 | -0.00021768 | 0.00006086 | 0.00012586 | -0.00006014 | 0.00003345 | -0.00010648 | -0.00000944 |
| 27 | 49 | 0.04157823 | 0.05114704 | -0.01973023 | -0.03579329 | 0.00511691 | 0.01301337 | 0.00077397 | -0.00149165 | 0.00038349 | -0.00038001 | -0.00011310 | -0.00010498 | 0.00027598 | -0.00018523 | 0.00016918 | -0.00024733 | 0.00006359 |
| 27 | 50 | 0.04138131 | 0.05003113 | -0.02105600 | -0.03552430 | 0.00605495 | 0.01307903 | 0.00070125 | -0.00169551 | -0.00001283 | -0.00056682 | -0.00010801 | 0.00000754 | 0.00016059 | -0.00011893 | 0.00009555 | -0.00018575 | 0.00002483 |
| 27 | 51 | 0.04118638 | 0.05002767 | -0.02051177 | -0.03525042 | 0.00583373 | 0.01254296 | 0.00067327 | -0.00134729 | 0.00047720 | -0.00046863 | -0.00008778 | -0.00011780 | 0.00030667 | -0.00020253 | 0.00021200 | -0.00029463 | 0.00008297 |
| 28 | 20 | 0.04939383 | 0.06483883 | -0.00864593 | -0.04301227 | -0.01036901 | 0.00690176 | 0.00351650 | -0.00243841 | 0.00052885 | 0.00081789 | 0.00043579 | -0.00051820 | 0.00053831 | -0.00035848 | 0.00034899 | -0.00051649 | 0.00012080 |
| 28 | 21 | 0.04784069 | 0.06401725 | -0.00585030 | -0.04068618 | -0.01079262 | 0.00652522 | 0.00414528 | -0.00211585 | 0.00014517 | 0.00059226 | 0.00054348 | -0.00054026 | 0.00050504 | -0.00036244 | 0.00036068 | -0.00054266 | 0.00012440 |
| 28 | 22 | 0.04433196 | 0.06747121 | 0.00248328 | -0.04061972 | -0.01338311 | 0.00789579 | 0.00521760 | -0.00160766 | 0.00046472 | 0.00098584 | 0.00016710 | -0.00074897 | 0.00081846 | -0.00016911 | 0.00048436 | -0.00063739 | 0.00023155 |
| 28 | 23 | 0.04519036 | 0.06682535 | 0.00010342 | -0.03990468 | -0.01076774 | 0.00890470 | 0.00471366 | -0.00174319 | 0.00034824 | 0.00088326 | 0.00011900 | -0.00068480 | 0.00075370 | -0.00014823 | 0.00043288 | -0.00067661 | 0.00012873 |
| 28 | 24 | 0.04515592 | 0.06663114 | -0.00132778 | -0.04082680 | -0.00972017 | 0.01012872 | 0.00475838 | -0.00205442 | 0.00039067 | 0.00087971 | 0.00005819 | -0.00072038 | 0.00078962 | -0.00015187 | 0.00041797 | -0.00055111 | 0.00025521 |
| 28 | 25 | 0.04528941 | 0.06694595 | -0.00063804 | -0.04007301 | -0.01042473 | 0.01004243 | 0.00494811 | -0.00152282 | 0.00038076 | 0.00088753 | -0.00000405 | -0.00058298 | 0.00060148 | -0.00009500 | 0.00030639 | -0.00048637 | 0.00013511 |
| 28 | 26 | 0.04507728 | 0.06711623 | -0.00024554 | -0.04009632 | -0.01150015 | 0.00998196 | 0.00520617 | -0.00110566 | 0.00039704 | 0.00092510 | -0.00009154 | -0.00045105 | 0.00043190 | -0.00002290 | 0.00018509 | -0.00025185 | 0.00014854 |
| 28 | 27 | 0.04528690 | 0.06736143 | 0.00014015 | -0.03928762 | -0.01167117 | 0.01010343 | 0.00539257 | -0.00052512 | 0.00036659 | 0.00084713 | -0.00013576 | -0.00033596 | 0.00027115 | 0.00001153 | 0.00009983 | -0.00022846 | 0.00001572 |
| 28 | 28 | 0.04523301 | 0.06712856 | -0.00027001 | -0.03914862 | -0.01182871 | 0.01041431 | 0.00541877 | -0.00034651 | 0.00027687 | 0.00086016 | -0.00022216 | -0.00019301 | 0.00009728 | 0.00008614 | -0.00002944 | 0.00000250 | 0.00001618 |
| 28 | 29 | 0.04544083 | 0.06647697 | -0.00132816 | -0.03871391 | -0.01039396 | 0.01099477 | 0.00537170 | -0.00023782 | 0.00032278 | 0.00072992 | -0.00020528 | -0.00019597 | 0.00008739 | 0.00005644 | -0.00001341 | -0.00008701 | -0.00004602 |
| 28 | 30 | 0.04536765 | 0.06541165 | -0.00292655 | -0.03863137 | -0.00926312 | 0.01128878 | 0.00505512 | -0.00053090 | 0.00026286 | 0.00070971 | -0.00023790 | -0.00015379 | 0.00005625 | 0.00008103 | -0.00005595 | 0.00004112 | 0.00001747 |
| 28 | 31 | 0.04536195 | 0.06473847 | -0.00363746 | -0.03808538 | -0.00776892 | 0.01195725 | 0.00487582 | -0.00054464 | 0.00037860 | 0.00062573 | -0.00023714 | -0.00015952 | 0.00005443 | 0.00004977 | -0.00003338 | -0.00003488 | -0.00002500 |
| 28 | 32 | 0.04533784 | 0.06372958 | -0.00526610 | -0.03773613 | -0.00654466 | 0.01193117 | 0.00443951 | -0.00078854 | 0.00026898 | 0.00054536 | -0.00025639 | -0.00012880 | 0.00005262 | 0.00005976 | -0.00006141 | 0.00006503 | 0.00002892 |
| 28 | 33 | 0.04512960 | 0.06305336 | -0.00559937 | -0.03699715 | -0.00500455 | 0.01260510 | 0.00409991 | -0.00092370 | 0.00045696 | 0.00050745 | -0.00027180 | -0.00013344 | 0.00005613 | 0.00002753 | -0.00003427 | 0.00000520 | 0.00000560 |
| 28 | 34 | 0.04520719 | 0.06212106 | -0.00733087 | -0.03651040 | -0.00376578 | 0.01230514 | 0.00359173 | -0.00110284 | 0.00029656 | 0.00038061 | -0.00027840 | -0.00011155 | 0.00007587 | 0.00002818 | -0.00005204 | 0.00007994 | 0.00004661 |
| 28 | 35 | 0.04485007 | 0.06145963 | -0.00732573 | -0.03552155 | -0.00221220 | 0.01289593 | 0.00308699 | -0.00134519 | 0.00054675 | 0.00038454 | -0.00030707 | -0.00011000 | 0.00007879 | -0.00000435 | -0.00002572 | 0.00004155 | 0.00004149 |
| 28 | 36 | 0.04505365 | 0.06061802 | -0.00919218 | -0.03504788 | -0.00105122 | 0.01239281 | 0.00256940 | -0.00144420 | 0.00034225 | 0.00022600 | -0.00030186 | -0.00009576 | 0.00011232 | -0.00000805 | -0.00003570 | 0.00009259 | 0.00006705 |
| 28 | 37 | 0.04463097 | 0.05998445 | -0.00893977 | -0.03377746 | 0.00046818 | 0.01280673 | 0.00191035 | -0.00177015 | 0.00063716 | 0.00026443 | -0.00033816 | -0.00008245 | 0.00010668 | -0.00004061 | -0.00001743 | 0.00008139 | 0.00007847 |
| 28 | 38 | 0.04495522 | 0.05924305 | -0.01092922 | -0.03345685 | 0.00147940 | 0.01219100 | 0.00144450 | -0.00177991 | 0.00039981 | 0.00008664 | -0.00032237 | -0.00007697 | 0.00014906 | -0.00004498 | -0.00001919 | 0.00010736 | 0.00008699 |
| 28 | 39 | 0.04455553 | 0.05863865 | -0.01054374 | -0.03189835 | 0.00291513 | 0.01233983 | 0.00064600 | -0.00216461 | 0.00072052 | 0.00014976 | -0.00035880 | -0.00004712 | 0.00012720 | -0.00007862 | -0.00001555 | 0.00012692 | 0.00011235 |
| 28 | 40 | 0.04481537 | 0.05788864 | -0.01241767 | -0.03261158 | 0.00260684 | 0.01189522 | 0.00066545 | -0.00176718 | 0.00040205 | 0.00008196 | -0.00038726 | 0.00003483 | 0.00006681 | 0.00000104 | -0.00006413 | 0.00014653 | 0.00001387 |
| 28 | 41 | 0.04426605 | 0.05721658 | -0.01207917 | -0.03242979 | 0.00276471 | 0.01222170 | 0.00014372 | -0.00188344 | 0.00068026 | 0.00025950 | -0.00047825 | 0.00013836 | -0.00003549 | 0.00003955 | -0.00009513 | 0.00014690 | -0.00006373 |
| 28 | 42 | 0.04457156 | 0.05630102 | -0.01429409 | -0.03277298 | 0.00305451 | 0.01195719 | 0.00039610 | -0.00163613 | 0.00028664 | 0.00014307 | -0.00043487 | 0.00015940 | -0.00003872 | 0.00007074 | -0.00012659 | 0.00016573 | -0.00008576 |
| 28 | 43 | 0.04401771 | 0.05563746 | -0.01440822 | -0.03321979 | 0.00338002 | 0.01248312 | -0.00008457 | -0.00185236 | 0.00055441 | 0.00026493 | -0.00049311 | 0.00020667 | -0.00006995 | 0.00006905 | -0.00011641 | 0.00011400 | -0.00013741 |
| 28 | 44 | 0.04430529 | 0.05474110 | -0.01650938 | -0.03316684 | 0.00393486 | 0.01223235 | 0.00036624 | -0.00166633 | 0.00013775 | 0.00012024 | -0.00040343 | 0.00018313 | -0.00003578 | 0.00007252 | -0.00013323 | 0.00013625 | -0.00009940 |
| 28 | 45 | 0.04375881 | 0.05418146 | -0.01688440 | -0.03406899 | 0.00437427 | 0.01292404 | 0.00000836 | -0.00187984 | 0.00040416 | 0.00016297 | -0.00042767 | 0.00017557 | -0.00000840 | 0.00003379 | -0.00008465 | 0.00004324 | -0.00012230 |
| 28 | 46 | 0.04399504 | 0.05340739 | -0.01857813 | -0.03363186 | 0.00490790 | 0.01255116 | 0.00051246 | -0.00171152 | 0.00000032 | 0.00002072 | -0.00032844 | 0.00014518 | 0.00002550 | 0.00003639 | -0.00010254 | 0.00008087 | -0.00006027 |
| 28 | 47 | 0.04348231 | 0.05298368 | -0.01908427 | -0.03480484 | 0.00540008 | 0.01332185 | 0.00027337 | -0.00183604 | 0.00026603 | -0.00003153 | -0.00032432 | 0.00009786 | 0.00008739 | -0.00003047 | -0.00002376 | -0.00004166 | -0.00006500 |
| 28 | 48 | 0.04365729 | 0.05233752 | -0.02031602 | -0.03404691 | 0.00579666 | 0.01278922 | 0.00069322 | -0.00169669 | -0.00010790 | -0.00014425 | -0.00024112 | 0.00008833 | 0.00009619 | -0.00001213 | -0.00005509 | 0.00001871 | -0.00000635 |
| 28 | 49 | 0.04320104 | 0.05201348 | -0.02093268 | -0.03536311 | 0.00630256 | 0.01354446 | 0.00052185 | -0.00169683 | 0.00014504 | -0.00029552 | -0.00021427 | 0.00001642 | 0.00017270 | -0.00009846 | 0.00004547 | -0.00012541 | -0.00000483 |
| 28 | 50 | 0.04331810 | 0.05146332 | -0.02177140 | -0.03436530 | 0.00657069 | 0.01289256 | 0.00079038 | -0.00161349 | -0.00019074 | -0.00036139 | -0.00015835 | 0.00003791 | 0.00014934 | -0.00005908 | -0.00000365 | -0.00004132 | 0.00003878 |

TABLE 3. Nuclear charge density distribution FB coefficients.

| $Z$ | $N$ | $a_1$ | $a_2$ | $a_3$ | $a_4$ | $a_5$ | $a_6$ | $a_7$ | $a_8$ | $a_9$ | $a_{10}$ | $a_{11}$ | $a_{12}$ | $a_{13}$ | $a_{14}$ | $a_{15}$ | $a_{16}$ | $a_{17}$ |
|---|---|---|---|---|---|---|---|---|---|---|---|---|---|---|---|---|---|---|
| 28 | 51 | 0.04274252 | 0.05098820 | -0.02149254 | -0.03480696 | 0.00676184 | 0.01288282 | 0.00039210 | -0.00156726 | 0.00022849 | -0.00034900 | -0.00019192 | -0.00000385 | 0.00021938 | -0.00010839 | 0.00008932 | -0.00017731 | 0.00001608 |
| 28 | 52 | 0.04279511 | 0.05028262 | -0.02157356 | -0.03298221 | 0.00668563 | 0.01144001 | 0.00037956 | -0.00149098 | 0.00003714 | -0.00021514 | -0.00019407 | 0.00004584 | 0.00019576 | -0.00002719 | 0.00002721 | -0.00008988 | 0.00002795 |
| 28 | 53 | 0.04231102 | 0.04979621 | -0.02119106 | -0.03307021 | 0.00718315 | 0.01145913 | -0.00015428 | -0.00141476 | 0.00053903 | -0.00016908 | -0.00025830 | 0.00004094 | 0.00022761 | -0.00005885 | 0.00010761 | -0.00021241 | -0.00002124 |
| 28 | 54 | 0.04253652 | 0.04913111 | -0.02165053 | -0.03143284 | 0.00716406 | 0.00999778 | -0.00020546 | -0.00128475 | 0.00029962 | -0.00008798 | -0.00023861 | 0.00007070 | 0.00021721 | 0.00001198 | 0.00005813 | -0.00015294 | -0.00002097 |
| 29 | 23 | 0.04609747 | 0.06781422 | 0.00039009 | -0.03847151 | -0.00966154 | 0.00905951 | 0.00510481 | -0.00142166 | 0.00029010 | 0.00094575 | 0.00010253 | -0.00074075 | 0.00083687 | -0.00009489 | 0.00047778 | -0.00079570 | 0.00009730 |
| 29 | 24 | 0.04602497 | 0.06702660 | -0.00072586 | -0.03872661 | -0.00838666 | 0.00954045 | 0.00428322 | -0.00196626 | 0.00036887 | 0.00083367 | 0.00005314 | -0.00066607 | 0.00076686 | -0.00012231 | 0.00044294 | -0.00073054 | 0.00009683 |
| 29 | 25 | 0.04616333 | 0.06742764 | -0.00040894 | -0.03815005 | -0.00919565 | 0.00972737 | 0.00521134 | -0.00103040 | 0.00037194 | 0.00089053 | 0.00003306 | -0.00062379 | 0.00066604 | -0.00006120 | 0.00036942 | -0.00062779 | 0.00007670 |
| 29 | 26 | 0.04605465 | 0.06759675 | 0.00014729 | -0.03820547 | -0.01009518 | 0.00941962 | 0.00461050 | -0.00096750 | 0.00044276 | 0.00087231 | -0.00010326 | -0.00038639 | 0.00039878 | 0.00000687 | 0.00021937 | -0.00044887 | -0.00004102 |
| 29 | 27 | 0.04619800 | 0.06769849 | -0.00018844 | -0.03788353 | -0.01028731 | 0.00979933 | 0.00545286 | -0.00010132 | 0.00043175 | 0.00081285 | -0.00008079 | -0.00039880 | 0.00036642 | 0.00001158 | 0.00019152 | -0.00039441 | -0.00003373 |
| 29 | 28 | 0.04638929 | 0.06760144 | -0.00053755 | -0.03773302 | -0.01012834 | 0.01013740 | 0.00470091 | -0.00018393 | 0.00039632 | 0.00077477 | -0.00023611 | -0.00011542 | 0.00004512 | 0.00010603 | 0.00000570 | -0.00019554 | -0.00019613 |
| 29 | 29 | 0.04641047 | 0.06677082 | -0.00203406 | -0.03764679 | -0.00889325 | 0.01086131 | 0.00533368 | 0.00011049 | 0.00042294 | 0.00069584 | -0.00014488 | -0.00026599 | 0.00019019 | 0.00004111 | 0.00008245 | -0.00024662 | -0.00008285 |
| 29 | 30 | 0.04648306 | 0.06577400 | -0.00334372 | -0.03743636 | -0.00743506 | 0.01140551 | 0.00442357 | -0.00043405 | 0.00039002 | 0.00063131 | -0.00024913 | -0.00009367 | 0.00001464 | 0.00008874 | -0.00002016 | -0.00012745 | -0.00015068 |
| 29 | 31 | 0.04642545 | 0.06509759 | -0.00441873 | -0.03696440 | -0.00613039 | 0.01202894 | 0.00482752 | -0.00026565 | 0.00046942 | 0.00062854 | -0.00018003 | -0.00022296 | 0.00014757 | 0.00004060 | 0.00005067 | -0.00017448 | -0.00005417 |
| 29 | 32 | 0.04638338 | 0.06399307 | -0.00575147 | -0.03662783 | -0.00456868 | 0.01232007 | 0.00383740 | -0.00076464 | 0.00042543 | 0.00047978 | -0.00026738 | -0.00008361 | 0.00002396 | 0.00005587 | -0.00002308 | -0.00007762 | -0.00009571 |
| 29 | 33 | 0.04629483 | 0.06348312 | -0.00642782 | -0.03579292 | -0.00327751 | 0.01279815 | 0.00403903 | -0.00070424 | 0.00055635 | 0.00055449 | -0.00021780 | -0.00018720 | 0.00013483 | 0.00002727 | 0.00003683 | -0.00011351 | -0.00001536 |
| 29 | 34 | 0.04617110 | 0.06231458 | -0.00779175 | -0.03533853 | -0.00166726 | 0.01280074 | 0.00297058 | -0.00115047 | 0.00049246 | 0.00033288 | -0.00028926 | -0.00007889 | 0.00006204 | 0.00001347 | -0.00001101 | -0.00003924 | -0.00003573 |
| 29 | 35 | 0.04612919 | 0.06195091 | -0.00819911 | -0.03421988 | -0.00044820 | 0.01311592 | 0.00300785 | -0.00117056 | 0.00066636 | 0.00047646 | -0.00025456 | -0.00015221 | 0.00013922 | 0.00000567 | 0.00003193 | -0.00005688 | 0.00002863 |
| 29 | 36 | 0.04595517 | 0.06077355 | -0.00957582 | -0.03368751 | 0.00110841 | 0.01282235 | 0.00189588 | -0.00155164 | 0.00057628 | 0.00019923 | -0.00031050 | -0.00007211 | 0.00011140 | -0.00003210 | 0.00000506 | -0.00000295 | 0.00002421 |
| 29 | 37 | 0.04603355 | 0.06050767 | -0.00989261 | -0.03239134 | 0.00223761 | 0.01297290 | 0.00179544 | -0.00162951 | 0.00078345 | 0.00039557 | -0.00028537 | -0.00011249 | 0.00014718 | -0.00002067 | 0.00002779 | 0.00000057 | 0.00007249 |
| 29 | 38 | 0.04583759 | 0.05938114 | -0.01123966 | -0.03183281 | 0.00362386 | 0.01240546 | 0.00069785 | -0.00193098 | 0.00066236 | 0.00008059 | -0.00032504 | -0.00005771 | 0.00015533 | -0.00007686 | 0.00001628 | 0.00003708 | 0.00007886 |
| 29 | 39 | 0.04607597 | 0.05914088 | -0.01164431 | -0.03048722 | 0.00468896 | 0.01241762 | 0.00047076 | -0.00205801 | 0.00089776 | 0.00031011 | -0.00030528 | -0.00006541 | 0.00014884 | -0.00005110 | 0.00002007 | 0.00005965 | 0.00011124 |
| 29 | 40 | 0.04569716 | 0.05802291 | -0.01288223 | -0.03123644 | 0.00462318 | 0.01196400 | -0.00024767 | -0.00196102 | 0.00071093 | 0.00009450 | -0.00039532 | 0.00005267 | 0.00008152 | -0.00003471 | -0.00002236 | 0.00007622 | 0.00001205 |
| 29 | 41 | 0.04576585 | 0.05782564 | -0.01344504 | -0.03171711 | 0.00434958 | 0.01229366 | -0.00019506 | -0.00186972 | 0.00089540 | 0.00037868 | -0.00040516 | 0.00008121 | 0.00002942 | 0.00002639 | -0.00002825 | 0.00005716 | -0.00004305 |
| 29 | 42 | 0.04544408 | 0.05646193 | -0.01512600 | -0.03216944 | 0.00491988 | 0.01212938 | -0.00064605 | -0.00186956 | 0.00065334 | 0.00015721 | -0.00045273 | 0.00017090 | -0.00001090 | 0.00002906 | -0.00006722 | 0.00007550 | -0.00009598 |
| 29 | 43 | 0.04548461 | 0.05633646 | -0.01594481 | -0.03308187 | 0.00476551 | 0.01257868 | -0.00048464 | -0.00184662 | 0.00081563 | 0.00033559 | -0.00040691 | 0.00012026 | 0.00002171 | 0.00002239 | -0.00002592 | 0.00001079 | -0.00009844 |
| 29 | 44 | 0.04519030 | 0.05496491 | -0.01766162 | -0.03322106 | 0.00572999 | 0.01262240 | -0.00064269 | -0.00188868 | 0.00054319 | 0.00011253 | -0.00042036 | 0.00018374 | 0.00000101 | 0.00002094 | -0.00006064 | 0.00003326 | -0.00011017 |
| 29 | 45 | 0.04519941 | 0.05493329 | -0.01854234 | -0.03434335 | 0.00562740 | 0.01306804 | -0.00038825 | -0.00181744 | 0.00071483 | 0.00018532 | -0.00033570 | 0.00006871 | 0.00009879 | -0.00003807 | 0.00002349 | -0.00006943 | -0.00007134 |
| 29 | 46 | 0.04492713 | 0.05370091 | -0.02002152 | -0.03419209 | 0.00671965 | 0.01320738 | -0.00037170 | -0.00186764 | 0.00042544 | -0.00003088 | -0.00033748 | 0.00013500 | 0.00006400 | -0.00002618 | -0.00002052 | -0.00003086 | -0.00007161 |
| 29 | 47 | 0.04490359 | 0.05374639 | -0.02084826 | -0.03536388 | 0.00658941 | 0.01351246 | -0.00009640 | -0.00168490 | 0.00062673 | -0.00005076 | -0.00023339 | -0.00002325 | 0.00020528 | -0.00011982 | 0.00009873 | -0.00016446 | -0.00000814 |
| 29 | 48 | 0.04466697 | 0.05266574 | -0.02207645 | -0.03499801 | 0.00767537 | 0.01368126 | -0.00006025 | -0.00175383 | 0.00031918 | -0.00025468 | -0.00023883 | 0.00007111 | 0.00012815 | -0.00008434 | 0.00003326 | -0.00010088 | -0.00002335 |
| 29 | 49 | 0.04460033 | 0.05275726 | -0.02277706 | -0.03611793 | 0.00746014 | 0.01376052 | 0.00018204 | -0.00144631 | 0.00055649 | -0.00034488 | -0.00013093 | -0.00011409 | 0.00029914 | -0.00019770 | 0.00018021 | -0.00026020 | 0.00005276 |
| 29 | 50 | 0.04441576 | 0.05177065 | -0.02388140 | -0.03562702 | 0.00852690 | 0.01395037 | 0.00012883 | -0.00156070 | 0.00022493 | -0.00053427 | -0.00014474 | 0.00001880 | 0.00016759 | -0.00013713 | 0.00008803 | -0.00016917 | 0.00000944 |
| 29 | 51 | 0.04424557 | 0.05173551 | -0.02349840 | -0.03555730 | 0.00809699 | 0.01317119 | 0.00004515 | -0.00128186 | 0.00066183 | -0.00040877 | -0.00010942 | -0.00013587 | 0.00034338 | -0.00021084 | 0.00022731 | -0.00031585 | 0.00007168 |
| 29 | 52 | 0.04385317 | 0.05062550 | -0.02327953 | -0.03380131 | 0.00864778 | 0.01234213 | -0.00024694 | -0.00140290 | 0.00049628 | -0.00034837 | -0.00019234 | 0.00002907 | 0.00021658 | -0.00009868 | 0.00011982 | -0.00021026 | 0.00000937 |
| 29 | 53 | 0.04400964 | 0.05055520 | -0.02341405 | -0.03365013 | 0.00884494 | 0.01188309 | -0.00050624 | -0.00116334 | 0.00095769 | -0.00022643 | -0.00016396 | -0.00008717 | 0.00033613 | -0.00016203 | 0.00023730 | -0.00033748 | 0.00003789 |
| 29 | 54 | 0.04356164 | 0.04945814 | -0.02297308 | -0.03170726 | 0.00915621 | 0.01069320 | -0.00079848 | -0.00112868 | 0.00080266 | -0.00018461 | -0.00024605 | 0.00006009 | 0.00023316 | -0.00005096 | 0.00014708 | -0.00026226 | -0.00003250 |
| 29 | 55 | 0.04398018 | 0.04934571 | -0.02352636 | -0.03147019 | 0.00989620 | 0.01058698 | -0.00114334 | -0.00094331 | 0.00126245 | -0.00007207 | -0.00021917 | -0.00001999 | 0.00029696 | -0.00010571 | 0.00023829 | -0.00035660 | -0.00002340 |
| 30 | 24 | 0.04652151 | 0.06760537 | 0.00001077 | -0.03756227 | -0.00778249 | 0.00878504 | 0.00381339 | -0.00184165 | 0.00056117 | 0.00083518 | -0.00000647 | -0.00061986 | 0.00078218 | -0.00009243 | 0.00047989 | -0.00073483 | 0.00008725 |
| 30 | 25 | 0.04667616 | 0.06790514 | 0.00062345 | -0.03682654 | -0.00858701 | 0.00864298 | 0.00396402 | -0.00145907 | 0.00044229 | 0.00086015 | -0.00004987 | -0.00047292 | 0.00057518 | -0.00002407 | 0.00035210 | -0.00066122 | -0.00004374 |
| 30 | 26 | 0.04678262 | 0.06789268 | -0.00005069 | -0.03723265 | -0.00892676 | 0.00894934 | 0.00384048 | -0.00106265 | 0.00060531 | 0.00089801 | -0.00017816 | -0.00030807 | 0.00038145 | 0.00005029 | 0.00023199 | -0.00042784 | -0.00006716 |
| 30 | 27 | 0.04703031 | 0.06808963 | 0.00008106 | -0.03663221 | -0.00915692 | 0.00915826 | 0.00393340 | -0.00068599 | 0.00047636 | 0.00084523 | -0.00021058 | -0.00016733 | 0.00018038 | 0.00009915 | 0.00011799 | -0.00037646 | -0.00021474 |
| 30 | 28 | 0.04734039 | 0.06762525 | -0.00168435 | -0.03706630 | -0.00850466 | 0.00996400 | 0.00362300 | -0.00051087 | 0.00054358 | 0.00084883 | -0.00033512 | 0.00000181 | -0.00001519 | 0.00017149 | -0.00001121 | -0.00014245 | -0.00024045 |
| 30 | 29 | 0.04746123 | 0.06695610 | -0.00257529 | -0.03662629 | -0.00731005 | 0.01058254 | 0.00374902 | -0.00050319 | 0.00042025 | 0.00068643 | -0.00027833 | -0.00001547 | -0.00002394 | 0.00013160 | -0.00000911 | -0.00020409 | -0.00027054 |
| 30 | 30 | 0.04756335 | 0.06574462 | -0.00486434 | -0.03672362 | -0.00562661 | 0.01116930 | 0.00337745 | -0.00068962 | 0.00044864 | 0.00061199 | -0.00032047 | 0.00000588 | -0.00002138 | 0.00013135 | -0.00002871 | -0.00008437 | -0.00018319 |
| 30 | 31 | 0.04754699 | 0.06504527 | -0.00555072 | -0.03616967 | -0.00433772 | 0.01183918 | 0.00337922 | -0.00076243 | 0.00039965 | 0.00049292 | -0.00027892 | -0.00001486 | -0.00002148 | 0.00009104 | -0.00002124 | -0.00013766 | -0.00019746 |
| 30 | 32 | 0.04761937 | 0.06392453 | -0.00767435 | -0.03593098 | -0.00265901 | 0.01198937 | 0.00286152 | -0.00093109 | 0.00039133 | 0.00038053 | -0.00030918 | -0.00000371 | 0.00001248 | 0.00007971 | -0.00002382 | -0.00004924 | -0.00011944 |
| 30 | 33 | 0.04745806 | 0.06323746 | -0.00804984 | -0.03513908 | -0.00130042 | 0.01259683 | 0.00269887 | -0.00108416 | 0.00042891 | 0.00030904 | -0.00028428 | -0.00002432 | 0.00002011 | 0.00003960 | -0.00001150 | -0.00009230 | -0.00011781 |
| 30 | 34 | 0.04755082 | 0.06223009 | -0.01006157 | -0.03472902 | 0.00022728 | 0.01235364 | 0.00211010 | -0.00120987 | 0.00037461 | 0.00017218 | -0.00030124 | -0.00002204 | 0.00007582 | 0.00002292 | -0.00000226 | -0.00003192 | -0.00005337 |
| 30 | 35 | 0.04728710 | 0.06159025 | -0.01011678 | -0.03362578 | 0.00160823 | 0.01279595 | 0.00176738 | -0.00143219 | 0.00049486 | 0.00014801 | -0.00029141 | -0.00003684 | 0.00008472 | -0.00001574 | 0.00000973 | -0.00005878 | -0.00003714 |
| 30 | 36 | 0.04743755 | 0.06069939 | -0.01208298 | -0.03323659 | 0.00288092 | 0.01226847 | 0.00119058 | -0.00149190 | 0.00039200 | -0.00000380 | -0.00029481 | -0.00004210 | 0.00015124 | -0.00003282 | 0.00002657 | -0.00002353 | 0.00000971 |
| 30 | 37 | 0.04714969 | 0.06012139 | -0.01190363 | -0.03180942 | 0.00423148 | 0.01246671 | 0.00067015 | -0.00176848 | 0.00057876 | 0.00001341 | -0.00029550 | -0.00004464 | 0.00015149 | -0.00006919 | 0.00003084 | -0.00002732 | 0.00003778 |
| 30 | 38 | 0.04736999 | 0.05933984 | -0.01386308 | -0.03159884 | 0.00521095 | 0.01179171 | 0.00017249 | -0.00175006 | 0.00043135 | -0.00014807 | -0.00028646 | -0.00005753 | 0.00022141 | -0.00008348 | 0.00005426 | -0.00001733 | 0.00006397 |
| 30 | 39 | 0.04713863 | 0.05880988 | -0.01359151 | -0.02989540 | 0.00649364 | 0.01170458 | -0.00051422 | -0.00207025 | 0.00066471 | -0.00010147 | -0.00029111 | -0.00004270 | 0.00020408 | -0.00011904 | 0.00004504 | 0.00000584 | 0.00010027 |
| 30 | 40 | 0.04726605 | 0.05793450 | -0.01568856 | -0.03101086 | 0.00628482 | 0.01139442 | -0.00061418 | -0.00176333 | 0.00045357 | -0.00015152 | -0.00032447 | 0.00002193 | 0.00016911 | -0.00005664 | 0.00002639 | 0.00001603 | 0.00002503 |
| 30 | 41 | 0.04689540 | 0.05728969 | -0.01573595 | -0.03094386 | 0.00644424 | 0.01169682 | -0.00113024 | -0.00192096 | 0.00067210 | 0.00001633 | -0.00038902 | 0.00011843 | 0.00006030 | -0.00002261 | -0.00001870 | 0.00003745 | -0.00003663 |
| 30 | 42 | 0.04703248 | 0.05634199 | -0.01795856 | -0.03167076 | 0.00673646 | 0.01158237 | -0.00086994 | -0.00170481 | 0.00039987 | -0.00008920 | -0.00035186 | 0.00012198 | 0.00007954 | -0.00000590 | -0.00002023 | 0.00004323 | -0.00003642 |
| 30 | 43 | 0.04665266 | 0.05577564 | -0.01828991 | -0.03217345 | 0.00698623 | 0.01213641 | -0.00128812 | -0.00193138 | 0.00060472 | 0.00003077 | -0.00039363 | 0.00017682 | 0.00002041 | -0.00000088 | -0.00003585 | 0.00002661 | -0.00008060 |
| 30 | 44 | 0.04676336 | 0.05490477 | -0.02023502 | -0.03242471 | 0.00749646 | 0.01202882 | -0.00079506 | -0.00172717 | 0.00031457 | -0.00010120 | -0.00032004 | 0.00014616 | 0.00006655 | -0.00000345 | -0.00002926 | 0.00003527 | -0.00003441 |
| 30 | 45 | 0.04640112 | 0.05446930 | -0.02075720 | -0.03335403 | 0.00785911 | 0.01282398 | -0.00107038 | -0.00193935 | 0.00051049 | -0.00005617 | -0.00033674 | 0.00016082 | 0.00004787 | -0.00002771 | -0.00001642 | -0.00001350 | -0.00006200 |
| 30 | 46 | 0.04646634 | 0.05371187 | -0.02225606 | -0.03313213 | 0.00837474 | 0.01257392 | -0.00054771 | -0.00172512 | 0.00023458 | -0.00018183 | -0.00026182 | 0.00013191 | 0.00008692 | -0.00002454 | -0.00001598 | 0.00000814 | -0.00000568 |
| 30 | 47 | 0.04615593 | 0.05339371 | -0.02296080 | -0.03437279 | 0.00882477 | 0.01350940 | -0.00072318 | -0.00186277 | 0.00042046 | -0.00023165 | -0.00025447 | 0.00011646 | 0.00009180 | -0.00007425 | 0.00002186 | -0.00006749 | -0.00002499 |
| 30 | 48 | 0.04616646 | 0.05272244 | -0.02402203 | -0.03374767 | 0.00925918 | 0.01307192 | -0.00031090 | -0.00165603 | 0.00017809 | -0.00032498 | -0.00019879 | 0.00011060 | 0.00010508 | -0.00005220 | 0.00000764 | -0.00002741 | 0.00002036 |
| 30 | 49 | 0.04592585 | 0.05246875 | -0.02493330 | -0.03521382 | 0.00974668 | 0.01402074 | -0.00046153 | -0.00169594 | 0.00034461 | -0.00047464 | -0.00017062 | 0.00007610 | 0.00011871 | -0.00012088 | 0.00006578 | -0.00012605 | 0.00000023 |
| 30 | 50 | 0.04588226 | 0.05185205 | -0.02562914 | -0.03427315 | 0.01010210 | 0.01343709 | -0.00020653 | -0.00152146 | 0.00014565 | -0.00051928 | -0.00014158 | 0.00009887 | 0.00010269 | -0.00007699 | 0.00003461 | -0.00006631 | 0.00002825 |
| 30 | 51 | 0.04551161 | 0.05147222 | -0.02548154 | -0.03459671 | 0.01016840 | 0.01331391 | -0.00052560 | -0.00149392 | 0.00042842 | -0.00052315 | -0.00014593 | 0.00005298 | 0.00016135 | -0.00012615 | 0.00010919 | -0.00018040 | 0.00001694 |
| 30 | 52 | 0.04535903 | 0.05071755 | -0.02517861 | -0.03256258 | 0.01002904 | 0.01167238 | -0.00046034 | -0.00131112 | 0.00031591 | -0.00038685 | -0.00014561 | 0.00006471 | 0.00018987 | -0.00005718 | 0.00008465 | -0.00013549 | 0.00004502 |
| 30 | 53 | 0.04508315 | 0.05031928 | -0.02491456 | -0.03243686 | 0.01039988 | 0.01150081 | -0.00090801 | -0.00121260 | 0.00069615 | -0.00035328 | -0.00017948 | 0.00005040 | 0.00021604 | -0.00008872 | 0.00014896 | -0.00023862 | 0.00001031 |
| 30 | 54 | 0.04506872 | 0.04957565 | -0.02497127 | -0.03068247 | 0.01026494 | 0.00991229 | -0.00087735 | -0.00100708 | 0.00052285 | -0.00027236 | -0.00015764 | 0.00004935 | 0.00024537 | -0.00002910 | 0.00013077 | -0.00021551 | 0.00002109 |
| 30 | 55 | 0.04490735 | 0.04914107 | -0.02461556 | -0.03002820 | 0.01097309 | 0.00967198 | -0.00144780 | -0.00082297 | 0.00098634 | -0.00021258 | -0.00021631 | 0.00006940 | 0.00023328 | -0.00004329 | 0.00018033 | -0.00030101 | -0.00003597 |
| 30 | 56 | 0.04501079 | 0.04844682 | -0.02500589 | -0.02868818 | 0.01079350 | 0.00823458 | -0.00142645 | -0.00063731 | 0.00074700 | -0.00018561 | -0.00017393 | 0.00005248 | 0.00026634 | 0.00000433 | 0.00016964 | -0.00029686 | -0.00003630 |
| 31 | 25 | 0.04755913 | 0.06826165 | 0.00075037 | -0.03468953 | -0.00734466 | 0.00819391 | 0.00417499 | -0.00110926 | 0.00029948 | 0.00088917 | -0.00001040 | -0.00048649 | 0.00061308 | 0.00003138 | 0.00037539 | -0.00075406 | -0.00009566 |
| 31 | 26 | 0.04754579 | 0.06859632 | 0.00106320 | -0.03521728 | -0.00817109 | 0.00810942 | 0.00330467 | -0.00120138 | 0.00049557 | 0.00088030 | -0.00016552 | -0.00025341 | 0.00036131 | 0.00006953 | 0.00024356 | -0.00055731 | -0.00018941 |
| 31 | 27 | 0.04786627 | 0.06829245 | -0.00026990 | -0.03509018 | -0.00772307 | 0.00881081 | 0.00391867 | -0.00047667 | 0.00039392 | 0.00082941 | -0.00015178 | -0.00021476 | 0.00026384 | 0.00011427 | 0.00018071 | -0.00049997 | -0.00024720 |
| 31 | 28 | 0.04827747 | 0.06824499 | -0.00139163 | -0.03561992 | -0.00731770 | 0.00948268 | 0.00286051 | -0.00072920 | 0.00048729 | 0.00079128 | -0.00032431 | 0.00006817 | -0.00005004 | 0.00017578 | 0.00000549 | -0.00027340 | -0.00038178 |
| 31 | 29 | 0.04834720 | 0.06712288 | -0.00324495 | -0.03541656 | -0.00581715 | 0.01033902 | 0.00362218 | -0.00036941 | 0.00037317 | 0.00067198 | -0.00021301 | -0.00007257 | 0.00006897 | 0.00013197 | 0.00006132 | -0.00032785 | -0.00029268 |
| 31 | 30 | 0.04857941 | 0.06620563 | -0.00496048 | -0.03554771 | -0.00427126 | 0.01109999 | 0.00269952 | -0.00091552 | 0.00040285 | 0.00055652 | -0.00030886 | 0.00006687 | -0.00006711 | 0.00013146 | -0.00002223 | -0.00018396 | -0.00029760 |
| 31 | 31 | 0.04858678 | 0.06531245 | -0.00628166 | -0.03491106 | -0.00286723 | 0.01167076 | 0.00327086 | -0.00063155 | 0.00035490 | 0.00052725 | -0.00021728 | -0.00006084 | 0.00005022 | 0.00010523 | 0.00003198 | -0.00024342 | -0.00022312 |
| 31 | 32 | 0.04866964 | 0.06424051 | -0.00803393 | -0.03485612 | -0.00110527 | 0.01221164 | 0.00223098 | -0.00115718 | 0.00037775 | 0.00032831 | -0.00029662 | 0.00005055 | -0.00003767 | 0.00007428 | -0.00002303 | -0.00012305 | -0.00020451 |
| 31 | 33 | 0.04865185 | 0.06359699 | -0.00882536 | -0.03381232 | 0.00012984 | 0.01246295 | 0.00262149 | -0.00094137 | 0.00040370 | 0.00039376 | -0.00022428 | -0.00005944 | 0.00006823 | 0.00006857 | 0.00002584 | -0.00018035 | -0.00014418 |
| 31 | 34 | 0.04859872 | 0.06245003 | -0.01051100 | -0.03355553 | 0.00194928 | 0.01267267 | 0.00148385 | -0.00142843 | 0.00040828 | 0.00012776 | -0.00028645 | 0.00002343 | 0.00002988 | 0.00001074 | -0.00000287 | -0.00008650 | -0.00010727 |
| 31 | 35 | 0.04863631 | 0.06201185 | -0.01097096 | -0.03224118 | 0.00300984 | 0.01265866 | 0.00171214 | -0.00127080 | 0.00050259 | 0.00027713 | -0.00023124 | -0.00006197 | 0.00010926 | 0.00002706 | 0.00003442 | -0.00013230 | -0.00006258 |
| 31 | 36 | 0.04847922 | 0.06087232 | -0.01250953 | -0.03181937 | 0.00470570 | 0.01249232 | 0.00053018 | -0.00169492 | 0.00047741 | -0.00003633 | -0.00027567 | -0.00000537 | 0.00011362 | -0.00005198 | 0.00002572 | -0.00006279 | -0.00001391 |
| 31 | 37 | 0.04864750 | 0.06054873 | -0.01292497 | -0.03043170 | 0.00565086 | 0.01231524 | 0.00060491 | -0.00159627 | 0.00063093 | 0.00017480 | -0.00023508 | -0.00006113 | 0.00015561 | -0.00001554 | 0.00004857 | -0.00009252 | 0.00001318 |
| 31 | 38 | 0.04842782 | 0.05948729 | -0.01425903 | -0.02989487 | 0.00707139 | 0.01178823 | -0.00055310 | -0.00193578 | 0.00056447 | -0.00016895 | -0.00026057 | -0.00002716 | 0.00019093 | -0.00010990 | 0.00005199 | -0.00004282 | 0.00006667 |
| 31 | 39 | 0.04875480 | 0.05916760 | -0.01489257 | -0.02865616 | 0.00799650 | 0.01158250 | -0.00062410 | -0.00190855 | 0.00077322 | 0.00007787 | -0.00023306 | -0.00005252 | 0.00019409 | -0.00005934 | 0.00006330 | -0.00005810 | 0.00007596 |
| 31 | 40 | 0.04834925 | 0.05809501 | -0.01623137 | -0.02954435 | 0.00799559 | 0.01124225 | -0.00150151 | -0.00197411 | 0.00062979 | -0.00015436 | -0.00030524 | 0.00005307 | 0.00014025 | -0.00008358 | 0.00002782 | -0.00000622 | 0.00003092 |
| 31 | 41 | 0.04852599 | 0.05780376 | -0.01715402 | -0.03023215 | 0.00768351 | 0.01153652 | -0.00138291 | -0.00183465 | 0.00078999 | 0.00016243 | -0.00031381 | 0.00008192 | 0.00007303 | 0.00000548 | 0.00001572 | -0.00003028 | -0.00004375 |
| 31 | 42 | 0.04812999 | 0.05658384 | -0.01877305 | -0.03091634 | 0.00820301 | 0.01152275 | -0.00188053 | -0.00196755 | 0.00060577 | -0.00008061 | -0.00034733 | 0.00015528 | 0.00004871 | -0.00003047 | -0.00001109 | 0.00001712 | -0.00004094 |
| 31 | 43 | 0.04827208 | 0.05639852 | -0.01980324 | -0.03192244 | 0.00799355 | 0.01197613 | -0.00163194 | -0.00186723 | 0.00074520 | 0.00013868 | -0.00030889 | 0.00012048 | 0.00004735 | -0.00000062 | 0.00001210 | -0.00004161 | -0.00007425 |
| 31 | 44 | 0.04789228 | 0.05525041 | -0.02131141 | -0.03228803 | 0.00885959 | 0.01220299 | -0.00178472 | -0.00200525 | 0.00053649 | -0.00010740 | -0.00032003 | 0.00017508 | 0.00003651 | -0.00003359 | -0.00001226 | 0.00000515 | -0.00004312 |
| 31 | 45 | 0.04799977 | 0.05514542 | -0.02238643 | -0.03347671 | 0.00873709 | 0.01270486 | -0.00145979 | -0.00186350 | 0.00068241 | 0.00001067 | -0.00024662 | 0.00008633 | 0.00008953 | -0.00005367 | 0.00004588 | -0.00008526 | -0.00004321 |
| 31 | 46 | 0.04765625 | 0.05414950 | -0.02362161 | -0.03350015 | 0.00976279 | 0.01303309 | -0.00144354 | -0.00197663 | 0.00046218 | -0.00022832 | -0.00025774 | 0.00015253 | 0.00005880 | -0.00006629 | 0.00001032 | -0.00002931 | -0.00001672 |
| 31 | 47 | 0.04772399 | 0.05408621 | -0.02470182 | -0.03478447 | 0.00965320 | 0.01344275 | -0.00112545 | -0.00175532 | 0.00063035 | -0.00020309 | -0.00016318 | 0.00002407 | 0.00015220 | -0.00012340 | 0.00010043 | -0.00014857 | 0.00000557 |
| 31 | 48 | 0.04743782 | 0.05321259 | -0.02571390 | -0.03452384 | 0.01072665 | 0.01376911 | -0.00111654 | -0.00185044 | 0.00040324 | -0.00042738 | -0.00018695 | 0.00012397 | 0.00007612 | -0.00010637 | 0.00004360 | -0.00007601 | 0.00000378 |
| 31 | 49 | 0.04744943 | 0.05316285 | -0.02673770 | -0.03583188 | 0.01056014 | 0.01399249 | -0.00085706 | -0.00154405 | 0.00059839 | -0.00047614 | -0.00008405 | -0.00003215 | 0.00020020 | -0.00018755 | 0.00016131 | -0.00022099 | 0.00004054 |
| 31 | 50 | 0.04723414 | 0.05234641 | -0.02765040 | -0.03536585 | 0.01164795 | 0.01428215 | -0.00096081 | -0.00164145 | 0.00036274 | -0.00067999 | -0.00012314 | 0.00010915 | 0.00006835 | -0.00013974 | 0.00007863 | -0.00012847 | 0.00000039 |
| 31 | 51 | 0.04713963 | 0.05216954 | -0.02747631 | -0.03522365 | 0.01117373 | 0.01337360 | -0.00093069 | -0.00132087 | 0.00070115 | -0.00053570 | -0.00005852 | -0.00005922 | 0.00024382 | -0.00019850 | 0.00020848 | -0.00027763 | 0.00005963 |
| 31 | 52 | 0.04670979 | 0.05119966 | -0.02693674 | -0.03319417 | 0.01169047 | 0.01237603 | -0.00118711 | -0.00137161 | 0.00057616 | -0.00051437 | -0.00013241 | 0.00007394 | 0.00015800 | -0.00011388 | 0.00012981 | -0.00019104 | 0.00002867 |
| 31 | 53 | 0.04692177 | 0.05101080 | -0.02718996 | -0.03291795 | 0.01174032 | 0.01172589 | -0.00130079 | -0.00105431 | 0.00095267 | -0.00036407 | -0.00008080 | -0.00005601 | 0.00027794 | -0.00015892 | 0.00023619 | -0.00031646 | 0.00005728 |
| 31 | 54 | 0.04640265 | 0.05001105 | -0.02641950 | -0.03072851 | 0.01202763 | 0.01040047 | -0.00156838 | -0.00097248 | 0.00082897 | -0.00036829 | -0.00014848 | 0.00006046 | 0.00021109 | -0.00007792 | 0.00017325 | -0.00026083 | 0.00001474 |
| 31 | 55 | 0.04685191 | 0.04980726 | -0.02702171 | -0.03038127 | 0.01252856 | 0.01009815 | -0.00174299 | -0.00069242 | 0.00121532 | -0.00021856 | -0.00010769 | -0.00003043 | 0.00027318 | -0.00010949 | 0.00025097 | -0.00034742 | 0.00002591 |
| 31 | 56 | 0.04630370 | 0.04880936 | -0.02610585 | -0.02808447 | 0.01261550 | 0.00846934 | -0.00206644 | -0.00048725 | 0.00109047 | -0.00025542 | -0.00016769 | 0.00007013 | 0.00022252 | -0.00003531 | 0.00020331 | -0.00032463 | -0.00003361 |
| 31 | 57 | 0.04687407 | 0.04858571 | -0.02694564 | -0.02779269 | 0.01343902 | 0.00861201 | -0.00219355 | -0.00028970 | 0.00146209 | -0.00011006 | -0.00013490 | 0.00001368 | 0.00023247 | -0.00005635 | 0.00025115 | -0.00036122 | -0.00002262 |
| 32 | 26 | 0.04838337 | 0.06881844 | 0.00035815 | -0.03447016 | -0.00666403 | 0.00779366 | 0.00235019 | -0.00135501 | 0.00076429 | 0.00091340 | -0.00026321 | -0.00018727 | 0.00037434 | 0.00009489 | 0.00026932 | -0.00051643 | -0.00017683 |
| 32 | 27 | 0.04871593 | 0.06890629 | 0.00000856 | -0.03412448 | -0.00679325 | 0.00816814 | 0.00223553 | -0.00123831 | 0.00052869 | 0.00087482 | -0.00029283 | -0.00001007 | 0.00012585 | 0.00015416 | 0.00011938 | -0.00042379 | -0.00034018 |
| 32 | 28 | 0.04935811 | 0.06814264 | -0.00316367 | -0.03517092 | -0.00531396 | 0.00941141 | 0.00163870 | -0.00109666 | 0.00072016 | 0.00085119 | -0.00043387 | 0.00015457 | -0.00005766 | 0.00021376 | 0.00001327 | -0.00021314 | -0.00037313 |
| 32 | 29 | 0.04945978 | 0.06743418 | -0.00403344 | -0.03471313 | -0.00427964 | 0.01009425 | 0.00180902 | -0.00119945 | 0.00045201 | 0.00066916 | -0.00035640 | 0.00015426 | -0.00010056 | 0.00017454 | -0.00001931 | -0.00022699 | -0.00039346 |
| 32 | 30 | 0.04974404 | 0.06605549 | -0.00701359 | -0.03504258 | -0.00214077 | 0.01091703 | 0.00145447 | -0.00123696 | 0.00056649 | 0.00054696 | -0.00039480 | 0.00014384 | -0.00006257 | 0.00015403 | -0.00000952 | -0.00013116 | -0.00028533 |
| 32 | 31 | 0.04976367 | 0.06531796 | -0.00770578 | -0.03440353 | -0.00102831 | 0.01158272 | 0.00154579 | -0.00136447 | 0.00036238 | 0.00039909 | -0.00032832 | 0.00013911 | -0.00009697 | 0.00011576 | -0.00003811 | -0.00014210 | -0.00029355 |
| 32 | 32 | 0.04992838 | 0.06407475 | -0.01031843 | -0.03436530 | 0.00098088 | 0.01186626 | 0.00104374 | -0.00140437 | 0.00045483 | 0.00026005 | -0.00035288 | 0.00011212 | -0.00001901 | 0.00008437 | -0.00000532 | -0.00008334 | -0.00018988 |

TABLE 3. Nuclear charge density distribution FB coefficients.

| $Z$ | $N$ | $a_1$ | $a_2$ | $a_3$ | $a_4$ | $a_5$ | $a_6$ | $a_7$ | $a_8$ | $a_9$ | $a_{10}$ | $a_{11}$ | $a_{12}$ | $a_{13}$ | $a_{14}$ | $a_{15}$ | $a_{16}$ | $a_{17}$ |
|---|---|---|---|---|---|---|---|---|---|---|---|---|---|---|---|---|---|---|
| 32 | 33 | 0.04984941 | 0.06336319 | -0.01069609 | -0.03340211 | 0.00214925 | 0.01238257 | 0.00100996 | -0.00154754 | 0.00033875 | 0.00015327 | -0.00029981 | 0.00010504 | -0.00004288 | 0.00004748 | -0.00002799 | -0.00009247 | -0.00018565 |
| 32 | 34 | 0.04994541 | 0.06229737 | -0.01295683 | -0.03316472 | 0.00384722 | 0.01216479 | 0.00043699 | -0.00157256 | 0.00039391 | 0.00001146 | -0.00031013 | 0.00006435 | 0.00006351 | 0.00001145 | 0.00002115 | -0.00006633 | -0.00009235 |
| 32 | 35 | 0.04979988 | 0.06164814 | -0.01301010 | -0.03181991 | 0.00503570 | 0.01242995 | 0.00024225 | -0.00172427 | 0.00037415 | -0.00005127 | -0.00027088 | 0.00005981 | 0.00004556 | -0.00002301 | 0.00000170 | -0.00007017 | -0.00007769 |
| 32 | 36 | 0.04988902 | 0.06075259 | -0.01503260 | -0.03159927 | 0.00633741 | 0.01186846 | -0.00031624 | -0.00171669 | 0.00037826 | -0.00019188 | -0.00026799 | 0.00001099 | 0.00016305 | -0.00005780 | 0.00005888 | -0.00006929 | -0.00000173 |
| 32 | 37 | 0.04974156 | 0.06016503 | -0.01488576 | -0.02992282 | 0.00751748 | 0.01183881 | -0.00069613 | -0.00187972 | 0.00044896 | -0.00021463 | -0.00024055 | 0.00001515 | 0.00014213 | -0.00008973 | 0.00003785 | -0.00006257 | 0.00001951 |
| 32 | 38 | 0.04986507 | 0.05941235 | -0.01677301 | -0.02988028 | 0.00842691 | 0.01111737 | -0.00117269 | -0.00182880 | 0.00039485 | -0.00035592 | -0.00022637 | -0.00003772 | 0.00025730 | -0.00011926 | 0.00009792 | -0.00008250 | 0.00007286 |
| 32 | 39 | 0.04977642 | 0.05884780 | -0.01662320 | -0.02801418 | 0.00959350 | 0.01082046 | -0.00174431 | -0.00201782 | 0.00054408 | -0.00035026 | -0.00020702 | -0.00002008 | 0.00022544 | -0.00015150 | 0.00007213 | -0.00006170 | 0.00009670 |
| 32 | 40 | 0.04980622 | 0.05801108 | -0.01878713 | -0.02950622 | 0.00935141 | 0.01061008 | -0.00190349 | -0.00182317 | 0.00042240 | -0.00036715 | -0.00023753 | 0.00001229 | 0.00022400 | -0.00010645 | 0.00008375 | -0.00005420 | 0.00006107 |
| 32 | 41 | 0.04960527 | 0.05736400 | -0.01912237 | -0.02946071 | 0.00944935 | 0.01085748 | -0.00237892 | -0.00197503 | 0.00056256 | -0.00021957 | -0.00028354 | 0.00012045 | 0.00008601 | -0.00006745 | 0.00001370 | -0.00000215 | 0.00000140 |
| 32 | 42 | 0.04959868 | 0.05652201 | -0.02121060 | -0.03054396 | 0.00964356 | 0.01086223 | -0.00212436 | -0.00181658 | 0.00039949 | -0.00029556 | -0.00025504 | 0.00009895 | 0.00013369 | -0.00006227 | 0.00004271 | -0.00000803 | 0.00002661 |
| 32 | 43 | 0.04939202 | 0.05601759 | -0.02169876 | -0.03098792 | 0.00979494 | 0.01140369 | -0.00247396 | -0.00203199 | 0.00051912 | -0.00018382 | -0.00028738 | 0.00017764 | 0.00003014 | -0.00004335 | -0.00000803 | 0.00001854 | -0.00002164 |
| 32 | 44 | 0.04934888 | 0.05525525 | -0.02344126 | -0.03154025 | 0.01023196 | 0.01142176 | -0.00199691 | -0.00184633 | 0.00034987 | -0.00028808 | -0.00023347 | 0.00013195 | 0.00009890 | -0.00005219 | 0.00002639 | 0.00000801 | 0.00003198 |
| 32 | 45 | 0.04916445 | 0.05489974 | -0.02408619 | -0.03237936 | 0.01053334 | 0.01227289 | -0.00220049 | -0.00205646 | 0.00045903 | -0.00024462 | -0.00024772 | 0.00017996 | 0.00002644 | -0.00005864 | -0.00000189 | 0.00000846 | -0.00000599 |
| 32 | 46 | 0.04908127 | 0.05420866 | -0.02540954 | -0.03240011 | 0.01104212 | 0.01213869 | -0.00171427 | -0.00183201 | 0.00031058 | -0.00034499 | -0.00019700 | 0.00014033 | 0.00008727 | -0.00005961 | 0.00002552 | 0.00000337 | 0.00004677 |
| 32 | 47 | 0.04894932 | 0.05396023 | -0.02626523 | -0.03356987 | 0.01148067 | 0.01318616 | -0.00183436 | -0.00198579 | 0.00041317 | -0.00039292 | -0.00019192 | 0.00016395 | 0.00003420 | -0.00009077 | 0.00002040 | -0.00002270 | 0.00001215 |
| 32 | 48 | 0.04882178 | 0.05330563 | -0.02720034 | -0.03313192 | 0.01196620 | 0.01284236 | -0.00146455 | -0.00174176 | 0.00030078 | -0.00046340 | -0.00015975 | 0.00014668 | 0.00007127 | -0.00007244 | 0.00003350 | -0.00001515 | 0.00004928 |
| 32 | 49 | 0.04875255 | 0.05310583 | -0.02829626 | -0.03456854 | 0.01246625 | 0.01392799 | -0.00158998 | -0.00181645 | 0.00039180 | -0.00060766 | -0.00013826 | 0.00015501 | 0.00002517 | -0.00012220 | 0.00004905 | -0.00006688 | 0.00000944 |
| 32 | 50 | 0.04858318 | 0.05246533 | -0.02890961 | -0.03376800 | 0.01290877 | 0.01341426 | -0.00136743 | -0.00157941 | 0.00032075 | -0.00063171 | -0.00012875 | 0.00016313 | 0.00003551 | -0.00008248 | 0.00004543 | -0.00004272 | 0.00002854 |
| 32 | 51 | 0.04839315 | 0.05210327 | -0.02891148 | -0.03389362 | 0.01293981 | 0.01322330 | -0.00161356 | -0.00156337 | 0.00046929 | -0.00065456 | -0.00010872 | 0.00012764 | 0.00006516 | -0.00012483 | 0.00009254 | -0.00012518 | 0.00002134 |
| 32 | 52 | 0.04809323 | 0.05129706 | -0.02840457 | -0.03183652 | 0.01278318 | 0.01146679 | -0.00151991 | -0.00130786 | 0.00041929 | -0.00052293 | -0.00009779 | 0.00009504 | 0.00014594 | -0.00007174 | 0.00010696 | -0.00012667 | 0.00006284 |
| 32 | 53 | 0.04799631 | 0.05090106 | -0.02828254 | -0.03142986 | 0.01311666 | 0.01117160 | -0.00184943 | -0.00117708 | 0.00066759 | -0.00051179 | -0.00010036 | 0.00008290 | 0.00015055 | -0.00009925 | 0.00014641 | -0.00019957 | 0.00003962 |
| 32 | 54 | 0.04778758 | 0.05011297 | -0.02804583 | -0.02971398 | 0.01288849 | 0.00952407 | -0.00179658 | -0.00093411 | 0.00055442 | -0.00043367 | -0.00007434 | 0.00004538 | 0.00022297 | -0.00005396 | 0.00016172 | -0.00021709 | 0.00005879 |
| 32 | 55 | 0.04778014 | 0.04967133 | -0.02779171 | -0.02872347 | 0.01352322 | 0.00910886 | -0.00220651 | -0.00067865 | 0.00089390 | -0.00039597 | -0.00009872 | 0.00006126 | 0.00019597 | -0.00006562 | 0.00018863 | -0.00027283 | 0.00002005 |
| 32 | 56 | 0.04765347 | 0.04893970 | -0.02782330 | -0.02751476 | 0.01317715 | 0.00770906 | -0.00217017 | -0.00050349 | 0.00070597 | -0.00036924 | -0.00005992 | 0.00001848 | 0.00025926 | -0.00002974 | 0.00020301 | -0.00029992 | 0.00002161 |
| 32 | 57 | 0.04770412 | 0.04844759 | -0.02743631 | -0.02599347 | 0.01405838 | 0.00719807 | -0.00262838 | -0.00013971 | 0.00111711 | -0.00031397 | -0.00010260 | 0.00006395 | 0.00019801 | -0.00002766 | 0.00021316 | -0.00032913 | -0.00002648 |
| 32 | 58 | 0.04764409 | 0.04779379 | -0.02770751 | -0.02544579 | 0.01350638 | 0.00617485 | -0.00256857 | -0.00007996 | 0.00084982 | -0.00032629 | -0.00005792 | 0.00001892 | 0.00024747 | 0.00000155 | 0.00022253 | -0.00035704 | -0.00003975 |
| 33 | 27 | 0.04953500 | 0.06887970 | -0.00059567 | -0.03234249 | -0.00513340 | 0.00781324 | 0.00210977 | -0.00116228 | 0.00032916 | 0.00088300 | -0.00024215 | -0.00002222 | 0.00017208 | 0.00019517 | 0.00013789 | -0.00049819 | -0.00037653 |
| 33 | 28 | 0.05019970 | 0.06865635 | -0.00292640 | -0.03368505 | -0.00425655 | 0.00888952 | 0.00076217 | -0.00156842 | 0.00054986 | 0.00081749 | -0.00041750 | 0.00024102 | -0.00011591 | 0.00021799 | -0.00000878 | -0.00027391 | -0.00048288 |
| 33 | 29 | 0.05026544 | 0.06742048 | -0.00470598 | -0.03317221 | -0.00270606 | 0.00969193 | 0.00158894 | -0.00118154 | 0.00029491 | 0.00068250 | -0.00029638 | 0.00012590 | -0.00003653 | 0.00020038 | 0.00001414 | -0.00031113 | -0.00041739 |
| 33 | 30 | 0.05071346 | 0.06643742 | -0.00709390 | -0.03371886 | -0.00098177 | 0.01069063 | 0.00065685 | -0.00168086 | 0.00039914 | 0.00051882 | -0.00037796 | 0.00023288 | -0.00014338 | 0.00016107 | -0.00004850 | -0.00016177 | -0.00038137 |
| 33 | 31 | 0.05073684 | 0.06547613 | -0.00825050 | -0.03273514 | 0.00031338 | 0.01105573 | 0.00137303 | -0.00129864 | 0.00021130 | 0.00047382 | -0.00027165 | 0.00012267 | -0.00005686 | 0.00016115 | -0.00002296 | -0.00021412 | -0.00032891 |
| 33 | 32 | 0.05099405 | 0.06432713 | -0.01062383 | -0.03307061 | 0.00226550 | 0.01185973 | 0.00032974 | -0.00178164 | 0.00030979 | 0.00023175 | -0.00033289 | 0.00020029 | -0.00011757 | 0.00009229 | -0.00005693 | -0.00008871 | -0.00026913 |
| 33 | 33 | 0.05100733 | 0.06364535 | -0.01121084 | -0.03167324 | 0.00329496 | 0.01178122 | 0.00090502 | -0.00141291 | 0.00021227 | 0.00028308 | -0.00024451 | 0.00009969 | -0.00003126 | 0.00011252 | -0.00002878 | -0.00015242 | -0.00023036 |
| 33 | 34 | 0.05107034 | 0.06245981 | -0.01334206 | -0.03173004 | 0.00524600 | 0.01221058 | -0.00022334 | -0.00186085 | 0.00029163 | -0.00001736 | -0.00028479 | 0.00014639 | -0.00004118 | 0.00001751 | -0.00003561 | -0.00005672 | -0.00015219 |
| 33 | 35 | 0.05114748 | 0.06197244 | -0.01366543 | -0.03014568 | 0.00608597 | 0.01182239 | 0.00018673 | -0.00152411 | 0.00029121 | 0.00011896 | -0.00021623 | 0.00006333 | 0.00002985 | 0.00005867 | -0.00000833 | -0.00012327 | -0.00013011 |
| 33 | 36 | 0.05105364 | 0.06085747 | -0.01542547 | -0.02992982 | 0.00782753 | 0.01180513 | -0.00097388 | -0.00191422 | 0.00033444 | -0.00022117 | -0.00023591 | 0.00008327 | 0.00006262 | -0.00005662 | 0.00000398 | -0.00005478 | -0.00004191 |
| 33 | 37 | 0.05125618 | 0.06043281 | -0.01587464 | -0.02846817 | 0.00861056 | 0.01131549 | -0.00074878 | -0.00164196 | 0.00043083 | -0.00002212 | -0.00018867 | 0.00002386 | 0.00010679 | 0.00000284 | 0.00002936 | -0.00011825 | -0.00003919 |
| 33 | 38 | 0.05106370 | 0.05945708 | -0.01721765 | -0.02801176 | 0.00999709 | 0.01086397 | -0.00188079 | -0.00195545 | 0.00041932 | -0.00038887 | -0.00018735 | 0.00002390 | 0.00016710 | -0.00012672 | 0.00005027 | -0.00007023 | 0.00005018 |
| 33 | 39 | 0.05139746 | 0.05896964 | -0.01808556 | -0.02697210 | 0.01085549 | 0.01050392 | -0.00182995 | -0.00178341 | 0.00061220 | -0.00015040 | -0.00016290 | -0.00001089 | 0.00018256 | -0.00005500 | 0.00007689 | -0.00012879 | 0.00003470 |
| 33 | 40 | 0.05105128 | 0.05808565 | -0.01936694 | -0.02786262 | 0.01077232 | 0.01022098 | -0.00275697 | -0.00195368 | 0.00048975 | -0.00038340 | -0.00020606 | 0.00007451 | 0.00013266 | -0.00011200 | 0.00003974 | -0.00003813 | 0.00003965 |
| 33 | 41 | 0.05129415 | 0.05767747 | -0.02058049 | -0.02869716 | 0.01044283 | 0.01046284 | -0.00257926 | -0.00179089 | 0.00060973 | -0.00004480 | -0.00022202 | 0.00011424 | 0.00004854 | 0.00000517 | 0.00001852 | -0.00005549 | -0.00004730 |
| 33 | 42 | 0.05088635 | 0.05674673 | -0.02194151 | -0.02947050 | 0.01075466 | 0.01051602 | -0.00309395 | -0.00200461 | 0.00047320 | -0.00028540 | -0.00024005 | 0.00016800 | 0.00003157 | -0.00005837 | -0.00000236 | 0.00001909 | -0.00000298 |
| 33 | 43 | 0.05109915 | 0.05645345 | -0.02318941 | -0.03051352 | 0.01056150 | 0.01096181 | -0.00278037 | -0.00187555 | 0.00056713 | -0.00003701 | -0.00021477 | 0.00015741 | -0.00000127 | 0.00000503 | 0.00000065 | -0.00002302 | -0.00005673 |
| 33 | 44 | 0.05067600 | 0.05562117 | -0.02437467 | -0.03100748 | 0.01121584 | 0.01128486 | -0.00296130 | -0.00207468 | 0.00042638 | -0.00028120 | -0.00022471 | 0.00019834 | -0.00000412 | -0.00005152 | -0.00001388 | 0.00003897 | 0.00000119 |
| 33 | 45 | 0.05085260 | 0.05539502 | -0.02566338 | -0.03219546 | 0.01116705 | 0.01183573 | -0.00258149 | -0.00191880 | 0.00052267 | -0.00013162 | -0.00016640 | 0.00014188 | 0.00000941 | -0.00003726 | 0.00001745 | -0.00002977 | -0.00002431 |
| 33 | 46 | 0.05046165 | 0.05468366 | -0.02661525 | -0.03235001 | 0.01205128 | 0.01227998 | -0.00260845 | -0.00206937 | 0.00038874 | -0.00037139 | -0.00018554 | 0.00019726 | -0.00000962 | -0.00007220 | -0.00000388 | 0.00002868 | 0.00001789 |
| 33 | 47 | 0.05058747 | 0.05448243 | -0.02792990 | -0.03365110 | 0.01205641 | 0.01278760 | -0.00225241 | -0.00186003 | 0.00050310 | -0.00031546 | -0.00010534 | 0.00010464 | 0.00004091 | -0.00009652 | 0.00005657 | -0.00006748 | 0.00001277 |
| 33 | 48 | 0.05026246 | 0.05384705 | -0.02871520 | -0.03349095 | 0.01305835 | 0.01322048 | -0.00228934 | -0.00195765 | 0.00037994 | -0.00054128 | -0.00014258 | 0.00019361 | -0.00001855 | -0.00010094 | 0.00001801 | -0.00000347 | 0.00001963 |
| 33 | 49 | 0.05031559 | 0.05364612 | -0.02999703 | -0.03486026 | 0.01302646 | 0.01357724 | -0.00201454 | -0.00169266 | 0.00051749 | -0.00056204 | -0.00005222 | 0.00007449 | 0.00006127 | -0.00015146 | 0.00010547 | -0.00012555 | 0.00002868 |
| 33 | 50 | 0.05007443 | 0.05303071 | -0.03071269 | -0.03444076 | 0.01407712 | 0.01393667 | -0.00215413 | -0.00175049 | 0.00040015 | -0.00076441 | -0.00010811 | 0.00020365 | -0.00004990 | -0.00012336 | 0.00004333 | -0.00004915 | -0.00000758 |
| 33 | 51 | 0.05005694 | 0.05264904 | -0.03081012 | -0.03421289 | 0.01367743 | 0.01298347 | -0.00203165 | -0.00142598 | 0.00060753 | -0.00062208 | -0.00002008 | 0.00004170 | 0.00010294 | -0.00016145 | 0.00015190 | -0.00018266 | 0.00004718 |
| 33 | 52 | 0.04964176 | 0.05181095 | -0.03008887 | -0.03204683 | 0.01416123 | 0.01188514 | -0.00227529 | -0.00139894 | 0.00052782 | -0.00063041 | -0.00007550 | 0.00013366 | 0.00005814 | -0.00010592 | 0.00010299 | -0.00012305 | 0.00003781 |
| 33 | 53 | 0.04988106 | 0.05143378 | -0.03049679 | -0.03165673 | 0.01416454 | 0.01114829 | -0.00221559 | -0.00103437 | 0.00078708 | -0.00047802 | -0.00000185 | 0.00000408 | 0.00016328 | -0.00013147 | 0.00019012 | -0.00023125 | 0.00007069 |
| 33 | 54 | 0.04934951 | 0.05055350 | -0.02951778 | -0.02933910 | 0.01444801 | 0.00975991 | -0.00250159 | -0.00091118 | 0.00070098 | -0.00051860 | -0.00004972 | 0.00008215 | 0.00013361 | -0.00008184 | 0.00015516 | -0.00020215 | 0.00004699 |
| 33 | 55 | 0.04976339 | 0.05017997 | -0.03018282 | -0.02893898 | 0.01478783 | 0.00937926 | -0.00244445 | -0.00056472 | 0.00099171 | -0.00035957 | 0.00000755 | -0.00001328 | 0.00018904 | -0.00009459 | 0.00021736 | -0.00027197 | 0.00006900 |
| 33 | 56 | 0.04917427 | 0.04929109 | -0.02902032 | -0.02653244 | 0.01486378 | 0.00773348 | -0.00280572 | -0.00035603 | 0.00089146 | -0.00043795 | -0.00003305 | 0.00005488 | 0.00016790 | -0.00005290 | 0.00019209 | -0.00027037 | 0.00002515 |
| 33 | 57 | 0.04964954 | 0.04891571 | -0.02988725 | -0.02634407 | 0.01544182 | 0.00785418 | -0.00268449 | -0.00010584 | 0.00119545 | -0.00027464 | 0.00000741 | -0.00001005 | 0.00017987 | -0.00005596 | 0.00023071 | -0.00029480 | 0.00004946 |
| 33 | 58 | 0.04905794 | 0.04804976 | -0.02861355 | -0.02394982 | 0.01526465 | 0.00600881 | -0.00312713 | 0.00016723 | 0.00107094 | -0.00038804 | -0.00002732 | 0.00005355 | 0.00015844 | -0.00002237 | 0.00020824 | -0.00031179 | -0.00001676 |
| 33 | 59 | 0.04950004 | 0.04764823 | -0.02965931 | -0.02419850 | 0.01596798 | 0.00673516 | -0.00287992 | 0.00025549 | 0.00138213 | -0.00022163 | -0.00000876 | 0.00001872 | 0.00013175 | -0.00001716 | 0.00022610 | -0.00028678 | 0.00001837 |
| 34 | 29 | 0.05152090 | 0.06754006 | -0.00630938 | -0.03280467 | -0.00087711 | 0.00958082 | -0.00038217 | -0.00206312 | 0.00048973 | 0.00067389 | -0.00044098 | 0.00032440 | -0.00016879 | 0.00020751 | -0.00004889 | -0.00019099 | -0.00046485 |
| 34 | 30 | 0.05192065 | 0.06607477 | -0.00962196 | -0.03330776 | 0.00138745 | 0.01042834 | -0.00064341 | -0.00193246 | 0.00066290 | 0.00051982 | -0.00046082 | 0.00027078 | -0.00009216 | 0.00017150 | -0.00000480 | -0.00013343 | -0.00033968 |
| 34 | 31 | 0.05199936 | 0.06531951 | -0.01029391 | -0.03248680 | 0.00237692 | 0.01107113 | -0.00054115 | -0.00210520 | 0.00033721 | 0.00035955 | -0.00038363 | 0.00029950 | -0.00017777 | 0.00014174 | -0.00007995 | -0.00008923 | -0.00035311 |
| 34 | 32 | 0.05222195 | 0.06402314 | -0.01313371 | -0.03266075 | 0.00441819 | 0.01135998 | -0.00091957 | -0.00197630 | 0.00050105 | 0.00019985 | -0.00038694 | 0.00022150 | -0.00005419 | 0.00009334 | -0.00000755 | -0.00007366 | -0.00022732 |
| 34 | 33 | 0.05224280 | 0.06330895 | -0.01344690 | -0.03144884 | 0.00539439 | 0.01176450 | -0.00087854 | -0.00209481 | 0.00025309 | 0.00007120 | -0.00031814 | 0.00024568 | -0.00012908 | 0.00006693 | -0.00007639 | -0.00003564 | -0.00023323 |
| 34 | 34 | 0.05233755 | 0.06224139 | -0.01579843 | -0.03142615 | 0.00705814 | 0.01152420 | -0.00133172 | -0.00197686 | 0.00039236 | -0.00007898 | -0.00030843 | 0.00014974 | 0.00002967 | 0.00001271 | 0.00001701 | -0.00005693 | -0.00011446 |
| 34 | 35 | 0.05232193 | 0.06157665 | -0.01581405 | -0.02982968 | 0.00803242 | 0.01161882 | -0.00141174 | -0.00204177 | 0.00024359 | -0.00017568 | -0.00024970 | 0.00017182 | -0.00003473 | -0.00001128 | -0.00004311 | -0.00002756 | -0.00011519 |
| 34 | 36 | 0.05236711 | 0.06072646 | -0.01781310 | -0.02980874 | 0.00926909 | 0.01102903 | -0.00187645 | -0.00193638 | 0.00033680 | -0.00031406 | -0.00023075 | 0.00006884 | 0.00013678 | -0.00006430 | 0.00005899 | -0.00007329 | -0.00001259 |
| 34 | 37 | 0.05235238 | 0.06007715 | -0.01772521 | -0.02796605 | 0.01027005 | 0.01082217 | -0.00213465 | -0.00197256 | 0.00029803 | -0.00038402 | -0.00018277 | 0.00009256 | 0.00007919 | -0.00008845 | 0.00000872 | -0.00005229 | -0.00001142 |
| 34 | 38 | 0.05241465 | 0.05940750 | -0.01950517 | -0.02808845 | 0.01110032 | 0.01008422 | -0.00254886 | -0.00187560 | 0.00032601 | -0.00051451 | -0.00015735 | -0.00000848 | 0.00024347 | -0.00013464 | 0.00010889 | -0.00011136 | 0.00006813 |
| 34 | 39 | 0.05242628 | 0.05871376 | -0.01954554 | -0.02622072 | 0.01216991 | 0.00967077 | -0.00301154 | -0.00192287 | 0.00040033 | -0.00056628 | -0.00011997 | 0.00002002 | 0.00018888 | -0.00016367 | 0.00006956 | -0.00009645 | 0.00006908 |
| 34 | 40 | 0.05241745 | 0.05806428 | -0.02157505 | -0.02787255 | 0.01182410 | 0.00947793 | -0.00319422 | -0.00181473 | 0.00034970 | -0.00054034 | -0.00014667 | 0.00001787 | 0.00021906 | -0.00013003 | 0.00010499 | -0.00008736 | 0.00007344 |
| 34 | 41 | 0.05234942 | 0.05740840 | -0.02208031 | -0.02779746 | 0.01181324 | 0.00964358 | -0.00360858 | -0.00193032 | 0.00041137 | -0.00041198 | -0.00018039 | 0.00014588 | 0.00004454 | -0.00008104 | 0.00000736 | -0.00000084 | 0.00000896 |
| 34 | 42 | 0.05224680 | 0.05675443 | -0.02394765 | -0.02909075 | 0.01187277 | 0.00968868 | -0.00339763 | -0.00182964 | 0.00034264 | -0.00044839 | -0.00016264 | 0.00009873 | 0.00012069 | -0.00008266 | 0.00006338 | -0.00001752 | 0.00005646 |
| 34 | 43 | 0.05217794 | 0.05629172 | -0.02451463 | -0.02940474 | 0.01191366 | 0.01016671 | -0.00368118 | -0.00202103 | 0.00037875 | -0.00034057 | -0.00018947 | 0.00020597 | -0.00002807 | -0.00004777 | -0.00002301 | 0.00005399 | 0.00000105 |
| 34 | 44 | 0.05202048 | 0.05566509 | -0.02608378 | -0.03020817 | 0.01227676 | 0.01026673 | -0.00327221 | -0.00187393 | 0.00031726 | -0.00041435 | -0.00015404 | 0.00014071 | 0.00006842 | -0.00006319 | 0.00004046 | 0.00002109 | 0.00006295 |
| 34 | 45 | 0.05197142 | 0.05536323 | -0.02678411 | -0.03086296 | 0.01251772 | 0.01109933 | -0.00341443 | -0.00207840 | 0.00034566 | -0.00036504 | -0.00016839 | 0.00022338 | -0.00005331 | -0.00005154 | -0.00002742 | 0.00007085 | 0.00001568 |
| 34 | 46 | 0.05177646 | 0.05474698 | -0.02801778 | -0.03115653 | 0.01303257 | 0.01108237 | -0.00299473 | -0.00187215 | 0.00030974 | -0.00044576 | -0.00013644 | 0.00016439 | 0.00003816 | -0.00006157 | 0.00003130 | 0.00003389 | 0.00007000 |
| 34 | 47 | 0.05176659 | 0.05454793 | -0.02892841 | -0.03212530 | 0.01346388 | 0.01215513 | -0.00306316 | -0.00203783 | 0.00034157 | -0.00048127 | -0.00013683 | 0.00022639 | -0.00006470 | -0.00007430 | -0.00001393 | 0.00005630 | 0.00002391 |
| 34 | 48 | 0.05154083 | 0.05392095 | -0.02984355 | -0.03196208 | 0.01401261 | 0.01194469 | -0.00273998 | -0.00178880 | 0.00033891 | -0.00054128 | -0.00011945 | 0.00018698 | 0.00000632 | -0.00006763 | 0.00003166 | 0.00002634 | 0.00006057 |
| 34 | 49 | 0.05157205 | 0.05376469 | -0.03098718 | -0.03320111 | 0.01453306 | 0.01307998 | -0.00282962 | -0.00188853 | 0.00037406 | -0.00066713 | -0.00010937 | 0.00023639 | -0.00008878 | -0.00009856 | 0.00000827 | 0.00001957 | 0.00000647 |
| 34 | 50 | 0.05132402 | 0.05311928 | -0.03163389 | -0.03266872 | 0.01506749 | 0.01269894 | -0.00262905 | -0.00162487 | 0.00040376 | -0.00068771 | -0.00010855 | 0.00021887 | -0.00004250 | -0.00007243 | 0.00003681 | 0.00000441 | 0.00002541 |
| 34 | 51 | 0.05128125 | 0.05273929 | -0.03172200 | -0.03248401 | 0.01510502 | 0.01243298 | -0.00282078 | -0.00160673 | 0.00043291 | -0.00071610 | -0.00007330 | 0.00020484 | -0.00005287 | -0.00010009 | 0.00004977 | -0.00003987 | 0.00001401 |
| 34 | 52 | 0.05091501 | 0.05189491 | -0.03117931 | -0.03058230 | 0.01497307 | 0.01069210 | -0.00273857 | -0.00131993 | 0.00041859 | -0.00060512 | -0.00004756 | 0.00012921 | 0.00007170 | -0.00006533 | 0.00009884 | -0.00008410 | 0.00006496 |
| 34 | 53 | 0.05096522 | 0.05145370 | -0.03117414 | -0.02984132 | 0.01532064 | 0.01031558 | -0.00294219 | -0.00115153 | 0.00053649 | -0.00061156 | -0.00002375 | 0.00012935 | 0.00004293 | -0.00008311 | 0.00010677 | -0.00011894 | 0.00004717 |
| 34 | 54 | 0.05062699 | 0.05066395 | -0.03074974 | -0.02831208 | 0.01503339 | 0.00873247 | -0.00292132 | -0.00092289 | 0.00046630 | -0.00054851 | 0.00000613 | 0.00005614 | 0.00015446 | -0.00005535 | 0.00015233 | -0.00017437 | 0.00007194 |
| 34 | 55 | 0.05073150 | 0.05015481 | -0.03062165 | -0.02701783 | 0.01566488 | 0.00824009 | -0.00313419 | -0.00060341 | 0.00067607 | -0.00053750 | 0.00001639 | 0.00007380 | 0.00010501 | -0.00006336 | 0.00015334 | -0.00019479 | 0.00004991 |
| 34 | 56 | 0.05043279 | 0.04945343 | -0.03033732 | -0.02605479 | 0.01515632 | 0.00698431 | -0.00316254 | -0.00050375 | 0.00052813 | -0.00051771 | 0.00004612 | 0.00000751 | 0.00019511 | -0.00004173 | 0.00018829 | -0.00024958 | 0.00004888 |
| 34 | 57 | 0.05053139 | 0.04887356 | -0.03009371 | -0.02434061 | 0.01600795 | 0.00641813 | -0.00336900 | -0.00006663 | 0.00082403 | -0.00049471 | 0.00004246 | 0.00004365 | 0.00012659 | -0.00004282 | 0.00018209 | -0.00025081 | 0.00002835 |
| 34 | 58 | 0.05031089 | 0.04824830 | -0.02998179 | -0.02408831 | 0.01516926 | 0.00562462 | -0.00340758 | -0.00013836 | 0.00058401 | -0.00050246 | 0.00006183 | -0.00000352 | 0.00017599 | -0.00001905 | 0.00019408 | -0.00028714 | -0.00000221 |
| 34 | 59 | 0.05033601 | 0.04759736 | -0.02967076 | -0.02217962 | 0.01616697 | 0.00505645 | -0.00357449 | 0.00036161 | 0.00095779 | -0.00047370 | 0.00004575 | 0.00004863 | 0.00009543 | -0.00001870 | 0.00018393 | -0.00026707 | -0.00001203 |
| 34 | 60 | 0.05057013 | 0.04668444 | -0.03029949 | -0.02253712 | 0.01493304 | 0.00472112 | -0.00339820 | 0.00034926 | 0.00066984 | -0.00047493 | 0.00001607 | 0.00008962 | 0.00000216 | 0.00005588 | 0.00012749 | -0.00021739 | -0.00009204 |
| 34 | 61 | 0.05054384 | 0.04588768 | -0.03014248 | -0.02062112 | 0.01602454 | 0.00417980 | -0.00340930 | 0.00092103 | 0.00110909 | -0.00045452 | -0.00000936 | 0.00016175 | -0.00009403 | 0.00005755 | 0.00011454 | -0.00016712 | -0.00007769 |
| 35 | 30 | 0.05285818 | 0.06629691 | -0.00977354 | -0.03163880 | 0.00255246 | 0.01004037 | -0.00155147 | -0.00247787 | 0.00042189 | 0.00052283 | -0.00044713 | 0.00039427 | -0.00021355 | 0.00019272 | -0.00009038 | -0.00010434 | -0.00043367 |
| 35 | 31 | 0.05285593 | 0.06543531 | -0.01042800 | -0.03021222 | 0.00358325 | 0.01009549 | -0.00075704 | -0.00204303 | 0.00011752 | 0.00048963 | -0.00033795 | 0.00030909 | -0.00015927 | 0.00021964 | -0.00009709 | -0.00013509 | -0.00039844 |
| 35 | 32 | 0.05327928 | 0.06418206 | -0.01330112 | -0.03089525 | 0.00554979 | 0.01104751 | -0.00170731 | -0.00240110 | 0.00027338 | 0.00020956 | -0.00036839 | 0.00034724 | -0.00020045 | 0.00012139 | -0.00010980 | -0.00002200 | -0.00031547 |
| 35 | 33 | 0.05329469 | 0.06355669 | -0.01350900 | -0.02915242 | 0.00629847 | 0.01061250 | -0.00103171 | -0.00194390 | 0.00006485 | 0.00025868 | -0.00027564 | 0.00026564 | -0.00013739 | 0.00016676 | -0.00010483 | -0.00007872 | -0.00029495 |
| 35 | 34 | 0.05349724 | 0.06232743 | -0.01598725 | -0.02951939 | 0.00818998 | 0.01120812 | -0.00201956 | -0.00225186 | 0.00019881 | -0.00007400 | -0.00028205 | 0.00027148 | -0.00013146 | 0.00004322 | -0.00009277 | 0.00000515 | -0.00019530 |
| 35 | 35 | 0.05358354 | 0.06181110 | -0.01615000 | -0.02777795 | 0.00881316 | 0.01048437 | -0.00152747 | -0.00181576 | 0.00011043 | 0.00004529 | -0.00021105 | 0.00019939 | -0.00007000 | 0.00010569 | -0.00007366 | -0.00007324 | -0.00019178 |
| 35 | 36 | 0.05359987 | 0.06072065 | -0.01808295 | -0.02778747 | 0.01045042 | 0.01062803 | -0.00252859 | -0.00207086 | 0.00020254 | -0.00032430 | -0.00019567 | 0.00017981 | -0.00002430 | -0.00003798 | -0.00004523 | -0.00001665 | -0.00008468 |
| 35 | 37 | 0.05378882 | 0.06016818 | -0.01861131 | -0.02642163 | 0.01112500 | 0.00990034 | -0.00225390 | -0.00170990 | 0.00024795 | -0.00015097 | -0.00015156 | 0.00012190 | 0.00002660 | 0.00003838 | -0.00001038 | -0.00011008 | -0.00009907 |
| 35 | 38 | 0.05368178 | 0.05927408 | -0.01996873 | -0.02607447 | 0.01239293 | 0.00958346 | -0.00323642 | -0.00190898 | 0.00027562 | -0.00054818 | -0.00011518 | 0.00008684 | 0.00009628 | -0.00012000 | 0.00002289 | -0.00007297 | 0.00000614 |
| 35 | 39 | 0.05394911 | 0.05858720 | -0.02108043 | -0.02538016 | 0.01324331 | 0.00913961 | -0.00314995 | -0.00167386 | 0.00045815 | -0.00033142 | -0.00010185 | 0.00004341 | 0.00013479 | -0.00003450 | 0.00007453 | -0.00017349 | -0.00002199 |
| 35 | 40 | 0.05374190 | 0.05797021 | -0.02216548 | -0.02607103 | 0.01299709 | 0.00887054 | -0.00400298 | -0.00183619 | 0.00034338 | -0.00056022 | -0.00011183 | 0.00011155 | 0.00007223 | -0.00011305 | 0.00002453 | -0.00004712 | 0.00001236 |
| 35 | 41 | 0.05398529 | 0.05748845 | -0.02351659 | -0.02696320 | 0.01269150 | 0.00900518 | -0.00382598 | -0.00170976 | 0.00042528 | -0.00019745 | -0.00014233 | 0.00016150 | -0.00001327 | 0.00002791 | -0.00000025 | -0.00005351 | -0.00007173 |
| 35 | 42 | 0.05364524 | 0.05688470 | -0.02455177 | -0.02763382 | 0.01272453 | 0.00904088 | -0.00431623 | -0.00190693 | 0.00032757 | -0.00042563 | -0.00014328 | 0.00020067 | -0.00003905 | -0.00005166 | -0.00002544 | 0.00004524 | -0.00000807 |
| 35 | 43 | 0.05385897 | 0.05651535 | -0.02589298 | -0.02862778 | 0.01260578 | 0.00941473 | -0.00399797 | -0.00182328 | 0.00037590 | -0.00014665 | -0.00013567 | 0.00021015 | -0.00008276 | 0.00003844 | -0.00003470 | 0.00002131 | -0.00006378 |
| 35 | 44 | 0.05346518 | 0.05598005 | -0.02679961 | -0.02913164 | 0.01298957 | 0.00974375 | -0.00420934 | -0.00200603 | 0.00029869 | -0.00037899 | -0.00014098 | 0.00023982 | -0.00009064 | -0.00003280 | -0.00004638 | 0.00009489 | 0.00000298 |
| 35 | 45 | 0.05363654 | 0.05567075 | -0.02816147 | -0.03021051 | 0.01306846 | 0.01026970 | -0.00382094 | -0.00191015 | 0.00034397 | -0.00019757 | -0.00010132 | 0.00020995 | -0.00009420 | 0.00000838 | -0.00003293 | 0.00004749 | -0.00002815 |
| 35 | 46 | 0.05326055 | 0.05519344 | -0.02895019 | -0.03046970 | 0.01375780 | 0.01077118 | -0.00389489 | -0.00203840 | 0.00029293 | -0.00043128 | -0.00012193 | 0.00025274 | -0.00011043 | -0.00004377 | -0.00004401 | 0.00010465 | 0.00001763 |
| 35 | 47 | 0.05336714 | 0.05490365 | -0.03032337 | -0.03164054 | 0.01392636 | 0.01129902 | -0.00353106 | -0.00190445 | 0.00035367 | -0.00034493 | -0.00006069 | 0.00019085 | -0.00008132 | -0.00004180 | -0.00000455 | 0.00003025 | 0.00000373 |

TABLE 3. Nuclear charge density distribution FB coefficients.

| $Z$ | $N$ | $a_1$ | $a_2$ | $a_3$ | $a_4$ | $a_5$ | $a_6$ | $a_7$ | $a_8$ | $a_9$ | $a_{10}$ | $a_{11}$ | $a_{12}$ | $a_{13}$ | $a_{14}$ | $a_{15}$ | $a_{16}$ | $a_{17}$ |
|---|---|---|---|---|---|---|---|---|---|---|---|---|---|---|---|---|---|---|
| 35 | 48 | 0.05305644 | 0.05444762 | -0.03104354 | -0.03164104 | 0.01480950 | 0.01183121 | -0.00359693 | -0.00196267 | 0.00032763 | -0.00057007 | -0.00010137 | 0.00026297 | -0.00012846 | -0.00006686 | -0.00002667 | 0.00008240 | 0.00001300 |
| 35 | 49 | 0.05307354 | 0.05415297 | -0.03237524 | -0.03288602 | 0.01495698 | 0.01224321 | -0.00332460 | -0.00178647 | 0.00041181 | -0.00056292 | -0.00003038 | 0.00017771 | -0.00007412 | -0.00009151 | 0.00003820 | -0.00001859 | 0.00001009 |
| 35 | 50 | 0.05285401 | 0.05368308 | -0.03308119 | -0.03264668 | 0.01593103 | 0.01271551 | -0.00345964 | -0.00178192 | 0.00039944 | -0.00076675 | -0.00008944 | 0.00028527 | -0.00016455 | -0.00008657 | -0.00000375 | 0.00003924 | -0.00002296 |
| 35 | 51 | 0.05287717 | 0.05313333 | -0.03330297 | -0.03222240 | 0.01566721 | 0.01173204 | -0.00328432 | -0.00149522 | 0.00047422 | -0.00062804 | 0.00001005 | 0.00014029 | -0.00003846 | -0.00010137 | 0.00008068 | -0.00007210 | 0.00002738 |
| 35 | 52 | 0.05256867 | 0.05237110 | -0.03264653 | -0.03012867 | 0.01609865 | 0.01066254 | -0.00351097 | -0.00137395 | 0.00042282 | -0.00066830 | -0.00001987 | 0.00019448 | -0.00006146 | -0.00007216 | 0.00005117 | -0.00003271 | 0.00002613 |
| 35 | 53 | 0.05277779 | 0.05182892 | -0.03309384 | -0.02960758 | 0.01613716 | 0.00991754 | -0.00329924 | -0.00103098 | 0.00055539 | -0.00052433 | 0.00006879 | 0.00007126 | 0.00003007 | -0.00008066 | 0.00011997 | -0.00012012 | 0.00006741 |
| 35 | 54 | 0.05232736 | 0.05103708 | -0.03212953 | -0.02734991 | 0.01640274 | 0.00860111 | -0.00361075 | -0.00085337 | 0.00049520 | -0.00060187 | 0.00004330 | 0.00011638 | 0.00001869 | -0.00005893 | 0.00010186 | -0.00010970 | 0.00004894 |
| 35 | 55 | 0.05261758 | 0.05050625 | -0.03276076 | -0.02697128 | 0.01666872 | 0.00825883 | -0.00334351 | -0.00054183 | 0.00067976 | -0.00044927 | 0.00011457 | 0.00001635 | 0.00007696 | -0.00006128 | 0.00015493 | -0.00016528 | 0.00008896 |
| 35 | 56 | 0.05208517 | 0.04971792 | -0.03155832 | -0.02463200 | 0.01672307 | 0.00674342 | -0.00375803 | -0.00032278 | 0.00059421 | -0.00056962 | 0.00009192 | 0.00005972 | 0.00006593 | -0.00004783 | 0.00013985 | -0.00017535 | 0.00004727 |
| 35 | 57 | 0.05234850 | 0.04919937 | -0.03233442 | -0.02465203 | 0.01712299 | 0.00694157 | -0.00343410 | -0.00014842 | 0.00082297 | -0.00040346 | 0.00013985 | -0.00001727 | 0.00009569 | -0.00004540 | 0.00017924 | -0.00019587 | 0.00009163 |
| 35 | 58 | 0.05180323 | 0.04842450 | -0.03099338 | -0.02235439 | 0.01688211 | 0.00529995 | -0.00393062 | 0.00009755 | 0.00069321 | -0.00056299 | 0.00011735 | 0.00003380 | 0.00006921 | -0.00003758 | 0.00015587 | -0.00021156 | 0.00002358 |
| 35 | 59 | 0.05203768 | 0.04787017 | -0.03195880 | -0.02290684 | 0.01730109 | 0.00605972 | -0.00352341 | 0.00009091 | 0.00096685 | -0.00037451 | 0.00012683 | -0.00000527 | 0.00005777 | -0.00002206 | 0.00017637 | -0.00018515 | 0.00006988 |
| 35 | 60 | 0.05168653 | 0.04694620 | -0.03084733 | -0.02073846 | 0.01668679 | 0.00436502 | -0.00394076 | 0.00046100 | 0.00079058 | -0.00055803 | 0.00009316 | 0.00008415 | -0.00003433 | 0.00000100 | 0.00012031 | -0.00016924 | -0.00002888 |
| 35 | 61 | 0.05262884 | 0.04577204 | -0.03314984 | -0.02116333 | 0.01734491 | 0.00512424 | -0.00322369 | 0.00077409 | 0.00116443 | -0.00037190 | 0.00004539 | 0.00016127 | -0.00020138 | 0.00007770 | 0.00007936 | -0.00001840 | 0.00002027 |
| 35 | 62 | 0.05351387 | 0.04365063 | -0.03442858 | -0.01810074 | 0.01740056 | 0.00286137 | -0.00340781 | 0.00208838 | 0.00107743 | -0.00064843 | 0.00004223 | 0.00035638 | -0.00050732 | 0.00015059 | -0.00005366 | 0.00011595 | -0.00008299 |
| 35 | 63 | 0.05426522 | 0.04261297 | -0.03697782 | -0.01837946 | 0.01904907 | 0.00374338 | -0.00307657 | 0.00259875 | 0.00160893 | -0.00064496 | 0.00004483 | 0.00043961 | -0.00067551 | 0.00016061 | -0.00007477 | 0.00029223 | 0.00005827 |
| 36 | 31 | 0.05416644 | 0.06504790 | -0.01293959 | -0.03004913 | 0.00576442 | 0.01006623 | -0.00266301 | -0.00274301 | 0.00035238 | 0.00038348 | -0.00043049 | 0.00044232 | -0.00024439 | 0.00017756 | -0.00012841 | -0.00002695 | -0.00039459 |
| 36 | 32 | 0.05442579 | 0.06379654 | -0.01570692 | -0.03039803 | 0.00756265 | 0.01024936 | -0.00286452 | -0.00244672 | 0.00055126 | 0.00020982 | -0.00040549 | 0.00031354 | -0.00008383 | 0.00011640 | -0.00002080 | -0.00005675 | -0.00025524 |
| 36 | 33 | 0.05453994 | 0.06310256 | -0.01590854 | -0.02891790 | 0.00836048 | 0.01052231 | -0.00278294 | -0.00252108 | 0.00021343 | 0.00007772 | -0.00033016 | 0.00037163 | -0.00020951 | 0.00010279 | -0.00013372 | 0.00002790 | -0.00027652 |
| 36 | 34 | 0.05465969 | 0.06208925 | -0.01822212 | -0.02912703 | 0.00982658 | 0.01024811 | -0.00306764 | -0.00225627 | 0.00039882 | -0.00008914 | -0.00029541 | 0.00022409 | -0.00001169 | 0.00003383 | -0.00000247 | -0.00004054 | -0.00014215 |
| 36 | 35 | 0.05475762 | 0.06139454 | -0.01822621 | -0.02737538 | 0.01059629 | 0.01021031 | -0.00307214 | -0.00224033 | 0.00015391 | -0.00020730 | -0.00022576 | 0.00027611 | -0.00012456 | 0.00002131 | -0.00010055 | 0.00002247 | -0.00016402 |
| 36 | 36 | 0.05481231 | 0.06062073 | -0.02017822 | -0.02757272 | 0.01170662 | 0.00963141 | -0.00341077 | -0.00203139 | 0.00030369 | -0.00036064 | -0.00018779 | 0.00012378 | 0.00008809 | -0.00004778 | 0.00004020 | -0.00006943 | -0.00004395 |
| 36 | 37 | 0.05489616 | 0.05985859 | -0.02023342 | -0.02576237 | 0.01253723 | 0.00934155 | -0.00358063 | -0.00196691 | 0.00017812 | -0.00047305 | -0.00012638 | 0.00016975 | -0.00000845 | -0.00006481 | -0.00003483 | -0.00003366 | -0.00006704 |
| 36 | 38 | 0.05496388 | 0.05929942 | -0.02192228 | -0.02603683 | 0.01329683 | 0.00863834 | -0.00392148 | -0.00181993 | 0.00026467 | -0.00061165 | -0.00008889 | 0.00002524 | 0.00019494 | -0.00012683 | 0.00009969 | -0.00013176 | 0.00003091 |
| 36 | 39 | 0.05501691 | 0.05841104 | -0.02223049 | -0.02440171 | 0.01426772 | 0.00822923 | -0.00428795 | -0.00175994 | 0.00027620 | -0.00072064 | -0.00003837 | 0.00006530 | 0.00011722 | -0.00015401 | 0.00005296 | -0.00012220 | 0.00000886 |
| 36 | 40 | 0.05504675 | 0.05804167 | -0.02398030 | -0.02593750 | 0.01382449 | 0.00796315 | -0.00449058 | -0.00169465 | 0.00027986 | -0.00065691 | -0.00006274 | 0.00003431 | 0.00017234 | -0.00012696 | 0.00010281 | -0.00011086 | 0.00004409 |
| 36 | 41 | 0.05504162 | 0.05737222 | -0.02454176 | -0.02579479 | 0.01370238 | 0.00801671 | -0.00484033 | -0.00176362 | 0.00027993 | -0.00053589 | -0.00008831 | 0.00018089 | -0.00003152 | -0.00006669 | -0.00001711 | 0.00000668 | -0.00002569 |
| 36 | 42 | 0.05492439 | 0.05694909 | -0.02615983 | -0.02710343 | 0.01362963 | 0.00800417 | -0.00471417 | -0.00170379 | 0.00028368 | -0.00053692 | -0.00008030 | 0.00011221 | 0.00006709 | -0.00007166 | 0.00005811 | -0.00001766 | 0.00003980 |
| 36 | 43 | 0.05491839 | 0.05651180 | -0.02670724 | -0.02721891 | 0.01357979 | 0.00834894 | -0.00494409 | -0.00186575 | 0.00025862 | -0.00041929 | -0.00010394 | 0.00024323 | -0.00011349 | -0.00002172 | -0.00005595 | 0.00009202 | -0.00002031 |
| 36 | 44 | 0.05472409 | 0.05603487 | -0.02815491 | -0.02818558 | 0.01385356 | 0.00846600 | -0.00464460 | -0.00176079 | 0.00027699 | -0.00047077 | -0.00008251 | 0.00015979 | 0.00000509 | -0.00004270 | 0.00003058 | 0.00004163 | 0.00005053 |
| 36 | 45 | 0.05472947 | 0.05576914 | -0.02880375 | -0.02855011 | 0.01404814 | 0.00916577 | -0.00474506 | -0.00195535 | 0.00024921 | -0.00039867 | -0.00009841 | 0.00026981 | -0.00014858 | -0.00001419 | -0.00006745 | 0.00013208 | -0.00000097 |
| 36 | 46 | 0.05449655 | 0.05523410 | -0.03004466 | -0.02913039 | 0.01454092 | 0.00926049 | -0.00441349 | -0.00178556 | 0.00029188 | -0.00047408 | -0.00007830 | 0.00019128 | -0.00003253 | -0.00003495 | 0.00001738 | 0.00006972 | 0.00005845 |
| 36 | 47 | 0.05452089 | 0.05507022 | -0.03088469 | -0.02975600 | 0.01497410 | 0.01021976 | -0.00444606 | -0.00195832 | 0.00027761 | -0.00047774 | -0.00008542 | 0.00028269 | -0.00016525 | -0.00003023 | -0.00005792 | 0.00013106 | 0.00000791 |
| 36 | 48 | 0.05426935 | 0.05447974 | -0.03190188 | -0.02996268 | 0.01555475 | 0.01019710 | -0.00416968 | -0.00173295 | 0.00034622 | -0.00054784 | -0.00007434 | 0.00022035 | -0.00006678 | -0.00003948 | 0.00001526 | 0.00007123 | 0.00004908 |
| 36 | 49 | 0.05430720 | 0.05435550 | -0.03295525 | -0.03083858 | 0.01611060 | 0.01124481 | -0.00422557 | -0.00185186 | 0.00034945 | -0.00063529 | -0.00007667 | 0.00030061 | -0.00018906 | -0.00005275 | -0.00003676 | 0.00009939 | -0.00001062 |
| 36 | 50 | 0.05405302 | 0.05371997 | -0.03376807 | -0.03072332 | 0.01670591 | 0.01109226 | -0.00403418 | -0.00159661 | 0.00043934 | -0.00067760 | -0.00007546 | 0.00025682 | -0.00011375 | -0.00004630 | 0.00001921 | 0.00005322 | 0.00001352 |
| 36 | 51 | 0.05409775 | 0.05330668 | -0.03382961 | -0.03009913 | 0.01677633 | 0.01069649 | -0.00418880 | -0.00156450 | 0.00037604 | -0.00068826 | -0.00003370 | 0.00026607 | -0.00016037 | -0.00005424 | -0.00000118 | 0.00004469 | -0.00000672 |
| 36 | 52 | 0.05376564 | 0.05245407 | -0.03337702 | -0.02852397 | 0.01665638 | 0.00914072 | -0.00414617 | -0.00128865 | 0.00036362 | -0.00061924 | 0.00000645 | 0.00015806 | -0.00001343 | -0.00003883 | 0.00006960 | -0.00002773 | 0.00004635 |
| 36 | 53 | 0.05390096 | 0.05194075 | -0.03342481 | -0.02741656 | 0.01704767 | 0.00869427 | -0.00422637 | -0.00108568 | 0.00037085 | -0.00062581 | 0.00004895 | 0.00017348 | -0.00007451 | -0.00004301 | 0.00004664 | -0.00002598 | 0.00002885 |
| 36 | 54 | 0.05352235 | 0.05120139 | -0.03288415 | -0.02620488 | 0.01669122 | 0.00733068 | -0.00429090 | -0.00093254 | 0.00031298 | -0.00059701 | 0.00008013 | 0.00007378 | 0.00005960 | -0.00003520 | 0.00011067 | -0.00010652 | 0.00005289 |
| 36 | 55 | 0.05366695 | 0.05058590 | -0.03288690 | -0.02470810 | 0.01736217 | 0.00685313 | -0.00429923 | -0.00057847 | 0.00040551 | -0.00060288 | 0.00011954 | 0.00009585 | -0.00001125 | -0.00003837 | 0.00008811 | -0.00009443 | 0.00004277 |
| 36 | 56 | 0.05329160 | 0.04997972 | -0.03229945 | -0.02402952 | 0.01666531 | 0.00583011 | -0.00447784 | -0.00062359 | 0.00026949 | -0.00060409 | 0.00013485 | 0.00001542 | 0.00009151 | -0.00003296 | 0.00013175 | -0.00016534 | 0.00003197 |
| 36 | 57 | 0.05334710 | 0.04926918 | -0.03225694 | -0.02235389 | 0.01754230 | 0.00537847 | -0.00442347 | -0.00017480 | 0.00045532 | -0.00061136 | 0.00016740 | 0.00004368 | 0.00001821 | -0.00003932 | 0.00011375 | -0.00014388 | 0.00003452 |
| 36 | 58 | 0.05315152 | 0.04869272 | -0.03181494 | -0.02223147 | 0.01638613 | 0.00473377 | -0.00464711 | -0.00039548 | 0.00022271 | -0.00061787 | 0.00015127 | 0.00001107 | 0.00004553 | -0.00001539 | 0.00011291 | -0.00017318 | -0.00002287 |
| 36 | 59 | 0.05303934 | 0.04789890 | -0.03175671 | -0.02062399 | 0.01737880 | 0.00437941 | -0.00452449 | 0.00008079 | 0.00050163 | -0.00062870 | 0.00017335 | 0.00004616 | -0.00002444 | -0.00002870 | 0.00010285 | -0.00014177 | -0.00000239 |
| 36 | 60 | 0.05407700 | 0.04647116 | -0.03308167 | -0.02019532 | 0.01612723 | 0.00356787 | -0.00444437 | 0.00039315 | 0.00029127 | -0.00061195 | 0.00009006 | 0.00017936 | -0.00025815 | 0.00009680 | -0.00001577 | -0.00001224 | -0.00012357 |
| 36 | 61 | 0.05395146 | 0.04544219 | -0.03342231 | -0.01864808 | 0.01729226 | 0.00322737 | -0.00417857 | 0.00098414 | 0.00064983 | -0.00065836 | 0.00011247 | 0.00023187 | -0.00034372 | 0.00007762 | -0.00002194 | 0.00004536 | -0.00007015 |
| 36 | 62 | 0.05516098 | 0.04369982 | -0.03579422 | -0.01831925 | 0.01716634 | 0.00295215 | -0.00423621 | 0.00207188 | 0.00074899 | -0.00080822 | 0.00004455 | 0.00049398 | -0.00078425 | 0.00019899 | -0.00020306 | 0.00030243 | -0.00013829 |
| 36 | 63 | 0.05449767 | 0.04285935 | -0.03583739 | -0.01747042 | 0.01832593 | 0.00313167 | -0.00392924 | 0.00250462 | 0.00109898 | -0.00095388 | 0.00009779 | 0.00051437 | -0.00082125 | 0.00012386 | -0.00018209 | 0.00035241 | -0.00003058 |
| 36 | 64 | 0.04419502 | 0.05037517 | -0.01942592 | -0.03254090 | 0.01001126 | 0.01478628 | -0.00320316 | -0.00226077 | 0.00055255 | -0.00087243 | -0.00035508 | 0.00052905 | -0.00059214 | -0.00007958 | -0.00023122 | 0.00035069 | -0.00012107 |
| 36 | 65 | 0.04370762 | 0.04933110 | -0.01984795 | -0.03158399 | 0.01109974 | 0.01471750 | -0.00299705 | -0.00188600 | 0.00085749 | -0.00005735 | -0.00028811 | 0.00053078 | -0.00061011 | -0.00017526 | -0.00020494 | 0.00040464 | -0.00000699 |
| 37 | 34 | 0.05578690 | 0.06209539 | -0.01811466 | -0.02665534 | 0.01076049 | 0.00951651 | -0.00375398 | -0.00246004 | 0.00016360 | -0.00004045 | -0.00027019 | 0.00037125 | -0.00020412 | 0.00008535 | -0.00014606 | 0.00004993 | -0.00024373 |
| 37 | 35 | 0.05585448 | 0.06155616 | -0.01822047 | -0.02497960 | 0.01123234 | 0.00854751 | -0.00332852 | -0.00204827 | 0.00001886 | 0.00007479 | -0.00020890 | 0.00031029 | -0.00014308 | 0.00016299 | -0.00012692 | -0.00003446 | -0.00024935 |
| 37 | 36 | 0.05602289 | 0.06047563 | -0.02027791 | -0.02519009 | 0.01266075 | 0.00887666 | -0.00405442 | -0.00208784 | 0.00012727 | -0.00034015 | -0.00015303 | 0.00025940 | -0.00010682 | -0.00000165 | -0.00009259 | 0.00000742 | -0.00014468 |
| 37 | 37 | 0.05615288 | 0.05981632 | -0.02091822 | -0.02410718 | 0.01326903 | 0.00798604 | -0.00387033 | -0.00177328 | 0.00013941 | -0.00018035 | -0.00012222 | 0.00020123 | -0.00003764 | 0.00008511 | -0.00004030 | -0.00010859 | -0.00016239 |
| 37 | 38 | 0.05619263 | 0.05895203 | -0.02236624 | -0.02391122 | 0.01437551 | 0.00789862 | -0.00459159 | -0.00177577 | 0.00017880 | -0.00063028 | -0.00004718 | 0.00014133 | 0.00001615 | -0.00009509 | -0.00000544 | -0.00008336 | -0.00006339 |
| 37 | 39 | 0.05632678 | 0.05813415 | -0.02359132 | -0.02362330 | 0.01520367 | 0.00738585 | -0.00458848 | -0.00161625 | 0.00035116 | -0.00041991 | -0.00005174 | 0.00008533 | 0.00009228 | -0.00000233 | 0.00007685 | -0.00021662 | -0.00008567 |
| 37 | 40 | 0.05632567 | 0.05773903 | -0.02452330 | -0.02399430 | 0.01483144 | 0.00714985 | -0.00525464 | -0.00163380 | 0.00023904 | -0.00066385 | -0.00002755 | 0.00014579 | -0.00000162 | -0.00009381 | 0.00000555 | -0.00006368 | -0.00004848 |
| 37 | 41 | 0.05647131 | 0.05730378 | -0.02570006 | -0.02476442 | 0.01456970 | 0.00705085 | -0.00517847 | -0.00163389 | 0.00030210 | -0.00025423 | -0.00007911 | 0.00019555 | -0.00005826 | 0.00006435 | -0.00001127 | -0.00006369 | -0.00011230 |
| 37 | 42 | 0.05629831 | 0.05693698 | -0.02656267 | -0.02522049 | 0.01436447 | 0.00704626 | -0.00558271 | -0.00169033 | 0.00023035 | -0.00048843 | -0.00005856 | 0.00023079 | -0.00011484 | -0.00002321 | -0.00005174 | 0.00005618 | -0.00005217 |
| 37 | 43 | 0.05639911 | 0.05661301 | -0.02767320 | -0.02597990 | 0.01433384 | 0.00719341 | -0.00535748 | -0.00174588 | 0.00025532 | -0.00015567 | -0.00007417 | 0.00024595 | -0.00013290 | 0.00008581 | -0.00005757 | 0.00004098 | -0.00008993 |
| 37 | 44 | 0.05614921 | 0.05625867 | -0.02853087 | -0.02643140 | 0.01445361 | 0.00748285 | -0.00556222 | -0.00180328 | 0.00022076 | -0.00039226 | -0.00006659 | 0.00027373 | -0.00017029 | 0.00000771 | -0.00007883 | 0.00012957 | -0.00003114 |
| 37 | 45 | 0.05619273 | 0.05598957 | -0.02962519 | -0.02719617 | 0.01466032 | 0.00780517 | -0.00525314 | -0.00185763 | 0.00023757 | -0.00015608 | -0.00005091 | 0.00025426 | -0.00015172 | 0.00006807 | -0.00006542 | 0.00009059 | -0.00004751 |
| 37 | 46 | 0.05594876 | 0.05562381 | -0.03052585 | -0.02757584 | 0.01512733 | 0.00833885 | -0.00534165 | -0.00187419 | 0.00023950 | -0.00039912 | -0.00006225 | 0.00029276 | -0.00019167 | 0.00000595 | -0.00008013 | 0.00015580 | -0.00001016 |
| 37 | 47 | 0.05591277 | 0.05536642 | -0.03160891 | -0.02838530 | 0.01545701 | 0.00870161 | -0.00504227 | -0.00189585 | 0.00026873 | -0.00026056 | -0.00002558 | 0.00024547 | -0.00014495 | 0.00002700 | -0.00004355 | 0.00008840 | -0.00001361 |
| 37 | 48 | 0.05572851 | 0.05497125 | -0.03257140 | -0.02865541 | 0.01618177 | 0.00936386 | -0.00509602 | -0.00184915 | 0.00030161 | -0.00050299 | -0.00005689 | 0.00030758 | -0.00020617 | -0.00001363 | -0.00006355 | 0.00014195 | -0.00000971 |
| 37 | 49 | 0.05559168 | 0.05469673 | -0.03360760 | -0.02952261 | 0.01651096 | 0.00965493 | -0.00487917 | -0.00182708 | 0.00035419 | -0.00044745 | -0.00001152 | 0.00024154 | -0.00014090 | -0.00001861 | -0.00000443 | 0.00004701 | -0.00000781 |
| 37 | 50 | 0.05549653 | 0.05426282 | -0.03463488 | -0.02966282 | 0.01737487 | 0.01033511 | -0.00495608 | -0.00171957 | 0.00040331 | -0.00067560 | -0.00005898 | 0.00033220 | -0.00023434 | -0.00003527 | -0.00003933 | 0.00010106 | -0.00004107 |
| 37 | 51 | 0.05546286 | 0.05365701 | -0.03464302 | -0.02884819 | 0.01726069 | 0.00924396 | -0.00479580 | -0.00153979 | 0.00037631 | -0.00051791 | 0.00003718 | 0.00020058 | -0.00011381 | -0.00002936 | 0.00003071 | 0.00000155 | 0.00000789 |
| 37 | 52 | 0.05538823 | 0.05287291 | -0.03439341 | -0.02710710 | 0.01760319 | 0.00840893 | -0.00496913 | -0.00129077 | 0.00031887 | -0.00061001 | 0.00003769 | 0.00023320 | -0.00015396 | -0.00002129 | -0.00000082 | 0.00004333 | -0.00000042 |
| 37 | 53 | 0.05544788 | 0.05228314 | -0.03454960 | -0.02634814 | 0.01770463 | 0.00762207 | -0.00469804 | -0.00108966 | 0.00034808 | -0.00045605 | 0.00012833 | 0.00011074 | -0.00005050 | -0.00001910 | 0.00006266 | -0.00003770 | 0.00005266 |
| 37 | 54 | 0.05521216 | 0.05148724 | -0.03394892 | -0.02444258 | 0.01790749 | 0.00661280 | -0.00499268 | -0.00081940 | 0.00027993 | -0.00059020 | 0.00012698 | 0.00014181 | -0.00008578 | -0.00001862 | 0.00003747 | -0.00002063 | 0.00002266 |
| 37 | 55 | 0.05523862 | 0.05093119 | -0.03419052 | -0.02400952 | 0.01812487 | 0.00626207 | -0.00465729 | -0.00071798 | 0.00037188 | -0.00042685 | 0.00020154 | 0.00002979 | 0.00000589 | -0.00002035 | 0.00009706 | -0.00008130 | 0.00008324 |
| 37 | 56 | 0.05490647 | 0.05014802 | -0.03331607 | -0.02204002 | 0.01811245 | 0.00514057 | -0.00507171 | -0.00044237 | 0.00026774 | -0.00060973 | 0.00019587 | 0.00006943 | -0.00004014 | -0.00002684 | 0.00006677 | -0.00007612 | 0.00002569 |
| 37 | 57 | 0.05481467 | 0.04963860 | -0.03356708 | -0.02208628 | 0.01833610 | 0.00525070 | -0.00474766 | -0.00055354 | 0.00042524 | -0.00041924 | 0.00024199 | -0.00002647 | 0.00004138 | -0.00002947 | 0.00012309 | -0.00011561 | 0.00009081 |
| 37 | 58 | 0.05450350 | 0.04882620 | -0.03261292 | -0.02018679 | 0.01800440 | 0.00409424 | -0.00520806 | -0.00025647 | 0.00025805 | -0.00064745 | 0.00022844 | 0.00003679 | -0.00004170 | -0.00003525 | 0.00007176 | -0.00010041 | 0.00000216 |
| 37 | 59 | 0.05452232 | 0.04823065 | -0.03307498 | -0.02046960 | 0.01813845 | 0.00437361 | -0.00486484 | -0.00049349 | 0.00050294 | -0.00040433 | 0.00021863 | -0.00000357 | -0.00001207 | -0.00001130 | 0.00010723 | -0.00009074 | 0.00006148 |
| 37 | 60 | 0.05473907 | 0.04696110 | -0.03294820 | -0.01848408 | 0.01759074 | 0.00309755 | -0.00509979 | 0.00014302 | 0.00029645 | -0.00066885 | 0.00018974 | 0.00013513 | -0.00022333 | 0.00002062 | -0.00000242 | -0.00000419 | -0.00005939 |
| 37 | 61 | 0.05584512 | 0.04558716 | -0.03518919 | -0.01746069 | 0.01857632 | 0.00235805 | -0.00475939 | 0.00063266 | 0.00079254 | -0.00046722 | 0.00015464 | 0.00021737 | -0.00035116 | 0.00009673 | -0.00000878 | 0.00011745 | 0.00003771 |
| 37 | 62 | 0.05599722 | 0.04395176 | -0.03587613 | -0.01606125 | 0.01853294 | 0.00180582 | -0.00483828 | 0.00176879 | 0.00070509 | -0.00087551 | 0.00016342 | 0.00042113 | -0.00069640 | 0.00011232 | -0.00015964 | 0.00028021 | -0.00005060 |
| 37 | 63 | 0.04405386 | 0.05257735 | -0.01760133 | -0.03216061 | 0.01127125 | 0.01474163 | -0.00354548 | -0.00368604 | 0.00074364 | -0.00061619 | -0.00029438 | 0.00029947 | -0.00017166 | -0.00021065 | -0.00003601 | 0.00021047 | 0.00010028 |
| 37 | 64 | 0.04443116 | 0.05070575 | -0.01898725 | -0.03084425 | 0.01124630 | 0.01403357 | -0.00374747 | -0.00263863 | 0.00057029 | -0.00101504 | -0.00023826 | 0.00045915 | -0.00048988 | -0.00020569 | -0.00017441 | 0.00033601 | 0.00000312 |
| 37 | 65 | 0.04315163 | 0.05156023 | -0.01709202 | -0.03082452 | 0.01153691 | 0.01373110 | -0.00470699 | -0.00394964 | 0.00095767 | -0.00077312 | -0.00026114 | 0.00033020 | -0.00019994 | -0.00029575 | -0.00001818 | 0.00023006 | 0.00015311 |
| 37 | 66 | 0.04379985 | 0.04962872 | -0.01899379 | -0.02992417 | 0.01121567 | 0.01310418 | -0.00465071 | -0.00290729 | 0.00069463 | -0.00113669 | -0.00019124 | 0.00046935 | -0.00051419 | -0.00028043 | -0.00016035 | 0.00035079 | 0.00005041 |
| 37 | 67 | 0.04279682 | 0.05016179 | -0.01773190 | -0.02945830 | 0.01182771 | 0.01269263 | -0.00505404 | -0.00385218 | 0.00088538 | -0.00082569 | -0.00021184 | 0.00035988 | -0.00027341 | -0.00027995 | -0.00003240 | 0.00023322 | 0.00013478 |
| 38 | 35 | 0.05702609 | 0.06112397 | -0.02006461 | -0.02429537 | 0.01280907 | 0.00813894 | -0.00464973 | -0.00225346 | 0.00013252 | -0.00015796 | -0.00019242 | 0.00034638 | -0.00018268 | 0.00006559 | -0.00013982 | 0.00003598 | -0.00022140 |
| 38 | 36 | 0.05715911 | 0.06044371 | -0.02198558 | -0.02470725 | 0.01377293 | 0.00761012 | -0.00489296 | -0.00197287 | 0.00030705 | -0.00034043 | -0.00013871 | 0.00015894 | 0.00005130 | -0.00001692 | 0.00002675 | -0.00009207 | -0.00009490 |
| 38 | 37 | 0.05728203 | 0.05953111 | -0.02227031 | -0.02313124 | 0.01447680 | 0.00734476 | -0.00500553 | -0.00184864 | 0.00012736 | -0.00048218 | -0.00006951 | 0.00022163 | -0.00007504 | -0.00002946 | -0.00006252 | -0.00004419 | -0.00013781 |
| 38 | 38 | 0.05743703 | 0.05909983 | -0.02389151 | -0.02353300 | 0.01516210 | 0.00671789 | -0.00530582 | -0.00167507 | 0.00024274 | -0.00064695 | -0.00002385 | 0.00004699 | 0.00014919 | -0.00010588 | 0.00009347 | -0.00017248 | -0.00003277 |
| 38 | 39 | 0.05746143 | 0.05799710 | -0.02449211 | -0.02230812 | 0.01603172 | 0.00642488 | -0.00558446 | -0.00156035 | 0.00020564 | -0.00079497 | 0.00003571 | 0.00009474 | 0.00005345 | -0.00013128 | 0.00004525 | -0.00016336 | -0.00007009 |
| 38 | 40 | 0.05760927 | 0.05794341 | -0.02587871 | -0.02348334 | 0.01553067 | 0.00599605 | -0.00583645 | -0.00150632 | 0.00024707 | -0.00071198 | 0.00001163 | 0.00004417 | 0.00012556 | -0.00011013 | 0.00010027 | -0.00015285 | -0.00001815 |
| 38 | 41 | 0.05756987 | 0.05724720 | -0.02640554 | -0.02323561 | 0.01534938 | 0.00591417 | -0.00611654 | -0.00152489 | 0.00021369 | -0.00058113 | -0.00000703 | 0.00020089 | -0.00009008 | -0.00003939 | -0.00003066 | -0.00001407 | -0.00008822 |
| 38 | 42 | 0.05753854 | 0.05708319 | -0.02774272 | -0.02434599 | 0.01514772 | 0.00572780 | -0.00612161 | -0.00148989 | 0.00026365 | -0.00056108 | -0.00000658 | 0.00011781 | 0.00002158 | -0.00004681 | 0.00005347 | -0.00004227 | -0.00001169 |
| 38 | 43 | 0.05749109 | 0.05664786 | -0.02818464 | -0.02420679 | 0.01506447 | 0.00587227 | -0.00630269 | -0.00161493 | 0.00021063 | -0.00041745 | -0.00002727 | 0.00026118 | -0.00016804 | 0.00001599 | -0.00007386 | 0.00009134 | -0.00007011 |
| 38 | 44 | 0.05736381 | 0.05635208 | -0.02950026 | -0.02517809 | 0.01520048 | 0.00589169 | -0.00616094 | -0.00155035 | 0.00027474 | -0.00045774 | -0.00001535 | 0.00016568 | -0.00003951 | -0.00000898 | 0.00002439 | 0.00003335 | 0.00000766 |
| 38 | 45 | 0.05731701 | 0.05609839 | -0.03000483 | -0.02517175 | 0.01538548 | 0.00633908 | -0.00623166 | -0.00172553 | 0.00022307 | -0.00034528 | -0.00003223 | 0.00028944 | -0.00019912 | 0.00003442 | -0.00008785 | 0.00014784 | -0.00004024 |
| 38 | 46 | 0.05714707 | 0.05567855 | -0.03126411 | -0.02595296 | 0.01577845 | 0.00647061 | -0.00603097 | -0.00160385 | 0.00030493 | -0.00042689 | -0.00001895 | 0.00019814 | -0.00007401 | 0.00000448 | 0.00001053 | 0.00007449 | 0.00002410 |
| 38 | 47 | 0.05710042 | 0.05552729 | -0.03193779 | -0.02613386 | 0.01623435 | 0.00716513 | -0.00603876 | -0.00177500 | 0.00027125 | -0.00037987 | -0.00003118 | 0.00030379 | -0.00020878 | 0.00002514 | -0.00007870 | 0.00015746 | -0.00002155 |
| 38 | 48 | 0.05691894 | 0.05500738 | -0.03309342 | -0.02670062 | 0.01676513 | 0.00731490 | -0.00584205 | -0.00159597 | 0.00037027 | -0.00047478 | -0.00002173 | 0.00022629 | -0.00010072 | 0.00000043 | 0.00000903 | 0.00008450 | 0.00002421 |
| 38 | 49 | 0.05686161 | 0.05489302 | -0.03396890 | -0.02710538 | 0.01737799 | 0.00812639 | -0.00586266 | -0.00172715 | 0.00036089 | -0.00050583 | -0.00003339 | 0.00032111 | -0.00022156 | 0.00000313 | -0.00005600 | 0.00013075 | -0.00003019 |
| 38 | 50 | 0.05669038 | 0.05430080 | -0.03500341 | -0.02745903 | 0.01796051 | 0.00824050 | -0.00570418 | -0.00151035 | 0.00047235 | -0.00058839 | -0.00002792 | 0.00025953 | -0.00013590 | -0.00001119 | 0.00001485 | 0.00007061 | -0.00000101 |
| 38 | 51 | 0.05673396 | 0.05384108 | -0.03494236 | -0.02633243 | 0.01809383 | 0.00766590 | -0.00582201 | -0.00146628 | 0.00034591 | -0.00056009 | 0.00001545 | 0.00028395 | -0.00020030 | 0.00000027 | -0.00002937 | 0.00008570 | -0.00002868 |
| 38 | 52 | 0.05653866 | 0.05304240 | -0.03457627 | -0.02513686 | 0.01791642 | 0.00640330 | -0.00587467 | -0.00125328 | 0.00030730 | -0.00054724 | 0.00006450 | 0.00015984 | -0.00005730 | -0.00000391 | 0.00004518 | 0.00000536 | 0.00001642 |
| 38 | 53 | 0.05665560 | 0.05243540 | -0.03462398 | -0.02372077 | 0.01837043 | 0.00590924 | -0.00583259 | -0.00103984 | 0.00023292 | -0.00053352 | 0.00011855 | 0.00018541 | -0.00013595 | 0.00000537 | 0.00000022 | 0.00003203 | -0.00000269 |
| 38 | 54 | 0.05634379 | 0.05182727 | -0.03390003 | -0.02277815 | 0.01787360 | 0.00479495 | -0.00606904 | -0.00102982 | 0.00015441 | -0.00055057 | 0.00014563 | 0.00007361 | 0.00000012 | -0.00000743 | 0.00006679 | -0.00005745 | 0.00001142 |
| 38 | 55 | 0.05641195 | 0.05108084 | -0.03402704 | -0.02127616 | 0.01858710 | 0.00444324 | -0.00588552 | -0.00069791 | 0.00015198 | -0.00055502 | 0.00020519 | 0.00009920 | -0.00008360 | -0.00000545 | 0.00002721 | -0.00002285 | 0.00000615 |
| 38 | 56 | 0.05609782 | 0.05064445 | -0.03303285 | -0.02062239 | 0.01764929 | 0.00348911 | -0.00631236 | -0.00093305 | -0.00000236 | -0.00058162 | 0.00020186 | 0.00001622 | 0.00002022 | -0.00001460 | 0.00006937 | -0.00010207 | -0.00002007 |
| 38 | 57 | 0.05598789 | 0.04978504 | -0.03320332 | -0.01929108 | 0.01852837 | 0.00335429 | -0.00603338 | -0.00056677 | 0.00008000 | -0.00060510 | 0.00026021 | 0.00004169 | -0.00005931 | -0.00002522 | 0.00004024 | -0.00006297 | -0.00000846 |
| 38 | 58 | 0.05613172 | 0.04923137 | -0.03250315 | -0.01860796 | 0.01712381 | 0.00228913 | -0.00646335 | -0.00081535 | -0.00013390 | -0.00059590 | 0.00020413 | 0.00003890 | -0.00006324 | 0.00001365 | 0.00002503 | -0.00008355 | -0.00008784 |
| 38 | 59 | 0.05580355 | 0.04826684 | -0.03273564 | -0.01760979 | 0.01805681 | 0.00239316 | -0.00611535 | -0.00047377 | 0.00002978 | -0.00064192 | 0.00025425 | 0.00007100 | -0.00014106 | -0.00001051 | 0.00000490 | -0.00003470 | -0.00005773 |

TABLE 3. Nuclear charge density distribution FB coefficients.

| $Z$ | $N$ | $a_1$ | $a_2$ | $a_3$ | $a_4$ | $a_5$ | $a_6$ | $a_7$ | $a_8$ | $a_9$ | $a_{10}$ | $a_{11}$ | $a_{12}$ | $a_{13}$ | $a_{14}$ | $a_{15}$ | $a_{16}$ | $a_{17}$ |
|---|---|---|---|---|---|---|---|---|---|---|---|---|---|---|---|---|---|---|
| 38 | 60 | 0.05764392 | 0.04654045 | -0.03454297 | -0.01537594 | 0.01736945 | 0.00014304 | -0.00630174 | 0.00041545 | 0.00007699 | -0.00061572 | 0.00014778 | 0.00027018 | -0.00043991 | 0.00014772 | -0.00011393 | 0.00011120 | -0.00015547 |
| 38 | 61 | 0.05718157 | 0.04539261 | -0.03512059 | -0.01464050 | 0.01849665 | 0.00043498 | -0.00590382 | 0.00083359 | 0.00030173 | -0.00075019 | 0.00021490 | 0.00031166 | -0.00053054 | 0.00009621 | -0.00012674 | 0.00018316 | -0.00008223 |
| 38 | 62 | 0.04574394 | 0.05353162 | -0.01701146 | -0.03036943 | 0.01013005 | 0.01271536 | -0.00488868 | -0.00379279 | 0.00008562 | -0.00071004 | -0.00029560 | 0.00036729 | -0.00028473 | -0.00013543 | -0.00013953 | 0.00018026 | -0.00010069 |
| 38 | 63 | 0.04518704 | 0.05225972 | -0.01775600 | -0.02977373 | 0.01124011 | 0.01303259 | -0.00459679 | -0.00346069 | 0.00027079 | -0.00091970 | -0.00020290 | 0.00038388 | -0.00035024 | -0.00022402 | -0.00014043 | 0.00025176 | -0.00000404 |
| 38 | 64 | 0.04539714 | 0.05207428 | -0.01783180 | -0.02890433 | 0.01086832 | 0.01158314 | -0.00565204 | -0.00370481 | 0.00029115 | -0.00083688 | -0.00021369 | 0.00038910 | -0.00034119 | -0.00018591 | -0.00011694 | 0.00017796 | -0.00003722 |
| 38 | 65 | 0.04470088 | 0.05076832 | -0.01839195 | -0.02843633 | 0.01168843 | 0.01183558 | -0.00547858 | -0.00349457 | 0.00045989 | -0.00107227 | -0.00013075 | 0.00039913 | -0.00039059 | -0.00029596 | -0.00011744 | 0.00025979 | 0.00006052 |
| 38 | 66 | 0.04525390 | 0.05070715 | -0.01892982 | -0.02759705 | 0.01139401 | 0.01045401 | -0.00625670 | -0.00377318 | 0.00037025 | -0.00092921 | -0.00011499 | 0.00036062 | -0.00034013 | -0.00024951 | -0.00008577 | 0.00016489 | 0.00002887 |
| 38 | 67 | 0.04432702 | 0.04931706 | -0.01919242 | -0.02727302 | 0.01204913 | 0.01099227 | -0.00571983 | -0.00342363 | 0.00039587 | -0.00108948 | -0.00008571 | 0.00042292 | -0.00045848 | -0.00027083 | -0.00012846 | 0.00026064 | 0.00004454 |
| 38 | 68 | 0.04470214 | 0.04917484 | -0.01963377 | -0.02672716 | 0.01184656 | 0.01011230 | -0.00602116 | -0.00350619 | 0.00021139 | -0.00087122 | -0.00012281 | 0.00042328 | -0.00046408 | -0.00012836 | -0.00011842 | 0.00015666 | -0.00006041 |
| 38 | 69 | 0.04387996 | 0.04787771 | -0.02011444 | -0.02646679 | 0.01262651 | 0.01074963 | -0.00543247 | -0.00319906 | 0.00019791 | -0.00103398 | -0.00007327 | 0.00047504 | -0.00057310 | -0.00016172 | -0.00016142 | 0.00024029 | -0.00004895 |
| 39 | 36 | 0.05825336 | 0.06014775 | -0.02191249 | -0.02201607 | 0.01460980 | 0.00652482 | -0.00552400 | -0.00196251 | 0.00012931 | -0.00028559 | -0.00010295 | 0.00029793 | -0.00014213 | 0.00003935 | -0.00010871 | -0.00001774 | -0.00020969 |
| 39 | 37 | 0.05826895 | 0.05947568 | -0.02253995 | -0.02125374 | 0.01516325 | 0.00549262 | -0.00556327 | -0.00183716 | 0.00013566 | -0.00010288 | -0.00009452 | 0.00023141 | -0.00003606 | 0.00013064 | -0.00003200 | -0.00014902 | -0.00021430 |
| 39 | 38 | 0.05851268 | 0.05855913 | -0.02422581 | -0.02127052 | 0.01611963 | 0.00573017 | -0.00596555 | -0.00159264 | 0.00015618 | -0.00063275 | 0.00001809 | 0.00016355 | -0.00002518 | -0.00006486 | -0.00000802 | -0.00012999 | -0.00013987 |
| 39 | 39 | 0.05842998 | 0.05778965 | -0.02517059 | -0.02121607 | 0.01684314 | 0.00505663 | -0.00617850 | -0.00167936 | 0.00032114 | -0.00038342 | -0.00001080 | 0.00008739 | 0.00010919 | 0.00002920 | 0.00010625 | -0.00028513 | -0.00013804 |
| 39 | 40 | 0.05870320 | 0.05745930 | -0.02623258 | -0.02133199 | 0.01646648 | 0.00495116 | -0.00656543 | -0.00141643 | 0.00020582 | -0.00068381 | 0.00004779 | 0.00015284 | -0.00003657 | -0.00006916 | 0.00000894 | -0.00011542 | -0.00012169 |
| 39 | 41 | 0.05862415 | 0.05724427 | -0.02677125 | -0.02168793 | 0.01623361 | 0.00442662 | -0.00670288 | -0.00164807 | 0.00028008 | -0.00018933 | -0.00002940 | 0.00018534 | -0.00002485 | 0.00009777 | 0.00001438 | -0.00012162 | -0.00014962 |
| 39 | 42 | 0.05872686 | 0.05693490 | -0.02780446 | -0.02196804 | 0.01590475 | 0.00443454 | -0.00694821 | -0.00143569 | 0.00021549 | -0.00047143 | 0.00001794 | 0.00023024 | -0.00013726 | 0.00000763 | -0.00005029 | 0.00001669 | -0.00011307 |
| 39 | 43 | 0.05858099 | 0.05682450 | -0.02823860 | -0.02223139 | 0.01592886 | 0.00414505 | -0.00692680 | -0.00172885 | 0.00025049 | -0.00004736 | -0.00002573 | 0.00023121 | -0.00008672 | 0.00012757 | -0.00003480 | -0.00000623 | -0.00011625 |
| 39 | 44 | 0.05860199 | 0.05648611 | -0.02937514 | -0.02264062 | 0.01584529 | 0.00438838 | -0.00706624 | -0.00154395 | 0.00022943 | -0.00032436 | 0.00000405 | 0.00026976 | -0.00018096 | 0.00004867 | -0.00007755 | 0.00010232 | -0.00007961 |
| 39 | 45 | 0.05838488 | 0.05641211 | -0.02978883 | -0.02285730 | 0.01612828 | 0.00428660 | -0.00694016 | -0.00183957 | 0.00024994 | 0.00000302 | -0.00000930 | 0.00023984 | -0.00009670 | 0.00012117 | -0.00004461 | 0.00005309 | -0.00006482 |
| 39 | 46 | 0.05840300 | 0.05601802 | -0.03109523 | -0.02336021 | 0.01637740 | 0.00481127 | -0.00700045 | -0.00164457 | 0.00026698 | -0.00027896 | -0.00000015 | 0.00028646 | -0.00018999 | 0.00005619 | -0.00007763 | 0.00013837 | -0.00004557 |
| 39 | 47 | 0.05809873 | 0.05593101 | -0.03151353 | -0.02358947 | 0.01681468 | 0.00479415 | -0.00685295 | -0.00190450 | 0.00029253 | -0.00005393 | 0.00000731 | 0.00023282 | -0.00008294 | 0.00008991 | -0.00002422 | 0.00005839 | -0.00002292 |
| 39 | 48 | 0.05816640 | 0.05547816 | -0.03299547 | -0.02416204 | 0.01735576 | 0.00555800 | -0.00686333 | -0.00167375 | 0.00034017 | -0.00034030 | -0.00000305 | 0.00029784 | -0.00018988 | 0.00004101 | -0.00005878 | 0.00013074 | -0.00003136 |
| 39 | 49 | 0.05775947 | 0.05534586 | -0.03339577 | -0.02443703 | 0.01782302 | 0.00553188 | -0.00676558 | -0.00187875 | 0.00038353 | -0.00020492 | 0.00001288 | 0.00023054 | -0.00007265 | 0.00004952 | 0.00001334 | 0.00002226 | -0.00001002 |
| 39 | 50 | 0.05790488 | 0.05484387 | -0.03502739 | -0.02505423 | 0.01854951 | 0.00644388 | -0.00675636 | -0.00160989 | 0.00044794 | -0.00048544 | -0.00001171 | 0.00031764 | -0.00020163 | 0.00001734 | -0.00003168 | 0.00009289 | -0.00004857 |
| 39 | 51 | 0.05767890 | 0.05432516 | -0.03442473 | -0.02367436 | 0.01855923 | 0.00513442 | -0.00669736 | -0.00164449 | 0.00036203 | -0.00027171 | 0.00006795 | 0.00018405 | -0.00004748 | 0.00003576 | 0.00004176 | -0.00001638 | 0.00000498 |
| 39 | 52 | 0.05794766 | 0.05343705 | -0.03483531 | -0.02247400 | 0.01877901 | 0.00468434 | -0.00679966 | -0.00123633 | 0.00026943 | -0.00044073 | 0.00009989 | 0.00021610 | -0.00014454 | 0.00002783 | -0.00001338 | 0.00005218 | -0.00002375 |
| 39 | 53 | 0.05769582 | 0.05297764 | -0.03422434 | -0.02128784 | 0.01889846 | 0.00369778 | -0.00662700 | -0.00134383 | 0.00023473 | -0.00023261 | 0.00017701 | 0.00007969 | 0.00001422 | 0.00003188 | 0.00006463 | -0.00004969 | 0.00004328 |
| 39 | 54 | 0.05780233 | 0.05208106 | -0.03428716 | -0.01996875 | 0.01898656 | 0.00319426 | -0.00685957 | -0.00093714 | 0.00012152 | -0.00045164 | 0.00019985 | 0.00012122 | -0.00009158 | 0.00001856 | 0.00000698 | 0.00000259 | -0.00001384 |
| 39 | 55 | 0.05739389 | 0.05171555 | -0.03355279 | -0.01916707 | 0.01908140 | 0.00254328 | -0.00669504 | -0.00125954 | 0.00014823 | -0.00022512 | 0.00026164 | -0.00001565 | 0.00007989 | 0.00000971 | 0.00009325 | -0.00009525 | 0.00006525 |
| 39 | 56 | 0.05741528 | 0.05080798 | -0.03338564 | -0.01783233 | 0.01895718 | 0.00204584 | -0.00701965 | -0.00085506 | -0.00001516 | -0.00050189 | 0.00027276 | 0.00004684 | -0.00005267 | -0.00000607 | 0.00002150 | -0.00004532 | -0.00002457 |
| 39 | 57 | 0.05685241 | 0.05052340 | -0.03248501 | -0.01735301 | 0.01891985 | 0.00155958 | -0.00696383 | -0.00144704 | 0.00008659 | -0.00022758 | 0.00030151 | -0.00007518 | 0.00012394 | -0.00001573 | 0.00011274 | -0.00013576 | 0.00005790 |
| 39 | 58 | 0.05701015 | 0.04947681 | -0.03246056 | -0.01606081 | 0.01851763 | 0.00107272 | -0.00721924 | -0.00094395 | -0.00014396 | -0.00055417 | 0.00029668 | 0.00002937 | -0.00007210 | -0.00001782 | 0.00000957 | -0.00006059 | -0.00006459 |
| 39 | 59 | 0.05697425 | 0.04891253 | -0.03217527 | -0.01513002 | 0.01846364 | 0.00002481 | -0.00713285 | -0.00136264 | 0.00011879 | -0.00021253 | 0.00026062 | -0.00000429 | 0.00001761 | 0.00002459 | 0.00007405 | -0.00008655 | 0.00001656 |
| 39 | 60 | 0.05790116 | 0.04715222 | -0.03356744 | -0.01335332 | 0.01832767 | -0.00080900 | -0.00709247 | -0.00021166 | -0.00005249 | -0.00060106 | 0.00025469 | 0.00019468 | -0.00033671 | 0.00006809 | -0.00008447 | 0.00007201 | -0.00011185 |
| 39 | 61 | 0.04564019 | 0.05570120 | -0.01469049 | -0.02951152 | 0.01061063 | 0.01192113 | -0.00543875 | -0.00555148 | 0.00001794 | -0.00027491 | -0.00023737 | 0.00012022 | 0.00015652 | -0.00021826 | 0.00002925 | 0.00001359 | 0.00004365 |
| 39 | 62 | 0.04594191 | 0.05396677 | -0.01615159 | -0.02848799 | 0.01098329 | 0.01179580 | -0.00550045 | -0.00434941 | -0.00004562 | -0.00076474 | -0.00018574 | 0.00030440 | -0.00019465 | -0.00022837 | -0.00010736 | 0.00015610 | -0.00003339 |
| 39 | 63 | 0.04498293 | 0.05396891 | -0.01552608 | -0.02788512 | 0.01187591 | 0.01088120 | -0.00636084 | -0.00517321 | 0.00041140 | -0.00053483 | -0.00016189 | 0.00019667 | 0.00004298 | -0.00029333 | 0.00004694 | 0.00005404 | 0.00013810 |
| 39 | 64 | 0.04566300 | 0.05201555 | -0.01762359 | -0.02681214 | 0.01204059 | 0.01044515 | -0.00631416 | -0.00406419 | 0.00022513 | -0.00097139 | -0.00008510 | 0.00033698 | -0.00027241 | -0.00029631 | -0.00008118 | 0.00016948 | 0.00005283 |
| 39 | 65 | 0.04470997 | 0.05202631 | -0.01688938 | -0.02596571 | 0.01283536 | 0.00938549 | -0.00729642 | -0.00512369 | 0.00056717 | -0.00070718 | -0.00005841 | 0.00018729 | 0.00002426 | -0.00037494 | 0.00007900 | 0.00005804 | 0.00020859 |
| 39 | 66 | 0.04534362 | 0.05032629 | -0.01889152 | -0.02559899 | 0.01255706 | 0.00929477 | -0.00693067 | -0.00411462 | 0.00031770 | -0.00110952 | 0.00001686 | 0.00030475 | -0.00026733 | -0.00037893 | -0.00004607 | 0.00016701 | 0.00013304 |
| 39 | 67 | 0.04445017 | 0.05035757 | -0.01816552 | -0.02474450 | 0.01331000 | 0.00859332 | -0.00740883 | -0.00494788 | 0.00043218 | -0.00076614 | 0.00000999 | 0.00021194 | -0.00006497 | -0.00035973 | 0.00004576 | 0.00008950 | 0.00019924 |
| 39 | 68 | 0.04501430 | 0.04884790 | -0.01986090 | -0.02457361 | 0.01307146 | 0.00896821 | -0.00665270 | -0.00385903 | 0.00007177 | -0.00102012 | 0.00002093 | 0.00037757 | -0.00041649 | -0.00024891 | -0.00011073 | 0.00018377 | 0.00003071 |
| 39 | 69 | 0.04427452 | 0.04881458 | -0.01964562 | -0.02399811 | 0.01378363 | 0.00837522 | -0.00693024 | -0.00459870 | 0.00011387 | -0.00075247 | 0.00004409 | 0.00027215 | -0.00021925 | -0.00025140 | -0.00002357 | 0.00010444 | 0.00010767 |
| 39 | 70 | 0.04465790 | 0.04740574 | -0.02099309 | -0.02371944 | 0.01380199 | 0.00872917 | -0.00641869 | -0.00364734 | -0.00011328 | -0.00096575 | 0.00003338 | 0.00042173 | -0.00052172 | -0.00014725 | -0.00014204 | 0.00016223 | -0.00006129 |
| 40 | 37 | 0.05942002 | 0.05894474 | -0.02517750 | -0.02171431 | 0.01640104 | 0.00601002 | -0.00580733 | -0.00154360 | 0.00013902 | -0.00050925 | -0.00003244 | 0.00024623 | -0.00011415 | 0.00001420 | -0.00005096 | -0.00011593 | -0.00020558 |
| 40 | 38 | 0.05961971 | 0.05870898 | -0.02658500 | -0.02230419 | 0.01695189 | 0.00570316 | -0.00617776 | -0.00154542 | 0.00025522 | -0.00062556 | 0.00000140 | 0.00006909 | 0.00010652 | -0.00006974 | 0.00009115 | -0.00022623 | -0.00010533 |
| 40 | 39 | 0.05978696 | 0.05741256 | -0.02769316 | -0.02130971 | 0.01803261 | 0.00552818 | -0.00627376 | -0.00124811 | 0.00019659 | -0.00085278 | 0.00007565 | 0.00010844 | 0.00000191 | -0.00009326 | 0.00005935 | -0.00023699 | -0.00014559 |
| 40 | 40 | 0.05995006 | 0.05759218 | -0.02855513 | -0.02210012 | 0.01732283 | 0.00498129 | -0.00664404 | -0.00132964 | 0.00024317 | -0.00071783 | 0.00005278 | 0.00005285 | 0.00007971 | -0.00008619 | 0.00009837 | -0.00020125 | -0.00008459 |
| 40 | 41 | 0.05999191 | 0.05672559 | -0.02938299 | -0.02165920 | 0.01728123 | 0.00458413 | -0.00676333 | -0.00109735 | 0.00019855 | -0.00066596 | 0.00006339 | 0.00019495 | -0.00013065 | -0.00001598 | -0.00000941 | -0.00008623 | -0.00014334 |
| 40 | 42 | 0.05997466 | 0.05682665 | -0.03021257 | -0.02248815 | 0.01687680 | 0.00435002 | -0.00691946 | -0.00121664 | 0.00025483 | -0.00059409 | 0.00005948 | 0.00011011 | -0.00001505 | -0.00003642 | 0.00005690 | -0.00008629 | -0.00006045 |
| 40 | 43 | 0.05998729 | 0.05624361 | -0.03091420 | -0.02214655 | 0.01682636 | 0.00403481 | -0.00702340 | -0.00112910 | 0.00019488 | -0.00050887 | 0.00006229 | 0.00024132 | -0.00019521 | 0.00002902 | -0.00004783 | 0.00002325 | -0.00010821 |
| 40 | 44 | 0.05987472 | 0.05621996 | -0.03177750 | -0.02295330 | 0.01679722 | 0.00411348 | -0.00703213 | -0.00122780 | 0.00026533 | -0.00050482 | 0.00006719 | 0.00014551 | -0.00006469 | -0.00000957 | 0.00003258 | -0.00000550 | -0.00002533 |
| 40 | 45 | 0.05985662 | 0.05585657 | -0.03247057 | -0.02273786 | 0.01692758 | 0.00401320 | -0.00709141 | -0.00123870 | 0.00020899 | -0.00042402 | 0.00006704 | 0.00025869 | -0.00020849 | 0.00003966 | -0.00005943 | 0.00008675 | -0.00006209 |
| 40 | 46 | 0.05970307 | 0.05569164 | -0.03336410 | -0.02347374 | 0.01721517 | 0.00432263 | -0.00701530 | -0.00127842 | 0.00029613 | -0.00047566 | 0.00007366 | 0.00016637 | -0.00008254 | -0.00000608 | 0.00002333 | 0.00004077 | 0.00000718 |
| 40 | 47 | 0.05964580 | 0.05548080 | -0.03412509 | -0.02342781 | 0.01758489 | 0.00447357 | -0.00703305 | -0.00133154 | 0.00025987 | -0.00043357 | 0.00007180 | 0.00026208 | -0.00019427 | 0.00002239 | -0.00004931 | 0.00010522 | -0.00002559 |
| 40 | 48 | 0.05948412 | 0.05517931 | -0.03502124 | -0.02405941 | 0.01807231 | 0.00489841 | -0.00692873 | -0.00130239 | 0.00036309 | -0.00051690 | 0.00007680 | 0.00018156 | -0.00008578 | -0.00002092 | 0.00002668 | 0.00005511 | 0.00002514 |
| 40 | 49 | 0.05937636 | 0.05505450 | -0.03588673 | -0.02423234 | 0.01860910 | 0.00524715 | -0.00694307 | -0.00135664 | 0.00035313 | -0.00053048 | 0.00007033 | 0.00026754 | -0.00017863 | -0.00000963 | -0.00002610 | 0.00008773 | -0.00001465 |
| 40 | 50 | 0.05923130 | 0.05463375 | -0.03677030 | -0.02473728 | 0.01919778 | 0.00569061 | -0.00685346 | -0.00127030 | 0.00046853 | -0.00062030 | 0.00007409 | 0.00020066 | -0.00009272 | -0.00004485 | 0.00003750 | 0.00004463 | 0.00001914 |
| 40 | 51 | 0.05926637 | 0.05412005 | -0.03672603 | -0.02341642 | 0.01922614 | 0.00471456 | -0.00699565 | -0.00118117 | 0.00031917 | -0.00056100 | 0.00012144 | 0.00021556 | -0.00013647 | -0.00001963 | -0.00000024 | 0.00004624 | -0.00000372 |
| 40 | 52 | 0.05914488 | 0.05345486 | -0.03608772 | -0.02223395 | 0.01897370 | 0.00371948 | -0.00721470 | -0.00115691 | 0.00025436 | -0.00054307 | 0.00016455 | 0.00008914 | -0.00000403 | -0.00003938 | 0.00005881 | -0.00001604 | 0.00003169 |
| 40 | 53 | 0.05920038 | 0.05279717 | -0.03621652 | -0.02085650 | 0.01931740 | 0.00296648 | -0.00714759 | -0.00091000 | 0.00015819 | -0.00050835 | 0.00022239 | 0.00010021 | -0.00005620 | -0.00001743 | 0.00002736 | -0.00000991 | 0.00002039 |
| 40 | 54 | 0.05894645 | 0.05233643 | -0.03500507 | -0.01969093 | 0.01867368 | 0.00198703 | -0.00759292 | -0.00114403 | 0.00004613 | -0.00049423 | 0.00023432 | -0.00000471 | 0.00006598 | -0.00003980 | 0.00007192 | -0.00007681 | 0.00002022 |
| 40 | 55 | 0.05888367 | 0.05158003 | -0.03522018 | -0.01848747 | 0.01926655 | 0.00156973 | -0.00735023 | -0.00081445 | 0.00002248 | -0.00049059 | 0.00029574 | 0.00000224 | 0.00001367 | -0.00002781 | 0.00005237 | -0.00007038 | 0.00002285 |
| 40 | 56 | 0.05877488 | 0.05116006 | -0.03385024 | -0.01727172 | 0.01817239 | 0.00045586 | -0.00791061 | -0.00119297 | -0.00014047 | -0.00044629 | 0.00026138 | -0.00004826 | 0.00007949 | -0.00002241 | 0.00006289 | -0.00011435 | -0.00002344 |
| 40 | 57 | 0.05849630 | 0.05035426 | -0.03405945 | -0.01641433 | 0.01888623 | 0.00039708 | -0.00758342 | -0.00087843 | -0.00008700 | -0.00047499 | 0.00031701 | -0.00004119 | 0.00003083 | -0.00002520 | 0.00005555 | -0.00010727 | -0.00000833 |
| 40 | 58 | 0.05946207 | 0.04933482 | -0.03399321 | -0.01443593 | 0.01773258 | -0.00152101 | -0.00787873 | -0.00064132 | -0.00008185 | -0.00034116 | 0.00019054 | 0.00007806 | -0.00011430 | 0.00009375 | -0.00001563 | -0.00003682 | -0.00010480 |
| 40 | 59 | 0.05912114 | 0.04839462 | -0.03440624 | -0.01374551 | 0.01855288 | -0.00143175 | -0.00752087 | -0.00028836 | 0.00003767 | -0.00042425 | 0.00024314 | 0.00010051 | -0.00017945 | 0.00007588 | -0.00002077 | -0.00001018 | -0.00006954 |
| 40 | 60 | 0.04717245 | 0.05674608 | -0.01540168 | -0.02990943 | 0.00988029 | 0.01131470 | -0.00615182 | -0.00497112 | -0.00001147 | -0.00028912 | -0.00034620 | 0.00021147 | 0.00004680 | -0.00013032 | -0.00004332 | 0.00001881 | -0.00009305 |
| 40 | 61 | 0.04674242 | 0.05568389 | -0.01603706 | -0.02930649 | 0.01087532 | 0.01149619 | -0.00589697 | -0.00461242 | 0.00011185 | -0.00047512 | -0.00026382 | 0.00022252 | -0.00001203 | -0.00018943 | -0.00003932 | 0.00005507 | -0.00002271 |
| 40 | 62 | 0.04699048 | 0.05487159 | -0.01701882 | -0.02811273 | 0.01140559 | 0.01014964 | -0.00683943 | -0.00447543 | 0.00034893 | -0.00043138 | -0.00025574 | 0.00026907 | -0.00006211 | -0.00015276 | -0.00001990 | 0.00002172 | -0.00002051 |
| 40 | 63 | 0.04650510 | 0.05373969 | -0.01759087 | -0.02754289 | 0.01218051 | 0.01021670 | -0.00670831 | -0.00422044 | 0.00044797 | -0.00065009 | -0.00017753 | 0.00027050 | -0.00010540 | -0.00023210 | -0.00001354 | 0.00006323 | 0.00005089 |
| 40 | 64 | 0.04683007 | 0.05324453 | -0.01850170 | -0.02692443 | 0.01227558 | 0.00933509 | -0.00731956 | -0.00440504 | 0.00044648 | -0.00053417 | -0.00015371 | 0.00025483 | -0.00009313 | -0.00019570 | 0.00000616 | 0.00000671 | 0.00003482 |
| 40 | 65 | 0.04617871 | 0.05210300 | -0.01884945 | -0.02641403 | 0.01282367 | 0.00927452 | -0.00730738 | -0.00422261 | 0.00056387 | -0.00077780 | -0.00008607 | 0.00025128 | -0.00011866 | -0.00029586 | 0.00001784 | 0.00005650 | 0.00011488 |
| 40 | 66 | 0.04670023 | 0.05184298 | -0.01989014 | -0.02599279 | 0.01287227 | 0.00862631 | -0.00762188 | -0.00435841 | 0.00049184 | -0.00064491 | -0.00005145 | 0.00021932 | -0.00009610 | -0.00025704 | 0.00003130 | 0.00000263 | 0.00010003 |
| 40 | 67 | 0.04584520 | 0.05059189 | -0.01987801 | -0.02538096 | 0.01321712 | 0.00855897 | -0.00738939 | -0.00405614 | 0.00048096 | -0.00080470 | -0.00003478 | 0.00026697 | -0.00018842 | -0.00026669 | 0.00000580 | 0.00006316 | 0.00010436 |
| 40 | 68 | 0.04619322 | 0.05026409 | -0.02063458 | -0.02503671 | 0.01321906 | 0.00814453 | -0.00738780 | -0.00404115 | 0.00031417 | -0.00058774 | -0.00005884 | 0.00027969 | -0.00022458 | -0.00012894 | -0.00000460 | 0.00000109 | 0.00001383 |
| 40 | 69 | 0.04549233 | 0.04908309 | -0.02088156 | -0.02437962 | 0.01376513 | 0.00811493 | -0.00714104 | -0.00374743 | 0.00027668 | -0.00076126 | -0.00002330 | 0.00032055 | -0.00031275 | -0.00015026 | -0.00003009 | 0.00004997 | 0.00001251 |
| 40 | 70 | 0.04573772 | 0.04873131 | -0.02161497 | -0.02417712 | 0.01385159 | 0.00774952 | -0.00722804 | -0.00374431 | 0.00019646 | -0.00057862 | -0.00005831 | 0.00032119 | -0.00032434 | -0.00002807 | -0.00001685 | -0.00002714 | -0.00006567 |
| 40 | 71 | 0.04515705 | 0.04763958 | -0.02207294 | -0.02353138 | 0.01453253 | 0.00773453 | -0.00696686 | -0.00346202 | 0.00013814 | -0.00076684 | -0.00000252 | 0.00034866 | -0.00039795 | -0.00006555 | -0.00003623 | 0.00000327 | -0.00007028 |
| 40 | 72 | 0.04532341 | 0.04729277 | -0.02277446 | -0.02347426 | 0.01465877 | 0.00740822 | -0.00716558 | -0.00345445 | 0.00015232 | -0.00061891 | -0.00004851 | 0.00034367 | -0.00039593 | 0.00004147 | -0.00000751 | -0.00007726 | -0.00013757 |
| 40 | 73 | 0.04482047 | 0.04630559 | -0.02337703 | -0.02289488 | 0.01541027 | 0.00740385 | -0.00689306 | -0.00319230 | 0.00007331 | -0.00082216 | 0.00002592 | 0.00035432 | -0.00044786 | -0.00001435 | -0.00001713 | -0.00007090 | -0.00014674 |
| 41 | 38 | 0.06060237 | 0.05765406 | -0.02852236 | -0.02150183 | 0.01825993 | 0.00570487 | -0.00609745 | -0.00124422 | 0.00014087 | -0.00070114 | 0.00002726 | 0.00019260 | -0.00008579 | -0.00000569 | 0.00000865 | -0.00021719 | -0.00020847 |
| 41 | 39 | 0.06048065 | 0.05694537 | -0.02925695 | -0.02141408 | 0.01894523 | 0.00510841 | -0.00609414 | -0.00146592 | 0.00025873 | -0.00051445 | -0.00000274 | 0.00010930 | 0.00005750 | 0.00007989 | 0.00014062 | -0.00039331 | -0.00018649 |
| 41 | 40 | 0.06102875 | 0.05650200 | -0.03072079 | -0.02137788 | 0.01873803 | 0.00501668 | -0.00655724 | -0.00098136 | 0.00016563 | -0.00079302 | 0.00008154 | 0.00016696 | -0.00010805 | -0.00002601 | 0.00002455 | -0.00019083 | -0.00017786 |
| 41 | 41 | 0.06085212 | 0.05620130 | -0.03105145 | -0.02144550 | 0.01832305 | 0.00398692 | -0.00659115 | -0.00124655 | 0.00021080 | -0.00039978 | 0.00003324 | 0.00016880 | -0.00005896 | 0.00011692 | 0.00007746 | -0.00024639 | -0.00016485 |
| 41 | 42 | 0.06121380 | 0.05588422 | -0.03233219 | -0.02152907 | 0.01810522 | 0.00404557 | -0.00687633 | -0.00082709 | 0.00015274 | -0.00066417 | 0.00010230 | 0.00021791 | -0.00020155 | 0.00001807 | -0.00002183 | -0.00006140 | -0.00014257 |
| 41 | 43 | 0.06097509 | 0.05571385 | -0.03262529 | -0.02169718 | 0.01784807 | 0.00316307 | -0.00688193 | -0.00119698 | 0.00017727 | -0.00031118 | 0.00007861 | 0.00018381 | -0.00010121 | 0.00011935 | 0.00004718 | -0.00014024 | -0.00010569 |
| 41 | 44 | 0.06123487 | 0.05547478 | -0.03387148 | -0.02189894 | 0.01785458 | 0.00354350 | -0.00703785 | -0.00083824 | 0.00014653 | -0.00057719 | 0.00012440 | 0.00023848 | -0.00023529 | 0.00003065 | -0.00004323 | 0.00002691 | -0.00008967 |
| 41 | 45 | 0.06091478 | 0.05538530 | -0.03413134 | -0.02213869 | 0.01779834 | 0.00282742 | -0.00700971 | -0.00124066 | 0.00017491 | -0.00028133 | 0.00012273 | 0.00016825 | -0.00008725 | 0.00008991 | 0.00004462 | -0.00007859 | -0.00003334 |
| 41 | 46 | 0.06113544 | 0.05517317 | -0.03543811 | -0.02245204 | 0.01814078 | 0.00358658 | -0.00707183 | -0.00091594 | 0.00017003 | -0.00056090 | 0.00014315 | 0.00023898 | -0.00022557 | 0.00001352 | -0.00004226 | 0.00007134 | -0.00003764 |
| 41 | 47 | 0.06071243 | 0.05512945 | -0.03562436 | -0.02273725 | 0.01823668 | 0.00300607 | -0.00702670 | -0.00129808 | 0.00021747 | -0.00032736 | 0.00015615 | 0.00014070 | -0.00004286 | 0.00003835 | 0.00006310 | -0.00006013 | 0.00002902 |
| 41 | 48 | 0.06092921 | 0.05489702 | -0.03704507 | -0.02316340 | 0.01889792 | 0.00409503 | -0.00703091 | -0.00098144 | 0.00023858 | -0.00062258 | 0.00015395 | 0.00023306 | -0.00019533 | -0.00002413 | -0.00002419 | 0.00007583 | -0.00000258 |
| 41 | 49 | 0.06039347 | 0.05486906 | -0.03712012 | -0.02348492 | 0.01904185 | 0.00358826 | -0.00700550 | -0.00131907 | 0.00030955 | -0.00044403 | 0.00017164 | 0.00011988 | 0.00000368 | -0.00002110 | 0.00009359 | -0.00007640 | 0.00006380 |
| 41 | 50 | 0.06062920 | 0.05457653 | -0.03868946 | -0.02401933 | 0.01993149 | 0.00489836 | -0.00699054 | -0.00099788 | 0.00035093 | -0.00074703 | 0.00015287 | 0.00023477 | -0.00016946 | -0.00006856 | 0.00000181 | 0.00005161 | 0.00000364 |
| 41 | 51 | 0.06027264 | 0.05408685 | -0.03782497 | -0.02271183 | 0.01959823 | 0.00305547 | -0.00708052 | -0.00120084 | 0.00030447 | -0.00046273 | 0.00022314 | 0.00004875 | 0.00007486 | -0.00004531 | 0.00012967 | -0.00011772 | 0.00009870 |
| 41 | 52 | 0.06066569 | 0.05327465 | -0.03829443 | -0.02143837 | 0.01988170 | 0.00286181 | -0.00728340 | -0.00078038 | 0.00017185 | -0.00063996 | 0.00025540 | 0.00010473 | -0.00006827 | -0.00005958 | 0.00003179 | -0.00000334 | 0.00004204 |
| 41 | 53 | 0.06017374 | 0.05295014 | -0.03719555 | -0.02031676 | 0.01971808 | 0.00144375 | -0.00728420 | -0.00111487 | 0.00021464 | -0.00034891 | 0.00030608 | -0.00008252 | 0.00019194 | -0.00004690 | 0.00017358 | -0.00017743 | 0.00014852 |
| 41 | 54 | 0.06042663 | 0.05210222 | -0.03728179 | -0.01890427 | 0.01975024 | 0.00116227 | -0.00759934 | -0.00071717 | 0.00003063 | -0.00055213 | 0.00032947 | -0.00001227 | 0.00003020 | -0.00005996 | 0.00006197 | -0.00006780 | 0.00005953 |
| 41 | 55 | 0.05978618 | 0.05196858 | -0.03594843 | -0.01806352 | 0.01966434 | 0.00012601 | -0.00761356 | -0.00125071 | 0.00018020 | -0.00023417 | 0.00034299 | -0.00018324 | 0.00029441 | -0.00005003 | 0.00021191 | -0.00023875 | 0.00016693 |
| 41 | 56 | 0.06004447 | 0.05097483 | -0.03594521 | -0.01660655 | 0.01938652 | -0.00022990 | -0.00791094 | -0.00079903 | -0.00006433 | -0.00046295 | 0.00034942 | -0.00007939 | 0.00008488 | -0.00004718 | 0.00007598 | -0.00011798 | 0.00004162 |
| 41 | 57 | 0.05975129 | 0.05083213 | -0.03497253 | -0.01574196 | 0.01931938 | -0.00133895 | -0.00789036 | -0.00124934 | 0.00026729 | -0.00008361 | 0.00028144 | -0.00016126 | 0.00026894 | 0.00000618 | 0.00019776 | -0.00023020 | 0.00012797 |
| 41 | 58 | 0.06038271 | 0.04937183 | -0.03561980 | -0.01406591 | 0.01892749 | -0.00198286 | -0.00793083 | -0.00037137 | 0.00008916 | -0.00031885 | 0.00025038 | 0.00003526 | -0.00006591 | 0.00006366 | 0.00001828 | -0.00005332 | -0.00002478 |
| 41 | 59 | 0.04811269 | 0.05840606 | -0.01625247 | -0.03062021 | 0.01092349 | 0.01090236 | -0.00618237 | -0.00573872 | 0.00024329 | 0.00009637 | -0.00033336 | -0.00003063 | 0.00046214 | -0.00017297 | 0.00015148 | -0.00015791 | 0.00008379 |
| 41 | 60 | 0.04755751 | 0.05717084 | -0.01627632 | -0.03013811 | 0.01088033 | 0.01129885 | -0.00627562 | -0.00490685 | 0.00022528 | -0.00024794 | -0.00032486 | 0.00016421 | 0.00013624 | -0.00017106 | 0.00000606 | -0.00000868 | -0.00000883 |
| 41 | 61 | 0.04702911 | 0.05734076 | -0.01618143 | -0.02973006 | 0.01211951 | 0.01064355 | -0.00699371 | -0.00540879 | 0.00077555 | -0.00004688 | -0.00033767 | 0.00007949 | 0.00034511 | -0.00020238 | 0.00016370 | -0.00012135 | 0.00014485 |
| 41 | 62 | 0.04734116 | 0.05524808 | -0.01785863 | -0.02826680 | 0.01243860 | 0.01008485 | -0.00708294 | -0.00440736 | 0.00063322 | -0.00037660 | -0.00025491 | 0.00022393 | 0.00003463 | -0.00018849 | 0.00003216 | -0.00000575 | 0.00005327 |
| 41 | 63 | 0.04682607 | 0.05561002 | -0.01751068 | -0.02791038 | 0.01338069 | 0.00948737 | -0.00791546 | -0.00523675 | 0.00100907 | -0.00016635 | -0.00026308 | 0.00009182 | 0.00030029 | -0.00024050 | 0.00019143 | -0.00012875 | 0.00018510 |
| 41 | 64 | 0.04700746 | 0.05365693 | -0.01907735 | -0.02715134 | 0.01324035 | 0.00932161 | -0.00766642 | -0.00435259 | 0.00078037 | -0.00047732 | -0.00017481 | 0.00021447 | 0.00000873 | -0.00023245 | 0.00006013 | -0.00001713 | 0.00010158 |
| 41 | 65 | 0.04653941 | 0.05405313 | -0.01867825 | -0.02645101 | 0.01423004 | 0.00843298 | -0.00861630 | -0.00517135 | 0.00113544 | -0.00030251 | -0.00017423 | 0.00006312 | 0.00030355 | -0.00030613 | 0.00022809 | -0.00013532 | 0.00024510 |

TABLE 3. Nuclear charge density distribution FB coefficients.

| $Z$ | $N$ | $a_1$ | $a_2$ | $a_3$ | $a_4$ | $a_5$ | $a_6$ | $a_7$ | $a_8$ | $a_9$ | $a_{10}$ | $a_{11}$ | $a_{12}$ | $a_{13}$ | $a_{14}$ | $a_{15}$ | $a_{16}$ | $a_{17}$ |
|---|---|---|---|---|---|---|---|---|---|---|---|---|---|---|---|---|---|---|
| 41 | 66 | 0.04675344 | 0.05224927 | -0.02028582 | -0.02619638 | 0.01378071 | 0.00854271 | -0.00807480 | -0.00430268 | 0.00086714 | -0.00059451 | -0.00008434 | 0.00017886 | 0.00001379 | -0.00029868 | 0.00009050 | -0.00001762 | 0.00016847 |
| 41 | 67 | 0.04627721 | 0.05251676 | -0.01980402 | -0.02534511 | 0.01455973 | 0.00759375 | -0.00866655 | -0.00487270 | 0.00103869 | -0.00037057 | -0.00011063 | 0.00007796 | 0.00022354 | -0.00028299 | 0.00021218 | -0.00012031 | 0.00023755 |
| 41 | 68 | 0.04633686 | 0.05070355 | -0.02107405 | -0.02516131 | 0.01404322 | 0.00791572 | -0.00781459 | -0.00390876 | 0.00065855 | -0.00055044 | -0.00007630 | 0.00023991 | -0.00012490 | -0.00016838 | 0.00005169 | -0.00002489 | 0.00007564 |
| 41 | 69 | 0.04602610 | 0.05093301 | -0.02107399 | -0.02457232 | 0.01483889 | 0.00713318 | -0.00823829 | -0.00439511 | 0.00077719 | -0.00039478 | -0.00007648 | 0.00013397 | 0.00007031 | -0.00017418 | 0.00016433 | -0.00012360 | 0.00014950 |
| 41 | 70 | 0.04599012 | 0.04920349 | -0.02213071 | -0.02421485 | 0.01462777 | 0.00739180 | -0.00761263 | -0.00354926 | 0.00049535 | -0.00055918 | -0.00005855 | 0.00027822 | -0.00022917 | -0.00006768 | 0.00003905 | -0.00006211 | -0.00000813 |
| 41 | 71 | 0.04581104 | 0.04939821 | -0.02254936 | -0.02384952 | 0.01543384 | 0.00668030 | -0.00789757 | -0.00393566 | 0.00057396 | -0.00047752 | -0.00003538 | 0.00015863 | -0.00003182 | -0.00009599 | 0.00015485 | -0.00017281 | 0.00007229 |
| 41 | 72 | 0.04569628 | 0.04779877 | -0.02338437 | -0.02341383 | 0.01541956 | 0.00692688 | -0.00750303 | -0.00320775 | 0.00039783 | -0.00062054 | -0.00003104 | 0.00029444 | -0.00029944 | -0.00000034 | 0.00005071 | -0.00012536 | -0.00008340 |
| 41 | 73 | 0.04561681 | 0.04795942 | -0.02414437 | -0.02321529 | 0.01624138 | 0.00621788 | -0.00768273 | -0.00348333 | 0.00044558 | -0.00061411 | 0.00000956 | 0.00015913 | -0.00009090 | -0.00004822 | 0.00017699 | -0.00025964 | 0.00000043 |
| 41 | 74 | 0.04543514 | 0.04651218 | -0.02477417 | -0.02280375 | 0.01631916 | 0.00651720 | -0.00749920 | -0.00288263 | 0.00036789 | -0.00072515 | 0.00000416 | 0.00029258 | -0.00034122 | 0.00003490 | 0.00008003 | -0.00020581 | -0.00015222 |
| 41 | 75 | 0.04541848 | 0.04663093 | -0.02578456 | -0.02271592 | 0.01715191 | 0.00578357 | -0.00759529 | -0.00304830 | 0.00038574 | -0.00078657 | 0.00005414 | 0.00014642 | -0.00012170 | -0.00002559 | 0.00021711 | -0.00036751 | -0.00007167 |
| 42 | 39 | 0.06177843 | 0.05637860 | -0.03165636 | -0.02123839 | 0.01992832 | 0.00548717 | -0.00645504 | -0.00118982 | 0.00011848 | -0.00081275 | 0.00006029 | 0.00015337 | -0.00006842 | -0.00001411 | 0.00005026 | -0.00029646 | -0.00022017 |
| 42 | 40 | 0.06197346 | 0.05681674 | -0.03194903 | -0.02173110 | 0.01884935 | 0.00479735 | -0.00706232 | -0.00153454 | 0.00017480 | -0.00058623 | 0.00003872 | 0.00009282 | 0.00001662 | -0.00003508 | 0.00006320 | -0.00020863 | -0.00013964 |
| 42 | 41 | 0.06217184 | 0.05553992 | -0.03344155 | -0.02096732 | 0.01936333 | 0.00424655 | -0.00679117 | -0.00080191 | 0.00010404 | -0.00073231 | 0.00010766 | 0.00020251 | -0.00018715 | 0.00002389 | -0.00000015 | -0.00015283 | -0.00018411 |
| 42 | 42 | 0.06216446 | 0.05603319 | -0.03359738 | -0.02160830 | 0.01859139 | 0.00396795 | -0.00721764 | -0.00122936 | 0.00017443 | -0.00056190 | 0.00009598 | 0.00011551 | -0.00006020 | -0.00002409 | 0.00003632 | -0.00009987 | -0.00008684 |
| 42 | 43 | 0.06238959 | 0.05502159 | -0.03504047 | -0.02098018 | 0.01895673 | 0.00334510 | -0.00697607 | -0.00061883 | 0.00008832 | -0.00067435 | 0.00015437 | 0.00022305 | -0.00024563 | 0.00003415 | -0.00002797 | -0.00004415 | -0.00012533 |
| 42 | 44 | 0.06225054 | 0.05548497 | -0.03514322 | -0.02168119 | 0.01858472 | 0.00347474 | -0.00726849 | -0.00106228 | 0.00017775 | -0.00056595 | 0.00014471 | 0.00012458 | -0.00009695 | -0.00003129 | 0.00002332 | -0.00002204 | -0.00002941 |
| 42 | 45 | 0.06247597 | 0.05470370 | -0.03662808 | -0.02128558 | 0.01895854 | 0.00295128 | -0.00705466 | -0.00057392 | 0.00008988 | -0.00066493 | 0.00019459 | 0.00022246 | -0.00025450 | 0.00001628 | -0.00003528 | 0.00002513 | -0.00006179 |
| 42 | 46 | 0.06224782 | 0.05507172 | -0.03669957 | -0.02195453 | 0.01896523 | 0.00339910 | -0.00725112 | -0.00097659 | 0.00020314 | -0.00061291 | 0.00018325 | 0.00012340 | -0.00010135 | -0.00005660 | 0.00002388 | 0.00002323 | 0.00002225 |
| 42 | 47 | 0.06242829 | 0.05448961 | -0.03822878 | -0.02184482 | 0.01942892 | 0.00308752 | -0.00705549 | -0.00058969 | 0.00012883 | -0.00071735 | 0.00022468 | 0.00020960 | -0.00022838 | -0.00002493 | -0.00002382 | 0.00005360 | -0.00000798 |
| 42 | 48 | 0.06214815 | 0.05471898 | -0.03828393 | -0.02241146 | 0.01972697 | 0.00372174 | -0.00719827 | -0.00091726 | 0.00026713 | -0.00070691 | 0.00021061 | 0.00011743 | -0.00008520 | -0.00009607 | 0.00003649 | 0.00003763 | 0.00005934 |
| 42 | 49 | 0.06223561 | 0.05429867 | -0.03981380 | -0.02261355 | 0.02026531 | 0.00365772 | -0.00702347 | -0.00061003 | 0.00021520 | -0.00082631 | 0.00024102 | 0.00019684 | -0.00018862 | -0.00007878 | 0.00000028 | 0.00004770 | 0.00002374 |
| 42 | 50 | 0.06194836 | 0.05436696 | -0.03989212 | -0.02303661 | 0.02075940 | 0.00435088 | -0.00715701 | -0.00085140 | 0.00037423 | -0.00083970 | 0.00022532 | 0.00011531 | -0.00006547 | -0.00014167 | 0.00005577 | 0.00002839 | 0.00007375 |
| 42 | 51 | 0.06214048 | 0.05356000 | -0.04039153 | -0.02172449 | 0.02069779 | 0.00294775 | -0.00721912 | -0.00053539 | 0.00018105 | -0.00082781 | 0.00029394 | 0.00012531 | -0.00011432 | -0.00010182 | 0.00002905 | 0.00001000 | 0.00005221 |
| 42 | 52 | 0.06189968 | 0.05327608 | -0.03887658 | -0.02023119 | 0.02032562 | 0.00210504 | -0.00771361 | -0.00090091 | 0.00015350 | -0.00068761 | 0.00030467 | -0.00000638 | 0.00004223 | -0.00013350 | 0.00007296 | -0.00002586 | 0.00009346 |
| 42 | 53 | 0.06199973 | 0.05241929 | -0.03945290 | -0.01908953 | 0.02046796 | 0.00095825 | -0.00767198 | -0.00052389 | 0.00002607 | -0.00068508 | 0.00037001 | -0.00000364 | 0.00000241 | -0.00009605 | 0.00005958 | -0.00005059 | 0.00008373 |
| 42 | 54 | 0.06172142 | 0.05218083 | -0.03763574 | -0.01764143 | 0.01972707 | 0.00014088 | -0.00821414 | -0.00103931 | -0.00002989 | -0.00051202 | 0.00033585 | -0.00008451 | 0.00011149 | -0.00010174 | 0.00007603 | -0.00007254 | 0.00007804 |
| 42 | 55 | 0.06162522 | 0.05137520 | -0.03805310 | -0.01670151 | 0.02005683 | -0.00065197 | -0.00810706 | -0.00066799 | -0.00006800 | -0.00052790 | 0.00038968 | -0.00008810 | 0.00008740 | -0.00007619 | 0.00007869 | -0.00010481 | 0.00007995 |
| 42 | 56 | 0.06186670 | 0.05086102 | -0.03690886 | -0.01529767 | 0.01898456 | -0.00174516 | -0.00836952 | -0.00081565 | 0.00004389 | -0.00025821 | 0.00023725 | -0.00000550 | 0.00001911 | 0.00002578 | 0.00002837 | -0.00004172 | 0.00000274 |
| 42 | 57 | 0.06168452 | 0.05009499 | -0.03721788 | -0.01441333 | 0.01945949 | -0.00231799 | -0.00824227 | -0.00043116 | 0.00009670 | -0.00030230 | 0.00027319 | 0.00000066 | -0.00000664 | 0.00003893 | 0.00003643 | -0.00006463 | 0.00001532 |
| 42 | 58 | 0.04862949 | 0.05917804 | -0.01657664 | -0.03199615 | 0.01028257 | 0.01174395 | -0.00663479 | -0.00562205 | 0.00017972 | 0.00000913 | -0.00041703 | 0.00013160 | 0.00025619 | -0.00014778 | 0.00001529 | -0.00002523 | -0.00003614 |
| 42 | 59 | 0.04842288 | 0.05840703 | -0.01696791 | -0.03107247 | 0.01100735 | 0.01130788 | -0.00657797 | -0.00520459 | 0.00027624 | -0.00011163 | -0.00036725 | 0.00013280 | 0.00023558 | -0.00016839 | 0.00003007 | -0.00003830 | 0.00000342 |
| 42 | 60 | 0.04848018 | 0.05721765 | -0.01833438 | -0.02986834 | 0.01225749 | 0.01049224 | -0.00738334 | -0.00491851 | 0.00066739 | -0.00003142 | -0.00035522 | 0.00021022 | 0.00013697 | -0.00012073 | 0.00003803 | -0.00002834 | 0.00001627 |
| 42 | 61 | 0.04821282 | 0.05652213 | -0.01853628 | -0.02906745 | 0.01280732 | 0.01011739 | -0.00739834 | -0.00460661 | 0.00075656 | -0.00018386 | -0.00031449 | 0.00020032 | 0.00013465 | -0.00016237 | 0.00005734 | -0.00004037 | 0.00005664 |
| 42 | 62 | 0.04824627 | 0.05557976 | -0.01968289 | -0.02851689 | 0.01344105 | 0.00987090 | -0.00787726 | -0.00474504 | 0.00081867 | -0.00006033 | -0.00027210 | 0.00021056 | 0.00009169 | -0.00012668 | 0.00005605 | -0.00004808 | 0.00004556 |
| 42 | 63 | 0.04786476 | 0.05497472 | -0.01967243 | -0.02788276 | 0.01379184 | 0.00947857 | -0.00797748 | -0.00449296 | 0.00094079 | -0.00024293 | -0.00024567 | 0.00019571 | 0.00010550 | -0.00018744 | 0.00008334 | -0.00005735 | 0.00009205 |
| 42 | 64 | 0.04807528 | 0.05419574 | -0.02090870 | -0.02743266 | 0.01428897 | 0.00935758 | -0.00821973 | -0.00461128 | 0.00090669 | -0.00011854 | -0.00018672 | 0.00019454 | 0.00006476 | -0.00015596 | 0.00006943 | -0.00005623 | 0.00007858 |
| 42 | 65 | 0.04761285 | 0.05361741 | -0.02080294 | -0.02688408 | 0.01447327 | 0.00882362 | -0.00839666 | -0.00439186 | 0.00105785 | -0.00033177 | -0.00016503 | 0.00016776 | 0.00010125 | -0.00023698 | 0.00011008 | -0.00006364 | 0.00014264 |
| 42 | 66 | 0.04802210 | 0.05304003 | -0.02204066 | -0.02643019 | 0.01484253 | 0.00872053 | -0.00844969 | -0.00443901 | 0.00097395 | -0.00020850 | -0.00010260 | 0.00016729 | 0.00005631 | -0.00020563 | 0.00008321 | -0.00005317 | 0.00012371 |
| 42 | 67 | 0.04728397 | 0.05222657 | -0.02166778 | -0.02590536 | 0.01472012 | 0.00803975 | -0.00843727 | -0.00410253 | 0.00100741 | -0.00036329 | -0.00011932 | 0.00017710 | 0.00003931 | -0.00020175 | 0.00010752 | -0.00006711 | 0.00013087 |
| 42 | 68 | 0.04741748 | 0.05146845 | -0.02260143 | -0.02557322 | 0.01488478 | 0.00800173 | -0.00823722 | -0.00404004 | 0.00082012 | -0.00017255 | -0.00010795 | 0.00021442 | -0.00005380 | -0.00007834 | 0.00006497 | -0.00007241 | 0.00004817 |
| 42 | 69 | 0.04687145 | 0.05069274 | -0.02252171 | -0.02494435 | 0.01503566 | 0.00737165 | -0.00822189 | -0.00367531 | 0.00083548 | -0.00036313 | -0.00010699 | 0.00022441 | -0.00007975 | -0.00008593 | 0.00008907 | -0.00009583 | 0.00004547 |
| 42 | 70 | 0.04693573 | 0.04995273 | -0.02346445 | -0.02475545 | 0.01529498 | 0.00740323 | -0.00808997 | -0.00365479 | 0.00071102 | -0.00020113 | -0.00010423 | 0.00024651 | -0.00014305 | 0.00002108 | 0.00006648 | -0.00011371 | -0.00002196 |
| 42 | 71 | 0.04655541 | 0.04924122 | -0.02363419 | -0.02407687 | 0.01565086 | 0.00680723 | -0.00807694 | -0.00327726 | 0.00071140 | -0.00042048 | -0.00008427 | 0.00025154 | -0.00016855 | -0.00000187 | 0.00009282 | -0.00014910 | -0.00003177 |
| 42 | 72 | 0.04655341 | 0.04854054 | -0.02456563 | -0.02404676 | 0.01595097 | 0.00688961 | -0.00803314 | -0.00328254 | 0.00066455 | -0.00028261 | -0.00009024 | 0.00026148 | -0.00020845 | 0.00008921 | 0.00008579 | -0.00017144 | -0.00008362 |
| 42 | 73 | 0.04631293 | 0.04790606 | -0.02493927 | -0.02335529 | 0.01644903 | 0.00631217 | -0.00802493 | -0.00290251 | 0.00064880 | -0.00052423 | -0.00005143 | 0.00025932 | -0.00022801 | 0.00004926 | 0.00011494 | -0.00022005 | -0.00010019 |
| 42 | 74 | 0.04625073 | 0.04724638 | -0.02585900 | -0.02348747 | 0.01675666 | 0.00644307 | -0.00807409 | -0.00292437 | 0.00068329 | -0.00040185 | -0.00006620 | 0.00026002 | -0.00025150 | 0.00012703 | 0.00011836 | -0.00023822 | -0.00013577 |
| 42 | 75 | 0.04611907 | 0.04668643 | -0.02638943 | -0.02281456 | 0.01734267 | 0.00589088 | -0.00806655 | -0.00255242 | 0.00064453 | -0.00065810 | -0.00001105 | 0.00025234 | -0.00026457 | 0.00007137 | 0.00014796 | -0.00029912 | -0.00016177 |
| 42 | 76 | 0.04601112 | 0.04605659 | -0.02731699 | -0.02309577 | 0.01765379 | 0.00607491 | -0.00820611 | -0.00258432 | 0.00075778 | -0.00054182 | -0.00003464 | 0.00024621 | -0.00027798 | 0.00013966 | 0.00015750 | -0.00030631 | -0.00018072 |
| 42 | 77 | 0.04594329 | 0.04555924 | -0.02794635 | -0.02248829 | 0.01827336 | 0.00559198 | -0.00818624 | -0.00223942 | 0.00068221 | -0.00080337 | 0.00003177 | 0.00023802 | -0.00028814 | 0.00007212 | 0.00018250 | -0.00037668 | -0.00022186 |
| 43 | 40 | 0.06284086 | 0.05530700 | -0.03415158 | -0.02057504 | 0.02030108 | 0.00444153 | -0.00683788 | -0.00111626 | 0.00005089 | -0.00068661 | 0.00006885 | 0.00020643 | -0.00015978 | 0.00005370 | 0.00001254 | -0.00024359 | -0.00023595 |
| 43 | 41 | 0.06257764 | 0.05483284 | -0.03441479 | -0.02028687 | 0.01999566 | 0.00324908 | -0.00665285 | -0.00127889 | 0.00007953 | -0.00041583 | 0.00005633 | 0.00019036 | -0.00009877 | 0.00016887 | 0.00008893 | -0.00031883 | -0.00019247 |
| 43 | 42 | 0.06323089 | 0.05461371 | -0.03585950 | -0.02020110 | 0.01994376 | 0.00329255 | -0.00703680 | -0.00072154 | 0.00003495 | -0.00067167 | 0.00014312 | 0.00021879 | -0.00023486 | 0.00005684 | -0.00001552 | -0.00012431 | -0.00017129 |
| 43 | 43 | 0.06295828 | 0.05424319 | -0.03624549 | -0.02017234 | 0.01979053 | 0.00225637 | -0.00686171 | -0.00097159 | 0.00005174 | -0.00043948 | 0.00015139 | 0.00016609 | -0.00012511 | 0.00013626 | 0.00008251 | -0.00022684 | -0.00010666 |
| 43 | 44 | 0.06351843 | 0.05421153 | -0.03750394 | -0.02015933 | 0.01981645 | 0.00253572 | -0.00711625 | -0.00048303 | 0.00002622 | -0.00069124 | 0.00020959 | 0.00021339 | -0.00026325 | 0.00003517 | -0.00002754 | -0.00003555 | -0.00009723 |
| 43 | 45 | 0.06321623 | 0.05389132 | -0.03801619 | -0.02036718 | 0.01981082 | 0.00165355 | -0.00694777 | -0.00076652 | 0.00004811 | -0.00049867 | 0.00023556 | 0.00012529 | -0.00010917 | 0.00008065 | 0.00008988 | -0.00016317 | -0.00001522 |
| 43 | 46 | 0.06368957 | 0.05398319 | -0.03917306 | -0.02045724 | 0.02007546 | 0.00228424 | -0.00711917 | -0.00035439 | 0.00004141 | -0.00075931 | 0.00026482 | 0.00019453 | -0.00025227 | -0.00000976 | -0.00002253 | 0.00001665 | -0.00002765 |
| 43 | 47 | 0.06331932 | 0.05369428 | -0.03971248 | -0.02083409 | 0.02015932 | 0.00152784 | -0.00695899 | -0.00063594 | 0.00007825 | -0.00059899 | 0.00030152 | 0.00007860 | -0.00006353 | 0.00000898 | 0.00010922 | -0.00013172 | 0.00006460 |
| 43 | 48 | 0.06370299 | 0.05383861 | -0.04083096 | -0.02105198 | 0.02071518 | 0.00252810 | -0.00707785 | -0.00028655 | 0.00009659 | -0.00087542 | 0.00030561 | 0.00017029 | -0.00021519 | -0.00007104 | -0.00000248 | 0.00003231 | 0.00002589 |
| 43 | 49 | 0.06323283 | 0.05357840 | -0.04127421 | -0.02152209 | 0.02080114 | 0.00185175 | -0.00693997 | -0.00055202 | 0.00015039 | -0.00073451 | 0.00034399 | 0.00003857 | -0.00000570 | -0.00006853 | 0.00013650 | -0.00013010 | 0.00011970 |
| 43 | 50 | 0.06353360 | 0.05370504 | -0.04242195 | -0.02188662 | 0.02161876 | 0.00317080 | -0.00703572 | -0.00024685 | 0.00019673 | -0.00102525 | 0.00032850 | 0.00015236 | -0.00017186 | -0.00013737 | 0.00002528 | 0.00002003 | 0.00005353 |
| 43 | 51 | 0.06313896 | 0.05299736 | -0.04175617 | -0.02071635 | 0.02119432 | 0.00120882 | -0.00715629 | -0.00053816 | 0.00014864 | -0.00072738 | 0.00039326 | -0.00004270 | 0.00008513 | -0.00010326 | 0.00017072 | -0.00016212 | 0.00016605 |
| 43 | 52 | 0.06351610 | 0.05258793 | -0.04160348 | -0.01914444 | 0.02137560 | 0.00101451 | -0.00756111 | -0.00025740 | 0.00002683 | -0.00086010 | 0.00040968 | 0.00001875 | -0.00005232 | -0.00013003 | 0.00005283 | -0.00003008 | 0.00009685 |
| 43 | 53 | 0.06289651 | 0.05211791 | -0.04070272 | -0.01843573 | 0.02106684 | -0.00038401 | -0.00768951 | -0.00076253 | 0.00009436 | -0.00053232 | 0.00042898 | -0.00015565 | 0.00020964 | -0.00009527 | 0.00020508 | -0.00021161 | 0.00020453 |
| 43 | 54 | 0.06317253 | 0.05160535 | -0.04025880 | -0.01676681 | 0.02091070 | -0.00074353 | -0.00810955 | -0.00047109 | -0.00007927 | -0.00065820 | 0.00043148 | -0.00007399 | 0.00004678 | -0.00010595 | 0.00007146 | -0.00007931 | 0.00010484 |
| 43 | 55 | 0.06267797 | 0.05132780 | -0.03947717 | -0.01639527 | 0.02071113 | -0.00174232 | -0.00816543 | -0.00096936 | 0.00017900 | -0.00028472 | 0.00036205 | -0.00016654 | 0.00024613 | -0.00003784 | 0.00020048 | -0.00021384 | 0.00018356 |
| 43 | 56 | 0.06299820 | 0.05058920 | -0.03912465 | -0.01483648 | 0.02016402 | -0.00231252 | -0.00838084 | -0.00043716 | 0.00007166 | -0.00037396 | 0.00030620 | -0.00000520 | -0.00000367 | 0.00000908 | 0.00003885 | -0.00005311 | 0.00004523 |
| 43 | 57 | 0.05080663 | 0.05951631 | -0.02005251 | -0.03177433 | 0.01194216 | 0.01066875 | -0.00658172 | -0.00579669 | 0.00017613 | 0.00008549 | -0.00032754 | -0.00004911 | 0.00051301 | -0.00017799 | 0.00016086 | -0.00016761 | 0.00010051 |
| 43 | 58 | 0.04935470 | 0.05937933 | -0.01813150 | -0.03214425 | 0.01127250 | 0.01160941 | -0.00674406 | -0.00547028 | 0.00026243 | -0.00007897 | -0.00039145 | 0.00012962 | 0.00027604 | -0.00017559 | 0.00003525 | -0.00003913 | 0.00000890 |
| 43 | 59 | 0.04903542 | 0.05934704 | -0.01857059 | -0.03185031 | 0.01278535 | 0.01124961 | -0.00729819 | -0.00569773 | 0.00081816 | 0.00009911 | -0.00042665 | 0.00007807 | 0.00043883 | -0.00017288 | 0.00017253 | -0.00013839 | 0.00013378 |
| 43 | 60 | 0.04915190 | 0.05754441 | -0.01968525 | -0.03000388 | 0.01327373 | 0.01037319 | -0.00761455 | -0.00479385 | 0.00080579 | -0.00009340 | -0.00035552 | 0.00020311 | 0.00018137 | -0.00014882 | 0.00006364 | -0.00004485 | 0.00005544 |
| 43 | 61 | 0.04885862 | 0.05784760 | -0.01967658 | -0.02984447 | 0.01441494 | 0.01024232 | -0.00823481 | -0.00541291 | 0.00115069 | 0.00008489 | -0.00038094 | 0.00009430 | 0.00040672 | -0.00017603 | 0.00020041 | -0.00015733 | 0.00015782 |
| 43 | 62 | 0.04878512 | 0.05604149 | -0.02071226 | -0.02868124 | 0.01444742 | 0.00978517 | -0.00821374 | -0.00462796 | 0.00102661 | -0.00010285 | -0.00029680 | 0.00019855 | 0.00015858 | -0.00015745 | 0.00008933 | -0.00006824 | 0.00008178 |
| 43 | 63 | 0.04860940 | 0.05652062 | -0.02059955 | -0.02827559 | 0.01555679 | 0.00938173 | -0.00897649 | -0.00526580 | 0.00135688 | 0.00003873 | -0.00031808 | 0.00007186 | 0.00041168 | -0.00020719 | 0.00023482 | -0.00018120 | 0.00019022 |
| 43 | 64 | 0.04852227 | 0.05473360 | -0.02171985 | -0.02756481 | 0.01529592 | 0.00920834 | -0.00865322 | -0.00447809 | 0.00117868 | -0.00014984 | -0.00022587 | 0.00017328 | 0.00015473 | -0.00019139 | 0.00011410 | -0.00008164 | 0.00011909 |
| 43 | 65 | 0.04844854 | 0.05527997 | -0.02157831 | -0.02681827 | 0.01636302 | 0.00831285 | -0.00955379 | -0.00507735 | 0.00151638 | -0.00005285 | -0.00023818 | 0.00002449 | 0.00044292 | -0.00026338 | 0.00027949 | -0.00019968 | 0.00025085 |
| 43 | 66 | 0.04838914 | 0.05361109 | -0.02270552 | -0.02652364 | 0.01582822 | 0.00845137 | -0.00896757 | -0.00430100 | 0.00129437 | -0.00023452 | -0.00014696 | 0.00013119 | 0.00017222 | -0.00024845 | 0.00014293 | -0.00008460 | 0.00017712 |
| 43 | 67 | 0.04819109 | 0.05392940 | -0.02248622 | -0.02578533 | 0.01651390 | 0.00733132 | -0.00962703 | -0.00469146 | 0.00148078 | -0.00010650 | -0.00018408 | 0.00002697 | 0.00038411 | -0.00023766 | 0.00027916 | -0.00019903 | 0.00024808 |
| 43 | 68 | 0.04786842 | 0.05207597 | -0.02330978 | -0.02561092 | 0.01581145 | 0.00761588 | -0.00877613 | -0.00382103 | 0.00113555 | -0.00020714 | -0.00014200 | 0.00018366 | 0.00004824 | -0.00012038 | 0.00011954 | -0.00010459 | 0.00009291 |
| 43 | 69 | 0.04787141 | 0.05238257 | -0.02356063 | -0.02507180 | 0.01659338 | 0.00669324 | -0.00930558 | -0.00412942 | 0.00128638 | -0.00014843 | -0.00015655 | 0.00008118 | 0.00023557 | -0.00013132 | 0.00024449 | -0.00021006 | 0.00016361 |
| 43 | 70 | 0.04748621 | 0.05059574 | -0.02422734 | -0.02473524 | 0.01618038 | 0.00693575 | -0.00861999 | -0.00337061 | 0.00100059 | -0.00024799 | -0.00012658 | 0.00022084 | -0.00005465 | -0.00002061 | 0.00011517 | -0.00014754 | 0.00001359 |
| 43 | 71 | 0.04764653 | 0.05089601 | -0.02486504 | -0.02437445 | 0.01702406 | 0.00613773 | -0.00903543 | -0.00359336 | 0.00112377 | -0.00025195 | -0.00012160 | 0.00011325 | 0.00012310 | -0.00005284 | 0.00023595 | -0.00025481 | 0.00008543 |
| 43 | 72 | 0.04721412 | 0.04922532 | -0.02538243 | -0.02396530 | 0.01681197 | 0.00636951 | -0.00853148 | -0.00295327 | 0.00091233 | -0.00034139 | -0.00010005 | 0.00023984 | -0.00013116 | 0.00004730 | 0.00012855 | -0.00020853 | -0.00005747 |
| 43 | 73 | 0.04749669 | 0.04951920 | -0.02632525 | -0.02374638 | 0.01769401 | 0.00564486 | -0.00884928 | -0.00309152 | 0.00100863 | -0.00039958 | -0.00008069 | 0.00012616 | 0.00004469 | -0.00000221 | 0.00024834 | -0.00032493 | 0.00001260 |
| 43 | 74 | 0.04702187 | 0.04797851 | -0.02671816 | -0.02334752 | 0.01759562 | 0.00589257 | -0.00851968 | -0.00256935 | 0.00087826 | -0.00046860 | -0.00006302 | 0.00024146 | -0.00018234 | 0.00008454 | 0.00015496 | -0.00027956 | -0.00011888 |
| 43 | 75 | 0.04739174 | 0.04825486 | -0.02788286 | -0.02322711 | 0.01849335 | 0.00521890 | -0.00874587 | -0.00262945 | 0.00094109 | -0.00056921 | -0.00003576 | 0.00012619 | -0.00000835 | 0.00002499 | 0.00027235 | -0.00040726 | -0.00005618 |
| 43 | 76 | 0.04688104 | 0.04683403 | -0.02820075 | -0.02290763 | 0.01845916 | 0.00551431 | -0.00857779 | -0.00221990 | 0.00089205 | -0.00061068 | -0.00001868 | 0.00023088 | -0.00021531 | 0.00009690 | 0.00018688 | -0.00035156 | -0.00017316 |
| 43 | 77 | 0.04729826 | 0.04707184 | -0.02950151 | -0.02284203 | 0.01934615 | 0.00488933 | -0.00870887 | -0.00221319 | 0.00090976 | -0.00074057 | 0.00000972 | 0.00012162 | -0.00004816 | 0.00003618 | 0.00029763 | -0.00048897 | -0.00012414 |
| 43 | 78 | 0.04676362 | 0.04575427 | -0.02980779 | -0.02266457 | 0.01936436 | 0.00527487 | -0.00869355 | -0.00190966 | 0.00093913 | -0.00075140 | 0.00002719 | 0.00021626 | -0.00024065 | 0.00009286 | 0.00021587 | -0.00041672 | -0.00022664 |
| 43 | 79 | 0.04718243 | 0.04593059 | -0.03115459 | -0.02261161 | 0.02021072 | 0.00469776 | -0.00872485 | -0.00185100 | 0.00089979 | -0.00089967 | 0.00005085 | 0.00012022 | -0.00008609 | 0.00003918 | 0.00031606 | -0.00056131 | -0.00019613 |
| 44 | 41 | 0.06380102 | 0.05430822 | -0.03637459 | -0.01971002 | 0.02057636 | 0.00327283 | -0.00720801 | -0.00114825 | -0.00001512 | -0.00055895 | 0.00008324 | 0.00024403 | -0.00022392 | 0.00010657 | -0.00001665 | -0.00019582 | -0.00023143 |
| 44 | 42 | 0.06400039 | 0.05518515 | -0.03623204 | -0.02032116 | 0.01951979 | 0.00282923 | -0.00786485 | -0.00176001 | 0.00011377 | -0.00030848 | 0.00006001 | 0.00015652 | -0.00008688 | 0.00003531 | 0.00000035 | -0.00009385 | -0.00010961 |
| 44 | 43 | 0.06421013 | 0.05378262 | -0.03805723 | -0.01934239 | 0.02048781 | 0.00231249 | -0.00731503 | -0.00073537 | -0.00002244 | -0.00061544 | 0.00017750 | 0.00022279 | -0.00025838 | 0.00007601 | -0.00002408 | -0.00010247 | -0.00014728 |
| 44 | 44 | 0.06419656 | 0.05473011 | -0.03771411 | -0.02010415 | 0.01975923 | 0.00228690 | -0.00788934 | -0.00143490 | 0.00012044 | -0.00041675 | 0.00015441 | 0.00011761 | -0.00008712 | -0.00001435 | 0.00001261 | -0.00003863 | -0.00002567 |
| 44 | 45 | 0.06457822 | 0.05349422 | -0.03978075 | -0.01932018 | 0.02064404 | 0.00173812 | -0.00731036 | -0.00042515 | -0.00001927 | -0.00071168 | 0.00026027 | 0.00019331 | -0.00025976 | 0.00002524 | -0.00002123 | -0.00003371 | -0.00006307 |
| 44 | 46 | 0.06438330 | 0.05441113 | -0.03928130 | -0.02011348 | 0.02025842 | 0.00204918 | -0.00783014 | -0.00115591 | 0.00014365 | -0.00055982 | 0.00023379 | 0.00008034 | -0.00006962 | -0.00007414 | 0.00003042 | -0.00000384 | 0.00004891 |
| 44 | 47 | 0.06483578 | 0.05333732 | -0.04154680 | -0.01964612 | 0.02113597 | 0.00163285 | -0.00724135 | -0.00019958 | 0.00000990 | -0.00085017 | 0.00032942 | 0.00015727 | -0.00023214 | -0.00004294 | -0.00000508 | 0.00000293 | 0.00001032 |
| 44 | 48 | 0.06449994 | 0.05415177 | -0.04092193 | -0.02035785 | 0.02105630 | 0.00216730 | -0.00773219 | -0.00090918 | 0.00019918 | -0.00073519 | 0.00029762 | 0.00004500 | -0.00003867 | -0.00014195 | 0.00005534 | 0.00000748 | 0.00010695 |
| 44 | 49 | 0.06491517 | 0.05323438 | -0.04327505 | -0.02027676 | 0.02192142 | 0.00197889 | -0.00714814 | -0.00003801 | 0.00007722 | -0.00102023 | 0.00038140 | 0.00012283 | -0.00018836 | -0.00012032 | 0.00002130 | 0.00000919 | 0.00006280 |
| 44 | 50 | 0.06450075 | 0.05389889 | -0.04258488 | -0.02082049 | 0.02209340 | 0.00262115 | -0.00763735 | -0.00068585 | 0.00029388 | -0.00093080 | 0.00034421 | 0.00001769 | -0.00000596 | -0.00021154 | 0.00008319 | 0.00000011 | 0.00014138 |
| 44 | 51 | 0.06487739 | 0.05266042 | -0.04366655 | -0.01929682 | 0.02227816 | 0.00125050 | -0.00738036 | -0.00001157 | 0.00004022 | -0.00102595 | 0.00043820 | 0.00004594 | -0.00011038 | -0.00015546 | 0.00004684 | -0.00001527 | 0.00010162 |
| 44 | 52 | 0.06441627 | 0.05285347 | -0.04156378 | -0.01824414 | 0.02162911 | 0.00068070 | -0.00811491 | -0.00083459 | 0.00009690 | -0.00073161 | 0.00039292 | -0.00007141 | 0.00006667 | -0.00018092 | 0.00008490 | -0.00002771 | 0.00015084 |
| 44 | 53 | 0.06459273 | 0.05172182 | -0.04244369 | -0.01694560 | 0.02183873 | -0.00051444 | -0.00796652 | -0.00026078 | -0.00008314 | -0.00081324 | 0.00046548 | -0.00004670 | -0.00001540 | -0.00013040 | 0.00006083 | -0.00005219 | 0.00011817 |
| 44 | 54 | 0.06423020 | 0.05187146 | -0.04059019 | -0.01655898 | 0.02075129 | -0.00090155 | -0.00837286 | -0.00087834 | 0.00011814 | -0.00043101 | 0.00029934 | -0.00002110 | 0.00002328 | -0.00006057 | 0.00004887 | -0.00000681 | 0.00008984 |
| 44 | 55 | 0.06425665 | 0.05090236 | -0.04119537 | -0.01538252 | 0.02099389 | -0.00193947 | -0.00834151 | -0.00040047 | 0.00003294 | -0.00050442 | 0.00034077 | 0.00001025 | -0.00004263 | -0.00001919 | 0.00003166 | -0.00002932 | 0.00006386 |
| 44 | 56 | 0.05039523 | 0.06090214 | -0.01923627 | -0.03446342 | 0.01125062 | 0.01306835 | -0.00668048 | -0.00612716 | 0.00020694 | -0.00002906 | -0.00043474 | 0.00012266 | 0.00030847 | -0.00020126 | 0.00003490 | -0.00000156 | 0.00001001 |
| 44 | 57 | 0.05032376 | 0.06010348 | -0.01967389 | -0.03334745 | 0.01166615 | 0.01223184 | -0.00674832 | -0.00569148 | 0.00019635 | -0.00013522 | -0.00040171 | 0.00015049 | 0.00026276 | -0.00018690 | 0.00002594 | -0.00001922 | 0.00000576 |
| 44 | 58 | 0.05031109 | 0.05892335 | -0.02116147 | -0.03208773 | 0.01349706 | 0.01155717 | -0.00760332 | -0.00527543 | 0.00077871 | 0.00001413 | -0.00040051 | 0.00022536 | 0.00017906 | -0.00013399 | 0.00005178 | -0.00000079 | 0.00004275 |
| 44 | 59 | 0.05015412 | 0.05831832 | -0.02126878 | -0.03109963 | 0.01381013 | 0.01088090 | -0.00770768 | -0.00495626 | 0.00078375 | -0.00010360 | -0.00038067 | 0.00023232 | 0.00017108 | -0.00014212 | 0.00005377 | -0.00002552 | 0.00004539 |
| 44 | 60 | 0.05004930 | 0.05725920 | -0.02241923 | -0.03040783 | 0.01496809 | 0.01084563 | -0.00823041 | -0.00501565 | 0.00096999 | 0.00006566 | -0.00033127 | 0.00023270 | 0.00013468 | -0.00010963 | 0.00006390 | -0.00002476 | 0.00005267 |
| 44 | 61 | 0.04978201 | 0.05685322 | -0.02221291 | -0.02960852 | 0.01515417 | 0.01026078 | -0.00836321 | -0.00474910 | 0.00103186 | -0.00006694 | -0.00032965 | 0.00022603 | 0.00015823 | -0.00013718 | 0.00007992 | -0.00005510 | 0.00006493 |

TABLE 3. Nuclear charge density distribution FB coefficients.

| $Z$ | $N$ | $a_1$ | $a_2$ | $a_3$ | $a_4$ | $a_5$ | $a_6$ | $a_7$ | $a_8$ | $a_9$ | $a_{10}$ | $a_{11}$ | $a_{12}$ | $a_{13}$ | $a_{14}$ | $a_{15}$ | $a_{16}$ | $a_{17}$ |
|---|---|---|---|---|---|---|---|---|---|---|---|---|---|---|---|---|---|---|
| 44 | 62 | 0.04982257 | 0.05586845 | -0.02343471 | -0.02898644 | 0.01610066 | 0.01033014 | -0.00867727 | -0.00480826 | 0.00110026 | 0.00008627 | -0.00026012 | 0.00022310 | 0.00010759 | -0.00011020 | 0.00007083 | -0.00004122 | 0.00006143 |
| 44 | 63 | 0.04950351 | 0.05559029 | -0.02308523 | -0.02832383 | 0.01617821 | 0.00970426 | -0.00884093 | -0.00455734 | 0.00121594 | -0.00006806 | -0.00026677 | 0.00019933 | 0.00016242 | -0.00015780 | 0.00010448 | -0.00007659 | 0.00009224 |
| 44 | 64 | 0.04971787 | 0.05472852 | -0.02432327 | -0.02768936 | 0.01690995 | 0.00975741 | -0.00897531 | -0.00455899 | 0.00121603 | 0.00006544 | -0.00018945 | 0.00020479 | 0.00009308 | -0.00013375 | 0.00007690 | -0.00004658 | 0.00008044 |
| 44 | 65 | 0.04936086 | 0.05451964 | -0.02392944 | -0.02714191 | 0.01686693 | 0.00899997 | -0.00917806 | -0.00433277 | 0.00136756 | -0.00011250 | -0.00019556 | 0.00015816 | 0.00018289 | -0.00020229 | 0.00013179 | -0.00008734 | 0.00013832 |
| 44 | 66 | 0.04972276 | 0.05384088 | -0.02508075 | -0.02654245 | 0.01736511 | 0.00906150 | -0.00914488 | -0.00428684 | 0.00131864 | 0.00000553 | -0.00012104 | 0.00017430 | 0.00009886 | -0.00017931 | 0.00008796 | -0.00004429 | 0.00011613 |
| 44 | 67 | 0.04902141 | 0.05331218 | -0.02452879 | -0.02617339 | 0.01695605 | 0.00810106 | -0.00923258 | -0.00396484 | 0.00136914 | -0.00012264 | -0.00015900 | 0.00015611 | 0.00014527 | -0.00016794 | 0.00013638 | -0.00009734 | 0.00013140 |
| 44 | 68 | 0.04901075 | 0.05235505 | -0.02539718 | -0.02576771 | 0.01717319 | 0.00818374 | -0.00899327 | -0.00383783 | 0.00121460 | 0.00005179 | -0.00012934 | 0.00020830 | 0.00001672 | -0.00005824 | 0.00007786 | -0.00006903 | 0.00005430 |
| 44 | 69 | 0.04853676 | 0.05183549 | -0.02520632 | -0.02532741 | 0.01703427 | 0.00727953 | -0.00910115 | -0.00347349 | 0.00124508 | -0.00012968 | -0.00015250 | 0.00019892 | 0.00003910 | -0.00005607 | 0.00012440 | -0.00012868 | 0.00005412 |
| 44 | 70 | 0.04846990 | 0.05091666 | -0.02607022 | -0.02503110 | 0.01735310 | 0.00743820 | -0.00889333 | -0.00340611 | 0.00113044 | 0.00002122 | -0.00012757 | 0.00023278 | -0.00005553 | 0.00003771 | 0.00008397 | -0.00011338 | -0.00000605 |
| 44 | 71 | 0.04819961 | 0.05044457 | -0.02616836 | -0.02454391 | 0.01744843 | 0.00661801 | -0.00900605 | -0.00301624 | 0.00114601 | -0.00020042 | -0.00013492 | 0.00022830 | -0.00004951 | 0.00002731 | 0.00012810 | -0.00017825 | -0.00001896 |
| 44 | 72 | 0.04806909 | 0.04957863 | -0.02702336 | -0.02438936 | 0.01781079 | 0.00681856 | -0.00885224 | -0.00299602 | 0.00108920 | -0.00006661 | -0.00011390 | 0.00024272 | -0.00011084 | 0.00010467 | 0.00010540 | -0.00017133 | -0.00005891 |
| 44 | 73 | 0.04797340 | 0.04917692 | -0.02734290 | -0.02388445 | 0.01807875 | 0.00608561 | -0.00896183 | -0.00260098 | 0.00108899 | -0.00031213 | -0.00010541 | 0.00024079 | -0.00011506 | 0.00008059 | 0.00014552 | -0.00023966 | -0.00008304 |
| 44 | 74 | 0.04777587 | 0.04835637 | -0.02820104 | -0.02388803 | 0.01844843 | 0.00631169 | -0.00886960 | -0.00261340 | 0.00109897 | -0.00018886 | -0.00008797 | 0.00023629 | -0.00014684 | 0.00014319 | 0.00013863 | -0.00023552 | -0.00010069 |
| 44 | 75 | 0.04782310 | 0.04802717 | -0.02868463 | -0.02338735 | 0.01882872 | 0.00566699 | -0.00896717 | -0.00222986 | 0.00107699 | -0.00044275 | -0.00006502 | 0.00023741 | -0.00015893 | 0.00010674 | 0.00017162 | -0.00030488 | -0.00013671 |
| 44 | 76 | 0.04756521 | 0.04723319 | -0.02957291 | -0.02354949 | 0.01919930 | 0.00591974 | -0.00893781 | -0.00226188 | 0.00115567 | -0.00032424 | -0.00005184 | 0.00021623 | -0.00016706 | 0.00015790 | 0.00017766 | -0.00029836 | -0.00013211 |
| 44 | 77 | 0.04771975 | 0.04695972 | -0.03017292 | -0.02306920 | 0.01964413 | 0.00537319 | -0.00901474 | -0.00190157 | 0.00110255 | -0.00057425 | -0.00001776 | 0.00022410 | -0.00018884 | 0.00011257 | 0.00019917 | -0.00036595 | -0.00018345 |
| 44 | 78 | 0.04742003 | 0.04617429 | -0.03112614 | -0.02337860 | 0.02003155 | 0.00566244 | -0.00904840 | -0.00194227 | 0.00124857 | -0.00045570 | -0.00001019 | 0.00018925 | -0.00017978 | 0.00015615 | 0.00021524 | -0.00035315 | -0.00015824 |
| 44 | 79 | 0.04763949 | 0.04592983 | -0.03179783 | -0.02293648 | 0.02050531 | 0.00523331 | -0.00910199 | -0.00161427 | 0.00115399 | -0.00069487 | 0.00002981 | 0.00020950 | -0.00021561 | 0.00010655 | 0.00022090 | -0.00041699 | -0.00023049 |
| 44 | 80 | 0.04732829 | 0.04514357 | -0.03285238 | -0.02337148 | 0.02093691 | 0.00556575 | -0.00919920 | -0.00165344 | 0.00136596 | -0.00057231 | 0.00003056 | 0.00016372 | -0.00019483 | 0.00014596 | 0.00024480 | -0.00039545 | -0.00018669 |
| 44 | 81 | 0.04755876 | 0.04490526 | -0.03354016 | -0.02299752 | 0.02140529 | 0.00527997 | -0.00923518 | -0.00137065 | 0.00121964 | -0.00079938 | 0.00007038 | 0.00020118 | -0.00024829 | 0.00009614 | 0.00023187 | -0.00045560 | -0.00028599 |
| 45 | 43 | 0.06444626 | 0.05280734 | -0.03929029 | -0.01908560 | 0.02090061 | 0.00154873 | -0.00715539 | -0.00128892 | -0.00007371 | -0.00035262 | 0.00012870 | 0.00022790 | -0.00019971 | 0.00021131 | 0.00005042 | -0.00023711 | -0.00015219 |
| 45 | 44 | 0.06510527 | 0.05306780 | -0.04017210 | -0.01871483 | 0.02098809 | 0.00150207 | -0.00760469 | -0.00082344 | -0.00005751 | -0.00056991 | 0.00020301 | 0.00022247 | -0.00026934 | 0.00008641 | -0.00002902 | -0.00008240 | -0.00011920 |
| 45 | 45 | 0.06489582 | 0.05247111 | -0.04116588 | -0.01902612 | 0.02124803 | 0.00094764 | -0.00720289 | -0.00087189 | -0.00005906 | -0.00050232 | 0.00025455 | 0.00013916 | -0.00015349 | 0.00012251 | 0.00008407 | -0.00019488 | -0.00003598 |
| 45 | 46 | 0.06553971 | 0.05287682 | -0.04197099 | -0.01874825 | 0.02140095 | 0.00111011 | -0.00753335 | -0.00044226 | -0.00004280 | -0.00073438 | 0.00030153 | 0.00016789 | -0.00024376 | 0.00000919 | -0.00001096 | -0.00003463 | -0.00002674 |
| 45 | 47 | 0.06530686 | 0.05228204 | -0.04307015 | -0.01925272 | 0.02173284 | 0.00066254 | -0.00712502 | -0.00048938 | -0.00003232 | -0.00067794 | 0.00036167 | 0.00005888 | -0.00009754 | 0.00002595 | 0.00011471 | -0.00016276 | 0.00006790 |
| 45 | 48 | 0.06587273 | 0.05276778 | -0.04382270 | -0.01910560 | 0.02208410 | 0.00113057 | -0.00739756 | -0.00012385 | -0.00000401 | -0.00093008 | 0.00038478 | 0.00011472 | -0.00020110 | -0.00007862 | 0.00001496 | -0.00001181 | 0.00004989 |
| 45 | 49 | 0.06557621 | 0.05217480 | -0.04489341 | -0.01972581 | 0.02239137 | 0.00074379 | -0.00699321 | -0.00016733 | 0.00001504 | -0.00086939 | 0.00044621 | -0.00000891 | -0.00003662 | -0.00007222 | 0.00014465 | -0.00014783 | 0.00014856 |
| 45 | 50 | 0.06602095 | 0.05267944 | -0.04561704 | -0.01974541 | 0.02298321 | 0.00155443 | -0.00724878 | 0.00012835 | 0.00006805 | -0.00114043 | 0.00044864 | 0.00007106 | -0.00015380 | -0.00016844 | 0.00004473 | -0.00001111 | 0.00010134 |
| 45 | 51 | 0.06555464 | 0.05173825 | -0.04533593 | -0.01901368 | 0.02271646 | 0.00022611 | -0.00723105 | -0.00019570 | 0.00000735 | -0.00087324 | 0.00049224 | -0.00007765 | 0.00003368 | -0.00011196 | 0.00016555 | -0.00015530 | 0.00019297 |
| 45 | 52 | 0.06580576 | 0.05177362 | -0.04454073 | -0.01741828 | 0.02264670 | -0.00009220 | -0.00780349 | -0.00012573 | -0.00007356 | -0.00094266 | 0.00048281 | -0.00001571 | -0.00007481 | -0.00014446 | 0.00005171 | -0.00003188 | 0.00012357 |
| 45 | 53 | 0.06528945 | 0.05110114 | -0.04433648 | -0.01741898 | 0.02235031 | -0.00090582 | -0.00784125 | -0.00057732 | 0.00002383 | -0.00063107 | 0.00043644 | -0.00008557 | 0.00006218 | -0.00005845 | 0.00015213 | -0.00014187 | 0.00017737 |
| 45 | 54 | 0.06537537 | 0.05109401 | -0.04328597 | -0.01614268 | 0.02180324 | -0.00130270 | -0.00821381 | -0.00038553 | -0.00000408 | -0.00064231 | 0.00036640 | 0.00003637 | -0.00009731 | -0.00004004 | 0.00002091 | -0.00000398 | 0.00007387 |
| 45 | 55 | 0.05352807 | 0.05947343 | -0.02520806 | -0.03314248 | 0.01342761 | 0.01139662 | -0.00647315 | -0.00566066 | -0.00007481 | -0.00020740 | -0.00026473 | 0.00003084 | 0.00035035 | -0.00018299 | 0.00010782 | -0.00009727 | 0.00007402 |
| 45 | 56 | 0.05125574 | 0.06062759 | -0.02142488 | -0.03468679 | 0.01210955 | 0.01310902 | -0.00663657 | -0.00590307 | 0.00009730 | -0.00023652 | -0.00040627 | 0.00018543 | 0.00021845 | -0.00019718 | 0.00000773 | 0.00001136 | -0.00000328 |
| 45 | 57 | 0.05107195 | 0.06017093 | -0.02247733 | -0.03447254 | 0.01361779 | 0.01273754 | -0.00716183 | -0.00593586 | 0.00055553 | -0.00010197 | -0.00044538 | 0.00018376 | 0.00030365 | -0.00015776 | 0.00011026 | -0.00006299 | 0.00007661 |
| 45 | 58 | 0.05116102 | 0.05888502 | -0.02313210 | -0.03235283 | 0.01436001 | 0.01161846 | -0.00769204 | -0.00511368 | 0.00070608 | -0.00018286 | -0.00039669 | 0.00028130 | 0.00011854 | -0.00013728 | 0.00003184 | 0.00000935 | 0.00002721 |
| 45 | 59 | 0.05095538 | 0.05882855 | -0.02349984 | -0.03219485 | 0.01556337 | 0.01165931 | -0.00818526 | -0.00556781 | 0.00093483 | -0.00005063 | -0.00041529 | 0.00020147 | 0.00028883 | -0.00013975 | 0.00013898 | -0.00008941 | 0.00009189 |
| 45 | 60 | 0.05082158 | 0.05743880 | -0.02407559 | -0.03069356 | 0.01584591 | 0.01089800 | -0.00843525 | -0.00486937 | 0.00096554 | -0.00011710 | -0.00034981 | 0.00027599 | 0.00010929 | -0.00012121 | 0.00005775 | -0.00002468 | 0.00003906 |
| 45 | 61 | 0.05071058 | 0.05768364 | -0.02414311 | -0.03033242 | 0.01704102 | 0.01085669 | -0.00897771 | -0.00536275 | 0.00119124 | -0.00001150 | -0.00036842 | 0.00017257 | 0.00031899 | -0.00014949 | 0.00017420 | -0.00012876 | 0.00010987 |
| 45 | 62 | 0.05054479 | 0.05620350 | -0.02486448 | -0.02923356 | 0.01702764 | 0.01029749 | -0.00897833 | -0.00464343 | 0.00116934 | -0.00008118 | -0.00029193 | 0.00024740 | 0.00012117 | -0.00013076 | 0.00008287 | -0.00005423 | 0.00005725 |
| 45 | 63 | 0.05055875 | 0.05663790 | -0.02480414 | -0.02860532 | 0.01819210 | 0.00990686 | -0.00959633 | -0.00510045 | 0.00142358 | -0.00001091 | -0.00030636 | 0.00012190 | 0.00036918 | -0.00018227 | 0.00021763 | -0.00016518 | 0.00014995 |
| 45 | 64 | 0.05039463 | 0.05516826 | -0.02558808 | -0.02788385 | 0.01788274 | 0.00960025 | -0.00935243 | -0.00437526 | 0.00135101 | -0.00008659 | -0.00022605 | 0.00020460 | 0.00014836 | -0.00016440 | 0.00011021 | -0.00007363 | 0.00009341 |
| 45 | 65 | 0.05047522 | 0.05571508 | -0.02543460 | -0.02705896 | 0.01890977 | 0.00873172 | -0.01006101 | -0.00480973 | 0.00163293 | -0.00004951 | -0.00023451 | 0.00005096 | 0.00044221 | -0.00023775 | 0.00027200 | -0.00019557 | 0.00021981 |
| 45 | 66 | 0.05034030 | 0.05434108 | -0.02621468 | -0.02669539 | 0.01834883 | 0.00875247 | -0.00958578 | -0.00410175 | 0.00150688 | -0.00013121 | -0.00015639 | 0.00014655 | 0.00019639 | -0.00022072 | 0.00014406 | -0.00008430 | 0.00015276 |
| 45 | 67 | 0.05021256 | 0.05459829 | -0.02607412 | -0.02606971 | 0.01891666 | 0.00761885 | -0.01018335 | -0.00438007 | 0.00166934 | -0.00006909 | -0.00019131 | 0.00003554 | 0.00042049 | -0.00021756 | 0.00028384 | -0.00020297 | 0.00023024 |
| 45 | 68 | 0.04973389 | 0.05291961 | -0.02660349 | -0.02586377 | 0.01817072 | 0.00781077 | -0.00948942 | -0.00357470 | 0.00142005 | -0.00009119 | -0.00015861 | 0.00018735 | 0.00010151 | -0.00010353 | 0.00012626 | -0.00010218 | 0.00008503 |
| 45 | 69 | 0.04985132 | 0.05318065 | -0.02697981 | -0.02544252 | 0.01883436 | 0.00683966 | -0.01002755 | -0.00379787 | 0.00154754 | -0.00010117 | -0.00017442 | 0.00008416 | 0.00029184 | -0.00012068 | 0.00025451 | -0.00021002 | 0.00015718 |
| 45 | 70 | 0.04930204 | 0.05153964 | -0.02732721 | -0.02507669 | 0.01834873 | 0.00703531 | -0.00941509 | -0.00308254 | 0.00133174 | -0.00012471 | -0.00015111 | 0.00022217 | 0.00001201 | -0.00000910 | 0.00012063 | -0.00013817 | 0.00001518 |
| 45 | 71 | 0.04962036 | 0.05181851 | -0.02810059 | -0.02481245 | 0.01909761 | 0.00620967 | -0.00987580 | -0.00323999 | 0.00143637 | -0.00019215 | -0.00015101 | 0.00012186 | 0.00018017 | -0.00004573 | 0.00023970 | -0.00023971 | 0.00008342 |
| 45 | 72 | 0.04901371 | 0.05026346 | -0.02830401 | -0.02438631 | 0.01879765 | 0.00642315 | -0.00936881 | -0.00263585 | 0.00126586 | -0.00020993 | -0.00013178 | 0.00024446 | -0.00006238 | 0.00005842 | 0.00012766 | -0.00018763 | -0.00005013 |
| 45 | 73 | 0.04949113 | 0.05056650 | -0.02936408 | -0.02423855 | 0.01960347 | 0.00571777 | -0.00975139 | -0.00273226 | 0.00134854 | -0.00031699 | -0.00012109 | 0.00014672 | 0.00009076 | 0.00000682 | 0.00023733 | -0.00028616 | 0.00001173 |
| 45 | 74 | 0.04882818 | 0.04911068 | -0.02947061 | -0.02384303 | 0.01941204 | 0.00595634 | -0.00934850 | -0.00224205 | 0.00123460 | -0.00032150 | -0.00009972 | 0.00025075 | -0.00011665 | 0.00009897 | 0.00014504 | -0.00024372 | -0.00010561 |
| 45 | 75 | 0.04942426 | 0.04943052 | -0.03071789 | -0.02376883 | 0.02024200 | 0.00535212 | -0.00965343 | -0.00228746 | 0.00128722 | -0.00045100 | -0.00008469 | 0.00015963 | 0.00002279 | 0.00003964 | 0.00024262 | -0.00034036 | -0.00005528 |
| 45 | 76 | 0.04870939 | 0.04806131 | -0.03079343 | -0.02347495 | 0.02011511 | 0.00562793 | -0.00934738 | -0.00190203 | 0.00123838 | -0.00043767 | -0.00005665 | 0.00024277 | -0.00015291 | 0.00011664 | 0.00016767 | -0.00029888 | -0.00015046 |
| 45 | 77 | 0.04938260 | 0.04837817 | -0.03213751 | -0.02342732 | 0.02093801 | 0.00511057 | -0.00957358 | -0.00190704 | 0.00124783 | -0.00057625 | -0.00004357 | 0.00016481 | -0.00003008 | 0.00005799 | 0.00024940 | -0.00039330 | -0.00011778 |
| 45 | 78 | 0.04863057 | 0.04707131 | -0.03226317 | -0.02328882 | 0.02087213 | 0.00544749 | -0.00936372 | -0.00161102 | 0.00127039 | -0.00054379 | -0.00000750 | 0.00022718 | -0.00017942 | 0.00011850 | 0.00018881 | -0.00034648 | -0.00018922 |
| 45 | 79 | 0.04933706 | 0.04736078 | -0.03361889 | -0.02321837 | 0.02165956 | 0.00499743 | -0.00951206 | -0.00158509 | 0.00122349 | -0.00068402 | -0.00000170 | 0.00016863 | -0.00007699 | 0.00006820 | 0.00025206 | -0.00043862 | -0.00017917 |
| 45 | 80 | 0.04857273 | 0.04609423 | -0.03387919 | -0.02327968 | 0.02168018 | 0.00543114 | -0.00940819 | -0.00136153 | 0.00132258 | -0.00063357 | 0.00004047 | 0.00021275 | -0.00020666 | 0.00011242 | 0.00020255 | -0.00038259 | -0.00022979 |
| 45 | 81 | 0.04926710 | 0.04633537 | -0.03516399 | -0.02313753 | 0.02240462 | 0.00501576 | -0.00948279 | -0.00131285 | 0.00120990 | -0.00077366 | 0.00003558 | 0.00017715 | -0.00012634 | 0.00007589 | 0.00024728 | -0.00047374 | -0.00024446 |
| 45 | 82 | 0.04852026 | 0.04509935 | -0.03563195 | -0.02344086 | 0.02254629 | 0.00559007 | -0.00950178 | -0.00114809 | 0.00138867 | -0.00070745 | 0.00007975 | 0.00020672 | -0.00024285 | 0.00010477 | 0.00020560 | -0.00040666 | -0.00028020 |
| 45 | 83 | 0.04894904 | 0.04475711 | -0.03611011 | -0.02208961 | 0.02279481 | 0.00436678 | -0.00944444 | -0.00102001 | 0.00110069 | -0.00083504 | 0.00008014 | 0.00018466 | -0.00019437 | 0.00011025 | 0.00023852 | -0.00050011 | -0.00028997 |
| 46 | 44 | 0.06545654 | 0.05415592 | -0.04008832 | -0.02040770 | 0.01989341 | 0.00176426 | -0.00838887 | -0.00220454 | 0.00014158 | -0.00009916 | 0.00002243 | 0.00022908 | -0.00014768 | 0.00009402 | -0.00003736 | 0.00000401 | -0.00006367 |
| 46 | 45 | 0.06590720 | 0.05247570 | -0.04230643 | -0.01856016 | 0.02143101 | 0.00099738 | -0.00787042 | -0.00096041 | -0.00006575 | -0.00053778 | 0.00021435 | 0.00022387 | -0.00027624 | 0.00009333 | -0.00003207 | -0.00006068 | -0.00009173 |
| 46 | 46 | 0.06588854 | 0.05373391 | -0.04191807 | -0.01991119 | 0.02083362 | 0.00134396 | -0.00841651 | -0.00175786 | 0.00015919 | -0.00033072 | 0.00017067 | 0.00011300 | -0.00007587 | -0.00001094 | 0.00001828 | -0.00000146 | 0.00004603 |
| 46 | 47 | 0.06638250 | 0.05235399 | -0.04415360 | -0.01864299 | 0.02206177 | 0.00076763 | -0.00774591 | -0.00052376 | -0.00004496 | -0.00075576 | 0.00033057 | 0.00014140 | -0.00022250 | -0.00000814 | 0.00000116 | -0.00003417 | 0.00000940 |
| 46 | 48 | 0.06616360 | 0.05351572 | -0.04360056 | -0.01992456 | 0.02175765 | 0.00134360 | -0.00830980 | -0.00137846 | 0.00018653 | -0.00058153 | 0.00028919 | 0.00001984 | -0.00000635 | -0.00011793 | 0.00006598 | -0.00000867 | 0.00013609 |
| 46 | 49 | 0.06677309 | 0.05226976 | -0.04606077 | -0.01901611 | 0.02288564 | 0.00088806 | -0.00755001 | -0.00013189 | -0.00000458 | -0.00099146 | 0.00042871 | 0.00007099 | -0.00016601 | -0.00011236 | 0.00003453 | -0.00002230 | 0.00008970 |
| 46 | 50 | 0.06638895 | 0.05330752 | -0.04535149 | -0.02016662 | 0.02281752 | 0.00161404 | -0.00815942 | -0.00100527 | 0.00024458 | -0.00083806 | 0.00038254 | -0.00004917 | 0.00004896 | -0.00021803 | 0.00010538 | -0.00001860 | 0.00019889 |
| 46 | 51 | 0.06678081 | 0.05180212 | -0.04650151 | -0.01826292 | 0.02321140 | 0.00040910 | -0.00770257 | -0.00011440 | -0.00004645 | -0.00001531 | 0.00048017 | 0.00001140 | -0.00011544 | -0.00014693 | 0.00004762 | -0.00002249 | 0.00012566 |
| 46 | 52 | 0.06607592 | 0.05252051 | -0.04442555 | -0.01887565 | 0.02212876 | 0.00053722 | -0.00834202 | -0.00117183 | 0.00018409 | -0.00058416 | 0.00031555 | -0.00001150 | 0.00000986 | -0.00011660 | 0.00006768 | 0.00001534 | 0.00015339 |
| 46 | 53 | 0.06629838 | 0.05121993 | -0.04528951 | -0.01718694 | 0.02244284 | -0.00054949 | -0.00810126 | -0.00045415 | -0.00002890 | -0.00074268 | 0.00037582 | 0.00006319 | -0.00014401 | -0.00005031 | 0.00001184 | 0.00001488 | 0.00007948 |
| 46 | 54 | 0.05201882 | 0.06203306 | -0.02260959 | -0.03756341 | 0.01217009 | 0.01497016 | -0.00656240 | -0.00673184 | 0.00012146 | -0.00016444 | -0.00044852 | 0.00014294 | 0.00030451 | -0.00024898 | 0.00003593 | 0.00003971 | 0.00004705 |
| 46 | 55 | 0.05207552 | 0.06101105 | -0.02321924 | -0.03618187 | 0.01248292 | 0.01410913 | -0.00649805 | -0.00615993 | -0.00001582 | -0.00033157 | -0.00041367 | 0.00022293 | 0.00017060 | -0.00020268 | -0.00001352 | 0.00004294 | -0.00001343 |
| 46 | 56 | 0.05209724 | 0.06011638 | -0.02472783 | -0.03494762 | 0.01464146 | 0.01324785 | -0.00760351 | -0.00573341 | 0.00071685 | -0.00011563 | -0.00043300 | 0.00028372 | 0.00013852 | -0.00015621 | 0.00003700 | 0.00006206 | 0.00005758 |
| 46 | 57 | 0.05208572 | 0.05931176 | -0.02506611 | -0.03375806 | 0.01483734 | 0.01250995 | -0.00762121 | -0.00531267 | 0.00059388 | -0.00028146 | -0.00041221 | 0.00033856 | 0.00005132 | -0.00013089 | 0.00000300 | 0.00005086 | 0.00000600 |
| 46 | 58 | 0.05196587 | 0.05841023 | -0.02611112 | -0.03283859 | 0.01640582 | 0.01227963 | -0.00841776 | -0.00535600 | 0.00091444 | -0.00004709 | -0.00036574 | 0.00030667 | 0.00007308 | -0.00011213 | 0.00004070 | 0.00004145 | 0.00004657 |
| 46 | 59 | 0.05182395 | 0.05786550 | -0.02609007 | -0.03192285 | 0.01646007 | 0.01166607 | -0.00845676 | -0.00502298 | 0.00084876 | -0.00021101 | -0.00036561 | 0.00033920 | 0.00003432 | -0.00010598 | 0.00002676 | 0.00001538 | 0.00000782 |
| 46 | 60 | 0.05178968 | 0.05697122 | -0.02712996 | -0.03104559 | 0.01778789 | 0.01155958 | -0.00905324 | -0.00507490 | 0.00105255 | 0.00001058 | -0.00029813 | 0.00030567 | 0.00003543 | -0.00009181 | 0.00004270 | 0.00001825 | 0.00003131 |
| 46 | 61 | 0.05158440 | 0.05663193 | -0.02688796 | -0.03030529 | 0.01777412 | 0.01097396 | -0.00908700 | -0.00476551 | 0.00105482 | -0.00015739 | -0.00030898 | 0.00031198 | 0.00004561 | -0.00010624 | 0.00005184 | -0.00002091 | 0.00001508 |
| 46 | 62 | 0.05168997 | 0.05577910 | -0.02795970 | -0.02942981 | 0.01883733 | 0.01085271 | -0.00950188 | -0.00475706 | 0.00118997 | 0.00003811 | -0.00023153 | 0.00029294 | 0.00001499 | -0.00009402 | 0.00004597 | 0.00000128 | 0.00002650 |
| 46 | 63 | 0.05144981 | 0.05560160 | -0.02757583 | -0.02880620 | 0.01877532 | 0.01022810 | -0.00952384 | -0.00445957 | 0.00125153 | -0.00013659 | -0.00024488 | 0.00026866 | 0.00007535 | -0.00013040 | 0.00007983 | -0.00004904 | 0.00004066 |
| 46 | 64 | 0.05167276 | 0.05485579 | -0.02859724 | -0.02802850 | 0.01949360 | 0.01010818 | -0.00975414 | -0.00441269 | 0.00132316 | 0.00003002 | -0.00016690 | 0.00026690 | 0.00001559 | -0.00011941 | 0.00005415 | -0.00000845 | 0.00004003 |
| 46 | 65 | 0.05139900 | 0.05479144 | -0.02812289 | -0.02748016 | 0.01938181 | 0.00937860 | -0.00978202 | -0.00413772 | 0.00143198 | -0.00015156 | -0.00017702 | 0.00020951 | 0.00012648 | -0.00017784 | 0.00011385 | -0.00006811 | 0.00009043 |
| 46 | 66 | 0.05173763 | 0.05417980 | -0.02909365 | -0.02684721 | 0.01979441 | 0.00929208 | -0.00985027 | -0.00406168 | 0.00145177 | -0.00001328 | -0.00010428 | 0.00022563 | 0.00004074 | -0.00016684 | 0.00007117 | -0.00001253 | 0.00007637 |
| 46 | 67 | 0.05106836 | 0.05379613 | -0.02844732 | -0.02645549 | 0.01942380 | 0.00843036 | -0.00981442 | -0.00369386 | 0.00149762 | -0.00014081 | -0.00014569 | 0.00019093 | 0.00012106 | -0.00015389 | 0.00012446 | -0.00007823 | 0.00009772 |
| 46 | 68 | 0.05099425 | 0.05284128 | -0.02916570 | -0.02598565 | 0.01961851 | 0.00841745 | -0.00970287 | -0.00352129 | 0.00143027 | 0.00004065 | -0.00011383 | 0.00024689 | -0.00001628 | -0.00006172 | 0.00006249 | -0.00002896 | 0.00003255 |
| 46 | 69 | 0.05053475 | 0.05247175 | -0.02889414 | -0.02562868 | 0.01941964 | 0.00757382 | -0.00975230 | -0.00315750 | 0.00144541 | -0.00013119 | -0.00014671 | 0.00022422 | 0.00004186 | -0.00005557 | 0.00011132 | -0.00009825 | 0.00003847 |
| 46 | 70 | 0.05040651 | 0.05154579 | -0.02956994 | -0.02521077 | 0.01972108 | 0.00763243 | -0.00962695 | -0.00303762 | 0.00140142 | 0.00002949 | -0.00011580 | 0.00026110 | -0.00006293 | 0.00002350 | 0.00006667 | -0.00006453 | -0.00001167 |
| 46 | 71 | 0.05016280 | 0.05121564 | -0.02963017 | -0.02489037 | 0.01969625 | 0.00687582 | -0.00971819 | -0.00267518 | 0.00139170 | -0.00017989 | -0.00013806 | 0.00025129 | -0.00003096 | 0.00002116 | 0.00010827 | -0.00013304 | -0.00002224 |
| 46 | 72 | 0.04996863 | 0.05033741 | -0.03027184 | -0.02456286 | 0.02005607 | 0.00697094 | -0.00959884 | -0.00260314 | 0.00138692 | -0.00003453 | -0.00010594 | 0.00026374 | -0.00009729 | 0.00008491 | 0.00008392 | -0.00011324 | -0.00005092 |
| 46 | 73 | 0.04992072 | 0.05007006 | -0.03058819 | -0.02428656 | 0.02017310 | 0.00634285 | -0.00969766 | -0.00225343 | 0.00135480 | -0.00026418 | -0.00011686 | 0.00026596 | -0.00008985 | 0.00007374 | 0.00011539 | -0.00017713 | -0.00007741 |
| 46 | 74 | 0.04965413 | 0.04923408 | -0.03122379 | -0.02408006 | 0.02055244 | 0.00644900 | -0.00959519 | -0.00221791 | 0.00139955 | -0.00012950 | -0.00008227 | 0.00025174 | -0.00011709 | 0.00012204 | 0.00011178 | -0.00016749 | -0.00008003 |
| 46 | 75 | 0.04976863 | 0.04903894 | -0.03171778 | -0.02385559 | 0.02076462 | 0.00596963 | -0.00967686 | -0.00189632 | 0.00134333 | -0.00035965 | -0.00008246 | 0.00026561 | -0.00013142 | 0.00010330 | 0.00013005 | -0.00022364 | -0.00012223 |
| 46 | 76 | 0.04943489 | 0.04822465 | -0.03239108 | -0.02378478 | 0.02115133 | 0.00607940 | -0.00959771 | -0.00188349 | 0.00144176 | -0.00023268 | -0.00004572 | 0.00022598 | -0.00012380 | 0.00013800 | 0.00014528 | -0.00021938 | -0.00009742 |
| 46 | 77 | 0.04967283 | 0.04809230 | -0.03299669 | -0.02361409 | 0.02141638 | 0.00575568 | -0.00964883 | -0.00160145 | 0.00135692 | -0.00044753 | -0.00003740 | 0.00025301 | -0.00015895 | 0.00011448 | 0.00014716 | -0.00026570 | -0.00015700 |
| 46 | 78 | 0.04929015 | 0.04727683 | -0.03375824 | -0.02368078 | 0.02182293 | 0.00587784 | -0.00959836 | -0.00159834 | 0.00150903 | -0.00032637 | -0.00000039 | 0.00019178 | -0.00012359 | 0.00013870 | 0.00017789 | -0.00026207 | -0.00010632 |
| 46 | 79 | 0.04961019 | 0.04718224 | -0.03442463 | -0.02355812 | 0.02211323 | 0.00570612 | -0.00962125 | -0.00135989 | 0.00139079 | -0.00051812 | 0.00001251 | 0.00023536 | -0.00018094 | 0.00011405 | 0.00016079 | -0.00029820 | -0.00018720 |
| 46 | 80 | 0.04920785 | 0.04635218 | -0.03532138 | -0.02375685 | 0.02256500 | 0.00585731 | -0.00960428 | -0.00135611 | 0.00159519 | -0.00040098 | 0.00004717 | 0.00015706 | -0.00012528 | 0.00013110 | 0.00020357 | -0.00029126 | -0.00011364 |
| 46 | 81 | 0.04956656 | 0.04626353 | -0.03600788 | -0.02367201 | 0.02286697 | 0.00582322 | -0.00961839 | -0.00115896 | 0.00144087 | -0.00057085 | 0.00005951 | 0.00022116 | -0.00020710 | 0.00010889 | 0.00016649 | -0.00031911 | -0.00022102 |
| 46 | 82 | 0.04918192 | 0.04542057 | -0.03707583 | -0.02399279 | 0.02338968 | 0.00601912 | -0.00963687 | -0.00114690 | 0.00169608 | -0.00045511 | 0.00008943 | 0.00012956 | -0.00013694 | 0.00012142 | 0.00021848 | -0.00030594 | -0.00012742 |
| 46 | 83 | 0.04920351 | 0.04482603 | -0.03675508 | -0.02275308 | 0.02311312 | 0.00517797 | -0.00960322 | -0.00088572 | 0.00138481 | -0.00059442 | 0.00009695 | 0.00022122 | -0.00026119 | 0.00014398 | 0.00016372 | -0.00034258 | -0.00025585 |
| 46 | 84 | 0.04859496 | 0.04365481 | -0.03670293 | -0.02182370 | 0.02282939 | 0.00441048 | -0.00944467 | -0.00072114 | 0.00147376 | -0.00045308 | 0.00012184 | 0.00014276 | -0.00021228 | 0.00019203 | 0.00019602 | -0.00032855 | -0.00016493 |
| 46 | 85 | 0.04861277 | 0.04287605 | -0.03666657 | -0.02071679 | 0.02270684 | 0.00367949 | -0.00940658 | -0.00043569 | 0.00119125 | -0.00061603 | 0.00012632 | 0.00023221 | -0.00034239 | 0.00021037 | 0.00014753 | -0.00036302 | -0.00029023 |
| 47 | 45 | 0.06558503 | 0.05183238 | -0.04326277 | -0.02008789 | 0.02123799 | 0.00143822 | -0.00758431 | -0.00167466 | -0.00013174 | -0.00032407 | 0.00010991 | 0.00028229 | -0.00027032 | 0.00022938 | 0.00000565 | -0.00013033 | -0.00012098 |
| 47 | 46 | 0.06658240 | 0.05204239 | -0.04442694 | -0.01899375 | 0.02177900 | 0.00084794 | -0.00808581 | -0.00113771 | -0.00004593 | -0.00051332 | 0.00020670 | 0.00023000 | -0.00027979 | 0.00010040 | -0.00003317 | -0.00003622 | -0.00006619 |
| 47 | 47 | 0.06648637 | 0.05133989 | -0.04574541 | -0.01939195 | 0.02232923 | 0.00068935 | -0.00760455 | -0.00101467 | -0.00007583 | -0.00053868 | 0.00028341 | 0.00013176 | -0.00017415 | 0.00010818 | 0.00006870 | -0.00012913 | 0.00001851 |
| 47 | 48 | 0.06710127 | 0.05193156 | -0.04633491 | -0.01905909 | 0.02260723 | 0.00073797 | -0.00791713 | -0.00065790 | -0.00002839 | -0.00076643 | 0.00034525 | 0.00011666 | -0.00019885 | -0.00002265 | 0.00001333 | -0.00002876 | 0.00004504 |
| 47 | 49 | 0.06698978 | 0.05120452 | -0.04767935 | -0.01953679 | 0.02312265 | 0.00060521 | -0.00740463 | -0.00052606 | -0.00004750 | -0.00076380 | 0.00041899 | 0.00001082 | -0.00008274 | -0.00001672 | 0.00011442 | -0.00012015 | 0.00013520 |
| 47 | 50 | 0.06753246 | 0.05184762 | -0.04826884 | -0.01942753 | 0.02352369 | 0.00094327 | -0.00766708 | -0.00021992 | 0.00000295 | -0.00102397 | 0.00045974 | 0.00002889 | -0.00013043 | -0.00014056 | 0.00005074 | -0.00002362 | 0.00012996 |
| 47 | 51 | 0.06704716 | 0.05090790 | -0.04822426 | -0.01917913 | 0.02334712 | 0.00033679 | -0.00756887 | -0.00054064 | -0.00004792 | -0.00077721 | 0.00044047 | -0.00001624 | -0.00006044 | -0.00004369 | 0.00011172 | -0.00009021 | 0.00016314 |

TABLE 3. Nuclear charge density distribution FB coefficients.

| $Z$ | $N$ | $a_1$ | $a_2$ | $a_3$ | $a_4$ | $a_5$ | $a_6$ | $a_7$ | $a_8$ | $a_9$ | $a_{10}$ | $a_{11}$ | $a_{12}$ | $a_{13}$ | $a_{14}$ | $a_{15}$ | $a_{16}$ | $a_{17}$ |
|---|---|---|---|---|---|---|---|---|---|---|---|---|---|---|---|---|---|---|
| 47 | 52 | 0.06701786 | 0.05132992 | -0.04712259 | -0.01849400 | 0.02285456 | 0.00022581 | -0.00804426 | -0.00062093 | -0.00003281 | -0.00078495 | 0.00036778 | 0.00008302 | -0.00016925 | -0.00005089 | 0.00000844 | 0.00002386 | 0.00008565 |
| 47 | 53 | 0.05569471 | 0.05909895 | -0.02998452 | -0.03477476 | 0.01465172 | 0.01243579 | -0.00631981 | -0.00568406 | -0.00030128 | -0.00041067 | -0.00022461 | 0.00011160 | 0.00019758 | -0.00016674 | 0.00005077 | -0.00002812 | 0.00004750 |
| 47 | 54 | 0.05274983 | 0.06130733 | -0.02493676 | -0.03781457 | 0.01270992 | 0.01511597 | -0.00639366 | -0.00648695 | -0.00012604 | -0.00038705 | -0.00042823 | 0.00025373 | 0.00013801 | -0.00020235 | -0.00003250 | 0.00006897 | -0.00002013 |
| 47 | 55 | 0.05269298 | 0.06049483 | -0.02636028 | -0.03743077 | 0.01413309 | 0.01456944 | -0.00691779 | -0.00634971 | 0.00019768 | -0.00031654 | -0.00045857 | 0.00030904 | 0.00013100 | -0.00013468 | 0.00002794 | 0.00003159 | 0.00001286 |
| 47 | 56 | 0.05286674 | 0.05966235 | -0.02690187 | -0.03530270 | 0.01516324 | 0.01346031 | -0.00755081 | -0.00558901 | 0.00046693 | -0.00035734 | -0.00043333 | 0.00039319 | -0.00000655 | -0.00012144 | -0.00002752 | 0.00009176 | -0.00001259 |
| 47 | 57 | 0.05279787 | 0.05914508 | -0.02773951 | -0.03496894 | 0.01630840 | 0.01331801 | -0.00805605 | -0.00587876 | 0.00054393 | -0.00030081 | -0.00042612 | 0.00035286 | 0.00007979 | -0.00010166 | 0.00004792 | 0.00000961 | 0.00000563 |
| 47 | 58 | 0.05271037 | 0.05820984 | -0.02803850 | -0.03326343 | 0.01694264 | 0.01251150 | -0.00846162 | -0.00524354 | 0.00070583 | -0.00030309 | -0.00038437 | 0.00040392 | -0.00003970 | -0.00008997 | -0.00000829 | 0.00005808 | -0.00002229 |
| 47 | 59 | 0.05267878 | 0.05801440 | -0.02851990 | -0.03292041 | 0.01801223 | 0.01243616 | -0.00895031 | -0.00561867 | 0.00076739 | -0.00026917 | -0.00037723 | 0.00033446 | 0.00009340 | -0.00009692 | 0.00007966 | -0.00003887 | -0.00000111 |
| 47 | 60 | 0.05253822 | 0.05696135 | -0.02890878 | -0.03148908 | 0.01838755 | 0.01172338 | -0.00917838 | -0.00495146 | 0.00089876 | -0.00025298 | -0.00032581 | 0.00038196 | -0.00003797 | -0.00008286 | 0.00001516 | 0.00001729 | -0.00002795 |
| 47 | 61 | 0.05259954 | 0.05700155 | -0.02918578 | -0.03098230 | 0.01945649 | 0.01149338 | -0.00963715 | -0.00529466 | 0.00099554 | -0.00024846 | -0.00031561 | 0.00028805 | 0.00013889 | -0.00011399 | 0.00012122 | -0.00008994 | 0.00001446 |
| 47 | 62 | 0.05244819 | 0.05590646 | -0.02965315 | -0.02986657 | 0.01951632 | 0.01089574 | -0.00969554 | -0.00461148 | 0.00109300 | -0.00022474 | -0.00025992 | 0.00034129 | -0.00001364 | -0.00009886 | 0.00004332 | -0.00001867 | -0.00001491 |
| 47 | 63 | 0.05256212 | 0.05614467 | -0.02970597 | -0.02919129 | 0.02052256 | 0.01040801 | -0.01011496 | -0.00491235 | 0.00123614 | -0.00024495 | -0.00024599 | 0.00021779 | 0.00021460 | -0.00015362 | 0.00017362 | -0.00013626 | 0.00006221 |
| 47 | 64 | 0.05242471 | 0.05507443 | -0.03022475 | -0.02844028 | 0.02024101 | 0.00999451 | -0.01000698 | -0.00424866 | 0.00128165 | -0.00022402 | -0.00019034 | 0.00028314 | 0.00003452 | -0.00013802 | 0.00007794 | -0.00004638 | 0.00002359 |
| 47 | 65 | 0.05253792 | 0.05545758 | -0.03005117 | -0.02760859 | 0.02114077 | 0.00915682 | -0.01042163 | -0.00451756 | 0.00148352 | -0.00025846 | -0.00017366 | 0.00012540 | 0.00032072 | -0.00021486 | 0.00023798 | -0.00017482 | 0.00014542 |
| 47 | 66 | 0.05244974 | 0.05445315 | -0.03063484 | -0.02723061 | 0.02057067 | 0.00900036 | -0.01015472 | -0.00389279 | 0.00146300 | -0.00025043 | -0.00012027 | 0.00020812 | 0.00010782 | -0.00019861 | 0.00012133 | -0.00006564 | 0.00009034 |
| 47 | 67 | 0.05229397 | 0.05457728 | -0.03043155 | -0.02657688 | 0.02115414 | 0.00804088 | -0.01050801 | -0.00402843 | 0.00158958 | -0.00026010 | -0.00013425 | 0.00008960 | 0.00033695 | -0.00020962 | 0.00025892 | -0.00018223 | 0.00017638 |
| 47 | 68 | 0.05186458 | 0.05320756 | -0.03086376 | -0.02632789 | 0.02047723 | 0.00812513 | -0.01007109 | -0.00328211 | 0.00146278 | -0.00021387 | -0.00012471 | 0.00023754 | 0.00003565 | -0.00010295 | 0.00010318 | -0.00006811 | 0.00004391 |
| 47 | 69 | 0.05193190 | 0.05337640 | -0.03113866 | -0.02593515 | 0.02107186 | 0.00725383 | -0.01045913 | -0.00344248 | 0.00154135 | -0.00027768 | -0.00012521 | 0.00012521 | 0.00024221 | -0.00013313 | 0.00023038 | -0.00017560 | 0.00012578 |
| 47 | 70 | 0.05141799 | 0.05200180 | -0.03135260 | -0.02550857 | 0.02063417 | 0.00737210 | -0.01003332 | -0.00274503 | 0.00144305 | -0.00022765 | -0.00012412 | 0.00026433 | -0.00003004 | -0.00002360 | 0.00009176 | -0.00008571 | -0.00000644 |
| 47 | 71 | 0.05170260 | 0.05221607 | -0.03203407 | -0.02532244 | 0.02125310 | 0.00661009 | -0.01041335 | -0.00289763 | 0.00148821 | -0.00033705 | -0.00011310 | 0.00015881 | 0.00015185 | -0.00007093 | 0.00020734 | -0.00018335 | 0.00006907 |
| 47 | 72 | 0.05110891 | 0.05088452 | -0.03207262 | -0.02481149 | 0.02100142 | 0.00677577 | -0.01001611 | -0.00228172 | 0.00141997 | -0.00028045 | -0.00011364 | 0.00028314 | -0.00008591 | 0.00003591 | 0.00008891 | -0.00011490 | -0.00005597 |
| 47 | 73 | 0.05158391 | 0.05114865 | -0.03305208 | -0.02477120 | 0.02163952 | 0.00612786 | -0.01036569 | -0.00241477 | 0.00143809 | -0.00041869 | -0.00009578 | 0.00018636 | 0.00007234 | -0.00002385 | 0.00019066 | -0.00020272 | 0.00000952 |
| 47 | 74 | 0.05090994 | 0.04987856 | -0.03298051 | -0.02428155 | 0.02150359 | 0.00634797 | -0.00999392 | -0.00189505 | 0.00140646 | -0.00035116 | -0.00009022 | 0.00028925 | -0.00012779 | 0.00007461 | 0.00009415 | -0.00014978 | -0.00009856 |
| 47 | 75 | 0.05153717 | 0.05018913 | -0.03414214 | -0.02432279 | 0.02214354 | 0.00580476 | -0.01030605 | -0.00200863 | 0.00139499 | -0.00050040 | -0.00007175 | 0.00020543 | 0.00000727 | 0.00000929 | 0.00017895 | -0.00022810 | -0.00004852 |
| 47 | 76 | 0.05078645 | 0.04897447 | -0.03404364 | -0.02394836 | 0.02207387 | 0.00609186 | -0.00994866 | -0.00158497 | 0.00140862 | -0.00041868 | -0.00005372 | 0.00028168 | -0.00015497 | 0.00009453 | 0.00010440 | -0.00018358 | -0.00013078 |
| 47 | 77 | 0.05152334 | 0.04931583 | -0.03528107 | -0.02400325 | 0.02269529 | 0.00563223 | -0.01022629 | -0.00168229 | 0.00135872 | -0.00056580 | -0.00004129 | 0.00021687 | -0.00004480 | 0.00003160 | 0.00016896 | -0.00025302 | -0.00010272 |
| 47 | 78 | 0.05070927 | 0.04813500 | -0.03525191 | -0.02381795 | 0.02267895 | 0.00600912 | -0.00987809 | -0.00134475 | 0.00142649 | -0.00046907 | -0.00000755 | 0.00026441 | -0.00017222 | 0.00010049 | 0.00011478 | -0.00021038 | -0.00015426 |
| 47 | 79 | 0.05151219 | 0.04848372 | -0.03647206 | -0.02381684 | 0.02326278 | 0.00559941 | -0.01013033 | -0.00142752 | 0.00132739 | -0.00060834 | -0.00000698 | 0.00022461 | -0.00008957 | 0.00004731 | 0.00015696 | -0.00027243 | -0.00015412 |
| 47 | 80 | 0.05065955 | 0.04731065 | -0.03661336 | -0.02387601 | 0.02332174 | 0.00609745 | -0.00980105 | -0.00116055 | 0.00145820 | -0.00049851 | 0.00004176 | 0.00024488 | -0.00018808 | 0.00009863 | 0.00012047 | -0.00022676 | -0.00017524 |
| 47 | 81 | 0.05148491 | 0.04764320 | -0.03773351 | -0.02375173 | 0.02384849 | 0.00569174 | -0.01003761 | -0.00122845 | 0.00130064 | -0.00063067 | 0.00002671 | 0.00023381 | -0.00013424 | 0.00006067 | 0.00014035 | -0.00028413 | -0.00020638 |
| 47 | 82 | 0.05062753 | 0.04645834 | -0.03813985 | -0.02409730 | 0.02402604 | 0.00634595 | -0.00975286 | -0.00101462 | 0.00150352 | -0.00051180 | 0.00008628 | 0.00023093 | -0.00021124 | 0.00009466 | 0.00011851 | -0.00023250 | -0.00020183 |
| 47 | 83 | 0.05120844 | 0.04627953 | -0.03840078 | -0.02263452 | 0.02414103 | 0.00491815 | -0.01006507 | -0.00090726 | 0.00122690 | -0.00068198 | 0.00005966 | 0.00023783 | -0.00018672 | 0.00009364 | 0.00013869 | -0.00032177 | -0.00025646 |
| 47 | 84 | 0.04999085 | 0.04464276 | -0.03782028 | -0.02194043 | 0.02364341 | 0.00462872 | -0.00978877 | -0.00055573 | 0.00136438 | -0.00054118 | 0.00010658 | 0.00023989 | -0.00027799 | 0.00016433 | 0.00011895 | -0.00028256 | -0.00024914 |
| 47 | 85 | 0.05073415 | 0.04435390 | -0.03851069 | -0.02043294 | 0.02398977 | 0.00323428 | -0.01001393 | -0.00037595 | 0.00107948 | -0.00077759 | 0.00009305 | 0.00023028 | -0.00024505 | 0.00013975 | 0.00014825 | -0.00037705 | -0.00029546 |
| 47 | 86 | 0.04948619 | 0.04277537 | -0.03771389 | -0.01985957 | 0.02315662 | 0.00298557 | -0.00960604 | 0.00000532 | 0.00120178 | -0.00058574 | 0.00012382 | 0.00024894 | -0.00035083 | 0.00022220 | 0.00010636 | -0.00031299 | -0.00029546 |
| 48 | 46 | 0.05599919 | 0.06154949 | -0.02762065 | -0.03743812 | 0.01281354 | 0.01387506 | -0.00707273 | -0.00721086 | -0.00014577 | 0.00009859 | -0.00050388 | 0.00025318 | 0.00019496 | -0.00009259 | -0.00002230 | 0.00002816 | -0.00004080 |
| 48 | 47 | 0.06686760 | 0.05199393 | -0.04605184 | -0.02033786 | 0.02182006 | 0.00129276 | -0.00820458 | -0.00146182 | -0.00000878 | -0.00048081 | 0.00016695 | 0.00024066 | -0.00026791 | 0.00010395 | -0.00003233 | -0.00000797 | -0.00004112 |
| 48 | 48 | 0.06484470 | 0.05487156 | -0.04270921 | -0.02507676 | 0.02004540 | 0.00421488 | -0.00830320 | -0.00308338 | 0.00021395 | -0.00025030 | 0.00001784 | 0.00012431 | 0.00001874 | -0.00003773 | 0.00004045 | 0.00002967 | 0.00010164 |
| 48 | 49 | 0.06768219 | 0.05165192 | -0.04841730 | -0.01997294 | 0.02299596 | 0.00100498 | -0.00802706 | -0.00084764 | 0.00000331 | -0.00075091 | 0.00034194 | 0.00009592 | -0.00017198 | -0.00003177 | 0.00002291 | -0.00001362 | 0.00008159 |
| 48 | 50 | 0.06726927 | 0.05304015 | -0.04777863 | -0.02208993 | 0.02271502 | 0.00210147 | -0.00844552 | -0.00176817 | 0.00028604 | -0.00055671 | 0.00027105 | -0.00001863 | 0.00005442 | -0.00013334 | 0.00009933 | 0.00000847 | 0.00021313 |
| 48 | 51 | 0.06755816 | 0.05147216 | -0.04871475 | -0.01997974 | 0.02305099 | 0.00098613 | -0.00803169 | -0.00086822 | -0.00001010 | -0.00076622 | 0.00034472 | 0.00009032 | -0.00016714 | -0.00004536 | 0.00001295 | 0.00002377 | 0.00009781 |
| 48 | 52 | 0.05322214 | 0.06278520 | -0.02588383 | -0.04113365 | 0.01255797 | 0.01686311 | -0.00656325 | -0.00757031 | 0.00001007 | -0.00017375 | -0.00048868 | 0.00014476 | 0.00034350 | -0.00027055 | 0.00004014 | 0.00006402 | 0.00009155 |
| 48 | 53 | 0.05328704 | 0.06156480 | -0.02649391 | -0.03951646 | 0.01278205 | 0.01606537 | -0.00633463 | -0.00687261 | -0.00021927 | -0.00039075 | -0.00045006 | 0.00027129 | 0.00012975 | -0.00019786 | -0.00004521 | 0.00008698 | -0.00001898 |
| 48 | 54 | 0.05339657 | 0.06105983 | -0.02798700 | -0.03842087 | 0.01513428 | 0.01513358 | -0.00762705 | -0.00651322 | 0.00056462 | -0.00016378 | -0.00049239 | 0.00033310 | 0.00013506 | -0.00016420 | 0.00001419 | 0.00012250 | 0.00008031 |
| 48 | 55 | 0.05348910 | 0.05998725 | -0.02853552 | -0.03694909 | 0.01530291 | 0.01439847 | -0.00750493 | -0.00594617 | 0.00034120 | -0.00038908 | -0.00046178 | 0.00043736 | -0.00004194 | -0.00010961 | -0.00005517 | 0.00012801 | -0.00002393 |
| 48 | 56 | 0.05348036 | 0.05935074 | -0.02953303 | -0.03575641 | 0.01727781 | 0.01394592 | -0.00858399 | -0.00598913 | 0.00074558 | -0.00013666 | -0.00042049 | 0.00037900 | 0.00003570 | -0.00011093 | 0.00000481 | 0.00011356 | 0.00004973 |
| 48 | 57 | 0.05344529 | 0.05853027 | -0.02978289 | -0.03468164 | 0.01725491 | 0.01337119 | -0.00847496 | -0.00554262 | 0.00055671 | -0.00036342 | -0.00040953 | 0.00046095 | -0.00009503 | -0.00007319 | -0.00004276 | 0.00009856 | -0.00004574 |
| 48 | 58 | 0.05344440 | 0.05787992 | -0.03064831 | -0.03350551 | 0.01897832 | 0.01303416 | -0.00938472 | -0.00561296 | 0.00086089 | -0.00010283 | -0.00034748 | 0.00039201 | -0.00002651 | -0.00008013 | -0.00000083 | 0.00008967 | 0.00001074 |
| 48 | 59 | 0.05335135 | 0.05726520 | -0.03073612 | -0.03273129 | 0.01884818 | 0.01251510 | -0.00926267 | -0.00521411 | 0.00072616 | -0.00033248 | -0.00034738 | 0.00044688 | -0.00010768 | -0.00006062 | -0.00002286 | 0.00005553 | -0.00006520 |
| 48 | 60 | 0.05343765 | 0.05660760 | -0.03159602 | -0.03153804 | 0.02031253 | 0.01214207 | -0.01000129 | -0.00521564 | 0.00098347 | -0.00008410 | -0.00027388 | 0.00038868 | -0.00006594 | -0.00006993 | -0.00000046 | 0.00006381 | -0.00001794 |
| 48 | 61 | 0.05332109 | 0.05617291 | -0.03157703 | -0.03098211 | 0.02011504 | 0.01161640 | -0.00985528 | -0.00484077 | 0.00090452 | -0.00031399 | -0.00027720 | 0.00041137 | -0.00009428 | -0.00007027 | 0.00000423 | 0.00001336 | -0.00006581 |
| 48 | 62 | 0.05347126 | 0.05557657 | -0.03235010 | -0.02990120 | 0.02118961 | 0.01123756 | -0.01039284 | -0.00480080 | 0.00110959 | -0.00008693 | -0.00020125 | 0.00036880 | -0.00008050 | -0.00008145 | 0.00000761 | 0.00004077 | -0.00002692 |
| 48 | 63 | 0.05334251 | 0.05529239 | -0.03224554 | -0.02947438 | 0.02095593 | 0.01064866 | -0.01023129 | -0.00444188 | 0.00108595 | -0.00031378 | -0.00020283 | 0.00035650 | -0.00005472 | -0.00010241 | 0.00003910 | -0.00002248 | -0.00003998 |
| 48 | 64 | 0.05355481 | 0.05479126 | -0.03293135 | -0.02858884 | 0.02162058 | 0.01029754 | -0.01057661 | -0.00438154 | 0.00123687 | -0.00011452 | -0.00013010 | 0.00033175 | -0.00006863 | -0.00011465 | 0.00002519 | 0.00002289 | -0.00001052 |
| 48 | 65 | 0.05340121 | 0.05462586 | -0.03273081 | -0.02821561 | 0.02136783 | 0.00959940 | -0.01041809 | -0.00404249 | 0.00126825 | -0.00033367 | -0.00012763 | 0.00028334 | 0.00001162 | -0.00015602 | 0.00008309 | -0.00004980 | 0.00001603 |
| 48 | 66 | 0.05369338 | 0.05422256 | -0.03339592 | -0.02755696 | 0.02170045 | 0.00930522 | -0.01061434 | -0.00397333 | 0.00137090 | -0.00016431 | -0.00006132 | 0.00027808 | -0.00003034 | -0.00016769 | 0.00005359 | 0.00001071 | 0.00003299 |
| 48 | 67 | 0.05317597 | 0.05377471 | -0.03302914 | -0.02720909 | 0.02139330 | 0.00861859 | -0.01041222 | -0.00350931 | 0.00138144 | -0.00033666 | -0.00009203 | 0.00025525 | 0.00001527 | -0.00014631 | 0.00009843 | -0.00005458 | 0.00003556 |
| 48 | 68 | 0.05310303 | 0.05299272 | -0.03355143 | -0.02665223 | 0.02161462 | 0.00845677 | -0.01048522 | -0.00331112 | 0.00142526 | -0.00014060 | -0.00006674 | 0.00030079 | -0.00009966 | -0.00008050 | 0.00004182 | 0.00001107 | -0.00000344 |
| 48 | 69 | 0.05271694 | 0.05262059 | -0.03338791 | -0.02632994 | 0.02146395 | 0.00783296 | -0.01036006 | -0.00289447 | 0.00140386 | -0.00033455 | -0.00009430 | 0.00028260 | -0.00005318 | -0.00006866 | 0.00008068 | -0.00005389 | -0.00000543 |
| 48 | 70 | 0.05259706 | 0.05182777 | -0.03387704 | -0.02581787 | 0.02174081 | 0.00767501 | -0.01042522 | -0.00274066 | 0.00146222 | -0.00015200 | -0.00006916 | 0.00031229 | -0.00014321 | -0.00000977 | 0.00004072 | -0.00000704 | -0.00003556 |
| 48 | 71 | 0.05236765 | 0.05154057 | -0.03391939 | -0.02554094 | 0.02174332 | 0.00718098 | -0.01034170 | -0.00236950 | 0.00141004 | -0.00036344 | -0.00009153 | 0.00030362 | -0.00010772 | -0.00000705 | 0.00006928 | -0.00006605 | -0.00004670 |
| 48 | 72 | 0.05219187 | 0.05076551 | -0.03437997 | -0.02509030 | 0.02206019 | 0.00701661 | -0.01039672 | -0.00225531 | 0.00149296 | -0.00019427 | -0.00006284 | 0.00030945 | -0.00016305 | 0.00004122 | 0.00005105 | -0.00003864 | -0.00005989 |
| 48 | 73 | 0.05212619 | 0.05056617 | -0.03461231 | -0.02488937 | 0.02217892 | 0.00669393 | -0.01032393 | -0.00193482 | 0.00141024 | -0.00041110 | -0.00007866 | 0.00031405 | -0.00014740 | 0.00003665 | 0.00006532 | -0.00008678 | -0.00008387 |
| 48 | 74 | 0.05188719 | 0.04981778 | -0.03506497 | -0.02452115 | 0.02251470 | 0.00651672 | -0.01036368 | -0.00185132 | 0.00152651 | -0.00025325 | -0.00004387 | 0.00029110 | -0.00016310 | 0.00007211 | 0.00007072 | -0.00007597 | -0.00007294 |
| 48 | 75 | 0.05197187 | 0.04970364 | -0.03544944 | -0.02442234 | 0.02269938 | 0.00638529 | -0.01027862 | -0.00159110 | 0.00141280 | -0.00045912 | -0.00005295 | 0.00031138 | -0.00017189 | 0.00006284 | 0.00006746 | -0.00010981 | -0.00011245 |
| 48 | 76 | 0.05167138 | 0.04897224 | -0.03593357 | -0.02415217 | 0.02304470 | 0.00619744 | -0.01029965 | -0.00152711 | 0.00156712 | -0.00031073 | -0.00001135 | 0.00025881 | -0.00014869 | 0.00008528 | 0.00009527 | -0.00011051 | -0.00007316 |
| 48 | 77 | 0.05187751 | 0.04893123 | -0.03642053 | -0.02416600 | 0.02325146 | 0.00625957 | -0.01019136 | -0.00133469 | 0.00142194 | -0.00049112 | -0.00001502 | 0.00029672 | -0.00018383 | 0.00007438 | 0.00007223 | -0.00012870 | -0.00013093 |
| 48 | 78 | 0.05153011 | 0.04819704 | -0.03698977 | -0.02400132 | 0.02361535 | 0.00607300 | -0.01019467 | -0.00127858 | 0.00161490 | -0.00035142 | 0.00003210 | 0.00021737 | -0.00012705 | 0.00008533 | 0.00011901 | -0.00013502 | -0.00006262 |
| 48 | 79 | 0.05181982 | 0.04820504 | -0.03753120 | -0.02411961 | 0.02381827 | 0.00631401 | -0.01006936 | -0.00115414 | 0.00143883 | -0.00049912 | 0.00003092 | 0.00027521 | -0.00018968 | 0.00007604 | 0.00007524 | -0.00013876 | -0.00014241 |
| 48 | 80 | 0.05145273 | 0.04745082 | -0.03824259 | -0.02406193 | 0.02422397 | 0.00614683 | -0.01006066 | -0.00109543 | 0.00166895 | -0.00036805 | 0.00008072 | 0.00017391 | -0.00010675 | 0.00007790 | 0.00013678 | -0.00014529 | -0.00004711 |
| 48 | 81 | 0.05178440 | 0.04747392 | -0.03879959 | -0.02426119 | 0.02441663 | 0.00653602 | -0.00994262 | -0.00103034 | 0.00146469 | -0.00048552 | 0.00007793 | 0.00025424 | -0.00019798 | 0.00007317 | 0.00007299 | -0.00013845 | -0.00015371 |
| 48 | 82 | 0.05143443 | 0.04669585 | -0.03970135 | -0.02430957 | 0.02489060 | 0.00640726 | -0.00992908 | -0.00096113 | 0.00173021 | -0.00036240 | 0.00012694 | 0.00013590 | -0.00009575 | 0.00006827 | 0.00014547 | -0.00014087 | -0.00003443 |
| 48 | 83 | 0.05145370 | 0.04624720 | -0.03926811 | -0.02326477 | 0.02455658 | 0.00574462 | -0.01000334 | -0.00073967 | 0.00143543 | -0.00049391 | 0.00010532 | 0.00024870 | -0.00023022 | 0.00010501 | 0.00007553 | -0.00017143 | -0.00019099 |
| 48 | 84 | 0.05086047 | 0.04512881 | -0.03908498 | -0.02200723 | 0.02426960 | 0.00444670 | -0.01001424 | -0.00043855 | 0.00159018 | -0.00039175 | 0.00013937 | 0.00014513 | -0.00014653 | 0.00013042 | 0.00014052 | -0.00020151 | -0.00009849 |
| 48 | 85 | 0.05089479 | 0.04450898 | -0.03894834 | -0.02108542 | 0.02412226 | 0.00389758 | -0.01005127 | -0.00016798 | 0.00133228 | -0.00055688 | 0.00011639 | 0.00025203 | -0.00028359 | 0.00016200 | 0.00008302 | -0.00023578 | -0.00024798 |
| 48 | 86 | 0.05044318 | 0.04351700 | -0.03874821 | -0.01985862 | 0.02362611 | 0.00266565 | -0.00986274 | 0.00016800 | 0.00143367 | -0.00043555 | 0.00015182 | 0.00015488 | -0.00020591 | 0.00018089 | 0.00012516 | -0.00024440 | -0.00016092 |
| 48 | 87 | 0.05046024 | 0.04271306 | -0.03889692 | -0.01905682 | 0.02359200 | 0.00219116 | -0.00983591 | 0.00046438 | 0.00120367 | -0.00061680 | 0.00012942 | 0.00024968 | -0.00033805 | 0.00020716 | 0.00007882 | -0.00027458 | -0.00029141 |
| 49 | 47 | 0.05562326 | 0.05995198 | -0.02959356 | -0.03790979 | 0.01348705 | 0.01441777 | -0.00635804 | -0.00695945 | -0.00052943 | -0.00015461 | -0.00042483 | 0.00027690 | 0.00010416 | -0.00001250 | -0.00001071 | -0.00005944 | -0.00009909 |
| 49 | 48 | 0.06486925 | 0.05378580 | -0.04408056 | -0.02519222 | 0.02016615 | 0.00444826 | -0.00792845 | -0.00277466 | -0.00004328 | -0.00040394 | 0.00001960 | 0.00024215 | -0.00016841 | 0.00005873 | -0.00002901 | 0.00002074 | -0.00001498 |
| 49 | 49 | 0.06476424 | 0.05329141 | -0.04496814 | -0.02542400 | 0.02064885 | 0.00452847 | -0.00751685 | -0.00268103 | -0.00012476 | -0.00042316 | 0.00009991 | 0.00013770 | -0.00006576 | 0.00004364 | 0.00003998 | -0.00002624 | 0.00006935 |
| 49 | 50 | 0.06795386 | 0.05169538 | -0.05000672 | -0.02155255 | 0.02305894 | 0.00171989 | -0.00803967 | -0.00117527 | 0.00004001 | -0.00068869 | 0.00030942 | 0.00008122 | -0.00013491 | -0.00003769 | 0.00002576 | 0.00001829 | 0.00012138 |
| 49 | 51 | 0.05718738 | 0.05899126 | -0.03347005 | -0.03655098 | 0.01519485 | 0.01325458 | -0.00638912 | -0.00600716 | -0.00039663 | -0.00035295 | -0.00023798 | 0.00013018 | 0.00017650 | -0.00013856 | 0.00002380 | 0.00001308 | 0.00006453 |
| 49 | 52 | 0.05371562 | 0.06183168 | -0.02782876 | -0.04120269 | 0.01272158 | 0.01692894 | -0.00630845 | -0.00729392 | -0.00028641 | -0.00034045 | -0.00047781 | 0.00027098 | 0.00015018 | -0.00019214 | -0.00004939 | 0.00009792 | -0.00000541 |
| 49 | 53 | 0.05366737 | 0.06095234 | -0.02904732 | -0.04041546 | 0.01404893 | 0.01622644 | -0.00682723 | -0.00706843 | -0.00008743 | -0.00030961 | -0.00050367 | 0.00037171 | 0.00008384 | -0.00011146 | -0.00003305 | 0.00010725 | -0.00000190 |
| 49 | 54 | 0.05396711 | 0.06033205 | -0.02990058 | -0.03864027 | 0.01525968 | 0.01528829 | -0.00748028 | -0.00636839 | 0.00022796 | -0.00036960 | -0.00049674 | 0.00046550 | -0.00004868 | -0.00009771 | -0.00007671 | 0.00015888 | -0.00002366 |
| 49 | 55 | 0.05402236 | 0.05953356 | -0.03086915 | -0.03773664 | 0.01648757 | 0.01486211 | -0.00802691 | -0.00648351 | 0.00019325 | -0.00037533 | -0.00046034 | 0.00045274 | -0.00002191 | -0.00006928 | -0.00003044 | 0.00009864 | -0.00003466 |
| 49 | 56 | 0.05403280 | 0.05887005 | -0.03125460 | -0.03613951 | 0.01738309 | 0.01419531 | -0.00850371 | -0.00591068 | 0.00041540 | -0.00037878 | -0.00044112 | 0.00050400 | -0.00012323 | -0.00005706 | -0.00007279 | 0.00013481 | -0.00005807 |
| 49 | 57 | 0.05408328 | 0.05833461 | -0.03196443 | -0.03552420 | 0.01840745 | 0.01393148 | -0.00900328 | -0.00616241 | 0.00033969 | -0.00041355 | -0.00040040 | 0.00045907 | -0.00005333 | -0.00005461 | -0.00000943 | 0.00004804 | -0.00007394 |
| 49 | 58 | 0.05401418 | 0.05759180 | -0.03227912 | -0.03399854 | 0.01913594 | 0.01330068 | -0.00934972 | -0.00554919 | 0.00055475 | -0.00037630 | -0.00037506 | 0.00049960 | -0.00015215 | -0.00004027 | -0.00005818 | 0.00009099 | -0.00009186 |
| 49 | 59 | 0.05414703 | 0.05722549 | -0.03293908 | -0.03349350 | 0.02006545 | 0.01296316 | -0.00978611 | -0.00579348 | 0.00049650 | -0.00044856 | -0.00032791 | 0.00043376 | -0.00004896 | -0.00005923 | 0.00002466 | -0.00001273 | -0.00009456 |
| 49 | 60 | 0.05404339 | 0.05646197 | -0.03321324 | -0.03210382 | 0.02056374 | 0.01236456 | -0.01000329 | -0.00514469 | 0.00070859 | -0.00037969 | -0.00029950 | 0.00047085 | -0.00015197 | -0.00004525 | -0.00003368 | 0.00004411 | -0.00010649 |
| 49 | 61 | 0.05423080 | 0.05625045 | -0.03376970 | -0.03165891 | 0.02135465 | 0.01186850 | -0.01035104 | -0.00535778 | 0.00068283 | -0.00048285 | -0.00024694 | 0.00038170 | -0.00001039 | -0.00008433 | 0.00007167 | -0.00007300 | -0.00008326 |
| 49 | 62 | 0.05411224 | 0.05552441 | -0.03399442 | -0.03050118 | 0.02154741 | 0.01135558 | -0.01043450 | -0.00471218 | 0.00087227 | -0.00039358 | -0.00021848 | 0.00042057 | -0.00012347 | -0.00007215 | 0.00000051 | 0.00000116 | -0.00009372 |
| 49 | 63 | 0.05431470 | 0.05545405 | -0.03438233 | -0.03004701 | 0.02219407 | 0.01062855 | -0.01070399 | -0.00488298 | 0.00090083 | -0.00051521 | -0.00016266 | 0.00030371 | 0.00006439 | -0.00013075 | 0.00013188 | -0.00012663 | -0.00003272 |
| 49 | 64 | 0.05420988 | 0.05479217 | -0.03459763 | -0.02919510 | 0.02206293 | 0.01025911 | -0.01066225 | -0.00427384 | 0.00104354 | -0.00042014 | -0.00013568 | 0.00035047 | -0.00006692 | -0.00012018 | 0.00004484 | -0.00003401 | -0.00004915 |
| 49 | 65 | 0.05437942 | 0.05484583 | -0.03475425 | -0.02865468 | 0.02260563 | 0.00925565 | -0.01089427 | -0.00440552 | 0.00115372 | -0.00054184 | -0.00008044 | 0.00020177 | 0.00017519 | -0.00019733 | 0.00020492 | -0.00016988 | 0.00005869 |
| 49 | 66 | 0.05432655 | 0.05424786 | -0.03504237 | -0.02815727 | 0.02218197 | 0.00907792 | -0.01074628 | -0.00385325 | 0.00122818 | -0.00045612 | -0.00005502 | 0.00026292 | 0.00001635 | -0.00018700 | 0.00009942 | -0.00005969 | 0.00002742 |
| 49 | 67 | 0.05423748 | 0.05410951 | -0.03513086 | -0.02767919 | 0.02262224 | 0.00815717 | -0.01092554 | -0.00384869 | 0.00129866 | -0.00056402 | -0.00003504 | 0.00015601 | 0.00020103 | -0.00020837 | 0.00023163 | -0.00017242 | 0.00010382 |
| 49 | 68 | 0.05392836 | 0.05313678 | -0.03542542 | -0.02726932 | 0.02218277 | 0.00827765 | -0.01067850 | -0.00315177 | 0.00128557 | -0.00046290 | -0.00005478 | 0.00029425 | -0.00006901 | -0.00011332 | 0.00007731 | -0.00004194 | -0.00000926 |
| 49 | 69 | 0.05397809 | 0.05311817 | -0.03577966 | -0.02700727 | 0.02262055 | 0.00744739 | -0.01090486 | -0.00323499 | 0.00130140 | -0.00059929 | -0.00002528 | 0.00018268 | 0.00012241 | -0.00015695 | 0.00020177 | -0.00014839 | 0.00007264 |
| 49 | 70 | 0.05360295 | 0.05208053 | -0.03591951 | -0.02645666 | 0.02235690 | 0.00756106 | -0.01066903 | -0.00255262 | 0.00132281 | -0.00048634 | -0.00005505 | 0.00032040 | -0.00013570 | -0.00005192 | 0.00005894 | -0.00003659 | -0.00004560 |
| 49 | 71 | 0.05381017 | 0.05217493 | -0.03651641 | -0.02637588 | 0.02279999 | 0.00684565 | -0.01090964 | -0.00269329 | 0.00129562 | -0.00064465 | -0.00001825 | 0.00020867 | 0.00005284 | -0.00011430 | 0.00017234 | -0.00013367 | 0.00003661 |
| 49 | 72 | 0.05336002 | 0.05112389 | -0.03650819 | -0.02573701 | 0.02269892 | 0.00698720 | -0.01067530 | -0.00205673 | 0.00134465 | -0.00052173 | -0.00005004 | 0.00033669 | -0.00018180 | -0.00000559 | 0.00004679 | -0.00004188 | -0.00007852 |
| 49 | 73 | 0.05372745 | 0.05132116 | -0.03729592 | -0.02579170 | 0.02313940 | 0.00638727 | -0.01091420 | -0.00223231 | 0.00128239 | -0.00069137 | -0.00001015 | 0.00023087 | -0.00000490 | -0.00008073 | 0.00014603 | -0.00012829 | -0.00000330 |
| 49 | 74 | 0.05319588 | 0.05028808 | -0.03718474 | -0.02515422 | 0.02315758 | 0.00659250 | -0.01065626 | -0.00166404 | 0.00135742 | -0.00055645 | -0.00003505 | 0.00034007 | -0.00020797 | 0.00002490 | 0.00004091 | -0.00005264 | -0.00010401 |
| 49 | 75 | 0.05370800 | 0.05057218 | -0.03808973 | -0.02528636 | 0.02357676 | 0.00608572 | -0.01089561 | -0.00186054 | 0.00126247 | -0.00072560 | 0.00000201 | 0.00024732 | -0.00004991 | -0.00005529 | 0.00012297 | -0.00012859 | -0.00004430 |
| 49 | 76 | 0.05309511 | 0.04956680 | -0.03795079 | -0.02475342 | 0.02366908 | 0.00639260 | -0.01058436 | -0.00137306 | 0.00136632 | -0.00057522 | -0.00000800 | 0.00033022 | -0.00021725 | 0.00004100 | 0.00003902 | -0.00006223 | -0.00011922 |
| 49 | 77 | 0.05372274 | 0.04991305 | -0.03889098 | -0.02489155 | 0.02404771 | 0.00593984 | -0.01083887 | -0.00158021 | 0.00123607 | -0.00073499 | 0.00001941 | 0.00025827 | -0.00008419 | -0.00003585 | 0.00010110 | -0.00012902 | -0.00008384 |
| 49 | 78 | 0.05303754 | 0.04892704 | -0.03881862 | -0.02455924 | 0.02419066 | 0.00638807 | -0.01045336 | -0.00117653 | 0.00137460 | -0.00056706 | 0.00002970 | 0.00031025 | -0.00021542 | 0.00004615 | 0.00003747 | -0.00006480 | -0.00012460 |
| 49 | 79 | 0.05374414 | 0.04930385 | -0.03971471 | -0.02462170 | 0.02451468 | 0.00593856 | -0.01074264 | -0.00138330 | 0.00120358 | -0.00071457 | 0.00004108 | 0.00026618 | -0.00011254 | -0.00001961 | 0.00007758 | -0.00012464 | -0.00012140 |

TABLE 3. Nuclear charge density distribution FB coefficients.

| $Z$ | $N$ | $a_1$ | $a_2$ | $a_3$ | $a_4$ | $a_5$ | $a_6$ | $a_7$ | $a_8$ | $a_9$ | $a_{10}$ | $a_{11}$ | $a_{12}$ | $a_{13}$ | $a_{14}$ | $a_{15}$ | $a_{16}$ | $a_{17}$ |
|---|---|---|---|---|---|---|---|---|---|---|---|---|---|---|---|---|---|---|
| 49 | 80 | 0.05300574 | 0.04831742 | -0.03981320 | -0.02456850 | 0.02471656 | 0.00656610 | -0.01028254 | -0.00105818 | 0.00138484 | -0.00053058 | 0.00007323 | 0.00028620 | -0.00021063 | 0.00004495 | 0.00003267 | -0.00005724 | -0.00012468 |
| 49 | 81 | 0.05375240 | 0.04869289 | -0.04059233 | -0.02447099 | 0.02497514 | 0.00606247 | -0.01062201 | -0.00125193 | 0.00116698 | -0.00066839 | 0.00006412 | 0.00027473 | -0.00014104 | -0.00000376 | 0.00005034 | -0.00011311 | -0.00015875 |
| 49 | 82 | 0.05298973 | 0.04768390 | -0.04096624 | -0.02475636 | 0.02527200 | 0.00690187 | -0.01011183 | -0.00099373 | 0.00140105 | -0.00047396 | 0.00011559 | 0.00026512 | -0.00021141 | 0.00004201 | 0.00002250 | -0.00003990 | -0.00012669 |
| 49 | 83 | 0.05350987 | 0.04754202 | -0.04100732 | -0.02327752 | 0.02523359 | 0.00519341 | -0.01069821 | -0.00090797 | 0.00113192 | -0.00071189 | 0.00008930 | 0.00026909 | -0.00016548 | 0.00002115 | 0.00005760 | -0.00016133 | -0.00020441 |
| 49 | 84 | 0.05239957 | 0.04606878 | -0.04050870 | -0.02248750 | 0.02492211 | 0.00486668 | -0.01038766 | -0.00042558 | 0.00134798 | -0.00055021 | 0.00012353 | 0.00026049 | -0.00024126 | 0.00009542 | 0.00004648 | -0.00013271 | -0.00019064 |
| 49 | 85 | 0.05305002 | 0.04580965 | -0.04100837 | -0.02105290 | 0.02517751 | 0.00332863 | -0.01076889 | -0.00026713 | 0.00108844 | -0.00086232 | 0.00011922 | 0.00024005 | -0.00018144 | 0.00004697 | 0.00010001 | -0.00026276 | -0.00024392 |
| 49 | 86 | 0.05190593 | 0.04438548 | -0.04028000 | -0.02038392 | 0.02448339 | 0.00298243 | -0.01038580 | 0.00022544 | 0.00127943 | -0.00064253 | 0.00013516 | 0.00024882 | -0.00027234 | 0.00013534 | 0.00006549 | -0.00020945 | -0.00024236 |
| 49 | 87 | 0.05262105 | 0.04400547 | -0.04120534 | -0.01902316 | 0.02497141 | 0.00163043 | -0.01056913 | 0.00037254 | 0.00102670 | -0.00097887 | 0.00014895 | 0.00020333 | -0.00019599 | 0.00006649 | 0.00013069 | -0.00033090 | -0.00025806 |
| 49 | 88 | 0.05152276 | 0.04266387 | -0.04034622 | -0.01852140 | 0.02396198 | 0.00132105 | -0.01010908 | 0.00087252 | 0.00118676 | -0.00070848 | 0.00015120 | 0.00022889 | -0.00030204 | 0.00016401 | 0.00007176 | -0.00025149 | -0.00026969 |
| 50 | 49 | 0.05770679 | 0.05959206 | -0.03385973 | -0.03752749 | 0.01476044 | 0.01360467 | -0.00680722 | -0.00635556 | -0.00028767 | -0.00023814 | -0.00034394 | 0.00021003 | 0.00012620 | -0.00010763 | -0.00002245 | 0.00004410 | 0.00001519 |
| 50 | 50 | 0.05415480 | 0.06342976 | -0.02851653 | -0.04445398 | 0.01251839 | 0.01845361 | -0.00659855 | -0.00845907 | -0.00004539 | 0.00000225 | -0.00054929 | 0.00009476 | 0.00045883 | -0.00027551 | 0.00006189 | 0.00007470 | 0.00017046 |
| 50 | 51 | 0.05406798 | 0.06215660 | -0.02889101 | -0.04279789 | 0.01254849 | 0.01769316 | -0.00630115 | -0.00773348 | -0.00032363 | -0.00023490 | -0.00051084 | 0.00024966 | 0.00020207 | -0.00018771 | -0.00004457 | 0.00010514 | 0.00002502 |
| 50 | 52 | 0.05417880 | 0.06203273 | -0.03026444 | -0.04210097 | 0.01488533 | 0.01698091 | -0.00763355 | -0.00753313 | 0.00040333 | -0.00003589 | -0.00058111 | 0.00033702 | 0.00021386 | -0.00016250 | 0.00000130 | 0.00017380 | 0.00013379 |
| 50 | 53 | 0.05432269 | 0.06074399 | -0.03094192 | -0.04031949 | 0.01504835 | 0.01611778 | -0.00746478 | -0.00684009 | 0.00013369 | -0.00029618 | -0.00053736 | 0.00047341 | -0.00002287 | -0.00008846 | -0.00009040 | 0.00018613 | -0.00000731 |
| 50 | 54 | 0.05447978 | 0.06033316 | -0.03199388 | -0.03890202 | 0.01738378 | 0.01557762 | -0.00873418 | -0.00688481 | 0.00055158 | -0.00007393 | -0.00050455 | 0.00041152 | 0.00007930 | -0.00010352 | -0.00002420 | 0.00017791 | 0.00008393 |
| 50 | 55 | 0.05449307 | 0.05926976 | -0.03241067 | -0.03759517 | 0.01733215 | 0.01495670 | -0.00854202 | -0.00633203 | 0.00029079 | -0.00034399 | -0.00047809 | 0.00052834 | -0.00011976 | -0.00004385 | -0.00009575 | 0.00016735 | -0.00005482 |
| 50 | 56 | 0.05458725 | 0.05886882 | -0.03317995 | -0.03616595 | 0.01942600 | 0.01453418 | -0.00967891 | -0.00643199 | 0.00062453 | -0.00009522 | -0.00042499 | 0.00044117 | -0.00000758 | -0.00006813 | -0.00004026 | 0.00015515 | 0.00002204 |
| 50 | 57 | 0.05454306 | 0.05797811 | -0.03350015 | -0.03526419 | 0.01924156 | 0.01403838 | -0.00944203 | -0.00594319 | 0.00039603 | -0.00037565 | -0.00040818 | 0.00053507 | -0.00016601 | -0.00002338 | -0.00008754 | 0.00012301 | -0.00010367 |
| 50 | 58 | 0.05468566 | 0.05754791 | -0.03423086 | -0.03377599 | 0.02113205 | 0.01354069 | -0.01043145 | -0.00596197 | 0.00071056 | -0.00012440 | -0.00034052 | 0.00044865 | -0.00006644 | -0.00005339 | -0.00004532 | 0.00012310 | -0.00002866 |
| 50 | 59 | 0.05462481 | 0.05681179 | -0.03451664 | -0.03321349 | 0.02084307 | 0.01309357 | -0.01014743 | -0.00551275 | 0.00051975 | -0.00040888 | -0.00032713 | 0.00051457 | -0.00018032 | -0.00002463 | -0.00006701 | 0.00007220 | -0.00013302 |
| 50 | 60 | 0.05478726 | 0.05641909 | -0.03511932 | -0.03181324 | 0.02236294 | 0.01255616 | -0.01093246 | -0.00547489 | 0.00080658 | -0.00016518 | -0.00025316 | 0.00043485 | -0.00009612 | -0.00006000 | -0.00003845 | 0.00008864 | -0.00005737 |
| 50 | 61 | 0.05473454 | 0.05581628 | -0.03540152 | -0.03149726 | 0.02199798 | 0.01207971 | -0.01062174 | -0.00505078 | 0.00065886 | -0.00044671 | -0.00023882 | 0.00047005 | -0.00016399 | -0.00004764 | -0.00003470 | 0.00002298 | -0.00013378 |
| 50 | 62 | 0.05490530 | 0.05550446 | -0.03584512 | -0.03028840 | 0.02307565 | 0.01154940 | -0.01118691 | -0.00498368 | 0.00090881 | -0.00021923 | -0.00016463 | 0.00040044 | -0.00009667 | -0.00008769 | -0.00001948 | 0.00005693 | -0.00005786 |
| 50 | 63 | 0.05486531 | 0.05501184 | -0.03612277 | -0.03012340 | 0.02265630 | 0.01097210 | -0.01087898 | -0.00457696 | 0.00081116 | -0.00049073 | -0.00014718 | 0.00040378 | -0.00011807 | -0.00009162 | 0.00000918 | -0.00001954 | -0.00010100 |
| 50 | 64 | 0.05505547 | 0.05479364 | -0.03645103 | -0.02915266 | 0.02332877 | 0.01048897 | -0.01125279 | -0.00450030 | 0.00102277 | -0.00028516 | -0.00007666 | 0.00034687 | -0.00006937 | -0.00013502 | 0.00001174 | 0.00003118 | -0.00002730 |
| 50 | 65 | 0.05501260 | 0.05439072 | -0.03669225 | -0.02905993 | 0.02286796 | 0.00976212 | -0.01097625 | -0.00411228 | 0.00098256 | -0.00053865 | -0.00005632 | 0.00031833 | -0.00004427 | -0.00015461 | 0.00006441 | -0.00005249 | -0.00003356 |
| 50 | 66 | 0.05524243 | 0.05425840 | -0.03698463 | -0.02834047 | 0.02323647 | 0.00936718 | -0.01121183 | -0.00404370 | 0.00115854 | -0.00035586 | 0.00000780 | 0.00027698 | -0.00001690 | -0.00019885 | 0.00005428 | 0.00001310 | 0.00003283 |
| 50 | 67 | 0.05494011 | 0.05361669 | -0.03716330 | -0.02819083 | 0.02284715 | 0.00874775 | -0.01094649 | -0.00351243 | 0.00111676 | -0.00057825 | -0.00001041 | 0.00028689 | -0.00004800 | -0.00015963 | 0.00008348 | -0.00005006 | -0.00000503 |
| 50 | 68 | 0.05488936 | 0.05313926 | -0.03737061 | -0.02745096 | 0.02324949 | 0.00857107 | -0.01111375 | -0.00328772 | 0.00126037 | -0.00038157 | 0.00000715 | 0.00030749 | -0.00011227 | -0.00013000 | 0.00003770 | 0.00003097 | -0.00000291 |
| 50 | 69 | 0.05468512 | 0.05258903 | -0.03771006 | -0.02738277 | 0.02294186 | 0.00800603 | -0.01092804 | -0.00283257 | 0.00118007 | -0.00061488 | -0.00000912 | 0.00031928 | -0.00013518 | -0.00010110 | 0.00005998 | -0.00002843 | -0.00003934 |
| 50 | 70 | 0.05459444 | 0.05207114 | -0.03787678 | -0.02666530 | 0.02336518 | 0.00776229 | -0.01111399 | -0.00263982 | 0.00134296 | -0.00041808 | 0.00000506 | 0.00032764 | -0.00018025 | -0.00007184 | 0.00003019 | 0.00003174 | -0.00003411 |
| 50 | 71 | 0.05449259 | 0.05163677 | -0.03832529 | -0.02666484 | 0.02316521 | 0.00734029 | -0.01096533 | -0.00226670 | 0.00122381 | -0.00065384 | -0.00000808 | 0.00034421 | -0.00020007 | -0.00005367 | 0.00004097 | -0.00001825 | -0.00007154 |
| 50 | 72 | 0.05436271 | 0.05110783 | -0.03847080 | -0.02598657 | 0.02359517 | 0.00702132 | -0.01116302 | -0.00210650 | 0.00140677 | -0.00046034 | 0.00000683 | 0.00033276 | -0.00021691 | -0.00002791 | 0.00003277 | 0.00001886 | -0.00005631 |
| 50 | 73 | 0.05436241 | 0.05080108 | -0.03897728 | -0.02604825 | 0.02350797 | 0.00681563 | -0.01100520 | -0.00181723 | 0.00125025 | -0.00068669 | -0.00000193 | 0.00035724 | -0.00024012 | -0.00001930 | 0.00002781 | -0.00001642 | -0.00009773 |
| 50 | 74 | 0.05419421 | 0.05027844 | -0.03913117 | -0.02543280 | 0.02392342 | 0.00642739 | -0.01119849 | -0.00168668 | 0.00145522 | -0.00049672 | 0.00001711 | 0.00032095 | -0.00022393 | 0.00000047 | 0.00004318 | 0.00000034 | -0.00006515 |
| 50 | 75 | 0.05428746 | 0.05009729 | -0.03965378 | -0.02556437 | 0.02392888 | 0.00647885 | -0.01099856 | -0.00148268 | 0.00126352 | -0.00070070 | 0.00001336 | 0.00035651 | -0.00025704 | 0.00000191 | 0.00001927 | -0.00001664 | -0.00011410 |
| 50 | 76 | 0.05408465 | 0.04958196 | -0.03985621 | -0.02503244 | 0.02431552 | 0.00603899 | -0.01116850 | -0.00137592 | 0.00149158 | -0.00051287 | 0.00003845 | 0.00029368 | -0.00020763 | 0.00001447 | 0.00005682 | -0.00001416 | -0.00005824 |
| 50 | 77 | 0.05425407 | 0.04951024 | -0.04036376 | -0.02524750 | 0.02437901 | 0.00635138 | -0.01091684 | -0.00125768 | 0.00126756 | -0.00068396 | 0.00003913 | 0.00034336 | -0.00025613 | 0.00001194 | 0.00001229 | -0.00001188 | -0.00011896 |
| 50 | 78 | 0.05402709 | 0.04898963 | -0.04066455 | -0.02481076 | 0.02473973 | 0.00588562 | -0.01104824 | -0.00116642 | 0.00151801 | -0.00049725 | 0.00007025 | 0.00025553 | -0.00017733 | 0.00001749 | 0.00006832 | -0.00001637 | -0.00003665 |
| 50 | 79 | 0.05424614 | 0.04899709 | -0.04113595 | -0.02511598 | 0.02482971 | 0.00643033 | -0.01076053 | -0.00112961 | 0.00126595 | -0.00063150 | 0.00007336 | 0.00032217 | -0.00024537 | 0.00001437 | 0.00000335 | 0.00000281 | -0.00011421 |
| 50 | 80 | 0.05401529 | 0.04845424 | -0.04159019 | -0.02477714 | 0.02518566 | 0.00596682 | -0.01084697 | -0.00104395 | 0.00153708 | -0.00044635 | 0.00010858 | 0.00021321 | -0.00014342 | 0.00001407 | 0.00007333 | -0.00000190 | -0.00000529 |
| 50 | 81 | 0.05425063 | 0.04849903 | -0.04201386 | -0.02516431 | 0.02528585 | 0.00669271 | -0.01055881 | -0.00107638 | 0.00126318 | -0.00054859 | 0.00011101 | 0.00029926 | -0.00023397 | 0.00001339 | -0.00001002 | 0.00002871 | -0.00010540 |
| 50 | 82 | 0.05404628 | 0.04792407 | -0.04267298 | -0.02492126 | 0.02566822 | 0.00625687 | -0.01060312 | -0.00098637 | 0.00155376 | -0.00036622 | 0.00014696 | 0.00017391 | -0.00011546 | 0.00000882 | 0.00006976 | 0.00002884 | 0.00002810 |
| 50 | 83 | 0.05396989 | 0.04746761 | -0.04229236 | -0.02411385 | 0.02538666 | 0.00579103 | -0.01067310 | -0.00077614 | 0.00125488 | -0.00055201 | 0.00013397 | 0.00028225 | -0.00023563 | 0.00003493 | 0.00000033 | -0.00001285 | -0.00013599 |
| 50 | 84 | 0.05353528 | 0.04645711 | -0.04219464 | -0.02273431 | 0.02509722 | 0.00404355 | -0.01091893 | -0.00039190 | 0.00148219 | -0.00046730 | 0.00016157 | 0.00016475 | -0.00013375 | 0.00004799 | 0.00009284 | -0.00007461 | -0.00004515 |
| 50 | 85 | 0.05344738 | 0.04587404 | -0.04200619 | -0.02197868 | 0.02505499 | 0.00373151 | -0.01089630 | -0.00012395 | 0.00123338 | -0.00067651 | 0.00014797 | 0.00026064 | -0.00024338 | 0.00006756 | 0.00003912 | -0.00012269 | -0.00019419 |
| 50 | 86 | 0.05313453 | 0.04494272 | -0.04198629 | -0.02079761 | 0.02450744 | 0.00208532 | -0.01096290 | 0.00024539 | 0.00139651 | -0.00057629 | 0.00018117 | 0.00014978 | -0.00015372 | 0.00007404 | 0.00011084 | -0.00016154 | -0.00010645 |
| 50 | 87 | 0.05300986 | 0.04423147 | -0.04196609 | -0.02007503 | 0.02464700 | 0.00190423 | -0.01081334 | 0.00055234 | 0.00119385 | -0.00079065 | 0.00016865 | 0.00022900 | -0.00024889 | 0.00008745 | 0.00007064 | -0.00020572 | -0.00022833 |
| 50 | 88 | 0.05285071 | 0.04341671 | -0.04207603 | -0.01915584 | 0.02392611 | 0.00045620 | -0.01074911 | 0.00084941 | 0.00129580 | -0.00065550 | 0.00020514 | 0.00013109 | -0.00017703 | 0.00008996 | 0.00011536 | -0.00021343 | -0.00014647 |
| 50 | 89 | 0.05266173 | 0.04259375 | -0.04220096 | -0.01849094 | 0.02418501 | 0.00038965 | -0.01048419 | 0.00115206 | 0.00113976 | -0.00085266 | 0.00019187 | 0.00019152 | -0.00025456 | 0.00009806 | 0.00008577 | -0.00024284 | -0.00023165 |
| 50 | 90 | 0.05266744 | 0.04194326 | -0.04244084 | -0.01787116 | 0.02339709 | -0.00079616 | -0.01038296 | 0.00133779 | 0.00120088 | -0.00067791 | 0.00022790 | 0.00011418 | -0.00020643 | 0.00009814 | 0.00010129 | -0.00021826 | -0.00016202 |
| 51 | 51 | 0.05451532 | 0.06119818 | -0.03071558 | -0.04228009 | 0.01350595 | 0.01647763 | -0.00731096 | -0.00772638 | -0.00023056 | -0.00015574 | -0.00048936 | 0.00035494 | 0.00007403 | -0.00008467 | -0.00005764 | 0.00011253 | -0.00004264 |
| 51 | 52 | 0.05505915 | 0.06071030 | -0.03207727 | -0.04074996 | 0.01478208 | 0.01542194 | -0.00792197 | -0.00698652 | 0.00010028 | -0.00027108 | -0.00046919 | 0.00045560 | -0.00005570 | -0.00008382 | -0.00008764 | 0.00016092 | -0.00004555 |
| 51 | 53 | 0.05524987 | 0.05949247 | -0.03352859 | -0.03925236 | 0.01655447 | 0.01478533 | -0.00859615 | -0.00684742 | 0.00011536 | -0.00028706 | -0.00044323 | 0.00050112 | -0.00009933 | -0.00002526 | -0.00007455 | 0.00014542 | -0.00006773 |
| 51 | 54 | 0.05543913 | 0.05898739 | -0.03419757 | -0.03769898 | 0.01757488 | 0.01401857 | -0.00909348 | -0.00627780 | 0.00033215 | -0.00034925 | -0.00040415 | 0.00053904 | -0.00018104 | -0.00003140 | -0.00009578 | 0.00016409 | -0.00007127 |
| 51 | 55 | 0.05543699 | 0.05815155 | -0.03498021 | -0.03678216 | 0.01886922 | 0.01382635 | -0.00964894 | -0.00644467 | 0.00024116 | -0.00038004 | -0.00037956 | 0.00053083 | -0.00015995 | -0.00000246 | -0.00006015 | 0.00010228 | -0.00011334 |
| 51 | 56 | 0.05547358 | 0.05761310 | -0.03536545 | -0.03524600 | 0.01969638 | 0.01313127 | -0.01003589 | -0.00588803 | 0.00043394 | -0.00039267 | -0.00033958 | 0.00055348 | -0.00023095 | -0.00000567 | -0.00008783 | 0.00012406 | -0.00011479 |
| 51 | 57 | 0.05553084 | 0.05695141 | -0.03608793 | -0.03459397 | 0.02082481 | 0.01296401 | -0.01048393 | -0.00606941 | 0.00034546 | -0.00046407 | -0.00030952 | 0.00052404 | -0.00017898 | 0.00000288 | -0.00003194 | 0.00003832 | -0.00015330 |
| 51 | 58 | 0.05550028 | 0.05639730 | -0.03633440 | -0.03309070 | 0.02146984 | 0.01227219 | -0.01076899 | -0.00548082 | 0.00054296 | -0.00043299 | -0.00026923 | 0.00054090 | -0.00024735 | 0.00000020 | -0.00006884 | 0.00007270 | -0.00014745 |
| 51 | 59 | 0.05562948 | 0.05588132 | -0.03702309 | -0.03264087 | 0.02240328 | 0.01201754 | -0.01108317 | -0.00563320 | 0.00047085 | -0.00054367 | -0.00023194 | 0.00049137 | -0.00016532 | -0.00000979 | 0.00000837 | -0.00003022 | -0.00016923 |
| 51 | 60 | 0.05555759 | 0.05536777 | -0.03713790 | -0.03126246 | 0.02279086 | 0.01135751 | -0.01125735 | -0.00503709 | 0.00066575 | -0.00047370 | -0.00019335 | 0.00050544 | -0.00023363 | -0.00001440 | -0.00003980 | 0.00002073 | -0.00015677 |
| 51 | 61 | 0.05573903 | 0.05498467 | -0.03775524 | -0.03094391 | 0.02351233 | 0.01093045 | -0.01144623 | -0.00514598 | 0.00062567 | -0.00061514 | -0.00014923 | 0.00043274 | -0.00011659 | -0.00004236 | 0.00006106 | -0.00009499 | -0.00014988 |
| 51 | 62 | 0.05565262 | 0.05454599 | -0.03776955 | -0.02976880 | 0.02360363 | 0.01034734 | -0.01151082 | -0.00457066 | 0.00080079 | -0.00051720 | -0.00011369 | 0.00044817 | -0.00019022 | -0.00004987 | -0.00000057 | -0.00002570 | -0.00013543 |
| 51 | 63 | 0.05585440 | 0.05428457 | -0.03826575 | -0.02948885 | 0.02414615 | 0.00967710 | -0.01161562 | -0.00463039 | 0.00081953 | -0.00067491 | -0.00006497 | 0.00034767 | -0.00003022 | -0.00009570 | 0.00012720 | -0.00015112 | -0.00008863 |
| 51 | 64 | 0.05578512 | 0.05392608 | -0.03824954 | -0.02857667 | 0.02394789 | 0.00921790 | -0.01158440 | -0.00409870 | 0.00095443 | -0.00056408 | -0.00003291 | 0.00037021 | -0.00011752 | -0.00010545 | 0.00004987 | -0.00006369 | -0.00007989 |
| 51 | 65 | 0.05595838 | 0.05377879 | -0.03855333 | -0.02824614 | 0.02438442 | 0.00828000 | -0.01166495 | -0.00412256 | 0.00106224 | -0.00071665 | 0.00001542 | 0.00023807 | 0.00009347 | -0.00016815 | 0.00020666 | -0.00019631 | 0.00001416 |
| 51 | 66 | 0.05594175 | 0.05348384 | -0.03860400 | -0.02764365 | 0.02393131 | 0.00798430 | -0.01155807 | -0.00364872 | 0.00113725 | -0.00060892 | 0.00004443 | 0.00027435 | -0.00001755 | -0.00017825 | 0.00011152 | -0.00009234 | 0.00000793 |
| 51 | 67 | 0.05586162 | 0.05315344 | -0.03888780 | -0.02732657 | 0.02435363 | 0.00720512 | -0.01162536 | -0.00354463 | 0.00121481 | -0.00075507 | 0.00006579 | 0.00018257 | 0.00012922 | -0.00019104 | 0.00023867 | -0.00019466 | 0.00007322 |
| 51 | 68 | 0.05560975 | 0.05244951 | -0.03904197 | -0.02673948 | 0.02396372 | 0.00725280 | -0.01143477 | -0.00290334 | 0.00120532 | -0.00064530 | 0.00006099 | 0.00029610 | -0.00010467 | -0.00012220 | 0.00009003 | -0.00006068 | -0.00001298 |
| 51 | 69 | 0.05564363 | 0.05227726 | -0.03952473 | -0.02664274 | 0.02435469 | 0.00654851 | -0.01157719 | -0.00292413 | 0.00122719 | -0.00080787 | 0.00008821 | 0.00019745 | 0.00005790 | -0.00015793 | 0.00021048 | -0.00015734 | 0.00006111 |
| 51 | 70 | 0.05534204 | 0.05146457 | -0.03957878 | -0.02591739 | 0.02414297 | 0.00659083 | -0.01138315 | -0.00227690 | 0.00124855 | -0.00068648 | 0.00007422 | 0.00031415 | -0.00017304 | -0.00007591 | 0.00006982 | -0.00004010 | -0.00003421 |
| 51 | 71 | 0.05550774 | 0.05144064 | -0.04024624 | -0.02601021 | 0.02450786 | 0.00598560 | -0.01157125 | -0.00239806 | 0.00122371 | -0.00085612 | 0.00010513 | 0.00021297 | -0.00000360 | -0.00013107 | 0.00017907 | -0.00012692 | 0.00004394 |
| 51 | 72 | 0.05514756 | 0.05057523 | -0.04018877 | -0.02518663 | 0.02447455 | 0.00606107 | -0.01136232 | -0.00177576 | 0.00126842 | -0.00072593 | 0.00008841 | 0.00032470 | -0.00021981 | -0.00004098 | 0.00005305 | -0.00003001 | -0.00005427 |
| 51 | 73 | 0.05545250 | 0.05068331 | -0.04100154 | -0.02541725 | 0.02481687 | 0.00556141 | -0.01157559 | -0.00197038 | 0.00120484 | -0.00089306 | 0.00011950 | 0.00022707 | -0.00005318 | -0.00010985 | 0.00014738 | -0.00010508 | 0.00002055 |
| 51 | 74 | 0.05502458 | 0.04980734 | -0.04085425 | -0.02457681 | 0.02492451 | 0.00571172 | -0.01132388 | -0.00139773 | 0.00127108 | -0.00075165 | 0.00010768 | 0.00032522 | -0.00024556 | -0.00001771 | 0.00003988 | -0.00002607 | -0.00007019 |
| 51 | 75 | 0.05546153 | 0.05002384 | -0.04175390 | -0.02487811 | 0.02524119 | 0.00529952 | -0.01155874 | -0.00164151 | 0.00117317 | -0.00090805 | 0.00013408 | 0.00023847 | -0.00009087 | -0.00009291 | 0.00011637 | -0.00008971 | -0.00000787 |
| 51 | 76 | 0.05496084 | 0.04915936 | -0.04157008 | -0.02412373 | 0.02543935 | 0.00556695 | -0.01123146 | -0.00113536 | 0.00126405 | -0.00075006 | 0.00013432 | 0.00031554 | -0.00025395 | -0.00000464 | 0.00002856 | -0.00002204 | -0.00007958 |
| 51 | 77 | 0.05550867 | 0.04945257 | -0.04248963 | -0.02441775 | 0.02572284 | 0.00520164 | -0.01150031 | -0.00140697 | 0.00113217 | -0.00089116 | 0.00015024 | 0.00024763 | -0.00011940 | -0.00007814 | 0.00008498 | -0.00007618 | -0.00003950 |
| 51 | 78 | 0.05493739 | 0.04860146 | -0.04234591 | -0.02385029 | 0.02597543 | 0.00562659 | -0.01107421 | -0.00097571 | 0.00125418 | -0.00071191 | 0.00016755 | 0.00029852 | -0.00025138 | 0.00000134 | 0.00001623 | -0.00001229 | -0.00008267 |
| 51 | 79 | 0.05556576 | 0.04893363 | -0.04321969 | -0.02405264 | 0.02621822 | 0.00525239 | -0.01139753 | -0.00125488 | 0.00108563 | -0.00083851 | 0.00016739 | 0.00025678 | -0.00014377 | -0.00006292 | 0.00005121 | -0.00005989 | -0.00007356 |
| 51 | 80 | 0.05493556 | 0.04808275 | -0.04320727 | -0.02375591 | 0.02651966 | 0.00586939 | -0.01087188 | -0.00089848 | 0.00124770 | -0.00063738 | 0.00020337 | 0.00027965 | -0.00024618 | 0.00000432 | 0.00000022 | 0.00000608 | -0.00008331 |
| 51 | 81 | 0.05561014 | 0.04841624 | -0.04397484 | -0.02378185 | 0.02671488 | 0.00542473 | -0.01126651 | -0.00116541 | 0.00103812 | -0.00075492 | 0.00018321 | 0.00026915 | -0.00017002 | -0.00004482 | 0.00001358 | -0.00003841 | -0.00011113 |
| 51 | 82 | 0.05494271 | 0.04754639 | -0.04419002 | -0.02381933 | 0.02709018 | 0.00625938 | -0.01066794 | -0.00087680 | 0.00125128 | -0.00053650 | 0.00023559 | 0.00026528 | -0.00024677 | 0.00000826 | -0.00002068 | 0.00003246 | -0.00008791 |
| 51 | 83 | 0.05538723 | 0.04736884 | -0.04430041 | -0.02257595 | 0.02696029 | 0.00453009 | -0.01134888 | -0.00083223 | 0.00101196 | -0.00079291 | 0.00020233 | 0.00025565 | -0.00017587 | -0.00002597 | 0.00002592 | -0.00009201 | -0.00015180 |
| 51 | 84 | 0.05437062 | 0.04602865 | -0.04372877 | -0.02163872 | 0.02671685 | 0.00410558 | -0.01099464 | -0.00028169 | 0.00122806 | -0.00064623 | 0.00024240 | 0.00023767 | -0.00023872 | 0.00004300 | 0.00002515 | -0.00008194 | -0.00014079 |
| 51 | 85 | 0.05492722 | 0.04574921 | -0.04425752 | -0.02044230 | 0.02685916 | 0.00258386 | -0.01143792 | -0.00018033 | 0.00099940 | -0.00096207 | 0.00022815 | 0.00020491 | -0.00015486 | -0.00001481 | 0.00008837 | -0.00021087 | -0.00017651 |
| 51 | 86 | 0.05387794 | 0.04443507 | -0.04348179 | -0.01961010 | 0.02625603 | 0.00211527 | -0.01102828 | 0.00038101 | 0.00119206 | -0.00075635 | 0.00025178 | 0.00020180 | -0.00023013 | 0.00006712 | 0.00006571 | -0.00017838 | -0.00017812 |
| 51 | 87 | 0.05448615 | 0.04407365 | -0.04438072 | -0.01850095 | 0.02663362 | 0.00083285 | -0.01126269 | 0.00045765 | 0.00098316 | -0.00108109 | 0.00025080 | 0.00014807 | -0.00013345 | -0.00000674 | 0.00013805 | -0.00029334 | -0.00017466 |
| 51 | 88 | 0.05347532 | 0.04281370 | -0.04349205 | -0.01782760 | 0.02572450 | 0.00038531 | -0.01078645 | 0.00102011 | 0.00114316 | -0.00082191 | 0.00026187 | 0.00016016 | -0.00022237 | 0.00008403 | 0.00009145 | -0.00023633 | -0.00019046 |
| 51 | 89 | 0.05405403 | 0.04243994 | -0.04464303 | -0.01690004 | 0.02630690 | -0.00062741 | -0.01092508 | 0.00098661 | 0.00098661 | -0.00111511 | 0.00026341 | 0.00009168 | -0.00011417 | 0.00000029 | 0.00016813 | -0.00032445 | -0.00014400 |
| 51 | 90 | 0.05314998 | 0.04125931 | -0.04373550 | -0.01641182 | 0.02518294 | -0.00099666 | -0.01039561 | 0.00153270 | 0.00110770 | -0.00081453 | 0.00026629 | 0.00011912 | -0.00021782 | 0.00009507 | 0.00009750 | -0.00024380 | -0.00017559 |
| 51 | 91 | 0.05361254 | 0.04098279 | -0.04494904 | -0.01576263 | 0.02593010 | -0.00174410 | -0.01059004 | 0.00133480 | 0.00104805 | -0.00106025 | 0.00026003 | 0.00004290 | -0.00009845 | 0.00000415 | 0.00017913 | -0.00030251 | -0.00008828 |
| 52 | 52 | 0.05639911 | 0.06010370 | -0.03517578 | -0.03938736 | 0.01744135 | 0.01377103 | -0.00973029 | -0.00681888 | 0.00059865 | -0.00010172 | -0.00036196 | 0.00041703 | -0.00002332 | -0.00009210 | -0.00001457 | 0.00014779 | 0.00003535 |
| 52 | 53 | 0.05641496 | 0.05881688 | -0.03581778 | -0.03801059 | 0.01751671 | 0.01318143 | -0.00951082 | -0.00615946 | 0.00036740 | -0.00034874 | -0.00033591 | 0.00054737 | -0.00024941 | -0.00001894 | -0.00010392 | 0.00016567 | -0.00010108 |
| 52 | 54 | 0.05656585 | 0.05848376 | -0.03666227 | -0.03645731 | 0.01984308 | 0.01267479 | -0.01072058 | -0.00631714 | 0.00067619 | -0.00014544 | -0.00028153 | 0.00045647 | -0.00011537 | -0.00005132 | -0.00003048 | 0.00014255 | 0.00000016 |
| 52 | 55 | 0.05647097 | 0.05733133 | -0.03716881 | -0.03542341 | 0.01987125 | 0.01225695 | -0.01050020 | -0.00573807 | 0.00048132 | -0.00040573 | -0.00027225 | 0.00057007 | -0.00030512 | 0.00001125 | -0.00009534 | 0.00013338 | -0.00013301 |
| 52 | 56 | 0.05658358 | 0.05707464 | -0.03766897 | -0.03391400 | 0.02182360 | 0.01179087 | -0.01151701 | -0.00589441 | 0.00073891 | -0.00018098 | -0.00021111 | 0.00047237 | -0.00017336 | -0.00002663 | -0.00003725 | 0.00011681 | -0.00004327 |
| 52 | 57 | 0.05644314 | 0.05604801 | -0.03813069 | -0.03317442 | 0.02182311 | 0.01146373 | -0.01127283 | -0.00535722 | 0.00058147 | -0.00045356 | -0.00021045 | 0.00056443 | -0.00032358 | 0.00002349 | -0.00007672 | 0.00008335 | -0.00016483 |
| 52 | 58 | 0.05658187 | 0.05585034 | -0.03844074 | -0.03173225 | 0.02338048 | 0.01094848 | -0.01207531 | -0.00544962 | 0.00081722 | -0.00021895 | -0.00014453 | 0.00047272 | -0.00020630 | -0.00001857 | -0.00003541 | 0.00008341 | -0.00007791 |
| 52 | 59 | 0.05643982 | 0.05495712 | -0.03888958 | -0.03123607 | 0.02333980 | 0.01064550 | -0.01179728 | -0.00494211 | 0.00069194 | -0.00049860 | -0.00014586 | 0.00053829 | -0.00031364 | 0.00001716 | -0.00005012 | 0.00003022 | -0.00017996 |
| 52 | 60 | 0.05660766 | 0.05484209 | -0.03902811 | -0.02993798 | 0.02444270 | 0.01009108 | -0.01237973 | -0.00498239 | 0.00090604 | -0.00026133 | -0.00007916 | 0.00045626 | -0.00021355 | -0.00002880 | -0.00002499 | 0.00004963 | -0.00009373 |
| 52 | 61 | 0.05648794 | 0.05408509 | -0.03946710 | -0.02961958 | 0.02435178 | 0.00974442 | -0.01207271 | -0.00449802 | 0.00081146 | -0.00054263 | -0.00007779 | 0.00049153 | -0.00027516 | -0.00000918 | -0.00001554 | -0.00001846 | -0.00016863 |
| 52 | 62 | 0.05669248 | 0.05405812 | -0.03948491 | -0.02850362 | 0.02502301 | 0.00917023 | -0.01246571 | -0.00450034 | 0.00100556 | -0.00031118 | -0.00001354 | 0.00042154 | -0.00019400 | -0.00005838 | -0.00000476 | 0.00001951 | -0.00008402 |
| 52 | 63 | 0.05659490 | 0.05343388 | -0.03988916 | -0.02829509 | 0.02487747 | 0.00871950 | -0.01214538 | -0.00403721 | 0.00094582 | -0.00058819 | -0.00000717 | 0.00042344 | -0.00020683 | -0.00005617 | 0.00002904 | -0.00005952 | -0.00012474 |
| 52 | 64 | 0.05684523 | 0.05348446 | -0.03985835 | -0.02737668 | 0.02520177 | 0.00815731 | -0.01240322 | -0.00401859 | 0.00112350 | -0.00036852 | 0.00005144 | 0.00036807 | -0.00014683 | -0.00010670 | 0.00002716 | -0.00000590 | -0.00004606 |
| 52 | 65 | 0.05674710 | 0.05298419 | -0.04017996 | -0.02722000 | 0.02500846 | 0.00756553 | -0.01209295 | -0.00358230 | 0.00110729 | -0.00063331 | 0.00006244 | 0.00033467 | -0.00010820 | -0.00012227 | 0.00008575 | -0.00009287 | -0.00004749 |
| 52 | 66 | 0.05705129 | 0.05309030 | -0.04017782 | -0.02650607 | 0.02510188 | 0.00706307 | -0.01227755 | -0.00356371 | 0.00127096 | -0.00042680 | 0.00011220 | 0.00029792 | -0.00007358 | -0.00017051 | 0.00007126 | -0.00002706 | 0.00001748 |
| 52 | 67 | 0.05663483 | 0.05231240 | -0.04045507 | -0.02628899 | 0.02495876 | 0.00659605 | -0.01194513 | -0.00297802 | 0.00123663 | -0.00067089 | 0.00010993 | 0.00028868 | -0.00008928 | -0.00013533 | 0.00010946 | -0.00008750 | -0.00000182 |
| 52 | 68 | 0.05660472 | 0.05197458 | -0.04047880 | -0.02555242 | 0.02508668 | 0.00632036 | -0.01204954 | -0.00279870 | 0.00134715 | -0.00045602 | 0.00013952 | 0.00030407 | -0.00015120 | -0.00011501 | 0.00005902 | -0.00000079 | 0.00000684 |
| 52 | 69 | 0.05628730 | 0.05131399 | -0.04088942 | -0.02539192 | 0.02505773 | 0.00593073 | -0.01179781 | -0.00228981 | 0.00128350 | -0.00071194 | 0.00013648 | 0.00029651 | -0.00015729 | -0.00009109 | 0.00009116 | -0.00005686 | -0.00000940 |

TABLE 3. Nuclear charge density distribution FB coefficients.

| $Z$ | $N$ | $a_1$ | $a_2$ | $a_3$ | $a_4$ | $a_5$ | $a_6$ | $a_7$ | $a_8$ | $a_9$ | $a_{10}$ | $a_{11}$ | $a_{12}$ | $a_{13}$ | $a_{14}$ | $a_{15}$ | $a_{16}$ | $a_{17}$ |
|---|---|---|---|---|---|---|---|---|---|---|---|---|---|---|---|---|---|---|
| 52 | 70 | 0.05622349 | 0.05091224 | -0.04091452 | -0.02466916 | 0.02524368 | 0.00561966 | -0.01190707 | -0.00215149 | 0.00139550 | -0.00050108 | 0.00016417 | 0.00030117 | -0.00020164 | -0.00007074 | 0.00005431 | 0.00000683 | -0.00000169 |
| 52 | 71 | 0.05601770 | 0.05038914 | -0.04143223 | -0.02457709 | 0.02533080 | 0.00538218 | -0.01170731 | -0.00173442 | 0.00130118 | -0.00075578 | 0.00016070 | 0.00029994 | -0.00020581 | -0.00005730 | 0.00007442 | -0.00003818 | -0.00001848 |
| 52 | 72 | 0.05592274 | 0.04995105 | -0.04147052 | -0.02387709 | 0.02557465 | 0.00502913 | -0.01181434 | -0.00162718 | 0.00142163 | -0.00055426 | 0.00018905 | 0.00028744 | -0.00022458 | -0.00003917 | 0.00005726 | -0.00000104 | -0.00000690 |
| 52 | 73 | 0.05583484 | 0.04957054 | -0.04207021 | -0.02387152 | 0.02575895 | 0.00500267 | -0.01163110 | -0.00131159 | 0.00129593 | -0.00079157 | 0.00018594 | 0.00029711 | -0.00023561 | -0.00003391 | 0.00005984 | -0.00002893 | -0.00002831 |
| 52 | 74 | 0.05570583 | 0.04911380 | -0.04213447 | -0.02321105 | 0.02604496 | 0.00460715 | -0.01172220 | -0.00122027 | 0.00143491 | -0.00060146 | 0.00021713 | 0.00026275 | -0.00022413 | -0.00001998 | 0.00006553 | -0.00001706 | -0.00000665 |
| 52 | 75 | 0.05573150 | 0.04886767 | -0.04278978 | -0.02330987 | 0.02629290 | 0.00482359 | -0.01152592 | -0.00101177 | 0.00127785 | -0.00080495 | 0.00021497 | 0.00028717 | -0.00024962 | -0.00001969 | 0.00004651 | -0.00002385 | -0.00003685 |
| 52 | 76 | 0.05556640 | 0.04839531 | -0.04290181 | -0.02270636 | 0.02660407 | 0.00439132 | -0.01158919 | -0.00092046 | 0.00144400 | -0.00062655 | 0.00024990 | 0.00022931 | -0.00020762 | -0.00001088 | 0.00007496 | -0.00003230 | 0.00000022 |
| 52 | 77 | 0.05568834 | 0.04826303 | -0.04358346 | -0.02291723 | 0.02688065 | 0.00485207 | -0.01136695 | -0.00081971 | 0.00125712 | -0.00078326 | 0.00024820 | 0.00027163 | -0.00025300 | -0.00001189 | 0.00003219 | -0.00001702 | -0.00004323 |
| 52 | 78 | 0.05549271 | 0.04776637 | -0.04377805 | -0.02238432 | 0.02721067 | 0.00439456 | -0.01139835 | -0.00071471 | 0.00145517 | -0.00061689 | 0.00028614 | 0.00019190 | -0.00018428 | -0.00000795 | 0.00008066 | -0.00003895 | 0.00001220 |
| 52 | 79 | 0.05568193 | 0.04771528 | -0.04445698 | -0.02269729 | 0.02749249 | 0.00507227 | -0.01115984 | -0.00071497 | 0.00124226 | -0.00072147 | 0.00028290 | 0.00025482 | -0.00025310 | -0.00000656 | 0.00001415 | -0.00000408 | -0.00004962 |
| 52 | 80 | 0.05547284 | 0.04718382 | -0.04477784 | -0.02224471 | 0.02784872 | 0.00460437 | -0.01116474 | -0.00058604 | 0.00147360 | -0.00056832 | 0.00032170 | 0.00015701 | -0.00016368 | -0.00000665 | 0.00007874 | -0.00003261 | 0.00002474 |
| 52 | 81 | 0.05569293 | 0.04717089 | -0.04543068 | -0.02263239 | 0.02813055 | 0.00544993 | -0.01094041 | -0.00067216 | 0.00124071 | -0.00062522 | 0.00031374 | 0.00024283 | -0.00025819 | 0.00000040 | -0.00000940 | 0.00001621 | -0.00006117 |
| 52 | 82 | 0.05549876 | 0.04660409 | -0.04592012 | -0.02226853 | 0.02852763 | 0.00498685 | -0.01092988 | -0.00051314 | 0.00150509 | -0.00048644 | 0.00035054 | 0.00013109 | -0.00015389 | -0.00000300 | 0.00006757 | -0.00001315 | 0.00003136 |
| 52 | 83 | 0.05540601 | 0.04614335 | -0.04571128 | -0.02157813 | 0.02827329 | 0.00460059 | -0.01102865 | -0.00040005 | 0.00122857 | -0.00061915 | 0.00032789 | 0.00022281 | -0.00025415 | 0.00002189 | 0.00000224 | -0.00002796 | -0.00009220 |
| 52 | 84 | 0.05494124 | 0.04517941 | -0.04530972 | -0.02009401 | 0.02794103 | 0.00285812 | -0.01111902 | 0.00005170 | 0.00143215 | -0.00056402 | 0.00035822 | 0.00009900 | -0.00013646 | 0.00002704 | 0.00010218 | -0.00011342 | -0.00001581 |
| 52 | 85 | 0.05487546 | 0.04458800 | -0.04535032 | -0.01949693 | 0.02781023 | 0.00251998 | -0.01118951 | 0.00020004 | 0.00118007 | -0.00072467 | 0.00033135 | 0.00018334 | -0.00023494 | 0.00004874 | 0.00004877 | -0.00013057 | -0.00012857 |
| 52 | 86 | 0.05449748 | 0.04368668 | -0.04494974 | -0.01808461 | 0.02730581 | 0.00094721 | -0.01104072 | 0.00066343 | 0.00134291 | -0.00062693 | 0.00036320 | 0.00006725 | -0.00012796 | 0.00005188 | 0.00012404 | -0.00019141 | -0.00005746 |
| 52 | 87 | 0.05443414 | 0.04297024 | -0.04521556 | -0.01757009 | 0.02725531 | 0.00063885 | -0.01105609 | 0.00082933 | 0.00111713 | -0.00079582 | 0.00033277 | 0.00013964 | -0.00021909 | 0.00007088 | 0.00008118 | -0.00020210 | -0.00014777 |
| 52 | 88 | 0.05416649 | 0.04218517 | -0.04484886 | -0.01632689 | 0.02666337 | -0.00066305 | -0.01074335 | 0.00124380 | 0.00125195 | -0.00063955 | 0.00036266 | 0.00003991 | -0.00013070 | 0.00007445 | 0.00012629 | -0.00023136 | -0.00008689 |
| 52 | 89 | 0.05407175 | 0.04138603 | -0.04529142 | -0.01593245 | 0.02666061 | -0.00093892 | -0.01071897 | 0.00138684 | 0.00106587 | -0.00079659 | 0.00032723 | 0.00009626 | -0.00020696 | 0.00008971 | 0.00009416 | -0.00022830 | -0.00014411 |
| 52 | 90 | 0.05392506 | 0.04078118 | -0.04495071 | -0.01492774 | 0.02607874 | -0.00191901 | -0.01036760 | 0.00171010 | 0.00119793 | -0.00058577 | 0.00035162 | 0.00002205 | -0.00014478 | 0.00009413 | 0.00010881 | -0.00022714 | -0.00010104 |
| 52 | 91 | 0.05376126 | 0.03997783 | -0.04548718 | -0.01471760 | 0.02610875 | -0.00214883 | -0.01036139 | 0.00178246 | 0.00107398 | -0.00072093 | 0.00030875 | 0.00005967 | -0.00019837 | 0.00010255 | 0.00009008 | -0.00020789 | -0.00011832 |
| 52 | 92 | 0.05374345 | 0.03958797 | -0.04515947 | -0.01395250 | 0.02561669 | -0.00282427 | -0.01008222 | 0.00201070 | 0.00122503 | -0.00047542 | 0.00032661 | 0.00001697 | -0.00016764 | 0.00010692 | 0.00007811 | -0.00018392 | -0.00009998 |
| 52 | 93 | 0.05347804 | 0.03886664 | -0.04569431 | -0.01398289 | 0.02567159 | -0.00299704 | -0.01016182 | 0.00197314 | 0.00118271 | -0.00059026 | 0.00027476 | 0.00003422 | -0.00019188 | 0.00010431 | 0.00007657 | -0.00015040 | -0.00007451 |
| 53 | 53 | 0.05727320 | 0.05749582 | -0.03826705 | -0.03731354 | 0.01902934 | 0.01238096 | -0.01045661 | -0.00606358 | 0.00039337 | -0.00032576 | -0.00029467 | 0.00059881 | -0.00035104 | 0.00006511 | -0.00010763 | 0.00014416 | -0.00019306 |
| 53 | 54 | 0.05750135 | 0.05717403 | -0.03878753 | -0.03572720 | 0.01976662 | 0.01140713 | -0.01084527 | -0.00551707 | 0.00053419 | -0.00041188 | -0.00021046 | 0.00058460 | -0.00038263 | 0.00002766 | -0.00010787 | 0.00014546 | -0.00016016 |
| 53 | 55 | 0.05733014 | 0.05615775 | -0.03950547 | -0.03494835 | 0.02134147 | 0.01157436 | -0.01137158 | -0.00569141 | 0.00051107 | -0.00042096 | -0.00023810 | 0.00060504 | -0.00037550 | 0.00008178 | -0.00007853 | 0.00008839 | -0.00022357 |
| 53 | 56 | 0.05743390 | 0.05581915 | -0.03978269 | -0.03340570 | 0.02187156 | 0.01064586 | -0.01165332 | -0.00515474 | 0.00063052 | -0.00046738 | -0.00015377 | 0.00058388 | -0.00040267 | 0.00004480 | -0.00008884 | 0.00009914 | -0.00018647 |
| 53 | 57 | 0.05733173 | 0.05498438 | -0.04040342 | -0.03282921 | 0.02327374 | 0.01081652 | -0.01203950 | -0.00531001 | 0.00061965 | -0.00051303 | -0.00018148 | 0.00058832 | -0.00036878 | 0.00008384 | -0.00004068 | 0.00002018 | -0.00024677 |
| 53 | 58 | 0.05736916 | 0.05465214 | -0.04052906 | -0.03136797 | 0.02358816 | 0.00990974 | -0.01222451 | -0.00477053 | 0.00073383 | -0.00051772 | -0.00009832 | 0.00056574 | -0.00039631 | 0.00004573 | -0.00006355 | 0.00004652 | -0.00020374 |
| 53 | 59 | 0.05735544 | 0.05399118 | -0.04107151 | -0.03094131 | 0.02477157 | 0.00998534 | -0.01244471 | -0.00488063 | 0.00073672 | -0.00059840 | -0.00012047 | 0.00055014 | -0.00033235 | 0.00006863 | 0.00000448 | -0.00004763 | -0.00024636 |
| 53 | 60 | 0.05736180 | 0.05370319 | -0.04108622 | -0.02962902 | 0.02483223 | 0.00911915 | -0.01254271 | -0.00435609 | 0.00084298 | -0.00056347 | -0.00004068 | 0.00052950 | -0.00036376 | 0.00002829 | -0.00003280 | -0.00000300 | -0.00019996 |
| 53 | 61 | 0.05742144 | 0.05321228 | -0.04151972 | -0.02927679 | 0.02579068 | 0.00901419 | -0.01260852 | -0.00440997 | 0.00087341 | -0.00067249 | -0.00005449 | 0.00048665 | -0.00026076 | 0.00003299 | 0.00005895 | -0.00010860 | -0.00021023 |
| 53 | 62 | 0.05743162 | 0.05298796 | -0.04148381 | -0.02816995 | 0.02558709 | 0.00821608 | -0.01263767 | -0.00391865 | 0.00096118 | -0.00060776 | 0.00002016 | 0.00047257 | -0.00030213 | -0.00000947 | 0.00000577 | -0.00004545 | -0.00016654 |
| 53 | 63 | 0.05752100 | 0.05266146 | -0.04174665 | -0.02780548 | 0.02636699 | 0.00786993 | -0.01259494 | -0.00391784 | 0.00104647 | -0.00073189 | 0.00001382 | 0.00039459 | -0.00014830 | -0.00002459 | 0.00012671 | -0.00016147 | -0.00013199 |
| 53 | 64 | 0.05756887 | 0.05249789 | -0.04174219 | -0.02695105 | 0.02591589 | 0.00717174 | -0.01257793 | -0.00347478 | 0.00110064 | -0.00065183 | 0.00008211 | 0.00039308 | -0.00020819 | -0.00006762 | 0.00005617 | -0.00008129 | -0.00009974 |
| 53 | 65 | 0.05761893 | 0.05232533 | -0.04174936 | -0.02650554 | 0.02660491 | 0.00657910 | -0.01249940 | -0.00344448 | 0.00127200 | -0.00076967 | 0.00007808 | 0.00027499 | 0.00000585 | -0.00010169 | 0.00021005 | -0.00020811 | -0.00001465 |
| 53 | 66 | 0.05773872 | 0.05220317 | -0.04187133 | -0.02593919 | 0.02594256 | 0.00601143 | -0.01245573 | -0.00305682 | 0.00127745 | -0.00068970 | 0.00013958 | 0.00029279 | -0.00008243 | -0.00014279 | 0.00012064 | -0.00011298 | -0.00000319 |
| 53 | 67 | 0.05747776 | 0.05182276 | -0.04187038 | -0.02553754 | 0.02657113 | 0.00557641 | -0.01237708 | -0.00289697 | 0.00142121 | -0.00079868 | 0.00012599 | 0.00020489 | 0.00006664 | -0.00013098 | 0.00024797 | -0.00020520 | 0.00006052 |
| 53 | 68 | 0.05731770 | 0.05120856 | -0.04218702 | -0.02499256 | 0.02593815 | 0.00534641 | -0.01220969 | -0.00230957 | 0.00133036 | -0.00072675 | 0.00017867 | 0.00029027 | -0.00014945 | -0.00010003 | 0.00010480 | -0.00007186 | 0.00000223 |
| 53 | 69 | 0.05717592 | 0.05100553 | -0.04240692 | -0.02483979 | 0.02650851 | 0.00494756 | -0.01227075 | -0.00230234 | 0.00142969 | -0.00084230 | 0.00016345 | 0.00019952 | 0.00001428 | -0.00010879 | 0.00022588 | -0.00015870 | 0.00007321 |
| 53 | 70 | 0.05696688 | 0.05026404 | -0.04263882 | -0.02412205 | 0.02611093 | 0.00477084 | -0.01204163 | -0.00169955 | 0.00134963 | -0.00077098 | 0.00021396 | 0.00028381 | -0.00019673 | -0.00006790 | 0.00009027 | -0.00004390 | 0.00000675 |
| 53 | 71 | 0.05696994 | 0.05020635 | -0.04311433 | -0.02420976 | 0.02660619 | 0.00441539 | -0.01221432 | -0.00181760 | 0.00140979 | -0.00088432 | 0.00019694 | 0.00019475 | -0.00003031 | -0.00009403 | 0.00019891 | -0.00011917 | 0.00008081 |
| 53 | 72 | 0.05671188 | 0.04940683 | -0.04322901 | -0.02335132 | 0.02646511 | 0.00434171 | -0.01192225 | -0.00123142 | 0.00133984 | -0.00081362 | 0.00024700 | 0.00027356 | -0.00022633 | -0.00004566 | 0.00007727 | -0.00002847 | 0.00000760 |
| 53 | 73 | 0.05687414 | 0.04946068 | -0.04395277 | -0.02362878 | 0.02689445 | 0.00404342 | -0.01217126 | -0.00144185 | 0.00136430 | -0.00091674 | 0.00022725 | 0.00019155 | -0.00006841 | -0.00008437 | 0.00016805 | -0.00008830 | 0.00007834 |
| 53 | 74 | 0.05655617 | 0.04865532 | -0.04394307 | -0.02271043 | 0.02696665 | 0.00410252 | -0.01180487 | -0.00089474 | 0.00131229 | -0.00084054 | 0.00028018 | 0.00025920 | -0.00024164 | -0.00003190 | 0.00006496 | -0.00002135 | 0.00000506 |
| 53 | 75 | 0.05687473 | 0.04878802 | -0.04486848 | -0.02309686 | 0.02735057 | 0.00386534 | -0.01210358 | -0.00116690 | 0.00130115 | -0.00092737 | 0.00025555 | 0.00019012 | -0.00010103 | -0.00007724 | 0.00013391 | -0.00006482 | 0.00006456 |
| 53 | 76 | 0.05648381 | 0.04800371 | -0.04475937 | -0.02222280 | 0.02756478 | 0.00407382 | -0.01164957 | -0.00067106 | 0.00127994 | -0.00083665 | 0.00031476 | 0.00024152 | -0.00024728 | -0.00002406 | 0.00005131 | -0.00001653 | 0.00000021 |
| 53 | 77 | 0.05693952 | 0.04818278 | -0.04581691 | -0.02262471 | 0.02792186 | 0.00388075 | -0.01198936 | -0.00097932 | 0.00123033 | -0.00090492 | 0.00028192 | 0.00019153 | -0.00013055 | -0.00006955 | 0.00009608 | -0.00004534 | 0.00003996 |
| 53 | 78 | 0.05646864 | 0.04742121 | -0.04566332 | -0.02189511 | 0.02821754 | 0.00425064 | -0.01144358 | -0.00053810 | 0.00125395 | -0.00079196 | 0.00034933 | 0.00022375 | -0.00024974 | -0.00001834 | 0.00003349 | -0.00000842 | -0.00000775 |
| 53 | 79 | 0.05703119 | 0.04761414 | -0.04677740 | -0.02221921 | 0.02855904 | 0.00406265 | -0.01183160 | -0.00086066 | 0.00116140 | -0.00084489 | 0.00030468 | 0.00019851 | -0.00016152 | -0.00005789 | 0.00005353 | -0.00002614 | 0.00000455 |
| 53 | 80 | 0.05648483 | 0.04686007 | -0.04665621 | -0.02171428 | 0.02890880 | 0.00460442 | -0.01120912 | -0.00047076 | 0.00124294 | -0.00070667 | 0.00037967 | 0.00021145 | -0.00025708 | -0.00001041 | 0.00000906 | 0.00000620 | -0.00002260 |
| 53 | 81 | 0.05711847 | 0.04703660 | -0.04775459 | -0.02187558 | 0.02923761 | 0.00436856 | -0.01165692 | -0.00078826 | 0.00110272 | -0.00075222 | 0.00032064 | 0.00021477 | -0.00019973 | -0.00003927 | 0.00000564 | -0.00000510 | -0.00004296 |
| 53 | 82 | 0.05651404 | 0.04627106 | -0.04775232 | -0.02165425 | 0.02964823 | 0.00509096 | -0.01099406 | -0.00044265 | 0.00125388 | -0.00059138 | 0.00040006 | 0.00021054 | -0.00027681 | 0.00000331 | -0.00002276 | 0.00002731 | -0.00004979 |
| 53 | 83 | 0.05693165 | 0.04596550 | -0.04819999 | -0.02068653 | 0.02950859 | 0.00353612 | -0.01171910 | -0.00050834 | 0.00104754 | -0.00077742 | 0.00033248 | 0.00019999 | -0.00020668 | -0.00001957 | 0.00001315 | -0.00005274 | -0.00008269 |
| 53 | 84 | 0.05595966 | 0.04474364 | -0.04736495 | -0.01958484 | 0.02923623 | 0.00298330 | -0.01128199 | 0.00006119 | 0.00118654 | -0.00068978 | 0.00040255 | 0.00016635 | -0.00024754 | 0.00003109 | 0.00002852 | -0.00007653 | -0.00008021 |
| 53 | 85 | 0.05650704 | 0.04435635 | -0.04818286 | -0.01863567 | 0.02924956 | 0.00158256 | -0.01179589 | 0.00004307 | 0.00097432 | -0.00091675 | 0.00034350 | 0.00014235 | -0.00017663 | -0.00000523 | 0.00007128 | -0.00015235 | -0.00009356 |
| 53 | 86 | 0.05549621 | 0.04311864 | -0.04717655 | -0.01758264 | 0.02871166 | 0.00102140 | -0.01127108 | 0.00063504 | 0.00109608 | -0.00076715 | 0.00040031 | 0.00012107 | -0.00022682 | 0.00005684 | 0.00006502 | -0.00015524 | -0.00010253 |
| 53 | 87 | 0.05609327 | 0.04272259 | -0.04826178 | -0.01674662 | 0.02887410 | -0.00017192 | -0.01162845 | 0.00058677 | 0.00091431 | -0.00098974 | 0.00034458 | 0.00008414 | -0.00014931 | 0.00001103 | 0.00011280 | -0.00021419 | -0.00008411 |
| 53 | 88 | 0.05511449 | 0.04148705 | -0.04717072 | -0.01577700 | 0.02811260 | -0.00068702 | -0.01100979 | 0.00119245 | 0.00100918 | -0.00078058 | 0.00038921 | 0.00007792 | -0.00021348 | 0.00008260 | 0.00008087 | -0.00019331 | -0.00010881 |
| 53 | 89 | 0.05566270 | 0.04122106 | -0.04833959 | -0.01520674 | 0.02841798 | -0.00163668 | -0.01135177 | 0.00103470 | 0.00091677 | -0.00097370 | 0.00032977 | 0.00003071 | -0.00012231 | 0.00002712 | 0.00013768 | -0.00023167 | -0.00005196 |
| 53 | 90 | 0.05478492 | 0.04000613 | -0.04725496 | -0.01433101 | 0.02751747 | -0.00205995 | -0.01066457 | 0.00163688 | 0.00098024 | -0.00071449 | 0.00036355 | 0.00004178 | -0.00020437 | 0.00010539 | 0.00007873 | -0.00018742 | -0.00009622 |
| 53 | 91 | 0.05519626 | 0.04000970 | -0.04830567 | -0.01414509 | 0.02793709 | -0.00278249 | -0.01114429 | 0.00132224 | 0.00102822 | -0.00087816 | 0.00029599 | -0.00001362 | -0.00009265 | 0.00003709 | 0.00015206 | -0.00021012 | 0.00000035 |
| 53 | 92 | 0.05447914 | 0.03882864 | -0.04731540 | -0.01334979 | 0.02701098 | -0.00308299 | -0.01043538 | 0.00190396 | 0.00106218 | -0.00058514 | 0.00032024 | 0.00001622 | -0.00019528 | 0.00011871 | 0.00006773 | -0.00014633 | -0.00006624 |
| 53 | 93 | 0.05469674 | 0.03915425 | -0.04812520 | -0.01356473 | 0.02749027 | -0.00364144 | -0.01113318 | 0.00142740 | 0.00126494 | -0.00072974 | 0.00024523 | -0.00004780 | -0.00005868 | 0.00003504 | 0.00016222 | -0.00015992 | 0.00006818 |
| 53 | 94 | 0.05418871 | 0.03801289 | -0.04730660 | -0.01281677 | 0.02665488 | -0.00381185 | -0.01045202 | 0.00198228 | 0.00127648 | -0.00042276 | 0.00026105 | 0.00000129 | -0.00018285 | 0.00011692 | 0.00005664 | -0.00008248 | -0.00002165 |
| 54 | 54 | 0.05863042 | 0.05692821 | -0.04068533 | -0.03464261 | 0.02147541 | 0.01009858 | -0.01218499 | -0.00549929 | 0.00081780 | -0.00023261 | -0.00010666 | 0.00050054 | -0.00031207 | 0.00000555 | -0.00004880 | 0.00012432 | -0.00009754 |
| 54 | 55 | 0.05844888 | 0.05573175 | -0.04121841 | -0.03373246 | 0.02164203 | 0.00982939 | -0.01192278 | -0.00489263 | 0.00068681 | -0.00047289 | -0.00010328 | 0.00060076 | -0.00048198 | 0.00006530 | -0.00010403 | 0.00011618 | -0.00021485 |
| 54 | 56 | 0.05856957 | 0.05561741 | -0.04145196 | -0.03234498 | 0.02327988 | 0.00934509 | -0.01279196 | -0.00511382 | 0.00087047 | -0.00027779 | -0.00004823 | 0.00050666 | -0.00034523 | 0.00002124 | -0.00004943 | 0.00009960 | -0.00012255 |
| 54 | 57 | 0.05833378 | 0.05449292 | -0.04197257 | -0.03161889 | 0.02352198 | 0.00914883 | -0.01253511 | -0.00453483 | 0.00078510 | -0.00052919 | -0.00005387 | 0.00058825 | -0.00047842 | 0.00007182 | -0.00007945 | 0.00006620 | -0.00023037 |
| 54 | 58 | 0.05851687 | 0.05447691 | -0.04202654 | -0.03033507 | 0.02473598 | 0.00863677 | -0.01318168 | -0.00470718 | 0.00094236 | -0.00031751 | 0.00000378 | 0.00050350 | -0.00035914 | 0.00002369 | -0.00004550 | 0.00007151 | -0.00014008 |
| 54 | 59 | 0.05826923 | 0.05345515 | -0.04253254 | -0.02976382 | 0.02500186 | 0.00845618 | -0.01290539 | -0.00415009 | 0.00088904 | -0.00057786 | -0.00000459 | 0.00056106 | -0.00045145 | 0.00006219 | -0.00005143 | 0.00001722 | -0.00023092 |
| 54 | 60 | 0.05852770 | 0.05354391 | -0.04246528 | -0.02862368 | 0.02577928 | 0.00790523 | -0.01335062 | -0.00427798 | 0.00102696 | -0.00035485 | 0.00005364 | 0.00048695 | -0.00035073 | 0.00001008 | -0.00003595 | 0.00004516 | -0.00014065 |
| 54 | 61 | 0.05829199 | 0.05265100 | -0.04293175 | -0.02816743 | 0.02601935 | 0.00767879 | -0.01304060 | -0.00374066 | 0.00099708 | -0.00062137 | 0.00004757 | 0.00051501 | -0.00039719 | 0.00003344 | -0.00001808 | -0.00002549 | -0.00020586 |
| 54 | 62 | 0.05862648 | 0.05283700 | -0.04280256 | -0.02718722 | 0.02640896 | 0.00709224 | -0.01333305 | -0.00383264 | 0.00112472 | -0.00039492 | 0.00010342 | 0.00045241 | -0.00031521 | -0.00002179 | -0.00001718 | 0.00002143 | -0.00011741 |
| 54 | 63 | 0.05840155 | 0.05208884 | -0.04318490 | -0.02680055 | 0.02659330 | 0.00676432 | -0.01299222 | -0.00331736 | 0.00111845 | -0.00066325 | 0.00010235 | 0.00044576 | -0.00030997 | -0.00001612 | 0.00002572 | -0.00006226 | -0.00014851 |
| 54 | 64 | 0.05880359 | 0.05234956 | -0.04305461 | -0.02598835 | 0.02669221 | 0.00617060 | -0.01319658 | -0.00338880 | 0.00124464 | -0.00043898 | 0.00015172 | 0.00039708 | -0.00024862 | -0.00007166 | 0.00001519 | -0.00000205 | -0.00006835 |
| 54 | 65 | 0.05856391 | 0.05175006 | -0.04329494 | -0.02563261 | 0.02681910 | 0.00570901 | -0.01284782 | -0.00290632 | 0.00127061 | -0.00070124 | 0.00015522 | 0.00035210 | -0.00018672 | -0.00008477 | 0.00008487 | -0.00009675 | -0.00005941 |
| 54 | 66 | 0.05902384 | 0.05205056 | -0.04322601 | -0.02499435 | 0.02674997 | 0.00516160 | -0.01303005 | -0.00297708 | 0.00140052 | -0.00048075 | 0.00019381 | 0.00032250 | -0.00015159 | -0.00013581 | 0.00006336 | -0.00002855 | 0.00000200 |
| 54 | 67 | 0.05841531 | 0.05116547 | -0.04340922 | -0.02463753 | 0.02680726 | 0.00482533 | -0.01261091 | -0.00233332 | 0.00138954 | -0.00073147 | 0.00020255 | 0.00029182 | -0.00014425 | -0.00010440 | 0.00011301 | -0.00008938 | 0.00000232 |
| 54 | 68 | 0.05850647 | 0.05097197 | -0.04344235 | -0.02402446 | 0.02669420 | 0.00452215 | -0.01266854 | -0.00223314 | 0.00143890 | -0.00051070 | 0.00024275 | 0.00030719 | -0.00021137 | -0.00009312 | 0.00005215 | 0.00000989 | 0.00001497 |
| 54 | 69 | 0.05798954 | 0.05021638 | -0.04375341 | -0.02373547 | 0.02685249 | 0.00422992 | -0.01235857 | -0.00167093 | 0.00141253 | -0.00077042 | 0.00024806 | 0.00027621 | -0.00019101 | -0.00007287 | 0.00009996 | -0.00005030 | 0.00001950 |
| 54 | 70 | 0.05805531 | 0.04994349 | -0.04382396 | -0.02310795 | 0.02685030 | 0.00393588 | -0.01240079 | -0.00161605 | 0.00144594 | -0.00056109 | 0.00028828 | 0.00028148 | -0.00024154 | -0.00006270 | 0.00005063 | 0.00002577 | 0.00002968 |
| 54 | 71 | 0.05765609 | 0.04933129 | -0.04427729 | -0.02293004 | 0.02708550 | 0.00374184 | -0.01218968 | -0.00115882 | 0.00140116 | -0.00081433 | 0.00028968 | 0.00025674 | -0.00021886 | -0.00005174 | 0.00008953 | -0.00002675 | 0.00003327 |
| 54 | 72 | 0.05770041 | 0.04900464 | -0.04437931 | -0.02227986 | 0.02721632 | 0.00345954 | -0.01220528 | -0.00112636 | 0.00143255 | -0.00062263 | 0.00033047 | 0.00024738 | -0.00024653 | -0.00004379 | 0.00005739 | 0.00002189 | 0.00004317 |
| 54 | 73 | 0.05743555 | 0.04853620 | -0.04497430 | -0.02224224 | 0.02750022 | 0.00341924 | -0.01206009 | -0.00078724 | 0.00136635 | -0.00085092 | 0.00032853 | 0.00023480 | -0.00023248 | -0.00003917 | 0.00008037 | -0.00001592 | 0.00004083 |
| 54 | 74 | 0.05745096 | 0.04817231 | -0.04509856 | -0.02157878 | 0.02775314 | 0.00314511 | -0.01203720 | -0.00075121 | 0.00141304 | -0.00067930 | 0.00037048 | 0.00020740 | -0.00023373 | -0.00003436 | 0.00006875 | 0.00000581 | 0.00005479 |
| 54 | 75 | 0.05732124 | 0.04783660 | -0.04581326 | -0.02168809 | 0.02806074 | 0.00330313 | -0.01191591 | -0.00053431 | 0.00132300 | -0.00086409 | 0.00036599 | 0.00021144 | -0.00023686 | -0.00003259 | 0.00007008 | -0.00001200 | 0.00004213 |
| 54 | 76 | 0.05729887 | 0.04743899 | -0.04596373 | -0.02103399 | 0.02840791 | 0.00302945 | -0.01185237 | -0.00047301 | 0.00139962 | -0.00071296 | 0.00040893 | 0.00016520 | -0.00021216 | -0.00003101 | 0.00007943 | -0.00001287 | 0.00006433 |
| 54 | 77 | 0.05728808 | 0.04721342 | -0.04676004 | -0.02127370 | 0.02872129 | 0.00340636 | -0.01172298 | -0.00037466 | 0.00128467 | -0.00083959 | 0.00040168 | 0.00018925 | -0.00023815 | -0.00002827 | 0.00005539 | -0.00000838 | 0.00003699 |
| 54 | 78 | 0.05722672 | 0.04677644 | -0.04695963 | -0.02065743 | 0.02913615 | 0.00312663 | -0.01162955 | -0.00027406 | 0.00140044 | -0.00070925 | 0.00044445 | 0.00012624 | -0.00019169 | -0.00002929 | 0.00008373 | -0.00002529 | 0.00007003 |
| 54 | 79 | 0.05730537 | 0.04662704 | -0.04779450 | -0.02099127 | 0.02944820 | 0.00371171 | -0.01148423 | -0.00028253 | 0.00126148 | -0.00077155 | 0.00043263 | 0.00017326 | -0.00024422 | -0.00002162 | 0.00003303 | 0.00000017 | 0.00002286 |
| 54 | 80 | 0.05721609 | 0.04614596 | -0.04807959 | -0.02044169 | 0.02991541 | 0.00342394 | -0.01138140 | -0.00013648 | 0.00142087 | -0.00066286 | 0.00047333 | 0.00009710 | -0.00018190 | -0.00002463 | 0.00007723 | -0.00002576 | 0.00006793 |
| 54 | 81 | 0.05734722 | 0.04603051 | -0.04891411 | -0.02082052 | 0.03023116 | 0.00417795 | -0.01123822 | -0.00023240 | 0.00126071 | -0.00066568 | 0.00045388 | 0.00016959 | -0.00026320 | -0.00000850 | 0.00000119 | 0.00001580 | -0.00000484 |
| 54 | 82 | 0.05725288 | 0.04551150 | -0.04932314 | -0.02036495 | 0.03074539 | 0.00388481 | -0.01114882 | -0.00004163 | 0.00146545 | -0.00057910 | 0.00049059 | 0.00008375 | -0.00019008 | -0.00001360 | 0.00005820 | -0.00001282 | 0.00005303 |
| 54 | 83 | 0.05709798 | 0.04494474 | -0.04935357 | -0.01975494 | 0.03047320 | 0.00341478 | -0.01130049 | -0.00000415 | 0.00122525 | -0.00064816 | 0.00046098 | 0.00014845 | -0.00026040 | 0.00001457 | 0.00000875 | -0.00002412 | -0.00003573 |
| 54 | 84 | 0.05672474 | 0.04402165 | -0.04887082 | -0.01824850 | 0.03024944 | 0.00190459 | -0.01126836 | 0.00042051 | 0.00134651 | -0.00062580 | 0.00049398 | 0.00003738 | -0.00015531 | 0.00001417 | 0.00009221 | -0.00009882 | 0.00002519 |
| 54 | 85 | 0.05662306 | 0.04330825 | -0.04918425 | -0.01773018 | 0.03003110 | 0.00142473 | -0.01141560 | 0.00048462 | 0.00110583 | -0.00072902 | 0.00045952 | 0.00009961 | -0.00023372 | 0.00004177 | 0.00004926 | -0.00010656 | -0.00005609 |
| 54 | 86 | 0.05631043 | 0.04245085 | -0.04863768 | -0.01624190 | 0.02967477 | 0.00013498 | -0.01112495 | 0.00092439 | 0.00120406 | -0.00063892 | 0.00048826 | 0.00000000 | -0.00014059 | 0.00004477 | 0.00010394 | -0.00015544 | -0.00000506 |
| 54 | 87 | 0.05623628 | 0.04162103 | -0.04917623 | -0.01581239 | 0.02948043 | -0.00035566 | -0.01124706 | 0.00100253 | 0.00097267 | -0.00075755 | 0.00044761 | 0.00005659 | -0.00022131 | 0.00007252 | 0.00006647 | -0.00015192 | -0.00006899 |
| 54 | 88 | 0.05599363 | 0.04090860 | -0.04856875 | -0.01446894 | 0.02907281 | -0.00134702 | -0.01080920 | 0.00139903 | 0.00107932 | -0.00058970 | 0.00046849 | -0.00002317 | -0.00014604 | 0.00007921 | 0.00009113 | -0.00017289 | -0.00003269 |
| 54 | 89 | 0.05590449 | 0.04004777 | -0.04922853 | -0.01417856 | 0.02888686 | -0.00183455 | -0.01093242 | 0.00145832 | 0.00088665 | -0.00070763 | 0.00041934 | 0.00002326 | -0.00021783 | 0.00010440 | 0.00006265 | -0.00015486 | -0.00006881 |
| 54 | 90 | 0.05574250 | 0.03954615 | -0.04856082 | -0.01305689 | 0.02852090 | -0.00250912 | -0.01049255 | 0.00177441 | 0.00103262 | -0.00047522 | 0.00043046 | -0.00002939 | -0.00016606 | 0.00011286 | 0.00006087 | -0.00015319 | -0.00005365 |
| 54 | 91 | 0.05559314 | 0.03877059 | -0.04921592 | -0.01297599 | 0.02834113 | -0.00298159 | -0.01068011 | 0.00177506 | 0.00091521 | -0.00058752 | 0.00037096 | 0.00000176 | -0.00021525 | 0.00012981 | 0.00004860 | -0.00012354 | -0.00005388 |
| 54 | 92 | 0.05553611 | 0.03846865 | -0.04854484 | -0.01204348 | 0.02810355 | -0.00338848 | -0.01034229 | 0.00201449 | 0.00111126 | -0.00031527 | 0.00037368 | -0.00001985 | -0.00019257 | 0.00013850 | 0.00002463 | -0.00010691 | -0.00006548 |
| 54 | 93 | 0.05529066 | 0.03787937 | -0.04908991 | -0.01222685 | 0.02792799 | -0.00384135 | -0.01065459 | 0.00192199 | 0.00109404 | -0.00042581 | 0.00030359 | -0.00000894 | -0.00020691 | 0.00014105 | 0.00003574 | -0.00007150 | -0.00002513 |
| 54 | 94 | 0.05503773 | 0.03783439 | -0.04809108 | -0.01209277 | 0.02738834 | -0.00336794 | -0.01007777 | 0.00188312 | 0.00119758 | -0.00013035 | 0.00031515 | -0.00002525 | -0.00019915 | 0.00013327 | -0.00001235 | -0.00006089 | -0.00006595 |
| 54 | 95 | 0.05462101 | 0.03763211 | -0.04827135 | -0.01277126 | 0.02705308 | -0.00366851 | -0.01045576 | 0.00164176 | 0.00126769 | -0.00026425 | 0.00024246 | -0.00005480 | -0.00015262 | 0.00010655 | 0.00003442 | -0.00003515 | 0.00002517 |
| 54 | 96 | 0.03569672 | 0.05574770 | -0.01248334 | -0.05440583 | 0.00141437 | 0.03026147 | 0.00025821 | -0.01438481 | -0.00131684 | 0.00304988 | -0.00057963 | -0.00066535 | 0.00103089 | 0.00006516 | -0.00010723 | -0.00022869 | 0.00014981 |

TABLE 3. Nuclear charge density distribution FB coefficients.

| $Z$ | $N$ | $a_1$ | $a_2$ | $a_3$ | $a_4$ | $a_5$ | $a_6$ | $a_7$ | $a_8$ | $a_9$ | $a_{10}$ | $a_{11}$ | $a_{12}$ | $a_{13}$ | $a_{14}$ | $a_{15}$ | $a_{16}$ | $a_{17}$ |
|---|---|---|---|---|---|---|---|---|---|---|---|---|---|---|---|---|---|---|
| 55 | 56 | 0.05931488 | 0.05448696 | -0.04317980 | -0.03194803 | 0.02318472 | 0.00838463 | -0.01274278 | -0.00425228 | 0.00084263 | -0.00053205 | -0.00001629 | 0.00060797 | -0.00055882 | 0.00009703 | -0.00009708 | 0.00008620 | -0.00026245 |
| 55 | 57 | 0.05905658 | 0.05355689 | -0.04357766 | -0.03126806 | 0.02492559 | 0.00886305 | -0.01302872 | -0.00437526 | 0.00091111 | -0.00054649 | -0.00008543 | 0.00062913 | -0.00052361 | 0.00015342 | -0.00005175 | 0.00001599 | -0.00032375 |
| 55 | 58 | 0.05920323 | 0.05336901 | -0.04375064 | -0.02999660 | 0.02485961 | 0.00776549 | -0.01315868 | -0.00389214 | 0.00094217 | -0.00058505 | 0.00002737 | 0.00058766 | -0.00053692 | 0.00009362 | -0.00007140 | 0.00003977 | -0.00026419 |
| 55 | 59 | 0.05903047 | 0.05261780 | -0.04403358 | -0.02938151 | 0.02642148 | 0.00816505 | -0.01327025 | -0.00396737 | 0.00102430 | -0.00062713 | -0.00003819 | 0.00058783 | -0.00046971 | 0.00013667 | -0.00000645 | -0.00004504 | -0.00030643 |
| 55 | 60 | 0.05917598 | 0.05246628 | -0.04416484 | -0.02826498 | 0.02614219 | 0.00710314 | -0.01334104 | -0.00350874 | 0.00104498 | -0.00062881 | 0.00007271 | 0.00055108 | -0.00048972 | 0.00007293 | -0.00004233 | -0.00000206 | -0.00024488 |
| 55 | 61 | 0.05907070 | 0.05191155 | -0.04427611 | -0.02766076 | 0.02747783 | 0.00733346 | -0.01327475 | -0.00352911 | 0.00114672 | -0.00069449 | 0.00001550 | 0.00051939 | -0.00037781 | 0.00009879 | 0.00004753 | -0.00009944 | -0.00025229 |
| 55 | 62 | 0.05924663 | 0.05180848 | -0.04443042 | -0.02674408 | 0.02699335 | 0.00632462 | -0.01331760 | -0.00310829 | 0.00115435 | -0.00066797 | 0.00012141 | 0.00049167 | -0.00040976 | 0.00003189 | -0.00000456 | -0.00003882 | -0.00019533 |
| 55 | 63 | 0.05915254 | 0.05145334 | -0.04429176 | -0.02609560 | 0.02812468 | 0.00632463 | -0.01312232 | -0.00308417 | 0.00129904 | -0.00074570 | 0.00007219 | 0.00041840 | -0.00023950 | 0.00003843 | 0.00011728 | -0.00015042 | -0.00015665 |
| 55 | 64 | 0.05938814 | 0.05139446 | -0.04454214 | -0.02541251 | 0.02746207 | 0.00539516 | -0.01316335 | -0.00271080 | 0.00128559 | -0.00070381 | 0.00017048 | 0.00040523 | -0.00029075 | -0.00002956 | 0.00004876 | -0.00007469 | -0.00011286 |
| 55 | 65 | 0.05922194 | 0.05122193 | -0.04407749 | -0.02470052 | 0.02846190 | 0.00516008 | -0.01293777 | -0.00268059 | 0.00150350 | -0.00077298 | 0.00012334 | 0.00028675 | -0.00005481 | -0.00004051 | 0.00020699 | -0.00020365 | -0.00002668 |
| 55 | 66 | 0.05954699 | 0.05119222 | -0.04449586 | -0.02426269 | 0.02767044 | 0.00434429 | -0.01298781 | -0.00235441 | 0.00146043 | -0.00072901 | 0.00021231 | 0.00029405 | -0.00013353 | -0.00010680 | 0.00012136 | -0.00011494 | -0.00000424 |
| 55 | 67 | 0.05904212 | 0.05079545 | -0.04407056 | -0.02373401 | 0.02842508 | 0.00422207 | -0.01278582 | -0.00220013 | 0.00163947 | -0.00078411 | 0.00016503 | 0.00020648 | 0.00002564 | -0.00007063 | 0.00024883 | -0.00020211 | 0.00005701 |
| 55 | 68 | 0.05907050 | 0.05025410 | -0.04472816 | -0.02334327 | 0.02758482 | 0.00375439 | -0.01264595 | -0.00164278 | 0.00148686 | -0.00075717 | 0.00026491 | 0.00027388 | -0.00018440 | -0.00007333 | 0.00010588 | -0.00006218 | 0.00002056 |
| 55 | 69 | 0.05869735 | 0.05000215 | -0.04463078 | -0.02314177 | 0.02820367 | 0.00355930 | -0.01267893 | -0.00166363 | 0.00163200 | -0.00080817 | 0.00020873 | 0.00018968 | -0.00001695 | -0.00005196 | 0.00022834 | -0.00014692 | 0.00008357 |
| 55 | 70 | 0.05866575 | 0.04935184 | -0.04517399 | -0.02252933 | 0.02766353 | 0.00321637 | -0.01241966 | -0.00108938 | 0.00147511 | -0.00079492 | 0.00031407 | 0.00024716 | -0.00021305 | -0.00005155 | 0.00009578 | -0.00002708 | 0.00004575 |
| 55 | 71 | 0.05845274 | 0.04919916 | -0.04545821 | -0.02263326 | 0.02814009 | 0.00296037 | -0.01264123 | -0.00124529 | 0.00159014 | -0.00083758 | 0.00025144 | 0.00016992 | -0.00005033 | -0.00004328 | 0.00020634 | -0.00010207 | 0.00010780 |
| 55 | 72 | 0.05837476 | 0.04851761 | -0.04584418 | -0.02182859 | 0.02793961 | 0.00279981 | -0.01227509 | -0.00068633 | 0.00143482 | -0.00083434 | 0.00035903 | 0.00021714 | -0.00022525 | -0.00003943 | 0.00008943 | -0.00000923 | 0.00006499 |
| 55 | 73 | 0.05833735 | 0.04842607 | -0.04649896 | -0.02215134 | 0.02832277 | 0.00253390 | -0.01261231 | -0.00093158 | 0.00152066 | -0.00086412 | 0.00029198 | 0.00015023 | -0.00007749 | -0.00004173 | 0.00018126 | -0.00006858 | 0.00012218 |
| 55 | 74 | 0.05820652 | 0.04776808 | -0.04670399 | -0.02124148 | 0.02840888 | 0.00257094 | -0.01214802 | -0.00040815 | 0.00138171 | -0.00086057 | 0.00040049 | 0.00018600 | -0.00022686 | -0.00003434 | 0.00008374 | -0.00000371 | 0.00007591 |
| 55 | 75 | 0.05834240 | 0.04770674 | -0.04767191 | -0.02167027 | 0.02876000 | 0.00234474 | -0.01253566 | -0.00070516 | 0.00143484 | -0.00087279 | 0.00032967 | 0.00013308 | -0.00010104 | -0.00004347 | 0.00015135 | -0.00004433 | 0.00012289 |
| 55 | 76 | 0.05814290 | 0.04709697 | -0.04770231 | -0.02076718 | 0.02903565 | 0.00256826 | -0.01198205 | -0.00022533 | 0.00133116 | -0.00085667 | 0.00043860 | 0.00015645 | -0.00022429 | -0.00003255 | 0.00007458 | -0.00000328 | 0.00007774 |
| 55 | 77 | 0.05843313 | 0.04703921 | -0.04890318 | -0.02119288 | 0.02939931 | 0.00240113 | -0.01238361 | -0.00054783 | 0.00134512 | -0.00084916 | 0.00036304 | 0.00012178 | -0.00012502 | -0.00004378 | 0.00011462 | -0.00002524 | 0.00010815 |
| 55 | 78 | 0.05815212 | 0.04647527 | -0.04879712 | -0.02040175 | 0.02977422 | 0.00279314 | -0.01175645 | -0.00011057 | 0.00129467 | -0.00081077 | 0.00047143 | 0.00013331 | -0.00022557 | -0.00002917 | 0.00005749 | -0.00000087 | 0.00006859 |
| 55 | 79 | 0.05856578 | 0.04639761 | -0.05014445 | -0.02073071 | 0.03017205 | 0.00266712 | -0.01216551 | -0.00043971 | 0.00126283 | -0.00078684 | 0.00038922 | 0.00012104 | -0.00015556 | -0.00003775 | 0.00006938 | -0.00000710 | 0.00007634 |
| 55 | 80 | 0.05820191 | 0.04586139 | -0.04996563 | -0.02013354 | 0.03059117 | 0.00321305 | -0.01149381 | -0.00003906 | 0.00127996 | -0.00072210 | 0.00049490 | 0.00012300 | -0.00023959 | -0.00001932 | 0.00002922 | 0.00000816 | 0.00004460 |
| 55 | 81 | 0.05870171 | 0.04574425 | -0.05137550 | -0.02028973 | 0.03102667 | 0.00308259 | -0.01192130 | -0.00035932 | 0.00119753 | -0.00069062 | 0.00040440 | 0.00013587 | -0.00019945 | -0.00002147 | 0.00001520 | 0.00001261 | 0.00002545 |
| 55 | 82 | 0.05826777 | 0.04521652 | -0.05120093 | -0.01994503 | 0.03147283 | 0.00377390 | -0.01124641 | 0.00001163 | 0.00129229 | -0.00060145 | 0.00050426 | 0.00013123 | -0.00027358 | 0.00000032 | -0.00001092 | 0.00002490 | 0.00000142 |
| 55 | 83 | 0.05857726 | 0.04459381 | -0.05204043 | -0.01906995 | 0.03139040 | 0.00235270 | -0.01192917 | -0.00012133 | 0.00110208 | -0.00070279 | 0.00041193 | 0.00012049 | -0.00021252 | 0.00000013 | 0.00001325 | -0.00002343 | -0.00001225 |
| 55 | 84 | 0.05778785 | 0.04357088 | -0.05109740 | -0.01792471 | 0.03115743 | 0.00178535 | -0.01148378 | 0.00040448 | 0.00115435 | -0.00067869 | 0.00050617 | 0.00007582 | -0.00023717 | 0.00002727 | 0.00003502 | -0.00006029 | -0.00001219 |
| 55 | 85 | 0.05822989 | 0.04291931 | -0.05218961 | -0.01706641 | 0.03111040 | 0.00050917 | -0.01197129 | 0.00032129 | 0.00094332 | -0.00081131 | 0.00041386 | 0.00006580 | -0.00019272 | 0.00002322 | 0.00005414 | -0.00009282 | -0.00001710 |
| 55 | 86 | 0.05740384 | 0.04183119 | -0.05114222 | -0.01593290 | 0.03070526 | -0.00002867 | -0.01141742 | 0.00085828 | 0.00097788 | -0.00072053 | 0.00049739 | 0.00003022 | -0.00022302 | 0.00006048 | 0.00005481 | -0.00011142 | -0.00002694 |
| 55 | 87 | 0.05787506 | 0.04130699 | -0.05228947 | -0.01525116 | 0.03069727 | -0.00111354 | -0.01181647 | 0.00075491 | 0.00081888 | -0.00084774 | 0.00039868 | 0.00002105 | -0.00018260 | 0.00005243 | 0.00007481 | -0.00012382 | -0.00001078 |
| 55 | 88 | 0.05708129 | 0.04016112 | -0.05122434 | -0.01414842 | 0.03016112 | -0.00157069 | -0.01115031 | 0.00129463 | 0.00082656 | -0.00069112 | 0.00047114 | -0.00000110 | -0.00022603 | 0.00009922 | 0.00004893 | -0.00012063 | -0.00003625 |
| 55 | 89 | 0.05747317 | 0.03995316 | -0.05220305 | -0.01382654 | 0.03019981 | -0.00245453 | -0.01162882 | 0.00110335 | 0.00080266 | -0.00080056 | 0.00036165 | -0.00001172 | -0.00017257 | 0.00008107 | 0.00008304 | -0.00011945 | 0.00001035 |
| 55 | 90 | 0.05677820 | 0.03876503 | -0.05120341 | -0.01275252 | 0.02961280 | -0.00279652 | -0.01088556 | 0.00163286 | 0.00078221 | -0.00058798 | 0.00042249 | -0.00001678 | -0.00023517 | 0.00013577 | 0.00002885 | -0.00009475 | -0.00003572 |
| 55 | 91 | 0.05701766 | 0.03897228 | -0.05189067 | -0.01286829 | 0.02970422 | -0.00352369 | -0.01157591 | 0.00131743 | 0.00093988 | -0.00068525 | 0.00030337 | -0.00003354 | -0.00015480 | 0.00010033 | 0.00008863 | -0.00009023 | 0.00004577 |
| 55 | 92 | 0.05647803 | 0.03776393 | -0.05102503 | -0.01180580 | 0.02916550 | -0.00374153 | -0.01081508 | 0.00182811 | 0.00089865 | -0.00043502 | 0.00035161 | -0.00001952 | -0.00023941 | 0.00016056 | 0.00000892 | -0.00004850 | -0.00002385 |
| 55 | 93 | 0.05464246 | 0.04056376 | -0.04752121 | -0.01730904 | 0.02578358 | -0.00033788 | -0.00946493 | 0.00000830 | 0.00058141 | -0.00061808 | 0.00032702 | -0.00030968 | 0.00014660 | -0.00005741 | 0.00016302 | -0.00020097 | 0.00019079 |
| 55 | 94 | 0.05584306 | 0.03736911 | -0.05021612 | -0.01202679 | 0.02833757 | -0.00373487 | -0.01059697 | 0.00164103 | 0.00103943 | -0.00026975 | 0.00028573 | -0.00005357 | -0.00020013 | 0.00014068 | -0.00000046 | -0.00001547 | 0.00000855 |
| 55 | 95 | 0.03553126 | 0.05849621 | -0.00973073 | -0.05684520 | -0.00096748 | 0.03139536 | 0.00178608 | -0.01473680 | -0.00192072 | 0.00235291 | -0.00036291 | -0.00100302 | 0.00139073 | -0.00018527 | 0.00013119 | -0.00056725 | 0.00029963 |
| 55 | 96 | 0.03614049 | 0.05676434 | -0.01152802 | -0.05430915 | 0.00104864 | 0.02998064 | 0.00103522 | -0.01381450 | -0.00150348 | 0.00248427 | -0.00040933 | -0.00092585 | 0.00127686 | -0.00010619 | 0.00008139 | -0.00047876 | 0.00024750 |
| 55 | 97 | 0.03767349 | 0.05123716 | -0.01859333 | -0.04584725 | 0.00457637 | 0.02485279 | -0.00192644 | -0.01262148 | -0.00134210 | 0.00264113 | -0.00027236 | -0.00037391 | 0.00049786 | -0.00008182 | -0.00029917 | 0.00015414 | 0.00019334 |
| 56 | 57 | 0.06014747 | 0.05344205 | -0.04472219 | -0.03029292 | 0.02445920 | 0.00707224 | -0.01331707 | -0.00359771 | 0.00099949 | -0.00058453 | 0.00005195 | 0.00061140 | -0.00061941 | 0.00012426 | -0.00009194 | 0.00006190 | -0.00030200 |
| 56 | 58 | 0.06044113 | 0.05350750 | -0.04477331 | -0.02921018 | 0.02540015 | 0.00653204 | -0.01389029 | -0.00384128 | 0.00108760 | -0.00039897 | 0.00009935 | 0.00054592 | -0.00051576 | 0.00007638 | -0.00007126 | 0.00007567 | -0.00021848 |
| 56 | 59 | 0.06007945 | 0.05245168 | -0.04514936 | -0.02844196 | 0.02596361 | 0.00650311 | -0.01353888 | -0.00323428 | 0.00109672 | -0.00063030 | 0.00009275 | 0.00058265 | -0.00057940 | 0.00011039 | -0.00006714 | 0.00002293 | -0.00028624 |
| 56 | 60 | 0.06049362 | 0.05263885 | -0.04513923 | -0.02749461 | 0.02650165 | 0.00592689 | -0.01392943 | -0.00344650 | 0.00116238 | -0.00042695 | 0.00014158 | 0.00052661 | -0.00049330 | 0.00005831 | -0.00006541 | 0.00006031 | -0.00020118 |
| 56 | 61 | 0.06010740 | 0.05168943 | -0.04543251 | -0.02676591 | 0.02709271 | 0.00585307 | -0.01354400 | -0.00285324 | 0.00119817 | -0.00066721 | 0.00013698 | 0.00053201 | -0.00050685 | 0.00007731 | -0.00003491 | -0.00001273 | -0.00024276 |
| 56 | 62 | 0.06062714 | 0.05198903 | -0.04540156 | -0.02596759 | 0.02727753 | 0.00523270 | -0.01380931 | -0.00303975 | 0.00125677 | -0.00045339 | 0.00018352 | 0.00048460 | -0.00043636 | 0.00002215 | -0.00004589 | 0.00004147 | -0.00015903 |
| 56 | 63 | 0.06021653 | 0.05117668 | -0.04555714 | -0.02526170 | 0.02783282 | 0.00505750 | -0.01338784 | -0.00246951 | 0.00131356 | -0.00069933 | 0.00018345 | 0.00045250 | -0.00039287 | 0.00002327 | 0.00001268 | -0.00004908 | -0.00016588 |
| 56 | 64 | 0.06081428 | 0.05155803 | -0.04554763 | -0.02461879 | 0.02777382 | 0.00442352 | -0.01360242 | -0.00264343 | 0.00138126 | -0.00047966 | 0.00022225 | 0.00041628 | -0.00033982 | -0.00003136 | -0.00000657 | 0.00001385 | -0.00009225 |
| 56 | 65 | 0.06035791 | 0.05089778 | -0.04551274 | -0.02393337 | 0.02826347 | 0.00411218 | -0.01317679 | -0.00211733 | 0.00146418 | -0.00072287 | 0.00022528 | 0.00034359 | -0.00023509 | -0.00004865 | 0.00008172 | -0.00009228 | -0.00005960 |
| 56 | 66 | 0.06100833 | 0.05131626 | -0.04557049 | -0.02344664 | 0.02809937 | 0.00352780 | -0.01341069 | -0.00229365 | 0.00155267 | -0.00049875 | 0.00025126 | 0.00032497 | -0.00020651 | -0.00009726 | 0.00005481 | -0.00002683 | -0.00000853 |
| 56 | 67 | 0.06017806 | 0.05037516 | -0.04551749 | -0.02288892 | 0.02831658 | 0.00332251 | -0.01288988 | -0.00160794 | 0.00156871 | -0.00073616 | 0.00026887 | 0.00026990 | -0.00016928 | -0.00007141 | 0.00011385 | -0.00008560 | 0.00001428 |
| 56 | 68 | 0.06047220 | 0.05029225 | -0.04575492 | -0.02250027 | 0.02798929 | 0.00302160 | -0.01294008 | -0.00160257 | 0.00154180 | -0.00052227 | 0.00031104 | 0.00029910 | -0.00025764 | -0.00006283 | 0.00003553 | 0.00002986 | 0.00001910 |
| 56 | 69 | 0.05972739 | 0.04946711 | -0.04588447 | -0.02208117 | 0.02823177 | 0.00276266 | -0.01259659 | -0.00101511 | 0.00155492 | -0.00076076 | 0.00032253 | 0.00024159 | -0.00020516 | -0.00004579 | 0.00009876 | -0.00003501 | 0.00004521 |
| 56 | 70 | 0.05999398 | 0.04929842 | -0.04616603 | -0.02162623 | 0.02808770 | 0.00253179 | -0.01259891 | -0.00105082 | 0.00150168 | -0.00056813 | 0.00036778 | 0.00025965 | -0.00027624 | -0.00004186 | 0.00003081 | 0.00005982 | 0.00005039 |
| 56 | 71 | 0.05937448 | 0.04859884 | -0.04652363 | -0.02139718 | 0.02832956 | 0.00226140 | -0.01242979 | -0.00058416 | 0.00150899 | -0.00079407 | 0.00037230 | 0.00020687 | -0.00022060 | -0.00003158 | 0.00009141 | -0.00000518 | 0.00007387 |
| 56 | 72 | 0.05961599 | 0.04836871 | -0.04681945 | -0.02084972 | 0.02841235 | 0.00211265 | -0.01236835 | -0.00062652 | 0.00144876 | -0.00062937 | 0.00041949 | 0.00021085 | -0.00026916 | -0.00003257 | 0.00003875 | 0.00006412 | 0.00007929 |
| 56 | 73 | 0.05914759 | 0.04779771 | -0.04740283 | -0.02081312 | 0.02865346 | 0.00190843 | -0.01232107 | -0.00028567 | 0.00144687 | -0.00082582 | 0.00041779 | 0.00016884 | -0.00022163 | -0.00002659 | 0.00008871 | 0.00000652 | 0.00009526 |
| 56 | 74 | 0.05935336 | 0.04752037 | -0.04769013 | -0.02018848 | 0.02894636 | 0.00182862 | -0.01219382 | -0.00030476 | 0.00140090 | -0.00069060 | 0.00046585 | 0.00015732 | -0.00024564 | -0.00003199 | 0.00005420 | 0.00005001 | 0.00010261 |
| 56 | 75 | 0.05904184 | 0.04707113 | -0.04845822 | -0.02031163 | 0.02919641 | 0.00177333 | -0.01219380 | -0.00008468 | 0.00138499 | -0.00083908 | 0.00045908 | 0.00013081 | -0.00021501 | -0.00002724 | 0.00008560 | 0.00000695 | 0.00010696 |
| 56 | 76 | 0.05919855 | 0.04674970 | -0.04873548 | -0.01965505 | 0.02964739 | 0.00173325 | -0.01201798 | -0.00006104 | 0.00137154 | -0.00073210 | 0.00050655 | 0.00010470 | -0.00021642 | -0.00003581 | 0.00006972 | 0.00002830 | 0.00011845 |
| 56 | 77 | 0.05903032 | 0.04640285 | -0.04962796 | -0.01988783 | 0.02991244 | 0.00188375 | -0.01200144 | 0.00004872 | 0.00133568 | -0.00081721 | 0.00049490 | 0.00009797 | -0.00020930 | -0.00002833 | 0.00007611 | 0.00000456 | 0.00010670 |
| 56 | 78 | 0.05913086 | 0.04603434 | -0.04991564 | -0.01925576 | 0.03046545 | 0.00185344 | -0.01180936 | 0.00012324 | 0.00136816 | -0.00073656 | 0.00053995 | 0.00006023 | -0.00019310 | -0.00003869 | 0.00007737 | 0.00001010 | 0.00012430 |
| 56 | 79 | 0.05907811 | 0.04575974 | -0.05086797 | -0.01953983 | 0.03074689 | 0.00222358 | -0.01174462 | 0.00013879 | 0.00130684 | -0.00075213 | 0.00052194 | 0.00007737 | -0.00021451 | -0.00002411 | 0.00005503 | 0.00000655 | 0.00009084 |
| 56 | 80 | 0.05912645 | 0.04534337 | -0.05120179 | -0.01898652 | 0.03136123 | 0.00218200 | -0.01157564 | 0.00026248 | 0.00139436 | -0.00069595 | 0.00056267 | 0.00003173 | -0.00018659 | -0.00003556 | 0.00007093 | 0.00000370 | 0.00011643 |
| 56 | 81 | 0.05915388 | 0.04510590 | -0.05215333 | -0.01926104 | 0.03165973 | 0.00274334 | -0.01146587 | 0.00020638 | 0.00130364 | -0.00064851 | 0.00053576 | 0.00007604 | -0.00023973 | -0.00001007 | 0.00001978 | 0.00001673 | 0.00005504 |
| 56 | 82 | 0.05916528 | 0.04464934 | -0.05257539 | -0.01883249 | 0.03231355 | 0.00268098 | -0.01135754 | 0.00036969 | 0.00145248 | -0.00061390 | 0.00057057 | 0.00002566 | -0.00020473 | -0.00002306 | 0.00004766 | 0.00001283 | 0.00009093 |
| 56 | 83 | 0.05896947 | 0.04392039 | -0.05282477 | -0.01812989 | 0.03203991 | 0.00207553 | -0.01147569 | 0.00040011 | 0.00123286 | -0.00061954 | 0.00053849 | 0.00005320 | -0.00024040 | 0.00001508 | 0.00002036 | -0.00001609 | 0.00002525 |
| 56 | 84 | 0.05872830 | 0.04302625 | -0.05242718 | -0.01673531 | 0.03195851 | 0.00087990 | -0.01140947 | 0.00071552 | 0.00125950 | -0.00062254 | 0.00057012 | -0.00002708 | -0.00016883 | 0.00001010 | 0.00007026 | -0.00005342 | 0.00007207 |
| 56 | 85 | 0.05859330 | 0.04213994 | -0.05295367 | -0.01608326 | 0.03172745 | 0.00024748 | -0.01151539 | 0.00078092 | 0.00102726 | -0.00066795 | 0.00053485 | 0.00000164 | -0.00021977 | 0.00004882 | 0.00004507 | -0.00007326 | 0.00001310 |
| 56 | 86 | 0.05838993 | 0.04135318 | -0.05241476 | -0.01474065 | 0.03150838 | -0.00069079 | -0.01122371 | 0.00108883 | 0.00104479 | -0.00058727 | 0.00055575 | -0.00006066 | -0.00016380 | 0.00005189 | 0.00006336 | -0.00008517 | 0.00004354 |
| 56 | 87 | 0.05828609 | 0.04038317 | -0.05310280 | -0.01417020 | 0.03127905 | -0.00133447 | -0.01131035 | 0.00117960 | 0.00081651 | -0.00065915 | 0.00051414 | -0.00003084 | -0.00022552 | 0.00009163 | 0.00003949 | -0.00008915 | -0.00000156 |
| 56 | 88 | 0.05811927 | 0.03979630 | -0.05241648 | -0.01301060 | 0.03101832 | -0.00197305 | -0.01093731 | 0.00143001 | 0.00087777 | -0.00049118 | 0.00052126 | -0.00006972 | -0.00018602 | 0.00009980 | 0.00003104 | -0.00008000 | 0.00000967 |
| 56 | 89 | 0.05799976 | 0.03886715 | -0.05311662 | -0.01259904 | 0.03076875 | -0.00261146 | -0.01104732 | 0.00151793 | 0.00069277 | -0.00057861 | 0.00046970 | -0.00004201 | -0.00024613 | 0.00013698 | 0.00001388 | -0.00006782 | -0.00001291 |
| 56 | 90 | 0.05789166 | 0.03849661 | -0.05235435 | -0.01164042 | 0.03059095 | -0.00297882 | -0.01073558 | 0.00169191 | 0.00082958 | -0.00034169 | 0.00046395 | -0.00005479 | -0.00022535 | 0.00014615 | -0.00001448 | -0.00004620 | -0.00002468 |
| 56 | 91 | 0.05771158 | 0.03773823 | -0.05293405 | -0.01146752 | 0.03031611 | -0.00360762 | -0.01093472 | 0.00174214 | 0.00072886 | -0.00044197 | 0.00040020 | -0.00003567 | -0.00026698 | 0.00017430 | -0.00001552 | -0.00002416 | -0.00001722 |
| 56 | 92 | 0.05744274 | 0.03749641 | -0.05201624 | -0.01104869 | 0.03000158 | -0.00322184 | -0.01042073 | 0.00168606 | 0.00081426 | -0.00014621 | 0.00040548 | -0.00004132 | -0.00026227 | 0.00016701 | -0.00007092 | -0.00000314 | -0.00005609 |
| 56 | 93 | 0.05712280 | 0.03715624 | -0.05219437 | -0.01136827 | 0.02955281 | -0.00373095 | -0.01068746 | 0.00163442 | 0.00080998 | -0.00028040 | 0.00033480 | -0.00005429 | -0.00025108 | 0.00016955 | -0.00003830 | 0.00001122 | -0.00000362 |
| 56 | 94 | 0.03771841 | 0.05528357 | -0.01456779 | -0.05126005 | 0.00370768 | 0.02934614 | 0.00051678 | -0.01350370 | -0.00165629 | 0.00288779 | -0.00035875 | -0.00064984 | 0.00093158 | 0.00008267 | -0.00011963 | -0.00028595 | 0.00008725 |
| 56 | 95 | 0.03721344 | 0.05645235 | -0.01261991 | -0.05267241 | 0.00207139 | 0.02953552 | 0.00122173 | -0.01336509 | -0.00173987 | 0.00240814 | -0.00029901 | -0.00089562 | 0.00118898 | -0.00008207 | 0.00005258 | -0.00048686 | 0.00020245 |
| 56 | 96 | 0.03877044 | 0.05086172 | -0.01942596 | -0.04403575 | 0.00655156 | 0.02535885 | -0.00181447 | -0.01247683 | -0.00139604 | 0.00325856 | -0.00036478 | -0.00012566 | 0.00026783 | 0.00017701 | -0.00046799 | 0.00022872 | -0.00004455 |
| 56 | 97 | 0.03851369 | 0.05090405 | -0.01890896 | -0.04406575 | 0.00592878 | 0.02462925 | -0.00185429 | -0.01203802 | -0.00128027 | 0.00284811 | -0.00030218 | -0.00028476 | 0.00039790 | 0.00003466 | -0.00035399 | 0.00015598 | 0.00007108 |
| 56 | 98 | 0.03911719 | 0.05023727 | -0.02015662 | -0.04255007 | 0.00737953 | 0.02414568 | -0.00277208 | -0.01212399 | -0.00097003 | 0.00316324 | -0.00041218 | -0.00011741 | 0.00025289 | 0.00011774 | -0.00044758 | 0.00025615 | -0.00002855 |
| 57 | 59 | 0.06066058 | 0.05186459 | -0.04571452 | -0.02765667 | 0.02750040 | 0.00662454 | -0.01355490 | -0.00295685 | 0.00129213 | -0.00062367 | 0.00001844 | 0.00060156 | -0.00056824 | 0.00019440 | -0.00002483 | -0.00002739 | -0.00034989 |
| 57 | 60 | 0.06098133 | 0.05174058 | -0.04617528 | -0.02684901 | 0.02690610 | 0.00536248 | -0.01367381 | -0.00256261 | 0.00124159 | -0.00066218 | 0.00014667 | 0.00056872 | -0.00060137 | 0.00012156 | -0.00006538 | 0.00001435 | -0.00029313 |
| 57 | 61 | 0.06067798 | 0.05120003 | -0.04577176 | -0.02579487 | 0.02864378 | 0.00591776 | -0.01340377 | -0.00255185 | 0.00139678 | -0.00067447 | 0.00006722 | 0.00052119 | -0.00045133 | 0.00015414 | 0.00003036 | -0.00007833 | -0.00027534 |
| 57 | 62 | 0.06105459 | 0.05111948 | -0.04631292 | -0.02516922 | 0.02791560 | 0.00469759 | -0.01351545 | -0.00218844 | 0.00134333 | -0.00069017 | 0.00019193 | 0.00049578 | -0.00049361 | 0.00007497 | -0.00002385 | -0.00002117 | -0.00022125 |
| 57 | 63 | 0.06071708 | 0.05077583 | -0.04561845 | -0.02409740 | 0.02939075 | 0.00500645 | -0.01314670 | -0.00217052 | 0.00152555 | -0.00070698 | 0.00011810 | 0.00040361 | -0.00028135 | 0.00009258 | 0.00010727 | -0.00013562 | -0.00016188 |
| 57 | 64 | 0.06116848 | 0.05073892 | -0.04627930 | -0.02365151 | 0.02858984 | 0.00386612 | -0.01327043 | -0.00183897 | 0.00147229 | -0.00070956 | 0.00023524 | 0.00038914 | -0.00033670 | 0.00000897 | 0.00004096 | -0.00006627 | -0.00011702 |
| 57 | 65 | 0.06071767 | 0.05056886 | -0.04527086 | -0.02264113 | 0.02982167 | 0.00390774 | -0.01294938 | -0.00187220 | 0.00170747 | -0.00071254 | 0.00015925 | 0.00025505 | -0.00006358 | 0.00001561 | 0.00020957 | -0.00020556 | -0.00002208 |
| 57 | 66 | 0.06126606 | 0.05056920 | -0.04607874 | -0.02234063 | 0.02903080 | 0.00290093 | -0.01307901 | -0.00156321 | 0.00165452 | -0.00071090 | 0.00026582 | 0.00025557 | -0.00013693 | -0.00006958 | 0.00013131 | -0.00012591 | 0.00000754 |
| 57 | 67 | 0.06051950 | 0.05014574 | -0.04529455 | -0.02177395 | 0.02971613 | 0.00296602 | -0.01285127 | -0.00149964 | 0.00181480 | -0.00069489 | 0.00019098 | 0.00017119 | 0.00002650 | -0.00000784 | 0.00025199 | -0.00020856 | 0.00005906 |
| 57 | 68 | 0.06081000 | 0.04966618 | -0.04637740 | -0.02153612 | 0.02880042 | 0.00235697 | -0.01270816 | -0.00092947 | 0.00163372 | -0.00071850 | 0.00031975 | 0.00023310 | -0.00018785 | -0.00003637 | 0.00010464 | -0.00005726 | 0.00003566 |
| 57 | 69 | 0.06021594 | 0.04931273 | -0.04608527 | -0.02138528 | 0.02925547 | 0.00220904 | -0.01280661 | -0.00105079 | 0.00176652 | -0.00068920 | 0.00023192 | 0.00015786 | -0.00002590 | 0.00001745 | 0.00022185 | -0.00014166 | 0.00008187 |
| 57 | 70 | 0.06042167 | 0.04876986 | -0.04698996 | -0.02088532 | 0.02870328 | 0.00179952 | -0.01250651 | -0.00047159 | 0.00157857 | -0.00073742 | 0.00037140 | 0.00020046 | -0.00021479 | -0.00001616 | 0.00008959 | -0.00001026 | 0.00006734 |
| 57 | 71 | 0.05999968 | 0.04844529 | -0.04720110 | -0.02105203 | 0.02899242 | 0.00149727 | -0.01283132 | -0.00071089 | 0.00169127 | -0.00069921 | 0.00027600 | 0.00013427 | -0.00006329 | 0.00002787 | 0.00019762 | -0.00008944 | 0.00010939 |
| 57 | 72 | 0.06014576 | 0.04791712 | -0.04788983 | -0.02033530 | 0.02883767 | 0.00133177 | -0.01240951 | -0.00015649 | 0.00150600 | -0.00076485 | 0.00041941 | 0.00016029 | -0.00022186 | -0.00000805 | 0.00008467 | 0.00001391 | 0.00009614 |
| 57 | 73 | 0.05990421 | 0.04759308 | -0.04855517 | -0.02067205 | 0.02907187 | 0.00097892 | -0.01283891 | -0.00045269 | 0.00159971 | -0.00071809 | 0.00032104 | 0.00010309 | -0.00008663 | 0.00002542 | 0.00017662 | -0.00005336 | 0.00013393 |
| 57 | 74 | 0.05999359 | 0.04712924 | -0.04900500 | -0.01984143 | 0.02924336 | 0.00105243 | -0.01232593 | 0.00005758 | 0.00143323 | -0.00078777 | 0.00046350 | 0.00011593 | -0.00021498 | -0.00000929 | 0.00008499 | 0.00001991 | 0.00011789 |
| 57 | 75 | 0.05992848 | 0.04678531 | -0.05004827 | -0.02021236 | 0.02951420 | 0.00074044 | -0.01276270 | -0.00025635 | 0.00150137 | -0.00072818 | 0.00036451 | 0.00006963 | -0.00010051 | 0.00001540 | 0.00015392 | -0.00003047 | 0.00014859 |
| 57 | 76 | 0.05994922 | 0.04640394 | -0.05025829 | -0.01938937 | 0.02989164 | 0.00101988 | -0.01218455 | 0.00020471 | 0.00137266 | -0.00078764 | 0.00050269 | 0.00007305 | -0.00020295 | -0.00001469 | 0.00008326 | 0.00001630 | 0.00012904 |
| 57 | 77 | 0.06004306 | 0.04602527 | -0.05159074 | -0.01969207 | 0.03024555 | 0.00079551 | -0.01257618 | -0.00010685 | 0.00140506 | -0.00071106 | 0.00040309 | 0.00004163 | -0.00011330 | 0.00000510 | 0.00012400 | -0.00001490 | 0.00014759 |
| 57 | 78 | 0.05998130 | 0.04571919 | -0.05158659 | -0.01898469 | 0.03071551 | 0.00124252 | -0.01196012 | 0.00030927 | 0.00133164 | -0.00074958 | 0.00053425 | 0.00003976 | -0.00019716 | -0.00001763 | 0.00007193 | 0.00001242 | 0.00012530 |
| 57 | 79 | 0.06020457 | 0.04529630 | -0.05311283 | -0.01914627 | 0.03116041 | 0.00110099 | -0.01229751 | 0.00001135 | 0.00131961 | -0.00065759 | 0.00043255 | 0.00002782 | -0.00013534 | 0.00000171 | 0.00008273 | -0.00000071 | 0.00012616 |
| 57 | 80 | 0.06005522 | 0.04504516 | -0.05294564 | -0.01863454 | 0.03164751 | 0.00168451 | -0.01167752 | 0.00039014 | 0.00131467 | -0.00067034 | 0.00055397 | 0.00002486 | -0.00020907 | -0.00001208 | 0.00004574 | 0.00001526 | 0.00010179 |
| 57 | 81 | 0.06037106 | 0.04457350 | -0.05456990 | -0.01860254 | 0.03216909 | 0.00158136 | -0.01197973 | 0.00011557 | 0.00125357 | -0.00057170 | 0.00044848 | 0.00003564 | -0.00017592 | 0.00001042 | 0.00002878 | 0.00001582 | 0.00008076 |
| 57 | 82 | 0.06014220 | 0.04435649 | -0.05430877 | -0.01833969 | 0.03264004 | 0.00228153 | -0.01139502 | 0.00046324 | 0.00132509 | -0.00055942 | 0.00055771 | 0.00003487 | -0.00024682 | 0.00000550 | 0.00000331 | 0.00002764 | 0.00005465 |
| 57 | 83 | 0.06030976 | 0.04333900 | -0.05544129 | -0.01730200 | 0.03265030 | 0.00095472 | -0.01190466 | 0.00033887 | 0.00111997 | -0.00057584 | 0.00045493 | 0.00001952 | -0.00019506 | 0.00003337 | 0.00001736 | -0.00000867 | 0.00004639 |
| 57 | 84 | 0.05978161 | 0.04255226 | -0.05453529 | -0.01624902 | 0.03248911 | 0.00044819 | -0.01153987 | 0.00077406 | 0.00110834 | -0.00060462 | 0.00055907 | -0.00002466 | -0.00021628 | 0.00003934 | 0.00003457 | -0.00003596 | 0.00004744 |

TABLE 3. Nuclear charge density distribution FB coefficients.

| $Z$ | $N$ | $a_1$ | $a_2$ | $a_3$ | $a_4$ | $a_5$ | $a_6$ | $a_7$ | $a_8$ | $a_9$ | $a_{10}$ | $a_{11}$ | $a_{12}$ | $a_{13}$ | $a_{14}$ | $a_{15}$ | $a_{16}$ | $a_{17}$ |
|---|---|---|---|---|---|---|---|---|---|---|---|---|---|---|---|---|---|---|
| 57 | 85 | 0.06004623 | 0.04161794 | -0.05571466 | -0.01529276 | 0.03242542 | -0.00072721 | -0.01188959 | 0.00070933 | 0.00088696 | -0.00065705 | 0.00045192 | -0.00002464 | -0.00019472 | 0.00006827 | 0.00003700 | -0.00004923 | 0.00003953 |
| 57 | 86 | 0.05949455 | 0.04072892 | -0.05476837 | -0.01421878 | 0.03217360 | -0.00116463 | -0.01140863 | 0.00112916 | 0.00085414 | -0.00061193 | 0.00054554 | -0.00006180 | -0.00021954 | 0.00008407 | 0.00003254 | -0.00006023 | 0.00003112 |
| 57 | 87 | 0.05974670 | 0.04009049 | -0.05575290 | -0.01356723 | 0.03203244 | -0.00216268 | -0.01176148 | 0.00105796 | 0.00071531 | -0.00067260 | 0.00042551 | -0.00004557 | -0.00021209 | 0.00011086 | 0.00003483 | -0.00005269 | 0.00003393 |
| 57 | 88 | 0.05923015 | 0.03910775 | -0.05483901 | -0.01247969 | 0.03174210 | -0.00248278 | -0.01116464 | 0.00145616 | 0.00065798 | -0.00055622 | 0.00050849 | -0.00007173 | -0.00024744 | 0.00013536 | 0.00000448 | -0.00004534 | 0.00001117 |
| 57 | 89 | 0.05938678 | 0.03889745 | -0.05550837 | -0.01226340 | 0.03157688 | -0.00332886 | -0.01168927 | 0.00132584 | 0.00068518 | -0.00061411 | 0.00037314 | -0.00004494 | -0.00023451 | 0.00015217 | 0.00002156 | -0.00002708 | 0.00003346 |
| 57 | 90 | 0.05895968 | 0.03785120 | -0.05468111 | -0.01116592 | 0.03131864 | -0.00351348 | -0.01102059 | 0.00169601 | 0.00060864 | -0.00044179 | 0.00044453 | -0.00005776 | -0.00028409 | 0.00018296 | -0.00003352 | -0.00000438 | -0.00000733 |
| 57 | 91 | 0.05715416 | 0.04019045 | -0.05125796 | -0.01593907 | 0.02787364 | -0.00038853 | -0.00950774 | 0.00016110 | 0.00019209 | -0.00056464 | 0.00041786 | -0.00029907 | 0.00002933 | 0.00000892 | 0.00006982 | -0.00012387 | 0.00015966 |
| 57 | 92 | 0.05842420 | 0.03705750 | -0.05405335 | -0.01073640 | 0.03064610 | -0.00374572 | -0.01074307 | 0.00165672 | 0.00061337 | -0.00029037 | 0.00038464 | -0.00005947 | -0.00029587 | 0.00019251 | -0.00007233 | 0.00003634 | -0.00001212 |
| 57 | 93 | 0.03760142 | 0.05786869 | -0.01186029 | -0.05373042 | 0.00097119 | 0.02968991 | 0.00203509 | -0.01389108 | -0.00228245 | 0.00232003 | -0.00017284 | -0.00089601 | 0.00115811 | -0.00011596 | 0.00003039 | -0.00051730 | 0.00020067 |
| 57 | 94 | 0.03833544 | 0.05619349 | -0.01375734 | -0.05106205 | 0.00300741 | 0.02907142 | 0.00138149 | -0.01293766 | -0.00195312 | 0.00234763 | -0.00018956 | -0.00085515 | 0.00108835 | -0.00005812 | 0.00001400 | -0.00047110 | 0.00016942 |
| 57 | 95 | 0.03935258 | 0.05174201 | -0.01895672 | -0.04419080 | 0.00568734 | 0.02499912 | -0.00109228 | -0.01173141 | -0.00165405 | 0.00251828 | -0.00011803 | -0.00035633 | 0.00039322 | -0.00001269 | -0.00031672 | 0.00005526 | 0.00007924 |
| 57 | 96 | 0.03936092 | 0.05124515 | -0.01901081 | -0.04328945 | 0.00643192 | 0.02466864 | -0.00137475 | -0.01158875 | -0.00145523 | 0.00272653 | -0.00020848 | -0.00029970 | 0.00037229 | 0.00005521 | -0.00034086 | 0.00008417 | 0.00003207 |
| 57 | 97 | 0.03953943 | 0.05103414 | -0.01953981 | -0.04248728 | 0.00643237 | 0.02356258 | -0.00205705 | -0.01131261 | -0.00123735 | 0.00240477 | -0.00013262 | -0.00037824 | 0.00040492 | -0.00008576 | -0.00028710 | 0.00009349 | 0.00012731 |
| 57 | 98 | 0.03962872 | 0.05068780 | -0.01963052 | -0.04181017 | 0.00728451 | 0.02338353 | -0.00242061 | -0.01124852 | -0.00099543 | 0.00262069 | -0.00025176 | -0.00031254 | 0.00038758 | -0.00001475 | -0.00030599 | 0.00010580 | 0.00006566 |
| 57 | 99 | 0.03950747 | 0.05088947 | -0.01950525 | -0.04162577 | 0.00685665 | 0.02254625 | -0.00292327 | -0.01112373 | -0.00077025 | 0.00228897 | -0.00015738 | -0.00045628 | 0.00049935 | -0.00017912 | -0.00021425 | 0.00008354 | 0.00020287 |
| 57 | 100 | 0.03973560 | 0.05054203 | -0.01986033 | -0.04098534 | 0.00792869 | 0.02238536 | -0.00344058 | -0.01108088 | -0.00047306 | 0.00251255 | -0.00030238 | -0.00036776 | 0.00046517 | -0.00010278 | -0.00023687 | 0.00009474 | 0.00012545 |
| 58 | 61 | 0.06190060 | 0.05122395 | -0.04680882 | -0.02511995 | 0.02773135 | 0.00432442 | -0.01355202 | -0.00187686 | 0.00136723 | -0.00067852 | 0.00019410 | 0.00053689 | -0.00059315 | 0.00012671 | -0.00006065 | 0.00000743 | -0.00028082 |
| 58 | 62 | 0.06260686 | 0.05153045 | -0.04705763 | -0.02456389 | 0.02774987 | 0.00362144 | -0.01386302 | -0.00215896 | 0.00136242 | -0.00048430 | 0.00022962 | 0.00051064 | -0.00054288 | 0.00007701 | -0.00008195 | 0.00006429 | -0.00020910 |
| 58 | 63 | 0.06198557 | 0.05072955 | -0.04678575 | -0.02340981 | 0.02864171 | 0.00360728 | -0.01326839 | -0.00152588 | 0.00147348 | -0.00069240 | 0.00024018 | 0.00043315 | -0.00043983 | 0.00006745 | -0.00000156 | -0.00003776 | -0.00017935 |
| 58 | 64 | 0.06276759 | 0.05112908 | -0.04705739 | -0.02300183 | 0.02848026 | 0.00290824 | -0.01355900 | -0.00180454 | 0.00148927 | -0.00048762 | 0.00026570 | 0.00041453 | -0.00040239 | 0.00001888 | -0.00002637 | 0.00002164 | -0.00011954 |
| 58 | 65 | 0.06206023 | 0.05045111 | -0.04660479 | -0.02190446 | 0.02927766 | 0.00271663 | -0.01301679 | -0.00124447 | 0.00162656 | -0.00068871 | 0.00027612 | 0.00029647 | -0.00023602 | -0.00000647 | 0.00008678 | -0.00010142 | -0.00005312 |
| 58 | 66 | 0.06288296 | 0.05090304 | -0.04691648 | -0.02162963 | 0.02910162 | 0.00210699 | -0.01335399 | -0.00152581 | 0.00168381 | -0.00047471 | 0.00028619 | 0.00029338 | -0.00021981 | -0.00004890 | 0.00005824 | -0.00004450 | -0.00001827 |
| 58 | 67 | 0.06186910 | 0.04993175 | -0.04662725 | -0.02087264 | 0.02935393 | 0.00196315 | -0.01275009 | -0.00083238 | 0.00170641 | -0.00067269 | 0.00031154 | 0.00021350 | -0.00015300 | -0.00002570 | 0.00012079 | -0.00009922 | 0.00002365 |
| 58 | 68 | 0.06240197 | 0.04990939 | -0.04720278 | -0.02074997 | 0.02892173 | 0.00172095 | -0.01283169 | -0.00091432 | 0.00161017 | -0.00048443 | 0.00034496 | 0.00027304 | -0.00027908 | -0.00001488 | 0.00001909 | 0.00003578 | 0.00000863 |
| 58 | 69 | 0.06148547 | 0.04901254 | -0.04722591 | -0.02025125 | 0.02907256 | 0.00138674 | -0.01249681 | -0.00034368 | 0.00163466 | -0.00067260 | 0.00036151 | 0.00019080 | -0.00020038 | 0.00000473 | 0.00009062 | -0.00003180 | 0.00004937 |
| 58 | 70 | 0.06196424 | 0.04891564 | -0.04779490 | -0.01999158 | 0.02891786 | 0.00129024 | -0.01249344 | -0.00045666 | 0.00151707 | -0.00051515 | 0.00040166 | 0.00023625 | -0.00030477 | 0.00000422 | -0.00000042 | 0.00008775 | 0.00004276 |
| 58 | 71 | 0.06118717 | 0.04810959 | -0.04815249 | -0.01975061 | 0.02899535 | 0.00083283 | -0.01239537 | -0.00001124 | 0.00154632 | -0.00068669 | 0.00040932 | 0.00015620 | -0.00022325 | 0.00002043 | 0.00007566 | 0.00001097 | 0.00007853 |
| 58 | 72 | 0.06161000 | 0.04795716 | -0.04868253 | -0.01933994 | 0.02915063 | 0.00087669 | -0.01230609 | -0.00012200 | 0.00142749 | -0.00056560 | 0.00045409 | 0.00018563 | -0.00030081 | 0.00000899 | -0.00000019 | 0.00010903 | 0.00007849 |
| 58 | 73 | 0.06099817 | 0.04725746 | -0.04932853 | -0.01929251 | 0.02921810 | 0.00041835 | -0.01234896 | 0.00021363 | 0.00146057 | -0.00070932 | 0.00045451 | 0.00011143 | -0.00022423 | 0.00002229 | 0.00007317 | 0.00002969 | 0.00010658 |
| 58 | 74 | 0.06135523 | 0.04705666 | -0.04981086 | -0.01877596 | 0.02964024 | 0.00056131 | -0.01219786 | 0.00012794 | 0.00136109 | -0.00062401 | 0.00050135 | 0.00012569 | -0.00027480 | 0.00000203 | 0.00001471 | 0.00010469 | 0.00011168 |
| 58 | 75 | 0.06091611 | 0.04646807 | -0.05066656 | -0.01884318 | 0.02974718 | 0.00022857 | -0.01226776 | 0.00037076 | 0.00138988 | -0.00072428 | 0.00049632 | 0.00006165 | -0.00021071 | 0.00001460 | 0.00007599 | 0.00003089 | 0.00012878 |
| 58 | 76 | 0.06119481 | 0.04621784 | -0.05111232 | -0.01829644 | 0.03035593 | 0.00041628 | -0.01209649 | 0.00032154 | 0.00132921 | -0.00067004 | 0.00054232 | 0.00006397 | -0.00023841 | -0.00001155 | 0.00003479 | 0.00008586 | 0.00013852 |
| 58 | 77 | 0.06091910 | 0.04573161 | -0.05209009 | -0.01840890 | 0.03051890 | 0.00030051 | -0.01210034 | 0.00048603 | 0.00134009 | -0.00071249 | 0.00053245 | 0.00001582 | -0.00019484 | 0.00000445 | 0.00007463 | 0.00002468 | 0.00013952 |
| 58 | 78 | 0.06110983 | 0.04542790 | -0.05252481 | -0.01791135 | 0.03123383 | 0.00048560 | -0.01195849 | 0.00047411 | 0.00133537 | -0.00068249 | 0.00057462 | 0.00001081 | -0.00020627 | -0.00002487 | 0.00004869 | 0.00006627 | 0.00015454 |
| 58 | 79 | 0.06097553 | 0.04502666 | -0.05353795 | -0.01800996 | 0.03144488 | 0.00062195 | -0.01184693 | 0.00057649 | 0.00131381 | -0.00066233 | 0.00055890 | -0.00001516 | -0.00019074 | -0.00000054 | 0.00006054 | 0.00002099 | 0.00013263 |
| 58 | 80 | 0.06107603 | 0.04466701 | -0.05399763 | -0.01762901 | 0.03220636 | 0.00077362 | -0.01178367 | 0.00059356 | 0.00137909 | -0.00064856 | 0.00059446 | -0.00002277 | -0.00019286 | -0.00003131 | 0.00004675 | 0.00005787 | 0.00015474 |
| 58 | 81 | 0.06105383 | 0.04433128 | -0.05496647 | -0.01766067 | 0.03244910 | 0.00114032 | -0.01155328 | 0.00065580 | 0.00131326 | -0.00057544 | 0.00057108 | -0.00002181 | -0.00021030 | 0.00000531 | 0.00002903 | 0.00002623 | 0.00010278 |
| 58 | 82 | 0.06107147 | 0.04391831 | -0.05549295 | -0.01744577 | 0.03322289 | 0.00124542 | -0.01161026 | 0.00068683 | 0.00145928 | -0.00056855 | 0.00059765 | -0.00002806 | -0.00020880 | -0.00002638 | 0.00002380 | 0.00006758 | 0.00013475 |
| 58 | 83 | 0.06093830 | 0.04305993 | -0.05583618 | -0.01643095 | 0.03294550 | 0.00055051 | -0.01149296 | 0.00083145 | 0.00120400 | -0.00053912 | 0.00057229 | -0.00004574 | -0.00021538 | 0.00003179 | 0.00002157 | 0.00000169 | 0.00007508 |
| 58 | 84 | 0.06075950 | 0.04218216 | -0.05560344 | -0.01534037 | 0.03295841 | -0.00038183 | -0.01158479 | 0.00093956 | 0.00117257 | -0.00054935 | 0.00059314 | -0.00007661 | -0.00018825 | 0.00001895 | 0.00002813 | 0.00001872 | 0.00010898 |
| 58 | 85 | 0.06067842 | 0.04119618 | -0.05615327 | -0.01429512 | 0.03277764 | -0.00109083 | -0.01144161 | 0.00114609 | 0.00092428 | -0.00055574 | 0.00056573 | -0.00008963 | -0.00021252 | 0.00007768 | 0.00002552 | -0.00003248 | 0.00005884 |
| 58 | 86 | 0.06050277 | 0.04049645 | -0.05568087 | -0.01337815 | 0.03260750 | -0.00174883 | -0.01138729 | 0.00120517 | 0.00089158 | -0.00048748 | 0.00057102 | -0.00009565 | -0.00020440 | 0.00007429 | 0.00000219 | 0.00000388 | 0.00006859 |
| 58 | 87 | 0.06044073 | 0.03949558 | -0.05628582 | -0.01239257 | 0.03245632 | -0.00245094 | -0.01123454 | 0.00145935 | 0.00067106 | -0.00052394 | 0.00053747 | -0.00010170 | -0.00024251 | 0.00013259 | -0.00000104 | -0.00002708 | 0.00003291 |
| 58 | 88 | 0.06027696 | 0.03901080 | -0.05564491 | -0.01170109 | 0.03226336 | -0.00283500 | -0.01118784 | 0.00143960 | 0.00070012 | -0.00037568 | 0.00052568 | -0.00008106 | -0.00025144 | 0.00013384 | -0.00004613 | 0.00002076 | 0.00001749 |
| 58 | 89 | 0.06019216 | 0.03812690 | -0.05616556 | -0.01089350 | 0.03209794 | -0.00351480 | -0.01107803 | 0.00170882 | 0.00054295 | -0.00043568 | 0.00048177 | -0.00008297 | -0.00029107 | 0.00018776 | -0.00004443 | 0.00000852 | 0.00000245 |
| 58 | 90 | 0.05989316 | 0.03760269 | -0.05552517 | -0.01043325 | 0.03188642 | -0.00329152 | -0.01084253 | 0.00151604 | 0.00053228 | -0.00019804 | 0.00048433 | -0.00004599 | -0.00032749 | 0.00017633 | -0.00012541 | 0.00006479 | -0.00004275 |
| 58 | 91 | 0.05971459 | 0.03711214 | -0.05567818 | -0.01011202 | 0.03155068 | -0.00385233 | -0.01078198 | 0.00173241 | 0.00046651 | -0.00029971 | 0.00043138 | -0.00006783 | -0.00033136 | 0.00020985 | -0.00009924 | 0.00005483 | -0.00001983 |
| 58 | 92 | 0.03970645 | 0.05532122 | -0.01629248 | -0.04864230 | 0.00528080 | 0.02837676 | 0.00052489 | -0.01287232 | -0.00192079 | 0.00277783 | -0.00015426 | -0.00058881 | 0.00079765 | 0.00009289 | -0.00016339 | -0.00025410 | 0.00006695 |
| 58 | 93 | 0.03944599 | 0.05605761 | -0.01481780 | -0.04951043 | 0.00384758 | 0.02855019 | 0.00144455 | -0.01254459 | -0.00209816 | 0.00231659 | -0.00009039 | -0.00080639 | 0.00098694 | -0.00003617 | -0.00002715 | -0.00043862 | 0.00014738 |
| 58 | 94 | 0.04015583 | 0.05194022 | -0.01924855 | -0.04316962 | 0.00732096 | 0.02546861 | -0.00120174 | -0.01192034 | -0.00161005 | 0.00304012 | -0.00019092 | -0.00017259 | 0.00027890 | 0.00017845 | -0.00041592 | 0.00010236 | -0.00007180 |
| 58 | 95 | 0.04019273 | 0.05169673 | -0.01912653 | -0.04268422 | 0.00685881 | 0.02472989 | -0.00101223 | -0.01126476 | -0.00158563 | 0.00263934 | -0.00012532 | -0.00030799 | 0.00034939 | 0.00006925 | -0.00033402 | 0.00003631 | 0.00000902 |
| 58 | 96 | 0.04034863 | 0.05128701 | -0.01969490 | -0.04178365 | 0.00815027 | 0.02439035 | -0.00209101 | -0.01155544 | -0.00120468 | 0.00294213 | -0.00024901 | -0.00016024 | 0.00026552 | 0.00013211 | -0.00039149 | 0.00010542 | -0.00008220 |
| 58 | 97 | 0.04040165 | 0.05113628 | -0.01961867 | -0.04128329 | 0.00769833 | 0.02353003 | -0.00200953 | -0.01091594 | -0.00115030 | 0.00252341 | -0.00017211 | -0.00031933 | 0.00036477 | 0.00000475 | -0.00029827 | 0.00004547 | 0.00002870 |
| 58 | 98 | 0.04042867 | 0.05091359 | -0.01995063 | -0.04084568 | 0.00888277 | 0.02342033 | -0.00308071 | -0.01132113 | -0.00071552 | 0.00283837 | -0.00031339 | -0.00017644 | 0.00029614 | 0.00006681 | -0.00034028 | 0.00009017 | -0.00006840 |
| 58 | 99 | 0.04046960 | 0.05093111 | -0.01979627 | -0.04042387 | 0.00837335 | 0.02252943 | -0.00302883 | -0.01071959 | -0.00064387 | 0.00241055 | -0.00022629 | -0.00036625 | 0.00043443 | -0.00007614 | -0.00023150 | 0.00002730 | 0.00007207 |
| 58 | 100 | 0.04046617 | 0.05060343 | -0.02025152 | -0.04002475 | 0.00958643 | 0.02236457 | -0.00420185 | -0.01113771 | -0.00016453 | 0.00274232 | -0.00038064 | -0.00019518 | 0.00033523 | -0.00000607 | -0.00028212 | 0.00008258 | -0.00003756 |
| 58 | 101 | 0.04049601 | 0.05077746 | -0.02000076 | -0.03966192 | 0.00902290 | 0.02146646 | -0.00413766 | -0.01056088 | -0.00008174 | 0.00231454 | -0.00028365 | -0.00041517 | 0.00051145 | -0.00015965 | -0.00015799 | 0.00001356 | 0.00012852 |
| 59 | 62 | 0.06280734 | 0.05086455 | -0.04704402 | -0.02319374 | 0.02844854 | 0.00334051 | -0.01318095 | -0.00118593 | 0.00146188 | -0.00067466 | 0.00023892 | 0.00047717 | -0.00054436 | 0.00012731 | -0.00004531 | -0.00000842 | -0.00024795 |
| 59 | 63 | 0.06221344 | 0.05051434 | -0.04582642 | -0.02174963 | 0.03002930 | 0.00376872 | -0.01269378 | -0.00122747 | 0.00163399 | -0.00060770 | 0.00015723 | 0.00035200 | -0.00028117 | 0.00015318 | 0.00009691 | -0.00013375 | -0.00016696 |
| 59 | 64 | 0.06286156 | 0.05046092 | -0.04688780 | -0.02149282 | 0.02926293 | 0.00252639 | -0.01286003 | -0.00089532 | 0.00158005 | -0.00066379 | 0.00028159 | 0.00033910 | -0.00034103 | 0.00005963 | 0.00003816 | -0.00007360 | -0.00012204 |
| 59 | 65 | 0.06211704 | 0.05023105 | -0.04551501 | -0.02037756 | 0.03044416 | 0.00260628 | -0.01260826 | -0.00108347 | 0.00179072 | -0.00057551 | 0.00018933 | 0.00018629 | -0.00003477 | 0.00008422 | 0.00021588 | -0.00022921 | -0.00002506 |
| 59 | 66 | 0.06285930 | 0.05024466 | -0.04662176 | -0.02009904 | 0.02986138 | 0.00154909 | -0.01272818 | -0.00073139 | 0.00177053 | -0.00062413 | 0.00030293 | 0.00017692 | -0.00009537 | -0.00001575 | 0.00015252 | -0.00016274 | 0.00001527 |
| 59 | 67 | 0.06193659 | 0.04972150 | -0.04577383 | -0.01970727 | 0.03020275 | 0.00158017 | -0.01264528 | -0.00084652 | 0.00185286 | -0.00052232 | 0.00020908 | 0.00010706 | 0.00005046 | 0.00007449 | 0.00025415 | -0.00023852 | 0.00004096 |
| 59 | 68 | 0.06252248 | 0.04930469 | -0.04722361 | -0.01948479 | 0.02942642 | 0.00098447 | -0.01240638 | -0.00020419 | 0.00167233 | -0.00060521 | 0.00034841 | 0.00017013 | -0.00016951 | 0.00002817 | 0.00010181 | -0.00007514 | 0.00002636 |
| 59 | 69 | 0.06177375 | 0.04877920 | -0.04696813 | -0.01948283 | 0.02954279 | 0.00074208 | -0.01265184 | -0.00047087 | 0.00173621 | -0.00048849 | 0.00024265 | 0.00011048 | -0.00003230 | 0.00011398 | 0.00020154 | -0.00015603 | 0.00004285 |
| 59 | 70 | 0.06224011 | 0.04835224 | -0.04818718 | -0.01902590 | 0.02914329 | 0.00038089 | -0.01227676 | 0.00015277 | 0.00155871 | -0.00060110 | 0.00039326 | 0.00014913 | -0.00021776 | 0.00005594 | 0.00006887 | -0.00001022 | 0.00004666 |
| 59 | 71 | 0.06166501 | 0.04780238 | -0.04845870 | -0.01922741 | 0.02917940 | -0.00002291 | -0.01269431 | -0.00017560 | 0.00161705 | -0.00048147 | 0.00028328 | 0.00009414 | -0.00008968 | 0.00013136 | 0.00016391 | -0.00009117 | 0.00006053 |
| 59 | 72 | 0.06203957 | 0.04743499 | -0.04942491 | -0.01861367 | 0.02915795 | -0.00014105 | -0.01224726 | 0.00039245 | 0.00145351 | -0.00061452 | 0.00043716 | 0.00011228 | -0.00023723 | 0.00006566 | 0.00005508 | 0.00002774 | 0.00007345 |
| 59 | 73 | 0.06163359 | 0.04684663 | -0.05014387 | -0.01885089 | 0.02924492 | -0.00058601 | -0.01269630 | 0.00006422 | 0.00150682 | -0.00049580 | 0.00032875 | 0.00005863 | -0.00011822 | 0.00012740 | 0.00014020 | -0.00004781 | 0.00008830 |
| 59 | 74 | 0.06192809 | 0.04657862 | -0.05084159 | -0.01818915 | 0.02952099 | -0.00048245 | -0.01222032 | 0.00056059 | 0.00137080 | -0.00063545 | 0.00047980 | 0.00006234 | -0.00023116 | 0.00005956 | 0.00005578 | 0.00004105 | 0.00010233 |
| 59 | 75 | 0.06168408 | 0.04594420 | -0.05192821 | -0.01834970 | 0.02973362 | -0.00087228 | -0.01260273 | 0.00026118 | 0.00140855 | -0.00051340 | 0.00037564 | 0.00001151 | -0.00012395 | 0.00010852 | 0.00012366 | -0.00002358 | 0.00011686 |
| 59 | 76 | 0.06189552 | 0.04578544 | -0.05235368 | -0.01775052 | 0.03018906 | -0.00058276 | -0.01212393 | 0.00068607 | 0.00131494 | -0.00064493 | 0.00051947 | 0.00000772 | -0.00021035 | 0.00004415 | 0.00006122 | 0.00003823 | 0.00012676 |
| 59 | 77 | 0.06179944 | 0.04510537 | -0.05371935 | -0.01777159 | 0.03054307 | -0.00086328 | -0.01239103 | 0.00042335 | 0.00132275 | -0.00051350 | 0.00041923 | -0.00003441 | -0.00012067 | 0.00008457 | 0.00010469 | -0.00001099 | 0.00013565 |
| 59 | 78 | 0.06191891 | 0.04504292 | -0.05388940 | -0.01732587 | 0.03106689 | -0.00042473 | -0.01193023 | 0.00078511 | 0.00128526 | -0.00062482 | 0.00055251 | -0.00003907 | -0.00019077 | 0.00002822 | 0.00005996 | 0.00003088 | 0.00013886 |
| 59 | 79 | 0.06194722 | 0.04432528 | -0.05543342 | -0.01717327 | 0.03153863 | -0.00059360 | -0.01208053 | 0.00056095 | 0.00125201 | -0.00048375 | 0.00045389 | -0.00006511 | -0.00012480 | 0.00006542 | 0.00007502 | -0.00000150 | 0.00013526 |
| 59 | 80 | 0.06197005 | 0.04433410 | -0.05539045 | -0.01694325 | 0.03205474 | -0.00003611 | -0.01166173 | 0.00086886 | 0.00128089 | -0.00056747 | 0.00057397 | -0.00006521 | -0.00018855 | 0.00001981 | 0.00004306 | 0.00002915 | 0.00013106 |
| 59 | 81 | 0.06209190 | 0.04359009 | -0.05700967 | -0.01659428 | 0.03260526 | -0.00013672 | -0.01172697 | 0.00068771 | 0.00120142 | -0.00042532 | 0.00047428 | -0.00006955 | -0.00014995 | 0.00005821 | 0.00003057 | 0.00001086 | 0.00010915 |
| 59 | 82 | 0.06202363 | 0.04364393 | -0.05681806 | -0.01661582 | 0.03307829 | 0.00052050 | -0.01137751 | 0.00094752 | 0.00130236 | -0.00047853 | 0.00057923 | -0.00006159 | -0.00021506 | 0.00002393 | 0.00000663 | 0.00003873 | 0.00009809 |
| 59 | 83 | 0.06206617 | 0.04235271 | -0.05794862 | -0.01523052 | 0.03314847 | -0.00069795 | -0.01158301 | 0.00091942 | 0.00104762 | -0.00043320 | 0.00048217 | -0.00008466 | -0.00017231 | 0.00007949 | 0.00001372 | -0.00000515 | 0.00008022 |
| 59 | 84 | 0.06179985 | 0.04177056 | -0.05720055 | -0.01439672 | 0.03303603 | -0.00115146 | -0.01140258 | 0.00123029 | 0.00100883 | -0.00049987 | 0.00057856 | -0.00011260 | -0.00020517 | 0.00007128 | 0.00001764 | -0.00000648 | 0.00008128 |
| 59 | 85 | 0.06187085 | 0.04071186 | -0.05816256 | -0.01321332 | 0.03299761 | -0.00220817 | -0.01154196 | 0.00126517 | 0.00078198 | -0.00049891 | 0.00047453 | -0.00010904 | -0.00019543 | 0.00012613 | 0.00001475 | -0.00002597 | 0.00006159 |
| 59 | 86 | 0.06159401 | 0.04002392 | -0.05738210 | -0.01234379 | 0.03283120 | -0.00255894 | -0.01124553 | 0.00153357 | 0.00071445 | -0.00049041 | 0.00055902 | -0.00012902 | -0.00023343 | 0.00012848 | -0.00000402 | -0.00001364 | 0.00005030 |
| 59 | 87 | 0.06162460 | 0.03932997 | -0.05807936 | -0.01154295 | 0.03272134 | -0.00346403 | -0.01150229 | 0.00157424 | 0.00060838 | -0.00050707 | 0.00043947 | -0.00009714 | -0.00024519 | 0.00017956 | -0.00000785 | -0.00001080 | 0.00003448 |
| 59 | 88 | 0.06137068 | 0.03857050 | -0.05729710 | -0.01066109 | 0.03256366 | -0.00366383 | -0.01110040 | 0.00179254 | 0.00052089 | -0.00043145 | 0.00051261 | -0.00010839 | -0.00028920 | 0.00018872 | -0.00004853 | 0.00001263 | 0.00000923 |
| 59 | 89 | 0.05957329 | 0.04023840 | -0.05412674 | -0.01433460 | 0.02937103 | -0.00082428 | -0.00933808 | 0.00055658 | -0.00002791 | -0.00049428 | 0.00051679 | -0.00032167 | -0.00003326 | 0.00004969 | 0.00001155 | -0.00008868 | 0.00014108 |
| 59 | 90 | 0.06095401 | 0.03733875 | -0.05696600 | -0.00950277 | 0.03217779 | -0.00411852 | -0.01081238 | 0.00187849 | 0.00036875 | -0.00031578 | 0.00047401 | -0.00007681 | -0.00035702 | 0.00022130 | -0.00011739 | 0.00006268 | -0.00003012 |
| 59 | 91 | 0.03967092 | 0.05757405 | -0.01387119 | -0.05068965 | 0.00268839 | 0.02964205 | 0.00202247 | -0.01305602 | -0.00236518 | 0.00237013 | -0.00000769 | -0.00078789 | 0.00095140 | -0.00006546 | -0.00005933 | -0.00043629 | 0.00015241 |
| 59 | 92 | 0.04046991 | 0.05610105 | -0.01569745 | -0.04810970 | 0.00457444 | 0.02795456 | 0.00132556 | -0.01221544 | -0.00213524 | 0.00232515 | -0.00001050 | -0.00074898 | 0.00089398 | -0.00001916 | -0.00006517 | -0.00039463 | 0.00013421 |
| 59 | 93 | 0.04099042 | 0.05258946 | -0.01936209 | -0.04300802 | 0.00654701 | 0.02511608 | -0.00055751 | -0.01116800 | -0.00178753 | 0.00249770 | 0.00000820 | -0.00033621 | 0.00031697 | 0.00002869 | -0.00033972 | 0.00001582 | 0.00002867 |
| 59 | 94 | 0.04093484 | 0.05229694 | -0.01918225 | -0.04231856 | 0.00720622 | 0.02476985 | -0.00087390 | -0.01109098 | -0.00162130 | 0.00260144 | -0.00006287 | -0.00030580 | 0.00033525 | 0.00007449 | -0.00032909 | 0.00001000 | -0.00000079 |
| 59 | 95 | 0.04111139 | 0.05190639 | -0.01976204 | -0.04155914 | 0.00724286 | 0.02395341 | -0.00142418 | -0.01079025 | -0.00145445 | 0.00236204 | -0.00001611 | -0.00034628 | 0.00031704 | -0.00003160 | -0.00031882 | 0.00004171 | 0.00004743 |
| 59 | 96 | 0.04107245 | 0.05169675 | -0.01960832 | -0.04101625 | 0.00805975 | 0.02367680 | -0.00185255 | -0.01076978 | -0.00121370 | 0.00248379 | -0.00011369 | -0.00030998 | 0.00034467 | 0.00001575 | -0.00029817 | 0.00001730 | 0.00000728 |
| 59 | 97 | 0.04106094 | 0.05164579 | -0.01972136 | -0.04067128 | 0.00774422 | 0.02299971 | -0.00229529 | -0.01058062 | -0.00104866 | 0.00222916 | -0.00004816 | -0.00040161 | 0.00038593 | -0.00011022 | -0.00025871 | 0.00002740 | 0.00008993 |
| 59 | 98 | 0.04108211 | 0.05140461 | -0.01976644 | -0.04015774 | 0.00877356 | 0.02270446 | -0.00288681 | -0.01057882 | -0.00073024 | 0.00237208 | -0.00017140 | -0.00034446 | 0.00040159 | -0.00005769 | -0.00023871 | 0.00000096 | 0.00003647 |
| 59 | 99 | 0.04094914 | 0.05145895 | -0.01964852 | -0.03987259 | 0.00823698 | 0.02197604 | -0.00325370 | -0.01040738 | -0.00057557 | 0.00211440 | -0.00008362 | -0.00046625 | 0.00047249 | -0.00019223 | -0.00018460 | 0.00000754 | 0.00014582 |
| 59 | 100 | 0.04104414 | 0.05117694 | -0.01992996 | -0.03939480 | 0.00944967 | 0.02165318 | -0.00401739 | -0.01042421 | -0.00018564 | 0.00227840 | -0.00023293 | -0.00038171 | 0.00046751 | -0.00013390 | -0.00017086 | -0.00001263 | 0.00007851 |
| 59 | 101 | 0.04079768 | 0.05129839 | -0.01957865 | -0.03912508 | 0.00870307 | 0.02088104 | -0.00425167 | -0.01024492 | -0.00005963 | 0.00201826 | -0.00012081 | -0.00053395 | 0.00056683 | -0.00027280 | -0.00010256 | -0.00001382 | 0.00021082 |
| 59 | 102 | 0.04097230 | 0.05098585 | -0.02010356 | -0.03869673 | 0.01006664 | 0.02053144 | -0.00519708 | -0.01028846 | 0.00039782 | 0.00220371 | -0.00029665 | -0.00041787 | 0.00053602 | -0.00020908 | -0.00009928 | -0.00002100 | 0.00013002 |
| 60 | 64 | 0.06456111 | 0.05093035 | -0.04753869 | -0.02092934 | 0.02890353 | 0.00160662 | -0.01304396 | -0.00089644 | 0.00149858 | -0.00046810 | 0.00029396 | 0.00038810 | -0.00043165 | 0.00008251 | -0.00004031 | 0.00000768 | -0.00015812 |
| 60 | 65 | 0.06365505 | 0.05022904 | -0.04674921 | -0.01952144 | 0.02973334 | 0.00137908 | -0.01240991 | -0.00034871 | 0.00165469 | -0.00059117 | 0.00031269 | 0.00021751 | -0.00020381 | 0.00005906 | 0.00009663 | -0.00013641 | -0.00006166 |
| 60 | 66 | 0.06456691 | 0.05066709 | -0.04728711 | -0.01944326 | 0.02972976 | 0.00084524 | -0.01287186 | -0.00071883 | 0.00170631 | -0.00041155 | 0.00030686 | 0.00023484 | -0.00020078 | 0.00001666 | 0.00006903 | -0.00008894 | -0.00004462 |
| 60 | 67 | 0.06348951 | 0.04962423 | -0.04699163 | -0.01860854 | 0.02974232 | 0.00056062 | -0.01225518 | -0.00006538 | 0.00169121 | -0.00053681 | 0.00033671 | 0.00013365 | -0.00011761 | 0.00005139 | 0.00012793 | -0.00014139 | 0.00000411 |
| 60 | 68 | 0.06420434 | 0.04962986 | -0.04785273 | -0.01866088 | 0.02945882 | 0.00049933 | -0.01237239 | -0.00019989 | 0.00155066 | -0.00040672 | 0.00035816 | 0.00023314 | -0.00028326 | 0.00005722 | 0.00000220 | 0.00001321 | -0.00003535 |
| 60 | 69 | 0.06326127 | 0.04862011 | -0.04800406 | -0.01816363 | 0.02928090 | -0.00006337 | -0.01207233 | 0.00032785 | 0.00154061 | -0.00051507 | 0.00037707 | 0.00013162 | -0.00019725 | 0.00009431 | 0.00007171 | -0.00005488 | 0.00000702 |
| 60 | 70 | 0.06386567 | 0.04857257 | -0.04877038 | -0.01803545 | 0.02933597 | 0.00005365 | -0.01210127 | 0.00016350 | 0.00139518 | -0.00042200 | 0.00040855 | 0.00021293 | -0.00033192 | 0.00008038 | -0.00004061 | 0.00008741 | -0.00001437 |
| 60 | 71 | 0.06308385 | 0.04763630 | -0.04930456 | -0.01778265 | 0.02908949 | -0.00065879 | -0.01202891 | 0.00059410 | 0.00140492 | -0.00051496 | 0.00041788 | 0.00011052 | -0.00024471 | 0.00011631 | 0.00003869 | 0.00000549 | 0.00002320 |
| 60 | 72 | 0.06357169 | 0.04754127 | -0.04997338 | -0.01750849 | 0.02945405 | -0.00042447 | -0.01200399 | 0.00041849 | 0.00127042 | -0.00046091 | 0.00045700 | 0.00017220 | -0.00034357 | 0.00008414 | -0.00005549 | 0.00012774 | 0.00001689 |
| 60 | 73 | 0.06296515 | 0.04671270 | -0.05078780 | -0.01738477 | 0.02925707 | -0.00112960 | -0.01203217 | 0.00078250 | 0.00130306 | -0.00053403 | 0.00045943 | 0.00006898 | -0.00025681 | 0.00011664 | 0.00002878 | 0.00003685 | 0.00005074 |
| 60 | 74 | 0.06332911 | 0.04656767 | -0.05137053 | -0.01703646 | 0.02984885 | -0.00085176 | -0.01199548 | 0.00061004 | 0.00119599 | -0.00051514 | 0.00050293 | 0.00011322 | -0.00032128 | 0.00007004 | -0.00004561 | 0.00013608 | 0.00005575 |
| 60 | 75 | 0.06290252 | 0.04586341 | -0.05236627 | -0.01695479 | 0.02976958 | -0.00140639 | -0.01200105 | 0.00092516 | 0.00123985 | -0.00055759 | 0.00050092 | 0.00001265 | -0.00023997 | 0.00009987 | 0.00003464 | 0.00004375 | 0.00008386 |

TABLE 3. Nuclear charge density distribution FB coefficients.

| $Z$ | $N$ | $a_1$ | $a_2$ | $a_3$ | $a_4$ | $a_5$ | $a_6$ | $a_7$ | $a_8$ | $a_9$ | $a_{10}$ | $a_{11}$ | $a_{12}$ | $a_{13}$ | $a_{14}$ | $a_{15}$ | $a_{16}$ | $a_{17}$ |
|---|---|---|---|---|---|---|---|---|---|---|---|---|---|---|---|---|---|---|
| 60 | 76 | 0.06313383 | 0.04566418 | -0.05287222 | -0.01661231 | 0.03048616 | -0.00114472 | -0.01199016 | 0.00076561 | 0.00117724 | -0.00056445 | 0.00054480 | 0.00004466 | -0.00027677 | 0.00004405 | -0.00002163 | 0.00012284 | 0.00009702 |
| 60 | 77 | 0.06288350 | 0.04508407 | -0.05396504 | -0.01652081 | 0.03054135 | -0.00144627 | -0.01188290 | 0.00103866 | 0.00121111 | -0.00056494 | 0.00053946 | -0.00004610 | -0.00020962 | 0.00007481 | 0.00004375 | 0.00003709 | 0.00011371 |
| 60 | 78 | 0.06297483 | 0.04482795 | -0.05439931 | -0.01625435 | 0.03128510 | -0.00123582 | -0.01192756 | 0.00089563 | 0.00120863 | -0.00058445 | 0.00057939 | -0.00001935 | -0.00022877 | 0.00001538 | 0.00000130 | 0.00010391 | 0.00013308 |
| 60 | 79 | 0.06288888 | 0.04436215 | -0.05551811 | -0.01612061 | 0.03146180 | -0.00124229 | -0.01166933 | 0.00113177 | 0.00121063 | -0.00054000 | 0.00057011 | -0.00009174 | -0.00018532 | 0.00005154 | 0.00004304 | 0.00002986 | 0.00013024 |
| 60 | 80 | 0.06283923 | 0.04404925 | -0.05588818 | -0.01598058 | 0.03215801 | -0.00109548 | -0.01179252 | 0.00100327 | 0.00128066 | -0.00055841 | 0.00060203 | -0.00006322 | -0.00019756 | -0.00000655 | 0.00000899 | 0.00009468 | 0.00015522 |
| 60 | 81 | 0.06289823 | 0.04368375 | -0.05697507 | -0.01577735 | 0.03243581 | -0.00082972 | -0.01139841 | 0.00121244 | 0.00123454 | -0.00047920 | 0.00058722 | -0.00011068 | -0.00018425 | 0.00003819 | 0.00002358 | 0.00003200 | 0.00012480 |
| 60 | 82 | 0.06271765 | 0.04331746 | -0.05729796 | -0.01579035 | 0.03304264 | -0.00074355 | -0.01161692 | 0.00109374 | 0.00138531 | -0.00048482 | 0.00060792 | -0.00007481 | -0.00019869 | -0.00001499 | -0.00000704 | 0.00010468 | 0.00015576 |
| 60 | 83 | 0.06281330 | 0.04243069 | -0.05784731 | -0.01448923 | 0.03291570 | -0.00139612 | -0.01127227 | 0.00139093 | 0.00109424 | -0.00045182 | 0.00059028 | -0.00013255 | -0.00019302 | 0.00006253 | 0.00001080 | 0.00001342 | 0.00010012 |
| 60 | 84 | 0.06250071 | 0.04162105 | -0.05737754 | -0.01374943 | 0.03269932 | -0.00223974 | -0.01152937 | 0.00130389 | 0.00099863 | -0.00048627 | 0.00060080 | -0.00010614 | -0.00020585 | 0.00004326 | -0.00001769 | 0.00005850 | 0.00010450 |
| 60 | 85 | 0.06264210 | 0.04067798 | -0.05804398 | -0.01231803 | 0.03279572 | -0.00285969 | -0.01116968 | 0.00169100 | 0.00076935 | -0.00046586 | 0.00057871 | -0.00015424 | -0.00021751 | 0.00012112 | -0.00000340 | -0.00000911 | 0.00006341 |
| 60 | 86 | 0.06228763 | 0.04006170 | -0.05730602 | -0.01187346 | 0.03238992 | -0.00344644 | -0.01139462 | 0.00151656 | 0.00067599 | -0.00044850 | 0.00057120 | -0.00009605 | -0.00025355 | 0.00010949 | -0.00005615 | 0.00004427 | 0.00003457 |
| 60 | 87 | 0.06244465 | 0.03917727 | -0.05797409 | -0.01047129 | 0.03262002 | -0.00401762 | -0.01105982 | 0.00196217 | 0.00052504 | -0.00044019 | 0.00054116 | -0.00013378 | -0.00027640 | 0.00018426 | -0.00004470 | 0.00000351 | 0.00001195 |
| 60 | 88 | 0.06196466 | 0.03828626 | -0.05738467 | -0.00989013 | 0.03224136 | -0.00418817 | -0.01111313 | 0.00167782 | 0.00035646 | -0.00032128 | 0.00055240 | -0.00003299 | -0.00037381 | 0.00017008 | -0.00015286 | 0.00008294 | -0.00005910 |
| 60 | 89 | 0.06208834 | 0.03774469 | -0.05780369 | -0.00892286 | 0.03242229 | -0.00461284 | -0.01082196 | 0.00211156 | 0.00031066 | -0.00034979 | 0.00051423 | -0.00008534 | -0.00037041 | 0.00022527 | -0.00012466 | 0.00005462 | -0.00004493 |
| 60 | 90 | 0.04110955 | 0.05615147 | -0.01694441 | -0.04696819 | 0.00601159 | 0.02710513 | -0.00024641 | -0.01257735 | -0.00178173 | 0.00278360 | -0.00003177 | -0.00048894 | 0.00070479 | 0.00006909 | -0.00019370 | -0.00018369 | 0.00007317 |
| 60 | 91 | 0.04132198 | 0.05636279 | -0.01633289 | -0.04700264 | 0.00516914 | 0.02727783 | 0.00092040 | -0.01199566 | -0.00202344 | 0.00238042 | 0.00004051 | -0.00068207 | 0.00081881 | -0.00001252 | -0.00009501 | -0.00034292 | 0.00012873 |
| 60 | 92 | 0.04103682 | 0.05359445 | -0.01874406 | -0.04340893 | 0.00758209 | 0.02535353 | -0.00163661 | -0.01196180 | -0.00141362 | 0.00298083 | -0.00010367 | -0.00017027 | 0.00032627 | 0.00012741 | -0.00037113 | 0.00006087 | -0.00005233 |
| 60 | 93 | 0.04150789 | 0.05306841 | -0.01914189 | -0.04229786 | 0.00747832 | 0.02475188 | -0.00108491 | -0.01109805 | -0.00150844 | 0.00262325 | -0.00003155 | -0.00028960 | 0.00033637 | 0.00006621 | -0.00032165 | 0.00000331 | 0.00000084 |
| 60 | 94 | 0.04105304 | 0.05283415 | -0.01915302 | -0.04223140 | 0.00854349 | 0.02450866 | -0.00257239 | -0.01171835 | -0.00101573 | 0.00291826 | -0.00017892 | -0.00014085 | 0.00030426 | 0.00009293 | -0.00035805 | 0.00007496 | -0.00007431 |
| 60 | 95 | 0.04155679 | 0.05240510 | -0.01955942 | -0.04112348 | 0.00838338 | 0.02378895 | -0.00208994 | -0.01084631 | -0.00111984 | 0.00251754 | -0.00008949 | -0.00028131 | 0.00033602 | 0.00001341 | -0.00030023 | 0.00001938 | 0.00000057 |
| 60 | 96 | 0.04097627 | 0.05227532 | -0.01941483 | -0.04133654 | 0.00941648 | 0.02364761 | -0.00363243 | -0.01155956 | -0.00054135 | 0.00284940 | -0.00025794 | -0.00013267 | 0.00031481 | 0.00004300 | -0.00032299 | 0.00007507 | -0.00007729 |
| 60 | 97 | 0.04148588 | 0.05200710 | -0.01973333 | -0.04030300 | 0.00915592 | 0.02287469 | -0.00316925 | -0.01070092 | -0.00065625 | 0.00241809 | -0.00015297 | -0.00029946 | 0.00037734 | -0.00005245 | -0.00025246 | 0.00001447 | 0.00001917 |
| 60 | 98 | 0.04085642 | 0.05178451 | -0.01968051 | -0.04052699 | 0.01021472 | 0.02265594 | -0.00481048 | -0.01143147 | -0.00001115 | 0.00278438 | -0.00033804 | -0.00012875 | 0.00033414 | -0.00001392 | -0.00027973 | 0.00007965 | -0.00006475 |
| 60 | 99 | 0.04136254 | 0.05168279 | -0.01988368 | -0.03956876 | 0.00986626 | 0.02185951 | -0.00434133 | -0.01058519 | -0.00013146 | 0.00233550 | -0.00021979 | -0.00032171 | 0.00042899 | -0.00012116 | -0.00019490 | 0.00001031 | 0.00005036 |
| 60 | 100 | 0.04070970 | 0.05135258 | -0.01994923 | -0.03978791 | 0.01090892 | 0.02155851 | -0.00605031 | -0.01132311 | 0.00055294 | 0.00272716 | -0.00041770 | -0.00012636 | 0.00035751 | -0.00007407 | -0.00023260 | 0.00009149 | -0.00003877 |
| 60 | 101 | 0.04120544 | 0.05140770 | -0.02002407 | -0.03889553 | 0.01049791 | 0.02076019 | -0.00555816 | -0.01048339 | 0.00043334 | 0.00227098 | -0.00028864 | -0.00034411 | 0.00048453 | -0.00018902 | -0.00013226 | 0.00000958 | 0.00009097 |
| 60 | 102 | 0.04055231 | 0.05097824 | -0.02021460 | -0.03910965 | 0.01149101 | 0.02039189 | -0.00730477 | -0.01122683 | 0.00113241 | 0.00267955 | -0.00049574 | -0.00012413 | 0.00038226 | -0.00013461 | -0.00018430 | 0.00011014 | -0.00000259 |
| 60 | 103 | 0.04103041 | 0.05116984 | -0.02015654 | -0.03826903 | 0.01104600 | 0.01960438 | -0.00678071 | -0.01038708 | 0.00101923 | 0.00222417 | -0.00035834 | -0.00036445 | 0.00054005 | -0.00025355 | -0.00006788 | 0.00001259 | 0.00013757 |
| 61 | 65 | 0.06348367 | 0.04989828 | -0.04539946 | -0.01809064 | 0.03021772 | 0.00105519 | -0.01208629 | -0.00038633 | 0.00166572 | -0.00035739 | 0.00021608 | 0.00010586 | -0.00000967 | 0.00018279 | 0.00021033 | -0.00027476 | -0.00007035 |
| 61 | 66 | 0.06435554 | 0.04995113 | -0.04656031 | -0.01766756 | 0.03000283 | 0.00010178 | -0.01207311 | 0.00005620 | 0.00169379 | -0.00046622 | 0.00032892 | 0.00008141 | -0.00004526 | 0.00007218 | 0.00016941 | -0.00022732 | -0.00001460 |
| 61 | 67 | 0.06335004 | 0.04924498 | -0.04600571 | -0.01760009 | 0.02990435 | -0.00003783 | -0.01226408 | -0.00026548 | 0.00168253 | -0.00027526 | 0.00022538 | 0.00003718 | 0.00005904 | 0.00018734 | 0.00024008 | -0.00028858 | -0.00002686 |
| 61 | 68 | 0.06420827 | 0.04888910 | -0.04763327 | -0.01720184 | 0.02940047 | -0.00052983 | -0.01181864 | 0.00049714 | 0.00149579 | -0.00043120 | 0.00036497 | 0.00009973 | -0.00015725 | 0.00013122 | 0.00008854 | -0.00012341 | -0.00003472 |
| 61 | 69 | 0.06338614 | 0.04816270 | -0.04760813 | -0.01732886 | 0.02923751 | -0.00086529 | -0.01221817 | 0.00011038 | 0.00148833 | -0.00023297 | 0.00025445 | 0.00005963 | -0.00005953 | 0.00023929 | 0.00015973 | -0.00019136 | -0.00005264 |
| 61 | 70 | 0.06408606 | 0.04783453 | -0.04899791 | -0.01681853 | 0.02902784 | -0.00117311 | -0.01172905 | 0.00080640 | 0.00131700 | -0.00041908 | 0.00040246 | 0.00009882 | -0.00023764 | 0.00016807 | 0.00003164 | -0.00004220 | -0.00003740 |
| 61 | 71 | 0.06342810 | 0.04708723 | -0.04939679 | -0.01695821 | 0.02893154 | -0.00160312 | -0.01218145 | 0.00043694 | 0.00132608 | -0.00022677 | 0.00029271 | 0.00005415 | -0.00013993 | 0.00026174 | 0.00010371 | -0.00011458 | -0.00005097 |
| 61 | 72 | 0.06398889 | 0.04684325 | -0.05053224 | -0.01642561 | 0.02899554 | -0.00174597 | -0.01172549 | 0.00103253 | 0.00118429 | -0.00043332 | 0.00044204 | 0.00007293 | -0.00027649 | 0.00017899 | 0.00000386 | 0.00000903 | -0.00002067 |
| 61 | 73 | 0.06347758 | 0.04607490 | -0.05127552 | -0.01646701 | 0.02903832 | -0.00219090 | -0.01211852 | 0.00072240 | 0.00120668 | -0.00025166 | 0.00033861 | 0.00001920 | -0.00017491 | 0.00025457 | 0.00007428 | -0.00006474 | -0.00002413 |
| 61 | 74 | 0.06391308 | 0.04594254 | -0.05214179 | -0.01599698 | 0.02931361 | -0.00218726 | -0.01173120 | 0.00120894 | 0.00110661 | -0.00046522 | 0.00048387 | 0.00002322 | -0.00027362 | 0.00016571 | 0.00000250 | 0.00002980 | 0.00001260 |
| 61 | 75 | 0.06353664 | 0.04515574 | -0.05316242 | -0.01589011 | 0.02951581 | -0.00257808 | -0.01199132 | 0.00096850 | 0.00112661 | -0.00029125 | 0.00038930 | -0.00003745 | -0.00017044 | 0.00022413 | 0.00006491 | -0.00004052 | 0.00001876 |
| 61 | 76 | 0.06385268 | 0.04513579 | -0.05375244 | -0.01555587 | 0.02991549 | -0.00244212 | -0.01167946 | 0.00135261 | 0.00107836 | -0.00049574 | 0.00052622 | -0.00004054 | -0.00024070 | 0.00013560 | 0.00001667 | 0.00002804 | 0.00005448 |
| 61 | 77 | 0.06360203 | 0.04433788 | -0.05497873 | -0.01528701 | 0.03026006 | -0.00272622 | -0.01177122 | 0.00117621 | 0.00107569 | -0.00032389 | 0.00043982 | -0.00009995 | -0.00014431 | 0.00018169 | 0.00006259 | -0.00003303 | 0.00006395 |
| 61 | 78 | 0.06379931 | 0.04441265 | -0.05530094 | -0.01514389 | 0.03069668 | -0.00246885 | -0.01153132 | 0.00147000 | 0.00108704 | -0.00050358 | 0.00056485 | -0.00010208 | -0.00019833 | 0.00009960 | 0.00003089 | 0.00001750 | 0.00009347 |
| 61 | 79 | 0.06366313 | 0.04361454 | -0.05665236 | -0.01471087 | 0.03114716 | -0.00263139 | -0.01146198 | 0.00135070 | 0.00104538 | -0.00033315 | 0.00048345 | -0.00014955 | -0.00011936 | 0.00013980 | 0.00005355 | -0.00003050 | 0.00009679 |
| 61 | 80 | 0.06374365 | 0.04375763 | -0.05673632 | -0.01479217 | 0.03155531 | -0.00226080 | -0.01129338 | 0.00156558 | 0.00112114 | -0.00047589 | 0.00059367 | -0.00014396 | -0.00016891 | 0.00006872 | 0.00003095 | 0.00001193 | 0.00011739 |
| 61 | 81 | 0.06370532 | 0.04296886 | -0.05813303 | -0.01418903 | 0.03207314 | -0.00233520 | -0.01110642 | 0.00150316 | 0.00103234 | -0.00031513 | 0.00051342 | -0.00017098 | -0.00011525 | 0.00010845 | 0.00002863 | -0.00002378 | 0.00010558 |
| 61 | 82 | 0.06367814 | 0.04315381 | -0.05802977 | -0.01450713 | 0.03242034 | -0.00185535 | -0.01101449 | 0.00164718 | 0.00117383 | -0.00041383 | 0.00060684 | -0.00015319 | -0.00016955 | 0.00005076 | 0.00000864 | 0.00002009 | 0.00011663 |
| 61 | 83 | 0.06365936 | 0.04187007 | -0.05891540 | -0.01285855 | 0.03255109 | -0.00289148 | -0.01094482 | 0.00175557 | 0.00088740 | -0.00034473 | 0.00052366 | -0.00018178 | -0.00013603 | 0.00012244 | 0.00001305 | -0.00003630 | 0.00008429 |
| 61 | 84 | 0.06354020 | 0.04143912 | -0.05819830 | -0.01227500 | 0.03232126 | -0.00338719 | -0.01096651 | 0.00194941 | 0.00082533 | -0.00045554 | 0.00060210 | -0.00018062 | -0.00018879 | 0.00010927 | 0.00000393 | -0.00002002 | 0.00007348 |
| 61 | 85 | 0.06352060 | 0.04040886 | -0.05898418 | -0.01086394 | 0.03254415 | -0.00424675 | -0.01100399 | 0.00210742 | 0.00064654 | -0.00041326 | 0.00050706 | -0.00017272 | -0.00018941 | 0.00017685 | -0.00000191 | -0.00004522 | 0.00004100 |
| 61 | 86 | 0.06336568 | 0.03993555 | -0.05811701 | -0.01031384 | 0.03221132 | -0.00461016 | -0.01091653 | 0.00223298 | 0.00054628 | -0.00046723 | 0.00057143 | -0.00016159 | -0.00024724 | 0.00017294 | -0.00003035 | -0.00002427 | 0.00001185 |
| 61 | 87 | 0.06161449 | 0.04099878 | -0.05536263 | -0.01277103 | 0.02968747 | -0.00196749 | -0.00906261 | 0.00120568 | -0.00004982 | -0.00043499 | 0.00061434 | -0.00036783 | -0.00001984 | 0.00005176 | 0.00000240 | -0.00011358 | 0.00013183 |
| 61 | 88 | 0.06306684 | 0.03832581 | -0.05811904 | -0.00839019 | 0.03221815 | -0.00537486 | -0.01079082 | 0.00244178 | 0.00029283 | -0.00040465 | 0.00055289 | -0.00009438 | -0.00036659 | 0.00022029 | -0.00011854 | 0.00002660 | -0.00006283 |
| 61 | 89 | 0.04129812 | 0.05801676 | -0.01516448 | -0.04843178 | 0.00404039 | 0.02815350 | 0.00116568 | -0.01246747 | -0.00194109 | 0.00254750 | 0.00007277 | -0.00068165 | 0.00083993 | -0.00005906 | -0.00009748 | -0.00036393 | 0.00014507 |
| 61 | 90 | 0.04195696 | 0.05684249 | -0.01676344 | -0.04632268 | 0.00563129 | 0.02652787 | 0.00015168 | -0.01191888 | -0.00173106 | 0.00248364 | 0.00005428 | -0.00060753 | 0.00076974 | -0.00002258 | -0.00011330 | -0.00028667 | 0.00013142 |
| 61 | 91 | 0.04214899 | 0.05406046 | -0.01936969 | -0.04278930 | 0.00714516 | 0.02491318 | -0.00102842 | -0.01109699 | -0.00143410 | 0.00264573 | 0.00004323 | -0.00030207 | 0.00032473 | 0.00001976 | -0.00032942 | 0.00000531 | 0.00002984 |
| 61 | 92 | 0.04188829 | 0.05399847 | -0.01905873 | -0.04267988 | 0.00769832 | 0.02464351 | -0.00171974 | -0.01128386 | -0.00120911 | 0.00270390 | -0.00003782 | -0.00026148 | 0.00035968 | 0.00003891 | -0.00030750 | 0.00001202 | 0.00001423 |
| 61 | 93 | 0.04217172 | 0.05326778 | -0.01983305 | -0.04163812 | 0.00792822 | 0.02409389 | -0.00191980 | -0.01087063 | -0.00117034 | 0.00252831 | 0.00000403 | -0.00027930 | 0.00029052 | -0.00002478 | -0.00033704 | 0.00005792 | 0.00002759 |
| 61 | 94 | 0.04184363 | 0.05326013 | -0.01952623 | -0.04164891 | 0.00869933 | 0.02382674 | -0.00278876 | -0.01112963 | -0.00082473 | 0.00262356 | -0.00010633 | -0.00023829 | 0.00034901 | -0.00000799 | -0.00029795 | 0.00004452 | 0.00000976 |
| 61 | 95 | 0.04204788 | 0.05278919 | -0.01993502 | -0.04083087 | 0.00856925 | 0.02332513 | -0.00286007 | -0.01075255 | -0.00083341 | 0.00241121 | -0.00003904 | -0.00029502 | 0.00031391 | -0.00008494 | -0.00030893 | 0.00007450 | 0.00004465 |
| 61 | 96 | 0.04168726 | 0.05274457 | -0.01976111 | -0.04088125 | 0.00956196 | 0.02299219 | -0.00393238 | -0.01105420 | -0.00037333 | 0.00254721 | -0.00017786 | -0.00023915 | 0.00037486 | -0.00006623 | -0.00026418 | 0.00005770 | 0.00002226 |
| 61 | 97 | 0.04184998 | 0.05240161 | -0.01994631 | -0.04008052 | 0.00916868 | 0.02242746 | -0.00388363 | -0.01065239 | -0.00042655 | 0.00230569 | -0.00008321 | -0.00032601 | 0.00036212 | -0.00015034 | -0.00026076 | 0.00007737 | 0.00007571 |
| 61 | 98 | 0.04147967 | 0.05230604 | -0.01994159 | -0.04017566 | 0.01032837 | 0.02202791 | -0.00514975 | -0.01099284 | 0.00013221 | 0.00248319 | -0.00025051 | -0.00024652 | 0.00041228 | -0.00012738 | -0.00021929 | 0.00006873 | 0.00004714 |
| 61 | 99 | 0.04160224 | 0.05206804 | -0.01991028 | -0.03936991 | 0.00971057 | 0.02141489 | -0.00494257 | -0.01055136 | 0.00002976 | 0.00221388 | -0.00012757 | -0.00036597 | 0.00042574 | -0.00021664 | -0.00019847 | 0.00007123 | 0.00011738 |
| 61 | 100 | 0.04124681 | 0.05192626 | -0.02009105 | -0.03951396 | 0.01098723 | 0.02095837 | -0.00639095 | -0.01092962 | 0.00067239 | 0.00243240 | -0.00032308 | -0.00025677 | 0.00045521 | -0.00018804 | -0.00016765 | 0.00007989 | 0.00008142 |
| 61 | 101 | 0.04132903 | 0.05176898 | -0.01985602 | -0.03869339 | 0.01019072 | 0.02031815 | -0.00600021 | -0.01043896 | 0.00051745 | 0.00213671 | -0.00017201 | -0.00040953 | 0.00049715 | -0.00028053 | -0.00012753 | 0.00005928 | 0.00016546 |
| 61 | 102 | 0.04101075 | 0.05159517 | -0.02022346 | -0.03888513 | 0.01154311 | 0.01981829 | -0.00761758 | -0.01085601 | 0.00122985 | 0.00239428 | -0.00039462 | -0.00026758 | 0.00049967 | -0.00024585 | -0.00011246 | 0.00009123 | 0.00012150 |
| 61 | 103 | 0.04105055 | 0.05149259 | -0.01980265 | -0.03804662 | 0.01061268 | 0.01916879 | -0.00703250 | -0.01031071 | 0.00102143 | 0.00207441 | -0.00021666 | -0.00045247 | 0.00057039 | -0.00033966 | -0.00005254 | 0.00004388 | 0.00021597 |
| 61 | 104 | 0.04078718 | 0.05130623 | -0.02034486 | -0.03828198 | 0.01200737 | 0.01864244 | -0.00880130 | -0.01076930 | 0.00178921 | 0.00236757 | -0.00046407 | -0.00027785 | 0.00054344 | -0.00029935 | -0.00005589 | 0.00010180 | 0.00016415 |
| 62 | 66 | 0.06599344 | 0.05028703 | -0.04700358 | -0.01689856 | 0.02997015 | -0.00034592 | -0.01203466 | 0.00006571 | 0.00147949 | -0.00033091 | 0.00033634 | 0.00017158 | -0.00017778 | 0.00009869 | 0.00006647 | -0.00016364 | -0.00012545 |
| 62 | 67 | 0.06504969 | 0.04913575 | -0.04712903 | -0.01620356 | 0.02950064 | -0.00102674 | -0.01156278 | 0.00063876 | 0.00143200 | -0.00034864 | 0.00035881 | 0.00005447 | -0.00009848 | 0.00016292 | 0.00011917 | -0.00021152 | -0.00007589 |
| 62 | 68 | 0.06576297 | 0.04910824 | -0.04795860 | -0.01619483 | 0.02956227 | -0.00079916 | -0.01159050 | 0.00051531 | 0.00121880 | -0.00032545 | 0.00038358 | 0.00018882 | -0.00028413 | 0.00014799 | -0.00002371 | -0.00005791 | -0.00014698 |
| 62 | 69 | 0.06500147 | 0.04800341 | -0.04856188 | -0.01578416 | 0.02898448 | -0.00170723 | -0.01140555 | 0.00100509 | 0.00120139 | -0.00032972 | 0.00039408 | 0.00007388 | -0.00021093 | 0.00021463 | 0.00003812 | -0.00011488 | -0.00010241 |
| 62 | 70 | 0.06552667 | 0.04794060 | -0.04919164 | -0.01562955 | 0.02928439 | -0.00138455 | -0.01137002 | 0.00084085 | 0.00099309 | -0.00034420 | 0.00043063 | 0.00018387 | -0.00035240 | 0.00017622 | -0.00008346 | 0.00001837 | -0.00014969 |
| 62 | 71 | 0.06494883 | 0.04694330 | -0.05013437 | -0.01536536 | 0.02877231 | -0.00236000 | -0.01136034 | 0.00128789 | 0.00102749 | -0.00034137 | 0.00043198 | 0.00006702 | -0.00027968 | 0.00023806 | -0.00001089 | -0.00004683 | -0.00010416 |
| 62 | 72 | 0.06528210 | 0.04685389 | -0.05057113 | -0.01513427 | 0.02922000 | -0.00203261 | -0.01130668 | 0.00109278 | 0.00083945 | -0.00039025 | 0.00047711 | 0.00015148 | -0.00037413 | 0.00017990 | -0.00010643 | 0.00005923 | -0.00012960 |
| 62 | 73 | 0.06488021 | 0.04599854 | -0.05173795 | -0.01491885 | 0.02888335 | -0.00294201 | -0.01136405 | 0.00151868 | 0.00092648 | -0.00037968 | 0.00047316 | 0.00003055 | -0.00029792 | 0.00023223 | -0.00002542 | -0.00001200 | -0.00007994 |
| 62 | 74 | 0.06502796 | 0.04588943 | -0.05198862 | -0.01467787 | 0.02939006 | -0.00266220 | -0.01131803 | 0.00130809 | 0.00077399 | -0.00045506 | 0.00052276 | 0.00009295 | -0.00035041 | 0.00016041 | -0.00009511 | 0.00006671 | -0.00008724 |
| 62 | 75 | 0.06478953 | 0.04518117 | -0.05329910 | -0.01446399 | 0.02926786 | -0.00340269 | -0.01134965 | 0.00171326 | 0.00089472 | -0.00042944 | 0.00051710 | -0.00003005 | -0.00027159 | 0.00020219 | -0.00001253 | -0.00000638 | -0.00003475 |
| 62 | 76 | 0.06476872 | 0.04505529 | -0.05337504 | -0.01426771 | 0.02976146 | -0.00317851 | -0.01132010 | 0.00150343 | 0.00079062 | -0.00051889 | 0.00056638 | 0.00001798 | -0.00029441 | 0.00012464 | -0.00006150 | 0.00005179 | -0.00002893 |
| 62 | 77 | 0.06467696 | 0.04448000 | -0.05476784 | -0.01404152 | 0.02984015 | -0.00367740 | -0.01125760 | 0.00187533 | 0.00091487 | -0.00046822 | 0.00056093 | -0.00010045 | -0.00021917 | 0.00015823 | 0.00001249 | -0.00001756 | 0.00002047 |
| 62 | 78 | 0.06451465 | 0.04433341 | -0.05469610 | -0.01392744 | 0.03027499 | -0.00348920 | -0.01124788 | 0.00168067 | 0.00086857 | -0.00055809 | 0.00060516 | -0.00005717 | -0.00022842 | 0.00008324 | -0.00002354 | 0.00003100 | 0.00003403 |
| 62 | 79 | 0.06454737 | 0.04387178 | -0.05611113 | -0.01368583 | 0.03051141 | -0.00371468 | -0.01106131 | 0.00200479 | 0.00096585 | -0.00047641 | 0.00059925 | -0.00016156 | -0.00016583 | 0.00011292 | 0.00003167 | -0.00002968 | 0.00007127 |
| 62 | 80 | 0.06427884 | 0.04369331 | -0.05594233 | -0.01367020 | 0.03087591 | -0.00353877 | -0.01107933 | 0.00183873 | 0.00098435 | -0.00055652 | 0.00063478 | -0.00011476 | -0.00017655 | 0.00004717 | 0.00000150 | 0.00001991 | 0.00008815 |
| 62 | 81 | 0.06440812 | 0.04333015 | -0.05731418 | -0.01340649 | 0.03121625 | -0.00350824 | -0.01078307 | 0.00210608 | 0.00103173 | -0.00044690 | 0.00062559 | -0.00019601 | -0.00013487 | 0.00007715 | 0.00003112 | -0.00003013 | 0.00010360 |
| 62 | 82 | 0.06407406 | 0.04310383 | -0.05712140 | -0.01348455 | 0.03153291 | -0.00333216 | -0.01084327 | 0.00198287 | 0.00112126 | -0.00051282 | 0.00065052 | -0.00014096 | -0.00015744 | 0.00002443 | 0.00000279 | 0.00002754 | 0.00012132 |
| 62 | 83 | 0.06428034 | 0.04227562 | -0.05789407 | -0.01220030 | 0.03149685 | -0.00409492 | -0.01063015 | 0.00230273 | 0.00088919 | -0.00045489 | 0.00062976 | -0.00021071 | -0.00014433 | 0.00009402 | 0.00002039 | -0.00004844 | 0.00008268 |
| 62 | 84 | 0.06393307 | 0.04159860 | -0.05708012 | -0.01163077 | 0.03115826 | -0.00472745 | -0.01087588 | 0.00217888 | 0.00073058 | -0.00055328 | 0.00062920 | -0.00013686 | -0.00020046 | 0.00009001 | -0.00001878 | -0.00001681 | 0.00003810 |
| 62 | 85 | 0.06413063 | 0.04081063 | -0.05780501 | -0.01020365 | 0.03141923 | -0.00538971 | -0.01067133 | 0.00259026 | 0.00059615 | -0.00050304 | 0.00060230 | -0.00019375 | -0.00019944 | 0.00015560 | -0.00000453 | -0.00006929 | 0.00001296 |
| 62 | 86 | 0.06374971 | 0.03947718 | -0.05761348 | -0.00897622 | 0.03129947 | -0.00590906 | -0.01090014 | 0.00242427 | 0.00035057 | -0.00046724 | 0.00062102 | -0.00003087 | -0.00037756 | 0.00016487 | -0.00013394 | 0.00002921 | -0.00008695 |
| 62 | 87 | 0.06389780 | 0.03904011 | -0.05804153 | -0.00794347 | 0.03165188 | -0.00635321 | -0.01073771 | 0.00284622 | 0.00033725 | -0.00046329 | 0.00058609 | -0.00010433 | -0.00034505 | 0.00020740 | -0.00010002 | -0.00001556 | -0.00007813 |
| 62 | 88 | 0.04229626 | 0.05750772 | -0.01752200 | -0.04640944 | 0.00626151 | 0.02569093 | -0.00174553 | -0.01249644 | -0.00118635 | 0.00288462 | -0.00001137 | -0.00039203 | 0.00067498 | 0.00000293 | -0.00019625 | -0.00011336 | 0.00010232 |
| 62 | 89 | 0.04244612 | 0.05747476 | -0.01713726 | -0.04605880 | 0.00597297 | 0.02574192 | -0.00090345 | -0.01195375 | -0.00128732 | 0.00261991 | 0.00003440 | -0.00053263 | 0.00074574 | -0.00004950 | -0.00012138 | -0.00023038 | 0.00014130 |
| 62 | 90 | 0.04194792 | 0.05557914 | -0.01874903 | -0.04435764 | 0.00767233 | 0.02489859 | -0.00291815 | -0.01221442 | -0.00077892 | 0.00304168 | -0.00010436 | -0.00016317 | 0.00042006 | 0.00003095 | -0.00031478 | 0.00005248 | 0.00000401 |
| 62 | 91 | 0.04218363 | 0.05502103 | -0.01905408 | -0.04334120 | 0.00787804 | 0.02443529 | -0.00265734 | -0.01156049 | -0.00076139 | 0.00282055 | -0.00007298 | -0.00023135 | 0.00040320 | -0.00000538 | -0.00028621 | 0.00002692 | 0.00003723 |
| 62 | 92 | 0.04188684 | 0.05471766 | -0.01930361 | -0.04333083 | 0.00886425 | 0.02429700 | -0.00387346 | -0.01207536 | -0.00038305 | 0.00302050 | -0.00019180 | -0.00012693 | 0.00039836 | 0.00000493 | -0.00031044 | 0.00007964 | -0.00001692 |
| 62 | 93 | 0.04207078 | 0.05420619 | -0.01961913 | -0.04242466 | 0.00901090 | 0.02376257 | -0.00378672 | -0.01149501 | -0.00037380 | 0.00276958 | -0.00015203 | -0.00019574 | 0.00038460 | -0.00004644 | -0.00028740 | 0.00007620 | 0.00003182 |
| 62 | 94 | 0.04174414 | 0.05397829 | -0.01971924 | -0.04242678 | 0.00997107 | 0.02358519 | -0.00493596 | -0.01196778 | 0.00007109 | 0.00298838 | -0.00027905 | -0.00010780 | 0.00040032 | -0.00003217 | -0.00028753 | 0.00009220 | -0.00002505 |
| 62 | 95 | 0.04185589 | 0.05357032 | -0.01995674 | -0.04168051 | 0.00999522 | 0.02301448 | -0.00497535 | -0.01147795 | 0.00006709 | 0.00271857 | -0.00023073 | -0.00018281 | 0.00039776 | -0.00009687 | -0.00026645 | 0.00010674 | 0.00004137 |
| 62 | 96 | 0.04156010 | 0.05329205 | -0.02009168 | -0.04154771 | 0.01095819 | 0.02268973 | -0.00608112 | -0.01185537 | 0.00056436 | 0.00295055 | -0.00036292 | -0.00009569 | 0.00041091 | -0.00007524 | -0.00025497 | 0.00010325 | -0.00002059 |
| 62 | 97 | 0.04159572 | 0.05300888 | -0.02021030 | -0.04096327 | 0.01084682 | 0.02210504 | -0.00620739 | -0.01145284 | 0.00054953 | 0.00267275 | -0.00030692 | -0.00017940 | 0.00042346 | -0.00015018 | -0.00023316 | 0.00013140 | 0.00006282 |
| 62 | 98 | 0.04135917 | 0.05266592 | -0.02043019 | -0.04070367 | 0.01179074 | 0.02164455 | -0.00725379 | -0.01173155 | 0.00107724 | 0.00291020 | -0.00044169 | -0.00008850 | 0.00042625 | -0.00012168 | -0.00021661 | 0.00011592 | -0.00000444 |
| 62 | 99 | 0.04132245 | 0.05251274 | -0.02041041 | -0.04026583 | 0.01155687 | 0.02106495 | -0.00743269 | -0.01140388 | 0.00105648 | 0.00263246 | -0.00037946 | -0.00018232 | 0.00045631 | -0.00020358 | -0.00019129 | 0.00015184 | 0.00009340 |
| 62 | 100 | 0.04116381 | 0.05210880 | -0.02074182 | -0.03990374 | 0.01246571 | 0.02049939 | -0.00841076 | -0.01159438 | 0.00159439 | 0.00286983 | -0.00051465 | -0.00008465 | 0.00044382 | -0.00016932 | -0.00017518 | 0.00013057 | 0.00002077 |
| 62 | 101 | 0.04106247 | 0.05207644 | -0.02058035 | -0.03958547 | 0.01213789 | 0.01993664 | -0.00861592 | -0.01132415 | 0.00157331 | 0.00259737 | -0.00044786 | -0.00018907 | 0.00049243 | -0.00025494 | -0.00014382 | 0.00016804 | 0.00012942 |
| 62 | 102 | 0.04098978 | 0.05162710 | -0.02102903 | -0.03915359 | 0.01299789 | 0.01930436 | -0.00952199 | -0.01144597 | 0.00210399 | 0.00283106 | -0.00058155 | -0.00008324 | 0.00046239 | -0.00021653 | -0.00013227 | 0.00014591 | 0.00005197 |
| 62 | 103 | 0.04083268 | 0.05169522 | -0.02073401 | -0.03892167 | 0.01261073 | 0.01876220 | -0.00973444 | -0.01121413 | 0.00208749 | 0.00256693 | -0.00051175 | -0.00019804 | 0.00052944 | -0.00030286 | -0.00009291 | 0.00017926 | 0.00016749 |
| 62 | 104 | 0.04084548 | 0.05122342 | -0.02128979 | -0.03845595 | 0.01340910 | 0.01810243 | -0.01056846 | -0.01129091 | 0.00259700 | 0.00279480 | -0.00064232 | -0.00008400 | 0.00048180 | -0.00026219 | -0.00008867 | 0.00016000 | 0.00008640 |
| 62 | 105 | 0.04079046 | 0.05089075 | -0.02160602 | -0.03812229 | 0.01304646 | 0.01759553 | -0.01030764 | -0.01050447 | 0.00249014 | 0.00243976 | -0.00059254 | -0.00010440 | 0.00042604 | -0.00028776 | -0.00010483 | 0.00021601 | 0.00012973 |
| 62 | 106 | 0.04104892 | 0.04986143 | -0.02289557 | -0.03716965 | 0.01404381 | 0.01675431 | -0.01058087 | -0.00987350 | 0.00289211 | 0.00253383 | -0.00072510 | 0.00008184 | 0.00026447 | -0.00018746 | -0.00013702 | 0.00018233 | -0.00002735 |
| 63 | 67 | 0.06476581 | 0.04842503 | -0.04656145 | -0.01542329 | 0.02934431 | -0.00180455 | -0.01187151 | 0.00027123 | 0.00134373 | 0.00001071 | 0.00024914 | -0.00001523 | 0.00002823 | 0.00031002 | 0.00019456 | -0.00034083 | -0.00015728 |

TABLE 3. Nuclear charge density distribution FB coefficients.

| $Z$ | $N$ | $a_1$ | $a_2$ | $a_3$ | $a_4$ | $a_5$ | $a_6$ | $a_7$ | $a_8$ | $a_9$ | $a_{10}$ | $a_{11}$ | $a_{12}$ | $a_{13}$ | $a_{14}$ | $a_{15}$ | $a_{16}$ | $a_{17}$ |
|---|---|---|---|---|---|---|---|---|---|---|---|---|---|---|---|---|---|---|
| 63 | 68 | 0.06579400 | 0.04811927 | -0.04808404 | -0.01478057 | 0.02897921 | -0.00225465 | -0.01113253 | 0.00115507 | 0.00107969 | -0.00024325 | 0.00039173 | 0.00003750 | -0.00016462 | 0.00025061 | 0.00005783 | -0.00020630 | -0.00016785 |
| 63 | 69 | 0.06496778 | 0.04725479 | -0.04841306 | -0.01493621 | 0.02885716 | -0.00258742 | -0.01171539 | 0.00072544 | 0.00110798 | 0.00002962 | 0.00027942 | 0.00001558 | -0.00010566 | 0.00036036 | 0.00009785 | -0.00024183 | -0.00020274 |
| 63 | 70 | 0.06579420 | 0.04699234 | -0.04969514 | -0.01432538 | 0.02862768 | -0.00297007 | -0.01103357 | 0.00150312 | 0.00086553 | -0.00025619 | 0.00042788 | 0.00004884 | -0.00026012 | 0.00028556 | -0.00001110 | -0.00012711 | -0.00019109 |
| 63 | 71 | 0.06511631 | 0.04618799 | -0.05026507 | -0.01436772 | 0.02866172 | -0.00333368 | -0.01157700 | 0.00115298 | 0.00094102 | 0.00000450 | 0.00031861 | 0.00001399 | -0.00018888 | 0.00037611 | 0.00003554 | -0.00016862 | -0.00021124 |
| 63 | 72 | 0.06574779 | 0.04601512 | -0.05127633 | -0.01385709 | 0.02855504 | -0.00365860 | -0.01101244 | 0.00179935 | 0.00074089 | -0.00030119 | 0.00046680 | 0.00002874 | -0.00030127 | 0.00028873 | -0.00004095 | -0.00008078 | -0.00018175 |
| 63 | 73 | 0.06519451 | 0.04527241 | -0.05202944 | -0.01375548 | 0.02872765 | -0.00403207 | -0.01144312 | 0.00154679 | 0.00084950 | -0.00005772 | 0.00036626 | -0.00002224 | -0.00021433 | 0.00035769 | 0.00001116 | -0.00012710 | -0.00018215 |
| 63 | 74 | 0.06564579 | 0.04520732 | -0.05275418 | -0.01339362 | 0.02871918 | -0.00428427 | -0.01101100 | 0.00205692 | 0.00070538 | -0.00036657 | 0.00050897 | -0.00002157 | -0.00028874 | 0.00026308 | -0.00003466 | -0.00006649 | -0.00014126 |
| 63 | 75 | 0.06520136 | 0.04451582 | -0.05365552 | -0.01314755 | 0.02900052 | -0.00463278 | -0.01127495 | 0.00189709 | 0.00082025 | -0.00014009 | 0.00042093 | -0.00008610 | -0.00018913 | 0.00031165 | 0.00001750 | -0.00011429 | -0.00012293 |
| 63 | 76 | 0.06549300 | 0.04455350 | -0.05409827 | -0.01297017 | 0.02905515 | -0.00477905 | -0.01096218 | 0.00227723 | 0.00073885 | -0.00043193 | 0.00055342 | -0.00009173 | -0.00023670 | 0.00021715 | -0.00000506 | -0.00007448 | -0.00007819 |
| 63 | 77 | 0.06514947 | 0.04389572 | -0.05511873 | -0.01258705 | 0.02941759 | -0.00505718 | -0.01102799 | 0.00219452 | 0.00082833 | -0.00021950 | 0.00047860 | -0.00016135 | -0.00013409 | 0.00024998 | 0.00003832 | -0.00011878 | -0.00004859 |
| 63 | 78 | 0.06530582 | 0.04401651 | -0.05530673 | -0.01261745 | 0.02949975 | -0.00506195 | -0.01081347 | 0.00245652 | 0.00081150 | -0.00047511 | 0.00059673 | -0.00016428 | -0.00016937 | 0.00016320 | 0.00002899 | -0.00008934 | -0.00000701 |
| 63 | 79 | 0.06505778 | 0.04337608 | -0.05640968 | -0.01209870 | 0.02991937 | -0.00524290 | -0.01068623 | 0.00243623 | 0.00084954 | -0.00027642 | 0.00053246 | -0.00022725 | -0.00007669 | 0.00018683 | 0.00005484 | -0.00012613 | 0.00002198 |
| 63 | 80 | 0.06510607 | 0.04355393 | -0.05639296 | -0.01234472 | 0.03000775 | -0.00508221 | -0.01055511 | 0.00259503 | 0.00089674 | -0.00048239 | 0.00063329 | -0.00022012 | -0.00011308 | 0.00011364 | 0.00004954 | -0.00009644 | 0.00005549 |
| 63 | 81 | 0.06494360 | 0.04292040 | -0.05752964 | -0.01168179 | 0.03046092 | -0.00518286 | -0.01027943 | 0.00263060 | 0.00086982 | -0.00030367 | 0.00057474 | -0.00026567 | -0.00004146 | 0.00013423 | 0.00005312 | -0.00012524 | 0.00007151 |
| 63 | 82 | 0.06491385 | 0.04312901 | -0.05737850 | -0.01213584 | 0.03055780 | -0.00484751 | -0.01022654 | 0.00270264 | 0.00097949 | -0.00045427 | 0.00065704 | -0.00024446 | -0.00008826 | 0.00007756 | 0.00004547 | -0.00008716 | 0.00009480 |
| 63 | 83 | 0.06484921 | 0.04200870 | -0.05811401 | -0.01045567 | 0.03084564 | -0.00578116 | -0.01020546 | 0.00289901 | 0.00077362 | -0.00036047 | 0.00058168 | -0.00026217 | -0.00006595 | 0.00013845 | 0.00003918 | -0.00013125 | 0.00005581 |
| 63 | 84 | 0.06480744 | 0.04172811 | -0.05733130 | -0.01018025 | 0.03048645 | -0.00621692 | -0.01038791 | 0.00298085 | 0.00069231 | -0.00053060 | 0.00062893 | -0.00022959 | -0.00013932 | 0.00013544 | 0.00002947 | -0.00011997 | 0.00002056 |
| 63 | 85 | 0.06298786 | 0.04240767 | -0.05487847 | -0.01178031 | 0.02853411 | -0.00382625 | -0.00881442 | 0.00196881 | 0.00016049 | -0.00036911 | 0.00069083 | -0.00041443 | 0.00006519 | 0.00000986 | 0.00003307 | -0.00017892 | 0.00013704 |
| 63 | 86 | 0.06467623 | 0.03978877 | -0.05791702 | -0.00759836 | 0.03096825 | -0.00741449 | -0.01071222 | 0.00327746 | 0.00045763 | -0.00050155 | 0.00060993 | -0.00011314 | -0.00031650 | 0.00019165 | -0.00007603 | -0.00005572 | -0.00008495 |
| 63 | 87 | 0.04222586 | 0.05914695 | -0.01577699 | -0.04764138 | 0.00479313 | 0.02639711 | -0.00075022 | -0.01232338 | -0.00099032 | 0.00284485 | 0.00003670 | -0.00057161 | 0.00083376 | -0.00011480 | -0.00008994 | -0.00029933 | 0.00016418 |
| 63 | 88 | 0.04294764 | 0.05815629 | -0.01757166 | -0.04601375 | 0.00618488 | 0.02498215 | -0.00198128 | -0.01200541 | -0.00080449 | 0.00275963 | 0.00000171 | -0.00046666 | 0.00073236 | -0.00008267 | -0.00012577 | -0.00017984 | 0.00015317 |
| 63 | 89 | 0.04263796 | 0.05606087 | -0.01904931 | -0.04386649 | 0.00738542 | 0.02429775 | -0.00275177 | -0.01151925 | -0.00050163 | 0.00293843 | -0.00003587 | -0.00025582 | 0.00043241 | -0.00005883 | -0.00027640 | 0.00000583 | 0.00006780 |
| 63 | 90 | 0.04257371 | 0.05604719 | -0.01919003 | -0.04401724 | 0.00797797 | 0.02415949 | -0.00359452 | -0.01179211 | -0.00028782 | 0.00293801 | -0.00011345 | -0.00020895 | 0.00045040 | -0.00005314 | -0.00026345 | 0.00003842 | 0.00006175 |
| 63 | 91 | 0.04253764 | 0.05513963 | -0.01973595 | -0.04303065 | 0.00843015 | 0.02381430 | -0.00381683 | -0.01150395 | -0.00022787 | 0.00287838 | -0.00010037 | -0.00019750 | 0.00037037 | -0.00009037 | -0.00031087 | 0.00009943 | 0.00005868 |
| 63 | 92 | 0.04243695 | 0.05516790 | -0.01987278 | -0.04316131 | 0.00925676 | 0.02361479 | -0.00474953 | -0.01177780 | 0.00010185 | 0.00290996 | -0.00019995 | -0.00016570 | 0.00042679 | -0.00008849 | -0.00027185 | 0.00009827 | 0.00005555 |
| 63 | 93 | 0.04230702 | 0.05441872 | -0.02010651 | -0.04235665 | 0.00934249 | 0.02328992 | -0.00492083 | -0.01155114 | 0.00009220 | 0.00281805 | -0.00016435 | -0.00016955 | 0.00035036 | -0.00013168 | -0.00031862 | 0.00016561 | 0.00006368 |
| 63 | 94 | 0.04220314 | 0.05442346 | -0.02033347 | -0.04239912 | 0.01038561 | 0.02295063 | -0.00594879 | -0.01178627 | 0.00053174 | 0.00287850 | -0.00028329 | -0.00014375 | 0.00043071 | -0.00013118 | -0.00026001 | 0.00013958 | 0.00006216 |
| 63 | 95 | 0.04199890 | 0.05377780 | -0.02032626 | -0.04167840 | 0.01015006 | 0.02258523 | -0.00606403 | -0.01158894 | 0.00045859 | 0.00275919 | -0.00022409 | -0.00016123 | 0.00035645 | -0.00017793 | -0.00030512 | 0.00021361 | 0.00008207 |
| 63 | 96 | 0.04192536 | 0.05374535 | -0.02068422 | -0.04163190 | 0.01135262 | 0.02209694 | -0.00716590 | -0.01176792 | 0.00099095 | 0.00284567 | -0.00036078 | -0.00013374 | 0.00044779 | -0.00017670 | -0.00023495 | 0.00017176 | 0.00008004 |
| 63 | 97 | 0.04164691 | 0.05320437 | -0.02044061 | -0.04099684 | 0.01083411 | 0.02171295 | -0.00719548 | -0.01159542 | 0.00085734 | 0.00270098 | -0.00027802 | -0.00016907 | 0.00038321 | -0.00022683 | -0.00027337 | 0.00024477 | 0.00011199 |
| 63 | 98 | 0.04164001 | 0.05313523 | -0.02095843 | -0.04086594 | 0.01214539 | 0.02108780 | -0.00835166 | -0.01170824 | 0.00146475 | 0.00281130 | -0.00043124 | -0.00013291 | 0.00047340 | -0.00022299 | -0.00019991 | 0.00019622 | 0.00010687 |
| 63 | 99 | 0.04128562 | 0.05268935 | -0.02049445 | -0.04031717 | 0.01140084 | 0.02071103 | -0.00827927 | -0.01155879 | 0.00127560 | 0.00264366 | -0.00032625 | -0.00018823 | 0.00042402 | -0.00027591 | -0.00022781 | 0.00026119 | 0.00014947 |
| 63 | 100 | 0.04137609 | 0.05259395 | -0.02118521 | -0.04010870 | 0.01277937 | 0.01997192 | -0.00947404 | -0.01160313 | 0.00194108 | 0.00277554 | -0.00049458 | -0.00013859 | 0.00050376 | -0.00026832 | -0.00015781 | 0.00021325 | 0.00013916 |
| 63 | 101 | 0.04094268 | 0.05222355 | -0.02052471 | -0.03964221 | 0.01186943 | 0.01962128 | -0.00929557 | -0.01147519 | 0.00170261 | 0.00258783 | -0.00036966 | -0.00021402 | 0.00047270 | -0.00032297 | -0.00017291 | 0.00026517 | 0.00019028 |
| 63 | 102 | 0.04115107 | 0.05211971 | -0.02138386 | -0.03936716 | 0.01328142 | 0.01879810 | -0.01051567 | -0.01145690 | 0.00241007 | 0.00273885 | -0.00055113 | -0.00014868 | 0.00053638 | -0.00031139 | -0.00011092 | 0.00022266 | 0.00017355 |
| 63 | 103 | 0.04063618 | 0.05179838 | -0.02055634 | -0.03897386 | 0.01226159 | 0.01848151 | -0.01023563 | -0.01134671 | 0.00212959 | 0.00253418 | -0.00040915 | -0.00024271 | 0.00052444 | -0.00036637 | -0.00011244 | 0.00025875 | 0.00023085 |
| 63 | 104 | 0.04097171 | 0.05170875 | -0.02156341 | -0.03864821 | 0.01367966 | 0.01760880 | -0.01146925 | -0.01127911 | 0.00286374 | 0.00270198 | -0.00060130 | -0.00016183 | 0.00056993 | -0.00035134 | -0.00006082 | 0.00022413 | 0.00020741 |
| 63 | 105 | 0.04058069 | 0.05111618 | -0.02132369 | -0.03829644 | 0.01252551 | 0.01740578 | -0.01068983 | -0.01065354 | 0.00246843 | 0.00239329 | -0.00047818 | -0.00016121 | 0.00043357 | -0.00035527 | -0.00012339 | 0.00029133 | 0.00019942 |
| 63 | 106 | 0.04120618 | 0.05053153 | -0.02316234 | -0.03751630 | 0.01421736 | 0.01636968 | -0.01153404 | -0.00987578 | 0.00318267 | 0.00244360 | -0.00070243 | 0.00002340 | 0.00034036 | -0.00027928 | -0.00012313 | 0.00026500 | 0.00008952 |
| 63 | 107 | 0.04077953 | 0.05013543 | -0.02282641 | -0.03747259 | 0.01280355 | 0.01623584 | -0.01075603 | -0.00935603 | 0.00274392 | 0.00213825 | -0.00058414 | 0.00002396 | 0.00021594 | -0.00029122 | -0.00018786 | 0.00033438 | 0.00008809 |
| 64 | 69 | 0.06649136 | 0.04698047 | -0.04935081 | -0.01345981 | 0.02861466 | -0.00353967 | -0.01086046 | 0.00162029 | 0.00074836 | -0.00016532 | 0.00042215 | 0.00002155 | -0.00021798 | 0.00031602 | 0.00000299 | -0.00021994 | -0.00027255 |
| 64 | 70 | 0.06685901 | 0.04688682 | -0.04974432 | -0.01325633 | 0.02906682 | -0.00299376 | -0.01064918 | 0.00147625 | 0.00051965 | -0.00026606 | 0.00045486 | 0.00014271 | -0.00034856 | 0.00026923 | -0.00010671 | -0.00011490 | -0.00032898 |
| 64 | 71 | 0.06647728 | 0.04598833 | -0.05092551 | -0.01299420 | 0.02841352 | -0.00429220 | -0.01082111 | 0.00197635 | 0.00061314 | -0.00021616 | 0.00045594 | 0.00001761 | -0.00027931 | 0.00032754 | -0.00004083 | -0.00016449 | -0.00027830 |
| 64 | 72 | 0.06669132 | 0.04593388 | -0.05113166 | -0.01280633 | 0.02889478 | -0.00377023 | -0.01063014 | 0.00178105 | 0.00042357 | -0.00033042 | 0.00049153 | 0.00011349 | -0.00036870 | 0.00027075 | -0.00012301 | -0.00007848 | -0.00030420 |
| 64 | 73 | 0.06639097 | 0.04520727 | -0.05233520 | -0.01254119 | 0.02840388 | -0.00501989 | -0.01082621 | 0.00229454 | 0.00057902 | -0.00029297 | 0.00049244 | -0.00001781 | -0.00028490 | 0.00030907 | -0.00004459 | -0.00013992 | -0.00024690 |
| 64 | 74 | 0.06649331 | 0.04516970 | -0.05241920 | -0.01237865 | 0.02890728 | -0.00452295 | -0.01066357 | 0.00207051 | 0.00042851 | -0.00041520 | 0.00052943 | 0.00005845 | -0.00034480 | 0.00024841 | -0.00010456 | -0.00006770 | -0.00024892 |
| 64 | 75 | 0.06623752 | 0.04461454 | -0.05356731 | -0.01212648 | 0.02853376 | -0.00566978 | -0.01081030 | 0.00257595 | 0.00062445 | -0.00037882 | 0.00053276 | -0.00007903 | -0.00024403 | 0.00026710 | -0.00001761 | -0.00013977 | -0.00018406 |
| 64 | 76 | 0.06627081 | 0.04456246 | -0.05360273 | -0.01198865 | 0.02907656 | -0.00517850 | -0.01067701 | 0.00234964 | 0.00051398 | -0.00050371 | 0.00056932 | -0.00001517 | -0.00028807 | 0.00020836 | -0.00006223 | -0.00007493 | -0.00017054 |
| 64 | 77 | 0.06603562 | 0.04415899 | -0.05464714 | -0.01177312 | 0.02876194 | -0.00615843 | -0.01070770 | 0.00281569 | 0.00071548 | -0.00045292 | 0.00057609 | -0.00015303 | -0.00017622 | 0.00021183 | 0.00002354 | -0.00015160 | -0.00010198 |
| 64 | 78 | 0.06603725 | 0.04405992 | -0.05470691 | -0.01165590 | 0.02937105 | -0.00565429 | -0.01060745 | 0.00261492 | 0.00065012 | -0.00057620 | 0.00061020 | -0.00009453 | -0.00021709 | 0.00015976 | -0.00001186 | -0.00008852 | -0.00008099 |
| 64 | 79 | 0.06581197 | 0.04378200 | -0.05561567 | -0.01149003 | 0.02906543 | -0.00640729 | -0.01048144 | 0.00301014 | 0.00081836 | -0.00049814 | 0.00061890 | -0.00022247 | -0.00010617 | 0.00015495 | 0.00006020 | -0.00016143 | -0.00001715 |
| 64 | 80 | 0.06581092 | 0.04360740 | -0.05576598 | -0.01138722 | 0.02976805 | -0.00588737 | -0.01042459 | 0.00286268 | 0.00080833 | -0.00061746 | 0.00064871 | -0.00016402 | -0.00015334 | 0.00011268 | 0.00003027 | -0.00009660 | 0.00000514 |
| 64 | 81 | 0.06559382 | 0.04343443 | -0.05651278 | -0.01126607 | 0.02943856 | -0.00638331 | -0.01014389 | 0.00316403 | 0.00091079 | -0.00050828 | 0.00065549 | -0.00027066 | -0.00005671 | 0.00010667 | 0.00007761 | -0.00015887 | 0.00005352 |
| 64 | 82 | 0.06561116 | 0.04316138 | -0.05681184 | -0.01116801 | 0.03025905 | -0.00586330 | -0.01014562 | 0.00309627 | 0.00097018 | -0.00062316 | 0.00067971 | -0.00020930 | -0.00011569 | 0.00007557 | 0.00005212 | -0.00009120 | 0.00007362 |
| 64 | 83 | 0.06547459 | 0.04260064 | -0.05697468 | -0.01023598 | 0.02965329 | -0.00696366 | -0.01011759 | 0.00336018 | 0.00082895 | -0.00054315 | 0.00064965 | -0.00026750 | -0.00007701 | 0.00011653 | 0.00006662 | -0.00016476 | 0.00003734 |
| 64 | 84 | 0.06575922 | 0.04056567 | -0.05831429 | -0.00771905 | 0.03076281 | -0.00772021 | -0.01067646 | 0.00346345 | 0.00060924 | -0.00053383 | 0.00067503 | -0.00003580 | -0.00038565 | 0.00016761 | -0.00010206 | -0.00000365 | -0.00008124 |
| 64 | 85 | 0.06549984 | 0.04045831 | -0.05802599 | -0.00728529 | 0.03038818 | -0.00844828 | -0.01072349 | 0.00371177 | 0.00062946 | -0.00051276 | 0.00062682 | -0.00011820 | -0.00029600 | 0.00017969 | -0.00005485 | -0.00008111 | -0.00008250 |
| 64 | 86 | 0.04384477 | 0.05886209 | -0.01850668 | -0.04611158 | 0.00630228 | 0.02451898 | -0.00294662 | -0.01232860 | -0.00060646 | 0.00300279 | -0.00001021 | -0.00032792 | 0.00063977 | -0.00004403 | -0.00019964 | -0.00005498 | 0.00013536 |
| 64 | 87 | 0.04356115 | 0.05881028 | -0.01806609 | -0.04598409 | 0.00625469 | 0.02429864 | -0.00284451 | -0.01200264 | -0.00039124 | 0.00288092 | -0.00002272 | -0.00041464 | 0.00071524 | -0.00010997 | -0.00013254 | -0.00013754 | 0.00016310 |
| 64 | 88 | 0.04347036 | 0.05746436 | -0.01934601 | -0.04498436 | 0.00755712 | 0.02434979 | -0.00387208 | -0.01222132 | -0.00021880 | 0.00312318 | -0.00010770 | -0.00017050 | 0.00047424 | -0.00003294 | -0.00027626 | 0.00005247 | 0.00005683 |
| 64 | 89 | 0.04314935 | 0.05701416 | -0.01943302 | -0.04452673 | 0.00794975 | 0.02386217 | -0.00429890 | -0.01190556 | 0.00010526 | 0.00303514 | -0.00014029 | -0.00019650 | 0.00048449 | -0.00009117 | -0.00024621 | 0.00004451 | 0.00008138 |
| 64 | 90 | 0.04344920 | 0.05658086 | -0.02006326 | -0.04403951 | 0.00892777 | 0.02394312 | -0.00477400 | -0.01209449 | 0.00016781 | 0.00311850 | -0.00019864 | -0.00013555 | 0.00045627 | -0.00005365 | -0.00027401 | 0.00007810 | 0.00003243 |
| 64 | 91 | 0.04302691 | 0.05609421 | -0.02023168 | -0.04369803 | 0.00936553 | 0.02342730 | -0.00545200 | -0.01190498 | 0.00049709 | 0.00302076 | -0.00023170 | -0.00014873 | 0.00045736 | -0.00012147 | -0.00025849 | 0.00010799 | 0.00007204 |
| 64 | 92 | 0.04332483 | 0.05574211 | -0.02064993 | -0.04310577 | 0.01027466 | 0.02337987 | -0.00579006 | -0.01197180 | 0.00061066 | 0.00310395 | -0.00028917 | -0.00011449 | 0.00045828 | -0.00008248 | -0.00025403 | 0.00008648 | 0.00001572 |
| 64 | 93 | 0.04280252 | 0.05526570 | -0.02081831 | -0.04290309 | 0.01064907 | 0.02284262 | -0.00664773 | -0.01191148 | 0.00092229 | 0.00300191 | -0.00031875 | -0.00012063 | 0.00045478 | -0.00015744 | -0.00025192 | 0.00015311 | 0.00007370 |
| 64 | 94 | 0.04313153 | 0.05492226 | -0.02115833 | -0.04214946 | 0.01150654 | 0.02259569 | -0.00688520 | -0.01182929 | 0.00108763 | 0.00308003 | -0.00037478 | -0.00010173 | 0.00047016 | -0.00011665 | -0.00022273 | 0.00008879 | 0.00000965 |
| 64 | 95 | 0.04252572 | 0.05448959 | -0.02127539 | -0.04207956 | 0.01175703 | 0.02204581 | -0.00785016 | -0.01188127 | 0.00136901 | 0.00297778 | -0.00039791 | -0.00010593 | 0.00046605 | -0.00019623 | -0.00023148 | 0.00018707 | 0.00008637 |
| 64 | 96 | 0.04290155 | 0.05414327 | -0.02159845 | -0.04120159 | 0.01255905 | 0.02162101 | -0.00800177 | -0.01166422 | 0.00157664 | 0.00304849 | -0.00045283 | -0.00009546 | 0.00048770 | -0.00015447 | -0.00018455 | 0.00009016 | 0.00001501 |
| 64 | 97 | 0.04223641 | 0.05377831 | -0.02163476 | -0.04124517 | 0.01266526 | 0.02107094 | -0.00900859 | -0.01180106 | 0.00182295 | 0.00294756 | -0.00046764 | -0.00010234 | 0.00048703 | -0.00023648 | -0.00020026 | 0.00021149 | 0.00010841 |
| 64 | 98 | 0.04266583 | 0.05342632 | -0.02198284 | -0.04028774 | 0.01341327 | 0.02051092 | -0.00909350 | -0.01147860 | 0.00206098 | 0.00301167 | -0.00052223 | -0.00009392 | 0.00050778 | -0.00019432 | -0.00014299 | 0.00009280 | 0.00003018 |
| 64 | 99 | 0.04196694 | 0.05313990 | -0.02192789 | -0.04041620 | 0.01338514 | 0.01997070 | -0.01009075 | -0.01166871 | 0.00227285 | 0.00291159 | -0.00052785 | -0.00010718 | 0.00051399 | -0.00027682 | -0.00016131 | 0.00022714 | 0.00013667 |
| 64 | 100 | 0.04244711 | 0.05278726 | -0.02232056 | -0.03942575 | 0.01407908 | 0.01932484 | -0.01012883 | -0.01127823 | 0.00252891 | 0.00297177 | -0.00058280 | -0.00009596 | 0.00052872 | -0.00023485 | -0.00010023 | 0.00009667 | 0.00005266 |
| 64 | 101 | 0.04173695 | 0.05257676 | -0.02217750 | -0.03960566 | 0.01394472 | 0.01879920 | -0.01108070 | -0.01149089 | 0.00271004 | 0.00287088 | -0.00057918 | -0.00011808 | 0.00054427 | -0.00031606 | -0.00011705 | 0.00023430 | 0.00016791 |
| 64 | 102 | 0.04225838 | 0.05223509 | -0.02261577 | -0.03862637 | 0.01458084 | 0.01811566 | -0.01108958 | -0.01107118 | 0.00297267 | 0.00293058 | -0.00063488 | -0.00010097 | 0.00055006 | -0.00027502 | -0.00005729 | 0.00010053 | 0.00007987 |
| 64 | 103 | 0.04155414 | 0.05208701 | -0.02239605 | -0.03882420 | 0.01437622 | 0.01760415 | -0.01197395 | -0.01127973 | 0.00312804 | 0.00282693 | -0.00062250 | -0.00013333 | 0.00057634 | -0.00035332 | -0.00006918 | 0.00023307 | 0.00019949 |
| 64 | 104 | 0.04210471 | 0.05177221 | -0.02286928 | -0.03789426 | 0.01494812 | 0.01692478 | -0.01196795 | -0.01086579 | 0.00338771 | 0.00288946 | -0.00067909 | -0.00010873 | 0.00057205 | -0.00031408 | -0.00001448 | 0.00010275 | 0.00010958 |
| 64 | 105 | 0.04163190 | 0.05132053 | -0.02328636 | -0.03785291 | 0.01481231 | 0.01639820 | -0.01240324 | -0.01044339 | 0.00348816 | 0.00267042 | -0.00069127 | -0.00005124 | 0.00047801 | -0.00033581 | -0.00007608 | 0.00025003 | 0.00015620 |
| 64 | 106 | 0.04240252 | 0.05055439 | -0.02444202 | -0.03652408 | 0.01562083 | 0.01559737 | -0.01199195 | -0.00938194 | 0.00371172 | 0.00261134 | -0.00076924 | 0.00005728 | 0.00035580 | -0.00024022 | -0.00006014 | 0.00011731 | -0.00001100 |
| 64 | 107 | 0.04194110 | 0.05023614 | -0.02482723 | -0.03658671 | 0.01542131 | 0.01509351 | -0.01247777 | -0.00897639 | 0.00382185 | 0.00238312 | -0.00079442 | 0.00012400 | 0.00026343 | -0.00026575 | -0.00012151 | 0.00026135 | 0.00003015 |
| 64 | 108 | 0.04268226 | 0.04939739 | -0.02592541 | -0.03511363 | 0.01639551 | 0.01423066 | -0.01206406 | -0.00792085 | 0.00404380 | 0.00231112 | -0.00085782 | 0.00020765 | 0.00016716 | -0.00017316 | -0.00008126 | 0.00009678 | -0.00013450 |
| 65 | 70 | 0.06709180 | 0.04587658 | -0.05087225 | -0.01235788 | 0.02854326 | -0.00483502 | -0.01088735 | 0.00202325 | 0.00058988 | -0.00010613 | 0.00043298 | 0.00000534 | -0.00024401 | 0.00035023 | -0.00003058 | -0.00024323 | -0.00036056 |
| 65 | 71 | 0.06661656 | 0.04446457 | -0.05244661 | -0.01140924 | 0.02926575 | -0.00549364 | -0.01172357 | 0.00189342 | 0.00079845 | 0.00023809 | 0.00034742 | 0.00003207 | -0.00028101 | 0.00042542 | -0.00006096 | -0.00018178 | -0.00038909 |
| 65 | 72 | 0.06705143 | 0.04513350 | -0.05224583 | -0.01192495 | 0.02842174 | -0.00560428 | -0.01089393 | 0.00239558 | 0.00055866 | -0.00019071 | 0.00046163 | -0.00001386 | -0.00026951 | 0.00034101 | -0.00004606 | -0.00020722 | -0.00033947 |
| 65 | 73 | 0.06663165 | 0.04394171 | -0.05370218 | -0.01094628 | 0.02911502 | -0.00629323 | -0.01157952 | 0.00236565 | 0.00080021 | 0.00012335 | 0.00038557 | -0.00000937 | -0.00027423 | 0.00039206 | -0.00006092 | -0.00016272 | -0.00034451 |
| 65 | 74 | 0.06694190 | 0.04460080 | -0.05340389 | -0.01151309 | 0.02840277 | -0.00633734 | -0.01089339 | 0.00273697 | 0.00060662 | -0.00029167 | 0.00049519 | -0.00006028 | -0.00024810 | 0.00030795 | -0.00002805 | -0.00019518 | -0.00028343 |
| 65 | 75 | 0.06655537 | 0.04358413 | -0.05474631 | -0.01051381 | 0.02901276 | -0.00704185 | -0.01137212 | 0.00279454 | 0.00084433 | -0.00000748 | 0.00043396 | -0.00007614 | -0.00022416 | 0.00033647 | -0.00003229 | -0.00016615 | -0.00026681 |
| 65 | 76 | 0.06677413 | 0.04422046 | -0.05438537 | -0.01114213 | 0.02845553 | -0.00697052 | -0.01082165 | 0.00304269 | 0.00070260 | -0.00039154 | 0.00053505 | -0.00012656 | -0.00019247 | 0.00025820 | 0.00001140 | -0.00019899 | -0.00020083 |
| 65 | 77 | 0.06640918 | 0.04332589 | -0.05562814 | -0.01012288 | 0.02896002 | -0.00764532 | -0.01105417 | 0.00316545 | 0.00089800 | -0.00013321 | 0.00049054 | -0.00015568 | -0.00014955 | 0.00026877 | 0.00000917 | -0.00018112 | -0.00016953 |
| 65 | 78 | 0.06656853 | 0.04392496 | -0.05524674 | -0.01082409 | 0.02857043 | -0.00742232 | -0.01062775 | 0.00330645 | 0.00081289 | -0.00047294 | 0.00058003 | -0.00019987 | -0.00012178 | 0.00020098 | 0.00005627 | -0.00020747 | -0.00010494 |
| 65 | 79 | 0.06622351 | 0.04310492 | -0.05639941 | -0.00977286 | 0.02897154 | -0.00801870 | -0.01060974 | 0.00347139 | 0.00093404 | -0.00023692 | 0.00054982 | -0.00023070 | -0.00007425 | 0.00020058 | 0.00004584 | -0.00019541 | -0.00007081 |
| 65 | 80 | 0.06635146 | 0.04365502 | -0.05604382 | -0.01055642 | 0.02875591 | -0.00762880 | -0.01029783 | 0.00352677 | 0.00091098 | -0.00052460 | 0.00062593 | -0.00026462 | -0.00005749 | 0.00014593 | 0.00009083 | -0.00021006 | -0.00001191 |
| 65 | 81 | 0.06602875 | 0.04287749 | -0.05709986 | -0.00945121 | 0.02906435 | -0.00812788 | -0.01006930 | 0.00371865 | 0.00093993 | -0.00031204 | 0.00060404 | -0.00028433 | -0.00002037 | 0.00014212 | 0.00006402 | -0.00020004 | 0.00001107 |
| 65 | 82 | 0.06614891 | 0.04336983 | -0.05682009 | -0.01032067 | 0.02902976 | -0.00757441 | -0.00986531 | 0.00371178 | 0.00098452 | -0.00054549 | 0.00066643 | -0.00030643 | -0.00001821 | 0.00010107 | 0.00010390 | -0.00020004 | 0.00006264 |
| 65 | 83 | 0.06410811 | 0.04378956 | -0.05439054 | -0.01146545 | 0.02697437 | -0.00563802 | -0.00870269 | 0.00259676 | 0.00052442 | -0.00024398 | 0.00073279 | -0.00043368 | 0.00013360 | -0.00003637 | 0.00005525 | -0.00020860 | 0.00015800 |
| 65 | 84 | 0.06638425 | 0.04097176 | -0.05843597 | -0.00691065 | 0.03000904 | -0.00941672 | -0.01075189 | 0.00414738 | 0.00081794 | -0.00050629 | 0.00064189 | -0.00011775 | -0.00029148 | 0.00017468 | -0.00004040 | -0.00008936 | -0.00007323 |
| 65 | 85 | 0.04317942 | 0.06033756 | -0.01639869 | -0.04745791 | 0.00482474 | 0.02484335 | -0.00240828 | -0.01226201 | -0.00012913 | 0.00313014 | -0.00001748 | -0.00048330 | 0.00083110 | -0.00017319 | -0.00009691 | -0.00023300 | 0.00017915 |
| 65 | 86 | 0.04427135 | 0.05940236 | -0.01857124 | -0.04589154 | 0.00620907 | 0.02370319 | -0.00343469 | -0.01194192 | -0.00008241 | 0.00297875 | -0.00003113 | -0.00037713 | 0.00069207 | -0.00012792 | -0.00014266 | -0.00010231 | 0.00017309 |
| 65 | 87 | 0.04335406 | 0.05798049 | -0.01897660 | -0.04498045 | 0.00706657 | 0.02354385 | -0.00418376 | -0.01180460 | 0.00035062 | 0.00319700 | -0.00010679 | -0.00023697 | 0.00053583 | -0.00014280 | -0.00023339 | 0.00000207 | 0.00010396 |
| 65 | 88 | 0.04387577 | 0.05788963 | -0.01973026 | -0.04484334 | 0.00779383 | 0.02356307 | -0.00473073 | -0.01191302 | 0.00038791 | 0.00310876 | -0.00014831 | -0.00019208 | 0.00050157 | -0.00011661 | -0.00023691 | 0.00004868 | 0.00009731 |
| 65 | 89 | 0.04322919 | 0.05700310 | -0.01990131 | -0.04431157 | 0.00839833 | 0.02327106 | -0.00536633 | -0.01187816 | 0.00066874 | 0.00317124 | -0.00018835 | -0.00016366 | 0.00046739 | -0.00016872 | -0.00027500 | 0.00010945 | 0.00009254 |
| 65 | 90 | 0.04379375 | 0.05695625 | -0.02063368 | -0.04403615 | 0.00931593 | 0.02322308 | -0.00586745 | -0.01190104 | 0.00078321 | 0.00310078 | -0.00024329 | -0.00014059 | 0.00047040 | -0.00014273 | -0.00025159 | 0.00011159 | 0.00008183 |
| 65 | 91 | 0.04298072 | 0.05611100 | -0.02057416 | -0.04367199 | 0.00964904 | 0.02292826 | -0.00658318 | -0.01199126 | 0.00101012 | 0.00315106 | -0.00026860 | -0.00011101 | 0.00042551 | -0.00019735 | -0.00029991 | 0.00019967 | 0.00009003 |
| 65 | 92 | 0.04359686 | 0.05607317 | -0.02134028 | -0.04321746 | 0.01074528 | 0.02271440 | -0.00705963 | -0.01188894 | 0.00121238 | 0.00308940 | -0.00033435 | -0.00010712 | 0.00046235 | -0.00017339 | -0.00024794 | 0.00015609 | 0.00007599 |
| 65 | 93 | 0.04264875 | 0.05525802 | -0.02107222 | -0.04296866 | 0.01077197 | 0.02238429 | -0.00780540 | -0.01207908 | 0.00137235 | 0.00312842 | -0.00034128 | -0.00007799 | 0.00040615 | -0.00022823 | -0.00030623 | 0.00027223 | 0.00009922 |
| 65 | 94 | 0.04333084 | 0.05522357 | -0.02190496 | -0.04235149 | 0.01200492 | 0.02197601 | -0.00826165 | -0.01183771 | 0.00166069 | 0.00307137 | -0.00041687 | -0.00008792 | 0.00046943 | -0.00020691 | -0.00022962 | 0.00018807 | 0.00008141 |
| 65 | 95 | 0.04227333 | 0.05446203 | -0.02142407 | -0.04221841 | 0.01172146 | 0.02163669 | -0.00897469 | -0.01211390 | 0.00174422 | 0.00309719 | -0.00040295 | -0.00006493 | 0.00040846 | -0.00026174 | -0.00029367 | 0.00032466 | 0.00012004 |
| 65 | 96 | 0.04303899 | 0.05442942 | -0.02235444 | -0.04146441 | 0.01305172 | 0.02103915 | -0.00941792 | -0.01173479 | 0.00211190 | 0.00304507 | -0.00048858 | -0.00008118 | 0.00048756 | -0.00024244 | -0.00019990 | 0.00020975 | 0.00009721 |
| 65 | 97 | 0.04189638 | 0.05373428 | -0.02167257 | -0.04144074 | 0.01249318 | 0.02072089 | -0.01005126 | -0.01208163 | 0.00211593 | 0.00305507 | -0.00045284 | -0.00006885 | 0.00042817 | -0.00029711 | -0.00026501 | 0.00035717 | 0.00014949 |
| 65 | 98 | 0.04275810 | 0.05370618 | -0.02271962 | -0.04057926 | 0.01388614 | 0.01995845 | -0.01049184 | -0.01157877 | 0.00255349 | 0.00301061 | -0.00054898 | -0.00008427 | 0.00051283 | -0.00027888 | -0.00016219 | 0.00022256 | 0.00012063 |
| 65 | 99 | 0.04155038 | 0.05307661 | -0.02185978 | -0.04065052 | 0.01310766 | 0.01968399 | -0.01101640 | -0.01197965 | 0.00248021 | 0.00300210 | -0.00049184 | -0.00008546 | 0.00046001 | -0.00033299 | -0.00022408 | 0.00037149 | 0.00018363 |

TABLE 3. Nuclear charge density distribution FB coefficients.

| $Z$ | $N$ | $a_1$ | $a_2$ | $a_3$ | $a_4$ | $a_5$ | $a_6$ | $a_7$ | $a_8$ | $a_9$ | $a_{10}$ | $a_{11}$ | $a_{12}$ | $a_{13}$ | $a_{14}$ | $a_{15}$ | $a_{16}$ | $a_{17}$ |
|---|---|---|---|---|---|---|---|---|---|---|---|---|---|---|---|---|---|---|
| 65 | 100 | 0.04251168 | 0.05306147 | -0.02302497 | -0.03971415 | 0.01453210 | 0.01879219 | -0.01146489 | -0.01137718 | 0.00297633 | 0.00296910 | -0.00059857 | -0.00009477 | 0.00054230 | -0.00031518 | -0.00011921 | 0.00022724 | 0.00014855 |
| 65 | 101 | 0.04125475 | 0.05248445 | -0.02201786 | -0.03985883 | 0.01359515 | 0.01857159 | -0.01186728 | -0.01181374 | 0.00283208 | 0.00293968 | -0.00052159 | -0.00011061 | 0.00049919 | -0.00036801 | -0.00017450 | 0.00036992 | 0.00021885 |
| 65 | 102 | 0.04231028 | 0.05249653 | -0.02328550 | -0.03888281 | 0.01502225 | 0.01759251 | -0.01233189 | -0.01114304 | 0.00337414 | 0.00292221 | -0.00063834 | -0.00011079 | 0.00057422 | -0.00035048 | -0.00007280 | 0.00022412 | 0.00017828 |
| 65 | 103 | 0.04101753 | 0.05195020 | -0.02216756 | -0.03907471 | 0.01398616 | 0.01742274 | -0.01261057 | -0.01159469 | 0.00316840 | 0.00286991 | -0.00054387 | -0.00014089 | 0.00054200 | -0.00040099 | -0.00011923 | 0.00035477 | 0.00025237 |
| 65 | 104 | 0.04215502 | 0.05200815 | -0.02350875 | -0.03809526 | 0.01538996 | 0.01640135 | -0.01309636 | -0.01089104 | 0.00374320 | 0.00287184 | -0.00066951 | -0.00013094 | 0.00060771 | -0.00038408 | -0.00002418 | 0.00021338 | 0.00020774 |
| 65 | 105 | 0.04108741 | 0.05130005 | -0.02296026 | -0.03819624 | 0.01425917 | 0.01627921 | -0.01293479 | -0.01078501 | 0.00345985 | 0.00269767 | -0.00059897 | -0.00007210 | 0.00045807 | -0.00038895 | -0.00012492 | 0.00036848 | 0.00021827 |
| 65 | 106 | 0.04250485 | 0.05099679 | -0.02503461 | -0.03683688 | 0.01593869 | 0.01514847 | -0.01315664 | -0.00942140 | 0.00407405 | 0.00258791 | -0.00077168 | 0.00004604 | 0.00038653 | -0.00031539 | -0.00007857 | 0.00023912 | 0.00008593 |
| 65 | 107 | 0.04143367 | 0.05048319 | -0.02437625 | -0.03712950 | 0.01453239 | 0.01503050 | -0.01293443 | -0.00937226 | 0.00372761 | 0.00240329 | -0.00069586 | 0.00009366 | 0.00025747 | -0.00033163 | -0.00017802 | 0.00039043 | 0.00010880 |
| 65 | 108 | 0.04283495 | 0.05001404 | -0.02646440 | -0.03552289 | 0.01657140 | 0.01383218 | -0.01324391 | -0.00795314 | 0.00441072 | 0.00228148 | -0.00087499 | 0.00021102 | 0.00018808 | -0.00024909 | -0.00010939 | 0.00022743 | -0.00004417 |
| 65 | 109 | 0.04177018 | 0.04967204 | -0.02571799 | -0.03598454 | 0.01487828 | 0.01366205 | -0.01294080 | -0.00793191 | 0.00399608 | 0.00207955 | -0.00079252 | 0.00024499 | 0.00008201 | -0.00027525 | -0.00020429 | 0.00036863 | -0.00000953 |
| 66 | 72 | 0.06927725 | 0.04149173 | -0.05728205 | -0.00620201 | 0.03155863 | -0.00760069 | -0.01157545 | 0.00325636 | 0.00039193 | -0.00003869 | 0.00050768 | 0.00040610 | -0.00086377 | 0.00044326 | -0.00038603 | 0.00011171 | -0.00057167 |
| 66 | 73 | 0.06777988 | 0.04408686 | -0.05415929 | -0.01046417 | 0.02891463 | -0.00713568 | -0.01131042 | 0.00291236 | 0.00068613 | -0.00015099 | 0.00044874 | 0.00000103 | -0.00031190 | 0.00035741 | -0.00006634 | -0.00020313 | -0.00038811 |
| 66 | 74 | 0.06908155 | 0.04123684 | -0.05804854 | -0.00607286 | 0.03135588 | -0.00826237 | -0.01161261 | 0.00354745 | 0.00045946 | -0.00016000 | 0.00053660 | 0.00034407 | -0.00081612 | 0.00041218 | -0.00034725 | 0.00011163 | -0.00049424 |
| 66 | 75 | 0.06767114 | 0.04379182 | -0.05504334 | -0.01009059 | 0.02883164 | -0.00784178 | -0.01125140 | 0.00327567 | 0.00077403 | -0.00027254 | 0.00048272 | -0.00005300 | -0.00027315 | 0.00031688 | -0.00003353 | -0.00019833 | -0.00031211 |
| 66 | 76 | 0.06882299 | 0.04111572 | -0.05865689 | -0.00596767 | 0.03123861 | -0.00882866 | -0.01158008 | 0.00382583 | 0.00057896 | -0.00028715 | 0.00057413 | 0.00026059 | -0.00073501 | 0.00036220 | -0.00028565 | 0.00009383 | -0.00039297 |
| 66 | 77 | 0.06750870 | 0.04358774 | -0.05579052 | -0.00974781 | 0.02878682 | -0.00843606 | -0.01108363 | 0.00360415 | 0.00088047 | -0.00038620 | 0.00052581 | -0.00012246 | -0.00020983 | 0.00026420 | 0.00001293 | -0.00020390 | -0.00021535 |
| 66 | 78 | 0.06850427 | 0.04107791 | -0.05914011 | -0.00592124 | 0.03117704 | -0.00923259 | -0.01142296 | 0.00408414 | 0.00072627 | -0.00040290 | 0.00061981 | 0.00016332 | -0.00063235 | 0.00030012 | -0.00021236 | 0.00006543 | -0.00027638 |
| 66 | 79 | 0.06731144 | 0.04341132 | -0.05646246 | -0.00944189 | 0.02879298 | -0.00884551 | -0.01077502 | 0.00389255 | 0.00097917 | -0.00047841 | 0.00057586 | -0.00019541 | -0.00013899 | 0.00020725 | 0.00005915 | -0.00021073 | -0.00011058 |
| 66 | 80 | 0.06813606 | 0.04107525 | -0.05953294 | -0.00595723 | 0.03115323 | -0.00940651 | -0.01110839 | 0.00431506 | 0.00087770 | -0.00049233 | 0.00067046 | 0.00006434 | -0.00052485 | 0.00023443 | -0.00014067 | 0.00003525 | -0.00015659 |
| 66 | 81 | 0.06710513 | 0.04321415 | -0.05711325 | -0.00916414 | 0.02887693 | -0.00901996 | -0.01033193 | 0.00414225 | 0.00105278 | -0.00054214 | 0.00062769 | -0.00025766 | -0.00007920 | 0.00015418 | 0.00009222 | -0.00021084 | -0.00001292 |
| 66 | 82 | 0.06773411 | 0.04107242 | -0.05986347 | -0.00608293 | 0.03116149 | -0.00930947 | -0.01064272 | 0.00451632 | 0.00101715 | -0.00054812 | 0.00072021 | -0.00002264 | -0.00043022 | 0.00017349 | -0.00008250 | 0.00001082 | -0.00004750 |
| 66 | 83 | 0.06728491 | 0.04129431 | -0.05910036 | -0.00642381 | 0.02985267 | -0.01032111 | -0.01079085 | 0.00458101 | 0.00100594 | -0.00049286 | 0.00065825 | -0.00011076 | -0.00030313 | 0.00017550 | -0.00003254 | -0.00008398 | -0.00005954 |
| 66 | 84 | 0.04539611 | 0.05995683 | -0.01950298 | -0.04573440 | 0.00636937 | 0.02360751 | -0.00373318 | -0.01213497 | -0.00015475 | 0.00311168 | 0.00000298 | -0.00029502 | 0.00060379 | -0.00006928 | -0.00020193 | -0.00000660 | 0.00018236 |
| 66 | 85 | 0.04501053 | 0.05992074 | -0.01905302 | -0.04574141 | 0.00610677 | 0.02318191 | -0.00381115 | -0.01185389 | 0.00013987 | 0.00305650 | -0.00002541 | -0.00035297 | 0.00066861 | -0.00013969 | -0.00015338 | -0.00007261 | 0.00018763 |
| 66 | 86 | 0.04515815 | 0.05898461 | -0.02014266 | -0.04518940 | 0.00735633 | 0.02375820 | -0.00444326 | -0.01209032 | 0.00017624 | 0.00319344 | -0.00009106 | -0.00018308 | 0.00048792 | -0.00006573 | -0.00025650 | 0.00006740 | 0.00011340 |
| 66 | 87 | 0.04466973 | 0.05865715 | -0.02005187 | -0.04500878 | 0.00755628 | 0.02325829 | -0.00495742 | -0.01185889 | 0.00057865 | 0.00316244 | -0.00014019 | -0.00019404 | 0.00050738 | -0.00013349 | -0.00023345 | 0.00005373 | 0.00011457 |
| 66 | 88 | 0.04526179 | 0.05816047 | -0.02100966 | -0.04435511 | 0.00873958 | 0.02350035 | -0.00525778 | -0.01192806 | 0.00054139 | 0.00318980 | -0.00018164 | -0.00014576 | 0.00046330 | -0.00008165 | -0.00025997 | 0.00009212 | 0.00007827 |
| 66 | 89 | 0.04464719 | 0.05773421 | -0.02104565 | -0.04423017 | 0.00913700 | 0.02300140 | -0.00606458 | -0.01182028 | 0.00097511 | 0.00315591 | -0.00023753 | -0.00013839 | 0.00046996 | -0.00015595 | -0.00025090 | 0.00011479 | 0.00009041 |
| 66 | 90 | 0.04522789 | 0.05731953 | -0.02175639 | -0.04345009 | 0.01021430 | 0.02308177 | -0.00621058 | -0.01175723 | 0.00097741 | 0.00318219 | -0.00027489 | -0.00012036 | 0.00045964 | -0.00010408 | -0.00024404 | 0.00009482 | 0.00004642 |
| 66 | 91 | 0.04449353 | 0.05682476 | -0.02186138 | -0.04340449 | 0.01068224 | 0.02256685 | -0.00724966 | -0.01177902 | 0.00141205 | 0.00314842 | -0.00033264 | -0.00009921 | 0.00045552 | -0.00018222 | -0.00024975 | 0.00015684 | 0.00007468 |
| 66 | 92 | 0.04508002 | 0.05645689 | -0.02240170 | -0.04247671 | 0.01164332 | 0.02242471 | -0.00726627 | -0.01156338 | 0.00145860 | 0.00316643 | -0.00036495 | -0.00010474 | 0.00047066 | -0.00013196 | -0.00021286 | 0.00008528 | 0.00002393 |
| 66 | 93 | 0.04424946 | 0.05592740 | -0.02252940 | -0.04251458 | 0.01208347 | 0.02188949 | -0.00845912 | -0.01170032 | 0.00187104 | 0.00313454 | -0.00041978 | -0.00007493 | 0.00045823 | -0.00021151 | -0.00023259 | 0.00018495 | 0.00007069 |
| 66 | 94 | 0.04485412 | 0.05560173 | -0.02295110 | -0.04147948 | 0.01292198 | 0.02154471 | -0.00836362 | -0.01134750 | 0.00195657 | 0.00314218 | -0.00044775 | -0.00009731 | 0.00049123 | -0.00016406 | -0.00017190 | 0.00007125 | 0.00001355 |
| 66 | 95 | 0.04396064 | 0.05507134 | -0.02306915 | -0.04159235 | 0.01327466 | 0.02099526 | -0.00963080 | -0.01157226 | 0.00233237 | 0.00311178 | -0.00049573 | -0.00006418 | 0.00047387 | -0.00024324 | -0.00020310 | 0.00020219 | 0.00007840 |
| 66 | 96 | 0.04459004 | 0.05478423 | -0.02341788 | -0.04049708 | 0.01399776 | 0.02049400 | -0.00944709 | -0.01111436 | 0.00244797 | 0.00311096 | -0.00052088 | -0.00009598 | 0.00051658 | -0.00019893 | -0.00012637 | 0.00005784 | 0.00001501 |
| 66 | 97 | 0.04366851 | 0.05427900 | -0.02350887 | -0.04066643 | 0.01424011 | 0.01993911 | -0.01072110 | -0.01139374 | 0.00278054 | 0.00307986 | -0.00055927 | -0.00006441 | 0.00049803 | -0.00027651 | -0.00016529 | 0.00021085 | 0.00009549 |
| 66 | 98 | 0.04432069 | 0.05402951 | -0.02381460 | -0.03955828 | 0.01485823 | 0.01933624 | -0.01047531 | -0.01087147 | 0.00291596 | 0.00307459 | -0.00058324 | -0.00009925 | 0.00054363 | -0.00023530 | -0.00007982 | 0.00004717 | 0.00002648 |
| 66 | 99 | 0.04340149 | 0.05356353 | -0.02387327 | -0.03975923 | 0.01499454 | 0.01878238 | -0.01170576 | -0.01117228 | 0.00320448 | 0.00303970 | -0.00061045 | -0.00007318 | 0.00052736 | -0.00031035 | -0.00012236 | 0.00021227 | 0.00011899 |
| 66 | 100 | 0.04406821 | 0.05335491 | -0.02414986 | -0.03868208 | 0.01551624 | 0.01813161 | -0.01142253 | -0.01062769 | 0.00334992 | 0.00303478 | -0.00063465 | -0.00010630 | 0.00057104 | -0.00027220 | -0.00003416 | 0.00003916 | 0.00004563 |
| 66 | 101 | 0.04317416 | 0.05293009 | -0.02417898 | -0.03888770 | 0.01556708 | 0.01758117 | -0.01257604 | -0.01092091 | 0.00359693 | 0.00299285 | -0.00065005 | -0.00008853 | 0.00055978 | -0.00034395 | -0.00007646 | 0.00020706 | 0.00014617 |
| 66 | 102 | 0.04384467 | 0.05276945 | -0.02442882 | -0.03787881 | 0.01599806 | 0.01692929 | -0.01227663 | -0.01039169 | 0.00374431 | 0.00299295 | -0.00067558 | -0.00011678 | 0.00059856 | -0.00030888 | 0.00000992 | 0.00003262 | 0.00007014 |
| 66 | 103 | 0.04299056 | 0.05237792 | -0.02443565 | -0.03806375 | 0.01599128 | 0.01638117 | -0.01333399 | -0.01065437 | 0.00395400 | 0.00294119 | -0.00067927 | -0.00010899 | 0.00059419 | -0.00037662 | -0.00002888 | 0.00019544 | 0.00017479 |
| 66 | 104 | 0.04365500 | 0.05227476 | -0.02465520 | -0.03715150 | 0.01633531 | 0.01576487 | -0.01303587 | -0.01017044 | 0.00409763 | 0.00295016 | -0.00070700 | -0.00013049 | 0.00062650 | -0.00034479 | 0.00005247 | 0.00002597 | 0.00009796 |
| 66 | 105 | 0.04310386 | 0.05166301 | -0.02529288 | -0.03704280 | 0.01641678 | 0.01517666 | -0.01370980 | -0.00978890 | 0.00428561 | 0.00277207 | -0.00074132 | -0.00003316 | 0.00050186 | -0.00035935 | -0.00003274 | 0.00020225 | 0.00012973 |
| 66 | 106 | 0.04398781 | 0.05120327 | -0.02614675 | -0.03577852 | 0.01700742 | 0.01446995 | -0.01311678 | -0.00869959 | 0.00444413 | 0.00266692 | -0.00080339 | 0.00003466 | 0.00041526 | -0.00027403 | 0.00000961 | 0.00003794 | -0.00002345 |
| 66 | 107 | 0.04346679 | 0.05074385 | -0.02671367 | -0.03574678 | 0.01698914 | 0.01390439 | -0.01380110 | -0.00832896 | 0.00462115 | 0.00247116 | -0.00084524 | 0.00013589 | 0.00029505 | -0.00029383 | -0.00007370 | 0.00020704 | 0.00000371 |
| 66 | 108 | 0.04428384 | 0.05018985 | -0.02752173 | -0.03439898 | 0.01774903 | 0.01317720 | -0.01321799 | -0.00726926 | 0.00478386 | 0.00236450 | -0.00089708 | 0.00018535 | 0.00022893 | -0.00020915 | -0.00001234 | 0.00001914 | -0.00014758 |
| 66 | 109 | 0.04378404 | 0.04986728 | -0.02799736 | -0.03444150 | 0.01760185 | 0.01261405 | -0.01388572 | -0.00690113 | 0.00494489 | 0.00215245 | -0.00094508 | 0.00028851 | 0.00011471 | -0.00023298 | -0.00009202 | 0.00017749 | -0.00012575 |
| 66 | 110 | 0.04451492 | 0.04925741 | -0.02873019 | -0.03305899 | 0.01851215 | 0.01191818 | -0.01330874 | -0.00591254 | 0.00510100 | 0.00204928 | -0.00098097 | 0.00031291 | 0.00007627 | -0.00015392 | -0.00001143 | -0.00002953 | -0.00026674 |
| 67 | 73 | 0.05263837 | 0.05838157 | -0.02707583 | -0.03904590 | 0.01244174 | 0.01967026 | -0.00561444 | -0.01131436 | -0.00090472 | 0.00292217 | 0.00022972 | -0.00052387 | 0.00068774 | -0.00016814 | -0.00034237 | 0.00011221 | 0.00023639 |
| 67 | 74 | 0.06840512 | 0.04314901 | -0.05672039 | -0.00911507 | 0.03085870 | -0.00735174 | -0.01183770 | 0.00338809 | 0.00102455 | -0.00015851 | 0.00045508 | 0.00008964 | -0.00043606 | 0.00033771 | -0.00011297 | -0.00009822 | -0.00037375 |
| 67 | 75 | 0.05221819 | 0.05857810 | -0.02709054 | -0.03920265 | 0.01182498 | 0.01885402 | -0.00534585 | -0.01075159 | -0.00064227 | 0.00279071 | 0.00025037 | -0.00059322 | 0.00078213 | -0.00021190 | -0.00025010 | 0.00005484 | 0.00031625 |
| 67 | 76 | 0.06823419 | 0.04295806 | -0.05748149 | -0.00888362 | 0.03066691 | -0.00802968 | -0.01173114 | 0.00369126 | 0.00111806 | -0.00026458 | 0.00049589 | 0.00001577 | -0.00036831 | 0.00028603 | -0.00006719 | -0.00010141 | -0.00027625 |
| 67 | 77 | 0.05172487 | 0.05880754 | -0.02693466 | -0.03937719 | 0.01116941 | 0.01806441 | -0.00498239 | -0.01019479 | -0.00039958 | 0.00264062 | 0.00027043 | -0.00067215 | 0.00089243 | -0.00025694 | -0.00014455 | -0.00002837 | 0.00039357 |
| 67 | 78 | 0.06802040 | 0.04283456 | -0.05812164 | -0.00868637 | 0.03051385 | -0.00851912 | -0.01146335 | 0.00396070 | 0.00121197 | -0.00035361 | 0.00054564 | -0.00006516 | -0.00028804 | 0.00022657 | -0.00001829 | -0.00010697 | -0.00016511 |
| 67 | 79 | 0.05118439 | 0.05902223 | -0.02666612 | -0.03959090 | 0.01049789 | 0.01741487 | -0.00450984 | -0.00966792 | -0.00019852 | 0.00249162 | 0.00028471 | -0.00074472 | 0.00099689 | -0.00029415 | -0.00004335 | -0.00012175 | 0.00045564 |
| 67 | 80 | 0.06778617 | 0.04272515 | -0.05869721 | -0.00852633 | 0.03041811 | -0.00876081 | -0.01102760 | 0.00419356 | 0.00128529 | -0.00041545 | 0.00059994 | -0.00014009 | -0.00021276 | 0.00016756 | 0.00002067 | -0.00010719 | -0.00005494 |
| 67 | 81 | 0.05062751 | 0.05919074 | -0.02633623 | -0.03985500 | 0.00983800 | 0.01699715 | -0.00393926 | -0.00918835 | -0.00005228 | 0.00235863 | 0.00028730 | -0.00079474 | 0.00107422 | -0.00031560 | 0.00003743 | -0.00021045 | 0.00048997 |
| 67 | 82 | 0.06755600 | 0.04259419 | -0.05925190 | -0.00839032 | 0.03040804 | -0.00873645 | -0.01046298 | 0.00439586 | 0.00132962 | -0.00044916 | 0.00065183 | -0.00019580 | -0.00015888 | 0.00011637 | 0.00003999 | -0.00009725 | 0.00003904 |
| 67 | 83 | 0.04414751 | 0.06195436 | -0.01657636 | -0.04865515 | 0.00469573 | 0.02494119 | -0.00280511 | -0.01256379 | 0.00022357 | 0.00336322 | -0.00003082 | -0.00051387 | 0.00092659 | -0.00021300 | -0.00007445 | -0.00023582 | 0.00023132 |
| 67 | 84 | 0.04546255 | 0.06083193 | -0.01918401 | -0.04667857 | 0.00618125 | 0.02347796 | -0.00388813 | -0.01208291 | 0.00021207 | 0.00313023 | -0.00002643 | -0.00039715 | 0.00075456 | -0.00016083 | -0.00012328 | -0.00009212 | 0.00023341 |
| 67 | 85 | 0.04443456 | 0.05941013 | -0.01945504 | -0.04585835 | 0.00706504 | 0.02339293 | -0.00462428 | -0.01189733 | 0.00077888 | 0.00341620 | -0.00011582 | -0.00024534 | 0.00059682 | -0.00017305 | -0.00020560 | -0.00001607 | 0.00013746 |
| 67 | 86 | 0.04500756 | 0.05929602 | -0.02028613 | -0.04576683 | 0.00756686 | 0.02346534 | -0.00510103 | -0.01199673 | 0.00069843 | 0.00325872 | -0.00014297 | -0.00018586 | 0.00052500 | -0.00013947 | -0.00022255 | 0.00004680 | 0.00013050 |
| 67 | 87 | 0.04443733 | 0.05846420 | -0.02053396 | -0.04520367 | 0.00843712 | 0.02311956 | -0.00579737 | -0.01188481 | 0.00113055 | 0.00337474 | -0.00020610 | -0.00016549 | 0.00052020 | -0.00019460 | -0.00024155 | 0.00007988 | 0.00011664 |
| 67 | 88 | 0.04508819 | 0.05839897 | -0.02137325 | -0.04498405 | 0.00910607 | 0.02318417 | -0.00619340 | -0.01190539 | 0.00110788 | 0.00324274 | -0.00024522 | -0.00012558 | 0.00047908 | -0.00015901 | -0.00023991 | 0.00010858 | 0.00010469 |
| 67 | 89 | 0.04432587 | 0.05757190 | -0.02140538 | -0.04457767 | 0.00981497 | 0.02287941 | -0.00700723 | -0.01193021 | 0.00149798 | 0.00334840 | -0.00029940 | -0.00009769 | 0.00045987 | -0.00021569 | -0.00027134 | 0.00016800 | 0.00009775 |
| 67 | 90 | 0.04504211 | 0.05752150 | -0.02227945 | -0.04415934 | 0.01066244 | 0.02279757 | -0.00735338 | -0.01181793 | 0.00155887 | 0.00322957 | -0.00034726 | -0.00008043 | 0.00045451 | -0.00018193 | -0.00024179 | 0.00015175 | 0.00008282 |
| 67 | 91 | 0.04411633 | 0.05667050 | -0.02212323 | -0.04385487 | 0.01112562 | 0.02247010 | -0.00824042 | -0.01196200 | 0.00188129 | 0.00332493 | -0.00038667 | -0.00004497 | 0.00041606 | -0.00023697 | -0.00028846 | 0.00024481 | 0.00008739 |
| 67 | 92 | 0.04489412 | 0.05663548 | -0.02303856 | -0.04324147 | 0.01211747 | 0.02217405 | -0.00855043 | -0.01169206 | 0.00203564 | 0.00321065 | -0.00044182 | -0.00004974 | 0.00044661 | -0.00020776 | -0.00022813 | 0.00018043 | 0.00007051 |
| 67 | 93 | 0.04383134 | 0.05578564 | -0.02268220 | -0.04305116 | 0.01228159 | 0.02184136 | -0.00943772 | -0.01194896 | 0.00226912 | 0.00329343 | -0.00046189 | -0.00001252 | 0.00039358 | -0.00026057 | -0.00028790 | 0.00030390 | 0.00008936 |
| 67 | 94 | 0.04467851 | 0.05576975 | -0.02365598 | -0.04226752 | 0.01337988 | 0.02131412 | -0.00972426 | -0.01151749 | 0.00251667 | 0.00318167 | -0.00052454 | -0.00003381 | 0.00045331 | -0.00023650 | -0.00020078 | 0.00019677 | 0.00006989 |
| 67 | 95 | 0.04350965 | 0.05494273 | -0.02310699 | -0.04219426 | 0.01324650 | 0.02100593 | -0.01054985 | -0.01187325 | 0.00265051 | 0.00324853 | -0.00052248 | -0.00000023 | 0.00039163 | -0.00028701 | -0.00026972 | 0.00034282 | 0.00010280 |
| 67 | 96 | 0.04443492 | 0.05495086 | -0.02415491 | -0.04127590 | 0.01441609 | 0.02026682 | -0.01082664 | -0.01129518 | 0.00298383 | 0.00314204 | -0.00059353 | -0.00003057 | 0.00047076 | -0.00026741 | -0.00016345 | 0.00020317 | 0.00007964 |
| 67 | 97 | 0.04318912 | 0.05415785 | -0.02343473 | -0.04131056 | 0.01402238 | 0.02000793 | -0.01154677 | -0.01172995 | 0.00301664 | 0.00318897 | -0.00056835 | -0.00000492 | 0.00040625 | -0.00031563 | -0.00023685 | 0.00036197 | 0.00012450 |
| 67 | 98 | 0.04419513 | 0.05419769 | -0.02456079 | -0.04029891 | 0.01523111 | 0.01909798 | -0.01182725 | -0.01103461 | 0.00342366 | 0.00309299 | -0.00064857 | -0.00003738 | 0.00049537 | -0.00029951 | -0.00011974 | 0.00020139 | 0.00009706 |
| 67 | 99 | 0.04289698 | 0.05343737 | -0.02369997 | -0.04042161 | 0.01463225 | 0.01890017 | -0.01241744 | -0.01152438 | 0.00336139 | 0.00311597 | -0.00060090 | -0.00002241 | 0.00043269 | -0.00034529 | -0.00019300 | 0.00036333 | 0.00015076 |
| 67 | 100 | 0.04397857 | 0.05352049 | -0.02489340 | -0.03936091 | 0.01585036 | 0.01787132 | -0.01271262 | -0.01075010 | 0.00382745 | 0.00303655 | -0.00069042 | -0.00005194 | 0.00052463 | -0.00033190 | -0.00007234 | 0.00019248 | 0.00011934 |
| 67 | 101 | 0.04264815 | 0.05278068 | -0.02392843 | -0.03954417 | 0.01510686 | 0.01773171 | -0.01316470 | -0.01126811 | 0.00368135 | 0.00303203 | -0.00062218 | -0.00004873 | 0.00046673 | -0.00037477 | -0.00014162 | 0.00034943 | 0.00017835 |
| 67 | 102 | 0.04379392 | 0.05292233 | -0.02516611 | -0.03847845 | 0.01630750 | 0.01663936 | -0.01348208 | -0.01045692 | 0.00419084 | 0.00297497 | -0.00072044 | -0.00007243 | 0.00055703 | -0.00036383 | -0.00002298 | 0.00017701 | 0.00014405 |
| 67 | 103 | 0.04244781 | 0.05218343 | -0.02413670 | -0.03869131 | 0.01547686 | 0.01654310 | -0.01379950 | -0.01097517 | 0.00397529 | 0.00294019 | -0.00063441 | -0.00008062 | 0.00050511 | -0.00040301 | -0.00008547 | 0.00032280 | 0.00020483 |
| 67 | 104 | 0.04364287 | 0.05240082 | -0.02538798 | -0.03766094 | 0.01663719 | 0.01544031 | -0.01414335 | -0.01016819 | 0.00451316 | 0.00291039 | -0.00074032 | -0.00009742 | 0.00059172 | -0.00039466 | 0.00002736 | 0.00015529 | 0.00016926 |
| 67 | 105 | 0.04253341 | 0.05156088 | -0.02489926 | -0.03775792 | 0.01572221 | 0.01536821 | -0.01408881 | -0.01013392 | 0.00425139 | 0.00274974 | -0.00068249 | -0.00001995 | 0.00042798 | -0.00039357 | -0.00008461 | 0.00032558 | 0.00017210 |
| 67 | 106 | 0.04397053 | 0.05151589 | -0.02678853 | -0.03642699 | 0.01711680 | 0.01418746 | -0.01424205 | -0.00873036 | 0.00485145 | 0.00261681 | -0.00084082 | 0.00006743 | 0.00038653 | -0.00033450 | -0.00001770 | 0.00017527 | 0.00005778 |
| 67 | 107 | 0.04286850 | 0.05086784 | -0.02618705 | -0.03667054 | 0.01594552 | 0.01411217 | -0.01412225 | -0.00874092 | 0.00452629 | 0.00244353 | -0.00077624 | 0.00013258 | 0.00024368 | -0.00034498 | -0.00012749 | 0.00033996 | 0.00007247 |
| 67 | 108 | 0.04426311 | 0.05066434 | -0.02807007 | -0.03517826 | 0.01766044 | 0.01291488 | -0.01435052 | -0.00732299 | 0.00517758 | 0.00230511 | -0.00094190 | 0.00022132 | 0.00020281 | -0.00027545 | -0.00004278 | 0.00016196 | -0.00006164 |
| 67 | 109 | 0.04317897 | 0.05018543 | -0.02738231 | -0.03554942 | 0.01621604 | 0.01277788 | -0.01414772 | -0.00734742 | 0.00478603 | 0.00211248 | -0.00086973 | 0.00027282 | 0.00008180 | -0.00029594 | -0.00014795 | 0.00031604 | -0.00003576 |
| 67 | 110 | 0.04449508 | 0.04986547 | -0.02918834 | -0.03395706 | 0.01823180 | 0.01166444 | -0.01442971 | -0.00597784 | 0.00547499 | 0.00198143 | -0.00103553 | 0.00035538 | 0.00004945 | -0.00022160 | -0.00004562 | 0.00011513 | -0.00018201 |
| 67 | 111 | 0.04343902 | 0.04953067 | -0.02844002 | -0.03442763 | 0.01651875 | 0.01141303 | -0.01413194 | -0.00598710 | 0.00501532 | 0.00176201 | -0.00095541 | 0.00039214 | -0.00004832 | -0.00025031 | -0.00014269 | 0.00025153 | -0.00014705 |
| 68 | 74 | 0.04990558 | 0.06007393 | -0.02300018 | -0.04317807 | 0.01029432 | 0.02481624 | -0.00575767 | -0.01395247 | -0.00110926 | 0.00379714 | -0.00005159 | 0.00003891 | 0.00019532 | 0.00006602 | -0.00070740 | 0.00051809 | 0.00001654 |
| 68 | 75 | 0.06506509 | 0.04643835 | -0.05183040 | -0.01558984 | 0.02750028 | -0.00035870 | -0.00975417 | 0.00011108 | 0.00054902 | 0.00040533 | 0.00059982 | -0.00016279 | -0.00011241 | 0.00001759 | -0.00017448 | 0.00003316 | -0.00002860 |
| 68 | 76 | 0.04888977 | 0.06057760 | -0.02204578 | -0.04436966 | 0.00930587 | 0.02483551 | -0.00536686 | -0.01372775 | -0.00083831 | 0.00369906 | -0.00004085 | -0.00006109 | 0.00033267 | 0.00002049 | -0.00058376 | 0.00041821 | 0.00009684 |
| 68 | 77 | 0.06430206 | 0.04699361 | -0.05128840 | -0.01648844 | 0.02655655 | -0.00031165 | -0.00943033 | 0.00003066 | 0.00067201 | 0.00038545 | 0.00062966 | -0.00027435 | 0.00003113 | -0.00005451 | -0.00010030 | 0.00001039 | 0.00010629 |
| 68 | 78 | 0.04794056 | 0.06102159 | -0.02117945 | -0.04537634 | 0.00846913 | 0.02470932 | -0.00493408 | -0.01336481 | -0.00053778 | 0.00355965 | -0.00002800 | -0.00016730 | 0.00047845 | -0.00002913 | -0.00044376 | 0.00028887 | 0.00016774 |
| 68 | 79 | 0.06343145 | 0.04766009 | -0.05047721 | -0.01756830 | 0.02554029 | 0.00005157 | -0.00890620 | -0.00011870 | 0.00077874 | 0.00039648 | 0.00065774 | -0.00038271 | 0.00017686 | -0.00012517 | -0.00003109 | -0.00001416 | 0.00023869 |
| 68 | 80 | 0.04709092 | 0.06135989 | -0.02047317 | -0.04620102 | 0.00778016 | 0.02450454 | -0.00444233 | -0.01287986 | -0.00022980 | 0.00339733 | -0.00001648 | -0.00026616 | 0.00061337 | -0.00007502 | -0.00030379 | 0.00014578 | 0.00022081 |
| 68 | 81 | 0.06247399 | 0.04838984 | -0.04944605 | -0.01883854 | 0.02446591 | 0.00077015 | -0.00820667 | -0.00033713 | 0.00085662 | 0.00044547 | 0.00067559 | -0.00047223 | 0.00030517 | -0.00018442 | 0.00002104 | -0.00003394 | 0.00035116 |
| 68 | 82 | 0.04637210 | 0.06155647 | -0.01998915 | -0.04684483 | 0.00724126 | 0.02429595 | -0.00389000 | -0.01229452 | 0.00006683 | 0.00323039 | -0.00001105 | -0.00034235 | 0.00071656 | -0.00010917 | -0.00018090 | 0.00000621 | 0.00024732 |
| 68 | 83 | 0.04600858 | 0.06179601 | -0.01937699 | -0.04751321 | 0.00650794 | 0.02387075 | -0.00392307 | -0.01228081 | 0.00029170 | 0.00320970 | -0.00002915 | -0.00045420 | 0.00085730 | -0.00018650 | -0.00009609 | -0.00010072 | 0.00028724 |
| 68 | 84 | 0.04556850 | 0.06009023 | -0.02047230 | -0.04639004 | 0.00761007 | 0.02432649 | -0.00471617 | -0.01224671 | 0.00040522 | 0.00334857 | -0.00009366 | -0.00012924 | 0.00047586 | -0.00008157 | -0.00026867 | 0.00007909 | 0.00011098 |
| 68 | 85 | 0.04533660 | 0.05984467 | -0.02057402 | -0.04633396 | 0.00779212 | 0.02368251 | -0.00522876 | -0.01207648 | 0.00082777 | 0.00336005 | -0.00013961 | -0.00017572 | 0.00053948 | -0.00014667 | -0.00021710 | 0.00004341 | 0.00014109 |
| 68 | 86 | 0.04582828 | 0.05929225 | -0.02158173 | -0.04565635 | 0.00873902 | 0.02381631 | -0.00561350 | -0.01204465 | 0.00076914 | 0.00332338 | -0.00018856 | -0.00008789 | 0.00044558 | -0.00009422 | -0.00026714 | 0.00011058 | 0.00008751 |
| 68 | 87 | 0.04550646 | 0.05895497 | -0.02176949 | -0.04554366 | 0.00925747 | 0.02328144 | -0.00634824 | -0.01192974 | 0.00125162 | 0.00332560 | -0.00024547 | -0.00011253 | 0.00048858 | -0.00016461 | -0.00022932 | 0.00010257 | 0.00011716 |
| 68 | 88 | 0.04598404 | 0.05852381 | -0.02253554 | -0.04486742 | 0.01008629 | 0.02338607 | -0.00655396 | -0.01183447 | 0.00120972 | 0.00330653 | -0.00029073 | -0.00005659 | 0.00043182 | -0.00011170 | -0.00025328 | 0.00011983 | 0.00005734 |
| 68 | 89 | 0.04556578 | 0.05811299 | -0.02277305 | -0.04474547 | 0.01078948 | 0.02289938 | -0.00750032 | -0.01179939 | 0.00171508 | 0.00330038 | -0.00035405 | -0.00006217 | 0.00045584 | -0.00018517 | -0.00023148 | 0.00014507 | 0.00009156 |
| 68 | 90 | 0.04603045 | 0.05771279 | -0.02338082 | -0.04393423 | 0.01152042 | 0.02281011 | -0.00757318 | -0.01158309 | 0.00170987 | 0.00328742 | -0.00039229 | -0.00003292 | 0.00042887 | -0.00013312 | -0.00022667 | 0.00011364 | 0.00002787 |
| 68 | 91 | 0.04552393 | 0.05724820 | -0.02363694 | -0.04383046 | 0.01227311 | 0.02232304 | -0.00868751 | -0.01163116 | 0.00220669 | 0.00327326 | -0.00045635 | -0.00002389 | 0.00043642 | -0.00020780 | -0.00022135 | 0.00017461 | 0.00007154 |
| 68 | 92 | 0.04597002 | 0.05688039 | -0.02409474 | -0.04290929 | 0.01288607 | 0.02202662 | -0.00864062 | -0.01129935 | 0.00223836 | 0.00326013 | -0.00048720 | -0.00001910 | 0.00043729 | -0.00015886 | -0.00018809 | 0.00009720 | 0.00000629 |
| 68 | 93 | 0.04539630 | 0.05638521 | -0.02434668 | -0.04283586 | 0.01359047 | 0.02151359 | -0.00985887 | -0.01141679 | 0.00270387 | 0.00323704 | -0.00054655 | -0.00000067 | 0.00043177 | -0.00023335 | -0.00019770 | 0.00019148 | 0.00006207 |
| 68 | 94 | 0.04582698 | 0.05605603 | -0.02467873 | -0.04185081 | 0.01408441 | 0.02104805 | -0.00971130 | -0.01099693 | 0.00276547 | 0.00322388 | -0.00057148 | -0.00001439 | 0.00045436 | -0.00018823 | -0.00014154 | 0.00007647 | -0.00000441 |
| 68 | 95 | 0.04521475 | 0.05555199 | -0.02491568 | -0.04180701 | 0.01468689 | 0.02050160 | -0.01096669 | -0.01115921 | 0.00318635 | 0.00318985 | -0.00062182 | 0.00000825 | 0.00043980 | -0.00026152 | -0.00016290 | 0.00019740 | 0.00006356 |

TABLE 3. Nuclear charge density distribution FB coefficients.

| $Z$ | $N$ | $a_1$ | $a_2$ | $a_3$ | $a_4$ | $a_5$ | $a_6$ | $a_7$ | $a_8$ | $a_9$ | $a_{10}$ | $a_{11}$ | $a_{12}$ | $a_{13}$ | $a_{14}$ | $a_{15}$ | $a_{16}$ | $a_{17}$ |
|---|---|---|---|---|---|---|---|---|---|---|---|---|---|---|---|---|---|---|
| 68 | 96 | 0.04563329 | 0.05526932 | -0.02514724 | -0.04080906 | 0.01507069 | 0.01992683 | -0.01074211 | -0.01069044 | 0.00326788 | 0.00318024 | -0.00064299 | -0.00001707 | 0.00047688 | -0.00022019 | -0.00009134 | 0.00005562 | -0.00000424 |
| 68 | 97 | 0.04501166 | 0.05477172 | -0.02536828 | -0.04078496 | 0.01555501 | 0.01934954 | -0.01197705 | -0.01086925 | 0.00363855 | 0.00313264 | -0.00068149 | 0.00000523 | 0.00045731 | -0.00029147 | -0.00012035 | 0.00019416 | 0.00007400 |
| 68 | 98 | 0.04541808 | 0.05454452 | -0.02551767 | -0.03982115 | 0.01583971 | 0.01872694 | -0.01170085 | -0.01039302 | 0.00373020 | 0.00313128 | -0.00070104 | -0.00002558 | 0.00050254 | -0.00025378 | -0.00004062 | 0.00003651 | 0.00000529 |
| 68 | 99 | 0.04481102 | 0.05405954 | -0.02572778 | -0.03980155 | 0.01621494 | 0.01812470 | -0.01287240 | -0.01056182 | 0.00405038 | 0.00306761 | -0.00072625 | -0.00000719 | 0.00048159 | -0.00032229 | -0.00007307 | 0.00018309 | 0.00009069 |
| 68 | 100 | 0.04520296 | 0.05389747 | -0.02580583 | -0.03890971 | 0.01641054 | 0.01750607 | -0.01256924 | -0.01011474 | 0.00414456 | 0.00307883 | -0.00074601 | -0.00003882 | 0.00053019 | -0.00028822 | 0.00000890 | 0.00001919 | 0.00002206 |
| 68 | 101 | 0.04462699 | 0.05342249 | -0.02601261 | -0.03887819 | 0.01669955 | 0.01688463 | -0.01364933 | -0.01025196 | 0.00441706 | 0.00299723 | -0.00075758 | -0.00002681 | 0.00051071 | -0.00035314 | -0.00002319 | 0.00016506 | 0.00011104 |
| 68 | 102 | 0.04500192 | 0.05333515 | -0.02602542 | -0.03808453 | 0.01681425 | 0.01630783 | -0.01334130 | -0.00986181 | 0.00450908 | 0.00302419 | -0.00077903 | -0.00005599 | 0.00055938 | -0.00032285 | 0.00005660 | 0.00000274 | 0.00004389 |
| 68 | 103 | 0.04446616 | 0.05286090 | -0.02623690 | -0.03802661 | 0.01704508 | 0.01567151 | -0.01431444 | -0.00995185 | 0.00473833 | 0.00292380 | -0.00077743 | -0.00005186 | 0.00054343 | -0.00038334 | 0.00002800 | 0.00014060 | 0.00013295 |
| 68 | 104 | 0.04482305 | 0.05285693 | -0.02618868 | -0.03734526 | 0.01708586 | 0.01516030 | -0.01401996 | -0.00963639 | 0.00482611 | 0.00296810 | -0.00080166 | -0.00007642 | 0.00058996 | -0.00035714 | 0.00010250 | -0.00001411 | 0.00006877 |
| 68 | 105 | 0.04453278 | 0.05219195 | -0.02699596 | -0.03705082 | 0.01737359 | 0.01447848 | -0.01465887 | -0.00911104 | 0.00505114 | 0.00274208 | -0.00083137 | 0.00001140 | 0.00046330 | -0.00037159 | 0.00003261 | 0.00013917 | 0.00009561 |
| 68 | 106 | 0.04505954 | 0.05188340 | -0.02751480 | -0.03611287 | 0.01762859 | 0.01391946 | -0.01409421 | -0.00824769 | 0.00515259 | 0.00268101 | -0.00089051 | 0.00006870 | 0.00040238 | -0.00029592 | 0.00007100 | -0.00000686 | -0.00003399 |
| 68 | 107 | 0.04478642 | 0.05138208 | -0.02824754 | -0.03588409 | 0.01782381 | 0.01325061 | -0.01477583 | -0.00774097 | 0.00537681 | 0.00244034 | -0.00092976 | 0.00016099 | 0.00028142 | -0.00031737 | 0.00000300 | 0.00014197 | -0.00000901 |
| 68 | 108 | 0.04525531 | 0.05096665 | -0.02871761 | -0.03489234 | 0.01825542 | 0.01271119 | -0.01418474 | -0.00692428 | 0.00545600 | 0.00237858 | -0.00097695 | 0.00020048 | 0.00023941 | -0.00023857 | 0.00005783 | -0.00002831 | -0.00014035 |
| 68 | 109 | 0.04499109 | 0.05061439 | -0.02935645 | -0.03473713 | 0.01832590 | 0.01203887 | -0.01488111 | -0.00642797 | 0.00567544 | 0.00212617 | -0.00102561 | 0.00029635 | 0.00012469 | -0.00026473 | -0.00000693 | 0.00011232 | -0.00011912 |
| 68 | 110 | 0.04538918 | 0.05012373 | -0.02975830 | -0.03371632 | 0.01892623 | 0.01156008 | -0.01426126 | -0.00568863 | 0.00572573 | 0.00206670 | -0.00105458 | 0.00031151 | 0.00010821 | -0.00018842 | 0.00006494 | -0.00007798 | -0.00024385 |
| 68 | 111 | 0.04512779 | 0.04990052 | -0.03029056 | -0.03363848 | 0.01884771 | 0.01087447 | -0.01493782 | -0.00519524 | 0.00593284 | 0.00180523 | -0.00111149 | 0.00041010 | 0.00000008 | -0.00021714 | 0.00000437 | 0.00005052 | -0.00022864 |
| 68 | 112 | 0.04544604 | 0.04936067 | -0.03061166 | -0.03260059 | 0.01960039 | 0.01047100 | -0.01430104 | -0.00455415 | 0.00595227 | 0.00175064 | -0.00111838 | 0.00039701 | 0.00001272 | -0.00014741 | 0.00009259 | -0.00015370 | -0.00033918 |
| 69 | 75 | 0.04685677 | 0.06137985 | -0.01933641 | -0.04731553 | 0.00959783 | 0.02768296 | -0.00597239 | -0.01500989 | 0.00006278 | 0.00445053 | -0.00015402 | -0.00009288 | 0.00043953 | -0.00006548 | -0.00057171 | 0.00038456 | 0.00008264 |
| 69 | 76 | 0.04915966 | 0.06180669 | -0.02176943 | -0.04558202 | 0.00990999 | 0.02482333 | -0.00570159 | -0.01365191 | -0.00029222 | 0.00369027 | -0.00006986 | -0.00029089 | 0.00064275 | -0.00009735 | -0.00041166 | 0.00027173 | 0.00022612 |
| 69 | 77 | 0.04627050 | 0.06154612 | -0.01918302 | -0.04771060 | 0.00886783 | 0.02686067 | -0.00550982 | -0.01432630 | 0.00035378 | 0.00425409 | -0.00011587 | -0.00019801 | 0.00057895 | -0.00012547 | -0.00042795 | 0.00024210 | 0.00015278 |
| 69 | 78 | 0.04825007 | 0.06223684 | -0.02096637 | -0.04655483 | 0.00891114 | 0.02460958 | -0.00522774 | -0.01331558 | -0.00002981 | 0.00356591 | -0.00005372 | -0.00038583 | 0.00077798 | -0.00014863 | -0.00028384 | 0.00015262 | 0.00029356 |
| 69 | 79 | 0.04575091 | 0.06164216 | -0.01906410 | -0.04796078 | 0.00819098 | 0.02605578 | -0.00494694 | -0.01358223 | 0.00060203 | 0.00404276 | -0.00008575 | -0.00028705 | 0.00069706 | -0.00017235 | -0.00029635 | 0.00009537 | 0.00019854 |
| 69 | 80 | 0.04742236 | 0.06258091 | -0.02025339 | -0.04741353 | 0.00799603 | 0.02444401 | -0.00462033 | -0.01289357 | 0.00020453 | 0.00342937 | -0.00004078 | -0.00046674 | 0.00089300 | -0.00019047 | -0.00016693 | 0.00002766 | 0.00033788 |
| 69 | 81 | 0.04532455 | 0.06164411 | -0.01902651 | -0.04808113 | 0.00758396 | 0.02536136 | -0.00430542 | -0.01281090 | 0.00079515 | 0.00383831 | -0.00007005 | -0.00034367 | 0.00077283 | -0.00019807 | -0.00019375 | -0.00003775 | 0.00021017 |
| 69 | 82 | 0.04671085 | 0.06280141 | -0.01970497 | -0.04814503 | 0.00719761 | 0.02438250 | -0.00390719 | -0.01240197 | 0.00040043 | 0.00329452 | -0.00003664 | -0.00051926 | 0.00096879 | -0.00021586 | -0.00007538 | -0.00008828 | 0.00034998 |
| 69 | 83 | 0.04544167 | 0.06038490 | -0.02058441 | -0.04658359 | 0.00844414 | 0.02397248 | -0.00488270 | -0.01198798 | 0.00115067 | 0.00371570 | -0.00010350 | -0.00025051 | 0.00065790 | -0.00018813 | -0.00019363 | -0.00001348 | 0.00017222 |
| 69 | 84 | 0.04567081 | 0.06032455 | -0.02087950 | -0.04668471 | 0.00827077 | 0.02401292 | -0.00527307 | -0.01208278 | 0.00096948 | 0.00347222 | -0.00013030 | -0.00016247 | 0.00054675 | -0.00015389 | -0.00022075 | 0.00004569 | 0.00014318 |
| 69 | 85 | 0.04557850 | 0.05938749 | -0.02183216 | -0.04575722 | 0.00968508 | 0.02325894 | -0.00612544 | -0.01179092 | 0.00154421 | 0.00361780 | -0.00019652 | -0.00017253 | 0.00058125 | -0.00020579 | -0.00020186 | 0.00005308 | 0.00015382 |
| 69 | 86 | 0.04590979 | 0.05940831 | -0.02220989 | -0.04586883 | 0.00963009 | 0.02336724 | -0.00647906 | -0.01187820 | 0.00140855 | 0.00341161 | -0.00023807 | -0.00009793 | 0.00049461 | -0.00017112 | -0.00022265 | 0.00009864 | 0.00012428 |
| 69 | 87 | 0.04564362 | 0.05856301 | -0.02281771 | -0.04514783 | 0.01094674 | 0.02287306 | -0.00732336 | -0.01170304 | 0.00193478 | 0.00354114 | -0.00030052 | -0.00010068 | 0.00051427 | -0.00022306 | -0.00021680 | 0.00011784 | 0.00012627 |
| 69 | 88 | 0.04606085 | 0.05859790 | -0.02331628 | -0.04511856 | 0.01109484 | 0.02291176 | -0.00765957 | -0.01171022 | 0.00188256 | 0.00336619 | -0.00035227 | -0.00004473 | 0.00045834 | -0.00019077 | -0.00022027 | 0.00013696 | 0.00009845 |
| 69 | 89 | 0.04565300 | 0.05772183 | -0.02370673 | -0.04445236 | 0.01223468 | 0.02248345 | -0.00852812 | -0.01162134 | 0.00233533 | 0.00348113 | -0.00040358 | -0.00003027 | 0.00044574 | -0.00023723 | -0.00023632 | 0.00018596 | 0.00009554 |
| 69 | 90 | 0.04612698 | 0.05775911 | -0.02429850 | -0.04423567 | 0.01257506 | 0.02234592 | -0.00885474 | -0.01150563 | 0.00238571 | 0.00332550 | -0.00046210 | 0.00000094 | 0.00042884 | -0.00021100 | -0.00021177 | 0.00016617 | 0.00007258 |
| 69 | 91 | 0.04558065 | 0.05686776 | -0.02445229 | -0.04366593 | 0.01342891 | 0.02196239 | -0.00970413 | -0.01152049 | 0.00273641 | 0.00342342 | -0.00049661 | 0.00002708 | 0.00038833 | -0.00025195 | -0.00024881 | 0.00024736 | 0.00007166 |
| 69 | 92 | 0.04609907 | 0.05690904 | -0.02511838 | -0.04325376 | 0.01392582 | 0.02158427 | -0.01002826 | -0.01125951 | 0.00289461 | 0.00327951 | -0.00056014 | 0.00003285 | 0.00041183 | -0.00023357 | -0.00019232 | 0.00018422 | 0.00005467 |
| 69 | 93 | 0.04542880 | 0.05601839 | -0.02504116 | -0.04280791 | 0.01445626 | 0.02125320 | -0.01081342 | -0.01138216 | 0.00312838 | 0.00335756 | -0.00057406 | 0.00006495 | 0.00034961 | -0.00026955 | -0.00024724 | 0.00029432 | 0.00005953 |
| 69 | 94 | 0.04599388 | 0.05607384 | -0.02577345 | -0.04222114 | 0.01506834 | 0.02062550 | -0.01113947 | -0.01097696 | 0.00338756 | 0.00322414 | -0.00064237 | 0.00004953 | 0.00040822 | -0.00025885 | -0.00016180 | 0.00019136 | 0.00004752 |
| 69 | 95 | 0.04522083 | 0.05519320 | -0.02549163 | -0.04190642 | 0.01529182 | 0.02035936 | -0.01182442 | -0.01119692 | 0.00350223 | 0.00327862 | -0.00063391 | 0.00008239 | 0.00033112 | -0.00029059 | -0.00022975 | 0.00032292 | 0.00005926 |
| 69 | 96 | 0.04583962 | 0.05527861 | -0.02628332 | -0.04118490 | 0.01597972 | 0.01951723 | -0.01215528 | -0.01066957 | 0.00384743 | 0.00315930 | -0.00070744 | 0.00005235 | 0.00041624 | -0.00028636 | -0.00012257 | 0.00018875 | 0.00005039 |
| 69 | 97 | 0.04498638 | 0.05440812 | -0.02583562 | -0.04099040 | 0.01594372 | 0.01931992 | -0.01271798 | -0.01096496 | 0.00385079 | 0.00318609 | -0.00067666 | 0.00008194 | 0.00033061 | -0.00031440 | -0.00019797 | 0.00033248 | 0.00006823 |
| 69 | 98 | 0.04566304 | 0.05454229 | -0.02667416 | -0.04018314 | 0.01667267 | 0.01832308 | -0.01305627 | -0.01035181 | 0.00426307 | 0.00308679 | -0.00075582 | 0.00004376 | 0.00043347 | -0.00031529 | -0.00007743 | 0.00017760 | 0.00006098 |
| 69 | 99 | 0.04475065 | 0.05367323 | -0.02610478 | -0.04008471 | 0.01643733 | 0.01818565 | -0.01348866 | -0.01069359 | 0.00416959 | 0.00308209 | -0.00070432 | 0.00006750 | 0.00034429 | -0.00033978 | -0.00015497 | 0.00032445 | 0.00008305 |
| 69 | 100 | 0.04548385 | 0.05387566 | -0.02697036 | -0.03924181 | 0.01717809 | 0.01710168 | -0.01383707 | -0.01003744 | 0.00462948 | 0.00300894 | -0.00078910 | 0.00002620 | 0.00045776 | -0.00034479 | -0.00002870 | 0.00015887 | 0.00007670 |
| 69 | 101 | 0.04453061 | 0.05299298 | -0.02632418 | -0.03920810 | 0.01680366 | 0.01700440 | -0.01414198 | -0.01039373 | 0.00445714 | 0.00297000 | -0.00071953 | 0.00004301 | 0.00036833 | -0.00036550 | -0.00010405 | 0.00030129 | 0.00010065 |
| 69 | 102 | 0.04531448 | 0.05328189 | -0.02719259 | -0.03837489 | 0.01753336 | 0.01589735 | -0.01450326 | -0.00973687 | 0.00494687 | 0.00292802 | -0.00080948 | 0.00000188 | 0.00048748 | -0.00037413 | 0.00002204 | 0.00013328 | 0.00009523 |
| 69 | 103 | 0.04433564 | 0.05236778 | -0.02651154 | -0.03837343 | 0.01707268 | 0.01581460 | -0.01468990 | -0.01007662 | 0.00471434 | 0.00285349 | -0.00072501 | 0.00001188 | 0.00039947 | -0.00039051 | -0.00004804 | 0.00026575 | 0.00011872 |
| 69 | 104 | 0.04516187 | 0.05275827 | -0.02735810 | -0.03758663 | 0.01777518 | 0.01473847 | -0.01506695 | -0.00945606 | 0.00521913 | 0.00284586 | -0.00081937 | -0.00002734 | 0.00052130 | -0.00040266 | 0.00007381 | 0.00010143 | 0.00011470 |
| 69 | 105 | 0.04435886 | 0.05179065 | -0.02717246 | -0.03751672 | 0.01723890 | 0.01469142 | -0.01496076 | -0.00928293 | 0.00497747 | 0.00265281 | -0.00076574 | 0.00005993 | 0.00033543 | -0.00038784 | -0.00003751 | 0.00026013 | 0.00009392 |
| 69 | 106 | 0.04536496 | 0.05200490 | -0.02855586 | -0.03652658 | 0.01811949 | 0.01358952 | -0.01517486 | -0.00814139 | 0.00554101 | 0.00255592 | -0.00091079 | 0.00011201 | 0.00034706 | -0.00035647 | 0.00004283 | 0.00011856 | 0.00002886 |
| 69 | 107 | 0.04457106 | 0.05122659 | -0.02827192 | -0.03658687 | 0.01738328 | 0.01354859 | -0.01503689 | -0.00801500 | 0.00525471 | 0.00235284 | -0.00085294 | 0.00018823 | 0.00018212 | -0.00035392 | -0.00006381 | 0.00026940 | 0.00001804 |
| 69 | 108 | 0.04553060 | 0.05128297 | -0.02963287 | -0.03547804 | 0.01853010 | 0.01244066 | -0.01529150 | -0.00687599 | 0.00583792 | 0.00225338 | -0.00100477 | 0.00024309 | 0.00019214 | -0.00030818 | 0.00002812 | 0.00010591 | -0.00006614 |
| 69 | 109 | 0.04475528 | 0.05067508 | -0.02927590 | -0.03566017 | 0.01756548 | 0.01234652 | -0.01510309 | -0.00676579 | 0.00550430 | 0.00203421 | -0.00094207 | 0.00030696 | 0.00004895 | -0.00031631 | -0.00007172 | 0.00024436 | -0.00006714 |
| 69 | 110 | 0.04564089 | 0.05060293 | -0.03055844 | -0.03446712 | 0.01898499 | 0.01132780 | -0.01537849 | -0.00568167 | 0.00609748 | 0.00194312 | -0.00109391 | 0.00035824 | 0.00006395 | -0.00026134 | 0.00003225 | 0.00006206 | -0.00016516 |
| 69 | 111 | 0.04489280 | 0.05014217 | -0.03015439 | -0.03475563 | 0.01778592 | 0.01113028 | -0.01512608 | -0.00556268 | 0.00571300 | 0.00170158 | -0.00102606 | 0.00040886 | -0.00005646 | -0.00027818 | -0.00005807 | 0.00018203 | -0.00015802 |
| 69 | 112 | 0.04568263 | 0.04996942 | -0.03131105 | -0.03350663 | 0.01945664 | 0.01026918 | -0.01540350 | -0.00457340 | 0.00630800 | 0.00162971 | -0.00117147 | 0.00045164 | -0.00003242 | -0.00021886 | 0.00005625 | -0.00001250 | -0.00026316 |
| 69 | 113 | 0.04496795 | 0.04963120 | -0.03088116 | -0.03388081 | 0.01803777 | 0.00993589 | -0.01507792 | -0.00443056 | 0.00586728 | 0.00136061 | -0.00109818 | 0.00048855 | -0.00012899 | -0.00024220 | -0.00002176 | 0.00008218 | -0.00025086 |
| 70 | 78 | 0.04670039 | 0.06135638 | -0.01991881 | -0.04666656 | 0.00962609 | 0.02762297 | -0.00518439 | -0.01386454 | 0.00004276 | 0.00396633 | -0.00012173 | 0.00007756 | 0.00025246 | -0.00003143 | -0.00056770 | 0.00037488 | 0.00000445 |
| 70 | 79 | 0.04618494 | 0.06236879 | -0.01902809 | -0.04851323 | 0.00855382 | 0.02702865 | -0.00501835 | -0.01376638 | 0.00045768 | 0.00390257 | -0.00010942 | -0.00020222 | 0.00061403 | -0.00014522 | -0.00033774 | 0.00014696 | 0.00015855 |
| 70 | 80 | 0.04633223 | 0.06123487 | -0.02019226 | -0.04664925 | 0.00930756 | 0.02660725 | -0.00480671 | -0.01293670 | 0.00040223 | 0.00372723 | -0.00009190 | -0.00000765 | 0.00035391 | -0.00007501 | -0.00042611 | 0.00023127 | 0.00005047 |
| 70 | 81 | 0.04580463 | 0.06231276 | -0.01915146 | -0.04863090 | 0.00806127 | 0.02630373 | -0.00444960 | -0.01296559 | 0.00070911 | 0.00370394 | -0.00008592 | -0.00026394 | 0.00069085 | -0.00017724 | -0.00022887 | 0.00002191 | 0.00018212 |
| 70 | 82 | 0.04607174 | 0.06102124 | -0.02061983 | -0.04655332 | 0.00907193 | 0.02570746 | -0.00434166 | -0.01195883 | 0.00074102 | 0.00350586 | -0.00007532 | -0.00006119 | 0.00041414 | -0.00010109 | -0.00031199 | 0.00010224 | 0.00006648 |
| 70 | 83 | 0.04602499 | 0.06076732 | -0.02115111 | -0.04679716 | 0.00904811 | 0.02457546 | -0.00515558 | -0.01199682 | 0.00112869 | 0.00359476 | -0.00011533 | -0.00014592 | 0.00054327 | -0.00016109 | -0.00023536 | 0.00005493 | 0.00013520 |
| 70 | 84 | 0.04639669 | 0.06010735 | -0.02209045 | -0.04584339 | 0.00982013 | 0.02444801 | -0.00560290 | -0.01172639 | 0.00110344 | 0.00344158 | -0.00017072 | -0.00001543 | 0.00038929 | -0.00011219 | -0.00029235 | 0.00012932 | 0.00005941 |
| 70 | 85 | 0.04631218 | 0.05977359 | -0.02265909 | -0.04591908 | 0.01027403 | 0.02355115 | -0.00651423 | -0.01173205 | 0.00158230 | 0.00350598 | -0.00022319 | -0.00007970 | 0.00049139 | -0.00017703 | -0.00022403 | 0.00009966 | 0.00012254 |
| 70 | 86 | 0.04666049 | 0.05939688 | -0.02329324 | -0.04532939 | 0.01080381 | 0.02364314 | -0.00676219 | -0.01154729 | 0.00152865 | 0.00339188 | -0.00028009 | 0.00001427 | 0.00038627 | -0.00013034 | -0.00026506 | 0.00013439 | 0.00004288 |
| 70 | 87 | 0.04653018 | 0.05898381 | -0.02387202 | -0.04523709 | 0.01161811 | 0.02291851 | -0.00777615 | -0.01153641 | 0.00206361 | 0.00343344 | -0.00034185 | -0.00002691 | 0.00045798 | -0.00019702 | -0.00021143 | 0.00012929 | 0.00010011 |
| 70 | 88 | 0.04687409 | 0.05864488 | -0.02441750 | -0.04461967 | 0.01204281 | 0.02292500 | -0.00790099 | -0.01130989 | 0.00201990 | 0.00335241 | -0.00039205 | 0.00004491 | 0.00037902 | -0.00014927 | -0.00023699 | 0.00013103 | 0.00001625 |
| 70 | 89 | 0.04669420 | 0.05816967 | -0.02498323 | -0.04441354 | 0.01305258 | 0.02229254 | -0.00901603 | -0.01130441 | 0.00257434 | 0.00337278 | -0.00045851 | 0.00002433 | 0.00042256 | -0.00021599 | -0.00020070 | 0.00015531 | 0.00007131 |
| 70 | 90 | 0.04699553 | 0.05785988 | -0.02539153 | -0.04375123 | 0.01334774 | 0.02215767 | -0.00901901 | -0.01102784 | 0.00254733 | 0.00331434 | -0.00049890 | 0.00006937 | 0.00037530 | -0.00017044 | -0.00020317 | 0.00011922 | -0.00001030 |
| 70 | 91 | 0.04677014 | 0.05733892 | -0.02593098 | -0.04347760 | 0.01440710 | 0.02154770 | -0.01021114 | -0.01103651 | 0.00308983 | 0.00331303 | -0.00056432 | 0.00006507 | 0.00039442 | -0.00023619 | -0.00018426 | 0.00017388 | 0.00004646 |
| 70 | 92 | 0.04701146 | 0.05706440 | -0.02618095 | -0.04278703 | 0.01456762 | 0.02127330 | -0.01010438 | -0.01072238 | 0.00307953 | 0.00327258 | -0.00059496 | 0.00008396 | 0.00037899 | -0.00019455 | -0.00016248 | 0.00010139 | -0.00002989 |
| 70 | 93 | 0.04675341 | 0.05651293 | -0.02669159 | -0.04247745 | 0.01557604 | 0.02064209 | -0.01133296 | -0.01074097 | 0.00358741 | 0.00324791 | -0.00065388 | 0.00009091 | 0.00037837 | -0.00025877 | -0.00015888 | 0.00018326 | 0.00003110 |
| 70 | 94 | 0.04693392 | 0.05628539 | -0.02678684 | -0.04179007 | 0.01561303 | 0.02026653 | -0.01113631 | -0.01041385 | 0.00358975 | 0.00322550 | -0.00067681 | 0.00008819 | 0.00039048 | -0.00022148 | -0.00011659 | 0.00008066 | -0.00003946 |
| 70 | 95 | 0.04666139 | 0.05571570 | -0.02727479 | -0.04146257 | 0.01651738 | 0.01959549 | -0.01235515 | -0.01042987 | 0.00404917 | 0.00317554 | -0.00072498 | 0.00010147 | 0.00037514 | -0.00028376 | -0.00012484 | 0.00018329 | 0.00002633 |
| 70 | 96 | 0.04678860 | 0.05554854 | -0.02723107 | -0.04081212 | 0.01644911 | 0.01916974 | -0.01209310 | -0.01011811 | 0.00405914 | 0.00317326 | -0.00074300 | 0.00008318 | 0.00040865 | -0.00025075 | -0.00006796 | 0.00005927 | -0.00003881 |
| 70 | 97 | 0.04651988 | 0.05496788 | -0.02770700 | -0.04047475 | 0.01723363 | 0.01845741 | -0.01326038 | -0.01011624 | 0.00446356 | 0.00309658 | -0.00077772 | 0.00009850 | 0.00038339 | -0.00031062 | -0.00008400 | 0.00017445 | 0.00003068 |
| 70 | 98 | 0.04660364 | 0.05487337 | -0.02754360 | -0.03988864 | 0.01707920 | 0.01802754 | -0.01295940 | -0.00984517 | 0.00447749 | 0.00311653 | -0.00079364 | 0.00007052 | 0.00043215 | -0.00028180 | -0.00001862 | 0.00003811 | -0.00002933 |
| 70 | 99 | 0.04635295 | 0.05428307 | -0.02801861 | -0.03954317 | 0.01775206 | 0.01728096 | -0.01404333 | -0.00981130 | 0.00482567 | 0.00301275 | -0.00081371 | 0.00008447 | 0.00040122 | -0.00033859 | -0.00003841 | 0.00015739 | 0.00004175 |
| 70 | 100 | 0.04640323 | 0.05427090 | -0.02775459 | -0.03903845 | 0.01752853 | 0.01688029 | -0.01372914 | -0.00959941 | 0.00484197 | 0.00305590 | -0.00082991 | 0.00005179 | 0.00045981 | -0.00031404 | 0.00003020 | 0.00001704 | -0.00001302 |
| 70 | 101 | 0.04617907 | 0.05366710 | -0.02823794 | -0.03868373 | 0.01811021 | 0.01610909 | -0.01470927 | -0.00952272 | 0.00513630 | 0.00292595 | -0.00083533 | 0.00006190 | 0.00042671 | -0.00036692 | 0.00001025 | 0.00013282 | 0.00005709 |
| 70 | 102 | 0.04620551 | 0.05374382 | -0.02789116 | -0.03826662 | 0.01783191 | 0.01575696 | -0.01440457 | -0.00938042 | 0.00515515 | 0.00299161 | -0.00085363 | 0.00002845 | 0.00049069 | -0.00034691 | 0.00007798 | -0.00000465 | 0.00000799 |
| 70 | 103 | 0.04601092 | 0.05311920 | -0.02838929 | -0.03790121 | 0.01834672 | 0.01497040 | -0.01527041 | -0.00925399 | 0.00540026 | 0.00283780 | -0.00084531 | 0.00003307 | 0.00045817 | -0.00039493 | 0.00006078 | 0.00010142 | 0.00007460 |
| 70 | 104 | 0.04602296 | 0.05328808 | -0.02797656 | -0.03756837 | 0.01802581 | 0.01467446 | -0.01499342 | -0.00918416 | 0.00542280 | 0.00292357 | -0.00086689 | 0.00000184 | 0.00052393 | -0.00037984 | 0.00012461 | -0.00002780 | 0.00003171 |
| 70 | 105 | 0.04600264 | 0.05252766 | -0.02900080 | -0.03707008 | 0.01856026 | 0.01389549 | -0.01556395 | -0.00850990 | 0.00567167 | 0.00265304 | -0.00088830 | 0.00007888 | 0.00039638 | -0.00039007 | 0.00007344 | 0.00009453 | 0.00004973 |
| 70 | 106 | 0.04614152 | 0.05244616 | -0.02909019 | -0.03656814 | 0.01842060 | 0.01358976 | -0.01503955 | -0.00794239 | 0.00571053 | 0.00264362 | -0.00094500 | 0.00012006 | 0.00036739 | -0.00032942 | 0.00010571 | -0.00002331 | -0.00004651 |
| 70 | 107 | 0.04612112 | 0.05186309 | -0.03002976 | -0.03614242 | 0.01886960 | 0.01283893 | -0.01567526 | -0.00730824 | 0.00596313 | 0.00236312 | -0.00097594 | 0.00019980 | 0.00024893 | -0.00034936 | 0.00005710 | 0.00009712 | -0.00002620 |
| 70 | 108 | 0.04622512 | 0.05164744 | -0.03009512 | -0.03557711 | 0.01891267 | 0.01253580 | -0.01510332 | -0.00676698 | 0.00596837 | 0.00235020 | -0.00102223 | 0.00022811 | 0.00023219 | -0.00028026 | 0.00010192 | -0.00004446 | -0.00012980 |
| 70 | 109 | 0.04619607 | 0.05122847 | -0.03093097 | -0.03524198 | 0.01923767 | 0.01179817 | -0.01577866 | -0.00616620 | 0.00622160 | 0.00206462 | -0.00106406 | 0.00031090 | 0.00012213 | -0.00030649 | 0.00005639 | 0.00007101 | -0.00010988 |
| 70 | 110 | 0.04625967 | 0.05090306 | -0.03096381 | -0.03461096 | 0.01947453 | 0.01152831 | -0.01515875 | -0.00567173 | 0.00618978 | 0.00204764 | -0.00109297 | 0.00031984 | 0.00012413 | -0.00023528 | 0.00011546 | -0.00009194 | -0.00021343 |
| 70 | 111 | 0.04621562 | 0.05062735 | -0.03168473 | -0.03437919 | 0.01964723 | 0.01079607 | -0.01583993 | -0.00509785 | 0.00643654 | 0.00176134 | -0.00114585 | 0.00040594 | 0.00002168 | -0.00026449 | 0.00007335 | 0.00001493 | -0.00019719 |
| 70 | 112 | 0.04623414 | 0.05021677 | -0.03167594 | -0.03367444 | 0.02007265 | 0.01056795 | -0.01518431 | -0.00466440 | 0.00636824 | 0.00174013 | -0.00115248 | 0.00039115 | 0.00004650 | -0.00019639 | 0.00014693 | -0.00016459 | -0.00029313 |
| 70 | 113 | 0.04617073 | 0.05005994 | -0.03227594 | -0.03355372 | 0.02007507 | 0.00984194 | -0.01583198 | -0.00411366 | 0.00659754 | 0.00145690 | -0.00121513 | 0.00048046 | -0.00004888 | -0.00022576 | 0.00010846 | -0.00007039 | -0.00028401 |
| 70 | 114 | 0.04614005 | 0.04958431 | -0.03221821 | -0.03276215 | 0.02067327 | 0.00964663 | -0.01516546 | -0.00374952 | 0.00649687 | 0.00143173 | -0.00119726 | 0.00044041 | -0.00000046 | -0.00016420 | 0.00019515 | -0.00025950 | -0.00036576 |
| 70 | 115 | 0.04605422 | 0.04952215 | -0.03269355 | -0.03275543 | 0.02049575 | 0.00893517 | -0.01573915 | -0.00322283 | 0.00669453 | 0.00115493 | -0.00126719 | 0.00053253 | -0.00008890 | -0.00019159 | 0.00016041 | -0.00018199 | -0.00036700 |
| 71 | 79 | 0.04714522 | 0.06052965 | -0.02231007 | -0.04544565 | 0.01242909 | 0.02557570 | -0.00636432 | -0.01280275 | 0.00141213 | 0.00427027 | -0.00012802 | -0.00011821 | 0.00053179 | -0.00016327 | -0.00036561 | 0.00024533 | 0.00015425 |
| 71 | 80 | 0.04666511 | 0.06133441 | -0.02099408 | -0.04656259 | 0.01065949 | 0.02617925 | -0.00542155 | -0.01266013 | 0.00106746 | 0.00390502 | -0.00010950 | -0.00008770 | 0.00048025 | -0.00014929 | -0.00034847 | 0.00018331 | 0.00011268 |
| 71 | 81 | 0.04680303 | 0.06051156 | -0.02255999 | -0.04574145 | 0.01175338 | 0.02480126 | -0.00579846 | -0.01200874 | 0.00166048 | 0.00407621 | -0.00011038 | -0.00017758 | 0.00060936 | -0.00019852 | -0.00026233 | 0.00011722 | 0.00017457 |
| 71 | 82 | 0.04641364 | 0.06121102 | -0.02133899 | -0.04667322 | 0.01015523 | 0.02547125 | -0.00481176 | -0.01180314 | 0.00131143 | 0.00371923 | -0.00009514 | -0.00012743 | 0.00052709 | -0.00016957 | -0.00025974 | 0.00007073 | 0.00011769 |
| 71 | 83 | 0.04674570 | 0.05999083 | -0.02333273 | -0.04546725 | 0.01190156 | 0.02389176 | -0.00616632 | -0.01147124 | 0.00196372 | 0.00390872 | -0.00015729 | -0.00016074 | 0.00060202 | -0.00021382 | -0.00021062 | 0.00007465 | 0.00016444 |
| 71 | 84 | 0.04673356 | 0.06007116 | -0.02307761 | -0.04565316 | 0.01125034 | 0.02395725 | -0.00637225 | -0.01146779 | 0.00177858 | 0.00360792 | -0.00020099 | -0.00005597 | 0.00047250 | -0.00018180 | -0.00023660 | 0.00010841 | 0.00010969 |
| 71 | 85 | 0.04688260 | 0.05917063 | -0.02441756 | -0.04486171 | 0.01290298 | 0.02301615 | -0.00748983 | -0.01126941 | 0.00236826 | 0.00376649 | -0.00026447 | -0.00010018 | 0.00055271 | -0.00022990 | -0.00018999 | 0.00010078 | 0.00014359 |
| 71 | 86 | 0.04698549 | 0.05928678 | -0.02439583 | -0.04505889 | 0.01241346 | 0.02304127 | -0.00777317 | -0.01126032 | 0.00226182 | 0.00350624 | -0.00032309 | -0.00000654 | 0.00044841 | -0.00020241 | -0.00020910 | 0.00012536 | 0.00009329 |
| 71 | 87 | 0.04703579 | 0.05836357 | -0.02546055 | -0.04420149 | 0.01403326 | 0.02233016 | -0.00877500 | -0.01107587 | 0.00278109 | 0.00364645 | -0.00037672 | -0.00002829 | 0.00048522 | -0.00024135 | -0.00018831 | 0.00014020 | 0.00010779 |
| 71 | 88 | 0.04722417 | 0.05849034 | -0.02564049 | -0.04432147 | 0.01374691 | 0.02225843 | -0.00911096 | -0.01101512 | 0.00277496 | 0.00342079 | -0.00044571 | 0.00004736 | 0.00041331 | -0.00022104 | -0.00019122 | 0.00014422 | 0.00006473 |
| 71 | 89 | 0.04715677 | 0.05755487 | -0.02640106 | -0.04347479 | 0.01515918 | 0.02170500 | -0.00999183 | -0.01088386 | 0.00318900 | 0.00354223 | -0.00048372 | 0.00004372 | 0.00041107 | -0.00025081 | -0.00019650 | 0.00018703 | 0.00006807 |
| 71 | 90 | 0.04739610 | 0.05768068 | -0.02672940 | -0.04346409 | 0.01506337 | 0.02146699 | -0.01036549 | -0.01073899 | 0.00329117 | 0.00334310 | -0.00055884 | 0.00009549 | 0.00037801 | -0.00023968 | -0.00017480 | 0.00016105 | 0.00003539 |
| 71 | 91 | 0.04720496 | 0.05674685 | -0.02718663 | -0.04269395 | 0.01617293 | 0.02103480 | -0.01111944 | -0.01068800 | 0.00358151 | 0.00344360 | -0.00057752 | 0.00010482 | 0.00034329 | -0.00026150 | -0.00020393 | 0.00023258 | 0.00003414 |
| 71 | 92 | 0.04747277 | 0.05687267 | -0.02761460 | -0.04253362 | 0.01622982 | 0.02058743 | -0.01151973 | -0.01044475 | 0.00378681 | 0.00326602 | -0.00065579 | 0.00013069 | 0.00035095 | -0.00026002 | -0.00015389 | 0.00017244 | 0.00001318 |
| 71 | 93 | 0.04716477 | 0.05594687 | -0.02779997 | -0.04187589 | 0.01701599 | 0.02025650 | -0.01214388 | -0.01048142 | 0.00395081 | 0.00334197 | -0.00065356 | 0.00014825 | 0.00029083 | -0.00027540 | -0.00020277 | 0.00026913 | 0.00001141 |
| 71 | 94 | 0.04745459 | 0.05608618 | -0.02829054 | -0.04157854 | 0.01718178 | 0.01960234 | -0.01255895 | -0.01014531 | 0.00424393 | 0.00318575 | -0.00073336 | 0.00015020 | 0.00033613 | -0.00028268 | -0.00012607 | 0.00017641 | 0.00000141 |
| 71 | 95 | 0.04704443 | 0.05516630 | -0.02825371 | -0.04104156 | 0.01767264 | 0.01935510 | -0.01305472 | -0.01025894 | 0.00429093 | 0.00323271 | -0.00071045 | 0.00017193 | 0.00025752 | -0.00029298 | -0.00018928 | 0.00029161 | 0.00000097 |

TABLE 3. Nuclear charge density distribution FB coefficients.

| $Z$ | $N$ | $a_1$ | $a_2$ | $a_3$ | $a_4$ | $a_5$ | $a_6$ | $a_7$ | $a_8$ | $a_9$ | $a_{10}$ | $a_{11}$ | $a_{12}$ | $a_{13}$ | $a_{14}$ | $a_{15}$ | $a_{16}$ | $a_{17}$ |
|---|---|---|---|---|---|---|---|---|---|---|---|---|---|---|---|---|---|---|
| 71 | 96 | 0.04736066 | 0.05534074 | -0.02877969 | -0.04064095 | 0.01790796 | 0.01853515 | -0.01347315 | -0.00985158 | 0.00465137 | 0.00310126 | -0.00079113 | 0.00015445 | 0.00033425 | -0.00030749 | -0.00009137 | 0.00017209 | -0.00000013 |
| 71 | 97 | 0.04686594 | 0.05441765 | -0.02857666 | -0.04021251 | 0.01815562 | 0.01834953 | -0.01384682 | -0.01001875 | 0.00459827 | 0.00311498 | -0.00074919 | 0.00017719 | 0.00024328 | -0.00031366 | -0.00016310 | 0.00029784 | 0.00000110 |
| 71 | 98 | 0.04721640 | 0.05465112 | -0.02911662 | -0.03975100 | 0.01842950 | 0.01742433 | -0.01426051 | -0.00957123 | 0.00500485 | 0.00301301 | -0.00083060 | 0.00014545 | 0.00034426 | -0.00033390 | -0.00005094 | 0.00015928 | 0.00000664 |
| 71 | 99 | 0.04665419 | 0.05371132 | -0.02880147 | -0.03940682 | 0.01849208 | 0.01727365 | -0.01452199 | -0.00976228 | 0.00487197 | 0.00299049 | -0.00077221 | 0.00016712 | 0.00024577 | -0.00033634 | -0.00012594 | 0.00028799 | 0.00000890 |
| 71 | 100 | 0.04704527 | 0.05402496 | -0.02933740 | -0.03892551 | 0.01878259 | 0.01630586 | -0.01492777 | -0.00930820 | 0.00530592 | 0.00292208 | -0.00085430 | 0.00012583 | 0.00036438 | -0.00036121 | -0.00000618 | 0.00013824 | 0.00001927 |
| 71 | 101 | 0.04643060 | 0.05305386 | -0.02895715 | -0.03863702 | 0.01871332 | 0.01616236 | -0.01508838 | -0.00949285 | 0.00511369 | 0.00286217 | -0.00078246 | 0.00014530 | 0.00026181 | -0.00035989 | -0.00008027 | 0.00026369 | 0.00002153 |
| 71 | 102 | 0.04686587 | 0.05346332 | -0.02947432 | -0.03816970 | 0.01900699 | 0.01520581 | -0.01548778 | -0.00906283 | 0.00556017 | 0.00282965 | -0.00086520 | 0.00009822 | 0.00039263 | -0.00038871 | 0.00004165 | 0.00010955 | 0.00003545 |
| 71 | 103 | 0.04621105 | 0.05244774 | -0.02906687 | -0.03790987 | 0.01884900 | 0.01504471 | -0.01555812 | -0.00921420 | 0.00532686 | 0.00273319 | -0.00078290 | 0.00011513 | 0.00028815 | -0.00038326 | -0.00002866 | 0.00022732 | 0.00003660 |
| 71 | 104 | 0.04669146 | 0.05296224 | -0.02955433 | -0.03748054 | 0.01913957 | 0.01413957 | -0.01595610 | -0.00883252 | 0.00577531 | 0.00273667 | -0.00086620 | 0.00006504 | 0.00042704 | -0.00041576 | 0.00009154 | 0.00007400 | 0.00005328 |
| 71 | 105 | 0.04615778 | 0.05194009 | -0.02960496 | -0.03721528 | 0.01893618 | 0.01405402 | -0.01580543 | -0.00853196 | 0.00555965 | 0.00253232 | -0.00081459 | 0.00014599 | 0.00024342 | -0.00038848 | -0.00000895 | 0.00021684 | 0.00002489 |
| 71 | 106 | 0.04676097 | 0.05235972 | -0.03053229 | -0.03668921 | 0.01935411 | 0.01320185 | -0.01604884 | -0.00771021 | 0.00605646 | 0.00245925 | -0.00094346 | 0.00017054 | 0.00029250 | -0.00038527 | 0.00007705 | 0.00008861 | -0.00000128 |
| 71 | 107 | 0.04624625 | 0.05150492 | -0.03053101 | -0.03653546 | 0.01902629 | 0.01312386 | -0.01590144 | -0.00745795 | 0.00581137 | 0.00224646 | -0.00088953 | 0.00024140 | 0.00012973 | -0.00037030 | -0.00001629 | 0.00022167 | -0.00002167 |
| 71 | 108 | 0.04679590 | 0.05177826 | -0.03140128 | -0.03591912 | 0.01962790 | 0.01225711 | -0.01615283 | -0.00663454 | 0.00630934 | 0.00217260 | -0.00102592 | 0.00027194 | 0.00017248 | -0.00034958 | 0.00007452 | 0.00007851 | -0.00006583 |
| 71 | 109 | 0.04630604 | 0.05107719 | -0.03136746 | -0.03588355 | 0.01913730 | 0.01214010 | -0.01598814 | -0.00640873 | 0.00602919 | 0.00194651 | -0.00096892 | 0.00033136 | 0.00003128 | -0.00034546 | -0.00001022 | 0.00019829 | -0.00007770 |
| 71 | 110 | 0.04678722 | 0.05121745 | -0.03214794 | -0.03517539 | 0.01995763 | 0.01132835 | -0.01623718 | -0.00561843 | 0.00652470 | 0.00187947 | -0.00110744 | 0.00036340 | 0.00007258 | -0.00031128 | 0.00008667 | 0.00004074 | -0.00013775 |
| 71 | 111 | 0.04632757 | 0.05064929 | -0.03210478 | -0.03525807 | 0.01928588 | 0.01113351 | -0.01604036 | -0.00540384 | 0.00620177 | 0.00163530 | -0.00104702 | 0.00041093 | -0.00004686 | -0.00031588 | 0.00001181 | 0.00014306 | -0.00014221 |
| 71 | 112 | 0.04672737 | 0.05067686 | -0.03275897 | -0.03445541 | 0.02033053 | 0.01042992 | -0.01627246 | -0.00467240 | 0.00669255 | 0.00158283 | -0.00118168 | 0.00043997 | -0.00000288 | -0.00027295 | 0.00011494 | -0.00002575 | -0.00021379 |
| 71 | 113 | 0.04630242 | 0.05021802 | -0.03272745 | -0.03464783 | 0.01948100 | 0.01013476 | -0.01603522 | -0.00446472 | 0.00631641 | 0.00131701 | -0.00111767 | 0.00047630 | -0.00010110 | -0.00028374 | 0.00005061 | 0.00005484 | -0.00021363 |
| 71 | 114 | 0.04660993 | 0.05015533 | -0.03322100 | -0.03374938 | 0.02072678 | 0.00956827 | -0.01623648 | -0.00380663 | 0.00680244 | 0.00128606 | -0.00124297 | 0.00049846 | -0.00005169 | -0.00023659 | 0.00015907 | -0.00011972 | -0.00029069 |
| 71 | 115 | 0.04622234 | 0.04978409 | -0.03321357 | -0.03403492 | 0.01972017 | 0.00917099 | -0.01595816 | -0.00361472 | 0.00636050 | 0.00099731 | -0.00117535 | 0.00052580 | -0.00013053 | -0.00025071 | 0.00010466 | -0.00006409 | -0.00029022 |
| 71 | 116 | 0.04642901 | 0.04964934 | -0.03352163 | -0.03304150 | 0.02112302 | 0.00874477 | -0.01611853 | -0.00303224 | 0.00684416 | 0.00099294 | -0.00128715 | 0.00053833 | -0.00007471 | -0.00020306 | 0.00021677 | -0.00023736 | -0.00036593 |
| 71 | 117 | 0.04607905 | 0.04934968 | -0.03353792 | -0.03339792 | 0.01998885 | 0.00826327 | -0.01580748 | -0.00287705 | 0.00632349 | 0.00068309 | -0.00121627 | 0.00056088 | -0.00013802 | -0.00021738 | 0.00016976 | -0.00020762 | -0.00037066 |
| 72 | 81 | 0.04716564 | 0.06086567 | -0.02243341 | -0.04529173 | 0.01248183 | 0.02582181 | -0.00549706 | -0.01163496 | 0.00164753 | 0.00386348 | -0.00011916 | -0.00003497 | 0.00042302 | -0.00016915 | -0.00032403 | 0.00017399 | 0.00008506 |
| 72 | 82 | 0.04717481 | 0.06063225 | -0.02259134 | -0.04483420 | 0.01220463 | 0.02600239 | -0.00478800 | -0.01086011 | 0.00155633 | 0.00354698 | -0.00012424 | 0.00008413 | 0.00024209 | -0.00012561 | -0.00035401 | 0.00017065 | -0.00001632 |
| 72 | 83 | 0.04719925 | 0.06031950 | -0.02345158 | -0.04506121 | 0.01259638 | 0.02467499 | -0.00599748 | -0.01106799 | 0.00200252 | 0.00371009 | -0.00017186 | -0.00002531 | 0.00043168 | -0.00018562 | -0.00026111 | 0.00012644 | 0.00008484 |
| 72 | 84 | 0.04740395 | 0.05996531 | -0.02387707 | -0.04449886 | 0.01278457 | 0.02461197 | -0.00626613 | -0.01072419 | 0.00196364 | 0.00345519 | -0.00023656 | 0.00010044 | 0.00027611 | -0.00014948 | -0.00029980 | 0.00015908 | -0.00001038 |
| 72 | 85 | 0.04744909 | 0.05952708 | -0.02485230 | -0.04455095 | 0.01353965 | 0.02340958 | -0.00757054 | -0.01086222 | 0.00248190 | 0.00358302 | -0.00029642 | 0.00001879 | 0.00042194 | -0.00020624 | -0.00021668 | 0.00012869 | 0.00007571 |
| 72 | 86 | 0.04766776 | 0.05926985 | -0.02523328 | -0.04409746 | 0.01364376 | 0.02342459 | -0.00773570 | -0.01056930 | 0.00242698 | 0.00338317 | -0.00035726 | 0.00012767 | 0.00029179 | -0.00017092 | -0.00025547 | 0.00014635 | -0.00002353 |
| 72 | 87 | 0.04772796 | 0.05873935 | -0.02622003 | -0.04391959 | 0.01471335 | 0.02237343 | -0.00906060 | -0.01062222 | 0.00299107 | 0.00347228 | -0.00042416 | 0.00007218 | 0.00039417 | -0.00022471 | -0.00018730 | 0.00013662 | 0.00005004 |
| 72 | 88 | 0.04790712 | 0.05854195 | -0.02651043 | -0.04355342 | 0.01467996 | 0.02238518 | -0.00911704 | -0.01037827 | 0.00292320 | 0.00332414 | -0.00047579 | 0.00015802 | 0.00029498 | -0.00019136 | -0.00021750 | 0.00013366 | -0.00004579 |
| 72 | 89 | 0.04797354 | 0.05794866 | -0.02745207 | -0.04317061 | 0.01593794 | 0.02144812 | -0.01042694 | -0.01035603 | 0.00350144 | 0.00337437 | -0.00054406 | 0.00012502 | 0.00035802 | -0.00024238 | -0.00016677 | 0.00014827 | 0.00001898 |
| 72 | 90 | 0.04807007 | 0.05779944 | -0.02759659 | -0.04288071 | 0.01574532 | 0.02141857 | -0.01037586 | -0.01016220 | 0.00342447 | 0.00327191 | -0.00058419 | 0.00018347 | 0.00029439 | -0.00021248 | -0.00018147 | 0.00012093 | -0.00006750 |
| 72 | 91 | 0.04813646 | 0.05716436 | -0.02847409 | -0.04234598 | 0.01706061 | 0.02053738 | -0.01165259 | -0.01008073 | 0.00398821 | 0.00328359 | -0.00064824 | 0.00016826 | 0.00032411 | -0.00026102 | -0.00014799 | 0.00015984 | -0.00000793 |
| 72 | 92 | 0.04813163 | 0.05706279 | -0.02844436 | -0.04212363 | 0.01671878 | 0.02046031 | -0.01150675 | -0.00993885 | 0.00390634 | 0.00322185 | -0.00067745 | 0.00019928 | 0.00029638 | -0.00023535 | -0.00014448 | 0.00010803 | -0.00008232 |
| 72 | 93 | 0.04819681 | 0.05640014 | -0.02926072 | -0.04149003 | 0.01799401 | 0.01958762 | -0.01273364 | -0.00981029 | 0.00443387 | 0.00319512 | -0.00073251 | 0.00019679 | 0.00029963 | -0.00028169 | -0.00012599 | 0.00016777 | -0.00002532 |
| 72 | 94 | 0.04809396 | 0.05635309 | -0.02905631 | -0.04133147 | 0.01752995 | 0.01947759 | -0.01251366 | -0.00972295 | 0.00435110 | 0.00317062 | -0.00075330 | 0.00020375 | 0.00030441 | -0.00026044 | -0.00010535 | 0.00009465 | -0.00008748 |
| 72 | 95 | 0.04816105 | 0.05567149 | -0.02982517 | -0.04064138 | 0.01871008 | 0.01858751 | -0.01367160 | -0.00955343 | 0.00482844 | 0.00310605 | -0.00079589 | 0.00020915 | 0.00028797 | -0.00030467 | -0.00009841 | 0.00016940 | -0.00003181 |
| 72 | 96 | 0.04797734 | 0.05568842 | -0.02946484 | -0.04054448 | 0.01815586 | 0.01846563 | -0.01340233 | -0.00952243 | 0.00474869 | 0.00311584 | -0.00081149 | 0.00019733 | 0.00031964 | -0.00028779 | -0.00006402 | 0.00008022 | -0.00008291 |
| 72 | 97 | 0.04805098 | 0.05499175 | -0.03020253 | -0.03982816 | 0.01922060 | 0.01755079 | -0.01447255 | -0.00931363 | 0.00516886 | 0.00301508 | -0.00083963 | 0.00020642 | 0.00028970 | -0.00032966 | -0.00006478 | 0.00016307 | -0.00002855 |
| 72 | 98 | 0.04780891 | 0.05508067 | -0.02971444 | -0.03978850 | 0.01860654 | 0.01743454 | -0.01417987 | -0.00933867 | 0.00509607 | 0.00305581 | -0.00085317 | 0.00018156 | 0.00034179 | -0.00031716 | -0.00002091 | 0.00006400 | -0.00007002 |
| 72 | 99 | 0.04789252 | 0.05436908 | -0.03043501 | -0.03906644 | 0.01955817 | 0.01649904 | -0.01514751 | -0.00909034 | 0.00545769 | 0.00292200 | -0.00086633 | 0.00019093 | 0.00030370 | -0.00035608 | -0.00002566 | 0.00014809 | -0.00001783 |
| 72 | 100 | 0.04761486 | 0.05453444 | -0.02985015 | -0.03907529 | 0.01891000 | 0.01639801 | -0.01485549 | -0.00916832 | 0.00539548 | 0.00298931 | -0.00088031 | 0.00015847 | 0.00036984 | -0.00034809 | 0.00002342 | 0.00004530 | -0.00005086 |
| 72 | 101 | 0.04770916 | 0.05380576 | -0.03056286 | -0.03836135 | 0.01976224 | 0.01545136 | -0.01571137 | -0.00888016 | 0.00570127 | 0.00282718 | -0.00087910 | 0.00016547 | 0.00032807 | -0.00038322 | 0.00001795 | 0.00012455 | -0.00000209 |
| 72 | 102 | 0.04741651 | 0.05404769 | -0.02991158 | -0.03840528 | 0.01910087 | 0.01536708 | -0.01544014 | -0.00900497 | 0.00565254 | 0.00291537 | -0.00089524 | 0.00013014 | 0.00040239 | -0.00038004 | 0.00006851 | 0.00002355 | -0.00002754 |
| 72 | 103 | 0.04751932 | 0.05329906 | -0.03062042 | -0.03771003 | 0.01987082 | 0.01442081 | -0.01618054 | -0.00867793 | 0.00590774 | 0.00273116 | -0.00088105 | 0.00013281 | 0.00036055 | -0.00041039 | 0.00006495 | 0.00009309 | 0.00001648 |
| 72 | 104 | 0.04722970 | 0.05361360 | -0.02993120 | -0.03777177 | 0.01921347 | 0.01434812 | -0.01594521 | -0.00884044 | 0.00587446 | 0.00283319 | -0.00090026 | 0.00009851 | 0.00043793 | -0.00041241 | 0.00011396 | -0.00000161 | -0.00000196 |
| 72 | 105 | 0.04743592 | 0.05279519 | -0.03108981 | -0.03707951 | 0.01997502 | 0.01350768 | -0.01642348 | -0.00806660 | 0.00612980 | 0.00254892 | -0.00091228 | 0.00015806 | 0.00032139 | -0.00041335 | 0.00008544 | 0.00008382 | 0.00000846 |
| 72 | 106 | 0.04723733 | 0.05290430 | -0.03085111 | -0.03705119 | 0.01946997 | 0.01346425 | -0.01595812 | -0.00777646 | 0.00611329 | 0.00256339 | -0.00096539 | 0.00018595 | 0.00031628 | -0.00037356 | 0.00010892 | 0.00000112 | -0.00005210 |
| 72 | 107 | 0.04743217 | 0.05226760 | -0.03193118 | -0.03645088 | 0.02016223 | 0.01267830 | -0.01651571 | -0.00707044 | 0.00637144 | 0.00227375 | -0.00098502 | 0.00024485 | 0.00021323 | -0.00038726 | 0.00008422 | 0.00008665 | -0.00003496 |
| 72 | 108 | 0.04722024 | 0.05221737 | -0.03168795 | -0.03632436 | 0.01983044 | 0.01258702 | -0.01599190 | -0.00676108 | 0.00632355 | 0.00227905 | -0.00103163 | 0.00026764 | 0.00021039 | -0.00033345 | 0.00011571 | -0.00001840 | -0.00010903 |
| 72 | 109 | 0.04739350 | 0.05175079 | -0.03267022 | -0.03584281 | 0.02040632 | 0.01184134 | -0.01660773 | -0.00611858 | 0.00658180 | 0.00199022 | -0.00106154 | 0.00032745 | 0.00011886 | -0.00035571 | 0.00009431 | 0.00006551 | -0.00008802 |
| 72 | 110 | 0.04717057 | 0.05155934 | -0.03242394 | -0.03559063 | 0.02028345 | 0.01172640 | -0.01602772 | -0.00580522 | 0.00649981 | 0.00198328 | -0.00109431 | 0.00033870 | 0.00012515 | -0.00029443 | 0.00013659 | -0.00006202 | -0.00016982 |
| 72 | 111 | 0.04731519 | 0.05124076 | -0.03329992 | -0.03524774 | 0.02070813 | 0.01101177 | -0.01667305 | -0.00522124 | 0.00675147 | 0.00170057 | -0.00113609 | 0.00040121 | 0.00004266 | -0.00032089 | 0.00011774 | 0.00001776 | -0.00014892 |
| 72 | 112 | 0.04708060 | 0.05093404 | -0.03304082 | -0.03484572 | 0.02080627 | 0.01088580 | -0.01604611 | -0.00491725 | 0.00663571 | 0.00167989 | -0.00114890 | 0.00039554 | 0.00006362 | -0.00025831 | 0.00017229 | -0.00012954 | -0.00023138 |
| 72 | 113 | 0.04719265 | 0.05073657 | -0.03380841 | -0.03465338 | 0.02105853 | 0.01020149 | -0.01668675 | -0.00438948 | 0.00686979 | 0.00140775 | -0.00120273 | 0.00046235 | -0.00001219 | -0.00028500 | 0.00015515 | -0.00005715 | -0.00021515 |
| 72 | 114 | 0.04694299 | 0.05034093 | -0.03352159 | -0.03408131 | 0.02137001 | 0.01006536 | -0.01603081 | -0.00410479 | 0.00672378 | 0.00137330 | -0.00119171 | 0.00043658 | 0.00002626 | -0.00022589 | 0.00022170 | -0.00021862 | -0.00029109 |
| 72 | 115 | 0.04702097 | 0.05023878 | -0.03417998 | -0.03404368 | 0.02144001 | 0.00941918 | -0.01663107 | -0.00363604 | 0.00692569 | 0.00111551 | -0.00125628 | 0.00050887 | -0.00004470 | -0.00024969 | 0.00020534 | -0.00015721 | -0.00028421 |
| 72 | 116 | 0.04675079 | 0.04977325 | -0.03385308 | -0.03328564 | 0.02194485 | 0.00926548 | -0.01597163 | -0.00337535 | 0.00675581 | 0.00106842 | -0.00122039 | 0.00046284 | 0.00001035 | -0.00019663 | 0.00028173 | -0.00032476 | -0.00034760 |
| 72 | 117 | 0.04679495 | 0.04974676 | -0.03439815 | -0.03339984 | 0.02183009 | 0.00867089 | -0.01649988 | -0.00297468 | 0.00690902 | 0.00082819 | -0.00129318 | 0.00054156 | -0.00005712 | -0.00021542 | 0.00026504 | -0.00027763 | -0.00035434 |
| 72 | 118 | 0.04649790 | 0.04921666 | -0.03402913 | -0.03244474 | 0.02250503 | 0.00848896 | -0.01586659 | -0.00273570 | 0.00672389 | 0.00077006 | -0.00123437 | 0.00047815 | 0.00000985 | -0.00016849 | 0.00034745 | -0.00044167 | -0.00040119 |
| 73 | 82 | 0.04773773 | 0.06052010 | -0.02385876 | -0.04425863 | 0.01422803 | 0.02566980 | -0.00543440 | -0.01055703 | 0.00225025 | 0.00380684 | -0.00013823 | 0.00001467 | 0.00036297 | -0.00018709 | -0.00029613 | 0.00015284 | 0.00004668 |
| 73 | 83 | 0.04816273 | 0.05962824 | -0.02585954 | -0.04363140 | 0.01596212 | 0.02416212 | -0.00709347 | -0.01054740 | 0.00284926 | 0.00402135 | -0.00021166 | -0.00006916 | 0.00052093 | -0.00022898 | -0.00021544 | 0.00013283 | 0.00011821 |
| 73 | 84 | 0.04795344 | 0.05971737 | -0.02526421 | -0.04374868 | 0.01499740 | 0.02408820 | -0.00714012 | -0.01034077 | 0.00272478 | 0.00365851 | -0.00026313 | 0.00005201 | 0.00037018 | -0.00020762 | -0.00023568 | 0.00014149 | 0.00004550 |
| 73 | 85 | 0.04833765 | 0.05881469 | -0.02699508 | -0.04300782 | 0.01677790 | 0.02291588 | -0.00858448 | -0.01029608 | 0.00325233 | 0.00384850 | -0.00032741 | -0.00000730 | 0.00047566 | -0.00023683 | -0.00018100 | 0.00013558 | 0.00008754 |
| 73 | 86 | 0.04823235 | 0.05893500 | -0.02670520 | -0.04320157 | 0.01599049 | 0.02275450 | -0.00880013 | -0.01011512 | 0.00322527 | 0.00352530 | -0.00039476 | 0.00010190 | 0.00035605 | -0.00022591 | -0.00019238 | 0.00013649 | 0.00002477 |
| 73 | 87 | 0.04853715 | 0.05802671 | -0.02807689 | -0.04235730 | 0.01767296 | 0.02187588 | -0.00997818 | -0.01005354 | 0.00364964 | 0.00369294 | -0.00044183 | 0.00006283 | 0.00041314 | -0.00024303 | -0.00016546 | 0.00015223 | 0.00004648 |
| 73 | 88 | 0.04851602 | 0.05816163 | -0.02805516 | -0.04257062 | 0.01707713 | 0.02161635 | -0.01032373 | -0.00987635 | 0.00372382 | 0.00340801 | -0.00052071 | 0.00015599 | 0.00032663 | -0.00024286 | -0.00016375 | 0.00013955 | -0.00000527 |
| 73 | 89 | 0.04871086 | 0.05726600 | -0.02902986 | -0.04169534 | 0.01853236 | 0.02097865 | -0.01123293 | -0.00983003 | 0.00402706 | 0.00355354 | -0.00054625 | 0.00013151 | 0.00034379 | -0.00024997 | -0.00016262 | 0.00017943 | 0.00000470 |
| 73 | 90 | 0.04874278 | 0.05740501 | -0.02920767 | -0.04187800 | 0.01810884 | 0.02059942 | -0.01166902 | -0.00963935 | 0.00419532 | 0.00330363 | -0.00063186 | 0.00020453 | 0.00029260 | -0.00026021 | -0.00014352 | 0.00014823 | -0.00003426 |
| 73 | 91 | 0.04881558 | 0.05653058 | -0.02980635 | -0.04103339 | 0.01927338 | 0.02014449 | -0.01233940 | -0.00962935 | 0.00437658 | 0.00342477 | -0.00063488 | 0.00019016 | 0.00027820 | -0.00025971 | -0.00016431 | 0.00021107 | -0.00002994 |
| 73 | 92 | 0.04887399 | 0.05667311 | -0.03011221 | -0.04115588 | 0.01898501 | 0.01963274 | -0.01283068 | -0.00941685 | 0.00462325 | 0.00320755 | -0.00072300 | 0.00024060 | 0.00026317 | -0.00027930 | -0.00012532 | 0.00015848 | -0.00005522 |
| 73 | 93 | 0.04882748 | 0.05581800 | -0.03039099 | -0.04037746 | 0.01985488 | 0.01930615 | -0.01330469 | -0.00944785 | 0.00469450 | 0.00330015 | -0.00070492 | 0.00023310 | 0.00022471 | -0.00027339 | -0.00016309 | 0.00024024 | -0.00005291 |
| 73 | 94 | 0.04889980 | 0.05597531 | -0.03076672 | -0.04043379 | 0.01966201 | 0.01867172 | -0.01381985 | -0.00921464 | 0.00499949 | 0.00311538 | -0.00079249 | 0.00026084 | 0.00024439 | -0.00030075 | -0.00010447 | 0.00016604 | -0.00006516 |
| 73 | 95 | 0.04874519 | 0.05512929 | -0.03079577 | -0.03973177 | 0.02027042 | 0.01842469 | -0.01413961 | -0.00927694 | 0.00497982 | 0.00317489 | -0.00075613 | 0.00025787 | 0.00018817 | -0.00029103 | -0.00015392 | 0.00026106 | -0.00006310 |
| 73 | 96 | 0.04883293 | 0.05532086 | -0.03120112 | -0.03973366 | 0.02014060 | 0.01769896 | -0.01465321 | -0.00903171 | 0.00532282 | 0.00302383 | -0.00084130 | 0.00026489 | 0.00023897 | -0.00032458 | -0.00007838 | 0.00016744 | -0.00006421 |
| 73 | 97 | 0.04858451 | 0.05446904 | -0.03104966 | -0.03910065 | 0.02053623 | 0.01748918 | -0.01485403 | -0.00910659 | 0.00523358 | 0.00304693 | -0.00079004 | 0.00026483 | 0.00017007 | -0.00031194 | -0.00013451 | 0.00026968 | -0.00006189 |
| 73 | 98 | 0.04869745 | 0.05471645 | -0.03146010 | -0.03906799 | 0.02044740 | 0.01671404 | -0.01534923 | -0.00886256 | 0.00559711 | 0.00293100 | -0.00087193 | 0.00025439 | 0.00024712 | -0.00035034 | -0.00004619 | 0.00016051 | -0.00005429 |
| 73 | 99 | 0.04836942 | 0.05384307 | -0.03118797 | -0.03848786 | 0.02067959 | 0.01650746 | -0.01545728 | -0.00892807 | 0.00545839 | 0.00291639 | -0.00080917 | 0.00025613 | 0.00016938 | -0.00033503 | -0.00010484 | 0.00026446 | -0.00005194 |
| 73 | 100 | 0.04851925 | 0.05416450 | -0.03159011 | -0.03844055 | 0.02061960 | 0.01572351 | -0.01592692 | -0.00869965 | 0.00582941 | 0.00283606 | -0.00088758 | 0.00023197 | 0.00026745 | -0.00037739 | -0.00000817 | 0.00014435 | -0.00003790 |
| 73 | 101 | 0.04812438 | 0.05325588 | -0.03124468 | -0.03789552 | 0.02073044 | 0.01549571 | -0.01595931 | -0.00873516 | 0.00565799 | 0.00278477 | -0.00081637 | 0.00023481 | 0.00018362 | -0.00035913 | -0.00006626 | 0.00024561 | -0.00003615 |
| 73 | 102 | 0.04832118 | 0.05366309 | -0.03163262 | -0.03784844 | 0.02069522 | 0.01473463 | -0.01640459 | -0.00853491 | 0.00602812 | 0.00273890 | -0.00089151 | 0.00020055 | 0.00029772 | -0.00040496 | 0.00003478 | 0.00011908 | -0.00001750 |
| 73 | 103 | 0.04787016 | 0.05270923 | -0.03124879 | -0.03732350 | 0.02071673 | 0.01447123 | -0.01637090 | -0.00852412 | 0.00583655 | 0.00265398 | -0.00081441 | 0.00020410 | 0.00020968 | -0.00038321 | -0.00002077 | 0.00021457 | -0.00001712 |
| 73 | 104 | 0.04812094 | 0.05320706 | -0.03162153 | -0.03728465 | 0.02070788 | 0.01375291 | -0.01679862 | -0.00836056 | 0.00620160 | 0.00263964 | -0.00088666 | 0.00016297 | 0.00033541 | -0.00043234 | 0.00008149 | 0.00008554 | 0.00000484 |
| 73 | 105 | 0.04775172 | 0.05227047 | -0.03169444 | -0.03682835 | 0.02073465 | 0.01362415 | -0.01660349 | -0.00798002 | 0.00603267 | 0.00246174 | -0.00083748 | 0.00021540 | 0.00018811 | -0.00039592 | 0.00000676 | 0.00020295 | -0.00001201 |
| 73 | 106 | 0.04808488 | 0.05272568 | -0.03245214 | -0.03679520 | 0.02082915 | 0.01304551 | -0.01687481 | -0.00745477 | 0.00642598 | 0.00237646 | -0.00094722 | 0.00023173 | 0.00024340 | -0.00041678 | 0.00008408 | 0.00009751 | -0.00001758 |
| 73 | 107 | 0.04774820 | 0.05192059 | -0.03254066 | -0.03643006 | 0.02080649 | 0.01291382 | -0.01671467 | -0.00712495 | 0.00623719 | 0.00219466 | -0.00089727 | 0.00027492 | 0.00011661 | -0.00039148 | 0.00001750 | 0.00020505 | -0.00002813 |
| 73 | 108 | 0.04801919 | 0.05225037 | -0.03319188 | -0.03632775 | 0.02099565 | 0.01231353 | -0.01696182 | -0.00657761 | 0.00662505 | 0.00210256 | -0.00101483 | 0.00030101 | 0.00015929 | -0.00039404 | 0.00009474 | 0.00009005 | -0.00005027 |
| 73 | 109 | 0.04771919 | 0.05156438 | -0.03331401 | -0.03607082 | 0.02088445 | 0.01215178 | -0.01681455 | -0.00628839 | 0.00640542 | 0.00191244 | -0.00096319 | 0.00033423 | 0.00005290 | -0.00037842 | 0.00003656 | 0.00018548 | -0.00005413 |
| 73 | 110 | 0.04792244 | 0.05177107 | -0.03384386 | -0.03586907 | 0.02122128 | 0.01156650 | -0.01704106 | -0.00573843 | 0.00679034 | 0.00181937 | -0.00108502 | 0.00036717 | 0.00008685 | -0.00036545 | 0.00011565 | 0.00005941 | -0.00009309 |
| 73 | 111 | 0.04766334 | 0.05118421 | -0.03402219 | -0.03573135 | 0.02099610 | 0.01135243 | -0.01689114 | -0.00548580 | 0.00652651 | 0.00161659 | -0.00103117 | 0.00039119 | -0.00000074 | -0.00035739 | 0.00006514 | 0.00014079 | -0.00009145 |
| 73 | 112 | 0.04779285 | 0.05128400 | -0.03440198 | -0.03540194 | 0.02151047 | 0.01081742 | -0.01709016 | -0.00494879 | 0.00691124 | 0.00152937 | -0.00115236 | 0.00042657 | 0.00002953 | -0.00033288 | 0.00014796 | 0.00000353 | -0.00014483 |
| 73 | 113 | 0.04757929 | 0.05077450 | -0.03465693 | -0.03538344 | 0.02116453 | 0.01054091 | -0.01693128 | -0.00473860 | 0.00658756 | 0.00131054 | -0.00109615 | 0.00044387 | -0.00004225 | -0.00032981 | 0.00010326 | 0.00006941 | -0.00014076 |
| 73 | 114 | 0.04762758 | 0.05079094 | -0.03485137 | -0.03490719 | 0.02185656 | 0.01008051 | -0.01708832 | -0.00422279 | 0.00697606 | 0.00123616 | -0.00121132 | 0.00047649 | -0.00001044 | -0.00029824 | 0.00019138 | -0.00007738 | -0.00020365 |
| 73 | 115 | 0.04746409 | 0.05034189 | -0.03519305 | -0.03499586 | 0.02140002 | 0.00974737 | -0.01692480 | -0.00407156 | 0.00657616 | 0.00099981 | -0.00115295 | 0.00049129 | -0.00007078 | -0.00029741 | 0.00014932 | -0.00002746 | -0.00020191 |
| 73 | 116 | 0.04742254 | 0.05029667 | -0.03517093 | -0.03436488 | 0.02224287 | 0.00936897 | -0.01702188 | -0.00357611 | 0.00697360 | 0.00094437 | -0.00125726 | 0.00051609 | -0.00003317 | -0.00026283 | 0.00024378 | -0.00018041 | -0.00026769 |
| 73 | 117 | 0.04731196 | 0.04990066 | -0.03559326 | -0.03453939 | 0.02169524 | 0.00899909 | -0.01686792 | -0.00350784 | 0.00648323 | 0.00069135 | -0.00119736 | 0.00053426 | -0.00008782 | -0.00026149 | 0.00019994 | -0.00014544 | -0.00027428 |
| 73 | 118 | 0.04717235 | 0.04980449 | -0.03533860 | -0.03375487 | 0.02264661 | 0.00869291 | -0.01688865 | -0.00302342 | 0.00689492 | 0.00065890 | -0.00128739 | 0.00054720 | -0.00004198 | -0.00022675 | 0.00030114 | -0.00029998 | -0.00033586 |
| 73 | 119 | 0.04711428 | 0.04946366 | -0.03581925 | -0.03398922 | 0.02202812 | 0.00831355 | -0.01676545 | -0.00306302 | 0.00630491 | 0.00039201 | -0.00122707 | 0.00057565 | -0.00009771 | -0.00022233 | 0.00025028 | -0.00027720 | -0.00035710 |
| 73 | 120 | 0.04687092 | 0.04931109 | -0.03533904 | -0.03305760 | 0.02304521 | 0.00805745 | -0.01670007 | -0.00257483 | 0.00673480 | 0.00038365 | -0.00130133 | 0.00057434 | -0.00004350 | -0.00018870 | 0.00035799 | -0.00042850 | -0.00040831 |
| 73 | 121 | 0.04686031 | 0.04903197 | -0.03584573 | -0.03332433 | 0.02237189 | 0.00769467 | -0.01663078 | -0.00274040 | 0.00604245 | 0.00010628 | -0.00124192 | 0.00061983 | -0.00010734 | -0.00017900 | 0.00029481 | -0.00041368 | -0.00044962 |
| 74 | 83 | 0.04853525 | 0.05984544 | -0.02578980 | -0.04287420 | 0.01660502 | 0.02496541 | -0.00659901 | -0.00975939 | 0.00297745 | 0.00373531 | -0.00022798 | 0.00009805 | 0.00028513 | -0.00020259 | -0.00026577 | 0.00016299 | -0.00000092 |
| 74 | 84 | 0.04864050 | 0.05955272 | -0.02622969 | -0.04285193 | 0.01603520 | 0.02458179 | -0.00698596 | -0.00946273 | 0.00289085 | 0.00341541 | -0.00029688 | 0.00024020 | 0.00010043 | -0.00017334 | -0.00030315 | 0.00018169 | -0.00011243 |
| 74 | 85 | 0.04877842 | 0.05907861 | -0.02718636 | -0.04229179 | 0.01749851 | 0.02340194 | -0.00836041 | -0.00952289 | 0.00347131 | 0.00357927 | -0.00036048 | 0.00014143 | 0.00028840 | -0.00022187 | -0.00020843 | 0.00014600 | -0.00001459 |
| 74 | 86 | 0.04886082 | 0.05892321 | -0.02749426 | -0.04236089 | 0.01695163 | 0.02325247 | -0.00858664 | -0.00931573 | 0.00336380 | 0.00331499 | -0.00041963 | 0.00026094 | 0.00013825 | -0.00020218 | -0.00024835 | 0.00015992 | -0.00011747 |
| 74 | 87 | 0.04905832 | 0.05832917 | -0.02857269 | -0.04172092 | 0.01846282 | 0.02203961 | -0.01003378 | -0.00930885 | 0.00395662 | 0.00344333 | -0.00049072 | 0.00019281 | 0.00027280 | -0.00023893 | -0.00016899 | 0.00013866 | -0.00004071 |
| 74 | 88 | 0.04906657 | 0.05827611 | -0.02872644 | -0.04192315 | 0.01784041 | 0.02204210 | -0.01012253 | -0.00921661 | 0.00382486 | 0.00324173 | -0.00053808 | 0.00028597 | 0.00016091 | -0.00022677 | -0.00020582 | 0.00014439 | -0.00013017 |
| 74 | 89 | 0.04931274 | 0.05760737 | -0.02980755 | -0.04114484 | 0.01938426 | 0.02086416 | -0.01152582 | -0.00911975 | 0.00440970 | 0.00332626 | -0.00060804 | 0.00024262 | 0.00024688 | -0.00025569 | -0.00014346 | 0.00014129 | -0.00006872 |
| 74 | 90 | 0.04921246 | 0.05762691 | -0.02979122 | -0.04148338 | 0.01864202 | 0.02095050 | -0.01150478 | -0.00914503 | 0.00425752 | 0.00318643 | -0.00064361 | 0.00030806 | 0.00017417 | -0.00024952 | -0.00017192 | 0.00013586 | -0.00014195 |
| 74 | 91 | 0.04949166 | 0.05691835 | -0.03080852 | -0.04056673 | 0.02017231 | 0.01982225 | -0.01280538 | -0.00895960 | 0.00481614 | 0.00322395 | -0.00070593 | 0.00028311 | 0.00021989 | -0.00027377 | -0.00012571 | 0.00015092 | -0.00009055 |

TABLE 3. Nuclear charge density distribution FB coefficients.

| $Z$ | $N$ | $a_1$ | $a_2$ | $a_3$ | $a_4$ | $a_5$ | $a_6$ | $a_7$ | $a_8$ | $a_9$ | $a_{10}$ | $a_{11}$ | $a_{12}$ | $a_{13}$ | $a_{14}$ | $a_{15}$ | $a_{16}$ | $a_{17}$ |
|---|---|---|---|---|---|---|---|---|---|---|---|---|---|---|---|---|---|---|
| 74 | 92 | 0.04927227 | 0.05699143 | -0.03062014 | -0.04101753 | 0.01931302 | 0.01994674 | -0.01270188 | -0.00908815 | 0.00465214 | 0.00314004 | -0.00073144 | 0.00032194 | 0.00018451 | -0.00027251 | -0.00014232 | 0.00013259 | -0.00014706 |
| 74 | 93 | 0.04956977 | 0.05626535 | -0.03155047 | -0.03999545 | 0.02077997 | 0.01885454 | -0.01387984 | -0.00882745 | 0.00517052 | 0.00313080 | -0.00078191 | 0.00030937 | 0.00019946 | -0.00029416 | -0.00010958 | 0.00016266 | -0.00010180 |
| 74 | 94 | 0.04924236 | 0.05638348 | -0.03120475 | -0.04052410 | 0.01983247 | 0.01899090 | -0.01371944 | -0.00903697 | 0.00500481 | 0.00309470 | -0.00080010 | 0.00032496 | 0.00019714 | -0.00029715 | -0.00011324 | 0.00013138 | -0.00014299 |
| 74 | 95 | 0.04954716 | 0.05565188 | -0.03205058 | -0.03943848 | 0.02120116 | 0.01791634 | -0.01477256 | -0.00871616 | 0.00547409 | 0.00304145 | -0.00083650 | 0.00031953 | 0.00019039 | -0.00031720 | -0.00009040 | 0.00017147 | -0.00010150 |
| 74 | 96 | 0.04913675 | 0.05581296 | -0.03157547 | -0.04001024 | 0.02020258 | 0.01805007 | -0.01457871 | -0.00898285 | 0.00531608 | 0.00304427 | -0.00085033 | 0.00031670 | 0.00021514 | -0.00032407 | -0.00008212 | 0.00012893 | -0.00012979 |
| 74 | 97 | 0.04944118 | 0.05508099 | -0.03234957 | -0.03889944 | 0.02145694 | 0.01698214 | -0.01551008 | -0.00861456 | 0.00573248 | 0.00295190 | -0.00087200 | 0.00031414 | 0.00019462 | -0.00034264 | -0.00006548 | 0.00017337 | -0.00009099 |
| 74 | 98 | 0.04897834 | 0.05528463 | -0.03178029 | -0.03948260 | 0.02044273 | 0.01710335 | -0.01530376 | -0.00891635 | 0.00558963 | 0.00298436 | -0.00088410 | 0.00029837 | 0.00023956 | -0.00035328 | -0.00004770 | 0.00012260 | -0.00010899 |
| 74 | 99 | 0.04927666 | 0.05455378 | -0.03249606 | -0.03837762 | 0.02158147 | 0.01604105 | -0.01611711 | -0.00851039 | 0.00595352 | 0.00285974 | -0.00089159 | 0.00029526 | 0.00021178 | -0.00036989 | -0.00003392 | 0.00016604 | -0.00007272 |
| 74 | 100 | 0.04879144 | 0.05479775 | -0.03186987 | -0.03894317 | 0.02058146 | 0.01613980 | -0.01591628 | -0.00882761 | 0.00583104 | 0.00291214 | -0.00090385 | 0.00027202 | 0.00026992 | -0.00038439 | -0.00000969 | 0.00011079 | -0.00008270 |
| 74 | 101 | 0.04907846 | 0.05406841 | -0.03253653 | -0.03786863 | 0.02161154 | 0.01509073 | -0.01661513 | -0.00839229 | 0.00614569 | 0.00276385 | -0.00089854 | 0.00026574 | 0.00024009 | -0.00039814 | 0.00000398 | 0.00014865 | -0.00004937 |
| 74 | 102 | 0.04859758 | 0.05434718 | -0.03188962 | -0.03838939 | 0.02064953 | 0.01515468 | -0.01643402 | -0.00870696 | 0.00604676 | 0.00282595 | -0.00091212 | 0.00024000 | 0.00030480 | -0.00041673 | 0.00003155 | 0.00009274 | -0.00005309 |
| 74 | 103 | 0.04886783 | 0.05362034 | -0.03251066 | -0.03736581 | 0.02158082 | 0.01413262 | -0.01702196 | -0.00825063 | 0.00631702 | 0.00266388 | -0.00089591 | 0.00022858 | 0.00027700 | -0.00042655 | 0.00004718 | 0.00012150 | -0.00002339 |
| 74 | 104 | 0.04841375 | 0.05392488 | -0.03187665 | -0.03781567 | 0.02067558 | 0.01414660 | -0.01687065 | -0.00854550 | 0.00624325 | 0.00272495 | -0.00091121 | 0.00020456 | 0.00034245 | -0.00044954 | 0.00007536 | 0.00006837 | -0.00002205 |
| 74 | 105 | 0.04872930 | 0.05318613 | -0.03289065 | -0.03693375 | 0.02159582 | 0.01335122 | -0.01722919 | -0.00776546 | 0.00649610 | 0.00248646 | -0.00091703 | 0.00023308 | 0.00026189 | -0.00043676 | 0.00007480 | 0.00011173 | -0.00001276 |
| 74 | 106 | 0.04833584 | 0.05331669 | -0.03266592 | -0.03735961 | 0.02083499 | 0.01345606 | -0.01685387 | -0.00765783 | 0.00642770 | 0.00246350 | -0.00096275 | 0.00026097 | 0.00025640 | -0.00042160 | 0.00008379 | 0.00006949 | -0.00004431 |
| 74 | 107 | 0.04864184 | 0.05274406 | -0.03364305 | -0.03659373 | 0.02170605 | 0.01272711 | -0.01730399 | -0.00697028 | 0.00667822 | 0.00222425 | -0.00097365 | 0.00028590 | 0.00019235 | -0.00042334 | 0.00008795 | 0.00011432 | -0.00002562 |
| 74 | 108 | 0.04824439 | 0.05270912 | -0.03340019 | -0.03688400 | 0.02109388 | 0.01274423 | -0.01685818 | -0.00679254 | 0.00658849 | 0.00218327 | -0.00101680 | 0.00031640 | 0.00017897 | -0.00039054 | 0.00010116 | 0.00005162 | -0.00007493 |
| 74 | 109 | 0.04852904 | 0.05229159 | -0.03432228 | -0.03626651 | 0.02186416 | 0.01207198 | -0.01738187 | -0.00619625 | 0.00683298 | 0.00194881 | -0.00103596 | 0.00034039 | 0.00012822 | -0.00040255 | 0.00010840 | 0.00009791 | -0.00004941 |
| 74 | 110 | 0.04813630 | 0.05210568 | -0.03406884 | -0.03637335 | 0.02145803 | 0.01201846 | -0.01687336 | -0.00596233 | 0.00671924 | 0.00188726 | -0.00106971 | 0.00036719 | 0.00011453 | -0.00035803 | 0.00012930 | 0.00001236 | -0.00011259 |
| 74 | 111 | 0.04839210 | 0.05181984 | -0.03493118 | -0.03592940 | 0.02208740 | 0.01139463 | -0.01744763 | -0.00545471 | 0.00695034 | 0.00166208 | -0.00109968 | 0.00039372 | 0.00007258 | -0.00037551 | 0.00013743 | 0.00005945 | -0.00008440 |
| 74 | 112 | 0.04800703 | 0.05151156 | -0.03465448 | -0.03581373 | 0.02191889 | 0.01128805 | -0.01688384 | -0.00517990 | 0.00681205 | 0.00157980 | -0.00111748 | 0.00041038 | 0.00006611 | -0.00032557 | 0.00016867 | -0.00004868 | -0.00015539 |
| 74 | 113 | 0.04823098 | 0.05132880 | -0.03545983 | -0.03555798 | 0.02238282 | 0.01071078 | -0.01748263 | -0.00476075 | 0.00701835 | 0.00136744 | -0.00115966 | 0.00044317 | 0.00002819 | -0.00034393 | 0.00017524 | -0.00000231 | -0.00012984 |
| 74 | 114 | 0.04785074 | 0.05093141 | -0.03513561 | -0.03519280 | 0.02245594 | 0.01056353 | -0.01687272 | -0.00445791 | 0.00685810 | 0.00126644 | -0.00115644 | 0.00044444 | 0.00003466 | -0.00029401 | 0.00021808 | -0.00012968 | -0.00020143 |
| 74 | 115 | 0.04804369 | 0.05082561 | -0.03588694 | -0.03512850 | 0.02274475 | 0.01003897 | -0.01746984 | -0.00413170 | 0.00702502 | 0.00106963 | -0.00121085 | 0.00048690 | -0.00000345 | -0.00030946 | 0.00022066 | -0.00008628 | -0.00018435 |
| 74 | 116 | 0.04766047 | 0.05036610 | -0.03549077 | -0.03449907 | 0.02304146 | 0.00985627 | -0.01682613 | -0.00380813 | 0.00684841 | 0.00095347 | -0.00118399 | 0.00046995 | 0.00001832 | -0.00026314 | 0.00027452 | -0.00022652 | -0.00024948 |
| 74 | 117 | 0.04782591 | 0.05032019 | -0.03618443 | -0.03461912 | 0.02315574 | 0.00939621 | -0.01739899 | -0.00358430 | 0.00696030 | 0.00077424 | -0.00124931 | 0.00052499 | -0.00002327 | -0.00027308 | 0.00027088 | -0.00018878 | -0.00024656 |
| 74 | 118 | 0.04742903 | 0.04981002 | -0.03570353 | -0.03372127 | 0.02364636 | 0.00917701 | -0.01673719 | -0.00323973 | 0.00677509 | 0.00064695 | -0.00119918 | 0.00048989 | 0.00001222 | -0.00023140 | 0.00033339 | -0.00033327 | -0.00029950 |
| 74 | 119 | 0.04757105 | 0.04981896 | -0.03632521 | -0.03400996 | 0.02359143 | 0.00879379 | -0.01727063 | -0.00313108 | 0.00681775 | 0.00048656 | -0.00127309 | 0.00055984 | -0.00003518 | -0.00023458 | 0.00032158 | -0.00030374 | -0.00031579 |
| 74 | 120 | 0.04714949 | 0.04924948 | -0.03576740 | -0.03284835 | 0.02424636 | 0.00853356 | -0.01660883 | -0.00275720 | 0.00663258 | 0.00035144 | -0.00120288 | 0.00050907 | 0.00000911 | -0.00019616 | 0.00038926 | -0.00044308 | -0.00035272 |
| 74 | 121 | 0.04727103 | 0.04931821 | -0.03629284 | -0.03328282 | 0.02402845 | 0.00823469 | -0.01709758 | -0.00277672 | 0.00659523 | 0.00020994 | -0.00128259 | 0.00059588 | -0.00004585 | -0.00019240 | 0.00036761 | -0.00042362 | -0.00039220 |
| 74 | 122 | 0.04681590 | 0.04866334 | -0.03568880 | -0.03186978 | 0.02482683 | 0.00792846 | -0.01645397 | -0.00235865 | 0.00641866 | 0.00006865 | -0.00119756 | 0.00053293 | 0.00000094 | -0.00015421 | 0.00043702 | -0.00054947 | -0.00041118 |
| 74 | 123 | 0.04691709 | 0.04880029 | -0.03608948 | -0.03242084 | 0.02445278 | 0.00771300 | -0.01690306 | -0.00251603 | 0.00629466 | -0.00005563 | -0.00128020 | 0.00063832 | -0.00006326 | -0.00014411 | 0.00040402 | -0.00054078 | -0.00047659 |
| 75 | 84 | 0.04940630 | 0.05914327 | -0.02791436 | -0.04154911 | 0.01892477 | 0.02409665 | -0.00795497 | -0.00896074 | 0.00370540 | 0.00364885 | -0.00032918 | 0.00019785 | 0.00018296 | -0.00020562 | -0.00023593 | 0.00016367 | -0.00007382 |
| 75 | 85 | 0.04993638 | 0.05842957 | -0.02962452 | -0.04078399 | 0.02080280 | 0.02293755 | -0.00961524 | -0.00911235 | 0.00414952 | 0.00387401 | -0.00040171 | 0.00012490 | 0.00031601 | -0.00021509 | -0.00017954 | 0.00016357 | -0.00002872 |
| 75 | 86 | 0.04964896 | 0.05843332 | -0.02919788 | -0.04088440 | 0.01988181 | 0.02260382 | -0.00968512 | -0.00873016 | 0.00418319 | 0.00348853 | -0.00045974 | 0.00024287 | 0.00018436 | -0.00022496 | -0.00018470 | 0.00014669 | -0.00009329 |
| 75 | 87 | 0.05009180 | 0.05772570 | -0.03055079 | -0.04014083 | 0.02146052 | 0.02166932 | -0.01103815 | -0.00888444 | 0.00450236 | 0.00369981 | -0.00050955 | 0.00018459 | 0.00027296 | -0.00022074 | -0.00015295 | 0.00016269 | -0.00006519 |
| 75 | 88 | 0.04989127 | 0.05776071 | -0.03040544 | -0.04032306 | 0.02075364 | 0.02129495 | -0.01128520 | -0.00856751 | 0.00462160 | 0.00335541 | -0.00058066 | 0.00028926 | 0.00017351 | -0.00024301 | -0.00015071 | 0.00014151 | -0.00011682 |
| 75 | 89 | 0.05021487 | 0.05707272 | -0.03134378 | -0.03958495 | 0.02202125 | 0.02058108 | -0.01229385 | -0.00871131 | 0.00482381 | 0.00354661 | -0.00060376 | 0.00023917 | 0.00022720 | -0.00022991 | -0.00013931 | 0.00017428 | -0.00009629 |
| 75 | 90 | 0.05008292 | 0.05712899 | -0.03142882 | -0.03983406 | 0.02147429 | 0.02015916 | -0.01268142 | -0.00846265 | 0.00500873 | 0.00324561 | -0.00068399 | 0.00032951 | 0.00015734 | -0.00026160 | -0.00012942 | 0.00014742 | -0.00013603 |
| 75 | 91 | 0.05027416 | 0.05646415 | -0.03196473 | -0.03910123 | 0.02244648 | 0.01962417 | -0.01337277 | -0.00858734 | 0.00511272 | 0.00340998 | -0.00068115 | 0.00028358 | 0.00018493 | -0.00024350 | -0.00013307 | 0.00019452 | -0.00011728 |
| 75 | 92 | 0.05018979 | 0.05653653 | -0.03221628 | -0.03939310 | 0.02201104 | 0.01915015 | -0.01385752 | -0.00840189 | 0.00534244 | 0.00315210 | -0.00076610 | 0.00035811 | 0.00014304 | -0.00028207 | -0.00011456 | 0.00016033 | -0.00014564 |
| 75 | 93 | 0.05025081 | 0.05588898 | -0.03240469 | -0.03866827 | 0.02272478 | 0.01873748 | -0.01429051 | -0.00849955 | 0.00537134 | 0.00328348 | -0.00074105 | 0.00031447 | 0.00015148 | -0.00026158 | -0.00012817 | 0.00021764 | -0.00012603 |
| 75 | 94 | 0.05020010 | 0.05598057 | -0.03276456 | -0.03898104 | 0.02236328 | 0.01821343 | -0.01483045 | -0.00836781 | 0.00562711 | 0.00306707 | -0.00082675 | 0.00037200 | 0.00013642 | -0.00030513 | -0.00010005 | 0.00017446 | -0.00014368 |
| 75 | 95 | 0.05014203 | 0.05533754 | -0.03267769 | -0.03826412 | 0.02286578 | 0.01786885 | -0.01506920 | -0.00842995 | 0.00560325 | 0.00316113 | -0.00078445 | 0.00033028 | 0.00013056 | -0.00028350 | -0.00011946 | 0.00023757 | -0.00012277 |
| 75 | 96 | 0.05012173 | 0.05545911 | -0.03310223 | -0.03858279 | 0.02255348 | 0.01730475 | -0.01562893 | -0.00834114 | 0.00587058 | 0.00298382 | -0.00086782 | 0.00037053 | 0.00014090 | -0.00033082 | -0.00008143 | 0.00018434 | -0.00013081 |
| 75 | 97 | 0.04995935 | 0.05480458 | -0.03281192 | -0.03787064 | 0.02289180 | 0.01698563 | -0.01572764 | -0.00835933 | 0.00581231 | 0.00303883 | -0.00081317 | 0.00033112 | 0.00012380 | -0.00030820 | -0.00010364 | 0.00024939 | -0.00010938 |
| 75 | 98 | 0.04997455 | 0.05497035 | -0.03327315 | -0.03818619 | 0.02261586 | 0.01639597 | -0.01628175 | -0.00830338 | 0.00608180 | 0.00289772 | -0.00089224 | 0.00035492 | 0.00015744 | -0.00035867 | -0.00005628 | 0.00018598 | -0.00010928 |
| 75 | 99 | 0.04972283 | 0.05428874 | -0.03284147 | -0.03747511 | 0.02283043 | 0.01607458 | -0.01627928 | -0.00827079 | 0.00600217 | 0.00291469 | -0.00082942 | 0.00031847 | 0.00013094 | -0.00033446 | -0.00007935 | 0.00024994 | -0.00008866 |
| 75 | 100 | 0.04978258 | 0.05451154 | -0.03332421 | -0.03778110 | 0.02258722 | 0.01547298 | -0.01681345 | -0.00823883 | 0.00626936 | 0.00280608 | -0.00090323 | 0.00032757 | 0.00018505 | -0.00038788 | -0.00002395 | 0.00017722 | -0.00008191 |
| 75 | 101 | 0.04945515 | 0.05379046 | -0.03280008 | -0.03706942 | 0.02270923 | 0.01513608 | -0.01673407 | -0.00815168 | 0.00617619 | 0.00278838 | -0.00083550 | 0.00029464 | 0.00015027 | -0.00036108 | -0.00004682 | 0.00023798 | -0.00006356 |
| 75 | 102 | 0.04956875 | 0.05407843 | -0.03329818 | -0.03735890 | 0.02250144 | 0.01453096 | -0.01724364 | -0.00813532 | 0.00644082 | 0.00270774 | -0.00090381 | 0.00029142 | 0.00022154 | -0.00041749 | 0.00001502 | 0.00015749 | -0.00005150 |
| 75 | 103 | 0.04917714 | 0.05331009 | -0.03271739 | -0.03664857 | 0.02255280 | 0.01417763 | -0.01710051 | -0.00799411 | 0.00633738 | 0.00266049 | -0.00083351 | 0.00026229 | 0.00017929 | -0.00038706 | -0.00000727 | 0.00021381 | -0.00003671 |
| 75 | 104 | 0.04935224 | 0.05366534 | -0.03323080 | -0.03691249 | 0.02238690 | 0.01357032 | -0.01758756 | -0.00798418 | 0.00660228 | 0.00260230 | -0.00089660 | 0.00024941 | 0.00026417 | -0.00044657 | 0.00005938 | 0.00012737 | -0.00002041 |
| 75 | 105 | 0.04901386 | 0.05291307 | -0.03312004 | -0.03632427 | 0.02251904 | 0.01343318 | -0.01731966 | -0.00756979 | 0.00650069 | 0.00248405 | -0.00085052 | 0.00025497 | 0.00018130 | -0.00040463 | 0.00002548 | 0.00020248 | -0.00001550 |
| 75 | 106 | 0.04924558 | 0.05324639 | -0.03399775 | -0.03666108 | 0.02247673 | 0.01304983 | -0.01764465 | -0.00727185 | 0.00676493 | 0.00235653 | -0.00094223 | 0.00028379 | 0.00021351 | -0.00044166 | 0.00007597 | 0.00013644 | -0.00001523 |
| 75 | 107 | 0.04894903 | 0.05258381 | -0.03397357 | -0.03614819 | 0.02260560 | 0.01288449 | -0.01743594 | -0.00690887 | 0.00664970 | 0.00224368 | -0.00089696 | 0.00028146 | 0.00014970 | -0.00040839 | 0.00004943 | 0.00020264 | -0.00000634 |
| 75 | 108 | 0.04911806 | 0.05281487 | -0.03470103 | -0.03643451 | 0.02259321 | 0.01249121 | -0.01770664 | -0.00656497 | 0.00690522 | 0.00209298 | -0.00099504 | 0.00032346 | 0.00016253 | -0.00042886 | 0.00009740 | 0.00013069 | -0.00002075 |
| 75 | 109 | 0.04886856 | 0.05222730 | -0.03478186 | -0.03600862 | 0.02269197 | 0.01229081 | -0.01753340 | -0.00625240 | 0.00676038 | 0.00198136 | -0.00094981 | 0.00031395 | 0.00011603 | -0.00040268 | 0.00007654 | 0.00018742 | -0.00000848 |
| 75 | 110 | 0.04897278 | 0.05235633 | -0.03535214 | -0.03620646 | 0.02276240 | 0.01189550 | -0.01776765 | -0.00587358 | 0.00701450 | 0.00181293 | -0.00105217 | 0.00036679 | 0.00011342 | -0.00040849 | 0.00012472 | 0.00010679 | -0.00003845 |
| 75 | 111 | 0.04877621 | 0.05182543 | -0.03555741 | -0.03587616 | 0.02280941 | 0.01165678 | -0.01761339 | -0.00561545 | 0.00682254 | 0.00169815 | -0.00100649 | 0.00035263 | 0.00008046 | -0.00038738 | 0.00010657 | 0.00015442 | -0.00002476 |
| 75 | 112 | 0.04881245 | 0.05186684 | -0.03594867 | -0.03594704 | 0.02300438 | 0.01127551 | -0.01781534 | -0.00521292 | 0.00708122 | 0.00151936 | -0.00110945 | 0.00041162 | 0.00006889 | -0.00038167 | 0.00015828 | 0.00006258 | -0.00006877 |
| 75 | 113 | 0.04867528 | 0.05137662 | -0.03629180 | -0.03571259 | 0.02298862 | 0.01100294 | -0.01767411 | -0.00502008 | 0.00682424 | 0.00139752 | -0.00106320 | 0.00039704 | 0.00004407 | -0.00036345 | 0.00013871 | 0.00010234 | -0.00005714 |
| 75 | 114 | 0.04863783 | 0.05135280 | -0.03647343 | -0.03562704 | 0.02332748 | 0.01065171 | -0.01783421 | -0.00460150 | 0.00709299 | 0.00121696 | -0.00116209 | 0.00045587 | 0.00003128 | -0.00035000 | 0.00019749 | -0.00000228 | -0.00011115 |
| 75 | 115 | 0.04856613 | 0.05089504 | -0.03695494 | -0.03547831 | 0.02324911 | 0.01035876 | -0.01771156 | -0.00449057 | 0.00675484 | 0.00108524 | -0.00111546 | 0.00044648 | 0.00000824 | -0.00033261 | 0.00017154 | 0.00003153 | -0.00010636 |
| 75 | 116 | 0.04844658 | 0.05082752 | -0.03689733 | -0.03522050 | 0.02372527 | 0.01004636 | -0.01781011 | -0.00405800 | 0.00703883 | 0.00091169 | -0.00120560 | 0.00049838 | 0.00000164 | -0.00031494 | 0.00024053 | -0.00008592 | -0.00016446 |
| 75 | 117 | 0.04844417 | 0.05040390 | -0.03750151 | -0.03513895 | 0.02359134 | 0.00975243 | -0.01772062 | -0.00404678 | 0.00660776 | 0.00076851 | -0.00115905 | 0.00050055 | -0.00002639 | -0.00029665 | 0.00020298 | -0.00005546 | -0.00017198 |
| 75 | 118 | 0.04823308 | 0.05030489 | -0.03718616 | -0.03470572 | 0.02417816 | 0.00947734 | -0.01773494 | -0.00359732 | 0.00691109 | 0.00060991 | -0.00123683 | 0.00053966 | -0.00002133 | -0.00027716 | 0.00028432 | -0.00018410 | -0.00022755 |
| 75 | 119 | 0.04829908 | 0.04992392 | -0.03788367 | -0.03466793 | 0.02399669 | 0.00920172 | -0.01769684 | -0.00369861 | 0.00638139 | 0.00045447 | -0.00119080 | 0.00055956 | -0.00006079 | -0.00025668 | 0.00023026 | -0.00015383 | -0.00025250 |
| 75 | 120 | 0.04798846 | 0.04979159 | -0.03731041 | -0.03406421 | 0.02465955 | 0.00895352 | -0.01761016 | -0.00322701 | 0.00670631 | 0.00031693 | -0.00125463 | 0.00058211 | -0.00004160 | -0.00023611 | 0.00032473 | -0.00029066 | -0.00029972 |
| 75 | 121 | 0.04811603 | 0.04946151 | -0.03806599 | -0.03404454 | 0.02443588 | 0.00870886 | -0.01763870 | -0.00344333 | 0.00607795 | 0.00014831 | -0.00120917 | 0.00062459 | -0.00009781 | -0.00021267 | 0.00025021 | -0.00025702 | -0.00034573 |
| 75 | 122 | 0.04770168 | 0.04928040 | -0.03725546 | -0.03327943 | 0.02514492 | 0.00847245 | -0.01744813 | -0.00294498 | 0.00642498 | 0.00003539 | -0.00126002 | 0.00062950 | -0.00006522 | -0.00019003 | 0.00035730 | -0.00039843 | -0.00038081 |
| 75 | 123 | 0.04787769 | 0.04900144 | -0.03803702 | -0.03325034 | 0.02488151 | 0.00826060 | -0.01755009 | -0.00326642 | 0.00570150 | -0.00014824 | -0.00121433 | 0.00069708 | -0.00014170 | -0.00016344 | 0.00025977 | -0.00035773 | -0.00044913 |
| 75 | 124 | 0.04736071 | 0.04874769 | -0.03702790 | -0.03233638 | 0.02561999 | 0.00802158 | -0.01727001 | -0.00273938 | 0.00607061 | -0.00023577 | -0.00125574 | 0.00068569 | -0.00009896 | -0.00013643 | 0.00037820 | -0.00050058 | -0.00047093 |
| 76 | 85 | 0.05031606 | 0.05844576 | -0.03021711 | -0.04050228 | 0.02093480 | 0.02301194 | -0.00955928 | -0.00829521 | 0.00437499 | 0.00357094 | -0.00043837 | 0.00031246 | 0.00005843 | -0.00019288 | -0.00021143 | 0.00016119 | -0.00016822 |
| 76 | 86 | 0.05022616 | 0.05838547 | -0.03053840 | -0.04135301 | 0.01961764 | 0.02262656 | -0.00995198 | -0.00834746 | 0.00417604 | 0.00328827 | -0.00049008 | 0.00042600 | -0.00008328 | -0.00019365 | -0.00025348 | 0.00018318 | -0.00024950 |
| 76 | 87 | 0.05052226 | 0.05783102 | -0.03128047 | -0.03980771 | 0.02186848 | 0.02165430 | -0.01117583 | -0.00810973 | 0.00480575 | 0.00341140 | -0.00055842 | 0.00035269 | 0.00006285 | -0.00021430 | -0.00016730 | 0.00014814 | -0.00018556 |
| 76 | 88 | 0.05036808 | 0.05787950 | -0.03146391 | -0.04079460 | 0.02057612 | 0.02152913 | -0.01141457 | -0.00828757 | 0.00458866 | 0.00318617 | -0.00059743 | 0.00044123 | -0.00004306 | -0.00022695 | -0.00021012 | 0.00017197 | -0.00024832 |
| 76 | 89 | 0.05069553 | 0.05726869 | -0.03221834 | -0.03928329 | 0.02260810 | 0.02046677 | -0.01263023 | -0.00802087 | 0.00518152 | 0.00328551 | -0.00066380 | 0.00038844 | 0.00006206 | -0.00023576 | -0.00013887 | 0.00014819 | -0.00019921 |
| 76 | 90 | 0.05045336 | 0.05738631 | -0.03226592 | -0.04038860 | 0.02130071 | 0.02051808 | -0.01275386 | -0.00831617 | 0.00495300 | 0.00311939 | -0.00069260 | 0.00045339 | -0.00000971 | -0.00025663 | -0.00017927 | 0.00017169 | -0.00024378 |
| 76 | 91 | 0.05080225 | 0.05675119 | -0.03296760 | -0.03888340 | 0.02313057 | 0.01942523 | -0.01387881 | -0.00800398 | 0.00550208 | 0.00318600 | -0.00074988 | 0.00041483 | 0.00006060 | -0.00025848 | -0.00012105 | 0.00015898 | -0.00020396 |
| 76 | 92 | 0.05046715 | 0.05690437 | -0.03289723 | -0.04007285 | 0.02179664 | 0.01957398 | -0.01392739 | -0.00839385 | 0.00527092 | 0.00307392 | -0.00077135 | 0.00045895 | 0.00001939 | -0.00028463 | -0.00015612 | 0.00017960 | -0.00023234 |
| 76 | 93 | 0.05082470 | 0.05626859 | -0.03350805 | -0.03856102 | 0.02344638 | 0.01848149 | -0.01492145 | -0.00802970 | 0.00577435 | 0.00310272 | -0.00081550 | 0.00042861 | 0.00006326 | -0.00028329 | -0.00010755 | 0.00017519 | -0.00019751 |
| 76 | 94 | 0.05040894 | 0.05643458 | -0.03334619 | -0.03978962 | 0.02209783 | 0.01866616 | -0.01492617 | -0.00848194 | 0.00554978 | 0.00303500 | -0.00083225 | 0.00045556 | 0.00004761 | -0.00031282 | -0.00013543 | 0.00019089 | -0.00021264 |
| 76 | 95 | 0.05076263 | 0.05581228 | -0.03385279 | -0.03827511 | 0.02358703 | 0.01758504 | -0.01577995 | -0.00806810 | 0.00600828 | 0.00302576 | -0.00086186 | 0.00042840 | 0.00007370 | -0.00031048 | -0.00009274 | 0.00019058 | -0.00018009 |
| 76 | 96 | 0.05029011 | 0.05597778 | -0.03363048 | -0.03949534 | 0.02224814 | 0.01776163 | -0.01576201 | -0.00854759 | 0.00579865 | 0.00299021 | -0.00087594 | 0.00044245 | 0.00007764 | -0.00034238 | -0.00011293 | 0.00020040 | -0.00018523 |
| 76 | 97 | 0.05062939 | 0.05537599 | -0.03403502 | -0.03799372 | 0.02359317 | 0.01669665 | -0.01648166 | -0.00809206 | 0.00621408 | 0.00294730 | -0.00089139 | 0.00041457 | 0.00009372 | -0.00033974 | -0.00007274 | 0.00019979 | -0.00015370 |
| 76 | 98 | 0.05012907 | 0.05553285 | -0.03378541 | -0.03916088 | 0.02229082 | 0.01683276 | -0.01645516 | -0.00856536 | 0.00602606 | 0.00293078 | -0.00090420 | 0.00042029 | 0.00011082 | -0.00037364 | -0.00008592 | 0.00020395 | -0.00015191 |
| 76 | 99 | 0.05044563 | 0.05495475 | -0.03409616 | -0.03769384 | 0.02350606 | 0.01579224 | -0.01705150 | -0.00807948 | 0.00640073 | 0.00286223 | -0.00090699 | 0.00038891 | 0.00012316 | -0.00037036 | -0.00004569 | 0.00019920 | -0.00012123 |
| 76 | 100 | 0.04994639 | 0.05509607 | -0.03385272 | -0.03876841 | 0.02226390 | 0.01586062 | -0.01702638 | -0.00851686 | 0.00623921 | 0.00285145 | -0.00091935 | 0.00039076 | 0.00014704 | -0.00040624 | -0.00005328 | 0.00019887 | -0.00011498 |
| 76 | 101 | 0.05023369 | 0.05454393 | -0.03407753 | -0.03735990 | 0.02336253 | 0.01486049 | -0.01750962 | -0.00801409 | 0.00657541 | 0.00276778 | -0.00091153 | 0.00035397 | 0.00016042 | -0.00040135 | -0.00001134 | 0.00018711 | -0.00008573 |
| 76 | 102 | 0.04976124 | 0.05466180 | -0.03387279 | -0.03830814 | 0.02219870 | 0.01483520 | -0.01749333 | -0.00838989 | 0.00644374 | 0.00274994 | -0.00092375 | 0.00035603 | 0.00018520 | -0.00043940 | -0.00001506 | 0.00018394 | -0.00007675 |
| 76 | 103 | 0.05001391 | 0.05413862 | -0.03401573 | -0.03698196 | 0.02319290 | 0.01389870 | -0.01787165 | -0.00788525 | 0.00674345 | 0.00266282 | -0.00090757 | 0.00031261 | 0.00020307 | -0.00043168 | 0.00002946 | 0.00016334 | -0.00004987 |
| 76 | 104 | 0.04958978 | 0.05422394 | -0.03388044 | -0.03777654 | 0.02211962 | 0.01375426 | -0.01786985 | -0.00817781 | 0.00664385 | 0.00262623 | -0.00091955 | 0.00031821 | 0.00022376 | -0.00047213 | 0.00002801 | 0.00015906 | -0.00003918 |
| 76 | 105 | 0.04984041 | 0.05373445 | -0.03435257 | -0.03666751 | 0.02316413 | 0.01319311 | -0.01803451 | -0.00747432 | 0.00689685 | 0.00249183 | -0.00092214 | 0.00029878 | 0.00021103 | -0.00044647 | 0.00006226 | 0.00015239 | -0.00002310 |
| 76 | 106 | 0.04944210 | 0.05368109 | -0.03454876 | -0.03745262 | 0.02228215 | 0.01325283 | -0.01779290 | -0.00742662 | 0.00677826 | 0.00237569 | -0.00095966 | 0.00034489 | 0.00017512 | -0.00045299 | 0.00004691 | 0.00015836 | -0.00003653 |
| 76 | 107 | 0.04969570 | 0.05331898 | -0.03505959 | -0.03648039 | 0.02328498 | 0.01273549 | -0.01806273 | -0.00682788 | 0.00702309 | 0.00224630 | -0.00096564 | 0.00032126 | 0.00017832 | -0.00044100 | 0.00008539 | 0.00015375 | -0.00001194 |
| 76 | 108 | 0.04929202 | 0.05312080 | -0.03518989 | -0.03711177 | 0.02252267 | 0.01270923 | -0.01773515 | -0.00667480 | 0.00689236 | 0.00210026 | -0.00100250 | 0.00037501 | 0.00012722 | -0.00042963 | 0.00007241 | 0.00014149 | -0.00004371 |
| 76 | 109 | 0.04953785 | 0.05287350 | -0.03572598 | -0.03630741 | 0.02343559 | 0.01223339 | -0.01809328 | -0.00618172 | 0.00712344 | 0.00197915 | -0.00101496 | 0.00035103 | 0.00014164 | -0.00042750 | 0.00011215 | 0.00014090 | -0.00001290 |
| 76 | 110 | 0.04914026 | 0.05254512 | -0.03579886 | -0.03672855 | 0.02286066 | 0.01212987 | -0.01769535 | -0.00593845 | 0.00697850 | 0.00180336 | -0.00104546 | 0.00040621 | 0.00008373 | -0.00040313 | 0.00010552 | 0.00010634 | -0.00006054 |
| 76 | 111 | 0.04937163 | 0.05238942 | -0.03635757 | -0.03611661 | 0.02364427 | 0.01169347 | -0.01812217 | -0.00554999 | 0.00718763 | 0.00169262 | -0.00106717 | 0.00038692 | 0.00010288 | -0.00040643 | 0.00014276 | 0.00011145 | -0.00002740 |
| 76 | 112 | 0.04898529 | 0.05196116 | -0.03635945 | -0.03628020 | 0.02330226 | 0.01152924 | -0.01766342 | -0.00523512 | 0.00702738 | 0.00149038 | -0.00108528 | 0.00043622 | 0.00004774 | -0.00037465 | 0.00014621 | 0.00005247 | -0.00008606 |
| 76 | 113 | 0.04920057 | 0.05187004 | -0.03694452 | -0.03587440 | 0.02393219 | 0.01113419 | -0.01813948 | -0.00495156 | 0.00720347 | 0.00139096 | -0.00111825 | 0.00042720 | 0.00006455 | -0.00037901 | 0.00017685 | 0.00006419 | -0.00005585 |
| 76 | 114 | 0.04882310 | 0.05137840 | -0.03684739 | -0.03574891 | 0.02383857 | 0.01092648 | -0.01762376 | -0.00458194 | 0.00702925 | 0.00116827 | -0.00111869 | 0.00046360 | 0.00002087 | -0.00034497 | 0.00019309 | -0.00001856 | -0.00011897 |
| 76 | 115 | 0.04902533 | 0.05132860 | -0.03746191 | -0.03554988 | 0.02430700 | 0.01058000 | -0.01813195 | -0.00440665 | 0.00715930 | 0.00108011 | -0.00116379 | 0.00047021 | 0.00002885 | -0.00034677 | 0.00021326 | -0.00000040 | -0.00009765 |
| 76 | 116 | 0.04864725 | 0.05080470 | -0.03723493 | -0.03512147 | 0.02444777 | 0.01034124 | -0.01755988 | -0.00399366 | 0.00697504 | 0.00084483 | -0.00114313 | 0.00048838 | 0.00000251 | -0.00031410 | 0.00024337 | -0.00010315 | -0.00015809 |
| 76 | 117 | 0.04884245 | 0.05078295 | -0.03787464 | -0.03511758 | 0.02475998 | 0.01005400 | -0.01808705 | -0.00393259 | 0.00704613 | 0.00076698 | -0.00119996 | 0.00051504 | -0.00000330 | -0.00031098 | 0.00024992 | -0.00007987 | -0.00015165 |
| 76 | 118 | 0.04844948 | 0.05024235 | -0.03749671 | -0.03438774 | 0.02510027 | 0.00978952 | -0.01745980 | -0.00348041 | 0.00685756 | 0.00052762 | -0.00115741 | 0.00051235 | -0.00001046 | -0.00028094 | 0.00029309 | -0.00019601 | -0.00020289 |
| 76 | 119 | 0.04864405 | 0.05024780 | -0.03814581 | -0.03455720 | 0.02526869 | 0.00957131 | -0.01799732 | -0.00354010 | 0.00685886 | 0.00045824 | -0.00122430 | 0.00056220 | -0.00003315 | -0.00027198 | 0.00028386 | -0.00016975 | -0.00021657 |
| 76 | 120 | 0.04822015 | 0.04968485 | -0.03761588 | -0.03353883 | 0.02576589 | 0.00927973 | -0.01732037 | -0.00304593 | 0.00667232 | 0.00022253 | -0.00116205 | 0.00053880 | -0.00002330 | -0.00024333 | 0.00033764 | -0.00029095 | -0.00025371 |
| 76 | 121 | 0.04841838 | 0.04972649 | -0.03824715 | -0.03385177 | 0.02580439 | 0.00913480 | -0.01786392 | -0.00323069 | 0.00659629 | 0.00015886 | -0.00123633 | 0.00061362 | -0.00006420 | -0.00022888 | 0.00031148 | -0.00026405 | -0.00029139 |
| 76 | 122 | 0.04794934 | 0.04911603 | -0.03758857 | -0.03256627 | 0.02642045 | 0.00881063 | -0.01714947 | -0.00268672 | 0.00641807 | -0.00006757 | -0.00115923 | 0.00057170 | -0.00004225 | -0.00019845 | 0.00037272 | -0.00038196 | -0.00031156 |

TABLE 3. Nuclear charge density distribution FB coefficients.

| $Z$ | $N$ | $a_1$ | $a_2$ | $a_3$ | $a_4$ | $a_5$ | $a_6$ | $a_7$ | $a_8$ | $a_9$ | $a_{10}$ | $a_{11}$ | $a_{12}$ | $a_{13}$ | $a_{14}$ | $a_{15}$ | $a_{16}$ | $a_{17}$ |
|---|---|---|---|---|---|---|---|---|---|---|---|---|---|---|---|---|---|---|
| 76 | 123 | 0.04815095 | 0.04920560 | -0.03816803 | -0.03298568 | 0.02634146 | 0.00873430 | -0.01769803 | -0.00299614 | 0.00626030 | -0.00012930 | -0.00123753 | 0.00067205 | -0.00010161 | -0.00017968 | 0.00032916 | -0.00035623 | -0.00037546 |
| 76 | 124 | 0.04762724 | 0.04851199 | -0.03742507 | -0.03146275 | 0.02704976 | 0.00837189 | -0.01696453 | -0.00239262 | 0.00609675 | -0.00034326 | -0.00115228 | 0.00061442 | -0.00007306 | -0.00014339 | 0.00039523 | -0.00046429 | -0.00037759 |
| 76 | 125 | 0.04782604 | 0.04865492 | -0.03791942 | -0.03194495 | 0.02686460 | 0.00834973 | -0.01751935 | -0.00282025 | 0.00585483 | -0.00040796 | -0.00123091 | 0.00074001 | -0.00015074 | -0.00012173 | 0.00033414 | -0.00044041 | -0.00046824 |
| 76 | 126 | 0.04724461 | 0.04784534 | -0.03714715 | -0.03022416 | 0.02764932 | 0.00794808 | -0.01678810 | -0.00214916 | 0.00571286 | -0.00060784 | -0.00114508 | 0.00066875 | -0.00011961 | -0.00007568 | 0.00040380 | -0.00053500 | -0.00045252 |
| 76 | 127 | 0.04749027 | 0.04772880 | -0.03788462 | -0.03039570 | 0.02750035 | 0.00739231 | -0.01767392 | -0.00238466 | 0.00565036 | -0.00075373 | -0.00119800 | 0.00076297 | -0.00019288 | -0.00010594 | 0.00034343 | -0.00047338 | -0.00049241 |
| 77 | 86 | 0.05119071 | 0.05783320 | -0.03255534 | -0.03989252 | 0.02245529 | 0.02173316 | -0.01140307 | -0.00787454 | 0.00494973 | 0.00352549 | -0.00055142 | 0.00043079 | -0.00007108 | -0.00016704 | -0.00019228 | 0.00015891 | -0.00027152 |
| 77 | 87 | 0.05164781 | 0.05740934 | -0.03373661 | -0.03936501 | 0.02375520 | 0.02069358 | -0.01291990 | -0.00821335 | 0.00523853 | 0.00375893 | -0.00060657 | 0.00035116 | 0.00005295 | -0.00015550 | -0.00014314 | 0.00016035 | -0.00023350 |
| 77 | 88 | 0.05133545 | 0.05732851 | -0.03336302 | -0.03924748 | 0.02326752 | 0.02052328 | -0.01286212 | -0.00776052 | 0.00531816 | 0.00337142 | -0.00065476 | 0.00046178 | -0.00006098 | -0.00019189 | -0.00015460 | 0.00015156 | -0.00028063 |
| 77 | 89 | 0.05164423 | 0.05690715 | -0.03418298 | -0.03883243 | 0.02412771 | 0.01959502 | -0.01403251 | -0.00807813 | 0.00551013 | 0.00359429 | -0.00067995 | 0.00038197 | 0.00004050 | -0.00017417 | -0.00011717 | 0.00015953 | -0.00024252 |
| 77 | 90 | 0.05142625 | 0.05688111 | -0.03401850 | -0.03879460 | 0.02383777 | 0.01946703 | -0.01413692 | -0.00774389 | 0.00563298 | 0.00325360 | -0.00074112 | 0.00048392 | -0.00005008 | -0.00021826 | -0.00013011 | 0.00015718 | -0.00028004 |
| 77 | 91 | 0.05158551 | 0.05645616 | -0.03448417 | -0.03842637 | 0.02434989 | 0.01864232 | -0.01497203 | -0.00800337 | 0.00575418 | 0.00345342 | -0.00073839 | 0.00040217 | 0.00003353 | -0.00019760 | -0.00009947 | 0.00016912 | -0.00023898 |
| 77 | 92 | 0.05144553 | 0.05647317 | -0.03449937 | -0.03847792 | 0.02417395 | 0.01852339 | -0.01521398 | -0.00778993 | 0.00590144 | 0.00316194 | -0.00080887 | 0.00049485 | -0.00003607 | -0.00024647 | -0.00011349 | 0.00017170 | -0.00026793 |
| 77 | 93 | 0.05146626 | 0.05603822 | -0.03464366 | -0.03810065 | 0.02443687 | 0.01778550 | -0.01575573 | -0.00796601 | 0.00597590 | 0.00332855 | -0.00078302 | 0.00041138 | 0.00003308 | -0.00022471 | -0.00008590 | 0.00018453 | -0.00022396 |
| 77 | 94 | 0.05138899 | 0.05608790 | -0.03481096 | -0.03823893 | 0.02431118 | 0.01764113 | -0.01610660 | -0.00786107 | 0.00613410 | 0.00308431 | -0.00085859 | 0.00049335 | -0.00001661 | -0.00027659 | -0.00009894 | 0.00018903 | -0.00024473 |
| 77 | 95 | 0.05128840 | 0.05563548 | -0.03468005 | -0.03781248 | 0.02441346 | 0.01697426 | -0.01640737 | -0.00794140 | 0.00617993 | 0.00321179 | -0.00081550 | 0.00040974 | 0.00004001 | -0.00025422 | -0.00007239 | 0.00020041 | -0.00019961 |
| 77 | 96 | 0.05126482 | 0.05571221 | -0.03497773 | -0.03802662 | 0.02429631 | 0.01677481 | -0.01683810 | -0.00792208 | 0.00634166 | 0.00300967 | -0.00089216 | 0.00047951 | 0.00000977 | -0.00030832 | -0.00008168 | 0.00020306 | -0.00021254 |
| 77 | 97 | 0.05106124 | 0.05523439 | -0.03462072 | -0.03752802 | 0.02430763 | 0.01616950 | -0.01694752 | -0.00790604 | 0.00636979 | 0.00309674 | -0.00083745 | 0.00039788 | 0.00005472 | -0.00028484 | -0.00005570 | 0.00021184 | -0.00016869 |
| 77 | 98 | 0.05108960 | 0.05533682 | -0.03503413 | -0.03780193 | 0.02417699 | 0.01589279 | -0.01743160 | -0.00794365 | 0.00653316 | 0.00292970 | -0.00091197 | 0.00045460 | 0.00004316 | -0.00034099 | -0.00005871 | 0.00020898 | -0.00017442 |
| 77 | 99 | 0.05079946 | 0.05482687 | -0.03449622 | -0.03722395 | 0.02414628 | 0.01534741 | -0.01739053 | -0.00784009 | 0.00654799 | 0.00297911 | -0.00085037 | 0.00037690 | 0.00007682 | -0.00031538 | -0.00003384 | 0.00021509 | -0.00013407 |
| 77 | 100 | 0.05088361 | 0.05495520 | -0.03501723 | -0.03753797 | 0.02399549 | 0.01497772 | -0.01790633 | -0.00790398 | 0.00671546 | 0.00283914 | -0.00092053 | 0.00042076 | 0.00008235 | -0.00037363 | -0.00002879 | 0.00020385 | -0.00013359 |
| 77 | 101 | 0.05052002 | 0.05440939 | -0.03433585 | -0.03688618 | 0.02395335 | 0.01449790 | -0.01774574 | -0.00772877 | 0.00671632 | 0.00285657 | -0.00085559 | 0.00034837 | 0.00010524 | -0.00034483 | -0.00000605 | 0.00020797 | -0.00009842 |
| 77 | 102 | 0.05066687 | 0.05456254 | -0.03496172 | -0.03721822 | 0.02378619 | 0.01402345 | -0.01827748 | -0.00778895 | 0.00689333 | 0.00273530 | -0.00092017 | 0.00038054 | 0.00012534 | -0.00040523 | 0.00000791 | 0.00018654 | -0.00009301 |
| 77 | 103 | 0.05023925 | 0.05398118 | -0.03416513 | -0.03650762 | 0.02374897 | 0.01362048 | -0.01801985 | -0.00756280 | 0.00687622 | 0.00272825 | -0.00085431 | 0.00031407 | 0.00013842 | -0.00037240 | 0.00002743 | 0.00018973 | -0.00006396 |
| 77 | 104 | 0.05045671 | 0.05415528 | -0.03489726 | -0.03683453 | 0.02357537 | 0.01303123 | -0.01855720 | -0.00759139 | 0.00706974 | 0.00261745 | -0.00091292 | 0.00033648 | 0.00016979 | -0.00043480 | 0.00005038 | 0.00015732 | -0.00005500 |
| 77 | 105 | 0.05005218 | 0.05358988 | -0.03453802 | -0.03623473 | 0.02373120 | 0.01295990 | -0.01817905 | -0.00719021 | 0.00701867 | 0.00256785 | -0.00086797 | 0.00029194 | 0.00016021 | -0.00039132 | 0.00006166 | 0.00017829 | -0.00003170 |
| 77 | 106 | 0.05029678 | 0.05374492 | -0.03560303 | -0.03663132 | 0.02373731 | 0.01266222 | -0.01854312 | -0.00699254 | 0.00718488 | 0.00239490 | -0.00094852 | 0.00034158 | 0.00015725 | -0.00043484 | 0.00007534 | 0.00016257 | -0.00003032 |
| 77 | 107 | 0.04994410 | 0.05322509 | -0.03541601 | -0.03612647 | 0.02388497 | 0.01251795 | -0.01825261 | -0.00664458 | 0.00712298 | 0.00235973 | -0.00090554 | 0.00029225 | 0.00016176 | -0.00039713 | 0.00009199 | 0.00017582 | -0.00000685 |
| 77 | 108 | 0.05012373 | 0.05330929 | -0.03626113 | -0.03644934 | 0.02391055 | 0.01225556 | -0.01852321 | -0.00638111 | 0.00727841 | 0.00214736 | -0.00099031 | 0.00035560 | 0.00013756 | -0.00042744 | 0.00010219 | 0.00015723 | -0.00001633 |
| 77 | 109 | 0.04983129 | 0.05281916 | -0.03626335 | -0.03603429 | 0.02404141 | 0.01204741 | -0.01829689 | -0.00608969 | 0.00718838 | 0.00212347 | -0.00094910 | 0.00030398 | 0.00015248 | -0.00039377 | 0.00012031 | 0.00016452 | 0.00000567 |
| 77 | 110 | 0.04994433 | 0.05283567 | -0.03688639 | -0.03626148 | 0.02412320 | 0.01180957 | -0.01850225 | -0.00576892 | 0.00734200 | 0.00187572 | -0.00103653 | 0.00037869 | 0.00011099 | -0.00041237 | 0.00013091 | 0.00013918 | -0.00001530 |
| 77 | 111 | 0.04971999 | 0.05236289 | -0.03708449 | -0.03592720 | 0.02423013 | 0.01155267 | -0.01832037 | -0.00553913 | 0.00720730 | 0.00185950 | -0.00099702 | 0.00032855 | 0.00013117 | -0.00038107 | 0.00014555 | 0.00014342 | 0.00000297 |
| 77 | 112 | 0.04976518 | 0.05232253 | -0.03747999 | -0.03603608 | 0.02440207 | 0.01133619 | -0.01847871 | -0.00517419 | 0.00736434 | 0.00158318 | -0.00108419 | 0.00041005 | 0.00007913 | -0.00039026 | 0.00016111 | 0.00010694 | -0.00002853 |
| 77 | 113 | 0.04961574 | 0.05186210 | -0.03786693 | -0.03576837 | 0.02448110 | 0.01105105 | -0.01832887 | -0.00501296 | 0.00717051 | 0.00157089 | -0.00104642 | 0.00036615 | 0.00009850 | -0.00035993 | 0.00016672 | 0.00011172 | -0.00001697 |
| 77 | 114 | 0.04959062 | 0.05177971 | -0.03802737 | -0.03574143 | 0.02476549 | 0.01085708 | -0.01844516 | -0.00461816 | 0.00733332 | 0.00127506 | -0.00112945 | 0.00044825 | 0.00004427 | -0.00036237 | 0.00019195 | 0.00006010 | -0.00005630 |
| 77 | 115 | 0.04952109 | 0.05133540 | -0.03858140 | -0.03552070 | 0.02481592 | 0.01056639 | -0.01832360 | -0.00453247 | 0.00706927 | 0.00126277 | -0.00109345 | 0.00041579 | 0.00005656 | -0.00033206 | 0.00018305 | 0.00006913 | -0.00005489 |
| 77 | 116 | 0.04942102 | 0.05122469 | -0.03850047 | -0.03534969 | 0.02521724 | 0.01039667 | -0.01839034 | -0.00412068 | 0.00723848 | 0.00095812 | -0.00116832 | 0.00049179 | 0.00000852 | -0.00033010 | 0.00022206 | -0.00000030 | -0.00009787 |
| 77 | 117 | 0.04943303 | 0.05080697 | -0.03918766 | -0.03515168 | 0.02524029 | 0.01012017 | -0.01830055 | -0.00411443 | 0.00689713 | 0.00094151 | -0.00113385 | 0.00047574 | 0.00000803 | -0.00029939 | 0.00019377 | 0.00001647 | -0.00011009 |
| 77 | 118 | 0.04925149 | 0.05067599 | -0.03886358 | -0.03483849 | 0.02574455 | 0.00997426 | -0.01830289 | -0.00369609 | 0.00707253 | 0.00063957 | -0.00119749 | 0.00053980 | -0.00002719 | -0.00029441 | 0.00024942 | -0.00007149 | -0.00015190 |
| 77 | 119 | 0.04934192 | 0.05029653 | -0.03964419 | -0.03463555 | 0.02574121 | 0.00972335 | -0.01825159 | -0.00376688 | 0.00665028 | 0.00061354 | -0.00116372 | 0.00054412 | -0.00004504 | -0.00026345 | 0.00019796 | -0.00004384 | -0.00018050 |
| 77 | 120 | 0.04907180 | 0.05014480 | -0.03908243 | -0.03418978 | 0.02632163 | 0.00959789 | -0.01817582 | -0.00335026 | 0.00683189 | 0.00032588 | -0.00121512 | 0.00059247 | -0.00006382 | -0.00025528 | 0.00027138 | -0.00014903 | -0.00021682 |
| 77 | 121 | 0.04923192 | 0.04980948 | -0.03991871 | -0.03395153 | 0.02629128 | 0.00937155 | -0.01816808 | -0.00348762 | 0.00632695 | 0.00028418 | -0.00118020 | 0.00061938 | -0.00010195 | -0.00022468 | 0.00019427 | -0.00010768 | -0.00026298 |
| 77 | 122 | 0.04886705 | 0.04962792 | -0.03913368 | -0.03338737 | 0.02691767 | 0.00926126 | -0.01801040 | -0.00307944 | 0.00651625 | 0.00002131 | -0.00122121 | 0.00065110 | -0.00010426 | -0.00021141 | 0.00028497 | -0.00022736 | -0.00029125 |
| 77 | 123 | 0.04908257 | 0.04933141 | -0.03999649 | -0.03308085 | 0.02685769 | 0.00904486 | -0.01804563 | -0.00326543 | 0.00592647 | -0.00004335 | -0.00118208 | 0.00070059 | -0.00016372 | -0.00018221 | 0.00018103 | -0.00016964 | -0.00035398 |
| 77 | 124 | 0.04861905 | 0.04910421 | -0.03901192 | -0.03241558 | 0.02750576 | 0.00894482 | -0.01781794 | -0.00287121 | 0.00612775 | -0.00027297 | -0.00121763 | 0.00071750 | -0.00015271 | -0.00016046 | 0.00028740 | -0.00030067 | -0.00037410 |
| 77 | 125 | 0.04887108 | 0.04882819 | -0.03988368 | -0.03200605 | 0.02741177 | 0.00871226 | -0.01788818 | -0.00308315 | 0.00544933 | -0.00036817 | -0.00117014 | 0.00078727 | -0.00023259 | -0.00013388 | 0.00015673 | -0.00022407 | -0.00045006 |
| 77 | 126 | 0.04830796 | 0.04853703 | -0.03873123 | -0.03126096 | 0.02806876 | 0.00862113 | -0.01761829 | -0.00270731 | 0.00567040 | -0.00055883 | -0.00120771 | 0.00079304 | -0.00021333 | -0.00009955 | 0.00027686 | -0.00036404 | -0.00046441 |
| 77 | 127 | 0.04856447 | 0.04798183 | -0.03979577 | -0.03046375 | 0.02793198 | 0.00778120 | -0.01792845 | -0.00258814 | 0.00520899 | -0.00074195 | -0.00113177 | 0.00080948 | -0.00027413 | -0.00012223 | 0.00015947 | -0.00025008 | -0.00046830 |
| 77 | 128 | 0.04800341 | 0.04726936 | -0.03893732 | -0.02918647 | 0.02875572 | 0.00700328 | -0.01796515 | -0.00184684 | 0.00573513 | -0.00098996 | -0.00114733 | 0.00075661 | -0.00023426 | -0.00013103 | 0.00030435 | -0.00036158 | -0.00040923 |
| 78 | 87 | 0.05192254 | 0.05741422 | -0.03459790 | -0.03958835 | 0.02357517 | 0.02042501 | -0.01331732 | -0.00769123 | 0.00544591 | 0.00352209 | -0.00066149 | 0.00053149 | -0.00017236 | -0.00014061 | -0.00017277 | 0.00015757 | -0.00036024 |
| 78 | 88 | 0.05160560 | 0.05760035 | -0.03449740 | -0.04077730 | 0.02220857 | 0.02047738 | -0.01341346 | -0.00797383 | 0.00519854 | 0.00327965 | -0.00068553 | 0.00059107 | -0.00023700 | -0.00018857 | -0.00021907 | 0.00020982 | -0.00036836 |
| 78 | 89 | 0.05199420 | 0.05700881 | -0.03517280 | -0.03905614 | 0.02419288 | 0.01933309 | -0.01460221 | -0.00764583 | 0.00575597 | 0.00337489 | -0.00074478 | 0.00055090 | -0.00015586 | -0.00016977 | -0.00013934 | 0.00015511 | -0.00035732 |
| 78 | 90 | 0.05165889 | 0.05721728 | -0.03509326 | -0.04036742 | 0.02286765 | 0.01954343 | -0.01464128 | -0.00802075 | 0.00551013 | 0.00317880 | -0.00076387 | 0.00059878 | -0.00020404 | -0.00022346 | -0.00018936 | 0.00021505 | -0.00035417 |
| 78 | 91 | 0.05200750 | 0.05665206 | -0.03559957 | -0.03870249 | 0.02456797 | 0.01837041 | -0.01569393 | -0.00767491 | 0.00602452 | 0.00326177 | -0.00081127 | 0.00055946 | -0.00013544 | -0.00020118 | -0.00011621 | 0.00016404 | -0.00034183 |
| 78 | 92 | 0.05164952 | 0.05684727 | -0.03555177 | -0.04008586 | 0.02326991 | 0.01865761 | -0.01570805 | -0.00811514 | 0.00578619 | 0.00310697 | -0.00082816 | 0.00059911 | -0.00017135 | -0.00025705 | -0.00016822 | 0.00022935 | -0.00033050 |
| 78 | 93 | 0.05195667 | 0.05632066 | -0.03588097 | -0.03846394 | 0.02472966 | 0.01748738 | -0.01659978 | -0.00773967 | 0.00626111 | 0.00317102 | -0.00086140 | 0.00055673 | -0.00011043 | -0.00023432 | -0.00009841 | 0.00017915 | -0.00031473 |
| 78 | 94 | 0.05158002 | 0.05647573 | -0.03587975 | -0.03986900 | 0.02345419 | 0.01777837 | -0.01661276 | -0.00821253 | 0.00603598 | 0.00304822 | -0.00087774 | 0.00059113 | -0.00013806 | -0.00029010 | -0.00015052 | 0.00024700 | -0.00029815 |
| 78 | 95 | 0.05184583 | 0.05599469 | -0.03603590 | -0.03827848 | 0.02472449 | 0.01663504 | -0.01734014 | -0.00780150 | 0.00647564 | 0.00309002 | -0.00089672 | 0.00054299 | -0.00008030 | -0.00026854 | -0.00008103 | 0.00019429 | -0.00027851 |
| 78 | 96 | 0.05146214 | 0.05609141 | -0.03609824 | -0.03965544 | 0.02347583 | 0.01686936 | -0.01736508 | -0.00827053 | 0.00626938 | 0.00298683 | -0.00091311 | 0.00057484 | -0.00010360 | -0.00032314 | -0.00013160 | 0.00026192 | -0.00025921 |
| 78 | 97 | 0.05168722 | 0.05565926 | -0.03609294 | -0.03809430 | 0.02460310 | 0.01577528 | -0.01793714 | -0.00782645 | 0.00667652 | 0.00300782 | -0.00091920 | 0.00051933 | -0.00004521 | -0.00030308 | -0.00006033 | 0.00020396 | -0.00023643 |
| 78 | 98 | 0.05131302 | 0.05568573 | -0.03623680 | -0.03939563 | 0.02339115 | 0.01590440 | -0.01797850 | -0.00825430 | 0.00649468 | 0.00291025 | -0.00093552 | 0.00055110 | -0.00006817 | -0.00035621 | -0.00010817 | 0.00026909 | -0.00021644 |
| 78 | 99 | 0.05149788 | 0.05530446 | -0.03608449 | -0.03787340 | 0.02441197 | 0.01488434 | -0.01840962 | -0.00778801 | 0.00687002 | 0.00291646 | -0.00093090 | 0.00048750 | -0.00000621 | -0.00033702 | -0.00003417 | 0.00020418 | -0.00019193 |
| 78 | 100 | 0.05115155 | 0.05525271 | -0.03632806 | -0.03905604 | 0.02324895 | 0.01486926 | -0.01846685 | -0.00813973 | 0.00671775 | 0.00281054 | -0.00094663 | 0.00052138 | -0.00003257 | -0.00038885 | -0.00007851 | 0.00026526 | -0.00017264 |
| 78 | 101 | 0.05129604 | 0.05492421 | -0.03604211 | -0.03759120 | 0.02418965 | 0.01395140 | -0.01877252 | -0.00766831 | 0.00706029 | 0.00281109 | -0.00093378 | 0.00044967 | 0.00003505 | -0.00036937 | -0.00000184 | 0.00019282 | -0.00014810 |
| 78 | 102 | 0.05099526 | 0.05478916 | -0.03640298 | -0.03861983 | 0.02308727 | 0.01376155 | -0.01884300 | -0.00791441 | 0.00694198 | 0.00268460 | -0.00094825 | 0.00048736 | 0.00000212 | -0.00042028 | -0.00004227 | 0.00024900 | -0.00013017 |
| 78 | 103 | 0.05109844 | 0.05451554 | -0.03599349 | -0.03723489 | 0.02396572 | 0.01297529 | -0.01903788 | -0.00745801 | 0.00724969 | 0.00268970 | -0.00092959 | 0.00040803 | 0.00007663 | -0.00039927 | 0.00003630 | 0.00016936 | -0.00010737 |
| 78 | 104 | 0.05085877 | 0.05429549 | -0.03648749 | -0.03808490 | 0.02293390 | 0.01258964 | -0.01911868 | -0.00757694 | 0.00716873 | 0.00253367 | -0.00094221 | 0.00045057 | 0.00003496 | -0.00044963 | -0.00000008 | 0.00022050 | -0.00009066 |
| 78 | 105 | 0.05092734 | 0.05408024 | -0.03635630 | -0.03693207 | 0.02394399 | 0.01229591 | -0.01912218 | -0.00703561 | 0.00739388 | 0.00252023 | -0.00094094 | 0.00038366 | 0.00009615 | -0.00041301 | 0.00007047 | 0.00015550 | -0.00007401 |
| 78 | 106 | 0.05067667 | 0.05375651 | -0.03712967 | -0.03782496 | 0.02308991 | 0.01217854 | -0.01900045 | -0.00692110 | 0.00725698 | 0.00230290 | -0.00097603 | 0.00045562 | 0.00001371 | -0.00043212 | 0.00002413 | 0.00021662 | -0.00007441 |
| 78 | 107 | 0.05075672 | 0.05362832 | -0.03706625 | -0.03674867 | 0.02411727 | 0.01192696 | -0.01907676 | -0.00645562 | 0.00747887 | 0.00229595 | -0.00097664 | 0.00038455 | 0.00008949 | -0.00040829 | 0.00009736 | 0.00015415 | -0.00005090 |
| 78 | 108 | 0.05048591 | 0.05321006 | -0.03770072 | -0.03751415 | 0.02333407 | 0.01175508 | -0.01887228 | -0.00624629 | 0.00732944 | 0.00204454 | -0.00101189 | 0.00046455 | -0.00000780 | -0.00041184 | 0.00005328 | 0.00019964 | -0.00006591 |
| 78 | 109 | 0.05057357 | 0.05315077 | -0.03771475 | -0.03655409 | 0.02432191 | 0.01153517 | -0.01901670 | -0.00586204 | 0.00754052 | 0.00204595 | -0.00101743 | 0.00039486 | 0.00007527 | -0.00039707 | 0.00012468 | 0.00014382 | -0.00003813 |
| 78 | 110 | 0.05029379 | 0.05265544 | -0.03821008 | -0.03713282 | 0.02368616 | 0.01132144 | -0.01874243 | -0.00557065 | 0.00737746 | 0.00176112 | -0.00104786 | 0.00047670 | -0.00002827 | -0.00038925 | 0.00008758 | 0.00016825 | -0.00006612 |
| 78 | 111 | 0.05038664 | 0.05264277 | -0.03830978 | -0.03632035 | 0.02458676 | 0.01112775 | -0.01894596 | -0.00526982 | 0.00756946 | 0.00177166 | -0.00106124 | 0.00041483 | 0.00005359 | -0.00037950 | 0.00015194 | 0.00012323 | -0.00003764 |
| 78 | 112 | 0.05010553 | 0.05209678 | -0.03865894 | -0.03666604 | 0.02415341 | 0.01088728 | -0.01861185 | -0.00491581 | 0.00739020 | 0.00145788 | -0.00108143 | 0.00049126 | -0.00004633 | -0.00036483 | 0.00012637 | 0.00012238 | -0.00007533 |
| 78 | 113 | 0.05020431 | 0.05210919 | -0.03884946 | -0.03601944 | 0.02493546 | 0.01072099 | -0.01886407 | -0.00469941 | 0.00755392 | 0.00147687 | -0.00110499 | 0.00044384 | 0.00002571 | -0.00035633 | 0.00017842 | 0.00009160 | -0.00005056 |
| 78 | 114 | 0.04992298 | 0.05154175 | -0.03903896 | -0.03610388 | 0.02472860 | 0.01046758 | -0.01847390 | -0.00430322 | 0.00735641 | 0.00114206 | -0.00110986 | 0.00050764 | -0.00006119 | -0.00033890 | 0.00016801 | 0.00006354 | -0.00009319 |
| 78 | 115 | 0.05003214 | 0.05156310 | -0.03932036 | -0.03562615 | 0.02538017 | 0.01033554 | -0.01876586 | -0.00417248 | 0.00748194 | 0.00116725 | -0.00114498 | 0.00048070 | -0.00000645 | -0.00032873 | 0.00020310 | 0.00004900 | -0.00007706 |
| 78 | 116 | 0.04974390 | 0.05099810 | -0.03933419 | -0.03544049 | 0.02539066 | 0.01007815 | -0.01831729 | -0.00375082 | 0.00726602 | 0.00082205 | -0.00113081 | 0.00052587 | -0.00007321 | -0.00031130 | 0.00020990 | -0.00000518 | -0.00011900 |
| 78 | 117 | 0.04987076 | 0.05102133 | -0.03969928 | -0.03512016 | 0.02591724 | 0.00998971 | -0.01864294 | -0.00370739 | 0.00734347 | 0.00084958 | -0.00117756 | 0.00052409 | -0.00004117 | -0.00029776 | 0.00022460 | -0.00000324 | -0.00011637 |
| 78 | 118 | 0.04956181 | 0.05046959 | -0.03952558 | -0.03467163 | 0.02610840 | 0.00973029 | -0.01813103 | -0.00327047 | 0.00711128 | 0.00050606 | -0.00114290 | 0.00054699 | -0.00008428 | -0.00028104 | 0.00024871 | -0.00007936 | -0.00015211 |
| 78 | 119 | 0.04971447 | 0.05049757 | -0.03995853 | -0.03448541 | 0.02652680 | 0.00969277 | -0.01848697 | -0.00331543 | 0.00713158 | 0.00053071 | -0.00119986 | 0.00057318 | -0.00007775 | -0.00026390 | 0.00024108 | -0.00006230 | -0.00016709 |
| 78 | 120 | 0.04936649 | 0.05095219 | -0.03959684 | -0.03379259 | 0.02684651 | 0.00942606 | -0.01791022 | -0.00286632 | 0.00688762 | 0.00020084 | -0.00114626 | 0.00057299 | -0.00009784 | -0.00024627 | 0.00028074 | -0.00015377 | -0.00019224 |
| 78 | 121 | 0.04955103 | 0.04999500 | -0.04007352 | -0.03370793 | 0.02717745 | 0.00944029 | -0.01829403 | -0.00299876 | 0.00684263 | 0.00021631 | -0.00121049 | 0.00062803 | -0.00011714 | -0.00022654 | 0.00025027 | -0.00012394 | -0.00022753 |
| 78 | 122 | 0.04914493 | 0.04943189 | -0.03953987 | -0.03279654 | 0.02757267 | 0.00915585 | -0.01765992 | -0.00253451 | 0.00659393 | -0.00008965 | -0.00114261 | 0.00060642 | -0.00011826 | -0.00020438 | 0.00030260 | -0.00022331 | -0.00023957 |
| 78 | 123 | 0.04936247 | 0.04950081 | -0.04003027 | -0.03277349 | 0.02783452 | 0.00921315 | -0.01806854 | -0.00275023 | 0.00647606 | -0.00009015 | -0.00120990 | 0.00068948 | -0.00016193 | -0.00018387 | 0.00024971 | -0.00018308 | -0.00029617 |
| 78 | 124 | 0.04888236 | 0.04888527 | -0.03935796 | -0.03167528 | 0.02826283 | 0.00889987 | -0.01739592 | -0.00226430 | 0.00623246 | -0.00036473 | -0.00113506 | 0.00064958 | -0.00014987 | -0.00015247 | 0.00031187 | -0.00028382 | -0.00029461 |
| 78 | 125 | 0.04912678 | 0.04898511 | -0.03983060 | -0.03166752 | 0.02846834 | 0.00898111 | -0.01782477 | -0.00255530 | 0.00603397 | -0.00038800 | -0.00120044 | 0.00075870 | -0.00021566 | -0.00013307 | 0.00023733 | -0.00023470 | -0.00037179 |
| 78 | 126 | 0.04856331 | 0.04828297 | -0.03906546 | -0.03042198 | 0.02890343 | 0.00863316 | -0.01714137 | -0.00204063 | 0.00580832 | -0.00062649 | -0.00112758 | 0.00070370 | -0.00019586 | -0.00008775 | 0.00030753 | -0.00033265 | -0.00035776 |
| 78 | 127 | 0.04881569 | 0.04807435 | -0.03973716 | -0.02996044 | 0.02910062 | 0.00798634 | -0.01785855 | -0.00199457 | 0.00583456 | -0.00075940 | -0.00116439 | 0.00077391 | -0.00025140 | -0.00011865 | 0.00024674 | -0.00026506 | -0.00038354 |
| 78 | 128 | 0.04825515 | 0.04697409 | -0.03924097 | -0.02813564 | 0.02971794 | 0.00689892 | -0.01750113 | -0.00113853 | 0.00590116 | -0.00104797 | -0.00107172 | 0.00066771 | -0.00020877 | -0.00011365 | 0.00033910 | -0.00034582 | -0.00030838 |
| 78 | 129 | 0.04850027 | 0.04683093 | -0.03984021 | -0.02780208 | 0.02964115 | 0.00631634 | -0.01802814 | -0.00102563 | 0.00591754 | -0.00117352 | -0.00110014 | 0.00073058 | -0.00026769 | -0.00014431 | 0.00027368 | -0.00026549 | -0.00032281 |
| 78 | 130 | 0.04798503 | 0.04570590 | -0.03939992 | -0.02591448 | 0.03035384 | 0.00528216 | -0.01767443 | -0.00024415 | 0.00596643 | -0.00141048 | -0.00100934 | 0.00063610 | -0.00023376 | -0.00013268 | 0.00034922 | -0.00033213 | -0.00025393 |
| 79 | 89 | 0.05255316 | 0.05710298 | -0.03637596 | -0.03886569 | 0.02529516 | 0.01836733 | -0.01602132 | -0.00785352 | 0.00621970 | 0.00373544 | -0.00077177 | 0.00047360 | -0.00005172 | -0.00013764 | -0.00006553 | 0.00012103 | -0.00033892 |
| 79 | 90 | 0.05240345 | 0.05695306 | -0.03631440 | -0.03881128 | 0.02501232 | 0.01833426 | -0.01610825 | -0.00760514 | 0.00619194 | 0.00340754 | -0.00082239 | 0.00059706 | -0.00018595 | -0.00016545 | -0.00011179 | 0.00015509 | -0.00039107 |
| 79 | 91 | 0.05240524 | 0.05674303 | -0.03645004 | -0.03851580 | 0.02537370 | 0.01745306 | -0.01676843 | -0.00775577 | 0.00643192 | 0.00357616 | -0.00080734 | 0.00047668 | -0.00004216 | -0.00016762 | -0.00003713 | 0.00011932 | -0.00031916 |
| 79 | 92 | 0.05235545 | 0.05665216 | -0.03656991 | -0.03853106 | 0.02522235 | 0.01743865 | -0.01700626 | -0.00762697 | 0.00643589 | 0.00329167 | -0.00086993 | 0.00059389 | -0.00015996 | -0.00020115 | -0.00008789 | 0.00016415 | -0.00036277 |
| 79 | 93 | 0.05223284 | 0.05640175 | -0.03644889 | -0.03825416 | 0.02533455 | 0.01662792 | -0.01738013 | -0.00768264 | 0.00662951 | 0.00343620 | -0.00083417 | 0.00047261 | -0.00003032 | -0.00019848 | -0.00001510 | 0.00012464 | -0.00029150 |
| 79 | 94 | 0.05226292 | 0.05635485 | -0.03672625 | -0.03833248 | 0.02525567 | 0.01659112 | -0.01773440 | -0.00765084 | 0.00666073 | 0.00319189 | -0.00090425 | 0.00058144 | -0.00013119 | -0.00023722 | -0.00006768 | 0.00017658 | -0.00032566 |
| 79 | 95 | 0.05203857 | 0.05606065 | -0.03638689 | -0.03802984 | 0.02520880 | 0.01585085 | -0.01787798 | -0.00761173 | 0.00681592 | 0.00330740 | -0.00085403 | 0.00046237 | -0.00001651 | -0.00022903 | 0.00000340 | 0.00013275 | -0.00025878 |
| 79 | 96 | 0.05213307 | 0.05604185 | -0.03680509 | -0.03815583 | 0.02516156 | 0.01574954 | -0.01831469 | -0.00764244 | 0.00687357 | 0.00309626 | -0.00092714 | 0.00056077 | -0.00010000 | -0.00027269 | -0.00004727 | 0.00018671 | -0.00028314 |
| 79 | 97 | 0.05182641 | 0.05570522 | -0.03628076 | -0.03780134 | 0.02502729 | 0.01508816 | -0.01827959 | -0.00752236 | 0.00699356 | 0.00318249 | -0.00086805 | 0.00044661 | -0.00000056 | -0.00025845 | 0.00002109 | 0.00013957 | -0.00022346 |
| 79 | 98 | 0.05197763 | 0.05570003 | -0.03683167 | -0.03795383 | 0.02498814 | 0.01488458 | -0.01876660 | -0.00757360 | 0.00707985 | 0.00299504 | -0.00094034 | 0.00053323 | -0.00006702 | -0.00030669 | -0.00002383 | 0.00018996 | -0.00023862 |
| 79 | 99 | 0.05160242 | 0.05532626 | -0.03614801 | -0.03753883 | 0.02481709 | 0.01431651 | -0.01859689 | -0.00739689 | 0.00716388 | 0.00305585 | -0.00087685 | 0.00042586 | 0.00001772 | -0.00028615 | 0.00004017 | 0.00014162 | -0.00018765 |
| 79 | 100 | 0.05181059 | 0.05532189 | -0.03683124 | -0.03769363 | 0.02477707 | 0.01398042 | -0.01910559 | -0.00742427 | 0.00728311 | 0.00288157 | -0.00094542 | 0.00050044 | -0.00003328 | -0.00033843 | 0.00000419 | 0.00018334 | -0.00019507 |
| 79 | 101 | 0.05137450 | 0.05491960 | -0.03600565 | -0.03722372 | 0.02460030 | 0.01352285 | -0.01883718 | -0.00722177 | 0.00732777 | 0.00292368 | -0.00088080 | 0.00040071 | 0.00003835 | -0.00031176 | 0.00006214 | 0.00013640 | -0.00015305 |
| 79 | 102 | 0.05164582 | 0.05490480 | -0.03682655 | -0.03735622 | 0.02456132 | 0.01303268 | -0.01934386 | -0.00718329 | 0.00748526 | 0.00275237 | -0.00094378 | 0.00046403 | 0.00000004 | -0.00036724 | 0.00003724 | 0.00016547 | -0.00015478 |
| 79 | 103 | 0.05115140 | 0.05448500 | -0.03586920 | -0.03684686 | 0.02439393 | 0.01270249 | -0.01900519 | -0.00698819 | 0.00748572 | 0.00278392 | -0.00088015 | 0.00037185 | 0.00006100 | -0.00033502 | 0.00008773 | 0.00012246 | -0.00012099 |
| 79 | 104 | 0.05149538 | 0.05445017 | -0.03683618 | -0.03693466 | 0.02436506 | 0.01204537 | -0.01949194 | -0.00684800 | 0.00768693 | 0.00260681 | -0.00093669 | 0.00042548 | 0.00003187 | -0.00039261 | 0.00007498 | 0.00013640 | -0.00011921 |
| 79 | 105 | 0.05100346 | 0.05403569 | -0.03633211 | -0.03657998 | 0.02439651 | 0.01206548 | -0.01909047 | -0.00659128 | 0.00761206 | 0.00262629 | -0.00089207 | 0.00034678 | 0.00008137 | -0.00034862 | 0.00011891 | 0.00011060 | -0.00009049 |
| 79 | 106 | 0.05136269 | 0.05394489 | -0.03770061 | -0.03678720 | 0.02449767 | 0.01164255 | -0.01945498 | -0.00628741 | 0.00776177 | 0.00240122 | -0.00096740 | 0.00041926 | 0.00002670 | -0.00038642 | 0.00010172 | 0.00013628 | -0.00009487 |

TABLE 3. Nuclear charge density distribution FB coefficients.

| $Z$ | $N$ | $a_1$ | $a_2$ | $a_3$ | $a_4$ | $a_5$ | $a_6$ | $a_7$ | $a_8$ | $a_9$ | $a_{10}$ | $a_{11}$ | $a_{12}$ | $a_{13}$ | $a_{14}$ | $a_{15}$ | $a_{16}$ | $a_{17}$ |
|---|---|---|---|---|---|---|---|---|---|---|---|---|---|---|---|---|---|---|
| 79 | 107 | 0.05091780 | 0.05356711 | -0.03735517 | -0.03648571 | 0.02457193 | 0.01161617 | -0.01912003 | -0.00606903 | 0.00768455 | 0.00243888 | -0.00092469 | 0.00033657 | 0.00008957 | -0.00034873 | 0.00014869 | 0.00010600 | -0.00006578 |
| 79 | 108 | 0.05121041 | 0.05342605 | -0.03848666 | -0.03664172 | 0.02463227 | 0.01121834 | -0.01940295 | -0.00571402 | 0.00781623 | 0.00217295 | -0.00100372 | 0.00042170 | 0.00001527 | -0.00037477 | 0.00012764 | 0.00013028 | -0.00007807 |
| 79 | 109 | 0.05082979 | 0.05306390 | -0.03833377 | -0.03638549 | 0.02474397 | 0.01115177 | -0.01912026 | -0.00553715 | 0.00772127 | 0.00222564 | -0.00096346 | 0.00033982 | 0.00008419 | -0.00034113 | 0.00017198 | 0.00009955 | -0.00005157 |
| 79 | 110 | 0.05104503 | 0.05289200 | -0.03919053 | -0.03646459 | 0.02480542 | 0.01078537 | -0.01933598 | -0.00513635 | 0.00784473 | 0.00192169 | -0.00104426 | 0.00043339 | -0.00000271 | -0.00035792 | 0.00015232 | 0.00011763 | -0.00007033 |
| 79 | 111 | 0.05074417 | 0.05253204 | -0.03924887 | -0.03624346 | 0.02494595 | 0.01068806 | -0.01909350 | -0.00500349 | 0.00771904 | 0.00198598 | -0.00100699 | 0.00035750 | 0.00006460 | -0.00032646 | 0.00018787 | 0.00009123 | -0.00004933 |
| 79 | 112 | 0.05087513 | 0.05234726 | -0.03980593 | -0.03622102 | 0.02505303 | 0.01036474 | -0.01925091 | -0.00456912 | 0.00783795 | 0.00164908 | -0.00108656 | 0.00045414 | -0.00002653 | -0.00033657 | 0.00017514 | 0.00009765 | -0.00007289 |
| 79 | 113 | 0.05066603 | 0.05198656 | -0.04007415 | -0.03602063 | 0.02521213 | 0.01024771 | -0.01903972 | -0.00448070 | 0.00767177 | 0.00172095 | -0.00105256 | 0.00038920 | 0.00003211 | -0.00030612 | 0.00019604 | 0.00008035 | -0.00006001 |
| 79 | 114 | 0.05070940 | 0.05180224 | -0.04032300 | -0.03587969 | 0.02540245 | 0.00998086 | -0.01914173 | -0.00403078 | 0.00778434 | 0.00135862 | -0.00112728 | 0.00048308 | -0.00005474 | -0.00031179 | 0.00019535 | 0.00006994 | -0.00008640 |
| 79 | 115 | 0.05059908 | 0.05144777 | -0.04077929 | -0.03567924 | 0.02556987 | 0.00985330 | -0.01895636 | -0.00398359 | 0.00757115 | 0.00143292 | -0.00109629 | 0.00043319 | -0.00001039 | -0.00028207 | 0.00019663 | 0.00006586 | -0.00008380 |
| 79 | 116 | 0.05055371 | 0.05127036 | -0.04072831 | -0.03541612 | 0.02586531 | 0.00965479 | -0.01900073 | -0.00353991 | 0.00767211 | 0.00105529 | -0.00116269 | 0.00051903 | -0.00008569 | -0.00028465 | 0.00021194 | 0.00003486 | -0.00011084 |
| 79 | 117 | 0.05054323 | 0.05093497 | -0.04133455 | -0.03518696 | 0.02603162 | 0.00951946 | -0.01883856 | -0.00352593 | 0.00740739 | 0.00112529 | -0.00113353 | 0.00048692 | -0.00005931 | -0.00025628 | 0.00018998 | 0.00004708 | -0.00011987 |
| 79 | 118 | 0.05040869 | 0.05076349 | -0.04100623 | -0.03481368 | 0.02643355 | 0.00939759 | -0.01882040 | -0.00311136 | 0.00749090 | 0.00074484 | -0.00118929 | 0.00056091 | -0.00011825 | -0.00025579 | 0.00022361 | -0.00000615 | -0.00014547 |
| 79 | 119 | 0.05049239 | 0.05045925 | -0.04171593 | -0.03451909 | 0.02658966 | 0.00924656 | -0.01868027 | -0.00311739 | 0.00717020 | 0.00080204 | -0.00115962 | 0.00054765 | -0.00011145 | -0.00023013 | 0.00017616 | 0.00002457 | -0.00016636 |
| 79 | 120 | 0.05026795 | 0.05028589 | -0.04114208 | -0.03406163 | 0.02708063 | 0.00920539 | -0.01859636 | -0.00275308 | 0.00723304 | 0.00043291 | -0.00120452 | 0.00060817 | -0.00015243 | -0.00022503 | 0.00022867 | -0.00005026 | -0.00018904 |
| 79 | 121 | 0.05043360 | 0.05001722 | -0.04190958 | -0.03365809 | 0.02721673 | 0.00901762 | -0.01847643 | -0.00276153 | 0.00684978 | 0.00046729 | -0.00117062 | 0.00061311 | -0.00016502 | -0.00020386 | 0.00015463 | 0.00000069 | -0.00022052 |
| 79 | 122 | 0.05011754 | 0.04982883 | -0.04112680 | -0.03315166 | 0.02776799 | 0.00905771 | -0.01833094 | -0.00246455 | 0.00689406 | 0.00012404 | -0.00120744 | 0.00066105 | -0.00018978 | -0.00019099 | 0.00022508 | -0.00009346 | -0.00024010 |
| 79 | 123 | 0.05034696 | 0.04958742 | -0.04191501 | -0.03259142 | 0.02787236 | 0.00879980 | -0.01822663 | -0.00245533 | 0.00643810 | 0.00012483 | -0.00116428 | 0.00068188 | -0.00022017 | -0.00017622 | 0.00012417 | -0.00002045 | -0.00027918 |
| 79 | 124 | 0.04993689 | 0.04936752 | -0.04096127 | -0.03207498 | 0.02845453 | 0.00891999 | -0.01803582 | -0.00223706 | 0.00647293 | -0.00017919 | -0.00119906 | 0.00072043 | -0.00023322 | -0.00015108 | 0.00021080 | -0.00013114 | -0.00029740 |
| 79 | 125 | 0.05020742 | 0.04913117 | -0.04174604 | -0.03131042 | 0.02851314 | 0.00855047 | -0.01793846 | -0.00219036 | 0.00593064 | -0.00022210 | -0.00114071 | 0.00075341 | -0.00027876 | -0.00014441 | 0.00008330 | -0.00003386 | -0.00033935 |
| 79 | 126 | 0.04970048 | 0.04886239 | -0.04065849 | -0.03082278 | 0.02910564 | 0.00875048 | -0.01773220 | -0.00205577 | 0.00597225 | -0.00047628 | -0.00118238 | 0.00078730 | -0.00028621 | -0.00010183 | 0.00018435 | -0.00015907 | -0.00036004 |
| 79 | 127 | 0.04991800 | 0.04830944 | -0.04158106 | -0.02961676 | 0.02898962 | 0.00754801 | -0.01786916 | -0.00158588 | 0.00568775 | -0.00061067 | -0.00109844 | 0.00076580 | -0.00031091 | -0.00013459 | 0.00008382 | -0.00005273 | -0.00034179 |
| 79 | 128 | 0.04934662 | 0.04762957 | -0.04070475 | -0.02864479 | 0.02961816 | 0.00694246 | -0.01795112 | -0.00103795 | 0.00608519 | -0.00090963 | -0.00112476 | 0.00073719 | -0.00028867 | -0.00012565 | 0.00022758 | -0.00018425 | -0.00030188 |
| 79 | 129 | 0.04954066 | 0.04718655 | -0.04151935 | -0.02770857 | 0.02925124 | 0.00590923 | -0.01792412 | -0.00061562 | 0.00577391 | -0.00000897 | -0.00104418 | 0.00071737 | -0.00031600 | -0.00014921 | 0.00012530 | -0.00007448 | -0.00028362 |
| 79 | 130 | 0.04902320 | 0.04644783 | -0.04073215 | -0.02654458 | 0.02997597 | 0.00525006 | -0.01799051 | -0.00003271 | 0.00617876 | -0.00028486 | -0.00106047 | 0.00068755 | -0.00029655 | -0.00014429 | 0.00025405 | -0.00018800 | -0.00023747 |
| 79 | 131 | 0.04917862 | 0.04610191 | -0.04144018 | -0.02587859 | 0.02939440 | 0.00437891 | -0.01783240 | 0.00033563 | 0.00584613 | -0.00035994 | -0.00098426 | 0.00066989 | -0.00032525 | -0.00015975 | 0.00015382 | -0.00007977 | -0.00022116 |
| 80 | 90 | 0.05231006 | 0.05768181 | -0.03627094 | -0.03972270 | 0.02463804 | 0.01891268 | -0.01621485 | -0.00782303 | 0.00620037 | 0.00337816 | -0.00085802 | 0.00062780 | -0.00020028 | -0.00022082 | -0.00015954 | 0.00024707 | -0.00035647 |
| 80 | 91 | 0.05253975 | 0.05715764 | -0.03663376 | -0.03823928 | 0.02595733 | 0.01763588 | -0.01719889 | -0.00748618 | 0.00667835 | 0.00344987 | -0.00088486 | 0.00059141 | -0.00013891 | -0.00018649 | -0.00006634 | 0.00014669 | -0.00037306 |
| 80 | 92 | 0.05228906 | 0.05734539 | -0.03660319 | -0.03934880 | 0.02502809 | 0.01806166 | -0.01708117 | -0.00778772 | 0.00648029 | 0.00325382 | -0.00090327 | 0.00062644 | -0.00017809 | -0.00025867 | -0.00013499 | 0.00025813 | -0.00032745 |
| 80 | 93 | 0.05245572 | 0.05687680 | -0.03676375 | -0.03796435 | 0.02607925 | 0.01680819 | -0.01788310 | -0.00743680 | 0.00691770 | 0.00332201 | -0.00091583 | 0.00058019 | -0.00011308 | -0.00022432 | -0.00004116 | 0.00015332 | -0.00033682 |
| 80 | 94 | 0.05224039 | 0.05698502 | -0.03686770 | -0.03902756 | 0.02522908 | 0.01721280 | -0.01777451 | -0.00771906 | 0.00674817 | 0.00313989 | -0.00093655 | 0.00061953 | -0.00015895 | -0.00029454 | -0.00011506 | 0.00027233 | -0.00029213 |
| 80 | 95 | 0.05235375 | 0.05657506 | -0.03684872 | -0.03773539 | 0.02606635 | 0.01600231 | -0.01841478 | -0.00735730 | 0.00714645 | 0.00320281 | -0.00093688 | 0.00056302 | -0.00008855 | -0.00026037 | -0.00001903 | 0.00016094 | -0.00029627 |
| 80 | 96 | 0.05217068 | 0.05658603 | -0.03708132 | -0.03870196 | 0.02528977 | 0.01633036 | -0.01830994 | -0.00758361 | 0.00700971 | 0.00302297 | -0.00095881 | 0.00060735 | -0.00014245 | -0.00032815 | -0.00009573 | 0.00028382 | -0.00025333 |
| 80 | 97 | 0.05223985 | 0.05623661 | -0.03690729 | -0.03750112 | 0.02596339 | 0.01518367 | -0.01881574 | -0.00722160 | 0.00736865 | 0.00308200 | -0.00094962 | 0.00054102 | -0.00006558 | -0.00029378 | 0.00000311 | 0.00016478 | -0.00025465 |
| 80 | 98 | 0.05208983 | 0.05614061 | -0.03726266 | -0.03832578 | 0.02526206 | 0.01539177 | -0.01870420 | -0.00735622 | 0.00726914 | 0.00289253 | -0.00097127 | 0.00059029 | -0.00012799 | -0.00035930 | -0.00007382 | 0.00028785 | -0.00021368 |
| 80 | 99 | 0.05212196 | 0.05585258 | -0.03695735 | -0.03722385 | 0.02581099 | 0.01433066 | -0.01910397 | -0.00701072 | 0.00758696 | 0.00295199 | -0.00095536 | 0.00051523 | -0.00004430 | -0.00032394 | 0.00002746 | 0.00016126 | -0.00021463 |
| 80 | 100 | 0.05200890 | 0.05564811 | -0.03742924 | -0.03786888 | 0.02519294 | 0.01438977 | -0.01897348 | -0.00702331 | 0.00752870 | 0.00274269 | -0.00097527 | 0.00056873 | -0.00011487 | -0.00038776 | -0.00004730 | 0.00028144 | -0.00017516 |
| 80 | 101 | 0.05200861 | 0.05542067 | -0.03701444 | -0.03688006 | 0.02564263 | 0.01343402 | -0.01929377 | -0.00671403 | 0.00780267 | 0.00280842 | -0.00095521 | 0.00048656 | -0.00002478 | -0.00035048 | 0.00005528 | 0.00014831 | -0.00017811 |
| 80 | 102 | 0.05193818 | 0.05511462 | -0.03759534 | -0.03731921 | 0.02512012 | 0.01333120 | -0.01913301 | -0.00658379 | 0.00778873 | 0.00257278 | -0.00097223 | 0.00054294 | -0.00010217 | -0.00041335 | -0.00001535 | 0.00026354 | -0.00013891 |
| 80 | 103 | 0.05190784 | 0.05494393 | -0.03709063 | -0.03645922 | 0.02548371 | 0.01249475 | -0.01939698 | -0.00632922 | 0.00801599 | 0.00265007 | -0.00095012 | 0.00045577 | -0.00000693 | -0.00037323 | 0.00008696 | 0.00012527 | -0.00014614 |
| 80 | 104 | 0.05188603 | 0.05455171 | -0.03777073 | -0.03668076 | 0.02507062 | 0.01223384 | -0.01919724 | -0.00604759 | 0.00804811 | 0.00238656 | -0.00096353 | 0.00051307 | -0.00008879 | -0.00043594 | 0.00002191 | 0.00023478 | -0.00010526 |
| 80 | 105 | 0.05182154 | 0.05441238 | -0.03762087 | -0.03612110 | 0.02549922 | 0.01181130 | -0.01937691 | -0.00580774 | 0.00815800 | 0.00246417 | -0.00095934 | 0.00043478 | -0.00000342 | -0.00037896 | 0.00011770 | 0.00010982 | -0.00012079 |
| 80 | 106 | 0.05177200 | 0.05393251 | -0.03864044 | -0.03654660 | 0.02509993 | 0.01174461 | -0.01910258 | -0.00543827 | 0.00809134 | 0.00216765 | -0.00099354 | 0.00051272 | -0.00011566 | -0.00041164 | 0.00004695 | 0.00022793 | -0.00009293 |
| 80 | 107 | 0.05172265 | 0.05384719 | -0.03854045 | -0.03595212 | 0.02563515 | 0.01137921 | -0.01929392 | -0.00521229 | 0.00821429 | 0.00224742 | -0.00099045 | 0.00043119 | -0.00001787 | -0.00036671 | 0.00014312 | 0.00010667 | -0.00010292 |
| 80 | 108 | 0.05163978 | 0.05332047 | -0.03942633 | -0.03640424 | 0.02514890 | 0.01122624 | -0.01901925 | -0.00483319 | 0.00811447 | 0.00192917 | -0.00102599 | 0.00051560 | -0.00014070 | -0.00038541 | 0.00007512 | 0.00021095 | -0.00008585 |
| 80 | 109 | 0.05160825 | 0.05327361 | -0.03938885 | -0.03579768 | 0.02574887 | 0.01091429 | -0.01921553 | -0.00462255 | 0.00824579 | 0.00201232 | -0.00102679 | 0.00043661 | -0.00003869 | -0.00034965 | 0.00016614 | 0.00009983 | -0.00009192 |
| 80 | 110 | 0.05149414 | 0.05272231 | -0.04011416 | -0.03621572 | 0.02525796 | 0.01070124 | -0.01894237 | -0.00424385 | 0.00811147 | 0.00167235 | -0.00105912 | 0.00052122 | -0.00016256 | -0.00035813 | 0.00010625 | 0.00018345 | -0.00008436 |
| 80 | 111 | 0.05148354 | 0.05269877 | -0.04015005 | -0.03562317 | 0.02587789 | 0.01043861 | -0.01913808 | -0.00404991 | 0.00824588 | 0.00175961 | -0.00106656 | 0.00045099 | -0.00006503 | -0.00032850 | 0.00018630 | 0.00008881 | -0.00008860 |
| 80 | 112 | 0.05134086 | 0.05214567 | -0.04069018 | -0.03594540 | 0.02546095 | 0.01019879 | -0.01886001 | -0.00368396 | 0.00807352 | 0.00140017 | -0.00109056 | 0.00052914 | -0.00018022 | -0.00033058 | 0.00013939 | 0.00014599 | -0.00008861 |
| 80 | 113 | 0.05135557 | 0.05213387 | -0.04080555 | -0.03538911 | 0.02606245 | 0.00998281 | -0.01905180 | -0.00350842 | 0.00820487 | 0.00149113 | -0.00110695 | 0.00047357 | -0.00009525 | -0.00030440 | 0.00020302 | 0.00007310 | -0.00009353 |
| 80 | 114 | 0.05118484 | 0.05159763 | -0.04114185 | -0.03556492 | 0.02577805 | 0.00974885 | -0.01875552 | -0.00316760 | 0.00799046 | 0.00111732 | -0.00111758 | 0.00053914 | -0.00019345 | -0.00030321 | 0.00017281 | 0.00010022 | -0.00009851 |
| 80 | 115 | 0.05123110 | 0.05159202 | -0.04133575 | -0.03505623 | 0.02633680 | 0.00957879 | -0.01894272 | -0.00301266 | 0.00811128 | 0.00120969 | -0.00114431 | 0.00050309 | -0.00012737 | -0.00027862 | 0.00021561 | 0.00005249 | -0.00010687 |
| 80 | 116 | 0.05102842 | 0.05108255 | -0.04145875 | -0.03505560 | 0.02621077 | 0.00937514 | -0.01861145 | -0.00270713 | 0.00785229 | 0.00082959 | -0.00113764 | 0.00055151 | -0.00020306 | -0.00027589 | 0.00020411 | 0.00004890 | -0.00011378 |
| 80 | 117 | 0.05111391 | 0.05108457 | -0.04172220 | -0.03459099 | 0.02671988 | 0.00925094 | -0.01879594 | -0.00257522 | 0.00795349 | 0.00091888 | -0.00117474 | 0.00053823 | -0.00015972 | -0.00025216 | 0.00022316 | 0.00002748 | -0.00012824 |
| 80 | 118 | 0.05086977 | 0.05059954 | -0.04163400 | -0.03440800 | 0.02674212 | 0.00908887 | -0.01841462 | -0.00231105 | 0.00765060 | 0.00054311 | -0.00114894 | 0.00056715 | -0.00021103 | -0.00024771 | 0.00023045 | -0.00000423 | -0.00013407 |
| 80 | 119 | 0.05100229 | 0.05061653 | -0.04195052 | -0.03396879 | 0.02720940 | 0.00900823 | -0.01859966 | -0.00220429 | 0.00772124 | 0.00062266 | -0.00119478 | 0.00057801 | -0.00019157 | -0.00022535 | 0.00022443 | -0.00000026 | -0.00015679 |
| 80 | 120 | 0.05070223 | 0.05013977 | -0.04166612 | -0.03361894 | 0.02734106 | 0.00888475 | -0.01816104 | -0.00198242 | 0.00737980 | 0.00026321 | -0.00115097 | 0.00058764 | -0.00022048 | -0.00021688 | 0.00024886 | -0.00005497 | -0.00015922 |
| 80 | 121 | 0.05088738 | 0.05018159 | -0.04201374 | -0.03317416 | 0.02778181 | 0.00883944 | -0.01834939 | -0.00190193 | 0.00740687 | 0.00032476 | -0.00120214 | 0.00062221 | -0.00022369 | -0.00019739 | 0.00021779 | -0.00002779 | -0.00019136 |
| 80 | 122 | 0.05051420 | 0.04968501 | -0.04156080 | -0.03268796 | 0.02797037 | 0.00874057 | -0.01785922 | -0.00171802 | 0.00703787 | -0.00000656 | -0.00114487 | 0.00061495 | -0.00023520 | -0.00018079 | 0.00025677 | -0.00009919 | -0.00018938 |
| 80 | 123 | 0.05075258 | 0.04975873 | -0.04191497 | -0.03219835 | 0.02839882 | 0.00871341 | -0.01805148 | -0.00166340 | 0.00700615 | 0.00002809 | -0.00119633 | 0.00067140 | -0.00025844 | -0.00016618 | 0.00020137 | -0.00005119 | -0.00023082 |
| 80 | 124 | 0.05029006 | 0.04920793 | -0.04133177 | -0.03161481 | 0.02859484 | 0.00862120 | -0.01753033 | -0.00150876 | 0.00662663 | -0.00026498 | -0.00113342 | 0.00065099 | -0.00025913 | -0.00013626 | 0.00025237 | -0.00013354 | -0.00022512 |
| 80 | 125 | 0.05057456 | 0.04931169 | -0.04166902 | -0.03103670 | 0.02901774 | 0.00858507 | -0.01772438 | -0.00147788 | 0.00651895 | -0.00026575 | -0.00117911 | 0.00072670 | -0.00029933 | -0.00012839 | 0.00017353 | -0.00006631 | -0.00027442 |
| 80 | 126 | 0.05001152 | 0.04867493 | -0.04100024 | -0.03039982 | 0.02918738 | 0.00848634 | -0.01720468 | -0.00134138 | 0.00615142 | -0.00051296 | -0.00112074 | 0.00069704 | -0.00029552 | -0.00007988 | 0.00023501 | -0.00015599 | -0.00026719 |
| 80 | 127 | 0.05027239 | 0.04845817 | -0.04151694 | -0.02933981 | 0.02950839 | 0.00753668 | -0.01769449 | -0.00089012 | 0.00630566 | -0.00062665 | -0.00114187 | 0.00073156 | -0.00032147 | -0.00011464 | 0.00018060 | -0.00009037 | -0.00027143 |
| 80 | 128 | 0.04966151 | 0.04743687 | -0.04111734 | -0.02828612 | 0.02971574 | 0.00656479 | -0.01758679 | -0.00043140 | 0.00626445 | -0.00090653 | -0.00107204 | 0.00065269 | -0.00028958 | -0.00009763 | 0.00027696 | -0.00018999 | -0.00021864 |
| 80 | 129 | 0.04991686 | 0.04728546 | -0.04154424 | -0.02731940 | 0.02979561 | 0.00569721 | -0.01784383 | 0.00011259 | 0.00641442 | -0.00101383 | -0.00108659 | 0.00068026 | -0.00032023 | -0.00012908 | 0.00022048 | -0.00011670 | -0.00021423 |
| 80 | 130 | 0.04934868 | 0.04626488 | -0.04120426 | -0.02626767 | 0.03011456 | 0.00480769 | -0.01778714 | 0.00045285 | 0.00635780 | -0.00123615 | -0.00101827 | 0.00060890 | -0.00028907 | -0.00011085 | 0.00030128 | -0.00020004 | -0.00016359 |
| 80 | 131 | 0.04959468 | 0.04618221 | -0.04156030 | -0.02542750 | 0.02994958 | 0.00400914 | -0.01783274 | 0.00106903 | 0.00649908 | -0.00133800 | -0.00102601 | 0.00062892 | -0.00032143 | -0.00013898 | 0.00024482 | -0.00012238 | -0.00015107 |
| 80 | 132 | 0.04907241 | 0.04517504 | -0.04126075 | -0.02438564 | 0.03037084 | 0.00322607 | -0.01781393 | 0.00129014 | 0.00642861 | -0.00149844 | -0.00096002 | 0.00056575 | -0.00029236 | -0.00011975 | 0.00030904 | -0.00018742 | -0.00010240 |
| 80 | 133 | 0.04930915 | 0.04515879 | -0.04156972 | -0.02368397 | 0.02997964 | 0.00247766 | -0.01768071 | 0.00196228 | 0.00655636 | -0.00159955 | -0.00096127 | 0.00057823 | -0.00032417 | -0.00014441 | 0.00025471 | -0.00010921 | -0.00008334 |
| 80 | 134 | 0.04883086 | 0.04417586 | -0.04129044 | -0.02266118 | 0.03048711 | 0.00180858 | -0.01769320 | 0.00206859 | 0.00647698 | -0.00169708 | -0.00089856 | 0.00052322 | -0.00029754 | -0.00012466 | 0.00030299 | -0.00015604 | -0.00003646 |
| 80 | 135 | 0.04906043 | 0.04421739 | -0.04157724 | -0.02209051 | 0.02990487 | 0.00108923 | -0.01741905 | 0.00278491 | 0.00658601 | -0.00180445 | -0.00089399 | 0.00052884 | -0.00032758 | -0.00014552 | 0.00025252 | -0.00008100 | -0.00001318 |
| 80 | 136 | 0.04862110 | 0.04326976 | -0.04129873 | -0.02109952 | 0.03047724 | 0.00053014 | -0.01746134 | 0.00278365 | 0.00650531 | -0.00184031 | -0.00083559 | 0.00048130 | -0.00030274 | -0.00012602 | 0.00028685 | -0.00011125 | 0.00003225 |
| 81 | 95 | 0.05214462 | 0.05697504 | -0.03604767 | -0.03698633 | 0.02692339 | 0.01572600 | -0.01832776 | -0.00689449 | 0.00764919 | 0.00341142 | -0.00089417 | 0.00041697 | 0.00009453 | -0.00025523 | 0.00010962 | 0.00004870 | -0.00022626 |
| 81 | 96 | 0.05230440 | 0.05688990 | -0.03644238 | -0.03671752 | 0.02712265 | 0.01562715 | -0.01868331 | -0.00680901 | 0.00771202 | 0.00319239 | -0.00096429 | 0.00051425 | -0.00000319 | -0.00029398 | 0.00004318 | 0.00013104 | -0.00024145 |
| 81 | 97 | 0.05200490 | 0.05658587 | -0.03601836 | -0.03672587 | 0.02678626 | 0.01502185 | -0.01856486 | -0.00666683 | 0.00782290 | 0.00325719 | -0.00090026 | 0.00040777 | 0.00009093 | -0.00027687 | 0.00012628 | 0.00004938 | -0.00020063 |
| 81 | 98 | 0.05223566 | 0.05649826 | -0.03652476 | -0.03637423 | 0.02708382 | 0.01485161 | -0.01889605 | -0.00651002 | 0.00794847 | 0.00303973 | -0.00096943 | 0.00049593 | 0.00000431 | -0.00032219 | 0.00006500 | 0.00012871 | -0.00020691 |
| 81 | 99 | 0.05188024 | 0.05615747 | -0.03601876 | -0.03644348 | 0.02662070 | 0.01429909 | -0.01873980 | -0.00640576 | 0.00798973 | 0.00310158 | -0.00090417 | 0.00039736 | 0.00008449 | -0.00029530 | 0.00014131 | 0.00004808 | -0.00017730 |
| 81 | 100 | 0.05217960 | 0.05605201 | -0.03663280 | -0.03598152 | 0.02701718 | 0.01403745 | -0.01901233 | -0.00613324 | 0.00818047 | 0.00287631 | -0.00096965 | 0.00047594 | 0.00000790 | -0.00034607 | 0.00008816 | 0.00011948 | -0.00017618 |
| 81 | 101 | 0.05176887 | 0.05568826 | -0.03605152 | -0.03612286 | 0.02644225 | 0.01354399 | -0.01886081 | -0.00610407 | 0.00814942 | 0.00294125 | -0.00090568 | 0.00038544 | 0.00007659 | -0.00031074 | 0.00015646 | 0.00004252 | -0.00015682 |
| 81 | 102 | 0.05213883 | 0.05555522 | -0.03677243 | -0.03552864 | 0.02694484 | 0.01317980 | -0.01904794 | -0.00567903 | 0.00840733 | 0.00270080 | -0.00096566 | 0.00045429 | 0.00000889 | -0.00036575 | 0.00011375 | 0.00010210 | -0.00014986 |
| 81 | 103 | 0.05167026 | 0.05518090 | -0.03611769 | -0.03575497 | 0.02626356 | 0.01275066 | -0.01893286 | -0.00575734 | 0.00830169 | 0.00277453 | -0.00090447 | 0.00037164 | 0.00006859 | -0.00032351 | 0.00017306 | 0.00003112 | -0.00013936 |
| 81 | 104 | 0.05211543 | 0.05501695 | -0.03694595 | -0.03501536 | 0.02688224 | 0.01228250 | -0.01901646 | -0.00515439 | 0.00862788 | 0.00251462 | -0.00095808 | 0.00043085 | 0.00000870 | -0.00038154 | 0.00014222 | 0.00007651 | -0.00012782 |
| 81 | 105 | 0.05161596 | 0.05465360 | -0.03670117 | -0.03539046 | 0.02637076 | 0.01216375 | -0.01888761 | -0.00524778 | 0.00842764 | 0.00259879 | -0.00091428 | 0.00035190 | 0.00006965 | -0.00032967 | 0.00019885 | 0.00002088 | -0.00011786 |
| 81 | 106 | 0.05206178 | 0.05443847 | -0.03787579 | -0.03467338 | 0.02716612 | 0.01193907 | -0.01881847 | -0.00448385 | 0.00870470 | 0.00230063 | -0.00098351 | 0.00041857 | -0.00000500 | -0.00037176 | 0.00016877 | 0.00007591 | -0.00010626 |
| 81 | 107 | 0.05160115 | 0.05410831 | -0.03776680 | -0.03508569 | 0.02671459 | 0.01178596 | -0.01874971 | -0.00460951 | 0.00850813 | 0.00240503 | -0.00094233 | 0.00033724 | 0.00006858 | -0.00032624 | 0.00022546 | 0.00001964 | -0.00009538 |
| 81 | 108 | 0.05199664 | 0.05385616 | -0.03875143 | -0.03437723 | 0.02738392 | 0.01154841 | -0.01863474 | -0.00382584 | 0.00875685 | 0.00207170 | -0.00101416 | 0.00041450 | -0.00002522 | -0.00035803 | 0.00019199 | 0.00007436 | -0.00008937 |
| 81 | 109 | 0.05159355 | 0.05354087 | -0.03880005 | -0.03479354 | 0.02701005 | 0.01137674 | -0.01860574 | -0.00397614 | 0.00855576 | 0.00219223 | -0.00097680 | 0.00033604 | 0.00005297 | -0.00031690 | 0.00024248 | 0.00002293 | -0.00008021 |
| 81 | 110 | 0.05192248 | 0.05327782 | -0.03955801 | -0.03410952 | 0.02755556 | 0.01112136 | -0.01846997 | -0.00319357 | 0.00877891 | 0.00182881 | -0.00104898 | 0.00041900 | -0.00005135 | -0.00034081 | 0.00021145 | 0.00007156 | -0.00007785 |
| 81 | 111 | 0.05159433 | 0.05296793 | -0.03977117 | -0.03449223 | 0.02727543 | 0.01094794 | -0.01845973 | -0.00335708 | 0.00856802 | 0.00196027 | -0.00101625 | 0.00034815 | 0.00002394 | -0.00030290 | 0.00024973 | 0.00003061 | -0.00007243 |
| 81 | 112 | 0.05184392 | 0.05271365 | -0.04027840 | -0.03384310 | 0.02771064 | 0.01067963 | -0.01832151 | -0.00260211 | 0.00876198 | 0.00157354 | -0.00108583 | 0.00043176 | -0.00008201 | -0.00032088 | 0.00022667 | 0.00006689 | -0.00007240 |
| 81 | 113 | 0.05160481 | 0.05241013 | -0.04064485 | -0.03415037 | 0.02753677 | 0.01051812 | -0.01830998 | -0.00276355 | 0.00853833 | 0.00170920 | -0.00105773 | 0.00037209 | -0.00001555 | -0.00028609 | 0.00024769 | 0.00004143 | -0.00007181 |
| 81 | 114 | 0.05176638 | 0.05217475 | -0.04089367 | -0.03354258 | 0.02788377 | 0.01025163 | -0.01817965 | -0.00206658 | 0.00869451 | 0.00130796 | -0.00112160 | 0.00045190 | -0.00011525 | -0.00029921 | 0.00023709 | 0.00005969 | -0.00007352 |
| 81 | 115 | 0.05162512 | 0.05188731 | -0.04138435 | -0.03372816 | 0.02782339 | 0.01010743 | -0.01814952 | -0.00220737 | 0.00845599 | 0.00143915 | -0.00109695 | 0.00040535 | -0.00006132 | -0.00026861 | 0.00023727 | 0.00005355 | -0.00007781 |
| 81 | 116 | 0.05169391 | 0.05167063 | -0.04138436 | -0.03316861 | 0.02810681 | 0.00986576 | -0.01802966 | -0.00159983 | 0.00856387 | 0.00103445 | -0.00115254 | 0.00047827 | -0.00014910 | -0.00027675 | 0.00024200 | 0.00004972 | -0.00008138 |
| 81 | 117 | 0.05165259 | 0.05141304 | -0.04195666 | -0.03318160 | 0.02815983 | 0.00973099 | -0.01796765 | -0.00169940 | 0.00830688 | 0.00115061 | -0.00112868 | 0.00044491 | -0.00010898 | -0.00025240 | 0.00021928 | 0.00006538 | -0.00008948 |
| 81 | 118 | 0.05162654 | 0.05120593 | -0.04173296 | -0.03268312 | 0.02840015 | 0.00954219 | -0.01785517 | -0.00121018 | 0.00835806 | 0.00075554 | -0.00117484 | 0.00050977 | -0.00018218 | -0.00025403 | 0.00024044 | 0.00003764 | -0.00009561 |
| 81 | 119 | 0.05167951 | 0.05098905 | -0.04233808 | -0.03246902 | 0.02855695 | 0.00939216 | -0.01775255 | -0.00124752 | 0.00807513 | 0.00084463 | -0.00114750 | 0.00048806 | -0.00015511 | -0.00023859 | 0.00019397 | 0.00007658 | -0.00010538 |
| 81 | 120 | 0.05155828 | 0.05077647 | -0.04192732 | -0.03205411 | 0.02876651 | 0.00928537 | -0.01764305 | -0.00089990 | 0.00806738 | 0.00047366 | -0.00118530 | 0.00054577 | -0.00021430 | -0.00023082 | 0.00023114 | 0.00002535 | -0.00011541 |
| 81 | 121 | 0.05169198 | 0.05060115 | -0.04251913 | -0.03155722 | 0.02900629 | 0.00907810 | -0.01749483 | -0.00085520 | 0.00774546 | 0.00052314 | -0.00114878 | 0.00053291 | -0.00019825 | -0.00022693 | 0.00016071 | 0.00008879 | -0.00012368 |
| 81 | 122 | 0.05147582 | 0.05036581 | -0.04196440 | -0.03125833 | 0.02918998 | 0.00908011 | -0.01738840 | -0.00066459 | 0.00768553 | 0.00019081 | -0.00118213 | 0.00058634 | -0.00024682 | -0.00020573 | 0.00021252 | 0.00001598 | -0.00013970 |
| 81 | 123 | 0.05166952 | 0.05021798 | -0.04250724 | -0.03042574 | 0.02948080 | 0.00875982 | -0.01719163 | -0.00052078 | 0.00730589 | 0.00018890 | -0.00112969 | 0.00057880 | -0.00023931 | -0.00021549 | 0.00011807 | 0.00010548 | -0.00014245 |
| 81 | 124 | 0.05135865 | 0.04994347 | -0.04185303 | -0.03028142 | 0.02964102 | 0.00889289 | -0.01709851 | -0.00049393 | 0.00721055 | -0.00009168 | -0.00116556 | 0.00063229 | -0.00028254 | -0.00017611 | 0.00018300 | 0.00001331 | -0.00016761 |
| 81 | 125 | 0.05158649 | 0.04979368 | -0.04232674 | -0.02906848 | 0.02994259 | 0.00839811 | -0.01685019 | -0.00023825 | 0.00675047 | -0.00015453 | -0.00109019 | 0.00062613 | -0.00028122 | -0.00020065 | 0.00006437 | 0.00013105 | -0.00016016 |
| 81 | 126 | 0.05118062 | 0.04946656 | -0.04161412 | -0.02911736 | 0.03008566 | 0.00867917 | -0.01679391 | -0.00037339 | 0.00664524 | -0.00037319 | -0.00113826 | 0.00068475 | -0.00032518 | -0.00013812 | 0.00014143 | 0.00002090 | -0.00019880 |
| 81 | 127 | 0.05131911 | 0.04899860 | -0.04218022 | -0.02743755 | 0.03022321 | 0.00727730 | -0.01678233 | 0.00035268 | 0.00647295 | -0.00052932 | -0.00104538 | 0.00063269 | -0.00030815 | -0.00018986 | 0.00005873 | 0.00012376 | -0.00015060 |
| 81 | 128 | 0.05080744 | 0.04830353 | -0.04167970 | -0.02723230 | 0.03027786 | 0.00671220 | -0.01707353 | 0.00059398 | 0.00675482 | -0.00076829 | -0.00109243 | 0.00063823 | -0.00032315 | -0.00014643 | 0.00018985 | -0.00002331 | -0.00015321 |
| 81 | 129 | 0.05093380 | 0.04791337 | -0.04218382 | -0.02579726 | 0.03027008 | 0.00552654 | -0.01693369 | 0.00124994 | 0.00654349 | -0.00089734 | -0.00100590 | 0.00059706 | -0.00031825 | -0.00018518 | 0.00010263 | 0.00008413 | -0.00011296 |
| 81 | 130 | 0.05046238 | 0.04720194 | -0.04174937 | -0.02549420 | 0.03034536 | 0.00488841 | -0.01721925 | 0.00150421 | 0.00683868 | -0.00110256 | -0.00104225 | 0.00059123 | -0.00032217 | -0.00015030 | 0.00022416 | -0.00004831 | -0.00010213 |
| 81 | 131 | 0.05056081 | 0.04686455 | -0.04220089 | -0.02429712 | 0.03020530 | 0.00388951 | -0.01698593 | 0.00208857 | 0.00658937 | -0.00121713 | -0.00096242 | 0.00056208 | -0.00032978 | -0.00017640 | 0.00013543 | 0.00005880 | -0.00007231 |
| 81 | 132 | 0.05014967 | 0.04617293 | -0.04182028 | -0.02391416 | 0.03029564 | 0.00322396 | -0.01723547 | 0.00234344 | 0.00689280 | -0.00137533 | -0.00098811 | 0.00054465 | -0.00032222 | -0.00014992 | 0.00024432 | -0.00005439 | -0.00004647 |

TABLE 3. Nuclear charge density distribution FB coefficients.

| $Z$ | $N$ | $a_1$ | $a_2$ | $a_3$ | $a_4$ | $a_5$ | $a_6$ | $a_7$ | $a_8$ | $a_9$ | $a_{10}$ | $a_{11}$ | $a_{12}$ | $a_{13}$ | $a_{14}$ | $a_{15}$ | $a_{16}$ | $a_{17}$ |
|---|---|---|---|---|---|---|---|---|---|---|---|---|---|---|---|---|---|---|
| 81 | 133 | 0.05020734 | 0.04586371 | -0.04222730 | -0.02293086 | 0.03004173 | 0.00237449 | -0.01694385 | 0.00286222 | 0.00660845 | -0.00148920 | -0.00091522 | 0.00052816 | -0.00034241 | -0.00016422 | 0.00015683 | 0.00004789 | -0.00002901 |
| 81 | 134 | 0.04987011 | 0.04521959 | -0.04188988 | -0.02248492 | 0.03014559 | 0.00171358 | -0.01714012 | 0.00310735 | 0.00691642 | -0.00159153 | -0.00093100 | 0.00049934 | -0.00032315 | -0.00014555 | 0.00025173 | -0.00004402 | 0.00001206 |
| 81 | 135 | 0.04987800 | 0.04491684 | -0.04225849 | -0.02168024 | 0.02979931 | 0.00097793 | -0.01681922 | 0.00357067 | 0.00660055 | -0.00171743 | -0.00086481 | 0.00049562 | -0.00035575 | -0.00014932 | 0.00016747 | 0.00005018 | 0.00001618 |
| 81 | 136 | 0.04962179 | 0.04433981 | -0.04195567 | -0.02118896 | 0.02991651 | 0.00033927 | -0.01695846 | 0.00379796 | 0.00691136 | -0.00175929 | -0.00087214 | 0.00045599 | -0.00032469 | -0.00013754 | 0.00024868 | -0.00002090 | 0.00007150 |
| 81 | 137 | 0.04957514 | 0.04402567 | -0.04229017 | -0.02052184 | 0.02950077 | -0.00031080 | -0.01662746 | 0.00421753 | 0.00656695 | -0.00190759 | -0.00081188 | 0.00046469 | -0.00036937 | -0.00013237 | 0.00016860 | 0.00006360 | 0.00006235 |
| 82 | 96 | 0.05226766 | 0.05752906 | -0.03601746 | -0.03615391 | 0.02799903 | 0.01633149 | -0.01825256 | -0.00627720 | 0.00807144 | 0.00309476 | -0.00101957 | 0.00050469 | 0.00003426 | -0.00037249 | 0.00000341 | 0.00025416 | -0.00013637 |
| 82 | 97 | 0.05221785 | 0.05719682 | -0.03581176 | -0.03545587 | 0.02828543 | 0.01531399 | -0.01866710 | -0.00601402 | 0.00833195 | 0.00315446 | -0.00099309 | 0.00045299 | 0.00009523 | -0.00032470 | 0.00011222 | 0.00008884 | -0.00018020 |
| 82 | 98 | 0.05228181 | 0.05701682 | -0.03623088 | -0.03552794 | 0.02823734 | 0.01556905 | -0.01833307 | -0.00575341 | 0.00837617 | 0.00289901 | -0.00102246 | 0.00050065 | 0.00001644 | -0.00039703 | 0.00002132 | 0.00025434 | -0.00011035 |
| 82 | 99 | 0.05221453 | 0.05673670 | -0.03596028 | -0.03499264 | 0.02835205 | 0.01459302 | -0.01870559 | -0.00554656 | 0.00858142 | 0.00296878 | -0.00099308 | 0.00044200 | 0.00008303 | -0.00034551 | 0.00013170 | 0.00008155 | -0.00015681 |
| 82 | 100 | 0.05231539 | 0.05645756 | -0.03647158 | -0.03486509 | 0.02843543 | 0.01476983 | -0.01830714 | -0.00515245 | 0.00867314 | 0.00269416 | -0.00101995 | 0.00049345 | -0.00000393 | -0.00041731 | 0.00004082 | 0.00024808 | -0.00008635 |
| 82 | 101 | 0.05223310 | 0.05622365 | -0.03615390 | -0.03448659 | 0.02840156 | 0.01383586 | -0.01866543 | -0.00501287 | 0.00882320 | 0.00277413 | -0.00098961 | 0.00042998 | 0.00006692 | -0.00036190 | 0.00015179 | 0.00006865 | -0.00013753 |
| 82 | 102 | 0.05236694 | 0.05586653 | -0.03673875 | -0.03417156 | 0.02860963 | 0.01393444 | -0.01820238 | -0.00449153 | 0.00895945 | 0.00248364 | -0.00101307 | 0.00048173 | -0.00002310 | -0.00043395 | 0.00006333 | 0.00023457 | -0.00006379 |
| 82 | 103 | 0.05227071 | 0.05566766 | -0.03638982 | -0.03393879 | 0.02844428 | 0.01304010 | -0.01856457 | -0.00442306 | 0.00905518 | 0.00257206 | -0.00098306 | 0.00041617 | 0.00004940 | -0.00037442 | 0.00017354 | 0.00004958 | -0.00012181 |
| 82 | 104 | 0.05243440 | 0.05526201 | -0.03702749 | -0.03346088 | 0.02876938 | 0.01307119 | -0.01804243 | -0.00379166 | 0.00923236 | 0.00227325 | -0.00100270 | 0.00046435 | -0.00003778 | -0.00044765 | 0.00008947 | 0.00021412 | -0.00004170 |
| 82 | 105 | 0.05228201 | 0.05509670 | -0.03694555 | -0.03337611 | 0.02869507 | 0.01251114 | -0.01833161 | -0.00373377 | 0.00921799 | 0.00236231 | -0.00099012 | 0.00039779 | 0.00003633 | -0.00037605 | 0.00019919 | 0.00003860 | -0.00010157 |
| 82 | 106 | 0.05237019 | 0.05468631 | -0.03776146 | -0.03302517 | 0.02904366 | 0.01277866 | -0.01773319 | -0.00309211 | 0.00928683 | 0.00205350 | -0.00102624 | 0.00044731 | -0.00005619 | -0.00043045 | 0.00011696 | 0.00021155 | -0.00001866 |
| 82 | 107 | 0.05224972 | 0.05453846 | -0.03775192 | -0.03285926 | 0.02909854 | 0.01224652 | -0.01800789 | -0.00298968 | 0.00930236 | 0.00213873 | -0.00101620 | 0.00038160 | 0.00002297 | -0.00036767 | 0.00022410 | 0.00004158 | -0.00007616 |
| 82 | 108 | 0.05229507 | 0.05412622 | -0.03842767 | -0.03262426 | 0.02927212 | 0.01244345 | -0.01744891 | -0.00241823 | 0.00931407 | 0.00181871 | -0.00105088 | 0.00043302 | -0.00007379 | -0.00041287 | 0.00014569 | 0.00020250 | 0.00000138 |
| 82 | 109 | 0.05221187 | 0.05399184 | -0.03849330 | -0.03238451 | 0.02943072 | 0.01194569 | -0.01769944 | -0.00227218 | 0.00935863 | 0.00190316 | -0.00104599 | 0.00037244 | 0.00000412 | -0.00035742 | 0.00024458 | 0.00004624 | -0.00005314 |
| 82 | 110 | 0.05221169 | 0.05358673 | -0.03901650 | -0.03224232 | 0.02947224 | 0.01207692 | -0.01719310 | -0.00178549 | 0.00930663 | 0.00157105 | -0.00107537 | 0.00042203 | -0.00009018 | -0.00039501 | 0.00017500 | 0.00018678 | 0.00001747 |
| 82 | 111 | 0.05217129 | 0.05346527 | -0.03915672 | -0.03194030 | 0.02970527 | 0.01161940 | -0.01741063 | -0.00159730 | 0.00937842 | 0.00165692 | -0.00107786 | 0.00037056 | -0.00001945 | -0.00034571 | 0.00026016 | 0.00005198 | -0.00003335 |
| 82 | 112 | 0.05212372 | 0.05307126 | -0.03952013 | -0.03185930 | 0.02966413 | 0.01169843 | -0.01696181 | -0.00120984 | 0.00925464 | 0.00131387 | -0.00109789 | 0.00041497 | -0.00010525 | -0.00037675 | 0.00020357 | 0.00016488 | 0.00002862 |
| 82 | 113 | 0.05213207 | 0.05296632 | -0.03972953 | -0.03150545 | 0.02994416 | 0.01128501 | -0.01714010 | -0.00098178 | 0.00935004 | 0.00140167 | -0.00110921 | 0.00037578 | -0.00004649 | -0.00033300 | 0.00027031 | 0.00005789 | -0.00001782 |
| 82 | 114 | 0.05203445 | 0.05258115 | -0.03993195 | -0.03145293 | 0.02986659 | 0.01133151 | -0.01674459 | -0.00070570 | 0.00914728 | 0.00105144 | -0.00111627 | 0.00041253 | -0.00011935 | -0.00035779 | 0.00022951 | 0.00013809 | 0.00003406 |
| 82 | 115 | 0.05209785 | 0.05250004 | -0.04019966 | -0.03105059 | 0.03017330 | 0.01096246 | -0.01688150 | -0.00044110 | 0.00925974 | 0.00113928 | -0.00113677 | 0.00038764 | -0.00007555 | -0.00031969 | 0.00027438 | 0.00006308 | -0.00000752 |
| 82 | 116 | 0.05194502 | 0.05211493 | -0.04024651 | -0.03100099 | 0.03009245 | 0.01099816 | -0.01652727 | -0.00028381 | 0.00897441 | 0.00078844 | -0.00112830 | 0.00041547 | -0.00013330 | -0.00033753 | 0.00025055 | 0.00010862 | 0.00003340 |
| 82 | 117 | 0.05206962 | 0.05206719 | -0.04055639 | -0.03054192 | 0.03041576 | 0.01066832 | -0.01662524 | 0.00001288 | 0.00909361 | 0.00087173 | -0.00115695 | 0.00040563 | -0.00010544 | -0.00030590 | 0.00027154 | 0.00006726 | -0.00000314 |
| 82 | 118 | 0.05185306 | 0.05166725 | -0.04045998 | -0.03048312 | 0.03034574 | 0.01071248 | -0.01629626 | 0.00005047 | 0.00872815 | 0.00052921 | -0.00113221 | 0.00042465 | -0.00014850 | -0.00031503 | 0.00026432 | 0.00007944 | 0.00002672 |
| 82 | 119 | 0.05204347 | 0.05166192 | -0.04079197 | -0.02994610 | 0.03068469 | 0.01040964 | -0.01636127 | 0.00037408 | 0.00883958 | 0.00060092 | -0.00116640 | 0.00042937 | -0.00013570 | -0.00029125 | 0.00026070 | 0.00007120 | -0.00000479 |
| 82 | 120 | 0.05175158 | 0.05122726 | -0.04057167 | -0.02988162 | 0.03062094 | 0.01047553 | -0.01604355 | 0.00029819 | 0.00840404 | 0.00027704 | -0.00112718 | 0.00044103 | -0.00016689 | -0.00028884 | 0.00026863 | 0.00005389 | 0.00001437 |
| 82 | 121 | 0.05200897 | 0.05126951 | -0.04090399 | -0.02923453 | 0.03097920 | 0.01017922 | -0.01608275 | 0.00064353 | 0.00848904 | 0.00032840 | -0.00116267 | 0.00045888 | -0.00016695 | -0.00027463 | 0.00024058 | 0.00007691 | -0.00001205 |
| 82 | 122 | 0.05162889 | 0.05077725 | -0.04058588 | -0.02918154 | 0.03090530 | 0.01027326 | -0.01577104 | 0.00046677 | 0.00800162 | 0.00003372 | -0.00111378 | 0.00046566 | -0.00019085 | -0.00025695 | 0.00026176 | 0.00003529 | -0.00000330 |
| 82 | 123 | 0.05194875 | 0.05086450 | -0.04089799 | -0.02838570 | 0.03128484 | 0.00995447 | -0.01579016 | 0.00082930 | 0.00803796 | 0.00005524 | -0.00114494 | 0.00049466 | -0.00020108 | -0.00025392 | 0.00020978 | 0.00008743 | -0.00002419 |
| 82 | 124 | 0.05146903 | 0.05029227 | -0.04051333 | -0.02837071 | 0.03118295 | 0.01007843 | -0.01549258 | 0.00056885 | 0.00752446 | -0.00020061 | -0.00109427 | 0.00049958 | -0.00022296 | -0.00021670 | 0.00024269 | 0.00002632 | -0.00002627 |
| 82 | 125 | 0.05183938 | 0.05041077 | -0.04078905 | -0.02738610 | 0.03157887 | 0.00970092 | -0.01549436 | 0.00094508 | 0.00748741 | -0.00021808 | -0.00111457 | 0.00053762 | -0.00024100 | -0.00022594 | 0.00016716 | 0.00010608 | -0.00004069 |
| 82 | 126 | 0.05125303 | 0.04974157 | -0.04037119 | -0.02744042 | 0.03143930 | 0.00985659 | -0.01523274 | 0.00062049 | 0.00697963 | -0.00042715 | -0.00107254 | 0.00054357 | -0.00026569 | -0.00016489 | 0.00021128 | 0.00002859 | -0.00005505 |
| 82 | 127 | 0.05157122 | 0.04958897 | -0.04074055 | -0.02594705 | 0.03178859 | 0.00857106 | -0.01551784 | 0.00144287 | 0.00722685 | -0.00055087 | -0.00107633 | 0.00054297 | -0.00026599 | -0.00020909 | 0.00016608 | 0.00009338 | -0.00003368 |
| 82 | 128 | 0.05091388 | 0.04857808 | -0.04061535 | -0.02579418 | 0.03159164 | 0.00780002 | -0.01576291 | 0.00140197 | 0.00706854 | -0.00077922 | -0.00103655 | 0.00051074 | -0.00026645 | -0.00016751 | 0.00025361 | -0.00001738 | -0.00002498 |
| 82 | 129 | 0.05121564 | 0.04849959 | -0.04084428 | -0.02432796 | 0.03183052 | 0.00669311 | -0.01578385 | 0.00230715 | 0.00730450 | -0.00089880 | -0.00103526 | 0.00050454 | -0.00026964 | -0.00020754 | 0.00020646 | 0.00005404 | 0.00000323 |
| 82 | 130 | 0.05060438 | 0.04747995 | -0.04086290 | -0.02431014 | 0.03160919 | 0.00588396 | -0.01617400 | 0.00211985 | 0.00713367 | -0.00106678 | -0.00099800 | 0.00047743 | -0.00026790 | -0.00016573 | 0.00028171 | -0.00004346 | 0.00001062 |
| 82 | 131 | 0.05088722 | 0.04746833 | -0.04097918 | -0.02290015 | 0.03173349 | 0.00494172 | -0.01595589 | 0.00308683 | 0.00734868 | -0.00118663 | -0.00099106 | 0.00046650 | -0.00027359 | -0.00020120 | 0.00023392 | 0.00003195 | 0.00004396 |
| 82 | 132 | 0.05032356 | 0.04645693 | -0.04109953 | -0.02298953 | 0.03149794 | 0.00412817 | -0.01645708 | 0.00276811 | 0.00717303 | -0.00129098 | -0.00095671 | 0.00044418 | -0.00027008 | -0.00016004 | 0.00029596 | -0.00005011 | 0.00005134 |
| 82 | 133 | 0.05058733 | 0.04649988 | -0.04113483 | -0.02164830 | 0.03151214 | 0.00332064 | -0.01603783 | 0.00378079 | 0.00735856 | -0.00141788 | -0.00094405 | 0.00042962 | -0.00027810 | -0.00019049 | 0.00024898 | 0.00002592 | 0.00008747 |
| 82 | 134 | 0.05006897 | 0.04551291 | -0.04131360 | -0.02181903 | 0.03127380 | 0.00253235 | -0.01661842 | 0.00334799 | 0.00718754 | -0.00145789 | -0.00091313 | 0.00041150 | -0.00027280 | -0.00015097 | 0.00029804 | -0.00003989 | 0.00009600 |
| 82 | 135 | 0.05031474 | 0.04559402 | -0.04130049 | -0.02054566 | 0.03118864 | 0.00182113 | -0.01604195 | 0.00439440 | 0.00733608 | -0.00159950 | -0.00089494 | 0.00039456 | -0.00028325 | -0.00017592 | 0.00025313 | 0.00003339 | 0.00013234 |
| 82 | 136 | 0.04983729 | 0.04464773 | -0.04149710 | -0.02077798 | 0.03095833 | 0.00108296 | -0.01667512 | 0.00386504 | 0.00718020 | -0.00157631 | -0.00086811 | 0.00037979 | -0.00027568 | -0.00013909 | 0.00029043 | -0.00001650 | 0.00014302 |
| 82 | 137 | 0.05006678 | 0.04474790 | -0.04146630 | -0.01956178 | 0.03078827 | 0.00042869 | -0.01598437 | 0.00493632 | 0.00728492 | -0.00173991 | -0.00084452 | 0.00036178 | -0.00028897 | -0.00015812 | 0.00024836 | 0.00005118 | 0.00017712 |
| 82 | 138 | 0.04962520 | 0.04385837 | -0.04164571 | -0.01984400 | 0.03057451 | -0.00023982 | -0.01664955 | 0.00432647 | 0.00715491 | -0.00165595 | -0.00082264 | 0.00034940 | -0.00027829 | -0.00012501 | 0.00027583 | 0.00001601 | 0.00019073 |
| 83 | 101 | 0.05244711 | 0.05572291 | -0.03625108 | -0.03468658 | 0.02824375 | 0.01368098 | -0.01872646 | -0.00521158 | 0.00857302 | 0.00288628 | -0.00092427 | 0.00043095 | 0.00001491 | -0.00026730 | 0.00017704 | -0.00003542 | -0.00024692 |
| 83 | 102 | 0.05305653 | 0.05551738 | -0.03721619 | -0.03376130 | 0.02904374 | 0.01333205 | -0.01861835 | -0.00448032 | 0.00885975 | 0.00259538 | -0.00095899 | 0.00047214 | -0.00004402 | -0.00032501 | 0.00014550 | 0.00001033 | -0.00022550 |
| 83 | 103 | 0.05242514 | 0.05518093 | -0.03644686 | -0.03428426 | 0.02818120 | 0.01293131 | -0.01871489 | -0.00478630 | 0.00872151 | 0.00269801 | -0.00092066 | 0.00041886 | 0.00000032 | -0.00027694 | 0.00019273 | -0.00005026 | -0.00023346 |
| 83 | 104 | 0.05310712 | 0.05495434 | -0.03750225 | -0.03321229 | 0.02911632 | 0.01248402 | -0.01849984 | -0.00386768 | 0.00908231 | 0.00238966 | -0.00094954 | 0.00044852 | -0.00004718 | -0.00033785 | 0.00017482 | -0.00002099 | -0.00020749 |
| 83 | 105 | 0.05243225 | 0.05467726 | -0.03704198 | -0.03381551 | 0.02842463 | 0.01240848 | -0.01857419 | -0.00419734 | 0.00885945 | 0.00251212 | -0.00092381 | 0.00039091 | 0.00000668 | -0.00028477 | 0.00022217 | -0.00006450 | -0.00020790 |
| 83 | 106 | 0.05309766 | 0.05445817 | -0.03834935 | -0.03270971 | 0.02955961 | 0.01223588 | -0.01821844 | -0.00313616 | 0.00918882 | 0.00218244 | -0.00096554 | 0.00042045 | -0.00004629 | -0.00033514 | 0.00020675 | -0.00002100 | -0.00017316 |
| 83 | 107 | 0.05246151 | 0.05421050 | -0.03800349 | -0.03332991 | 0.02892500 | 0.01211832 | -0.01832242 | -0.00347335 | 0.00896929 | 0.00231853 | -0.00094013 | 0.00035823 | 0.00002183 | -0.00028850 | 0.00025636 | -0.00006898 | -0.00017267 |
| 83 | 108 | 0.05307200 | 0.05395393 | -0.03915270 | -0.03226510 | 0.02992070 | 0.01193790 | -0.01795239 | -0.00241817 | 0.00927145 | 0.00196088 | -0.00098619 | 0.00040109 | -0.00005335 | -0.00032927 | 0.00023419 | -0.00001927 | -0.00014238 |
| 83 | 109 | 0.05249644 | 0.05371363 | -0.03895002 | -0.03286456 | 0.02936461 | 0.01179141 | -0.01806273 | -0.00275064 | 0.00904736 | 0.00210601 | -0.00096208 | 0.00033916 | 0.00002109 | -0.00028709 | 0.00027990 | -0.00006639 | -0.00014350 |
| 83 | 110 | 0.05303226 | 0.05344590 | -0.03990297 | -0.03186950 | 0.03020958 | 0.01159529 | -0.01770656 | -0.00172735 | 0.00932294 | 0.00172564 | -0.00101059 | 0.00039114 | -0.00006837 | -0.00032023 | 0.00025642 | -0.00001590 | -0.00011596 |
| 83 | 111 | 0.05253817 | 0.05320002 | -0.03985834 | -0.03240685 | 0.02975147 | 0.01143364 | -0.01780058 | -0.00204068 | 0.00908871 | 0.00187430 | -0.00098842 | 0.00033392 | 0.00000503 | -0.00028136 | 0.00029220 | -0.00005650 | -0.00012043 |
| 83 | 112 | 0.05298318 | 0.05294001 | -0.04058914 | -0.03150301 | 0.03044724 | 0.01122358 | -0.01747930 | -0.00107954 | 0.00933210 | 0.00147783 | -0.00103682 | 0.00039074 | -0.00009055 | -0.00030825 | 0.00027269 | -0.00001146 | -0.00009487 |
| 83 | 113 | 0.05258806 | 0.05268693 | -0.04069897 | -0.03193303 | 0.03010286 | 0.01105758 | -0.01753645 | -0.00135712 | 0.00908407 | 0.00162316 | -0.00101651 | 0.00034151 | -0.00002396 | -0.00027268 | 0.00029337 | -0.00004029 | -0.00010341 |
| 83 | 114 | 0.05293087 | 0.05244328 | -0.04119774 | -0.03113457 | 0.03066318 | 0.01084576 | -0.01726291 | -0.00049125 | 0.00928467 | 0.00121894 | -0.00106202 | 0.00039939 | -0.00011835 | -0.00029376 | 0.00028227 | -0.00000685 | -0.00008009 |
| 83 | 115 | 0.05264676 | 0.05219169 | -0.04143881 | -0.03140746 | 0.03044309 | 0.01067865 | -0.01726652 | -0.00071482 | 0.00901978 | 0.00135241 | -0.00104236 | 0.00035985 | -0.00006208 | -0.00026270 | 0.00028405 | -0.00001964 | -0.00009227 |
| 83 | 116 | 0.05288063 | 0.05196224 | -0.04171284 | -0.03072460 | 0.03088936 | 0.01048645 | -0.01704556 | 0.00002253 | 0.00916493 | 0.00095075 | -0.00108268 | 0.00041616 | -0.00014984 | -0.00027723 | 0.00028447 | -0.00000296 | -0.00007237 |
| 83 | 117 | 0.05271218 | 0.05172706 | -0.04204519 | -0.03078542 | 0.03079710 | 0.01030953 | -0.01698389 | -0.00012823 | 0.00887869 | 0.00106214 | -0.00106102 | 0.00038625 | -0.00010491 | -0.00025299 | 0.00026501 | 0.00000336 | -0.00008658 |
| 83 | 118 | 0.05283429 | 0.05150060 | -0.04211733 | -0.03023018 | 0.03115170 | 0.01016463 | -0.01681412 | 0.00045086 | 0.00895793 | 0.00067514 | -0.00109514 | 0.00043996 | -0.00018320 | -0.00025892 | 0.00027857 | -0.00000013 | -0.00007206 |
| 83 | 119 | 0.05277771 | 0.05129616 | -0.04249090 | -0.03001986 | 0.03118149 | 0.00995398 | -0.01668077 | 0.00039088 | 0.00864225 | 0.00075315 | -0.00106732 | 0.00041808 | -0.00014858 | -0.00024439 | 0.00023674 | 0.00002748 | -0.00008560 |
| 83 | 120 | 0.05278768 | 0.05105559 | -0.04239592 | -0.02961124 | 0.03146225 | 0.00988602 | -0.01655841 | 0.00078907 | 0.00865178 | 0.00039402 | -0.00109634 | 0.00046996 | -0.00021747 | -0.00023849 | 0.00026373 | 0.00000235 | -0.00007897 |
| 83 | 121 | 0.05283024 | 0.05088815 | -0.04275989 | -0.02907016 | 0.03159638 | 0.00960207 | -0.01635125 | 0.00083558 | 0.00829351 | 0.00042725 | -0.00105683 | 0.00045338 | -0.00019085 | -0.00023658 | 0.00019904 | 0.00005320 | -0.00008822 |
| 83 | 122 | 0.05272889 | 0.05061444 | -0.04253944 | -0.02883605 | 0.03181494 | 0.00963871 | -0.01627535 | 0.00103993 | 0.00823945 | 0.00010916 | -0.00108459 | 0.00050590 | -0.00025300 | -0.00021466 | 0.00023892 | 0.00000643 | -0.00009255 |
| 83 | 123 | 0.05284965 | 0.05047645 | -0.04285168 | -0.02790981 | 0.03202283 | 0.00922947 | -0.01599478 | 0.00120513 | 0.00782048 | 0.00008754 | -0.00102697 | 0.00049127 | -0.00023177 | -0.00022766 | 0.00015100 | 0.00008288 | -0.00009310 |
| 83 | 124 | 0.05263798 | 0.05015203 | -0.04254900 | -0.02788462 | 0.03218780 | 0.00939367 | -0.01597264 | 0.00121331 | 0.00772009 | -0.00017785 | -0.00106030 | 0.00054820 | -0.00029166 | -0.00018496 | 0.00020313 | 0.00001502 | -0.00011214 |
| 83 | 125 | 0.05280988 | 0.05002050 | -0.04278304 | -0.02653118 | 0.03242791 | 0.00880268 | -0.01561921 | 0.00150461 | 0.00721923 | -0.00026165 | -0.00097793 | 0.00053193 | -0.00027353 | -0.00021414 | 0.00009149 | 0.00012013 | -0.00009908 |
| 83 | 126 | 0.05248855 | 0.04963157 | -0.04243793 | -0.02675002 | 0.03255011 | 0.00911155 | -0.01566969 | 0.00132431 | 0.00709946 | -0.00046561 | -0.00102635 | 0.00059766 | -0.00033645 | -0.00014581 | 0.00015567 | 0.00003127 | -0.00013741 |
| 83 | 127 | 0.05257454 | 0.04922725 | -0.04271139 | -0.02492275 | 0.03268777 | 0.00765341 | -0.01557265 | 0.00206973 | 0.00691176 | -0.00063237 | -0.00093164 | 0.00053808 | -0.00030109 | -0.00020111 | 0.00008203 | 0.00011696 | -0.00008775 |
| 83 | 128 | 0.05214164 | 0.04850922 | -0.04256523 | -0.02496658 | 0.03273313 | 0.00714011 | -0.01602962 | 0.00221775 | 0.00719529 | -0.00084417 | -0.00098550 | 0.00055531 | -0.00033448 | -0.00014924 | 0.00020276 | -0.00001611 | -0.00009761 |
| 83 | 129 | 0.05220686 | 0.04817283 | -0.04276061 | -0.02336469 | 0.03273813 | 0.00590237 | -0.01580453 | 0.00289836 | 0.00697050 | -0.00098678 | -0.00089846 | 0.00050803 | -0.00031284 | -0.00019095 | 0.00012450 | 0.00007407 | -0.00005761 |
| 83 | 130 | 0.05181535 | 0.04744209 | -0.04269819 | -0.02335136 | 0.03276679 | 0.00529127 | -0.01627347 | 0.00303981 | 0.00725766 | -0.00116372 | -0.00094109 | 0.00051277 | -0.00033271 | -0.00014778 | 0.00023679 | -0.00004676 | -0.00005385 |
| 83 | 131 | 0.05184800 | 0.04714711 | -0.04283808 | -0.02197356 | 0.03265349 | 0.00424559 | -0.01595806 | 0.00365462 | 0.00699860 | -0.00129372 | -0.00086220 | 0.00047939 | -0.00032609 | -0.00017606 | 0.00015636 | 0.00004402 | -0.00002585 |
| 83 | 132 | 0.05151356 | 0.04643787 | -0.04283316 | -0.02190687 | 0.03266145 | 0.00358144 | -0.01640026 | 0.00378306 | 0.00728380 | -0.00142486 | -0.00089320 | 0.00047101 | -0.00033164 | -0.00014172 | 0.00025761 | -0.00006077 | -0.00000695 |
| 83 | 133 | 0.05150500 | 0.04615959 | -0.04293913 | -0.02073811 | 0.03244662 | 0.00269073 | -0.01603367 | 0.00433645 | 0.00699448 | -0.00155427 | -0.00082277 | 0.00045246 | -0.00034073 | -0.00015722 | 0.00017723 | 0.00002709 | 0.00000737 |
| 83 | 134 | 0.05123696 | 0.04549848 | -0.04296565 | -0.02061932 | 0.03243684 | 0.00200877 | -0.01642178 | 0.00444815 | 0.00727460 | -0.00163276 | -0.00084252 | 0.00043091 | -0.00033154 | -0.00013143 | 0.00026635 | -0.00006004 | 0.00004164 |
| 83 | 135 | 0.05118223 | 0.04521573 | -0.04305683 | -0.01963552 | 0.03213801 | 0.00123601 | -0.01603788 | 0.00494721 | 0.00695883 | -0.00177239 | -0.00078042 | 0.00042744 | -0.00035648 | -0.00013526 | 0.00018760 | 0.00002236 | 0.00004160 |
| 83 | 136 | 0.05098408 | 0.04462241 | -0.04309093 | -0.01946623 | 0.03211755 | 0.00056029 | -0.01635752 | 0.00504068 | 0.00723348 | -0.00179516 | -0.00079004 | 0.00039312 | -0.00033243 | -0.00011744 | 0.00026496 | -0.00004758 | 0.00009038 |
| 83 | 137 | 0.05088178 | 0.04431816 | -0.04318299 | -0.01863765 | 0.03175233 | -0.00012566 | -0.01598062 | 0.00549352 | 0.00689398 | -0.00195370 | -0.00073553 | 0.00040436 | -0.00037289 | -0.00011100 | 0.00018862 | 0.00002812 | 0.00007626 |
| 83 | 138 | 0.05075233 | 0.04380645 | -0.04320464 | -0.01842284 | 0.03172899 | -0.00078139 | -0.01622945 | 0.00556844 | 0.00716525 | -0.00192061 | -0.00073674 | 0.00035806 | -0.00033409 | -0.00010039 | 0.00025574 | -0.00002670 | 0.00013785 |
| 83 | 139 | 0.05060410 | 0.04346766 | -0.04330945 | -0.01771627 | 0.03131503 | -0.00140329 | -0.01587295 | 0.00598344 | 0.00680325 | -0.00210439 | -0.00068852 | 0.00038316 | -0.00038944 | -0.00008527 | 0.00018177 | 0.00004228 | 0.00011081 |
| 83 | 140 | 0.05053901 | 0.04304681 | -0.04330356 | -0.01746627 | 0.03129465 | -0.00203330 | -0.01605840 | 0.00603953 | 0.00707504 | -0.00201728 | -0.00068349 | 0.00032592 | -0.00033626 | -0.00008097 | 0.00024093 | -0.00000053 | 0.00018292 |
| 83 | 141 | 0.05034871 | 0.04266398 | -0.04342916 | -0.01684663 | 0.03085004 | -0.00260571 | -0.01572569 | 0.00642514 | 0.00669053 | -0.00223053 | -0.00063984 | 0.00036366 | -0.00040560 | -0.00005885 | 0.00016870 | 0.00006263 | 0.00014477 |
| 84 | 102 | 0.05419917 | 0.05503595 | -0.03910990 | -0.03354107 | 0.02973181 | 0.01351931 | -0.01817670 | -0.00390867 | 0.00878020 | 0.00231940 | -0.00092312 | 0.00053460 | -0.00018672 | -0.00036006 | 0.00006303 | 0.00009216 | -0.00024232 |
| 84 | 103 | 0.05392506 | 0.05476883 | -0.03854023 | -0.03312038 | 0.02977980 | 0.01279163 | -0.01857395 | -0.00394609 | 0.00890397 | 0.00242085 | -0.00093359 | 0.00049946 | -0.00013338 | -0.00029344 | 0.00014512 | -0.00004476 | -0.00029299 |
| 84 | 104 | 0.05426211 | 0.05443644 | -0.03945634 | -0.03286821 | 0.02992351 | 0.01253865 | -0.01806566 | -0.00322833 | 0.00903329 | 0.00210556 | -0.00090976 | 0.00050039 | -0.00017216 | -0.00037672 | 0.00010405 | 0.00004840 | -0.00021628 |
| 84 | 105 | 0.05395392 | 0.05427850 | -0.03912249 | -0.03261147 | 0.03006442 | 0.01223026 | -0.01840670 | -0.00328906 | 0.00907191 | 0.00222163 | -0.00093214 | 0.00046517 | -0.00012148 | -0.00030178 | 0.00018141 | -0.00006592 | -0.00026173 |
| 84 | 106 | 0.05421290 | 0.05396577 | -0.04023484 | -0.03252605 | 0.03019130 | 0.01221522 | -0.01790084 | -0.00257690 | 0.00912052 | 0.00190522 | -0.00092409 | 0.00047633 | -0.00017829 | -0.00036620 | 0.00013484 | 0.00005056 | -0.00018065 |
| 84 | 107 | 0.05393993 | 0.05385438 | -0.03993487 | -0.03217673 | 0.03047287 | 0.01195333 | -0.01821069 | -0.00260397 | 0.00918061 | 0.00201976 | -0.00094528 | 0.00043497 | -0.00011328 | -0.00030271 | 0.00021503 | -0.00006483 | -0.00022117 |
| 84 | 108 | 0.05413339 | 0.05348972 | -0.04096439 | -0.03223411 | 0.03041144 | 0.01183501 | -0.01776748 | -0.00193976 | 0.00918688 | 0.00168727 | -0.00094128 | 0.00045825 | -0.00018704 | -0.00035359 | 0.00016595 | 0.00004812 | -0.00014902 |
| 84 | 109 | 0.05390105 | 0.05341109 | -0.04071014 | -0.03180321 | 0.03080102 | 0.01161827 | -0.01803445 | -0.00193099 | 0.00926425 | 0.00180143 | -0.00096290 | 0.00041522 | -0.00011470 | -0.00029956 | 0.00024387 | -0.00006221 | -0.00018546 |
| 84 | 110 | 0.05402766 | 0.05300621 | -0.04163926 | -0.03196713 | 0.03060834 | 0.01141104 | -0.01765946 | -0.00132839 | 0.00922220 | 0.00145259 | -0.00095999 | 0.00044646 | -0.00019818 | -0.00033857 | 0.00019661 | 0.00004065 | -0.00012249 |
| 84 | 111 | 0.05384134 | 0.05294761 | -0.04144313 | -0.03147370 | 0.03106545 | 0.01123513 | -0.01787455 | -0.00128241 | 0.00931215 | 0.00156688 | -0.00098372 | 0.00040667 | -0.00012608 | -0.00029198 | 0.00026677 | -0.00005815 | -0.00015576 |
| 84 | 112 | 0.05390198 | 0.05251440 | -0.04225194 | -0.03169320 | 0.03081020 | 0.01096691 | -0.01756241 | -0.00075617 | 0.00921301 | 0.00120331 | -0.00097812 | 0.00044102 | -0.00021135 | -0.00032085 | 0.00022553 | 0.00002814 | -0.00010213 |
| 84 | 113 | 0.05376767 | 0.05246618 | -0.04212489 | -0.03115880 | 0.03129389 | 0.01082511 | -0.01772092 | -0.00067293 | 0.00930981 | 0.00131695 | -0.00100538 | 0.00040929 | -0.00014674 | -0.00027999 | 0.00028265 | -0.00005320 | -0.00013325 |
| 84 | 114 | 0.05376314 | 0.05201492 | -0.04279194 | -0.03137675 | 0.03104378 | 0.01053229 | -0.01745630 | -0.00023679 | 0.00914410 | 0.00094271 | -0.00099307 | 0.00044179 | -0.00022616 | -0.00030020 | 0.00025101 | 0.00001103 | -0.00008881 |
| 84 | 115 | 0.05368811 | 0.05197250 | -0.04274116 | -0.03081801 | 0.03152102 | 0.01041611 | -0.01755876 | -0.00011816 | 0.00924018 | 0.00105297 | -0.00102465 | 0.00042226 | -0.00017501 | -0.00026394 | 0.00029056 | -0.00004827 | -0.00011891 |
| 84 | 116 | 0.05361662 | 0.05150934 | -0.04324586 | -0.03098221 | 0.03132902 | 0.01013580 | -0.01732003 | 0.00021770 | 0.00900080 | 0.00067478 | -0.00100222 | 0.00044857 | -0.00024242 | -0.00027623 | 0.00027115 | -0.00000968 | -0.00008312 |
| 84 | 117 | 0.05360940 | 0.05147417 | -0.04327253 | -0.03040382 | 0.03178090 | 0.01003547 | -0.01737179 | 0.00036771 | 0.00908581 | 0.00077664 | -0.00103781 | 0.00044420 | -0.00020859 | -0.00024435 | 0.00028981 | -0.00004433 | -0.00011338 |
| 84 | 118 | 0.05346444 | 0.05099836 | -0.04359905 | -0.03047763 | 0.03167428 | 0.00979714 | -0.01713718 | 0.00059909 | 0.00877131 | 0.00040378 | -0.00100358 | 0.00046140 | -0.00026039 | -0.00024824 | 0.00028401 | -0.00003241 | -0.00008544 |
| 84 | 119 | 0.05353388 | 0.05097749 | -0.04369649 | -0.02986839 | 0.03209742 | 0.00970124 | -0.01714651 | 0.00077464 | 0.00883144 | 0.00049000 | -0.00104143 | 0.00047362 | -0.00024526 | -0.00022146 | 0.00027987 | -0.00004187 | -0.00011685 |
| 84 | 120 | 0.05330358 | 0.05047933 | -0.04383855 | -0.02983778 | 0.03207418 | 0.00952066 | -0.01690147 | 0.00090470 | 0.00844888 | 0.00013355 | -0.00099647 | 0.00048067 | -0.00028104 | -0.00021502 | 0.00028789 | -0.00005516 | -0.00009603 |
| 84 | 121 | 0.05345672 | 0.05048274 | -0.04399218 | -0.02917076 | 0.03247645 | 0.00941476 | -0.01687682 | 0.00109892 | 0.00846661 | 0.00019536 | -0.00103315 | 0.00050938 | -0.00028370 | -0.00019484 | 0.00026019 | -0.00004036 | -0.00012909 |
| 84 | 122 | 0.05312536 | 0.04994383 | -0.04395664 | -0.02904631 | 0.03251098 | 0.00929297 | -0.01662005 | 0.00113836 | 0.00803309 | -0.00013296 | -0.00098197 | 0.00050721 | -0.00030607 | -0.00017467 | 0.00028147 | -0.00007564 | -0.00011516 |

TABLE 3. Nuclear charge density distribution FB coefficients.

| $Z$ | $N$ | $a_1$ | $a_2$ | $a_3$ | $a_4$ | $a_5$ | $a_6$ | $a_7$ | $a_8$ | $a_9$ | $a_{10}$ | $a_{11}$ | $a_{12}$ | $a_{13}$ | $a_{14}$ | $a_{15}$ | $a_{16}$ | $a_{17}$ |
|---|---|---|---|---|---|---|---|---|---|---|---|---|---|---|---|---|---|---|
| 84 | 123 | 0.05336438 | 0.04997963 | -0.04414610 | -0.02828302 | 0.03290330 | 0.00915769 | -0.01656768 | 0.00134442 | 0.00798769 | -0.00010477 | -0.00101258 | 0.00055112 | -0.00032411 | -0.00016296 | 0.00023020 | -0.00003805 | -0.00014958 |
| 84 | 124 | 0.05291573 | 0.04937654 | -0.04395393 | -0.02809673 | 0.03295904 | 0.00908599 | -0.01631329 | 0.00130995 | 0.00753009 | -0.00039400 | -0.00096303 | 0.00054203 | -0.00033769 | -0.00012477 | 0.00026410 | -0.00009170 | -0.00014319 |
| 84 | 125 | 0.05323473 | 0.04944507 | -0.04415679 | -0.02719355 | 0.03334704 | 0.00889577 | -0.01623644 | 0.00152204 | 0.00739900 | -0.00040779 | -0.00098177 | 0.00059914 | -0.00036824 | -0.00012305 | 0.00018942 | -0.00003227 | -0.00017783 |
| 84 | 126 | 0.05265667 | 0.04875613 | -0.04384033 | -0.02699283 | 0.03339075 | 0.00886471 | -0.01601087 | 0.00143355 | 0.00695150 | -0.00064887 | -0.00094411 | 0.00058583 | -0.00037812 | -0.00006253 | 0.00023591 | -0.00010164 | -0.00018046 |
| 84 | 127 | 0.05298806 | 0.04858435 | -0.04421468 | -0.02564525 | 0.03372284 | 0.00775014 | -0.01629870 | 0.00203121 | 0.00711876 | -0.00075507 | -0.00094417 | 0.00060452 | -0.00039094 | -0.00010173 | 0.00018916 | -0.00005112 | -0.00017410 |
| 84 | 128 | 0.05237098 | 0.04758536 | -0.04419048 | -0.02521356 | 0.03375959 | 0.00683026 | -0.01661387 | 0.00216970 | 0.00702975 | -0.00100052 | -0.00090497 | 0.00054875 | -0.00037087 | -0.00006876 | 0.00027595 | -0.00014301 | -0.00014161 |
| 84 | 129 | 0.05268033 | 0.04748045 | -0.04439994 | -0.02388639 | 0.03391814 | 0.00583142 | -0.01664856 | 0.00287293 | 0.00719613 | -0.00110431 | -0.00090144 | 0.00056135 | -0.00038758 | -0.00010363 | 0.00022799 | -0.00008759 | -0.00012983 |
| 84 | 130 | 0.05210402 | 0.04647929 | -0.04447773 | -0.02354237 | 0.03397074 | 0.00493671 | -0.01704526 | 0.00286362 | 0.00707641 | -0.00129286 | -0.00086112 | 0.00051075 | -0.00036478 | -0.00007099 | 0.00030149 | -0.00016635 | -0.00009793 |
| 84 | 131 | 0.05239051 | 0.04643602 | -0.04455866 | -0.02227499 | 0.03395388 | 0.00404453 | -0.01685174 | 0.00365058 | 0.00723505 | -0.00139610 | -0.00085406 | 0.00051789 | -0.00038445 | -0.00010088 | 0.00025349 | -0.00010826 | -0.00008208 |
| 84 | 132 | 0.05185489 | 0.04544604 | -0.04469822 | -0.02199643 | 0.03402692 | 0.00320278 | -0.01730025 | 0.00350808 | 0.00709084 | -0.00152621 | -0.00081265 | 0.00047245 | -0.00035986 | -0.00006954 | 0.00031316 | -0.00017238 | -0.00005010 |
| 84 | 133 | 0.05212107 | 0.04545541 | -0.04469273 | -0.02081367 | 0.03384643 | 0.00239848 | -0.01691648 | 0.00435915 | 0.00723504 | -0.00163301 | -0.00080257 | 0.00047518 | -0.00038198 | -0.00009375 | 0.00026626 | -0.00011414 | -0.00003211 |
| 84 | 134 | 0.05162216 | 0.04448943 | -0.04485339 | -0.02058086 | 0.03394133 | 0.00162724 | -0.01739276 | 0.00410122 | 0.00707563 | -0.00170539 | -0.00076047 | 0.00043445 | -0.00035582 | -0.00006477 | 0.00031284 | -0.00016365 | 0.00000057 |
| 84 | 135 | 0.05187213 | 0.04453905 | -0.04480381 | -0.01949230 | 0.03361958 | 0.00088622 | -0.01686359 | 0.00499978 | 0.00719913 | -0.00182125 | -0.00074808 | 0.00043411 | -0.00038032 | -0.00008265 | 0.00026804 | -0.00010773 | 0.00001847 |
| 84 | 136 | 0.05140420 | 0.04361035 | -0.04494838 | -0.01929237 | 0.03373407 | 0.00019498 | -0.01734994 | 0.00464459 | 0.00703574 | -0.00183798 | -0.00070601 | 0.00039724 | -0.00035216 | -0.00005712 | 0.00030329 | -0.00014380 | 0.00005245 |
| 84 | 137 | 0.05164210 | 0.04368519 | -0.04489297 | -0.01829403 | 0.03329998 | -0.00050806 | -0.01672013 | 0.00557714 | 0.00713254 | -0.00196877 | -0.00069192 | 0.00039535 | -0.00037934 | -0.00006814 | 0.00026114 | -0.00009221 | 0.00006808 |
| 84 | 138 | 0.05119941 | 0.04280767 | -0.04499063 | -0.01812291 | 0.03342808 | -0.00111579 | -0.01720533 | 0.00514124 | 0.00697721 | -0.00193261 | -0.00065092 | 0.00036124 | -0.00034827 | -0.00004707 | 0.00028746 | -0.00011679 | 0.00010388 |
| 84 | 139 | 0.05142879 | 0.04289118 | -0.04496108 | -0.01720029 | 0.03291341 | -0.00180276 | -0.01651390 | 0.00609724 | 0.00704134 | -0.00208375 | -0.00063536 | 0.00035928 | -0.00037879 | -0.00005088 | 0.00024801 | -0.00007080 | 0.00011538 |
| 84 | 140 | 0.05100652 | 0.04207889 | -0.04498861 | -0.01706279 | 0.03304620 | -0.00232741 | -0.01699277 | 0.00559442 | 0.00690586 | -0.00199765 | -0.00059673 | 0.00032681 | -0.00034369 | -0.00003509 | 0.00026808 | -0.00008622 | 0.00015337 |
| 84 | 141 | 0.05123011 | 0.04215399 | -0.04500914 | -0.01619387 | 0.03248274 | -0.00301472 | -0.01626974 | 0.00656604 | 0.00693138 | -0.00217364 | -0.00057947 | 0.00032614 | -0.00037838 | -0.00003156 | 0.00023084 | -0.00004633 | 0.00015940 |
| 84 | 142 | 0.05082472 | 0.04142039 | -0.04495110 | -0.01610225 | 0.03260912 | -0.00345911 | -0.01674217 | 0.00600691 | 0.00682657 | -0.00204046 | -0.00054471 | 0.00029432 | -0.00033822 | -0.00002163 | 0.00024729 | -0.00005498 | 0.00019971 |
| 84 | 143 | 0.05104436 | 0.04147046 | -0.04503857 | -0.01526057 | 0.03202682 | -0.00415732 | -0.01600776 | 0.00698882 | 0.00680773 | -0.00224474 | -0.00052503 | 0.00029600 | -0.00037791 | -0.00001082 | 0.00021144 | -0.00002103 | 0.00019950 |
| 85 | 106 | 0.05477029 | 0.05366231 | -0.04079855 | -0.03209450 | 0.03109494 | 0.01200526 | -0.01840110 | -0.00276529 | 0.00910063 | 0.00207047 | -0.00092027 | 0.00047333 | -0.00017536 | -0.00027802 | 0.00018938 | -0.00009590 | -0.00029211 |
| 85 | 107 | 0.05389635 | 0.05352891 | -0.03983949 | -0.03236467 | 0.03060069 | 0.01203915 | -0.01848599 | -0.00311482 | 0.00892597 | 0.00220654 | -0.00091757 | 0.00043216 | -0.00011403 | -0.00022904 | 0.00022275 | -0.00011943 | -0.00029314 |
| 85 | 108 | 0.05473606 | 0.05331999 | -0.04154675 | -0.03171164 | 0.03148922 | 0.01170052 | -0.01830967 | -0.00213542 | 0.00921499 | 0.00187299 | -0.00093226 | 0.00044301 | -0.00015981 | -0.00028313 | 0.00022404 | -0.00009231 | -0.00024538 |
| 85 | 109 | 0.05392325 | 0.05318896 | -0.04070091 | -0.03192832 | 0.03105588 | 0.01168983 | -0.01834308 | -0.00245089 | 0.00902304 | 0.00201224 | -0.00092477 | 0.00039403 | -0.00008514 | -0.00023858 | 0.00026039 | -0.00012528 | -0.00024706 |
| 85 | 110 | 0.05467038 | 0.05294602 | -0.04226229 | -0.03137998 | 0.03181214 | 0.01133398 | -0.01823212 | -0.00151303 | 0.00930061 | 0.00165616 | -0.00094794 | 0.00042431 | -0.00015528 | -0.00028321 | 0.00025353 | -0.00008748 | -0.00020455 |
| 85 | 111 | 0.05394248 | 0.05279237 | -0.04156652 | -0.03151317 | 0.03145347 | 0.01128703 | -0.01819175 | -0.00178243 | 0.00908105 | 0.00179301 | -0.00093563 | 0.00037209 | -0.00007523 | -0.00024157 | 0.00028682 | -0.00012498 | -0.00020927 |
| 85 | 112 | 0.05457937 | 0.05253293 | -0.04294506 | -0.03107467 | 0.03208486 | 0.01092045 | -0.01815518 | -0.00090848 | 0.00934306 | 0.00141990 | -0.00096544 | 0.00041796 | -0.00016255 | -0.00027753 | 0.00027628 | -0.00008147 | -0.00017116 |
| 85 | 113 | 0.05395875 | 0.05234311 | -0.04242192 | -0.03109457 | 0.03181111 | 0.01084444 | -0.01802637 | -0.00112136 | 0.00908778 | 0.00154811 | -0.00094823 | 0.00036701 | -0.00008462 | -0.00023812 | 0.00030052 | -0.00011815 | -0.00018079 |
| 85 | 114 | 0.05447187 | 0.05207862 | -0.04358909 | -0.03075923 | 0.03233811 | 0.01048515 | -0.01806067 | -0.00033522 | 0.00932447 | 0.00116486 | -0.00098195 | 0.00042374 | -0.00018116 | -0.00026586 | 0.00029086 | -0.00007488 | -0.00014662 |
| 85 | 115 | 0.05397748 | 0.05185232 | -0.04324291 | -0.03063475 | 0.03215591 | 0.01038332 | -0.01783778 | -0.00048263 | 0.00902738 | 0.00127743 | -0.00095938 | 0.00037794 | -0.00011144 | -0.00022901 | 0.00030073 | -0.00010558 | -0.00016228 |
| 85 | 116 | 0.05435760 | 0.05158712 | -0.04418067 | -0.03038787 | 0.03260712 | 0.01005818 | -0.01792904 | 0.00019172 | 0.00922537 | 0.00089238 | -0.00099394 | 0.00044052 | -0.00020937 | -0.00024847 | 0.00029621 | -0.00006874 | -0.00013200 |
| 85 | 117 | 0.05400279 | 0.05133585 | -0.04399664 | -0.03008497 | 0.03251865 | 0.00992618 | -0.01761525 | 0.00011756 | 0.00888149 | 0.00098150 | -0.00096489 | 0.00040261 | -0.00015187 | -0.00021555 | 0.00028758 | -0.00008899 | -0.00015388 |
| 85 | 118 | 0.05424421 | 0.05106697 | -0.04469856 | -0.02991078 | 0.03292316 | 0.00966623 | -0.01774375 | 0.00065826 | 0.00902744 | 0.00060445 | -0.00099780 | 0.00046663 | -0.00024463 | -0.00022581 | 0.00029175 | -0.00006420 | -0.00012799 |
| 85 | 119 | 0.05403487 | 0.05080931 | -0.04464592 | -0.02939212 | 0.03292421 | 0.00948922 | -0.01734924 | 0.00066410 | 0.00863140 | 0.00066186 | -0.00096020 | 0.00043789 | -0.00020110 | -0.00019898 | 0.00026174 | -0.00007022 | -0.00015512 |
| 85 | 120 | 0.05413401 | 0.05052716 | -0.04511722 | -0.02928137 | 0.03330419 | 0.00932376 | -0.01749606 | 0.00105424 | 0.00871655 | 0.00030369 | -0.00099075 | 0.00050032 | -0.00028442 | -0.00019803 | 0.00027721 | -0.00006197 | -0.00013482 |
| 85 | 121 | 0.05406709 | 0.05028170 | -0.04515589 | -0.02850850 | 0.03338121 | 0.00907512 | -0.01703444 | 0.00114580 | 0.00826116 | 0.00032142 | -0.00094141 | 0.00048058 | -0.00025459 | -0.00017985 | 0.00022401 | -0.00005028 | -0.00016487 |
| 85 | 122 | 0.05402128 | 0.04997144 | -0.04541245 | -0.02846413 | 0.03374821 | 0.00902686 | -0.01718864 | 0.00137620 | 0.00828571 | -0.00000668 | -0.00097172 | 0.00054040 | -0.00032718 | -0.00016452 | 0.00025247 | -0.00006175 | -0.00015235 |
| 85 | 123 | 0.05408424 | 0.04974960 | -0.04550177 | -0.02740203 | 0.03387587 | 0.00867003 | -0.01667296 | 0.00155758 | 0.00776121 | -0.00003521 | -0.00090649 | 0.00052809 | -0.00030940 | -0.00015747 | 0.00017496 | -0.00002859 | -0.00018148 |
| 85 | 124 | 0.05389082 | 0.04939342 | -0.04556787 | -0.02744048 | 0.03423246 | 0.00875280 | -0.01683706 | 0.00162864 | 0.00773689 | -0.00032305 | -0.00094206 | 0.00058646 | -0.00037294 | -0.00012352 | 0.00021752 | -0.00006204 | -0.00018024 |
| 85 | 125 | 0.05406219 | 0.04919444 | -0.04567480 | -0.02606363 | 0.03437416 | 0.00824682 | -0.01627645 | 0.00190120 | 0.00713163 | -0.00040184 | -0.00085618 | 0.00057891 | -0.00036478 | -0.00012956 | 0.00011494 | -0.00000299 | -0.00020315 |
| 85 | 126 | 0.05371879 | 0.04877442 | -0.04557984 | -0.02621165 | 0.03471987 | 0.00846678 | -0.01646805 | 0.00182314 | 0.00708137 | -0.00064158 | -0.00090567 | 0.00063867 | -0.00042320 | -0.00007222 | 0.00017257 | -0.00006050 | -0.00021805 |
| 85 | 127 | 0.05387015 | 0.04836633 | -0.04577295 | -0.02445896 | 0.03475586 | 0.00710847 | -0.01627624 | 0.00246678 | 0.00681713 | -0.00077756 | -0.00081203 | 0.00058847 | -0.00039600 | -0.00011096 | 0.00010483 | -0.00001056 | -0.00019812 |
| 85 | 128 | 0.05344588 | 0.04763608 | -0.04587476 | -0.02436620 | 0.03511316 | 0.00652098 | -0.01695662 | 0.00265750 | 0.00717117 | -0.00101559 | -0.00086709 | 0.00059936 | -0.00042280 | -0.00007546 | 0.00021383 | -0.00009903 | -0.00017611 |
| 85 | 129 | 0.05355205 | 0.04731692 | -0.04590275 | -0.02283823 | 0.03493237 | 0.00534794 | -0.01659793 | 0.00326777 | 0.00688617 | -0.00112794 | -0.00078050 | 0.00055521 | -0.00040308 | -0.00010420 | 0.00014495 | -0.00004806 | -0.00016258 |
| 85 | 130 | 0.05318378 | 0.04654927 | -0.04611306 | -0.02263205 | 0.03532417 | 0.00468488 | -0.01727365 | 0.00344741 | 0.00721992 | -0.00133594 | -0.00082257 | 0.00055860 | -0.00042229 | -0.00007395 | 0.00024210 | -0.00012355 | -0.00013064 |
| 85 | 131 | 0.05323688 | 0.04629803 | -0.04601316 | -0.02134291 | 0.03496275 | 0.00368830 | -0.01679642 | 0.00401491 | 0.00692111 | -0.00143358 | -0.00074482 | 0.00052313 | -0.00041191 | -0.00009277 | 0.00017383 | -0.00007352 | -0.00012576 |
| 85 | 132 | 0.05293553 | 0.04552094 | -0.04629942 | -0.02102746 | 0.03536335 | 0.00297918 | -0.01741901 | 0.00418299 | 0.00722562 | -0.00160219 | -0.00077221 | 0.00051725 | -0.00042184 | -0.00006801 | 0.00025750 | -0.00013430 | -0.00008249 |
| 85 | 133 | 0.05293261 | 0.04531781 | -0.04611010 | -0.01997474 | 0.03486193 | 0.00214116 | -0.01687542 | 0.00470255 | 0.00691982 | -0.00169475 | -0.00070505 | 0.00049283 | -0.00042261 | -0.00007718 | 0.00019110 | -0.00008662 | -0.00008811 |
| 85 | 134 | 0.05270229 | 0.04455383 | -0.04644040 | -0.01955651 | 0.03525183 | 0.00140599 | -0.01741038 | 0.00486071 | 0.00719054 | -0.00181835 | -0.00071704 | 0.00047628 | -0.00042157 | -0.00005801 | 0.00026141 | -0.00013313 | -0.00003311 |
| 85 | 135 | 0.05264435 | 0.04438087 | -0.04619730 | -0.01872366 | 0.03465134 | 0.00070495 | -0.01684836 | 0.00533040 | 0.00688304 | -0.00191486 | -0.00066164 | 0.00046474 | -0.00043498 | -0.00005807 | 0.00019749 | -0.00008839 | -0.00005034 |
| 85 | 136 | 0.05248363 | 0.04364840 | -0.04654239 | -0.01821315 | 0.03501688 | -0.00004529 | -0.01727696 | 0.00548141 | 0.00712015 | -0.00199105 | -0.00065860 | 0.00043651 | -0.00042139 | -0.00004445 | 0.00025605 | -0.00012288 | 0.00001588 |
| 85 | 137 | 0.05237447 | 0.04348946 | -0.04627608 | -0.01757288 | 0.03435533 | -0.00063023 | -0.01673414 | 0.00590177 | 0.00681376 | -0.00209920 | -0.00061528 | 0.00043897 | -0.00044847 | -0.00003621 | 0.00019451 | -0.00008081 | -0.00001316 |
| 85 | 138 | 0.05227822 | 0.04280403 | -0.04661095 | -0.01698591 | 0.03468711 | -0.00139126 | -0.01705246 | 0.00604816 | 0.00702158 | -0.00212798 | -0.00059861 | 0.00039858 | -0.00042103 | -0.00002796 | 0.00024399 | -0.00010670 | 0.00006300 |
| 85 | 139 | 0.05212336 | 0.04264447 | -0.04634605 | -0.01650371 | 0.03399808 | -0.00187782 | -0.01655328 | 0.00642183 | 0.00671631 | -0.00225380 | -0.00056671 | 0.00041541 | -0.00046233 | -0.00001243 | 0.00018410 | -0.00006626 | 0.00002280 |
| 85 | 140 | 0.05208461 | 0.04201955 | -0.04665074 | -0.01586152 | 0.03428936 | -0.00264914 | -0.01676934 | 0.00656478 | 0.00690215 | -0.00223655 | -0.00053864 | 0.00036292 | -0.00042021 | -0.00000924 | 0.00022766 | -0.00008748 | 0.00010708 |
| 85 | 141 | 0.05189040 | 0.04184600 | -0.04640593 | -0.01549888 | 0.03360140 | -0.00305121 | -0.01632520 | 0.00689628 | 0.00659557 | -0.00238452 | -0.00051663 | 0.00039381 | -0.00047580 | 0.00001246 | 0.00016829 | -0.00004719 | 0.00005706 |
| 85 | 142 | 0.05190147 | 0.04129330 | -0.04666585 | -0.01482739 | 0.03384665 | -0.00383360 | -0.01645539 | 0.00703492 | 0.00676839 | -0.00232327 | -0.00047995 | 0.00032982 | -0.00041872 | 0.00001107 | 0.00020908 | -0.00006754 | 0.00014731 |
| 85 | 143 | 0.05167449 | 0.04109367 | -0.04645457 | -0.01454453 | 0.03318382 | -0.00416179 | -0.01606671 | 0.00733050 | 0.00645650 | -0.00249657 | -0.00046569 | 0.00037389 | -0.00048820 | 0.00003774 | 0.00014895 | -0.00002578 | 0.00008928 |
| 85 | 144 | 0.05172798 | 0.04062323 | -0.04665999 | -0.01387274 | 0.03337763 | -0.00495560 | -0.01613228 | 0.00746173 | 0.00662558 | -0.00239347 | -0.00042348 | 0.00029944 | -0.00041649 | 0.00003236 | 0.00018976 | -0.00004855 | 0.00018324 |
| 86 | 107 | 0.05549584 | 0.05314796 | -0.04235832 | -0.03154988 | 0.03227979 | 0.01185391 | -0.01844837 | -0.00227405 | 0.00918653 | 0.00193163 | -0.00091397 | 0.00047154 | -0.00020172 | -0.00027109 | 0.00019990 | -0.00011194 | -0.00029660 |
| 86 | 108 | 0.05571853 | 0.05297549 | -0.04344002 | -0.03176723 | 0.03215104 | 0.01149614 | -0.01822667 | -0.00161872 | 0.00922913 | 0.00161302 | -0.00088189 | 0.00045629 | -0.00022291 | -0.00033800 | 0.00018702 | -0.00004294 | -0.00021245 |
| 86 | 109 | 0.05543230 | 0.05289290 | -0.04301126 | -0.03117776 | 0.03269062 | 0.01152712 | -0.01845717 | -0.00169536 | 0.00930562 | 0.00173428 | -0.00092510 | 0.00044234 | -0.00017893 | -0.00028073 | 0.00023501 | -0.00010533 | -0.00024391 |
| 86 | 110 | 0.05557885 | 0.05263363 | -0.04402532 | -0.03143746 | 0.03248099 | 0.01108056 | -0.01826366 | -0.00104424 | 0.00931209 | 0.00138728 | -0.00089652 | 0.00044121 | -0.00021922 | -0.00033284 | 0.00021877 | -0.00003943 | -0.00017019 |
| 86 | 111 | 0.05533378 | 0.05259019 | -0.04363752 | -0.03083763 | 0.03303855 | 0.01113643 | -0.01846393 | -0.00111734 | 0.00938834 | 0.00151428 | -0.00093830 | 0.00042549 | -0.00016877 | -0.00028415 | 0.00026432 | -0.00009777 | -0.00019817 |
| 86 | 112 | 0.05540867 | 0.05224931 | -0.04456831 | -0.03108736 | 0.03280853 | 0.01063196 | -0.01828147 | -0.00048172 | 0.00934743 | 0.00114105 | -0.00090989 | 0.00043396 | -0.00022185 | -0.00032243 | 0.00024775 | -0.00004000 | -0.00013517 |
| 86 | 113 | 0.05520811 | 0.05222581 | -0.04424309 | -0.03050121 | 0.03334442 | 0.01069915 | -0.01844698 | -0.00054881 | 0.00941681 | 0.00127152 | -0.00095120 | 0.00042160 | -0.00017244 | -0.00028025 | 0.00028590 | -0.00008936 | -0.00016125 |
| 86 | 114 | 0.05521574 | 0.05181341 | -0.04506811 | -0.03068658 | 0.03315110 | 0.01017546 | -0.01825449 | 0.00005652 | 0.00931637 | 0.00087736 | -0.00091923 | 0.00043415 | -0.00023086 | -0.00030588 | 0.00027220 | -0.00004476 | -0.00010871 |
| 86 | 115 | 0.05506549 | 0.05179369 | -0.04482609 | -0.03013039 | 0.03363725 | 0.01024183 | -0.01838303 | -0.00000258 | 0.00937063 | 0.00100684 | -0.00096064 | 0.00043024 | -0.00018960 | -0.00026857 | 0.00029814 | -0.00008081 | -0.00013475 |
| 86 | 116 | 0.05500741 | 0.05132232 | -0.04551780 | -0.03020357 | 0.03352399 | 0.00973821 | -0.01816054 | 0.00055657 | 0.00920086 | 0.00060034 | -0.00092193 | 0.00044114 | -0.00024581 | -0.00028254 | 0.00029043 | -0.00005367 | -0.00009190 |
| 86 | 117 | 0.05491638 | 0.05129674 | -0.04537433 | -0.02967977 | 0.03394888 | 0.00979432 | -0.01825185 | 0.00050608 | 0.00922918 | 0.00072203 | -0.00096313 | 0.00045002 | -0.00021839 | -0.00024926 | 0.00030004 | -0.00007334 | -0.00011987 |
| 86 | 118 | 0.05478885 | 0.05077739 | -0.04590474 | -0.02960827 | 0.03393724 | 0.00934261 | -0.01798632 | 0.00100544 | 0.00898655 | 0.00031470 | -0.00091627 | 0.00045440 | -0.00026610 | -0.00025176 | 0.00030096 | -0.00006642 | -0.00008567 |
| 86 | 119 | 0.05476858 | 0.05074500 | -0.04586591 | -0.02910218 | 0.03430636 | 0.00938157 | -0.01804115 | 0.00096237 | 0.00897504 | 0.00041983 | -0.00095562 | 0.00047905 | -0.00025594 | -0.00022271 | 0.00029122 | -0.00006835 | -0.00011740 |
| 86 | 120 | 0.05456137 | 0.05018270 | -0.04621287 | -0.02887593 | 0.03439213 | 0.00899988 | -0.01773170 | 0.00139406 | 0.00866566 | 0.00002515 | -0.00090213 | 0.00047374 | -0.00029135 | -0.00021268 | 0.00030272 | -0.00008249 | -0.00009090 |
| 86 | 121 | 0.05462385 | 0.05015095 | -0.04627297 | -0.02835641 | 0.03472291 | 0.00901548 | -0.01775067 | 0.00135560 | 0.00859736 | 0.00010382 | -0.00093652 | 0.00051552 | -0.00029942 | -0.00018897 | 0.00027181 | -0.00006681 | -0.00012771 |
| 86 | 122 | 0.05432118 | 0.04954223 | -0.04642631 | -0.02799121 | 0.03487926 | 0.00870641 | -0.01741111 | 0.00171968 | 0.00823898 | -0.00026417 | -0.00088135 | 0.00049950 | -0.00032178 | -0.00016405 | 0.00029518 | -0.00010101 | -0.00010842 |
| 86 | 123 | 0.05447528 | 0.04952313 | -0.04656839 | -0.02741544 | 0.03519195 | 0.00869011 | -0.01739395 | 0.00168247 | 0.00809492 | -0.00022171 | -0.00090652 | 0.00055828 | -0.00034707 | -0.00014731 | 0.00024225 | -0.00006868 | -0.00015088 |
| 86 | 124 | 0.05405898 | 0.04885708 | -0.04653355 | -0.02695143 | 0.03537910 | 0.00844505 | -0.01705103 | 0.00198678 | 0.00771611 | -0.00055006 | -0.00085765 | 0.00053235 | -0.00035824 | -0.00010423 | 0.00027843 | -0.00012089 | -0.00013889 |
| 86 | 125 | 0.05430641 | 0.04886075 | -0.04673282 | -0.02627230 | 0.03568748 | 0.00838328 | -0.01699685 | 0.00194828 | 0.00747717 | -0.00055207 | -0.00086888 | 0.00060695 | -0.00039866 | -0.00009595 | 0.00020320 | -0.00007275 | -0.00018673 |
| 86 | 126 | 0.05376082 | 0.04812397 | -0.04653030 | -0.02576730 | 0.03586553 | 0.00819097 | -0.01668435 | 0.00220572 | 0.00711365 | -0.00083027 | -0.00083606 | 0.00057281 | -0.00040186 | -0.00003146 | 0.00025316 | -0.00014085 | -0.00018252 |
| 86 | 127 | 0.05409379 | 0.04793458 | -0.04697701 | -0.02469655 | 0.03622000 | 0.00727135 | -0.01709568 | 0.00246160 | 0.00717643 | -0.00090793 | -0.00083202 | 0.00061724 | -0.00042907 | -0.00006664 | 0.00019976 | -0.00009224 | -0.00019016 |
| 86 | 128 | 0.05355185 | 0.04695355 | -0.04702498 | -0.02391706 | 0.03649046 | 0.00628588 | -0.01734705 | 0.00287818 | 0.00716501 | -0.00117546 | -0.00079910 | 0.00054238 | -0.00040007 | -0.00003427 | 0.00028413 | -0.00017021 | -0.00014373 |
| 86 | 129 | 0.05387231 | 0.04680799 | -0.04734551 | -0.02289018 | 0.03664202 | 0.00542359 | -0.01755828 | 0.00324062 | 0.00723412 | -0.00125210 | -0.00079302 | 0.00058326 | -0.00043499 | -0.00006437 | 0.00022889 | -0.00011750 | -0.00014972 |
| 86 | 130 | 0.05335362 | 0.04583474 | -0.04742028 | -0.02214207 | 0.03692172 | 0.00450084 | -0.01780178 | 0.00352575 | 0.00717264 | -0.00146860 | -0.00075458 | 0.00051097 | -0.00040064 | -0.00003292 | 0.00030074 | -0.00018439 | -0.00010167 |
| 86 | 131 | 0.05365848 | 0.04572961 | -0.04763790 | -0.02118424 | 0.03686503 | 0.00368452 | -0.01782362 | 0.00398264 | 0.00724428 | -0.00154535 | -0.00074655 | 0.00054760 | -0.00044086 | -0.00005741 | 0.00024523 | -0.00013051 | -0.00010657 |
| 86 | 132 | 0.05316335 | 0.04477487 | -0.04771594 | -0.02046805 | 0.03715869 | 0.00284993 | -0.01804686 | 0.00414181 | 0.00713827 | -0.00170965 | -0.00070267 | 0.00047874 | -0.00040279 | -0.00002805 | 0.00030415 | -0.00018481 | -0.00005692 |
| 86 | 133 | 0.05345332 | 0.04470466 | -0.04786139 | -0.01959552 | 0.03690403 | 0.00206583 | -0.01790058 | 0.00468056 | 0.00720805 | -0.00178915 | -0.00069317 | 0.00051092 | -0.00044639 | -0.00004629 | 0.00024986 | -0.00013261 | -0.00006171 |
| 86 | 134 | 0.05297910 | 0.04377898 | -0.04791820 | -0.01891062 | 0.03721280 | 0.00133140 | -0.01809964 | 0.00472333 | 0.00706710 | -0.00190227 | -0.00064454 | 0.00044593 | -0.00040557 | -0.00002025 | 0.00029661 | -0.00017423 | -0.00001055 |
| 86 | 135 | 0.05325688 | 0.04373619 | -0.04802499 | -0.01812945 | 0.03678399 | 0.00056447 | -0.01781520 | 0.00533166 | 0.00713066 | -0.00198835 | -0.00063433 | 0.00047396 | -0.00045124 | -0.00003160 | 0.00024483 | -0.00012627 | -0.00001654 |
| 86 | 136 | 0.05279966 | 0.04285084 | -0.04803724 | -0.01747692 | 0.03710454 | -0.00006774 | -0.01799169 | 0.00526956 | 0.00696697 | -0.00205245 | -0.00058207 | 0.00041278 | -0.00040793 | -0.00001009 | 0.00028109 | -0.00015617 | 0.00003612 |
| 86 | 137 | 0.05306862 | 0.04282633 | -0.04813775 | -0.01678335 | 0.03653494 | -0.00083162 | -0.01760337 | 0.00593598 | 0.00702009 | -0.00214954 | -0.00057202 | 0.00043744 | -0.00045502 | -0.00001400 | 0.00023273 | -0.00011448 | 0.00002745 |
| 86 | 138 | 0.05262450 | 0.04199334 | -0.04808513 | -0.01616767 | 0.03685957 | -0.00136599 | -0.01776213 | 0.00578082 | 0.00684690 | -0.00216719 | -0.00051753 | 0.00037953 | -0.00040883 | 0.00000192 | 0.00026073 | -0.00013433 | 0.00008175 |
| 86 | 139 | 0.05288780 | 0.04197673 | -0.04820763 | -0.01554976 | 0.03618747 | -0.00213780 | -0.01730362 | 0.00649477 | 0.00688525 | -0.00227969 | -0.00050831 | 0.00040195 | -0.00045733 | 0.00000582 | 0.00021621 | -0.00010009 | 0.00006898 |
| 86 | 140 | 0.05245366 | 0.04120835 | -0.04807423 | -0.01497963 | 0.03650501 | -0.00258229 | -0.01745118 | 0.00625752 | 0.00671572 | -0.00225337 | -0.00045317 | 0.00034648 | -0.00040748 | 0.00001528 | 0.00023839 | -0.00011195 | 0.00012513 |
| 86 | 141 | 0.05271381 | 0.04118844 | -0.04824152 | -0.01441939 | 0.03576969 | -0.00336840 | -0.01695153 | 0.00700961 | 0.00673460 | -0.00238521 | -0.00044505 | 0.00036797 | -0.00045789 | 0.00002714 | 0.00019752 | -0.00008550 | 0.00010712 |
| 86 | 142 | 0.05228763 | 0.04049646 | -0.04801634 | -0.01390741 | 0.03606660 | -0.00373255 | -0.01709533 | 0.00669970 | 0.00658095 | -0.00231713 | -0.00039097 | 0.00031402 | -0.00040346 | 0.00002959 | 0.00021633 | -0.00009152 | 0.00016529 |
| 86 | 143 | 0.05254627 | 0.04046158 | -0.04824532 | -0.01338278 | 0.03530532 | -0.00453477 | -0.01657660 | 0.00748193 | 0.00657517 | -0.00247152 | -0.00038372 | 0.00033587 | -0.00045664 | 0.00004935 | 0.00017843 | -0.00007238 | 0.00014127 |
| 86 | 144 | 0.05212720 | 0.03985676 | -0.04792232 | -0.01294468 | 0.03556693 | -0.00482829 | -0.01672452 | 0.00710695 | 0.00644808 | -0.00236352 | -0.00033246 | 0.00028261 | -0.00039677 | 0.00004453 | 0.00019608 | -0.00007459 | 0.00020150 |
| 86 | 145 | 0.05238524 | 0.03979520 | -0.04822423 | -0.01243134 | 0.03481328 | -0.00564464 | -0.01620112 | 0.00791297 | 0.00641226 | -0.00254293 | -0.00032534 | 0.00030594 | -0.00045368 | 0.00007192 | 0.00016010 | -0.00006176 | 0.00017115 |
| 87 | 110 | 0.05600139 | 0.05258887 | -0.04420502 | -0.03045267 | 0.03409335 | 0.01143259 | -0.01858662 | -0.00124301 | 0.00944534 | 0.00159188 | -0.00091930 | 0.00043095 | -0.00016808 | -0.00029672 | 0.00024887 | -0.00010807 | -0.00021759 |
| 87 | 111 | 0.05522162 | 0.05240627 | -0.04324574 | -0.03033807 | 0.03365822 | 0.01135020 | -0.01851643 | -0.00151101 | 0.00916004 | 0.00169903 | -0.00091281 | 0.00040398 | -0.00012061 | -0.00023679 | 0.00026649 | -0.00013802 | -0.00024311 |
| 87 | 112 | 0.05587291 | 0.05234502 | -0.04473529 | -0.03005961 | 0.03447181 | 0.01100961 | -0.01865428 | -0.00069907 | 0.00951149 | 0.00136239 | -0.00092825 | 0.00041705 | -0.00015478 | -0.00030237 | 0.00027668 | -0.00009808 | -0.00016863 |
| 87 | 113 | 0.05518963 | 0.05210847 | -0.04396874 | -0.02988594 | 0.03405990 | 0.01085111 | -0.01850364 | -0.00092051 | 0.00918044 | 0.00146440 | -0.00091301 | 0.00038501 | -0.00010313 | -0.00024369 | 0.00029203 | -0.00013549 | -0.00019747 |
| 87 | 114 | 0.05571758 | 0.05201467 | -0.04526187 | -0.02965234 | 0.03480344 | 0.01053599 | -0.01867148 | -0.00015729 | 0.00950957 | 0.00110769 | -0.00093432 | 0.00041623 | -0.00015741 | -0.00029888 | 0.00029570 | -0.00008759 | -0.00012977 |
| 87 | 115 | 0.05514948 | 0.05171354 | -0.04471264 | -0.02940075 | 0.03443146 | 0.01031007 | -0.01843887 | -0.00032860 | 0.00912601 | 0.00119876 | -0.00091045 | 0.00038410 | -0.00010842 | -0.00024139 | 0.00030290 | -0.00012707 | -0.00016372 |
| 87 | 116 | 0.05554561 | 0.05158950 | -0.04578562 | -0.02919534 | 0.03511366 | 0.01003913 | -0.01861288 | 0.00036975 | 0.00941850 | 0.00082934 | -0.00093436 | 0.00042776 | -0.00017547 | -0.00028569 | 0.00030434 | -0.00007741 | -0.00010265 |
| 87 | 117 | 0.05511034 | 0.05122755 | -0.04546040 | -0.02884093 | 0.03480155 | 0.00975603 | -0.01830487 | 0.00024915 | 0.00897683 | 0.00090316 | -0.00090169 | 0.00040019 | -0.00013482 | -0.00023026 | 0.00029790 | -0.00011360 | -0.00014309 |
| 87 | 118 | 0.05536710 | 0.05107274 | -0.04629493 | -0.02864685 | 0.03543201 | 0.00954964 | -0.01846047 | 0.00086638 | 0.00921880 | 0.00053003 | -0.00092531 | 0.00044997 | -0.00020674 | -0.00026291 | 0.00030181 | -0.00006897 | -0.00008849 |
| 87 | 119 | 0.05507899 | 0.05066765 | -0.04618068 | -0.02815895 | 0.03519817 | 0.00921708 | -0.01808991 | 0.00079534 | 0.00871442 | 0.00058010 | -0.00088309 | 0.00043065 | -0.00017824 | -0.00021133 | 0.00027718 | -0.00009690 | -0.00013604 |
| 87 | 120 | 0.05518936 | 0.05047715 | -0.04676666 | -0.02796448 | 0.03578443 | 0.00909327 | -0.01820861 | 0.00131721 | 0.00889618 | 0.00021345 | -0.00090521 | 0.00048084 | -0.00024795 | -0.00023092 | 0.00028812 | -0.00006395 | -0.00008819 |

TABLE 3. Nuclear charge density distribution FB coefficients.

| $Z$ | $N$ | $a_1$ | $a_2$ | $a_3$ | $a_4$ | $a_5$ | $a_6$ | $a_7$ | $a_8$ | $a_9$ | $a_{10}$ | $a_{11}$ | $a_{12}$ | $a_{13}$ | $a_{14}$ | $a_{15}$ | $a_{16}$ | $a_{17}$ |
|---|---|---|---|---|---|---|---|---|---|---|---|---|---|---|---|---|---|---|
| 87 | 121 | 0.05505655 | 0.05005595 | -0.04683296 | -0.02731039 | 0.03563849 | 0.00871182 | -0.01779116 | 0.00129450 | 0.00832538 | 0.00023376 | -0.00085186 | 0.00047197 | -0.00023331 | -0.00018567 | 0.00024196 | -0.00007902 | -0.00014217 |
| 87 | 122 | 0.05501333 | 0.04981969 | -0.04717052 | -0.02711385 | 0.03618416 | 0.00868293 | -0.01786696 | 0.00171134 | 0.00844529 | -0.00011583 | -0.00087408 | 0.00051863 | -0.00029597 | -0.00018978 | 0.00026391 | -0.00006362 | -0.00010241 |
| 87 | 123 | 0.05503505 | 0.04941115 | -0.04737651 | -0.02626473 | 0.03611977 | 0.00824327 | -0.01741680 | 0.00173687 | 0.00780502 | -0.00012994 | -0.00080720 | 0.00052072 | -0.00029503 | -0.00015358 | 0.00019404 | -0.00006125 | -0.00016042 |
| 87 | 124 | 0.05483133 | 0.04911461 | -0.04747651 | -0.02607718 | 0.03662530 | 0.00831467 | -0.01745986 | 0.00204597 | 0.00787222 | -0.00045273 | -0.00083450 | 0.00056235 | -0.00034881 | -0.00013876 | 0.00023028 | -0.00006832 | -0.00013151 |
| 87 | 125 | 0.05499644 | 0.04874092 | -0.04777983 | -0.02501421 | 0.03661751 | 0.00779882 | -0.01698604 | 0.00212061 | 0.00715998 | -0.00050363 | -0.00075112 | 0.00057426 | -0.00036010 | -0.00011411 | 0.00013544 | -0.00004355 | -0.00018927 |
| 87 | 126 | 0.05462626 | 0.04836762 | -0.04766220 | -0.02485856 | 0.03708311 | 0.00797030 | -0.01702170 | 0.00232724 | 0.00719424 | -0.00079197 | -0.00079152 | 0.00061172 | -0.00040610 | -0.00007628 | 0.00018856 | -0.00007714 | -0.00017544 |
| 87 | 127 | 0.05484961 | 0.04786238 | -0.04810092 | -0.02349078 | 0.03711088 | 0.00671503 | -0.01703646 | 0.00267058 | 0.00684245 | -0.00087488 | -0.00070941 | 0.00058973 | -0.00040024 | -0.00008818 | 0.00012232 | -0.00005105 | -0.00019206 |
| 87 | 128 | 0.05445165 | 0.04720937 | -0.04819274 | -0.02302619 | 0.03773989 | 0.00614630 | -0.01764625 | 0.00307727 | 0.00726218 | -0.00115351 | -0.00076018 | 0.00058692 | -0.00041983 | -0.00007197 | 0.00021742 | -0.00010109 | -0.00014168 |
| 87 | 129 | 0.05461655 | 0.04680773 | -0.04840414 | -0.02188403 | 0.03747823 | 0.00503328 | -0.01747440 | 0.00341375 | 0.00690729 | -0.00121476 | -0.00068359 | 0.00056471 | -0.00041557 | -0.00007857 | 0.00015358 | -0.00007790 | -0.00016160 |
| 87 | 130 | 0.05428083 | 0.04608441 | -0.04863348 | -0.02126251 | 0.03817962 | 0.00440413 | -0.01806123 | 0.00380691 | 0.00727820 | -0.00146911 | -0.00072033 | 0.00056020 | -0.00043453 | -0.00006227 | 0.00023352 | -0.00011415 | -0.00010586 |
| 87 | 131 | 0.05438012 | 0.04577476 | -0.04864542 | -0.02035576 | 0.03767131 | 0.00343495 | -0.01774400 | 0.00412864 | 0.00693044 | -0.00151602 | -0.00065156 | 0.00054010 | -0.00043302 | -0.00006394 | 0.00017363 | -0.00009523 | -0.00013076 |
| 87 | 132 | 0.05411346 | 0.04500029 | -0.04898757 | -0.01959338 | 0.03840766 | 0.00276313 | -0.01826262 | 0.00450716 | 0.00724197 | -0.00173794 | -0.00067172 | 0.00053153 | -0.00044912 | -0.00004804 | 0.00023772 | -0.00011746 | -0.00006832 |
| 87 | 133 | 0.05414659 | 0.04477042 | -0.04883591 | -0.01891866 | 0.03770421 | 0.00193384 | -0.01784847 | 0.00480743 | 0.00691007 | -0.00177799 | -0.00061327 | 0.00051630 | -0.00045222 | -0.00004482 | 0.00018246 | -0.00010309 | -0.00009986 |
| 87 | 134 | 0.05394898 | 0.04396272 | -0.04926248 | -0.01803418 | 0.03844265 | 0.00122807 | -0.01826739 | 0.00517288 | 0.00715790 | -0.00196282 | -0.00061532 | 0.00050126 | -0.00046271 | -0.00003013 | 0.00023192 | -0.00011320 | -0.00003001 |
| 87 | 135 | 0.05392020 | 0.04379987 | -0.04898643 | -0.01757569 | 0.03759921 | 0.00053074 | -0.01780546 | 0.00544628 | 0.00684818 | -0.00200305 | -0.00056938 | 0.00049361 | -0.00047257 | -0.00002193 | 0.00018119 | -0.00010273 | -0.00006950 |
| 87 | 136 | 0.05378692 | 0.04297694 | -0.04946755 | -0.01659105 | 0.03831235 | -0.00020704 | -0.01810798 | 0.00580159 | 0.00703416 | -0.00214887 | -0.00055300 | 0.00046989 | -0.00047450 | -0.00000936 | 0.00021868 | -0.00010417 | 0.00000787 |
| 87 | 137 | 0.05370317 | 0.04286745 | -0.04910561 | -0.01632277 | 0.03738282 | -0.00078243 | -0.01764174 | 0.00604431 | 0.00674967 | -0.00219548 | -0.00052099 | 0.00047213 | -0.00049317 | 0.00000387 | 0.00017174 | -0.00009618 | -0.00004030 |
| 87 | 138 | 0.05362696 | 0.04204802 | -0.04961206 | -0.01526329 | 0.03804880 | -0.00155380 | -0.01782485 | 0.00639225 | 0.00688101 | -0.00230218 | -0.00048711 | 0.00043797 | -0.00048378 | 0.00001343 | 0.00020078 | -0.00009322 | 0.00004411 |
| 87 | 139 | 0.05349626 | 0.04197719 | -0.04919934 | -0.01515200 | 0.03708208 | -0.00201787 | -0.01738762 | 0.00660242 | 0.00662124 | -0.00236042 | -0.00046940 | 0.00045175 | -0.00051294 | 0.00003160 | 0.00015641 | -0.00008580 | -0.00001278 |
| 87 | 140 | 0.05346909 | 0.04118054 | -0.04970469 | -0.01404600 | 0.03768409 | -0.00282494 | -0.01745935 | 0.00694433 | 0.00670887 | -0.00242873 | -0.00042002 | 0.00040605 | -0.00049006 | 0.00003746 | 0.00018075 | -0.00008278 | 0.00007774 |
| 87 | 141 | 0.05329945 | 0.04113269 | -0.04927133 | -0.01405436 | 0.03672186 | -0.00318795 | -0.01707237 | 0.00712232 | 0.00647013 | -0.00250289 | -0.00041588 | 0.00043226 | -0.00053089 | 0.00006027 | 0.00013745 | -0.00007381 | 0.00001274 |
| 87 | 142 | 0.05331362 | 0.04037795 | -0.04975310 | -0.01293233 | 0.03724713 | -0.00403133 | -0.01704867 | 0.00745738 | 0.00652701 | -0.00253377 | -0.00035385 | 0.00037464 | -0.00049310 | 0.00006203 | 0.00016064 | -0.00007461 | 0.00010807 |
| 87 | 143 | 0.05311236 | 0.04033690 | -0.04932392 | -0.01302169 | 0.03632341 | -0.00430276 | -0.01672134 | 0.00760581 | 0.00630320 | -0.00262733 | -0.00036154 | 0.00041343 | -0.00054622 | 0.00008902 | 0.00011684 | -0.00006203 | 0.00003608 |
| 87 | 144 | 0.05316109 | 0.03964205 | -0.04976416 | -0.01191479 | 0.03676222 | -0.00518079 | -0.01662292 | 0.00793088 | 0.00634270 | -0.00262157 | -0.00029021 | 0.00034423 | -0.00049295 | 0.00008661 | 0.00014184 | -0.00006974 | 0.00013471 |
| 87 | 145 | 0.05293469 | 0.03959176 | -0.04935884 | -0.01204744 | 0.03590390 | -0.00536944 | -0.01635480 | 0.00805452 | 0.00612645 | -0.00273731 | -0.00030732 | 0.00039509 | -0.00055846 | 0.00011714 | 0.00009609 | -0.00005181 | 0.00005719 |
| 87 | 146 | 0.05301226 | 0.03897281 | -0.04974411 | -0.01098619 | 0.03624849 | -0.00627806 | -0.01620442 | 0.00836453 | 0.00616108 | -0.00269552 | -0.00023022 | 0.00031525 | -0.00048996 | 0.00011076 | 0.00012515 | -0.00006854 | 0.00015751 |
| 88 | 113 | 0.05630181 | 0.05216768 | -0.04559711 | -0.02907179 | 0.03593940 | 0.01082159 | -0.01880079 | -0.00025041 | 0.00962653 | 0.00119057 | -0.00091224 | 0.00040206 | -0.00012286 | -0.00033154 | 0.00028931 | -0.00009475 | -0.00012492 |
| 88 | 114 | 0.05632431 | 0.05184619 | -0.04644064 | -0.02906800 | 0.03580579 | 0.01004610 | -0.01867628 | 0.00049471 | 0.00956374 | 0.00078687 | -0.00086097 | 0.00038427 | -0.00014837 | -0.00035784 | 0.00030125 | -0.00008009 | -0.00006253 |
| 88 | 115 | 0.05612588 | 0.05184721 | -0.04606437 | -0.02857698 | 0.03626150 | 0.01028546 | -0.01882977 | 0.00027175 | 0.00957397 | 0.00091969 | -0.00090919 | 0.00040513 | -0.00012817 | -0.00032615 | 0.00030418 | -0.00008292 | -0.00008669 |
| 88 | 116 | 0.05608875 | 0.05139282 | -0.04685446 | -0.02845740 | 0.03617702 | 0.00951007 | -0.01863193 | 0.00102522 | 0.00944430 | 0.00049697 | -0.00085502 | 0.00039176 | -0.00016493 | -0.00033443 | 0.00031593 | -0.00007912 | -0.00003786 |
| 88 | 117 | 0.05593241 | 0.05140962 | -0.04654460 | -0.02802487 | 0.03654778 | 0.00972588 | -0.01875574 | 0.00078537 | 0.00942066 | 0.00062563 | -0.00089791 | 0.00041952 | -0.00014985 | -0.00030942 | 0.00030800 | -0.00007166 | -0.00006072 |
| 88 | 118 | 0.05583255 | 0.05084177 | -0.04724839 | -0.02776988 | 0.03653670 | 0.00899372 | -0.01848244 | 0.00152528 | 0.00922352 | 0.00019791 | -0.00084081 | 0.00040496 | -0.00019015 | -0.00030041 | 0.00032268 | -0.00008179 | -0.00002420 |
| 88 | 119 | 0.05573067 | 0.05085982 | -0.04702622 | -0.02737951 | 0.03682810 | 0.00917665 | -0.01856492 | 0.00127467 | 0.00915053 | 0.00031216 | -0.00087631 | 0.00044336 | -0.00018523 | -0.00028150 | 0.00030033 | -0.00006260 | -0.00004825 |
| 88 | 120 | 0.05555994 | 0.05019911 | -0.04760966 | -0.02698367 | 0.03690174 | 0.00852174 | -0.01822913 | 0.00198220 | 0.00889488 | -0.00010507 | -0.00081878 | 0.00042314 | -0.00022251 | -0.00025538 | 0.00032108 | -0.00008870 | -0.00002276 |
| 88 | 121 | 0.05552734 | 0.05021375 | -0.04748547 | -0.02660527 | 0.03713081 | 0.00866660 | -0.01825991 | 0.00172395 | 0.00875433 | -0.00001609 | -0.00084386 | 0.00047467 | -0.00023074 | -0.00024279 | 0.00028166 | -0.00005773 | -0.00005044 |
| 88 | 122 | 0.05527249 | 0.04947678 | -0.04792092 | -0.02608299 | 0.03728366 | 0.00810787 | -0.01788913 | 0.00238748 | 0.00846024 | -0.00040748 | -0.00079126 | 0.00044610 | -0.00026073 | -0.00019889 | 0.00031136 | -0.00010053 | -0.00003496 |
| 88 | 123 | 0.05532321 | 0.04949282 | -0.04789125 | -0.02567576 | 0.03747265 | 0.00821146 | -0.01786105 | 0.00212232 | 0.00823329 | -0.00035401 | -0.00080235 | 0.00051196 | -0.00028310 | -0.00019338 | 0.00025326 | -0.00005881 | -0.00006838 |
| 88 | 124 | 0.05496771 | 0.04868934 | -0.04816338 | -0.02506296 | 0.03768330 | 0.00775232 | -0.01749212 | 0.00273912 | 0.00793052 | -0.00070585 | -0.00076226 | 0.00047420 | -0.00030417 | -0.00013036 | 0.00029439 | -0.00011776 | -0.00006208 |
| 88 | 125 | 0.05511043 | 0.04871661 | -0.04821197 | -0.02458243 | 0.03785115 | 0.00781010 | -0.01740294 | 0.00246733 | 0.00760060 | -0.00069633 | -0.00075612 | 0.00055459 | -0.00034036 | -0.00013267 | 0.00021684 | -0.00006658 | -0.00010288 |
| 88 | 126 | 0.05463834 | 0.04785026 | -0.04832044 | -0.02393202 | 0.03808907 | 0.00744369 | -0.01707385 | 0.00304191 | 0.00732412 | -0.00099769 | -0.00073690 | 0.00050800 | -0.00035286 | -0.00004911 | 0.00027144 | -0.00014029 | -0.00010492 |
| 88 | 127 | 0.05493209 | 0.04772418 | -0.04867238 | -0.02306992 | 0.03846720 | 0.00674181 | -0.01753332 | 0.00297668 | 0.00728298 | -0.00104886 | -0.00072091 | 0.00057018 | -0.00038138 | -0.00009346 | 0.00020882 | -0.00008273 | -0.00011310 |
| 88 | 128 | 0.05450571 | 0.04666875 | -0.04899329 | -0.02209871 | 0.03889057 | 0.00566484 | -0.01778566 | 0.00365328 | 0.00733374 | -0.00132961 | -0.00070623 | 0.00049041 | -0.00036382 | -0.00004084 | 0.00029112 | -0.00015922 | -0.00007565 |
| 88 | 129 | 0.05480631 | 0.04656149 | -0.04928391 | -0.02128809 | 0.03913799 | 0.00502059 | -0.01811120 | 0.00368059 | 0.00730784 | -0.00138084 | -0.00068965 | 0.00055371 | -0.00040486 | -0.00008018 | 0.00022438 | -0.00009502 | -0.00008576 |
| 88 | 130 | 0.05437792 | 0.04551727 | -0.04956305 | -0.02032979 | 0.03947712 | 0.00398310 | -0.01827550 | 0.00424927 | 0.00728912 | -0.00161677 | -0.00066563 | 0.00047229 | -0.00037869 | -0.00002803 | 0.00029691 | -0.00016557 | -0.00004479 |
| 88 | 131 | 0.05468100 | 0.04542310 | -0.04980169 | -0.01957596 | 0.03958197 | 0.00337911 | -0.01846366 | 0.00436855 | 0.00727520 | -0.00167015 | -0.00064847 | 0.00053524 | -0.00042950 | -0.00006156 | 0.00022782 | -0.00009841 | -0.00005732 |
| 88 | 132 | 0.05425108 | 0.04440462 | -0.05002656 | -0.01865094 | 0.03984839 | 0.00240993 | -0.01854241 | 0.00482478 | 0.00719415 | -0.00185966 | -0.00061507 | 0.00045328 | -0.00039587 | -0.00001201 | 0.00029045 | -0.00016128 | -0.00001262 |
| 88 | 133 | 0.05455438 | 0.04431716 | -0.05022799 | -0.01795597 | 0.03980824 | 0.00182881 | -0.01859490 | 0.00503419 | 0.00718806 | -0.00191780 | -0.00059752 | 0.00051445 | -0.00045368 | -0.00003882 | 0.00022083 | -0.00009494 | -0.00002806 |
| 88 | 134 | 0.05412285 | 0.04333905 | -0.05038650 | -0.01707921 | 0.04001577 | 0.00094556 | -0.01860239 | 0.00537711 | 0.00705623 | -0.00206138 | -0.00055563 | 0.00043311 | -0.00041374 | 0.00000610 | 0.00027415 | -0.00014916 | 0.00002025 |
| 88 | 135 | 0.05442520 | 0.04325191 | -0.05056874 | -0.01644171 | 0.03983821 | 0.00037013 | -0.01852841 | 0.00567348 | 0.00705401 | -0.00212735 | -0.00053818 | 0.00049137 | -0.00047601 | -0.00001316 | 0.00020590 | -0.00008726 | 0.00000127 |
| 88 | 136 | 0.05399207 | 0.04232858 | -0.05064942 | -0.01562365 | 0.03999990 | -0.00041821 | -0.01848471 | 0.00590510 | 0.00688539 | -0.00222662 | -0.00048927 | 0.00041159 | -0.00043068 | 0.00002536 | 0.00025097 | -0.00013248 | 0.00005306 |
| 88 | 137 | 0.05429296 | 0.04223598 | -0.05083151 | -0.01503908 | 0.03970130 | -0.00100372 | -0.01830098 | 0.00628369 | 0.00688381 | -0.00230372 | -0.00047267 | 0.00046628 | -0.00049530 | 0.00001436 | 0.00018584 | -0.00007816 | 0.00002978 |
| 88 | 138 | 0.05385860 | 0.04138102 | -0.05082402 | -0.01428695 | 0.03982734 | -0.00169432 | -0.01822692 | 0.00640819 | 0.00669309 | -0.00236074 | -0.00041854 | 0.00038859 | -0.00044510 | 0.00004502 | 0.00022404 | -0.00011453 | 0.00008500 |
| 88 | 139 | 0.05415781 | 0.04127785 | -0.05102439 | -0.01374809 | 0.03943049 | -0.00230271 | -0.01795549 | 0.00686266 | 0.00668944 | -0.00245216 | -0.00040363 | 0.00043958 | -0.00051059 | 0.00004274 | 0.00016343 | -0.00007015 | 0.00005663 |
| 88 | 140 | 0.05372302 | 0.04050330 | -0.05091995 | -0.01306742 | 0.03952695 | -0.00289679 | -0.01786932 | 0.00688577 | 0.00649073 | -0.00246897 | -0.00034615 | 0.00036412 | -0.00045570 | 0.00006452 | 0.00019627 | -0.00009821 | 0.00011533 |
| 88 | 141 | 0.05402044 | 0.04038498 | -0.05115540 | -0.01256511 | 0.03905821 | -0.00353660 | -0.01753436 | 0.00740826 | 0.00648240 | -0.00257747 | -0.00033368 | 0.00041176 | -0.00052131 | 0.00007118 | 0.00014100 | -0.00006513 | 0.00008115 |
| 88 | 142 | 0.05358635 | 0.03970082 | -0.05094726 | -0.01196084 | 0.03912689 | -0.00403813 | -0.01745037 | 0.00733675 | 0.00628841 | -0.00255595 | -0.00027467 | 0.00033834 | -0.00046160 | 0.00008345 | 0.00017002 | -0.00008571 | 0.00014344 |
| 88 | 143 | 0.05388193 | 0.03956288 | -0.05123223 | -0.01148460 | 0.03861343 | -0.00471314 | -0.01707499 | 0.00791837 | 0.00627233 | -0.00268372 | -0.00026507 | 0.00038338 | -0.00052725 | 0.00009904 | 0.00012030 | -0.00006427 | 0.00010291 |
| 88 | 144 | 0.05344984 | 0.03897692 | -0.05091596 | -0.01096199 | 0.03865223 | -0.00512794 | -0.01700357 | 0.00775951 | 0.00609405 | -0.00262548 | -0.00020628 | 0.00031158 | -0.00046243 | 0.00010159 | 0.00014698 | -0.00007843 | 0.00016888 |
| 88 | 145 | 0.05374350 | 0.03881452 | -0.05126226 | -0.01050028 | 0.03812021 | -0.00583738 | -0.01660733 | 0.00839106 | 0.00606648 | -0.00277413 | -0.00019955 | 0.00035501 | -0.00052863 | 0.00012589 | 0.00010239 | -0.00006798 | 0.00012168 |
| 88 | 146 | 0.05331489 | 0.03833249 | -0.05083587 | -0.01006536 | 0.03812394 | -0.00617257 | -0.01655587 | 0.00815207 | 0.00591302 | -0.00268057 | -0.00014261 | 0.00028433 | -0.00045835 | 0.00011883 | 0.00012810 | -0.00007694 | 0.00019129 |
| 88 | 147 | 0.05360646 | 0.03814030 | -0.05125254 | -0.00960587 | 0.03759728 | -0.00691201 | -0.01615332 | 0.00882478 | 0.00586955 | -0.00285122 | -0.00013830 | 0.00032721 | -0.00052596 | 0.00015143 | 0.00008773 | -0.00007614 | 0.00013737 |
| 89 | 116 | 0.05646230 | 0.05166137 | -0.04677065 | -0.02740166 | 0.03747389 | 0.00978574 | -0.01896456 | 0.00071834 | 0.00955877 | 0.00070497 | -0.00087256 | 0.00039355 | -0.00009845 | -0.00035233 | 0.00030893 | -0.00008087 | -0.00004387 |
| 89 | 117 | 0.05598960 | 0.05124906 | -0.04632877 | -0.02694521 | 0.03724470 | 0.00955671 | -0.01872640 | 0.00051957 | 0.00906841 | 0.00076050 | -0.00084351 | 0.00038637 | -0.00007914 | -0.00028537 | 0.00029382 | -0.00010661 | -0.00009214 |
| 89 | 118 | 0.05625209 | 0.05119732 | -0.04721466 | -0.02674905 | 0.03770842 | 0.00916031 | -0.01884969 | 0.00122684 | 0.00933333 | 0.00039555 | -0.00084916 | 0.00040989 | -0.00012491 | -0.00033056 | 0.00030723 | -0.00006892 | -0.00001998 |
| 89 | 119 | 0.05590163 | 0.05073892 | -0.04699685 | -0.02623490 | 0.03754811 | 0.00889296 | -0.01855536 | 0.00104886 | 0.00878648 | 0.00044059 | -0.00080900 | 0.00040510 | -0.00010893 | -0.00026842 | 0.00027961 | -0.00008998 | -0.00006999 |
| 89 | 120 | 0.05603034 | 0.05060954 | -0.04766923 | -0.02601074 | 0.03792707 | 0.00855905 | -0.01859966 | 0.00171621 | 0.00899057 | 0.00007072 | -0.00081543 | 0.00043377 | -0.00016435 | -0.00029660 | 0.00029435 | -0.00005931 | -0.00000923 |
| 89 | 121 | 0.05582086 | 0.05012327 | -0.04767093 | -0.02541781 | 0.03785415 | 0.00827016 | -0.01825701 | 0.00155276 | 0.00837991 | 0.00009987 | -0.00076303 | 0.00043548 | -0.00015579 | -0.00024101 | 0.00024961 | -0.00007148 | -0.00006278 |
| 89 | 122 | 0.05580379 | 0.04991654 | -0.04811273 | -0.02516012 | 0.03816153 | 0.00801345 | -0.01822692 | 0.00217037 | 0.00852700 | -0.00026440 | -0.00077245 | 0.00046350 | -0.00021321 | -0.00025084 | 0.00027132 | -0.00005449 | -0.00001328 |
| 89 | 123 | 0.05574778 | 0.04943273 | -0.04830809 | -0.02446601 | 0.03818174 | 0.00771181 | -0.01784749 | 0.00201616 | 0.00784831 | -0.00025567 | -0.00070636 | 0.00047436 | -0.00021413 | -0.00020408 | 0.00020624 | -0.00005404 | -0.00007094 |
| 89 | 124 | 0.05557363 | 0.04914296 | -0.04851530 | -0.02417931 | 0.03843276 | 0.00754071 | -0.01776270 | 0.00257826 | 0.00794985 | -0.00060480 | -0.00072363 | 0.00049810 | -0.00026842 | -0.00019341 | 0.00023995 | -0.00005677 | -0.00003387 |
| 89 | 125 | 0.05567298 | 0.04869519 | -0.04886387 | -0.02336725 | 0.03853193 | 0.00722317 | -0.01735898 | 0.00243138 | 0.00720333 | -0.00061934 | -0.00064241 | 0.00051923 | -0.00027932 | -0.00015777 | 0.00015269 | -0.00003995 | -0.00009448 |
| 89 | 126 | 0.05533268 | 0.04831216 | -0.04884552 | -0.02306685 | 0.03874220 | 0.00714038 | -0.01725066 | 0.00293734 | 0.00727758 | -0.00094578 | -0.00067451 | 0.00053743 | -0.00032826 | -0.00012384 | 0.00020245 | -0.00006744 | -0.00007237 |
| 89 | 127 | 0.05556385 | 0.04778029 | -0.04939622 | -0.02197928 | 0.03909486 | 0.00621782 | -0.01744655 | 0.00295630 | 0.00688985 | -0.00097332 | -0.00060363 | 0.00053836 | -0.00032652 | -0.00012471 | 0.00013668 | -0.00004428 | -0.00010231 |
| 89 | 128 | 0.05524797 | 0.04712451 | -0.04962892 | -0.02130061 | 0.03963288 | 0.00545289 | -0.01798829 | 0.00360826 | 0.00731404 | -0.00128797 | -0.00065277 | 0.00053074 | -0.00035840 | -0.00010703 | 0.00021797 | -0.00007958 | -0.00005375 |
| 89 | 129 | 0.05542627 | 0.04670697 | -0.04992903 | -0.02042802 | 0.03971583 | 0.00467775 | -0.01800980 | 0.00362719 | 0.00694342 | -0.00129488 | -0.00058680 | 0.00052896 | -0.00035625 | -0.00010726 | 0.00015557 | -0.00005874 | -0.00008467 |
| 89 | 130 | 0.05516302 | 0.04594213 | -0.05032319 | -0.01957969 | 0.04029674 | 0.00382500 | -0.01850186 | 0.00427432 | 0.00728932 | -0.00159218 | -0.00062079 | 0.00052272 | -0.00039172 | -0.00008364 | 0.00022110 | -0.00008331 | -0.00003502 |
| 89 | 131 | 0.05528177 | 0.04563330 | -0.05038508 | -0.01892261 | 0.04013603 | 0.00319313 | -0.01837802 | 0.00429110 | 0.00694490 | -0.00158560 | -0.00056191 | 0.00052004 | -0.00038995 | -0.00008352 | 0.00016336 | -0.00006647 | -0.00006866 |
| 89 | 132 | 0.05507461 | 0.04477511 | -0.05092366 | -0.01793080 | 0.04073253 | 0.00227299 | -0.01878253 | 0.00492878 | 0.00720462 | -0.00185809 | -0.00057780 | 0.00051242 | -0.00042597 | -0.00005534 | 0.00021340 | -0.00008052 | -0.00001588 |
| 89 | 133 | 0.05513291 | 0.04456647 | -0.05077216 | -0.01748120 | 0.04036260 | 0.00177648 | -0.01854805 | 0.00494102 | 0.00689318 | -0.00184445 | -0.00052839 | 0.00051116 | -0.00042617 | -0.00005447 | 0.00016076 | -0.00006834 | -0.00005401 |
| 89 | 134 | 0.05498049 | 0.04363434 | -0.05143037 | -0.01637307 | 0.04095261 | 0.00080322 | -0.01884179 | 0.00556684 | 0.00706617 | -0.00208789 | -0.00052436 | 0.00049930 | -0.00045919 | -0.00002373 | 0.00019713 | -0.00007370 | 0.00000350 |
| 89 | 135 | 0.05498153 | 0.04351459 | -0.05109907 | -0.01611512 | 0.04041383 | 0.00043065 | -0.01853466 | 0.00557201 | 0.00679204 | -0.00207286 | -0.00048673 | 0.00050196 | -0.00046328 | -0.00002126 | 0.00014943 | -0.00006602 | -0.00004069 |
| 89 | 136 | 0.05487952 | 0.04253161 | -0.05184642 | -0.01491797 | 0.04098005 | -0.00058578 | -0.01870793 | 0.00618478 | 0.00688425 | -0.00228532 | -0.00046221 | 0.00048321 | -0.00048962 | 0.00000972 | 0.00017502 | -0.00006554 | 0.00002267 |
| 89 | 137 | 0.05482892 | 0.04248730 | -0.05137380 | -0.01482935 | 0.04031586 | -0.00084837 | -0.01836607 | 0.00618054 | 0.00664908 | -0.00227383 | -0.00043822 | 0.00049208 | -0.00049953 | 0.00001478 | 0.00013172 | -0.00006162 | -0.00002876 |
| 89 | 138 | 0.05477158 | 0.04147915 | -0.05217674 | -0.01357033 | 0.04084479 | -0.00190024 | -0.01842012 | 0.00677937 | 0.00667142 | -0.00245472 | -0.00039388 | 0.00046422 | -0.00051582 | 0.00004377 | 0.00014993 | -0.00005858 | 0.00004107 |
| 89 | 139 | 0.05467597 | 0.04149515 | -0.05160237 | -0.01362396 | 0.04009850 | -0.00206791 | -0.01807781 | 0.00676394 | 0.00647422 | -0.00245107 | -0.00038460 | 0.00048117 | -0.00053319 | 0.00005224 | 0.00011021 | -0.00005723 | -0.00001821 |
| 89 | 140 | 0.05465730 | 0.04048845 | -0.05242737 | -0.01232999 | 0.04057940 | -0.00314804 | -0.01802172 | 0.00734740 | 0.00644077 | -0.00260030 | -0.00032220 | 0.00044265 | -0.00053670 | 0.00007738 | 0.00012445 | -0.00005490 | 0.00005821 |
| 89 | 141 | 0.05452352 | 0.04054854 | -0.05178920 | -0.01249622 | 0.03979154 | -0.00323544 | -0.01770661 | 0.00731988 | 0.00627794 | -0.00260825 | -0.00032776 | 0.00046898 | -0.00056282 | 0.00008980 | 0.00008732 | -0.00005462 | -0.00000896 |
| 89 | 142 | 0.05453793 | 0.03956914 | -0.05260513 | -0.01119361 | 0.04021548 | -0.00433647 | -0.01755466 | 0.00788558 | 0.00620419 | -0.00272572 | -0.00024992 | 0.00041894 | -0.00055163 | 0.00010976 | 0.00010067 | -0.00005586 | 0.00007368 |
| 89 | 143 | 0.05437234 | 0.03965639 | -0.05193752 | -0.01144197 | 0.03942183 | -0.00435668 | -0.01728585 | 0.00784628 | 0.00606989 | -0.00274851 | -0.00026947 | 0.00045533 | -0.00058738 | 0.00012637 | 0.00006498 | -0.00005500 | -0.00000088 |
| 89 | 144 | 0.05441494 | 0.03872792 | -0.05271733 | -0.01015622 | 0.03978087 | -0.00547108 | -0.01705547 | 0.00839068 | 0.00597130 | -0.00283393 | -0.00017940 | 0.00039364 | -0.00056045 | 0.00014038 | 0.00008002 | -0.00006212 | 0.00008723 |
| 89 | 145 | 0.05422335 | 0.03882526 | -0.05205016 | -0.01045690 | 0.03901187 | -0.00543497 | -0.01684306 | 0.00834131 | 0.00585809 | -0.00287435 | -0.00021124 | 0.00044025 | -0.00060635 | 0.00016109 | 0.00004458 | -0.00005900 | 0.00000621 |
| 89 | 146 | 0.05428996 | 0.03796838 | -0.05277162 | -0.00921229 | 0.03929850 | -0.00655533 | -0.01655329 | 0.00885982 | 0.00574897 | -0.00292726 | -0.00011244 | 0.00036742 | -0.00056346 | 0.00016894 | 0.00006328 | -0.00007370 | 0.00009872 |
| 89 | 147 | 0.05407741 | 0.03805898 | -0.05212989 | -0.00953713 | 0.03857934 | -0.00647151 | -0.01639913 | 0.00880352 | 0.00564858 | -0.00298762 | -0.00015425 | 0.00042390 | -0.00061967 | 0.00019344 | 0.00002692 | -0.00006677 | 0.00001246 |
| 89 | 148 | 0.05416453 | 0.03729084 | -0.05277590 | -0.00835621 | 0.03878598 | -0.00759107 | -0.01606938 | 0.00929073 | 0.00554133 | -0.00300752 | -0.00005024 | 0.00034092 | -0.00056132 | 0.00019533 | 0.00005067 | -0.00009009 | 0.00010809 |
| 90 | 118 | 0.05670312 | 0.05085554 | -0.04826217 | -0.02575918 | 0.03841012 | 0.00811302 | -0.01884298 | 0.00208730 | 0.00928644 | 0.00004477 | -0.00076735 | 0.00034571 | -0.00010246 | -0.00034169 | 0.00034264 | -0.00011571 | 0.00002851 |
| 90 | 119 | 0.05652221 | 0.05092559 | -0.04784901 | -0.02545339 | 0.03848054 | 0.00830608 | -0.01890461 | 0.00168464 | 0.00913585 | 0.00014268 | -0.00078866 | 0.00040074 | -0.00010755 | -0.00034277 | 0.00030098 | -0.00007173 | 0.00001268 |
| 90 | 120 | 0.05640930 | 0.05020982 | -0.04863406 | -0.02491440 | 0.03866721 | 0.00753722 | -0.01859108 | 0.00260138 | 0.00895244 | -0.00026607 | -0.00073982 | 0.00036100 | -0.00013928 | -0.00029248 | 0.00033794 | -0.00011375 | 0.00003699 |
| 90 | 121 | 0.05628165 | 0.05030328 | -0.04827104 | -0.02463200 | 0.03863647 | 0.00768322 | -0.01857552 | 0.00218285 | 0.00872950 | -0.00018858 | -0.00074484 | 0.00042257 | -0.00014957 | -0.00030274 | 0.00028364 | -0.00006159 | 0.00002247 |
| 90 | 122 | 0.05609247 | 0.04946533 | -0.04899090 | -0.02401374 | 0.03891848 | 0.00703748 | -0.01823194 | 0.00307774 | 0.00852726 | -0.00057205 | -0.00070840 | 0.00037862 | -0.00018177 | -0.00023141 | 0.00032649 | -0.00011601 | 0.00003405 |
| 90 | 123 | 0.05603352 | 0.04957453 | -0.04869187 | -0.02372592 | 0.03880632 | 0.00713919 | -0.01812320 | 0.00264881 | 0.00821479 | -0.00052438 | -0.00069453 | 0.00044858 | -0.00019984 | -0.00025048 | 0.00025753 | -0.00005667 | 0.00001776 |
| 90 | 124 | 0.05575680 | 0.04863985 | -0.04931782 | -0.02305100 | 0.03918405 | 0.00662694 | -0.01779537 | 0.00350798 | 0.00801945 | -0.00087037 | -0.00067671 | 0.00039884 | -0.00022854 | -0.00015862 | 0.00030953 | -0.00012447 | 0.00001749 |
| 90 | 125 | 0.05577967 | 0.04876548 | -0.04908397 | -0.02272343 | 0.03901497 | 0.00668997 | -0.01758852 | 0.00307116 | 0.00760314 | -0.00086060 | -0.00064224 | 0.00047845 | -0.00025575 | -0.00018612 | 0.00022486 | -0.00005986 | -0.00000396 |
| 90 | 126 | 0.05540272 | 0.04775422 | -0.04959558 | -0.02202669 | 0.03947219 | 0.00630469 | -0.01731903 | 0.00388957 | 0.00744382 | -0.00115961 | -0.00064919 | 0.00042248 | -0.00027888 | -0.00007423 | 0.00028863 | -0.00014074 | -0.00001471 |
| 90 | 127 | 0.05562383 | 0.04773774 | -0.04972024 | -0.02132506 | 0.03964961 | 0.00567952 | -0.01772231 | 0.00358340 | 0.00728428 | -0.00119730 | -0.00061007 | 0.00049580 | -0.00030312 | -0.00013834 | 0.00021335 | -0.00007077 | -0.00001700 |
| 90 | 128 | 0.05532478 | 0.04654682 | -0.05044961 | -0.02028540 | 0.04033467 | 0.00460678 | -0.01806261 | 0.00445556 | 0.00740587 | -0.00147304 | -0.00062744 | 0.00041820 | -0.00030264 | -0.00005012 | 0.00029882 | -0.00015610 | -0.00000145 |
| 90 | 129 | 0.05557296 | 0.04652596 | -0.05058317 | -0.01962465 | 0.04051344 | 0.00407180 | -0.01839136 | 0.00422262 | 0.00727093 | -0.00151083 | -0.00058805 | 0.00049703 | -0.00034229 | -0.00011062 | 0.00021712 | -0.00007605 | -0.00000782 |
| 90 | 130 | 0.05524560 | 0.04534840 | -0.05121021 | -0.01860312 | 0.04098533 | 0.00299026 | -0.01858318 | 0.00501138 | 0.00730870 | -0.00174816 | -0.00059398 | 0.00041344 | -0.00033077 | -0.00002161 | 0.00029636 | -0.00016111 | 0.00001277 |
| 90 | 131 | 0.05551799 | 0.04531105 | -0.05135953 | -0.01797573 | 0.04115315 | 0.00252539 | -0.01882800 | 0.00485897 | 0.00719420 | -0.00178970 | -0.00055456 | 0.00049652 | -0.00038419 | -0.00007689 | 0.00020975 | -0.00007489 | 0.00000119 |
| 90 | 132 | 0.05516234 | 0.04417096 | -0.05187039 | -0.01700244 | 0.04142654 | 0.00146627 | -0.01888124 | 0.00555247 | 0.00715777 | -0.00198635 | -0.00054852 | 0.00040762 | -0.00036143 | 0.00000924 | 0.00028295 | -0.00015770 | 0.00002802 |
| 90 | 133 | 0.05545592 | 0.04410587 | -0.05204301 | -0.01640104 | 0.04157177 | 0.00105142 | -0.01903186 | 0.00548729 | 0.00705854 | -0.00203513 | -0.00050930 | 0.00049326 | -0.00042643 | -0.00003919 | 0.00019326 | -0.00006952 | 0.00001038 |

TABLE 3. Nuclear charge density distribution FB coefficients.

| $Z$ | $N$ | $a_1$ | $a_2$ | $a_3$ | $a_4$ | $a_5$ | $a_6$ | $a_7$ | $a_8$ | $a_9$ | $a_{10}$ | $a_{11}$ | $a_{12}$ | $a_{13}$ | $a_{14}$ | $a_{15}$ | $a_{16}$ | $a_{17}$ |
|---|---|---|---|---|---|---|---|---|---|---|---|---|---|---|---|---|---|---|
| 90 | 134 | 0.05507386 | 0.04302653 | -0.05242805 | -0.01549866 | 0.04167035 | 0.00003816 | -0.01897103 | 0.00607608 | 0.00696154 | -0.00219056 | -0.00049189 | 0.00040028 | -0.00039290 | 0.00004078 | 0.00026079 | -0.00014826 | 0.00004404 |
| 90 | 135 | 0.05538502 | 0.04292483 | -0.05263097 | -0.01491656 | 0.04178403 | -0.00034639 | -0.01902053 | 0.00610334 | 0.00687284 | -0.00224991 | -0.00045326 | 0.00048667 | -0.00046689 | 0.00000063 | 0.00017014 | -0.00006246 | 0.00001975 |
| 90 | 136 | 0.05498014 | 0.04192700 | -0.05288475 | -0.01410023 | 0.04173627 | -0.00129699 | -0.01887781 | 0.00658066 | 0.00673082 | -0.00236462 | -0.00042590 | 0.00039112 | -0.00042348 | 0.00007172 | 0.00023246 | -0.00013546 | 0.00006045 |
| 90 | 137 | 0.05530478 | 0.04178317 | -0.05312357 | -0.01353150 | 0.04181367 | -0.00166984 | -0.01882538 | 0.00670330 | 0.00664917 | -0.00243751 | -0.00038850 | 0.00047649 | -0.00050372 | 0.00004096 | 0.00014311 | -0.00005615 | 0.00002908 |
| 90 | 138 | 0.05488178 | 0.04088365 | -0.05324417 | -0.01280988 | 0.04164878 | -0.00254624 | -0.01863416 | 0.00706531 | 0.00647763 | -0.00251257 | -0.00035306 | 0.00037987 | -0.00045154 | 0.00010118 | 0.00020069 | -0.00012198 | 0.00007678 |
| 90 | 139 | 0.05521560 | 0.04069581 | -0.05352314 | -0.01224918 | 0.04168993 | -0.00292392 | -0.01848602 | 0.00728339 | 0.00640103 | -0.00260150 | -0.00031775 | 0.00046276 | -0.00053540 | 0.00008052 | 0.00011484 | -0.00005273 | 0.00003808 |
| 90 | 140 | 0.05477977 | 0.03990667 | -0.05351153 | -0.01162600 | 0.04143457 | -0.00371878 | -0.01827607 | 0.00752930 | 0.00621416 | -0.00263828 | -0.00027622 | 0.00036635 | -0.00047560 | 0.00012858 | 0.00016810 | -0.00011027 | 0.00009261 |
| 90 | 141 | 0.05511852 | 0.03967614 | -0.05383367 | -0.01106874 | 0.04144378 | -0.00411463 | -0.01804471 | 0.00783970 | 0.00614171 | -0.00274508 | -0.00024395 | 0.00044569 | -0.00056080 | 0.00011830 | 0.00008765 | -0.00005382 | 0.00004646 |
| 90 | 142 | 0.05467514 | 0.03900458 | -0.05369293 | -0.01054425 | 0.04112017 | -0.00482396 | -0.01783938 | 0.00797168 | 0.00595163 | -0.00274507 | -0.00019823 | 0.00035050 | -0.00049448 | 0.00015360 | 0.00013700 | -0.00010231 | 0.00010756 |
| 90 | 143 | 0.05501498 | 0.03873476 | -0.05406071 | -0.00998663 | 0.04110485 | -0.00524752 | -0.01754156 | 0.00836830 | 0.00588290 | -0.00287087 | -0.00016989 | 0.00042570 | -0.00057929 | 0.00015365 | 0.00006336 | -0.00006039 | 0.00005400 |
| 90 | 144 | 0.05456888 | 0.03818370 | -0.05379509 | -0.00955892 | 0.04072987 | -0.00587022 | -0.01735703 | 0.00839112 | 0.00569951 | -0.00283565 | -0.00012170 | 0.00033238 | -0.00050735 | 0.00017611 | 0.00010920 | -0.00009953 | 0.00012133 |
| 90 | 145 | 0.05490653 | 0.03787892 | -0.05421090 | -0.00899796 | 0.04069902 | -0.00632693 | -0.01701131 | 0.00886543 | 0.00563376 | -0.00298092 | -0.00009793 | 0.00040334 | -0.00059070 | 0.00018615 | 0.00004314 | -0.00007274 | 0.00006054 |
| 90 | 146 | 0.05446190 | 0.03744787 | -0.05382526 | -0.00866399 | 0.04028428 | -0.00686429 | -0.01685723 | 0.00878588 | 0.00546485 | -0.00291213 | -0.00004877 | 0.00031223 | -0.00051386 | 0.00019615 | 0.00008595 | -0.00010270 | 0.00013372 |
| 90 | 147 | 0.05479480 | 0.03711211 | -0.05429192 | -0.00809745 | 0.04024734 | -0.00735586 | -0.01648165 | 0.00932780 | 0.00540059 | -0.00307681 | -0.00002989 | 0.00037928 | -0.00059538 | 0.00021565 | 0.00002757 | -0.00009063 | 0.00006599 |
| 90 | 148 | 0.05435497 | 0.03679834 | -0.05379122 | -0.00785395 | 0.03979973 | -0.00781129 | -0.01636262 | 0.00915399 | 0.00525222 | -0.00297611 | 0.00001896 | 0.00029043 | -0.00051408 | 0.00021386 | 0.00006789 | -0.00011199 | 0.00014455 |
| 90 | 149 | 0.05468127 | 0.03643437 | -0.05431214 | -0.00728010 | 0.03976573 | -0.00833621 | -0.01597302 | 0.00975280 | 0.00518696 | -0.00315982 | 0.00003304 | 0.00035420 | -0.00059402 | 0.00024220 | 0.00001669 | -0.00011341 | 0.00007028 |
| 91 | 120 | 0.05675131 | 0.05059806 | -0.04844273 | -0.02417031 | 0.03887335 | 0.00722822 | -0.01890015 | 0.00216129 | 0.00883441 | -0.00012685 | -0.00071928 | 0.00039114 | -0.00009880 | -0.00034350 | 0.00028967 | -0.00008049 | 0.00003520 |
| 91 | 121 | 0.05645277 | 0.05016007 | -0.04821644 | -0.02344118 | 0.03908114 | 0.00716036 | -0.01853302 | 0.00193554 | 0.00825125 | -0.00010682 | -0.00066386 | 0.00040175 | -0.00008654 | -0.00030170 | 0.00024019 | -0.00006921 | 0.00000317 |
| 91 | 122 | 0.05649351 | 0.04994927 | -0.04883198 | -0.02328986 | 0.03897511 | 0.00662137 | -0.01848171 | 0.00267468 | 0.00838180 | -0.00045802 | -0.00066874 | 0.00040902 | -0.00014144 | -0.00029787 | 0.00026925 | -0.00006994 | 0.00004497 |
| 91 | 123 | 0.05633153 | 0.04952596 | -0.04880274 | -0.02256731 | 0.03921719 | 0.00656714 | -0.01804928 | 0.00242818 | 0.00773505 | -0.00044152 | -0.00059697 | 0.00042236 | -0.00013130 | -0.00026627 | 0.00020434 | -0.00004899 | 0.00001411 |
| 91 | 124 | 0.05622647 | 0.04919866 | -0.04922923 | -0.02235844 | 0.03909264 | 0.00611794 | -0.01794735 | 0.00315657 | 0.00783761 | -0.00078786 | -0.00061501 | 0.00042966 | -0.00019112 | -0.00024011 | 0.00024175 | -0.00006495 | 0.00004074 |
| 91 | 125 | 0.05621389 | 0.04881685 | -0.04937372 | -0.02162772 | 0.03937015 | 0.00608716 | -0.01746326 | 0.00288254 | 0.00712380 | -0.00077901 | -0.00052666 | 0.00044883 | -0.00018570 | -0.00021995 | 0.00015901 | -0.00003285 | 0.00000824 |
| 91 | 126 | 0.05595260 | 0.04837198 | -0.04960895 | -0.02136622 | 0.03925421 | 0.00572990 | -0.01734502 | 0.00359548 | 0.00721502 | -0.00111372 | -0.00056311 | 0.00045347 | -0.00024584 | -0.00017035 | 0.00020950 | -0.00006886 | 0.00001926 |
| 91 | 127 | 0.05612842 | 0.04790279 | -0.05005444 | -0.02039503 | 0.03996452 | 0.00518366 | -0.01755483 | 0.00338806 | 0.00682854 | -0.00110404 | -0.00049136 | 0.00046693 | -0.00023371 | -0.00018219 | 0.00014176 | -0.00003248 | 0.00000043 |
| 91 | 128 | 0.05592771 | 0.04714828 | -0.05061648 | -0.01970976 | 0.04028191 | 0.00414598 | -0.01815258 | 0.00420624 | 0.00721482 | -0.00143190 | -0.00055106 | 0.00046209 | -0.00028779 | -0.00013920 | 0.00021477 | -0.00007720 | 0.00001936 |
| 91 | 129 | 0.05607131 | 0.04679661 | -0.05083336 | -0.01893115 | 0.04083401 | 0.00378965 | -0.01823001 | 0.00398624 | 0.00686643 | -0.00140035 | -0.00048428 | 0.00047413 | -0.00027598 | -0.00015468 | 0.00014906 | -0.00003856 | 0.00000132 |
| 91 | 130 | 0.05589846 | 0.04590321 | -0.05155111 | -0.01808511 | 0.04110011 | 0.00260987 | -0.01873716 | 0.00482087 | 0.00715026 | -0.00171863 | -0.00052774 | 0.00046964 | -0.00033410 | -0.00010086 | 0.00020887 | -0.00007900 | 0.00001887 |
| 91 | 131 | 0.05600533 | 0.04566252 | -0.05154437 | -0.01749351 | 0.04150042 | 0.00243073 | -0.01870311 | 0.00459082 | 0.00684419 | -0.00167307 | -0.00046799 | 0.00048225 | -0.00032440 | -0.00011945 | 0.00014578 | -0.00004015 | -0.00000127 |
| 91 | 132 | 0.05586157 | 0.04465095 | -0.05240127 | -0.01651682 | 0.04170407 | 0.00113581 | -0.01908889 | 0.00543411 | 0.00702395 | -0.00197461 | -0.00049204 | 0.00047480 | -0.00038220 | -0.00005769 | 0.00019353 | -0.00007627 | 0.00001851 |
| 91 | 133 | 0.05592998 | 0.04451002 | -0.05218659 | -0.01610140 | 0.04196090 | 0.00111700 | -0.01896328 | 0.00519737 | 0.00676158 | -0.00192169 | -0.00044135 | 0.00049001 | -0.00037664 | -0.00007800 | 0.00013321 | -0.00003876 | -0.00000643 |
| 91 | 134 | 0.05581520 | 0.04340767 | -0.05315907 | -0.01502408 | 0.04210031 | -0.00026816 | -0.01921504 | 0.00604142 | 0.00684301 | -0.00220198 | -0.00044418 | 0.00047669 | -0.00042983 | -0.00001192 | 0.00017093 | -0.00007122 | 0.00001866 |
| 91 | 135 | 0.05584516 | 0.04335222 | -0.05276035 | -0.01476949 | 0.04222493 | -0.00014746 | -0.01901804 | 0.00580179 | 0.00662370 | -0.00214723 | -0.00040442 | 0.00049631 | -0.00043019 | -0.00003214 | 0.00011335 | -0.00003629 | -0.00001333 |
| 91 | 136 | 0.05575872 | 0.04219070 | -0.05381982 | -0.01361990 | 0.04230506 | -0.00159937 | -0.01913760 | 0.00663866 | 0.00661805 | -0.00240362 | -0.00038554 | 0.00047482 | -0.00047497 | 0.00003453 | 0.00014350 | -0.00006608 | 0.00001938 |
| 91 | 137 | 0.05575118 | 0.04220528 | -0.05326626 | -0.01350727 | 0.04231255 | -0.00136298 | -0.01889081 | 0.00639989 | 0.00644006 | -0.00235165 | -0.00035832 | 0.00050023 | -0.00048250 | 0.00001623 | 0.00008872 | -0.00003470 | -0.00002125 |
| 91 | 138 | 0.05569240 | 0.04101737 | -0.05438161 | -0.01231130 | 0.04234173 | -0.00285914 | -0.01888923 | 0.00722178 | 0.00636184 | -0.00258254 | -0.00031844 | 0.00046897 | -0.00051588 | 0.00008008 | 0.00011372 | -0.00006293 | 0.00002054 |
| 91 | 139 | 0.05564893 | 0.04108701 | -0.05370499 | -0.01231932 | 0.04225084 | -0.00253217 | -0.01861569 | 0.00698707 | 0.00622284 | -0.00253727 | -0.00030491 | 0.00050108 | -0.00053124 | 0.00006523 | 0.00006193 | -0.00003578 | -0.00002959 |
| 91 | 140 | 0.05561712 | 0.03990379 | -0.05484482 | -0.01110021 | 0.04223784 | -0.00405111 | -0.01850843 | 0.00778667 | 0.00608785 | -0.00274153 | -0.00024570 | 0.00045920 | -0.00055114 | 0.00012354 | 0.00008395 | -0.00006348 | 0.00002196 |
| 91 | 141 | 0.05553979 | 0.04001481 | -0.05407742 | -0.01120650 | 0.04207003 | -0.00365799 | -0.01823133 | 0.00755839 | 0.00598495 | -0.00270624 | -0.00024645 | 0.00049844 | -0.00057442 | 0.00011319 | 0.00003536 | -0.00004086 | -0.00003780 |
| 91 | 142 | 0.05553411 | 0.03886381 | -0.05521196 | -0.00998492 | 0.04202195 | -0.00517981 | -0.01803494 | 0.00832918 | 0.00580881 | -0.00288292 | -0.00017026 | 0.00044571 | -0.00057968 | 0.00016408 | 0.00005621 | -0.00006898 | 0.00002344 |
| 91 | 143 | 0.05542531 | 0.03900365 | -0.05438508 | -0.01016718 | 0.04179975 | -0.00474256 | -0.01777556 | 0.00810891 | 0.00573836 | -0.00286029 | -0.00018524 | 0.00049214 | -0.00061065 | 0.00015878 | 0.00001092 | -0.00005070 | -0.00004546 |
| 91 | 144 | 0.05544468 | 0.03790807 | -0.05548722 | -0.00896132 | 0.04172097 | -0.00624970 | -0.01750603 | 0.00884524 | 0.00553568 | -0.00300850 | -0.00009488 | 0.00042889 | -0.00060086 | 0.00020117 | 0.00003202 | -0.00008012 | 0.00002483 |
| 91 | 145 | 0.05530723 | 0.03806499 | -0.05463048 | -0.00919864 | 0.04146646 | -0.00578664 | -0.01728179 | 0.00863392 | 0.00549310 | -0.00300069 | -0.00012334 | 0.00048229 | -0.00063917 | 0.00020105 | -0.00001010 | -0.00006548 | -0.00005222 |
| 91 | 146 | 0.05535014 | 0.03704362 | -0.05567649 | -0.00802438 | 0.04135829 | -0.00726458 | -0.01695392 | 0.00933105 | 0.00527688 | -0.00311963 | -0.00002189 | 0.00040925 | -0.00061453 | 0.00023457 | 0.00001239 | -0.00009702 | 0.00002602 |
| 91 | 147 | 0.05518717 | 0.03720594 | -0.05481714 | -0.00829767 | 0.04109189 | -0.00678990 | -0.01677685 | 0.00912948 | 0.00525661 | -0.00312833 | -0.00006246 | 0.00046923 | -0.00065983 | 0.00023945 | -0.00002698 | -0.00008493 | -0.00005792 |
| 91 | 148 | 0.05525185 | 0.03627368 | -0.05578690 | -0.00716881 | 0.04095269 | -0.00822755 | -0.01640448 | 0.00978334 | 0.00503809 | -0.00321734 | 0.00004697 | 0.00038737 | -0.00062096 | 0.00026424 | -0.00000225 | -0.00011935 | 0.00002691 |
| 91 | 149 | 0.05506660 | 0.03642955 | -0.05494960 | -0.00746117 | 0.04069276 | -0.00775130 | -0.01628078 | 0.00959251 | 0.00503383 | -0.00324384 | -0.00000385 | 0.00045348 | -0.00067299 | 0.00027376 | -0.00003953 | -0.00010845 | -0.00006250 |
| 91 | 150 | 0.05515097 | 0.03559800 | -0.05582680 | -0.00638987 | 0.04051821 | -0.00914108 | -0.01587698 | 0.01019948 | 0.00482235 | -0.00330247 | 0.00011051 | 0.00036396 | -0.00062080 | 0.00029034 | -0.00001191 | -0.00014642 | 0.00002741 |
| 92 | 123 | 0.05667321 | 0.04956795 | -0.04934261 | -0.02201488 | 0.03901749 | 0.00547630 | -0.01830829 | 0.00319185 | 0.00797623 | -0.00072871 | -0.00059224 | 0.00039164 | -0.00013801 | -0.00028226 | 0.00025317 | -0.00008394 | 0.00005896 |
| 92 | 124 | 0.05648387 | 0.04856698 | -0.05028729 | -0.02127948 | 0.03984768 | 0.00515453 | -0.01803769 | 0.00424280 | 0.00794702 | -0.00104067 | -0.00060621 | 0.00032421 | -0.00016005 | -0.00016451 | 0.00032337 | -0.00015815 | 0.00006291 |
| 92 | 125 | 0.05639139 | 0.04880870 | -0.04972240 | -0.02109219 | 0.03909610 | 0.00504360 | -0.01769916 | 0.00369019 | 0.00742460 | -0.00104658 | -0.00053891 | 0.00040607 | -0.00018561 | -0.00022019 | 0.00022592 | -0.00007906 | 0.00005599 |
| 92 | 126 | 0.05611562 | 0.04769802 | -0.05062780 | -0.02036028 | 0.04003476 | 0.00485426 | -0.01750792 | 0.00469889 | 0.00743259 | -0.00131792 | -0.00057953 | 0.00033682 | -0.00020801 | -0.00008190 | 0.00030486 | -0.00016374 | 0.00004684 |
| 92 | 127 | 0.05624434 | 0.04778713 | -0.05046437 | -0.01982290 | 0.03970534 | 0.00410210 | -0.01780920 | 0.00421605 | 0.00712352 | -0.00135903 | -0.00051129 | 0.00042140 | -0.00023323 | -0.00016692 | 0.00021315 | -0.00008529 | 0.00004383 |
| 92 | 128 | 0.05606399 | 0.04646590 | -0.05162155 | -0.01874937 | 0.04087315 | 0.00321663 | -0.01825223 | 0.00523732 | 0.00735042 | -0.00161046 | -0.00056759 | 0.00034251 | -0.00024055 | -0.00004134 | 0.00030885 | -0.00018173 | 0.00004235 |
| 92 | 129 | 0.05623290 | 0.04653214 | -0.05152688 | -0.01824236 | 0.04066519 | 0.00258191 | -0.01852408 | 0.00480744 | 0.00707333 | -0.00164951 | -0.00049809 | 0.00043503 | -0.00028102 | -0.00012511 | 0.00020950 | -0.00009121 | 0.00003462 |
| 92 | 130 | 0.05600451 | 0.04522941 | -0.05253030 | -0.01718933 | 0.04151868 | 0.00165027 | -0.01878016 | 0.00576785 | 0.00721006 | -0.00186978 | -0.00054266 | 0.00034698 | -0.00027645 | 0.00000262 | 0.00030206 | -0.00019110 | 0.00003978 |
| 92 | 131 | 0.05621316 | 0.04525408 | -0.05251757 | -0.01670038 | 0.04142587 | 0.00111292 | -0.01901419 | 0.00540243 | 0.00696105 | -0.00191142 | -0.00047254 | 0.00044663 | -0.00033172 | -0.00007759 | 0.00019644 | -0.00009223 | 0.00002572 |
| 92 | 132 | 0.05593690 | 0.04400364 | -0.05334543 | -0.01569927 | 0.04197675 | 0.00016583 | -0.01909532 | 0.00628556 | 0.00701792 | -0.00209792 | -0.00050429 | 0.00034975 | -0.00031411 | 0.00004744 | 0.00028598 | -0.00019336 | 0.00003937 |
| 92 | 133 | 0.05618357 | 0.04397008 | -0.05342470 | -0.01521791 | 0.04198724 | -0.00029396 | -0.01927856 | 0.00599638 | 0.00679197 | -0.00214657 | -0.00043412 | 0.00045516 | -0.00038308 | -0.00002687 | 0.00017581 | -0.00009026 | 0.00001778 |
| 92 | 134 | 0.05586233 | 0.04280322 | -0.05406291 | -0.01429301 | 0.04225976 | -0.00123128 | -0.01921172 | 0.00678716 | 0.00678269 | -0.00229783 | -0.00045311 | 0.00035057 | -0.00035216 | 0.00009116 | 0.00026237 | -0.00019029 | 0.00004103 |
| 92 | 135 | 0.05614385 | 0.04269866 | -0.05423983 | -0.01381089 | 0.04235853 | -0.00163240 | -0.01933048 | 0.00658498 | 0.00657487 | -0.00235751 | -0.00038351 | 0.00045998 | -0.00043313 | 0.00002485 | 0.00014967 | -0.00008725 | 0.00001105 |
| 92 | 136 | 0.05578272 | 0.04164172 | -0.05468213 | -0.01297896 | 0.04238531 | -0.00254016 | -0.01915138 | 0.00727060 | 0.00651461 | -0.00247281 | -0.00039070 | 0.00034933 | -0.00038936 | 0.00013236 | 0.00023320 | -0.00018384 | 0.00004439 |
| 92 | 137 | 0.05609458 | 0.04145842 | -0.05495751 | -0.01248978 | 0.04255659 | -0.00289996 | -0.01919448 | 0.00716407 | 0.00632098 | -0.00254694 | -0.00032243 | 0.00046080 | -0.00048017 | 0.00007576 | 0.00012019 | -0.00008507 | 0.00000551 |
| 92 | 138 | 0.05570013 | 0.04053124 | -0.05520488 | -0.01176075 | 0.04237437 | -0.00376338 | -0.01894173 | 0.00773474 | 0.00622471 | -0.00262613 | -0.00031936 | 0.00034597 | -0.00042455 | 0.00017018 | 0.00020053 | -0.00017602 | 0.00004897 |
| 92 | 139 | 0.05603671 | 0.04026697 | -0.05557484 | -0.01125969 | 0.04260366 | -0.00409723 | -0.01890278 | 0.00772957 | 0.00604280 | -0.00271739 | -0.00025332 | 0.00045757 | -0.00052274 | 0.00012449 | 0.00008949 | -0.00008540 | 0.00000098 |
| 92 | 140 | 0.05561637 | 0.03948228 | -0.05563439 | -0.01063811 | 0.04224945 | -0.00490589 | -0.01861300 | 0.00817896 | 0.00592409 | -0.00276074 | -0.00024180 | 0.00034046 | -0.00045665 | 0.00020418 | 0.00016644 | -0.00016876 | 0.00005424 |
| 92 | 141 | 0.05597146 | 0.03914006 | -0.05609128 | -0.01012148 | 0.04252493 | -0.00522666 | -0.01849140 | 0.00827738 | 0.00575293 | -0.00287097 | -0.00017902 | 0.00045040 | -0.00055962 | 0.00017007 | 0.00005957 | -0.00008961 | -0.00000278 |
| 92 | 142 | 0.05553281 | 0.03850354 | -0.05597468 | -0.00960777 | 0.04203303 | -0.00597388 | -0.01819582 | 0.00860285 | 0.00562322 | -0.00287918 | -0.00016083 | 0.00033277 | -0.00048462 | 0.00023424 | 0.00013286 | -0.00016377 | 0.00005976 |
| 92 | 143 | 0.05589993 | 0.03809081 | -0.05650821 | -0.00907266 | 0.04234592 | -0.00629167 | -0.01799669 | 0.00880351 | 0.00546302 | -0.00300934 | -0.00010240 | 0.00043953 | -0.00058986 | 0.00021186 | 0.00003213 | -0.00009866 | -0.00000597 |
| 92 | 144 | 0.05545026 | 0.03760184 | -0.05623017 | -0.00866455 | 0.04174625 | -0.00697392 | -0.01771934 | 0.00900593 | 0.00533128 | -0.00298349 | -0.00007919 | 0.00032287 | -0.00050761 | 0.00026044 | 0.00010148 | -0.00016246 | 0.00006513 |
| 92 | 145 | 0.05582315 | 0.03712920 | -0.05682879 | -0.00810875 | 0.04209053 | -0.00729600 | -0.01745241 | 0.00930411 | 0.00518294 | -0.00313368 | -0.00002611 | 0.00042528 | -0.00061287 | 0.00024952 | 0.00000847 | -0.00011309 | -0.00000875 |
| 92 | 146 | 0.05536909 | 0.03678212 | -0.05640545 | -0.00780216 | 0.04140768 | -0.00791230 | -0.01720970 | 0.00938751 | 0.00505581 | -0.00307523 | 0.00000075 | 0.00031081 | -0.00052494 | 0.00028298 | 0.00007366 | -0.00016584 | 0.00007008 |
| 92 | 147 | 0.05574200 | 0.03626174 | -0.05705778 | -0.00722441 | 0.04177938 | -0.00824322 | -0.01688783 | 0.00977569 | 0.00492025 | -0.00324486 | 0.00004764 | 0.00040810 | -0.00062843 | 0.00028292 | -0.00001056 | -0.00013295 | -0.00001125 |
| 92 | 148 | 0.05528926 | 0.03604736 | -0.05650534 | -0.00701410 | 0.04103268 | -0.00879467 | -0.01668910 | 0.00974663 | 0.00480238 | -0.00315556 | 0.00007699 | 0.00029667 | -0.00053622 | 0.00030215 | 0.00005034 | -0.00017448 | 0.00007441 |
| 92 | 149 | 0.05565730 | 0.03549146 | -0.05720135 | -0.00641424 | 0.04142899 | -0.00913666 | -0.01632662 | 0.01021518 | 0.00468001 | -0.00334349 | 0.00011715 | 0.00038850 | -0.00063671 | 0.00031213 | -0.00002457 | -0.00015793 | -0.00001353 |
| 92 | 150 | 0.05521062 | 0.03539863 | -0.05653485 | -0.00629439 | 0.04063303 | -0.00962588 | -0.01617555 | 0.01008206 | 0.00457454 | -0.00322532 | 0.00014806 | 0.00028066 | -0.00054135 | 0.00031828 | 0.00003209 | -0.00018857 | 0.00007800 |
| 92 | 151 | 0.05556975 | 0.03481806 | -0.05726698 | -0.00567342 | 0.04105152 | -0.00997940 | -0.01578674 | 0.01062017 | 0.00446487 | -0.00343009 | 0.00018128 | 0.00036706 | -0.00063822 | 0.00033738 | -0.00003358 | -0.00018739 | -0.00001571 |
| 93 | 126 | 0.05653707 | 0.04842331 | -0.05017481 | -0.01995049 | 0.03890217 | 0.00399604 | -0.01738662 | 0.00424296 | 0.00700483 | -0.00129325 | -0.00047016 | 0.00037747 | -0.00018143 | -0.00019226 | 0.00021186 | -0.00009822 | 0.00006496 |
| 93 | 127 | 0.05662034 | 0.04803343 | -0.05042656 | -0.01908990 | 0.03958276 | 0.00356640 | -0.01753099 | 0.00390251 | 0.00660111 | -0.00127359 | -0.00038694 | 0.00040399 | -0.00016510 | -0.00022407 | 0.00013673 | -0.00004557 | 0.00005930 |
| 93 | 128 | 0.05653725 | 0.04717823 | -0.05133022 | -0.01843010 | 0.03996323 | 0.00249906 | -0.01820992 | 0.00481072 | 0.00696924 | -0.00158477 | -0.00046660 | 0.00039546 | -0.00022870 | -0.00014822 | 0.00021078 | -0.00010892 | 0.00004771 |
| 93 | 129 | 0.05661229 | 0.04689516 | -0.05141501 | -0.01774397 | 0.04061819 | 0.00229073 | -0.01828082 | 0.00443384 | 0.00661976 | -0.00153962 | -0.00038784 | 0.00042363 | -0.00021395 | -0.00018730 | 0.00013553 | -0.00004905 | 0.00004396 |
| 93 | 130 | 0.05652776 | 0.04589644 | -0.05242756 | -0.01693163 | 0.04084425 | 0.00104261 | -0.01882128 | 0.00538566 | 0.00687328 | -0.00184913 | -0.00045115 | 0.00041162 | -0.00027950 | -0.00009763 | 0.00020060 | -0.00011440 | 0.00003120 |
| 93 | 131 | 0.05659246 | 0.04570757 | -0.05235264 | -0.01641184 | 0.04147299 | 0.00104068 | -0.01883557 | 0.00497764 | 0.00657772 | -0.00178682 | -0.00037888 | 0.00044373 | -0.00026931 | -0.00014243 | 0.00012521 | -0.00004947 | 0.00002532 |
| 93 | 132 | 0.05650758 | 0.04459537 | -0.05345263 | -0.01547620 | 0.04154042 | -0.00036118 | -0.01921405 | 0.00596312 | 0.00672062 | -0.00208805 | -0.00042269 | 0.00042483 | -0.00033163 | -0.00004322 | 0.00018291 | -0.00011629 | 0.00001631 |
| 93 | 133 | 0.05656001 | 0.04448381 | -0.05323066 | -0.01511113 | 0.04213713 | -0.00017577 | -0.01918054 | 0.00553210 | 0.00647549 | -0.00201587 | -0.00035862 | 0.00046269 | -0.00032871 | -0.00009140 | 0.00010727 | -0.00004850 | 0.00000467 |
| 93 | 134 | 0.05647713 | 0.04329390 | -0.05439459 | -0.01408142 | 0.04205554 | -0.00170356 | -0.01939451 | 0.00653877 | 0.00651823 | -0.00230398 | -0.00038126 | 0.00043442 | -0.00038330 | 0.00001259 | 0.00015941 | -0.00011628 | 0.00000343 |
| 93 | 135 | 0.05651494 | 0.04324127 | -0.05404141 | -0.01385665 | 0.04261161 | -0.00135373 | -0.01931723 | 0.00609474 | 0.00631843 | -0.00222814 | -0.00032681 | 0.00047919 | -0.00038960 | -0.00003632 | 0.00008365 | -0.00004790 | -0.00001685 |
| 93 | 136 | 0.05643758 | 0.04201094 | -0.05524575 | -0.01276035 | 0.04240068 | -0.00297911 | -0.01937979 | 0.00710850 | 0.00627553 | -0.00249951 | -0.00032796 | 0.00044009 | -0.00043300 | 0.00006782 | 0.00013188 | -0.00011597 | -0.00000744 |
| 93 | 137 | 0.05645805 | 0.04200017 | -0.05477837 | -0.01265912 | 0.04290780 | -0.00249078 | -0.01926186 | 0.00666203 | 0.00611572 | -0.00242532 | -0.00028429 | 0.00049224 | -0.00044950 | 0.00002067 | 0.00005649 | -0.00004929 | -0.00003831 |
| 93 | 138 | 0.05639045 | 0.04076451 | -0.05600103 | -0.01152127 | 0.04259254 | -0.00418528 | -0.01919511 | 0.00766833 | 0.00600341 | -0.00267710 | -0.00026471 | 0.00044181 | -0.00047945 | 0.00012093 | 0.00010211 | -0.00011687 | -0.00001652 |
| 93 | 139 | 0.05639073 | 0.04078191 | -0.05543651 | -0.01152506 | 0.04304545 | -0.00358598 | -0.01904193 | 0.00722930 | 0.00587910 | -0.00260895 | -0.00023279 | 0.00050115 | -0.00050614 | 0.00007762 | 0.00002797 | -0.00005407 | -0.00005901 |
| 93 | 140 | 0.05633722 | 0.03957094 | -0.05665768 | -0.01036798 | 0.04265163 | -0.00532191 | -0.01887088 | 0.00821442 | 0.00571334 | -0.00283880 | -0.00019401 | 0.00043969 | -0.00052155 | 0.00017087 | 0.00007187 | -0.00012035 | -0.00002417 |
| 93 | 141 | 0.05631460 | 0.03960708 | -0.05601230 | -0.01045731 | 0.04304952 | -0.00463906 | -0.01869150 | 0.00779100 | 0.00562133 | -0.00278024 | -0.00017455 | 0.00050555 | -0.00055754 | 0.00013280 | 0.00000011 | -0.00006325 | -0.00007836 |
| 93 | 142 | 0.05627910 | 0.03844448 | -0.05721482 | -0.00930052 | 0.04260035 | -0.00639049 | -0.01843975 | 0.00874302 | 0.00541662 | -0.00298622 | -0.00011861 | 0.00043394 | -0.00055833 | 0.00021692 | 0.00004278 | -0.00012750 | -0.00003072 |
| 93 | 143 | 0.05623146 | 0.03849383 | -0.05650418 | -0.00945584 | 0.04294687 | -0.00564986 | -0.01824666 | 0.00834117 | 0.00535469 | -0.00293994 | -0.00011204 | 0.00050530 | -0.00060216 | 0.00018488 | -0.00002542 | -0.00007739 | -0.00009597 |
| 93 | 144 | 0.05621701 | 0.03739686 | -0.05767314 | -0.00831604 | 0.04246096 | -0.00739365 | -0.01793411 | 0.00925053 | 0.00512347 | -0.00312054 | -0.00004123 | 0.00042482 | -0.00058900 | 0.00025865 | 0.00001624 | -0.00013912 | -0.00003648 |
| 93 | 145 | 0.05614299 | 0.03745659 | -0.05691256 | -0.00851896 | 0.04276336 | -0.00661805 | -0.01774178 | 0.00887390 | 0.00509000 | -0.00308836 | -0.00004762 | 0.00050054 | -0.00063899 | 0.00023285 | -0.00004735 | -0.00009663 | -0.00011152 |
| 93 | 146 | 0.05615150 | 0.03643705 | -0.05803484 | -0.00740996 | 0.04225426 | -0.00833452 | -0.01738396 | 0.00973353 | 0.00484258 | -0.00324254 | 0.00003564 | 0.00041262 | -0.00061302 | 0.00029588 | -0.00000663 | -0.00015562 | -0.00004164 |
| 93 | 147 | 0.05605066 | 0.03650549 | -0.05723969 | -0.00764411 | 0.04252180 | -0.00754314 | -0.01720683 | 0.00938380 | 0.00483585 | -0.00322555 | 0.00001662 | 0.00049157 | -0.00066751 | 0.00027613 | -0.00006493 | -0.00012070 | -0.00012487 |
| 93 | 148 | 0.05608299 | 0.03557096 | -0.05830351 | -0.00657669 | 0.04199797 | -0.00921651 | -0.01681547 | 0.01018891 | 0.00458058 | -0.00335273 | 0.00010991 | 0.00039768 | -0.00063011 | 0.00032859 | -0.00002508 | -0.00017710 | -0.00004634 |
| 93 | 149 | 0.05595576 | 0.03564618 | -0.05748966 | -0.00682849 | 0.04224072 | -0.00842466 | -0.01666610 | 0.00986634 | 0.00459832 | -0.00335142 | 0.00007901 | 0.00047889 | -0.00068777 | 0.00031445 | -0.00007780 | -0.00014906 | -0.00013597 |
| 93 | 150 | 0.05601164 | 0.03480160 | -0.05848385 | -0.00581081 | 0.04170642 | -0.01004305 | -0.01625004 | 0.01061396 | 0.00434198 | -0.00345142 | 0.00017997 | 0.00038040 | -0.00064028 | 0.00035692 | -0.00003875 | -0.00020328 | -0.00005065 |
| 93 | 151 | 0.05585933 | 0.03488025 | -0.05766789 | -0.00606956 | 0.04193421 | -0.00926234 | -0.01613789 | 0.01031804 | 0.00438113 | -0.00346588 | 0.00013834 | 0.00046310 | -0.00070019 | 0.00034785 | -0.00008604 | -0.00018096 | -0.00014492 |
| 93 | 152 | 0.05593766 | 0.03412899 | -0.05858198 | -0.00510733 | 0.04139002 | -0.01081760 | -0.01570420 | 0.01100647 | 0.00412916 | -0.00353890 | 0.00024471 | 0.00036124 | -0.00064388 | 0.00038110 | -0.00004761 | -0.00023365 | -0.00005465 |
| 94 | 127 | 0.05682360 | 0.04782195 | -0.05101252 | -0.01863344 | 0.03910466 | 0.00240488 | -0.01781949 | 0.00482360 | 0.00684665 | -0.00152460 | -0.00043411 | 0.00035290 | -0.00017907 | -0.00016934 | 0.00021238 | -0.00012962 | 0.00005873 |
| 94 | 128 | 0.05674742 | 0.04639444 | -0.05259089 | -0.01754375 | 0.04086533 | 0.00179406 | -0.01834794 | 0.00597318 | 0.00720645 | -0.00173920 | -0.00052706 | 0.00026682 | -0.00018278 | -0.00001327 | 0.00032308 | -0.00023527 | 0.00005208 |
| 94 | 129 | 0.05681977 | 0.04655631 | -0.05218252 | -0.01719315 | 0.04006826 | 0.00096963 | -0.01853198 | 0.00537761 | 0.00676712 | -0.00178933 | -0.00042823 | 0.00037215 | -0.00022845 | -0.00011693 | 0.00020522 | -0.00013984 | 0.00003532 |
| 94 | 130 | 0.05668614 | 0.04514802 | -0.05358161 | -0.01611818 | 0.04145971 | 0.00028497 | -0.01885920 | 0.00648568 | 0.00703568 | -0.00198100 | -0.00051033 | 0.00027557 | -0.00022098 | 0.00004277 | 0.00031503 | -0.00025098 | 0.00003633 |
| 94 | 131 | 0.05680366 | 0.04526033 | -0.05329151 | -0.01578018 | 0.04086403 | -0.00042294 | -0.01903726 | 0.00593660 | 0.00663271 | -0.00202934 | -0.00040964 | 0.00038844 | -0.00027940 | -0.00006012 | 0.00019086 | -0.00014602 | 0.00001432 |
| 94 | 132 | 0.05661357 | 0.04390974 | -0.05448265 | -0.01474826 | 0.04189322 | -0.00115194 | -0.01917409 | 0.00698468 | 0.00681854 | -0.00219578 | -0.00047939 | 0.00028178 | -0.00025954 | 0.00009800 | 0.00029944 | -0.00026027 | 0.00002469 |
| 94 | 133 | 0.05677639 | 0.04395274 | -0.05432743 | -0.01441240 | 0.04149218 | -0.00176300 | -0.01933661 | 0.00649630 | 0.00644883 | -0.00224707 | -0.00037784 | 0.00040109 | -0.00033025 | -0.00000146 | 0.00017075 | -0.00014957 | -0.00000374 |

TABLE 3. Nuclear charge density distribution FB coefficients.

| $Z$ | $N$ | $a_1$ | $a_2$ | $a_3$ | $a_4$ | $a_5$ | $a_6$ | $a_7$ | $a_8$ | $a_9$ | $a_{10}$ | $a_{11}$ | $a_{12}$ | $a_{13}$ | $a_{14}$ | $a_{15}$ | $a_{16}$ | $a_{17}$ |
|---|---|---|---|---|---|---|---|---|---|---|---|---|---|---|---|---|---|---|
| 94 | 134 | 0.05653318 | 0.04269464 | -0.05529063 | -0.01344624 | 0.04217868 | -0.00251054 | -0.01930686 | 0.00746658 | 0.00656334 | -0.00238642 | -0.00043480 | 0.00028555 | -0.00029761 | 0.00015039 | 0.00027760 | -0.00026437 | 0.00001697 |
| 94 | 135 | 0.05674005 | 0.04265217 | -0.05528139 | -0.01310430 | 0.04195944 | -0.00304309 | -0.01944129 | 0.00705274 | 0.00622318 | -0.00244517 | -0.00033336 | 0.00040984 | -0.00037975 | 0.00005690 | 0.00014631 | -0.00015187 | -0.00001882 |
| 94 | 136 | 0.05644875 | 0.04151573 | -0.05600554 | -0.01222067 | 0.04233210 | -0.00378784 | -0.01927706 | 0.00792920 | 0.00627922 | -0.00255585 | -0.00037791 | 0.00028708 | -0.00033456 | 0.00019862 | 0.00025086 | -0.00026465 | 0.00001262 |
| 94 | 137 | 0.05669687 | 0.04137624 | -0.05614737 | -0.01186651 | 0.04227798 | -0.00425822 | -0.01937021 | 0.00760216 | 0.00596488 | -0.00262612 | -0.00027760 | 0.00041475 | -0.00042690 | 0.00011334 | 0.00011899 | -0.00015422 | -0.00003122 |
| 94 | 138 | 0.05636371 | 0.04038377 | -0.05662980 | -0.01107647 | 0.04237127 | -0.00498361 | -0.01910764 | 0.00837145 | 0.00597559 | -0.00270688 | -0.00031074 | 0.00028665 | -0.00036988 | 0.00024195 | 0.00022066 | -0.00026256 | 0.00001093 |
| 94 | 139 | 0.05664898 | 0.04014088 | -0.05692146 | -0.01070567 | 0.04246370 | -0.00540565 | -0.01914773 | 0.00814099 | 0.00568389 | -0.00279206 | -0.00021258 | 0.00041600 | -0.00047087 | 0.00016669 | 0.00009019 | -0.00015785 | -0.00004143 |
| 94 | 140 | 0.05628090 | 0.03930753 | -0.05716714 | -0.01001543 | 0.04231457 | -0.00609966 | -0.01882321 | 0.00879298 | 0.00566171 | -0.00284199 | -0.00023568 | 0.00028451 | -0.00040303 | 0.00028013 | 0.00018845 | -0.00025959 | 0.00001115 |
| 94 | 141 | 0.05659799 | 0.03896030 | -0.05760159 | -0.00962466 | 0.04253504 | -0.00648460 | -0.01880149 | 0.00866586 | 0.00539035 | -0.00294469 | -0.00014075 | 0.00041389 | -0.00051091 | 0.00021620 | 0.00006132 | -0.00016385 | -0.00004999 |
| 94 | 142 | 0.05620230 | 0.03829390 | -0.05762186 | -0.00903658 | 0.04218015 | -0.00713944 | -0.01844865 | 0.00919396 | 0.00534625 | -0.00296331 | -0.00015532 | 0.00028083 | -0.00043349 | 0.00031324 | 0.00015567 | -0.00025713 | 0.00001252 |
| 94 | 143 | 0.05654505 | 0.03784664 | -0.05818695 | -0.00862317 | 0.04251136 | -0.00749589 | -0.01836036 | 0.00917354 | 0.00509397 | -0.00308527 | -0.00006471 | 0.00040871 | -0.00054631 | 0.00026141 | 0.00003367 | -0.00017313 | -0.00005739 |
| 94 | 144 | 0.05612902 | 0.03734832 | -0.05799820 | -0.00813676 | 0.04198499 | -0.00810735 | -0.01800798 | 0.00957469 | 0.00503701 | -0.00307259 | -0.00007220 | 0.00027573 | -0.00046071 | 0.00034156 | 0.00012367 | -0.00025648 | 0.00001440 |
| 94 | 145 | 0.05649070 | 0.03680997 | -0.05867790 | -0.00769834 | 0.04241173 | -0.00844147 | -0.01785268 | 0.00966094 | 0.00480353 | -0.00321464 | 0.00001303 | 0.00040070 | -0.00057643 | 0.00030209 | 0.00000840 | -0.00018637 | -0.00006403 |
| 94 | 146 | 0.05606137 | 0.03647488 | -0.05830004 | -0.00731121 | 0.04174445 | -0.00900828 | -0.01752338 | 0.00993545 | 0.00474068 | -0.00317119 | 0.00001130 | 0.00026924 | -0.00048412 | 0.00036550 | 0.00009367 | -0.00025871 | 0.00001628 |
| 94 | 147 | 0.05643510 | 0.03585815 | -0.05907571 | -0.00684571 | 0.04225363 | -0.00932400 | -0.01730472 | 0.01012514 | 0.00452649 | -0.00333332 | 0.00009015 | 0.00039012 | -0.00060078 | 0.00033819 | -0.00001353 | -0.00020397 | -0.00007014 |
| 94 | 148 | 0.05599906 | 0.03567656 | -0.05853076 | -0.00655419 | 0.04147168 | -0.00984720 | -0.01701457 | 0.01027632 | 0.00446263 | -0.00326014 | 0.00009311 | 0.00026136 | -0.00050324 | 0.00038550 | 0.00006667 | -0.00026465 | 0.00001780 |
| 94 | 149 | 0.05637811 | 0.03499670 | -0.05938260 | -0.00605988 | 0.04205218 | -0.01014663 | -0.01673953 | 0.01056343 | 0.00426860 | -0.00344158 | 0.00016471 | 0.00037721 | -0.00061900 | 0.00036975 | -0.00003145 | -0.00022603 | -0.00007589 |
| 94 | 150 | 0.05594124 | 0.03495535 | -0.05869322 | -0.00585956 | 0.04117735 | -0.01062884 | -0.01649839 | 0.01059709 | 0.00420685 | -0.00334017 | 0.00017152 | 0.00025209 | -0.00051765 | 0.00040197 | 0.00004345 | -0.00027486 | 0.00001872 |
| 94 | 151 | 0.05631937 | 0.03422866 | -0.05960186 | -0.00533535 | 0.04181946 | -0.01091274 | -0.01617634 | 0.01097339 | 0.00403387 | -0.00353956 | 0.00023521 | 0.00036225 | -0.00063099 | 0.00039693 | -0.00004498 | -0.00025237 | -0.00008132 |
| 94 | 152 | 0.05588687 | 0.03431218 | -0.05878998 | -0.00522130 | 0.04086949 | -0.01135757 | -0.01598865 | 0.01089725 | 0.00397597 | -0.00341176 | 0.00024520 | 0.00024144 | -0.00052711 | 0.00041532 | 0.00002450 | -0.00028959 | 0.00001891 |
| 94 | 153 | 0.05625856 | 0.03355471 | -0.05973780 | -0.00466713 | 0.04156433 | -0.01162583 | -0.01563023 | 0.01135303 | 0.00382450 | -0.00362734 | 0.00030056 | 0.00034555 | -0.00063688 | 0.00041998 | -0.00005403 | -0.00028258 | -0.00008649 |
| 95 | 128 | 0.05709879 | 0.04721183 | -0.05183679 | -0.01746928 | 0.03919360 | 0.00092024 | -0.01817115 | 0.00537354 | 0.00665038 | -0.00173187 | -0.00040413 | 0.00032997 | -0.00017996 | -0.00013497 | 0.00021000 | -0.00016627 | 0.00003907 |
| 95 | 129 | 0.05708002 | 0.04699527 | -0.05170662 | -0.01689886 | 0.03953644 | 0.00061329 | -0.01820876 | 0.00488866 | 0.00625268 | -0.00169665 | -0.00031137 | 0.00037967 | -0.00017610 | -0.00019384 | 0.00011987 | -0.00009031 | 0.00003691 |
| 95 | 130 | 0.05708375 | 0.04593920 | -0.05299907 | -0.01611229 | 0.04005425 | -0.00045265 | -0.01877076 | 0.00591310 | 0.00653382 | -0.00197163 | -0.00039484 | 0.00034818 | -0.00022874 | -0.00007689 | 0.00019895 | -0.00017683 | 0.00001249 |
| 95 | 131 | 0.05707183 | 0.04579627 | -0.05277630 | -0.01569222 | 0.04048096 | -0.00053960 | -0.01880418 | 0.00537243 | 0.00619859 | -0.00191339 | -0.00030716 | 0.00040450 | -0.00023023 | -0.00014406 | 0.00010635 | -0.00009161 | 0.00000904 |
| 95 | 132 | 0.05705595 | 0.04464598 | -0.05409747 | -0.01478557 | 0.04076210 | -0.00178261 | -0.01917440 | 0.00645510 | 0.00636942 | -0.00218939 | -0.00037252 | 0.00036263 | -0.00027740 | -0.00001655 | 0.00018245 | -0.00018425 | -0.00001018 |
| 95 | 133 | 0.05704904 | 0.04455522 | -0.05379742 | -0.01450560 | 0.04126545 | -0.00166729 | -0.01920764 | 0.00586933 | 0.00609167 | -0.00211531 | -0.00029130 | 0.00042682 | -0.00028689 | -0.00008932 | 0.00008750 | -0.00009229 | -0.00001854 |
| 95 | 134 | 0.05701827 | 0.04335021 | -0.05512281 | -0.01350386 | 0.04132123 | -0.00306125 | -0.01938904 | 0.00699572 | 0.00616340 | -0.00238783 | -0.00033722 | 0.00037311 | -0.00032488 | 0.00004384 | 0.00016166 | -0.00018966 | -0.00002884 |
| 95 | 135 | 0.05701362 | 0.04329034 | -0.05475969 | -0.01335191 | 0.04188645 | -0.00276498 | -0.01941780 | 0.00637869 | 0.00593636 | -0.00230463 | -0.00026349 | 0.00044564 | -0.00034419 | -0.00003162 | 0.00006482 | -0.00009381 | -0.00004499 |
| 95 | 136 | 0.05697379 | 0.04206865 | -0.05606883 | -0.01227906 | 0.04174058 | -0.00428185 | -0.01942864 | 0.00753142 | 0.00592335 | -0.00256955 | -0.00028986 | 0.00037973 | -0.00037045 | 0.00010255 | 0.00013769 | -0.00019418 | -0.00004383 |
| 95 | 137 | 0.05696795 | 0.04202120 | -0.05565458 | -0.01224135 | 0.04234886 | -0.00382859 | -0.01944471 | 0.00689851 | 0.00573983 | -0.00248339 | -0.00022442 | 0.00046039 | -0.00040047 | 0.00002723 | 0.00003980 | -0.00009746 | -0.00006985 |
| 95 | 138 | 0.05692540 | 0.04081643 | -0.05693149 | -0.01111988 | 0.04203255 | -0.00543964 | -0.01931223 | 0.00805894 | 0.00565767 | -0.00273680 | -0.00023199 | 0.00038282 | -0.00041359 | 0.00015835 | 0.00011163 | -0.00019885 | -0.00005574 |
| 95 | 139 | 0.05691445 | 0.04076744 | -0.05647514 | -0.01118104 | 0.04266478 | -0.00485464 | -0.01930758 | 0.00742539 | 0.00551117 | -0.00265318 | -0.00017554 | 0.00047080 | -0.00045431 | 0.00008564 | 0.00001388 | -0.00010423 | -0.00009291 |
| 95 | 140 | 0.05687542 | 0.03960685 | -0.05770848 | -0.01003175 | 0.04221193 | -0.00653167 | -0.01906234 | 0.00857529 | 0.00537509 | -0.00289145 | -0.00016567 | 0.00038278 | -0.00045387 | 0.00021044 | 0.00008456 | -0.00020465 | -0.00006530 |
| 95 | 141 | 0.05685528 | 0.03954783 | -0.05721607 | -0.01017517 | 0.04285152 | -0.00584020 | -0.01903194 | 0.00795484 | 0.00526054 | -0.00281501 | -0.00011886 | 0.00047687 | -0.00050443 | 0.00014227 | -0.00001162 | -0.00011486 | -0.00011408 |
| 95 | 142 | 0.05682544 | 0.03845149 | -0.05839849 | -0.00901701 | 0.04229481 | -0.00755678 | -0.01870339 | 0.00907767 | 0.00508428 | -0.00303490 | -0.00009320 | 0.00038001 | -0.00049084 | 0.00025836 | 0.00005755 | -0.00021250 | -0.00007322 |
| 95 | 143 | 0.05679215 | 0.03837936 | -0.05787363 | -0.00922528 | 0.04292941 | -0.00678282 | -0.01864675 | 0.00848158 | 0.00499828 | -0.00296935 | -0.00005666 | 0.00047872 | -0.00054972 | 0.00019604 | -0.00003551 | -0.00012979 | -0.00013335 |
| 95 | 144 | 0.05677637 | 0.03736037 | -0.05900103 | -0.00807536 | 0.04229756 | -0.00851529 | -0.01826045 | 0.00956343 | 0.00479342 | -0.00316818 | -0.00001698 | 0.00037484 | -0.00052401 | 0.00030184 | 0.00003167 | -0.00022317 | -0.00008011 |
| 95 | 145 | 0.05672634 | 0.03727667 | -0.05844561 | -0.00833103 | 0.04291966 | -0.00768061 | -0.01818169 | 0.00899998 | 0.00473421 | -0.00311614 | 0.00000872 | 0.00047656 | -0.00058927 | 0.00024610 | -0.00005676 | -0.00014919 | -0.00015070 |
| 95 | 146 | 0.05672846 | 0.03634194 | -0.05951614 | -0.00720419 | 0.04223607 | -0.00940866 | -0.01775793 | 0.01003003 | 0.00450993 | -0.00329192 | 0.00006069 | 0.00036752 | -0.00055290 | 0.00034083 | 0.00000786 | -0.00023728 | -0.00008642 |
| 95 | 147 | 0.05665869 | 0.03625146 | -0.05893133 | -0.00749077 | 0.04284241 | -0.00853229 | -0.01766491 | 0.00950452 | 0.00447687 | -0.00325498 | 0.00007512 | 0.00047068 | -0.00062242 | 0.00029185 | -0.00007462 | -0.00017298 | -0.00016612 |
| 95 | 148 | 0.05668144 | 0.03540305 | -0.05994429 | -0.00639939 | 0.04212479 | -0.01023916 | -0.01721853 | 0.01047499 | 0.00424007 | -0.00340650 | 0.00013771 | 0.00035824 | -0.00057702 | 0.00037534 | -0.00001306 | -0.00025521 | -0.00009246 |
| 95 | 149 | 0.05658972 | 0.03531217 | -0.05933174 | -0.00670215 | 0.04271545 | -0.00933719 | -0.01712141 | 0.00999006 | 0.00423316 | -0.00338527 | 0.00014067 | 0.00046139 | -0.00064875 | 0.00033291 | -0.00008858 | -0.00020082 | -0.00017957 |
| 95 | 150 | 0.05663468 | 0.03454884 | -0.06028654 | -0.00565594 | 0.04197615 | -0.01100959 | -0.01666247 | 0.01089598 | 0.00398879 | -0.00351206 | 0.00021232 | 0.00034718 | -0.00059600 | 0.00040548 | -0.00003048 | -0.00027711 | -0.00009839 |
| 95 | 151 | 0.05651966 | 0.03446386 | -0.05964925 | -0.00596273 | 0.04255323 | -0.01009527 | -0.01657197 | 0.01045220 | 0.00400807 | -0.00350639 | 0.00020385 | 0.00044910 | -0.00066816 | 0.00036913 | -0.00009842 | -0.00023221 | -0.00019107 |
| 95 | 152 | 0.05658739 | 0.03378258 | -0.06054458 | -0.00496856 | 0.04180020 | -0.01172313 | -0.01610679 | 0.01129088 | 0.00375962 | -0.00360859 | 0.00028309 | 0.00033447 | -0.00060960 | 0.00043140 | -0.00004399 | -0.00030288 | -0.00010426 |
| 95 | 153 | 0.05644859 | 0.03370829 | -0.05988774 | -0.00527024 | 0.04236676 | -0.01080719 | -0.01603288 | 0.01088736 | 0.00380466 | -0.00361781 | 0.00026358 | 0.00043425 | -0.00068085 | 0.00040055 | -0.00010417 | -0.00026651 | -0.00020065 |
| 95 | 154 | 0.05653875 | 0.03310564 | -0.06072105 | -0.00433232 | 0.04160430 | -0.01238314 | -0.01556522 | 0.01165781 | 0.00355460 | -0.00369604 | 0.00034901 | 0.00032029 | -0.00061779 | 0.00045331 | -0.00005346 | -0.00033223 | -0.00011007 |
| 96 | 135 | 0.05723463 | 0.04280403 | -0.05583099 | -0.01274023 | 0.04105918 | -0.00424280 | -0.01938205 | 0.00745775 | 0.00587738 | -0.00251092 | -0.00029987 | 0.00033974 | -0.00031396 | 0.00008948 | 0.00015657 | -0.00023434 | -0.00005470 |
| 96 | 136 | 0.05702953 | 0.04154435 | -0.05694269 | -0.01178070 | 0.04190800 | -0.00483146 | -0.01926476 | 0.00852449 | 0.00605619 | -0.00261813 | -0.00037797 | 0.00020921 | -0.00026742 | 0.00026133 | 0.00027978 | -0.00035974 | -0.00002355 |
| 96 | 137 | 0.05718176 | 0.04155641 | -0.05675366 | -0.01159373 | 0.04143166 | -0.00540744 | -0.01936775 | 0.00797147 | 0.00562989 | -0.00267784 | -0.00024993 | 0.00034352 | -0.00035458 | 0.00014686 | 0.00013395 | -0.00024081 | -0.00006801 |
| 96 | 138 | 0.05693209 | 0.04043154 | -0.05760911 | -0.01070921 | 0.04196320 | -0.00599753 | -0.01913463 | 0.00894374 | 0.00575275 | -0.00276296 | -0.00031640 | 0.00020827 | -0.00029864 | 0.00030764 | 0.00025376 | -0.00036278 | -0.00002759 |
| 96 | 139 | 0.05712802 | 0.04034087 | -0.05759928 | -0.01050736 | 0.04169750 | -0.00651147 | -0.01921636 | 0.00847607 | 0.00536284 | -0.00283248 | -0.00019041 | 0.00034429 | -0.00039295 | 0.00020074 | 0.00010975 | -0.00024736 | -0.00007814 |
| 96 | 140 | 0.05683929 | 0.03936865 | -0.05819686 | -0.00970647 | 0.04194256 | -0.00708876 | -0.01889863 | 0.00934144 | 0.00544050 | -0.00289372 | -0.00024609 | 0.00020604 | -0.00032830 | 0.00034802 | 0.00022542 | -0.00036375 | -0.00002871 |
| 96 | 141 | 0.05707569 | 0.03916792 | -0.05836685 | -0.00948573 | 0.04187004 | -0.00755189 | -0.01894805 | 0.00896913 | 0.00508374 | -0.00297641 | -0.00012327 | 0.00034257 | -0.00042894 | 0.00025059 | 0.00008478 | -0.00025479 | -0.00008595 |
| 96 | 142 | 0.05675375 | 0.03836042 | -0.05871136 | -0.00877381 | 0.04186066 | -0.00810650 | -0.01857662 | 0.00971815 | 0.00512643 | -0.00301226 | -0.00016925 | 0.00020292 | -0.00035631 | 0.00038265 | 0.00019576 | -0.00036378 | -0.00002788 |
| 96 | 143 | 0.05702622 | 0.03804670 | -0.05905598 | -0.00853094 | 0.04196309 | -0.00852731 | -0.01858402 | 0.00944846 | 0.00479983 | -0.00311085 | -0.00005066 | 0.00033882 | -0.00046234 | 0.00029617 | 0.00005986 | -0.00026390 | -0.00009228 |
| 96 | 144 | 0.05667709 | 0.03741028 | -0.05915783 | -0.00791052 | 0.04173096 | -0.00905332 | -0.01818772 | 0.01007469 | 0.00481674 | -0.00312020 | -0.00008810 | 0.00019918 | -0.00038250 | 0.00041195 | 0.00016574 | -0.00036395 | -0.00002595 |
| 96 | 145 | 0.05698024 | 0.03698526 | -0.05966640 | -0.00764280 | 0.04199016 | -0.00943775 | -0.01814569 | 0.00991196 | 0.00451794 | -0.00323664 | 0.00002527 | 0.00033338 | -0.00049285 | 0.00033742 | 0.00003585 | -0.00027536 | -0.00009781 |
| 96 | 146 | 0.05660995 | 0.03652073 | -0.05954068 | -0.00711416 | 0.04156541 | -0.00993259 | -0.01774978 | 0.01041203 | 0.00451679 | -0.00321887 | -0.00000478 | 0.00019498 | -0.00040664 | 0.00043642 | 0.00013631 | -0.00036530 | -0.00002367 |
| 96 | 147 | 0.05693769 | 0.03599057 | -0.06019794 | -0.00681916 | 0.04196421 | -0.01028432 | -0.01765386 | 0.01035760 | 0.00424418 | -0.00335434 | 0.00010246 | 0.00032650 | -0.00052007 | 0.00037435 | 0.00001354 | -0.00028975 | -0.00010305 |
| 96 | 148 | 0.05655221 | 0.03569364 | -0.05986344 | -0.00638097 | 0.04137432 | -0.01074821 | -0.01727909 | 0.01073100 | 0.00423105 | -0.00330933 | 0.00007877 | 0.00019037 | -0.00042839 | 0.00045661 | 0.00010834 | -0.00036870 | -0.00002161 |
| 96 | 149 | 0.05689799 | 0.03506856 | -0.06065039 | -0.00605635 | 0.04189682 | -0.01106898 | -0.01712802 | 0.01078335 | 0.00398378 | -0.00346422 | 0.00017903 | 0.00031831 | -0.00054357 | 0.00040704 | -0.00000636 | -0.00030745 | -0.00010834 |
| 96 | 150 | 0.05650301 | 0.03493047 | -0.06012861 | -0.00570624 | 0.04116625 | -0.01150422 | -0.01679006 | 0.01103222 | 0.00396304 | -0.00339240 | 0.00016086 | 0.00018532 | -0.00044738 | 0.00047304 | 0.00008260 | -0.00037491 | -0.00002020 |
| 96 | 151 | 0.05686012 | 0.03422406 | -0.06102376 | -0.00534993 | 0.04179796 | -0.01179421 | -0.01658562 | 0.01118722 | 0.00374090 | -0.00356638 | 0.00025338 | 0.00030889 | -0.00056295 | 0.00043563 | -0.00002329 | -0.00032866 | -0.00011385 |
| 96 | 152 | 0.05646106 | 0.03423235 | -0.06033770 | -0.00508467 | 0.04094801 | -0.01220458 | -0.01629508 | 0.01131601 | 0.00371540 | -0.00346867 | 0.00024006 | 0.00017975 | -0.00046324 | 0.00048619 | 0.00005970 | -0.00038445 | -0.00001971 |
| 96 | 153 | 0.05682294 | 0.03346055 | -0.06131839 | -0.00469498 | 0.04167565 | -0.01246289 | -0.01604174 | 0.01156730 | 0.00351859 | -0.00366076 | 0.00032421 | 0.00029830 | -0.00057787 | 0.00046024 | -0.00003684 | -0.00035337 | -0.00011962 |
| 96 | 154 | 0.05642475 | 0.03360018 | -0.06049151 | -0.00451093 | 0.04072457 | -0.01285296 | -0.01580453 | 0.01158234 | 0.00348990 | -0.00353851 | 0.00031524 | 0.00017355 | -0.00047564 | 0.00049647 | 0.00004013 | -0.00039769 | -0.00002029 |
| 96 | 155 | 0.05678521 | 0.03278010 | -0.06153523 | -0.00408677 | 0.04153583 | -0.01307813 | -0.01550876 | 0.01192187 | 0.00331876 | -0.00374723 | 0.00039053 | 0.00028659 | -0.00058816 | 0.00048107 | -0.00004681 | -0.00038139 | -0.00012563 |
| 96 | 156 | 0.05639235 | 0.03303454 | -0.06059043 | -0.00397991 | 0.04049915 | -0.01345274 | -0.01532680 | 0.01183089 | 0.00328749 | -0.00360215 | 0.00038554 | 0.00016661 | -0.00048433 | 0.00050421 | 0.00002419 | -0.00041476 | -0.00002198 |
| 97 | 136 | 0.05742826 | 0.04232144 | -0.05645326 | -0.01210670 | 0.04073081 | -0.00530773 | -0.01932668 | 0.00788189 | 0.00559899 | -0.00261801 | -0.00026484 | 0.00030089 | -0.00029706 | 0.00013321 | 0.00015574 | -0.00028188 | -0.00007859 |
| 97 | 137 | 0.05734557 | 0.04231538 | -0.05589323 | -0.01214494 | 0.04111943 | -0.00499606 | -0.01944521 | 0.00706690 | 0.00536493 | -0.00252545 | -0.00018056 | 0.00040542 | -0.00034050 | 0.00003330 | 0.00003728 | -0.00016090 | -0.00010289 |
| 97 | 138 | 0.05736789 | 0.04111844 | -0.05734204 | -0.01103268 | 0.04105825 | -0.00642031 | -0.01926886 | 0.00837309 | 0.00534903 | -0.00277184 | -0.00021274 | 0.00030160 | -0.00033233 | 0.00018818 | 0.00013496 | -0.00029000 | -0.00008936 |
| 97 | 139 | 0.05728999 | 0.04109606 | -0.05678811 | -0.01116558 | 0.04155083 | -0.00596627 | -0.01939559 | 0.00754153 | 0.00515998 | -0.00267491 | -0.00013447 | 0.00041255 | -0.00038407 | 0.00008773 | 0.00001654 | -0.00016875 | -0.00012161 |
| 97 | 140 | 0.05730945 | 0.03994840 | -0.05816100 | -0.01001305 | 0.04129856 | -0.00747437 | -0.01909139 | 0.00885445 | 0.00508466 | -0.00291505 | -0.00015204 | 0.00029992 | -0.00036571 | 0.00023933 | 0.00011288 | -0.00029802 | -0.00009703 |
| 97 | 141 | 0.05723337 | 0.03989863 | -0.05761896 | -0.01022690 | 0.04187258 | -0.00690105 | -0.01921306 | 0.00802407 | 0.00493400 | -0.00281954 | -0.00008028 | 0.00041593 | -0.00042525 | 0.00014068 | -0.00000403 | -0.00017965 | -0.00013808 |
| 97 | 142 | 0.05725507 | 0.03881929 | -0.05891038 | -0.00905185 | 0.04146357 | -0.00846698 | -0.01881194 | 0.00932412 | 0.00481224 | -0.00304903 | -0.00008459 | 0.00029644 | -0.00039727 | 0.00028639 | 0.00009011 | -0.00030664 | -0.00010256 |
| 97 | 143 | 0.05717752 | 0.03873594 | -0.05838220 | -0.00933198 | 0.04209803 | -0.00779616 | -0.01891786 | 0.00851086 | 0.00469440 | -0.00295998 | -0.00001994 | 0.00041601 | -0.00046371 | 0.00019151 | -0.00002378 | -0.00019395 | -0.00015281 |
| 97 | 144 | 0.05720598 | 0.03773816 | -0.05959042 | -0.00815110 | 0.04156490 | -0.00939669 | -0.01844868 | 0.00978045 | 0.00453780 | -0.00317481 | -0.00001231 | 0.00029163 | -0.00042698 | 0.00032928 | 0.00006730 | -0.00031652 | -0.00010684 |
| 97 | 145 | 0.05712349 | 0.03761987 | -0.05907479 | -0.00848225 | 0.04224196 | -0.00864806 | -0.01853206 | 0.00899760 | 0.00444870 | -0.00309633 | 0.00004454 | 0.00041323 | -0.00049905 | 0.00023965 | -0.00004211 | -0.00021182 | -0.00016623 |
| 97 | 146 | 0.05716258 | 0.03671130 | -0.06020119 | -0.00731078 | 0.04161356 | -0.01026340 | -0.01801975 | 0.01022186 | 0.00426694 | -0.00329317 | 0.00006289 | 0.00028584 | -0.00045466 | 0.00036804 | 0.00004513 | -0.00032831 | -0.00011059 |
| 97 | 147 | 0.05707172 | 0.03656115 | -0.05969407 | -0.00767768 | 0.04231931 | -0.00945406 | -0.01807811 | 0.00947966 | 0.00420406 | -0.00322827 | 0.00011118 | 0.00040795 | -0.00053084 | 0.00028467 | -0.00005845 | -0.00023329 | -0.00017860 |
| 97 | 148 | 0.05712457 | 0.03574440 | -0.06074223 | -0.00652916 | 0.04161983 | -0.01106805 | -0.01754273 | 0.01064678 | 0.00400463 | -0.00340459 | 0.00013919 | 0.00027926 | -0.00047999 | 0.00040276 | 0.00002429 | -0.00034254 | -0.00011435 |
| 97 | 149 | 0.05702205 | 0.03556912 | -0.06023800 | -0.00691726 | 0.04234426 | -0.01021239 | -0.01757769 | 0.00995233 | 0.00396693 | -0.00335519 | 0.00017816 | 0.00040045 | -0.00055861 | 0.00032618 | -0.00007229 | -0.00025827 | -0.00019008 |
| 97 | 150 | 0.05709110 | 0.03484259 | -0.06121292 | -0.00580318 | 0.04159272 | -0.01181232 | -0.01703413 | 0.01105356 | 0.00375513 | -0.00350936 | 0.00021494 | 0.00027200 | -0.00050259 | 0.00043355 | 0.00000537 | -0.00035961 | -0.00011848 |
| 97 | 151 | 0.05697405 | 0.03465134 | -0.06070532 | -0.00619927 | 0.04232936 | -0.01092214 | -0.01705063 | 0.01041101 | 0.00374268 | -0.00347628 | 0.00024392 | 0.00039097 | -0.00058196 | 0.00036387 | -0.00008327 | -0.00028651 | -0.00020069 |
| 97 | 152 | 0.05706087 | 0.03401033 | -0.06161235 | -0.00512888 | 0.04153990 | -0.01249846 | -0.01650896 | 0.01144056 | 0.00352189 | -0.00360756 | 0.00028870 | 0.00026405 | -0.00052204 | 0.00046054 | -0.00001108 | -0.00037977 | -0.00012315 |
| 97 | 153 | 0.05692694 | 0.03381332 | -0.06109565 | -0.00552173 | 0.04228492 | -0.01158337 | -0.01651411 | 0.01085154 | 0.00353542 | -0.00359071 | 0.00030720 | 0.00037969 | -0.00060057 | 0.00039756 | -0.00009113 | -0.00031763 | -0.00021039 |
| 97 | 154 | 0.05703244 | 0.03325126 | -0.06193971 | -0.00450188 | 0.04146754 | -0.01312900 | -0.01598037 | 0.01180608 | 0.00330752 | -0.00369913 | 0.00035931 | 0.00025540 | -0.00053795 | 0.00048387 | -0.00002468 | -0.00040308 | -0.00012843 |
| 97 | 155 | 0.05687998 | 0.03305838 | -0.06140980 | -0.00488272 | 0.04221887 | -0.01219683 | -0.01598230 | 0.01127032 | 0.00334792 | -0.00369772 | 0.00036702 | 0.00036683 | -0.00061432 | 0.00042716 | -0.00009580 | -0.00035118 | -0.00021911 |
| 97 | 156 | 0.05700428 | 0.03256803 | -0.06219456 | -0.00391777 | 0.04138015 | -0.01370672 | -0.01545944 | 0.01214851 | 0.00311371 | -0.00378392 | 0.00042583 | 0.00024600 | -0.00055004 | 0.00050365 | -0.00003517 | -0.00042945 | -0.00013426 |
| 97 | 157 | 0.05683235 | 0.03238754 | -0.06164977 | -0.00428054 | 0.04213663 | -0.01276416 | -0.01546628 | 0.01166441 | 0.00318164 | -0.00379666 | 0.00042276 | 0.00035259 | -0.00062322 | 0.00045270 | -0.00009735 | -0.00038663 | -0.00022680 |
| 98 | 139 | 0.05753571 | 0.04075058 | -0.05784497 | -0.01057848 | 0.04064044 | -0.00732816 | -0.01914047 | 0.00873885 | 0.00508348 | -0.00285373 | -0.00017757 | 0.00025490 | -0.00030452 | 0.00022566 | 0.00014019 | -0.00034051 | -0.00010705 |
| 98 | 140 | 0.05733680 | 0.03958746 | -0.05888763 | -0.00964950 | 0.04135117 | -0.00782951 | -0.01885458 | 0.00982148 | 0.00525609 | -0.00292782 | -0.00026182 | 0.00011509 | -0.00024503 | 0.00040430 | 0.00026948 | -0.00046934 | -0.00006269 |
| 98 | 141 | 0.05747356 | 0.03963125 | -0.05863115 | -0.00961975 | 0.04085587 | -0.00833589 | -0.01894411 | 0.00919730 | 0.00482430 | -0.00298676 | -0.00011620 | 0.00025090 | -0.00033305 | 0.00027370 | 0.00012026 | -0.00034968 | -0.00011206 |
| 98 | 142 | 0.05723873 | 0.03861238 | -0.05941952 | -0.00876204 | 0.04128879 | -0.00882808 | -0.01858462 | 0.01017521 | 0.00495022 | -0.00304181 | -0.00019027 | 0.00011057 | -0.00026833 | 0.00043968 | 0.00024396 | -0.00047206 | -0.00006105 |
| 98 | 143 | 0.05741764 | 0.03855022 | -0.05935632 | -0.00871355 | 0.04101245 | -0.00928420 | -0.01865911 | 0.00964404 | 0.00456044 | -0.00311163 | -0.00004891 | 0.00024577 | -0.00036035 | 0.00031775 | 0.00009964 | -0.00035911 | -0.00011517 |
| 98 | 144 | 0.05715152 | 0.03768695 | -0.05989281 | -0.00793302 | 0.04119289 | -0.00975890 | -0.01825117 | 0.01050868 | 0.00464827 | -0.00314585 | -0.00011359 | 0.00010605 | -0.00029072 | 0.00046949 | 0.00021736 | -0.00047364 | -0.00005775 |
| 98 | 145 | 0.05736902 | 0.03751273 | -0.06002130 | -0.00786172 | 0.04111985 | -0.01017178 | -0.01830092 | 0.01007788 | 0.00429687 | -0.00322927 | 0.00002260 | 0.00023998 | -0.00038649 | 0.00035784 | 0.00007884 | -0.00036942 | -0.00011729 |
| 98 | 146 | 0.05707636 | 0.03681220 | -0.06031290 | -0.00716207 | 0.04107290 | -0.01062421 | -0.01786846 | 0.01082314 | 0.00435448 | -0.00324122 | -0.00003355 | 0.00010177 | -0.00031220 | 0.00049425 | 0.00019037 | -0.00047498 | -0.00005366 |
| 98 | 147 | 0.05732789 | 0.03652363 | -0.06062632 | -0.00706445 | 0.04118694 | -0.01099855 | -0.01788479 | 0.01049770 | 0.00403810 | -0.00334040 | 0.00009666 | 0.00023384 | -0.00041136 | 0.00039405 | 0.00005842 | -0.00038119 | -0.00011912 |
| 98 | 148 | 0.05701350 | 0.03598871 | -0.06068404 | -0.00644747 | 0.04093671 | -0.01142693 | -0.01744944 | 0.01111988 | 0.00407229 | -0.00332905 | 0.00004816 | 0.00009790 | -0.00033267 | 0.00051452 | 0.00016365 | -0.00047689 | -0.00004948 |
| 98 | 149 | 0.05729375 | 0.03558750 | -0.06117089 | -0.00632040 | 0.04122159 | -0.01176534 | -0.01742543 | 0.01090233 | 0.00378813 | -0.00344550 | 0.00017170 | 0.00022750 | -0.00043474 | 0.00042654 | 0.00003895 | -0.00039495 | -0.00012123 |
| 98 | 150 | 0.05696244 | 0.03521687 | -0.06100932 | -0.00578632 | 0.04079074 | -0.01217044 | -0.01700579 | 0.01140004 | 0.00380445 | -0.00341028 | 0.00013006 | 0.00009446 | -0.00035195 | 0.00053085 | 0.00013785 | -0.00048016 | -0.00004579 |
| 98 | 151 | 0.05726557 | 0.03470869 | -0.06165393 | -0.00562701 | 0.04123063 | -0.01247374 | -0.01693675 | 0.01129052 | 0.00355039 | -0.00354489 | 0.00024627 | 0.00022103 | -0.00045625 | 0.00045543 | 0.00002098 | -0.00041112 | -0.00012398 |
| 98 | 152 | 0.05692209 | 0.03449702 | -0.06129055 | -0.00517488 | 0.04064002 | -0.01285822 | -0.01654775 | 0.01166457 | 0.00355304 | -0.00348567 | 0.00021083 | 0.00009138 | -0.00036975 | 0.00054377 | 0.00011354 | -0.00048544 | -0.00004302 |
| 98 | 153 | 0.05724185 | 0.03389131 | -0.06207396 | -0.00498086 | 0.04121969 | -0.01312574 | -0.01643147 | 0.01166090 | 0.00332764 | -0.00363868 | 0.00031913 | 0.00021439 | -0.00047550 | 0.00048085 | 0.00000499 | -0.00042998 | -0.00012756 |
| 98 | 154 | 0.05689086 | 0.03382967 | -0.06152841 | -0.00460886 | 0.04048823 | -0.01349366 | -0.01608425 | 0.01191410 | 0.00331958 | -0.00355583 | 0.00028933 | 0.00008855 | -0.00038577 | 0.00055374 | 0.00009124 | -0.00049327 | -0.00004145 |
| 98 | 155 | 0.05722093 | 0.03313899 | -0.06242922 | -0.00437808 | 0.04119317 | -0.01372362 | -0.01592099 | 0.01201201 | 0.00312202 | -0.00372686 | 0.00038921 | 0.00020748 | -0.00049206 | 0.00050291 | -0.00000862 | -0.00045168 | -0.00013205 |
| 98 | 156 | 0.05686693 | 0.03321548 | -0.06172268 | -0.00408376 | 0.04033790 | -0.01407988 | -0.01562284 | 0.01214893 | 0.00310506 | -0.00362115 | 0.00036463 | 0.00008578 | -0.00039966 | 0.00056114 | 0.00007138 | -0.00050405 | -0.00004127 |
| 98 | 157 | 0.05720111 | 0.03245480 | -0.06271815 | -0.00381467 | 0.04115427 | -0.01426983 | -0.01541506 | 0.01234238 | 0.00293502 | -0.00380930 | 0.00045567 | 0.00020020 | -0.00050557 | 0.00052170 | -0.00001958 | -0.00047618 | -0.00013739 |

TABLE 3. Nuclear charge density distribution FB coefficients.

| $Z$ | $N$ | $a_1$ | $a_2$ | $a_3$ | $a_4$ | $a_5$ | $a_6$ | $a_7$ | $a_8$ | $a_9$ | $a_{10}$ | $a_{11}$ | $a_{12}$ | $a_{13}$ | $a_{14}$ | $a_{15}$ | $a_{16}$ | $a_{17}$ |
|---|---|---|---|---|---|---|---|---|---|---|---|---|---|---|---|---|---|---|
| 98 | 158 | 0.05684829 | 0.03265529 | -0.06187241 | -0.00359509 | 0.04019036 | -0.01461973 | -0.01516986 | 0.01236906 | 0.00291006 | -0.00368188 | 0.00043599 | 0.00008289 | -0.00041112 | 0.00056628 | 0.00005426 | -0.00051804 | -0.00004253 |
| 99 | 140 | 0.05768811 | 0.04044474 | -0.05827508 | -0.01021362 | 0.04019416 | -0.00814182 | -0.01898939 | 0.00907247 | 0.00483383 | -0.00292573 | -0.00014391 | 0.00020465 | -0.00027226 | 0.00025909 | 0.00014893 | -0.00039150 | -0.00012104 |
| 99 | 141 | 0.05752037 | 0.04051061 | -0.05747427 | -0.01050046 | 0.04048352 | -0.00770348 | -0.01923417 | 0.00802812 | 0.00463346 | -0.00281595 | -0.00005309 | 0.00033543 | -0.00033707 | 0.00013342 | 0.00001631 | -0.00025077 | -0.00015510 |
| 99 | 142 | 0.05762322 | 0.03937857 | -0.05902497 | -0.00931056 | 0.04038539 | -0.00910646 | -0.01877982 | 0.00950878 | 0.00458129 | -0.00304975 | -0.00008234 | 0.00019857 | -0.00029632 | 0.00030392 | 0.00013106 | -0.00040153 | -0.00012346 |
| 99 | 143 | 0.05746275 | 0.03940728 | -0.05827112 | -0.00966382 | 0.04079755 | -0.00856473 | -0.01903298 | 0.00846695 | 0.00442236 | -0.00293962 | 0.00000435 | 0.00033178 | -0.00036498 | 0.00017838 | 0.00000160 | -0.00026426 | -0.00016311 |
| 99 | 144 | 0.05756634 | 0.03834757 | -0.05972212 | -0.00845462 | 0.04053252 | -0.01001359 | -0.01849343 | 0.00993342 | 0.00432692 | -0.00316638 | -0.00001565 | 0.00019199 | -0.00031969 | 0.00034495 | 0.00011245 | -0.00041151 | -0.00012427 |
| 99 | 145 | 0.05741099 | 0.03833209 | -0.05901792 | -0.00886093 | 0.04104671 | -0.00938702 | -0.01874163 | 0.00891098 | 0.00420352 | -0.00306104 | 0.00006628 | 0.00032627 | -0.00039162 | 0.00022164 | -0.00001258 | -0.00028027 | -0.00017001 |
| 99 | 146 | 0.05751834 | 0.03735555 | -0.06036771 | -0.00764753 | 0.04064360 | -0.01086211 | -0.01814329 | 0.01034559 | 0.00407480 | -0.00327645 | 0.00005470 | 0.00018535 | -0.00034251 | 0.00038227 | 0.00009350 | -0.00042198 | -0.00012430 |
| 99 | 147 | 0.05736548 | 0.03729263 | -0.05971229 | -0.00809333 | 0.04124076 | -0.01016673 | -0.01837584 | 0.00935711 | 0.00398209 | -0.00318029 | 0.00013108 | 0.00031940 | -0.00041702 | 0.00026299 | -0.00002593 | -0.00029892 | -0.00017641 |
| 99 | 148 | 0.05747924 | 0.03640610 | -0.06096220 | -0.00688957 | 0.04072563 | -0.01165206 | -0.01774216 | 0.01074455 | 0.00382854 | -0.00338066 | 0.00012726 | 0.00017890 | -0.00036473 | 0.00041602 | 0.00007469 | -0.00043351 | -0.00012427 |
| 99 | 149 | 0.05732597 | 0.03629646 | -0.06035143 | -0.00736153 | 0.04138930 | -0.01090115 | -0.01795174 | 0.00980188 | 0.00376305 | -0.00329711 | 0.00019721 | 0.00031154 | -0.00044099 | 0.00030219 | -0.00003809 | -0.00032028 | -0.00018274 |
| 99 | 150 | 0.05744847 | 0.03550286 | -0.06150510 | -0.00617969 | 0.04078454 | -0.01238433 | -0.01730228 | 0.01112950 | 0.00359128 | -0.00347953 | 0.00020070 | 0.00017278 | -0.00038616 | 0.00044634 | 0.00005650 | -0.00044660 | -0.00012471 |
| 99 | 151 | 0.05729169 | 0.03535092 | -0.06093229 | -0.00666513 | 0.04150128 | -0.01158846 | -0.01748521 | 0.01024156 | 0.00355099 | -0.00341098 | 0.00026326 | 0.00030290 | -0.00046323 | 0.00033898 | -0.00004870 | -0.00034433 | -0.00018924 |
| 99 | 152 | 0.05742478 | 0.03464948 | -0.06199508 | -0.00551573 | 0.04082527 | -0.01306042 | -0.01683526 | 0.01149952 | 0.00336567 | -0.00357341 | 0.00027378 | 0.00016701 | -0.00040648 | 0.00047339 | 0.00003943 | -0.00046169 | -0.00012601 |
| 99 | 153 | 0.05726150 | 0.03446284 | -0.06145177 | -0.00600309 | 0.04158448 | -0.01222773 | -0.01699126 | 0.01067236 | 0.00334992 | -0.00352123 | 0.00032798 | 0.00029362 | -0.00048332 | 0.00037309 | -0.00005746 | -0.00037096 | -0.00019598 |
| 99 | 154 | 0.05740663 | 0.03384959 | -0.06243021 | -0.00489477 | 0.04085167 | -0.01368221 | -0.01635180 | 0.01185352 | 0.00315390 | -0.00366249 | 0.00034542 | 0.00016151 | -0.00042529 | 0.00049728 | 0.00002391 | -0.00047910 | -0.00012836 |
| 99 | 155 | 0.05723403 | 0.03363817 | -0.06190711 | -0.00537401 | 0.04164531 | -0.01281878 | -0.01648354 | 0.01109053 | 0.00316304 | -0.00362708 | 0.00039038 | 0.00028372 | -0.00050085 | 0.00040427 | -0.00006409 | -0.00039996 | -0.00020291 |
| 99 | 156 | 0.05739215 | 0.03310674 | -0.06280810 | -0.00431339 | 0.04086660 | -0.01425171 | -0.01586158 | 0.01219029 | 0.00295764 | -0.00374682 | 0.00041468 | 0.00015613 | -0.00044219 | 0.00051810 | 0.00001033 | -0.00049904 | -0.00013186 |
| 99 | 157 | 0.05720788 | 0.03288168 | -0.06229607 | -0.00477644 | 0.04168864 | -0.01336219 | -0.01597391 | 0.01149257 | 0.00299277 | -0.00372776 | 0.00044965 | 0.00027322 | -0.00051546 | 0.00043229 | -0.00006843 | -0.00043104 | -0.00020990 |
| 99 | 158 | 0.05737950 | 0.03242417 | -0.06312628 | -0.00376806 | 0.04087194 | -0.01477110 | -0.01537314 | 0.01250851 | 0.00277811 | -0.00382630 | 0.00048077 | 0.00015073 | -0.00045676 | 0.00053594 | -0.00000104 | -0.00052156 | -0.00013648 |
| 99 | 159 | 0.05718167 | 0.03219663 | -0.06261742 | -0.00420895 | 0.04171772 | -0.01385923 | -0.01547222 | 0.01187541 | 0.00284058 | -0.00382255 | 0.00050523 | 0.00026212 | -0.00052688 | 0.00045703 | -0.00007043 | -0.00046383 | -0.00021677 |
| 100 | 141 | 0.05782724 | 0.04019141 | -0.05864460 | -0.00992186 | 0.03973203 | -0.00887270 | -0.01882121 | 0.00937789 | 0.00459957 | -0.00298975 | -0.00011145 | 0.00015206 | -0.00023676 | 0.00028861 | 0.00016042 | -0.00044229 | -0.00013165 |
| 100 | 142 | 0.05768252 | 0.03895327 | -0.05990195 | -0.00892542 | 0.04061504 | -0.00933538 | -0.01849058 | 0.01057611 | 0.00480431 | -0.00306168 | -0.00021000 | 0.00001354 | -0.00017967 | 0.00048454 | 0.00029377 | -0.00057572 | -0.00008812 |
| 100 | 143 | 0.05776051 | 0.03917887 | -0.05935657 | -0.00906983 | 0.03989972 | -0.00979708 | -0.01860278 | 0.00979310 | 0.00435441 | -0.00310583 | -0.00005012 | 0.00014421 | -0.00025681 | 0.00033033 | 0.00014445 | -0.00045299 | -0.00013174 |
| 100 | 144 | 0.05758226 | 0.03806922 | -0.06037333 | -0.00813298 | 0.04052266 | -0.01025224 | -0.01820789 | 0.01088964 | 0.00451190 | -0.00316206 | -0.00013907 | 0.00000719 | -0.00019744 | 0.00051481 | 0.00027110 | -0.00057940 | -0.00008360 |
| 100 | 145 | 0.05770319 | 0.03819826 | -0.06002337 | -0.00826017 | 0.04003642 | -0.01066567 | -0.01831797 | 0.01019670 | 0.00410989 | -0.00321503 | 0.00001560 | 0.00013641 | -0.00027667 | 0.00036847 | 0.00012766 | -0.00046331 | -0.00013047 |
| 100 | 146 | 0.05749516 | 0.03722857 | -0.06079937 | -0.00738950 | 0.04041811 | -0.01110626 | -0.01787880 | 0.01118406 | 0.00422741 | -0.00325377 | -0.00006426 | 0.00000156 | -0.00021493 | 0.00054000 | 0.00024744 | -0.00058174 | -0.00007783 |
| 100 | 147 | 0.05765595 | 0.03725223 | -0.06064662 | -0.00749443 | 0.04014873 | -0.01147770 | -0.01797784 | 0.01058819 | 0.00386931 | -0.00331813 | 0.00008445 | 0.00012905 | -0.00029649 | 0.00040316 | 0.00011040 | -0.00047378 | -0.00012863 |
| 100 | 148 | 0.05742193 | 0.03643092 | -0.06118513 | -0.00669479 | 0.04030773 | -0.01189973 | -0.01751374 | 0.01146076 | 0.00395352 | -0.00333789 | 0.00001310 | -0.00000313 | -0.00023222 | 0.00056062 | 0.00022329 | -0.00058347 | -0.00007154 |
| 100 | 149 | 0.05761874 | 0.03634340 | -0.06122686 | -0.00677289 | 0.04024216 | -0.01223341 | -0.01759298 | 0.01096717 | 0.00363551 | -0.00341582 | 0.00015522 | 0.00012236 | -0.00031626 | 0.00043457 | 0.00009306 | -0.00048491 | -0.00012688 |
| 100 | 150 | 0.05736253 | 0.03567569 | -0.06153449 | -0.00604748 | 0.04019640 | -0.01263552 | -0.01712204 | 0.01172113 | 0.00369227 | -0.00341546 | 0.00009175 | -0.00000679 | -0.00024929 | 0.00057718 | 0.00019913 | -0.00058529 | -0.00006537 |
| 100 | 151 | 0.05759087 | 0.03547450 | -0.06176371 | -0.00609469 | 0.04032108 | -0.01293381 | -0.01717356 | 0.01133315 | 0.00341091 | -0.00350865 | 0.00022678 | 0.00011644 | -0.00033582 | 0.00046285 | 0.00007606 | -0.00049721 | -0.00012575 |
| 100 | 152 | 0.05731627 | 0.03496231 | -0.06185009 | -0.00544521 | 0.04008770 | -0.01331677 | -0.01671184 | 0.01196645 | 0.00344513 | -0.00348736 | 0.00017059 | -0.00000940 | -0.00026601 | 0.00059018 | 0.00017543 | -0.00058785 | -0.00005984 |
| 100 | 153 | 0.05757106 | 0.03464852 | -0.06225573 | -0.00545797 | 0.04038881 | -0.01358040 | -0.01672915 | 0.01168550 | 0.00319749 | -0.00359703 | 0.00029810 | 0.00011128 | -0.00035489 | 0.00048814 | 0.00005984 | -0.00051114 | -0.00012562 |
| 100 | 154 | 0.05728192 | 0.03429038 | -0.06213323 | -0.00488497 | 0.03998414 | -0.01394670 | -0.01629039 | 0.01219780 | 0.00321310 | -0.00355439 | 0.00024865 | -0.00001105 | -0.00028218 | 0.00060010 | 0.00015264 | -0.00059174 | -0.00005535 |
| 100 | 155 | 0.05755765 | 0.03386861 | -0.06270071 | -0.00486017 | 0.04044778 | -0.01417496 | -0.01626868 | 0.01202340 | 0.00299689 | -0.00368122 | 0.00036824 | 0.00010676 | -0.00037310 | 0.00051056 | 0.00004480 | -0.00052704 | -0.00012670 |
| 100 | 156 | 0.05725783 | 0.03365982 | -0.06238420 | -0.00436322 | 0.03988712 | -0.01452834 | -0.01586380 | 0.01241602 | 0.00299684 | -0.00361717 | 0.00032507 | -0.00001190 | -0.00029754 | 0.00060734 | 0.00013117 | -0.00059751 | -0.00005218 |
| 100 | 157 | 0.05754872 | 0.03313805 | -0.06309587 | -0.00429832 | 0.04049948 | -0.01471937 | -0.01580028 | 0.01234589 | 0.00281036 | -0.00376135 | 0.00043638 | 0.00010272 | -0.00039006 | 0.00053019 | 0.00003129 | -0.00054516 | -0.00012910 |
| 100 | 158 | 0.05724208 | 0.03307092 | -0.06260218 | -0.00387622 | 0.03979729 | -0.01506449 | -0.01543745 | 0.01262166 | 0.00279666 | -0.00367620 | 0.00039913 | -0.00001214 | -0.00031179 | 0.00061226 | 0.00011138 | -0.00060555 | -0.00005053 |
| 100 | 159 | 0.05754225 | 0.03246013 | -0.06343808 | -0.00376926 | 0.04054464 | -0.01521554 | -0.01533125 | 0.01265182 | 0.00263884 | -0.00383741 | 0.00050183 | 0.00009896 | -0.00040533 | 0.00054709 | 0.00001962 | -0.00056563 | -0.00013281 |
| 100 | 160 | 0.05723264 | 0.03252439 | -0.06278564 | -0.00342020 | 0.03971455 | -0.01555751 | -0.01501585 | 0.01281497 | 0.00261272 | -0.00373181 | 0.00047020 | -0.00001198 | -0.00032464 | 0.00061514 | 0.00009359 | -0.00061621 | -0.00005050 |
| 101 | 144 | 0.05788693 | 0.03902142 | -0.05963798 | -0.00888383 | 0.03940889 | -0.01041801 | -0.01841658 | 0.01005405 | 0.00414228 | -0.00315642 | -0.00001931 | 0.00008895 | -0.00021569 | 0.00035337 | 0.00015967 | -0.00050360 | -0.00013747 |
| 101 | 145 | 0.05763620 | 0.03918446 | -0.05859375 | -0.00936053 | 0.03960589 | -0.00990529 | -0.01879515 | 0.00880086 | 0.00397104 | -0.00303350 | 0.00008115 | 0.00023062 | -0.00028577 | 0.00020152 | 0.00002398 | -0.00035336 | -0.00017273 |
| 101 | 146 | 0.05782957 | 0.03809048 | -0.06027335 | -0.00811708 | 0.03953424 | -0.01125031 | -0.01813546 | 0.01043775 | 0.00390759 | -0.00325893 | 0.00004511 | 0.00008015 | -0.00023246 | 0.00038883 | 0.00014453 | -0.00051413 | -0.00013441 |
| 101 | 147 | 0.05758981 | 0.03820808 | -0.05930111 | -0.00863611 | 0.03986170 | -0.01066630 | -0.01851897 | 0.00920464 | 0.00377423 | -0.00313901 | 0.00014277 | 0.00022105 | -0.00030263 | 0.00023748 | 0.00001413 | -0.00036969 | -0.00017292 |
| 101 | 148 | 0.05778321 | 0.03719011 | -0.06087227 | -0.00738996 | 0.03964545 | -0.01202803 | -0.01780644 | 0.01080969 | 0.00367833 | -0.00335565 | 0.00011219 | 0.00007221 | -0.00024961 | 0.00042110 | 0.00012878 | -0.00052448 | -0.00013093 |
| 101 | 149 | 0.05755255 | 0.03725605 | -0.05997381 | -0.00793891 | 0.04008622 | -0.01138588 | -0.01818256 | 0.00961157 | 0.00357775 | -0.00324303 | 0.00020621 | 0.00021136 | -0.00031968 | 0.00027229 | 0.00000456 | -0.00038771 | -0.00017319 |
| 101 | 150 | 0.05774777 | 0.03632208 | -0.06143549 | -0.00670263 | 0.03974685 | -0.01275163 | -0.01743841 | 0.01116967 | 0.00345667 | -0.00344724 | 0.00018094 | 0.00006534 | -0.00026715 | 0.00045034 | 0.00011278 | -0.00053514 | -0.00012768 |
| 101 | 151 | 0.05752380 | 0.03633331 | -0.06060939 | -0.00726957 | 0.04028512 | -0.01206182 | -0.01779700 | 0.01001937 | 0.00338486 | -0.00334550 | 0.00027035 | 0.00020184 | -0.00033690 | 0.00030585 | -0.00000450 | -0.00040758 | -0.00017403 |
| 101 | 152 | 0.05772242 | 0.03548838 | -0.06196268 | -0.00605431 | 0.03984153 | -0.01342230 | -0.01703972 | 0.01151747 | 0.00324442 | -0.00353430 | 0.00025040 | 0.00005960 | -0.00028495 | 0.00047671 | 0.00009689 | -0.00054662 | -0.00012513 |
| 101 | 153 | 0.05750249 | 0.03544525 | -0.06120453 | -0.00662803 | 0.04046331 | -0.01269268 | -0.01737338 | 0.01042539 | 0.00319860 | -0.00344617 | 0.00033414 | 0.00019264 | -0.00035407 | 0.00033800 | -0.00001276 | -0.00042941 | -0.00017575 |
| 101 | 154 | 0.05770593 | 0.03469134 | -0.06245245 | -0.00544335 | 0.03993157 | -0.01404157 | -0.01661824 | 0.01185272 | 0.00304298 | -0.00361729 | 0.00031972 | 0.00005494 | -0.00030276 | 0.00050035 | 0.00008148 | -0.00055936 | -0.00012365 |
| 101 | 155 | 0.05748711 | 0.03459763 | -0.06175530 | -0.00601364 | 0.04062487 | -0.01327766 | -0.01692254 | 0.01082673 | 0.00302170 | -0.00354462 | 0.00039668 | 0.00018382 | -0.00037085 | 0.00036849 | -0.00001994 | -0.00045324 | -0.00017847 |
| 101 | 156 | 0.05769657 | 0.03393360 | -0.06290247 | -0.00486746 | 0.04001819 | -0.01461121 | -0.01618130 | 0.01217487 | 0.00285350 | -0.00369656 | 0.00038812 | 0.00005126 | -0.00032024 | 0.00052137 | 0.00006692 | -0.00057374 | -0.00012348 |
| 101 | 157 | 0.05747598 | 0.03379608 | -0.06225745 | -0.00542545 | 0.04077283 | -0.01381657 | -0.01645466 | 0.01122034 | 0.00285643 | -0.00364035 | 0.00045720 | 0.00017530 | -0.00038681 | 0.00039707 | -0.00002580 | -0.00047902 | -0.00018219 |
| 101 | 158 | 0.05769238 | 0.03321810 | -0.06330971 | -0.00432401 | 0.04010182 | -0.01513296 | -0.01573563 | 0.01248317 | 0.00267686 | -0.00377230 | 0.00045490 | 0.00004835 | -0.00033702 | 0.00053984 | 0.00005356 | -0.00059004 | -0.00012474 |
| 101 | 159 | 0.05746730 | 0.03304599 | -0.06270678 | -0.00486236 | 0.04090929 | -0.01430968 | -0.01597908 | 0.01160312 | 0.00270463 | -0.00373274 | 0.00051507 | 0.00016700 | -0.00040152 | 0.00042346 | -0.00003013 | -0.00050662 | -0.00018679 |
| 101 | 160 | 0.05769129 | 0.03254800 | -0.06367070 | -0.00381021 | 0.04018211 | -0.01560856 | -0.01528736 | 0.01277668 | 0.00251368 | -0.00384459 | 0.00051943 | 0.00004601 | -0.00035269 | 0.00055582 | 0.00004167 | -0.00060847 | -0.00012744 |
| 101 | 161 | 0.05745927 | 0.03235207 | -0.06309945 | -0.00432328 | 0.04103530 | -0.01475781 | -0.01550401 | 0.01197210 | 0.00256762 | -0.00382117 | 0.00056981 | 0.00015879 | -0.00041454 | 0.00044740 | -0.00003279 | -0.00053582 | -0.00019208 |
| 102 | 146 | 0.05788130 | 0.03768009 | -0.06112888 | -0.00772441 | 0.03971469 | -0.01142499 | -0.01779700 | 0.01150615 | 0.00412100 | -0.00326305 | -0.00009068 | -0.00009745 | -0.00012258 | 0.00057532 | 0.00030227 | -0.00068172 | -0.00009818 |
| 102 | 147 | 0.05794640 | 0.03801417 | -0.06048290 | -0.00801417 | 0.03903394 | -0.01177637 | -0.01794827 | 0.01065986 | 0.00371850 | -0.00329907 | 0.00007305 | 0.00002412 | -0.00018801 | 0.00040657 | 0.00016240 | -0.00056364 | -0.00013661 |
| 102 | 148 | 0.05779578 | 0.03691995 | -0.06150780 | -0.00705405 | 0.03960390 | -0.01220995 | -0.01748087 | 0.01176646 | 0.00385578 | -0.00334432 | -0.00001903 | -0.00010375 | -0.00013622 | 0.00059665 | 0.00028125 | -0.00068446 | -0.00009057 |
| 102 | 149 | 0.05790103 | 0.03715846 | -0.06105633 | -0.00732337 | 0.03914201 | -0.01252158 | -0.01763093 | 0.01101337 | 0.00350014 | -0.00338995 | 0.00013816 | 0.00001572 | -0.00020282 | 0.00043663 | 0.00014804 | -0.00057382 | -0.00013178 |
| 102 | 150 | 0.05772498 | 0.03619615 | -0.06185684 | -0.00642504 | 0.03950081 | -0.01293887 | -0.01713886 | 0.01201020 | 0.00360276 | -0.00341860 | 0.00005449 | -0.00010865 | -0.00015015 | 0.00061389 | 0.00025964 | -0.00068628 | -0.00008265 |
| 102 | 151 | 0.05786718 | 0.03633087 | -0.06160047 | -0.00666851 | 0.03924816 | -0.01321485 | -0.01727972 | 0.01135537 | 0.00329012 | -0.00347589 | 0.00020469 | 0.00000870 | -0.00021835 | 0.00046388 | 0.00013326 | -0.00058399 | -0.00012724 |
| 102 | 152 | 0.05766867 | 0.03550738 | -0.06217933 | -0.00583604 | 0.03940821 | -0.01361469 | -0.01677763 | 0.01223875 | 0.00336296 | -0.00348686 | 0.00012903 | -0.00011208 | -0.00016434 | 0.00062747 | 0.00023781 | -0.00068777 | -0.00007496 |
| 102 | 153 | 0.05784409 | 0.03553272 | -0.06211509 | -0.00604877 | 0.03935453 | -0.01385750 | -0.01690152 | 0.01168586 | 0.00308971 | -0.00355750 | 0.00027189 | 0.00000310 | -0.00023450 | 0.00048850 | 0.00011839 | -0.00059465 | -0.00012345 |
| 102 | 154 | 0.05762606 | 0.03485246 | -0.06247758 | -0.00528500 | 0.03932777 | -0.01424057 | -0.01640285 | 0.01245337 | 0.00313690 | -0.00354996 | 0.00020381 | -0.00011406 | -0.00017873 | 0.00063783 | 0.00021610 | -0.00068949 | -0.00006792 |
| 102 | 155 | 0.05783042 | 0.03476573 | -0.06259894 | -0.00546261 | 0.03946231 | -0.01445119 | -0.01650273 | 0.01200468 | 0.00289986 | -0.00363530 | 0.00033905 | -0.00000112 | -0.00025107 | 0.00051059 | 0.00010376 | -0.00060623 | -0.00012076 |
| 102 | 156 | 0.05759593 | 0.03423047 | -0.06275262 | -0.00476926 | 0.03926026 | -0.01481960 | -0.01601942 | 0.01265521 | 0.00292482 | -0.00360871 | 0.00027815 | -0.00011470 | -0.00019316 | 0.00064535 | 0.00019485 | -0.00069199 | -0.00006190 |
| 102 | 157 | 0.05782454 | 0.03403199 | -0.06304970 | -0.00490794 | 0.03957170 | -0.01499767 | -0.01608928 | 0.01231150 | 0.00272126 | -0.00370970 | 0.00040552 | -0.00000412 | -0.00026774 | 0.00053028 | 0.00008971 | -0.00061914 | -0.00011941 |
| 102 | 158 | 0.05757662 | 0.03364079 | -0.06300445 | -0.00428583 | 0.03920569 | -0.01535461 | -0.01563150 | 0.01284519 | 0.00272670 | -0.00366378 | 0.00035143 | -0.00011414 | -0.00020744 | 0.00065039 | 0.00017439 | -0.00069574 | -0.00005716 |
| 102 | 159 | 0.05782443 | 0.03333406 | -0.06346430 | -0.00438237 | 0.03968227 | -0.01549861 | -0.01566670 | 0.01260579 | 0.00255445 | -0.00378098 | 0.00047071 | -0.00000607 | -0.00028416 | 0.00054764 | 0.00007654 | -0.00063369 | -0.00011953 |
| 102 | 160 | 0.05756627 | 0.03308329 | -0.06323212 | -0.00383154 | 0.03916335 | -0.01584815 | -0.01524266 | 0.01302403 | 0.00254238 | -0.00371574 | 0.00042311 | -0.00011258 | -0.00022131 | 0.00065328 | 0.00015504 | -0.00070118 | -0.00005391 |
| 102 | 161 | 0.05782802 | 0.03267479 | -0.06383900 | -0.00388339 | 0.03979297 | -0.01595563 | -0.01524003 | 0.01288683 | 0.00239978 | -0.00384930 | 0.00053408 | -0.00000723 | -0.00029995 | 0.00056271 | 0.00006454 | -0.00065011 | -0.00012118 |
| 102 | 162 | 0.05756289 | 0.03255834 | -0.06343391 | -0.00340317 | 0.03913205 | -0.01630235 | -0.01485591 | 0.01319221 | 0.00237161 | -0.00376503 | 0.00049270 | -0.00011024 | -0.00023452 | 0.00065427 | 0.00013705 | -0.00070866 | -0.00005228 |
| 103 | 148 | 0.05805443 | 0.03796114 | -0.06066062 | -0.00794211 | 0.03854178 | -0.01225143 | -0.01775861 | 0.01086562 | 0.00354133 | -0.00333615 | 0.00009951 | -0.00003094 | -0.00014406 | 0.00042216 | 0.00018077 | -0.00061161 | -0.00013754 |
| 103 | 149 | 0.05772999 | 0.03823971 | -0.05938049 | -0.00855630 | 0.03864871 | -0.01170428 | -0.01826084 | 0.00943490 | 0.00339146 | -0.00320631 | 0.00021101 | 0.00011152 | -0.00020873 | 0.00024519 | 0.00004914 | -0.00045925 | -0.00016884 |
| 103 | 150 | 0.05801008 | 0.03714866 | -0.06120803 | -0.00728569 | 0.03864490 | -0.01296575 | -0.01745308 | 0.01120184 | 0.00333337 | -0.00342169 | 0.00016249 | -0.00003972 | -0.00015681 | 0.00045023 | 0.00016766 | -0.00062158 | -0.00013160 |
| 103 | 151 | 0.05770031 | 0.03737779 | -0.06001589 | -0.00792268 | 0.03888424 | -0.01237089 | -0.01795562 | 0.00980535 | 0.00321757 | -0.00329737 | 0.00027192 | 0.00010008 | -0.00021932 | 0.00027425 | 0.00004225 | -0.00047595 | -0.00016444 |
| 103 | 152 | 0.05797772 | 0.03636068 | -0.06173171 | -0.00666185 | 0.03875294 | -0.01363020 | -0.01711831 | 0.01152692 | 0.00313440 | -0.00350241 | 0.00022670 | -0.00004687 | -0.00017054 | 0.00047569 | 0.00015401 | -0.00063128 | -0.00012600 |
| 103 | 153 | 0.05768077 | 0.03653390 | -0.06062954 | -0.00731073 | 0.03911168 | -0.01299606 | -0.01761008 | 0.01017765 | 0.00304867 | -0.00338701 | 0.00033306 | 0.00008961 | -0.00023108 | 0.00030270 | 0.00003540 | -0.00049372 | -0.00016090 |
| 103 | 154 | 0.05795656 | 0.03559795 | -0.06223155 | -0.00606970 | 0.03886729 | -0.01424629 | -0.01675994 | 0.01184101 | 0.00294523 | -0.00357892 | 0.00029151 | -0.00005235 | -0.00018518 | 0.00049869 | 0.00014010 | -0.00064116 | -0.00012115 |
| 103 | 155 | 0.05767023 | 0.03571155 | -0.06121859 | -0.00672027 | 0.03933328 | -0.01357891 | -0.01723169 | 0.01055017 | 0.00288655 | -0.00347525 | 0.00039372 | 0.00008024 | -0.00024388 | 0.00033045 | 0.00002881 | -0.00051277 | -0.00015856 |
| 103 | 156 | 0.05794533 | 0.03486165 | -0.06270643 | -0.00550770 | 0.03898841 | -0.01481578 | -0.01638317 | 0.01214412 | 0.00276641 | -0.00365179 | 0.00035639 | -0.00005625 | -0.00020053 | 0.00051936 | 0.00012623 | -0.00065166 | -0.00011739 |
| 103 | 157 | 0.05766708 | 0.03491495 | -0.06177915 | -0.00615078 | 0.03955057 | -0.01411901 | -0.01682777 | 0.01092093 | 0.00273287 | -0.00356197 | 0.00045325 | 0.00007194 | -0.00025746 | 0.00035734 | 0.00002272 | -0.00053328 | -0.00015761 |
| 103 | 158 | 0.05794238 | 0.03415335 | -0.06315420 | -0.00497390 | 0.03911587 | -0.01534044 | -0.01599275 | 0.01243613 | 0.00259831 | -0.00372148 | 0.00042078 | -0.00005869 | -0.00021633 | 0.00053780 | 0.00011269 | -0.00066318 | -0.00011496 |
| 103 | 159 | 0.05766947 | 0.03414887 | -0.06230655 | -0.00560140 | 0.03976412 | -0.01461624 | -0.01640548 | 0.01128769 | 0.00258901 | -0.00364694 | 0.00051110 | 0.00006461 | -0.00027146 | 0.00038313 | 0.00001738 | -0.00055534 | -0.00015811 |
| 103 | 160 | 0.05794579 | 0.03347512 | -0.06357181 | -0.00446608 | 0.03924856 | -0.01582197 | -0.01559312 | 0.01271670 | 0.00244112 | -0.00378833 | 0.00048421 | -0.00005987 | -0.00023226 | 0.00055409 | 0.00009977 | -0.00067606 | -0.00011402 |
| 103 | 161 | 0.05767542 | 0.03341833 | -0.06279563 | -0.00507121 | 0.03997385 | -0.01507079 | -0.01597162 | 0.01164799 | 0.00245616 | -0.00372981 | 0.00056682 | 0.00005811 | -0.00028545 | 0.00040757 | 0.00001299 | -0.00057896 | -0.00015999 |
| 103 | 162 | 0.05795340 | 0.03282950 | -0.06395557 | -0.00398193 | 0.03938477 | -0.01626183 | -0.01518835 | 0.01298529 | 0.00229495 | -0.00385258 | 0.00054622 | -0.00006004 | -0.00024795 | 0.00056827 | 0.00008774 | -0.00069059 | -0.00011462 |
| 103 | 163 | 0.05768289 | 0.03272837 | -0.06324121 | -0.00455936 | 0.04017882 | -0.01548312 | -0.01553250 | 0.01199923 | 0.00233521 | -0.00381018 | 0.00062002 | 0.00005226 | -0.00029899 | 0.00043035 | 0.00000971 | -0.00060409 | -0.00016313 |
| 104 | 149 | 0.05815423 | 0.03792486 | -0.06081323 | -0.00789335 | 0.03806245 | -0.01268202 | -0.01756841 | 0.01105713 | 0.00337504 | -0.00337066 | 0.00012457 | -0.00008449 | -0.00010117 | 0.00043602 | 0.00019923 | -0.00065783 | -0.00013756 |
| 104 | 150 | 0.05805915 | 0.03670870 | -0.06207437 | -0.00685783 | 0.03879434 | -0.01312231 | -0.01707350 | 0.01227373 | 0.00352553 | -0.00342260 | 0.00002286 | -0.00020509 | -0.00005959 | 0.00064251 | 0.00031547 | -0.00077993 | -0.00009855 |
| 104 | 151 | 0.05811092 | 0.03715387 | -0.06133464 | -0.00726959 | 0.03815923 | -0.01336684 | -0.01727451 | 0.01137719 | 0.00317693 | -0.00345132 | 0.00018532 | -0.00009359 | -0.00011211 | 0.00046230 | 0.00018726 | -0.00066761 | -0.00013074 |
| 104 | 152 | 0.05799120 | 0.03605334 | -0.06238713 | -0.00628733 | 0.03869585 | -0.01379218 | -0.01675626 | 0.01248904 | 0.00329295 | -0.00348832 | 0.00009186 | -0.00021003 | -0.00007090 | 0.00065713 | 0.00029613 | -0.00078176 | -0.00008966 |
| 104 | 153 | 0.05807989 | 0.03640434 | -0.06183701 | -0.00667550 | 0.03826667 | -0.01400378 | -0.01695554 | 0.01168639 | 0.00298835 | -0.00352722 | 0.00024711 | -0.00010085 | -0.00012424 | 0.00048615 | 0.00017465 | -0.00067687 | -0.00012428 |
| 104 | 154 | 0.05793781 | 0.03542724 | -0.06268080 | -0.00575091 | 0.03861597 | -0.01441328 | -0.01642553 | 0.01269004 | 0.00307378 | -0.00354832 | 0.00016167 | -0.00021325 | -0.00008277 | 0.00066843 | 0.00027642 | -0.00078293 | -0.00008100 |
| 104 | 155 | 0.05806042 | 0.03567655 | -0.06232040 | -0.00611006 | 0.03838567 | -0.01459454 | -0.01661609 | 0.01198500 | 0.00280976 | -0.00359899 | 0.00030948 | -0.00010628 | -0.00013748 | 0.00050769 | 0.00016166 | -0.00068604 | -0.00011855 |
| 104 | 156 | 0.05789818 | 0.03482882 | -0.06295733 | -0.00524654 | 0.03855521 | -0.01498878 | -0.01608525 | 0.01287794 | 0.00286793 | -0.00360344 | 0.00023180 | -0.00021479 | -0.00009514 | 0.00067676 | 0.00025659 | -0.00078393 | -0.00007295 |
| 104 | 157 | 0.05705125 | 0.03497116 | -0.06278392 | -0.00557177 | 0.03851618 | -0.01514099 | -0.01626036 | 0.01227319 | 0.00264138 | -0.00366721 | 0.00037198 | -0.00010993 | -0.00015167 | 0.00052705 | 0.00014853 | -0.00069555 | -0.00011388 |
| 104 | 158 | 0.05787111 | 0.03425673 | -0.06321758 | -0.00477177 | 0.03851326 | -0.01552171 | -0.01573865 | 0.01305386 | 0.00267507 | -0.00365448 | 0.00030178 | -0.00021474 | -0.00010791 | 0.00068245 | 0.00023689 | -0.00078523 | -0.00006580 |
| 104 | 159 | 0.05805079 | 0.03428925 | -0.06322557 | -0.00505869 | 0.03865726 | -0.01564496 | -0.01589214 | 0.01255098 | 0.00248326 | -0.00373240 | 0.00043420 | -0.00011194 | -0.00016657 | 0.00054432 | 0.00013552 | -0.00070579 | -0.00011050 |
| 104 | 160 | 0.05785503 | 0.03371000 | -0.06346152 | -0.00432394 | 0.03848920 | -0.01601480 | -0.01538842 | 0.01321875 | 0.00249475 | -0.00370213 | 0.00037120 | -0.00021325 | -0.00012093 | 0.00068580 | 0.00021759 | -0.00078727 | -0.00005979 |
| 104 | 161 | 0.05805712 | 0.03363245 | -0.06364248 | -0.00456872 | 0.03880736 | -0.01610810 | -0.01551491 | 0.01281820 | 0.00233534 | -0.00379495 | 0.00049574 | -0.00011252 | -0.00018189 | 0.00055959 | 0.00012291 | -0.00071714 | -0.00010856 |
| 104 | 162 | 0.05784823 | 0.03318806 | -0.06368822 | -0.00390020 | 0.03848165 | -0.01647047 | -0.01503679 | 0.01337339 | 0.00232642 | -0.00374700 | 0.00043967 | -0.00021049 | -0.00013400 | 0.00068709 | 0.00019893 | -0.00079046 | -0.00005514 |
| 104 | 163 | 0.05806824 | 0.03300287 | -0.06403100 | -0.00409968 | 0.03896437 | -0.01653185 | -0.01513190 | 0.01307451 | 0.00219750 | -0.00385516 | 0.00055620 | -0.00011190 | -0.00019731 | 0.00057290 | 0.00011093 | -0.00072988 | -0.00010817 |
| 104 | 164 | 0.05784884 | 0.03269085 | -0.06389605 | -0.00349779 | 0.03848881 | -0.01689062 | -0.01468563 | 0.01351842 | 0.00216956 | -0.00378962 | 0.00050683 | -0.00020668 | -0.00014691 | 0.00068656 | 0.00018114 | -0.00079515 | -0.00005197 |
| 105 | 150 | 0.05863682 | 0.03706823 | -0.06231559 | -0.00730142 | 0.03964620 | -0.01274234 | -0.01720262 | 0.01185309 | 0.00359425 | -0.00342065 | 0.00014008 | -0.00007648 | -0.00013546 | 0.00049195 | 0.00020636 | -0.00067819 | -0.00014416 |
| 105 | 151 | 0.05839914 | 0.03734474 | -0.06112634 | -0.00789235 | 0.04001257 | -0.01211275 | -0.01767001 | 0.01048346 | 0.00350855 | -0.00328867 | 0.00025283 | 0.00006843 | -0.00020069 | 0.00031123 | 0.00007855 | -0.00052836 | -0.00017540 |

TABLE 3. Nuclear charge density distribution FB coefficients.

| $Z$ | $N$ | $a_1$ | $a_2$ | $a_3$ | $a_4$ | $a_5$ | $a_6$ | $a_7$ | $a_8$ | $a_9$ | $a_{10}$ | $a_{11}$ | $a_{12}$ | $a_{13}$ | $a_{14}$ | $a_{15}$ | $a_{16}$ | $a_{17}$ |
|---|---|---|---|---|---|---|---|---|---|---|---|---|---|---|---|---|---|---|
| 105 | 152 | 0.05858916 | 0.03629013 | -0.06280605 | -0.00668008 | 0.03971418 | -0.01342460 | -0.01687772 | 0.01215608 | 0.00338424 | -0.00350323 | 0.00020631 | -0.00008239 | -0.00014908 | 0.00051525 | 0.00019225 | -0.00068888 | -0.00013937 |
| 105 | 153 | 0.05836279 | 0.03650329 | -0.06171116 | -0.00727477 | 0.04020572 | -0.01275266 | -0.01733075 | 0.01083950 | 0.00332945 | -0.00337976 | 0.00031531 | 0.00005927 | -0.00021298 | 0.00033866 | 0.00007106 | -0.00054699 | -0.00017297 |
| 105 | 154 | 0.05855114 | 0.03553757 | -0.06326939 | -0.00608813 | 0.03978966 | -0.01405641 | -0.01652833 | 0.01244726 | 0.00318392 | -0.00358128 | 0.00027332 | -0.00008671 | -0.00016357 | 0.00053596 | 0.00017779 | -0.00069953 | -0.00013519 |
| 105 | 155 | 0.05833297 | 0.03568458 | -0.06226649 | -0.00667740 | 0.04039175 | -0.01334852 | -0.01695820 | 0.01119358 | 0.00315647 | -0.00346905 | 0.00037756 | 0.00005103 | -0.00022613 | 0.00036512 | 0.00006374 | -0.00056665 | -0.00017147 |
| 105 | 156 | 0.05852172 | 0.03481175 | -0.06370502 | -0.00552456 | 0.03987296 | -0.01463945 | -0.01615964 | 0.01272679 | 0.00299394 | -0.00365534 | 0.00034050 | -0.00008950 | -0.00017875 | 0.00055421 | 0.00016328 | -0.00071055 | -0.00013195 |
| 105 | 157 | 0.05830853 | 0.03489293 | -0.06278907 | -0.00610022 | 0.04057217 | -0.01389980 | -0.01655972 | 0.01154394 | 0.00299138 | -0.00355644 | 0.00043892 | 0.00004370 | -0.00023989 | 0.00039046 | 0.00005684 | -0.00058749 | -0.00017108 |
| 105 | 158 | 0.05849951 | 0.03411416 | -0.06411144 | -0.00498788 | 0.03996367 | -0.01517546 | -0.01577653 | 0.01299472 | 0.00281474 | -0.00372587 | 0.00040729 | -0.00009091 | -0.00019437 | 0.00057015 | 0.00014900 | -0.00072235 | -0.00012989 |
| 105 | 159 | 0.05828805 | 0.03413298 | -0.06327499 | -0.00554293 | 0.04074771 | -0.01440638 | -0.01614252 | 0.01188863 | 0.00283576 | -0.00364174 | 0.00049879 | 0.00003721 | -0.00025396 | 0.00041449 | 0.00005060 | -0.00060962 | -0.00017191 |
| 105 | 160 | 0.05848286 | 0.03344680 | -0.06448636 | -0.00447629 | 0.04006062 | -0.01566615 | -0.01538347 | 0.01325090 | 0.00264662 | -0.00379324 | 0.00047316 | -0.00009114 | -0.00021013 | 0.00058388 | 0.00013527 | -0.00073527 | -0.00012916 |
| 105 | 161 | 0.05827001 | 0.03340966 | -0.06372008 | -0.00500502 | 0.04091850 | -0.01486851 | -0.01571343 | 0.01222553 | 0.00269089 | -0.00372468 | 0.00055664 | 0.00003144 | -0.00026794 | 0.00043699 | 0.00004523 | -0.00063306 | -0.00017392 |
| 105 | 162 | 0.05846998 | 0.03281202 | -0.06482681 | -0.00398779 | 0.04016223 | -0.01611312 | -0.01498462 | 0.01349506 | 0.00248974 | -0.00385772 | 0.00053759 | -0.00009039 | -0.00022570 | 0.00059551 | 0.00012234 | -0.00074959 | -0.00012987 |
| 105 | 163 | 0.05825285 | 0.03272774 | -0.06412015 | -0.00448600 | 0.04108396 | -0.01528681 | -0.01527887 | 0.01255247 | 0.00255785 | -0.00380489 | 0.00061204 | 0.00002624 | -0.00028146 | 0.00045771 | 0.00004090 | -0.00065775 | -0.00017702 |
| 105 | 164 | 0.05845900 | 0.03221250 | -0.06512943 | -0.00352052 | 0.04026645 | -0.01651769 | -0.01458380 | 0.01372677 | 0.00234427 | -0.00391947 | 0.00060012 | -0.00008893 | -0.00024075 | 0.00060509 | 0.00011046 | -0.00076552 | -0.00013202 |
| 105 | 165 | 0.05823511 | 0.03209164 | -0.06447134 | -0.00398553 | 0.04124308 | -0.01566214 | -0.01484464 | 0.01286734 | 0.00243744 | -0.00388196 | 0.00066463 | 0.00002146 | -0.00029416 | 0.00047646 | 0.00003774 | -0.00068359 | -0.00018108 |
| 106 | 152 | 0.05876324 | 0.03545041 | -0.06412923 | -0.00596190 | 0.04106892 | -0.01324897 | -0.01643416 | 0.01333186 | 0.00381808 | -0.00348732 | 0.00009083 | -0.00018666 | -0.00011233 | 0.00069554 | 0.00031876 | -0.00081371 | -0.00010935 |
| 106 | 153 | 0.05902340 | 0.03544580 | -0.06411671 | -0.00605480 | 0.04123373 | -0.01342839 | -0.01643844 | 0.01288848 | 0.00358385 | -0.00355182 | 0.00023349 | -0.00006381 | -0.00019251 | 0.00056236 | 0.00019458 | -0.00071396 | -0.00015633 |
| 106 | 154 | 0.05869515 | 0.03481334 | -0.06437545 | -0.00541553 | 0.04092498 | -0.01391077 | -0.01607845 | 0.01350594 | 0.00357110 | -0.00355477 | 0.00017060 | -0.00018693 | -0.00012555 | 0.00070100 | 0.00029622 | -0.00081667 | -0.00010250 |
| 106 | 155 | 0.05897335 | 0.03470206 | -0.06452342 | -0.00546896 | 0.04126323 | -0.01405183 | -0.01605976 | 0.01315723 | 0.00337123 | -0.00363189 | 0.00030541 | -0.00006599 | -0.00020832 | 0.00057926 | 0.00017867 | -0.00072638 | -0.00015420 |
| 106 | 156 | 0.05863764 | 0.03420708 | -0.06459560 | -0.00489955 | 0.04079809 | -0.01452061 | -0.01571058 | 0.01366799 | 0.00333821 | -0.00361731 | 0.00025000 | -0.00018581 | -0.00013914 | 0.00070376 | 0.00027392 | -0.00081997 | -0.00009662 |
| 106 | 157 | 0.05892890 | 0.03399117 | -0.06489337 | -0.00491134 | 0.04129653 | -0.01462367 | -0.01566418 | 0.01341343 | 0.00316973 | -0.00370799 | 0.00037683 | -0.00006706 | -0.00022431 | 0.00059367 | 0.00016312 | -0.00073960 | -0.00015314 |
| 106 | 158 | 0.05858947 | 0.03363131 | -0.06479046 | -0.00441210 | 0.04068777 | -0.01508136 | -0.01533512 | 0.01381889 | 0.00311956 | -0.00367557 | 0.00032834 | -0.00018348 | -0.00015289 | 0.00070419 | 0.00025218 | -0.00082412 | -0.00009198 |
| 106 | 159 | 0.05888871 | 0.03331551 | -0.06522499 | -0.00438056 | 0.04133319 | -0.01514566 | -0.01525738 | 0.01365698 | 0.00298007 | -0.00378044 | 0.00044705 | -0.00006719 | -0.00024019 | 0.00060574 | 0.00014825 | -0.00075397 | -0.00015332 |
| 106 | 160 | 0.05854913 | 0.03308610 | -0.06495997 | -0.00395109 | 0.04059297 | -0.01559572 | -0.01495601 | 0.01395941 | 0.00291511 | -0.00373011 | 0.00040502 | -0.00018013 | -0.00016654 | 0.00070265 | 0.00023134 | -0.00082955 | -0.00008877 |
| 106 | 161 | 0.05885134 | 0.03267761 | -0.06551617 | -0.00387501 | 0.04137218 | -0.01561955 | -0.01484457 | 0.01388766 | 0.00280283 | -0.00384946 | 0.00051544 | -0.00006659 | -0.00025563 | 0.00061560 | 0.00013436 | -0.00076974 | -0.00015483 |
| 106 | 162 | 0.05851502 | 0.03257189 | -0.06510331 | -0.00351417 | 0.04051219 | -0.01606603 | -0.01457667 | 0.01409011 | 0.00272471 | -0.00378138 | 0.00047949 | -0.00017600 | -0.00017985 | 0.00069942 | 0.00021167 | -0.00083661 | -0.00008714 |
| 106 | 163 | 0.05881530 | 0.03208023 | -0.06576452 | -0.00339303 | 0.04141219 | -0.01604703 | -0.01443058 | 0.01410513 | 0.00263847 | -0.00391513 | 0.00058142 | -0.00006547 | -0.00027031 | 0.00062338 | 0.00012171 | -0.00078709 | -0.00015770 |
| 106 | 164 | 0.05848548 | 0.03208954 | -0.06521903 | -0.00309909 | 0.04044360 | -0.01649432 | -0.01420006 | 0.01421137 | 0.00254817 | -0.00382974 | 0.00055128 | -0.00017130 | -0.00019255 | 0.00069476 | 0.00019344 | -0.00084560 | -0.00008718 |
| 106 | 165 | 0.05877927 | 0.03152613 | -0.06596776 | -0.00293310 | 0.04145158 | -0.01642970 | -0.01401971 | 0.01430896 | 0.00248731 | -0.00397748 | 0.00064449 | -0.00006403 | -0.00028396 | 0.00062918 | 0.00011050 | -0.00080609 | -0.00016191 |
| 106 | 166 | 0.05845892 | 0.03164034 | -0.06530529 | -0.00270381 | 0.04038508 | -0.01688223 | -0.01382878 | 0.01432339 | 0.00238528 | -0.00387542 | 0.00061993 | -0.00016625 | -0.00020440 | 0.00068887 | 0.00017686 | -0.00085674 | -0.00008892 |
| 106 | 167 | 0.05874212 | 0.03101796 | -0.06612381 | -0.00249409 | 0.04148847 | -0.01676920 | -0.01361589 | 0.01449872 | 0.00234966 | -0.00403639 | 0.00070420 | -0.00006244 | -0.00029633 | 0.00063309 | 0.00010089 | -0.00082676 | -0.00016742 |
| 107 | 153 | 0.05942948 | 0.03545146 | -0.06407262 | -0.00634072 | 0.04375976 | -0.01183202 | -0.01653549 | 0.01233975 | 0.00406279 | -0.00339385 | 0.00027713 | 0.00011137 | -0.00030757 | 0.00044415 | 0.00009561 | -0.00060128 | -0.00023164 |
| 107 | 154 | 0.05940660 | 0.03465125 | -0.06522684 | -0.00540395 | 0.04268499 | -0.01337248 | -0.01596039 | 0.01356243 | 0.00377448 | -0.00359661 | 0.00026650 | -0.00003947 | -0.00024056 | 0.00060192 | 0.00019504 | -0.00074314 | -0.00018063 |
| 107 | 155 | 0.05934945 | 0.03462915 | -0.06451107 | -0.00571983 | 0.04379665 | -0.01245831 | -0.01611520 | 0.01266646 | 0.00385000 | -0.00349347 | 0.00034579 | 0.00010481 | -0.00032339 | 0.00046836 | 0.00008595 | -0.00062509 | -0.00023425 |
| 107 | 156 | 0.05933985 | 0.03392921 | -0.06556074 | -0.00482915 | 0.04265408 | -0.01398463 | -0.01555567 | 0.01380536 | 0.00354984 | -0.00367841 | 0.00034244 | -0.00004045 | -0.00025655 | 0.00061469 | 0.00017823 | -0.00075768 | -0.00018028 |
| 107 | 157 | 0.05926973 | 0.03385340 | -0.06489393 | -0.00512386 | 0.04381648 | -0.01303390 | -0.01567499 | 0.01298229 | 0.00364665 | -0.00359003 | 0.00041303 | 0.00009827 | -0.00033841 | 0.00049029 | 0.00007724 | -0.00065004 | -0.00023731 |
| 107 | 158 | 0.05927604 | 0.03324704 | -0.06584955 | -0.00428268 | 0.04262329 | -0.01454248 | -0.01513801 | 0.01403500 | 0.00333762 | -0.00375613 | 0.00041692 | -0.00004075 | -0.00027222 | 0.00062506 | 0.00016225 | -0.00077338 | -0.00018107 |
| 107 | 159 | 0.05919010 | 0.03312904 | -0.06522007 | -0.00455337 | 0.04382234 | -0.01355929 | -0.01522444 | 0.01328541 | 0.00345526 | -0.00368304 | 0.00047800 | 0.00009179 | -0.00035236 | 0.00050984 | 0.00006974 | -0.00067608 | -0.00024089 |
| 107 | 160 | 0.05921416 | 0.03260752 | -0.06609223 | -0.00376336 | 0.04259251 | -0.01504800 | -0.01471390 | 0.01425120 | 0.00313889 | -0.00382989 | 0.00048915 | -0.00004054 | -0.00028726 | 0.00063319 | 0.00014741 | -0.00079049 | -0.00018311 |
| 107 | 161 | 0.05911038 | 0.03246006 | -0.06548899 | -0.00400867 | 0.04381657 | -0.01403565 | -0.01477220 | 0.01357416 | 0.00327788 | -0.00377198 | 0.00053999 | 0.00008535 | -0.00036501 | 0.00052695 | 0.00006362 | -0.00070312 | -0.00024498 |
| 107 | 162 | 0.05915324 | 0.03201345 | -0.06628761 | -0.00326993 | 0.04256124 | -0.01550323 | -0.01428927 | 0.01445368 | 0.00295449 | -0.00389973 | 0.00055845 | -0.00003998 | -0.00030136 | 0.00063924 | 0.00013400 | -0.00080914 | -0.00018646 |
| 107 | 163 | 0.05903044 | 0.03184946 | -0.06570084 | -0.00349008 | 0.04380071 | -0.01446476 | -0.01432578 | 0.01384710 | 0.00311597 | -0.00385635 | 0.00059844 | 0.00007900 | -0.00037619 | 0.00054161 | 0.00005899 | -0.00073099 | -0.00024957 |
| 107 | 164 | 0.05909240 | 0.03146742 | -0.06643466 | -0.00280123 | 0.04252856 | -0.01591030 | -0.01386939 | 0.01464212 | 0.00278504 | -0.00396560 | 0.00062423 | -0.00003923 | -0.00031429 | 0.00064338 | 0.00012222 | -0.00082940 | -0.00019110 |
| 107 | 165 | 0.05895022 | 0.03129905 | -0.06585664 | -0.00299799 | 0.04377551 | -0.01484899 | -0.01389146 | 0.01410305 | 0.00297045 | -0.00393567 | 0.00065292 | 0.00007274 | -0.00038581 | 0.00055384 | 0.00005592 | -0.00075950 | -0.00025461 |
| 107 | 166 | 0.05903100 | 0.03097169 | -0.06653275 | -0.00235644 | 0.04249318 | -0.01627145 | -0.01345894 | 0.01481616 | 0.00263101 | -0.00402736 | 0.00068601 | -0.00003839 | -0.00032588 | 0.00064574 | 0.00011220 | -0.00085123 | -0.00019701 |
| 107 | 167 | 0.05886981 | 0.03080944 | -0.06595819 | -0.00253299 | 0.04374092 | -0.01519122 | -0.01347425 | 0.01434113 | 0.00284175 | -0.00400952 | 0.00070316 | 0.00006661 | -0.00039386 | 0.00056376 | 0.00005438 | -0.00078843 | -0.00026009 |
| 107 | 168 | 0.05896864 | 0.03052800 | -0.06658185 | -0.00193519 | 0.04245349 | -0.01658916 | -0.01306193 | 0.01497548 | 0.00249264 | -0.00408482 | 0.00074340 | -0.00003755 | -0.00033603 | 0.00064647 | 0.00010403 | -0.00087453 | -0.00020413 |
| 108 | 155 | 0.05973445 | 0.03393577 | -0.06610625 | -0.00474317 | 0.04403217 | -0.01325597 | -0.01545254 | 0.01416912 | 0.00395625 | -0.00363710 | 0.00030372 | -0.00001106 | -0.00029158 | 0.00063324 | 0.00019477 | -0.00077683 | -0.00021147 |
| 108 | 156 | 0.05930198 | 0.03376123 | -0.06580466 | -0.00451934 | 0.04302080 | -0.01385312 | -0.01532146 | 0.01432918 | 0.00384585 | -0.00360580 | 0.00027170 | -0.00014278 | -0.00019325 | 0.00071245 | 0.00028909 | -0.00086026 | -0.00013574 |
| 108 | 157 | 0.05964742 | 0.03324856 | -0.06635460 | -0.00418559 | 0.04392697 | -0.01385516 | -0.01502790 | 0.01438384 | 0.00372083 | -0.00372014 | 0.00038222 | -0.00001179 | -0.00030669 | 0.00064195 | 0.00017772 | -0.00079378 | -0.00021261 |
| 108 | 158 | 0.05922646 | 0.03319132 | -0.06592288 | -0.00402697 | 0.04281877 | -0.01444221 | -0.01492141 | 0.01445499 | 0.00360096 | -0.00367109 | 0.00035612 | -0.00013965 | -0.00020667 | 0.00070869 | 0.00026615 | -0.00086798 | -0.00013343 |
| 108 | 159 | 0.05956152 | 0.03260841 | -0.06655153 | -0.00365692 | 0.04381924 | -0.01439805 | -0.01459607 | 0.01458478 | 0.00349976 | -0.00379877 | 0.00045807 | -0.00001219 | -0.00032104 | 0.00064847 | 0.00016198 | -0.00081212 | -0.00021496 |
| 108 | 160 | 0.05915574 | 0.03265855 | -0.06600650 | -0.00356095 | 0.04262879 | -0.01497838 | -0.01451911 | 0.01457115 | 0.00337235 | -0.00373214 | 0.00043745 | -0.00013595 | -0.00021970 | 0.00070350 | 0.00024466 | -0.00087741 | -0.00013283 |
| 108 | 161 | 0.05947632 | 0.03201817 | -0.06669721 | -0.00315638 | 0.04370945 | -0.01488709 | -0.01416423 | 0.01477170 | 0.00329435 | -0.00387293 | 0.00053047 | -0.00001239 | -0.00033437 | 0.00065298 | 0.00014783 | -0.00083198 | -0.00021857 |
| 108 | 162 | 0.05908874 | 0.03216436 | -0.06605551 | -0.00311985 | 0.04244977 | -0.01546434 | -0.01411950 | 0.01467791 | 0.00316040 | -0.00378916 | 0.00051501 | -0.00013189 | -0.00023206 | 0.00069716 | 0.00022493 | -0.00088884 | -0.00013401 |
| 108 | 163 | 0.05939150 | 0.03148022 | -0.06679212 | -0.00268329 | 0.04359754 | -0.01532498 | -0.01373878 | 0.01494433 | 0.00310556 | -0.00394250 | 0.00059875 | -0.00001247 | -0.00034647 | 0.00065569 | 0.00013549 | -0.00085338 | -0.00022346 |
| 108 | 164 | 0.05902451 | 0.03171037 | -0.06606952 | -0.00270243 | 0.04228017 | -0.01590263 | -0.01372692 | 0.01477531 | 0.00296537 | -0.00384231 | 0.00058827 | -0.00012766 | -0.00024350 | 0.00068988 | 0.00020718 | -0.00090244 | -0.00013700 |
| 108 | 165 | 0.05930696 | 0.03099642 | -0.06683717 | -0.00223720 | 0.04348293 | -0.01571466 | -0.01332522 | 0.01510240 | 0.00293409 | -0.00400729 | 0.00066240 | -0.00001249 | -0.00035722 | 0.00065678 | 0.00012507 | -0.00087626 | -0.00022963 |
| 108 | 166 | 0.05896223 | 0.03129816 | -0.06604827 | -0.00230772 | 0.04211805 | -0.01629572 | -0.01334514 | 0.01486327 | 0.00278734 | -0.00389163 | 0.00065679 | -0.00012339 | -0.00025385 | 0.00068186 | 0.00019159 | -0.00091828 | -0.00014178 |
| 108 | 167 | 0.05922270 | 0.03056788 | -0.06683385 | -0.00181812 | 0.04336443 | -0.01605937 | -0.01292819 | 0.01524564 | 0.00278024 | -0.00406709 | 0.00072103 | -0.00001246 | -0.00036661 | 0.00065642 | 0.00011665 | -0.00090051 | -0.00023704 |
| 108 | 168 | 0.05890136 | 0.03092923 | -0.06599169 | -0.00193534 | 0.04196112 | -0.01664604 | -0.01297742 | 0.01494157 | 0.00262631 | -0.00393709 | 0.00072023 | -0.00011918 | -0.00026301 | 0.00067326 | 0.00017825 | -0.00093632 | -0.00014833 |
| 108 | 169 | 0.05913905 | 0.03019496 | -0.06678446 | -0.00142657 | 0.04324036 | -0.01636264 | -0.01255151 | 0.01537385 | 0.00264404 | -0.00412166 | 0.00077439 | -0.00001235 | -0.00037470 | 0.00065483 | 0.00011020 | -0.00092591 | -0.00024566 |
| 109 | 156 | 0.06000676 | 0.03332536 | -0.06673995 | -0.00409635 | 0.04523545 | -0.01308703 | -0.01493217 | 0.01470193 | 0.00413052 | -0.00367257 | 0.00034237 | 0.00001995 | -0.00034450 | 0.00065681 | 0.00019517 | -0.00081554 | -0.00024840 |
| 109 | 157 | 0.06000494 | 0.03317752 | -0.06611270 | -0.00409094 | 0.04665135 | -0.01193470 | -0.01478806 | 0.01420678 | 0.00430753 | -0.00359820 | 0.00037745 | 0.00016213 | -0.00044960 | 0.00057625 | 0.00010376 | -0.00072900 | -0.00032766 |
| 109 | 158 | 0.05989772 | 0.03268520 | -0.06689560 | -0.00356376 | 0.04504369 | -0.01367400 | -0.01449717 | 0.01488698 | 0.00388661 | -0.00375604 | 0.00042151 | 0.00001870 | -0.00035786 | 0.00066190 | 0.00017859 | -0.00083506 | -0.00025082 |
| 109 | 159 | 0.05986959 | 0.03252865 | -0.06627136 | -0.00355377 | 0.04643995 | -0.01250081 | -0.01434286 | 0.01444799 | 0.00408293 | -0.00369541 | 0.00044543 | 0.00015398 | -0.00046032 | 0.00058949 | 0.00009567 | -0.00075719 | -0.00033142 |
| 109 | 160 | 0.05978927 | 0.03209876 | -0.06699679 | -0.00306152 | 0.04484810 | -0.01420406 | -0.01406253 | 0.01505786 | 0.00365943 | -0.00383457 | 0.00049671 | 0.00001757 | -0.00037019 | 0.00066513 | 0.00016375 | -0.00085604 | -0.00025451 |
| 109 | 161 | 0.05973557 | 0.03194579 | -0.06637110 | -0.00304963 | 0.04621696 | -0.01301763 | -0.01390821 | 0.01467153 | 0.00387597 | -0.00378694 | 0.00050922 | 0.00014576 | -0.00046930 | 0.00060022 | 0.00008937 | -0.00078601 | -0.00033543 |
| 109 | 162 | 0.05968169 | 0.03156826 | -0.06704555 | -0.00258942 | 0.04464963 | -0.01468044 | -0.01363579 | 0.01521433 | 0.00345040 | -0.00390794 | 0.00056725 | 0.00001653 | -0.00038133 | 0.00066670 | 0.00015085 | -0.00087849 | -0.00025949 |
| 109 | 163 | 0.05960388 | 0.03142890 | -0.06641664 | -0.00257908 | 0.04598559 | -0.01348822 | -0.01349136 | 0.01487692 | 0.00368810 | -0.00387221 | 0.00056827 | 0.00013761 | -0.00047662 | 0.00060865 | 0.00008489 | -0.00081526 | -0.00033981 |
| 109 | 164 | 0.05957537 | 0.03109512 | -0.06704428 | -0.00214746 | 0.04444845 | -0.01510679 | -0.01322340 | 0.01535614 | 0.00326042 | -0.00397588 | 0.00063255 | 0.00001558 | -0.00039121 | 0.00066683 | 0.00014004 | -0.00090233 | -0.00026579 |
| 109 | 165 | 0.05947542 | 0.03097650 | -0.06641310 | -0.00214268 | 0.04574786 | -0.01391616 | -0.01309772 | 0.01506397 | 0.00351993 | -0.00395078 | 0.00062227 | 0.00012972 | -0.00048246 | 0.00061505 | 0.00008217 | -0.00084472 | -0.00034462 |
| 109 | 166 | 0.05947083 | 0.03067991 | -0.06699608 | -0.00173602 | 0.04424404 | -0.01548701 | -0.01283061 | 0.01548308 | 0.00308994 | -0.00403816 | 0.00069225 | 0.00001476 | -0.00039985 | 0.00066575 | 0.00013135 | -0.00092741 | -0.00027339 |
| 109 | 167 | 0.05935091 | 0.03058601 | -0.06636579 | -0.00174076 | 0.04550463 | -0.01430534 | -0.01273098 | 0.01523277 | 0.00337137 | -0.00402238 | 0.00067108 | 0.00012222 | -0.00048701 | 0.00061968 | 0.00008112 | -0.00087418 | -0.00034992 |
| 109 | 168 | 0.05936864 | 0.03032221 | -0.06690452 | -0.00135579 | 0.04403507 | -0.01582539 | -0.01246160 | 0.01559500 | 0.00293894 | -0.00409454 | 0.00074615 | 0.00001414 | -0.00040737 | 0.00066367 | 0.00012474 | -0.00095352 | -0.00028226 |
| 109 | 169 | 0.05923089 | 0.03025392 | -0.06628015 | -0.00137358 | 0.04525602 | -0.01465970 | -0.01239329 | 0.01538361 | 0.00324173 | -0.00408683 | 0.00071476 | 0.00011523 | -0.00049052 | 0.00062283 | 0.00008156 | -0.00090342 | -0.00035574 |
| 109 | 170 | 0.05926944 | 0.03002069 | -0.06677373 | -0.00100787 | 0.04381964 | -0.01612644 | -0.01211946 | 0.01569184 | 0.00280702 | -0.00414482 | 0.00079419 | 0.00001382 | -0.00041398 | 0.00066079 | 0.00012012 | -0.00098042 | -0.00029237 |
| 110 | 157 | 0.06022844 | 0.03284051 | -0.06713307 | -0.00349860 | 0.04624926 | -0.01288745 | -0.01442761 | 0.01515489 | 0.00429941 | -0.00370178 | 0.00037875 | 0.00005263 | -0.00039904 | 0.00067421 | 0.00019776 | -0.00085985 | -0.00029145 |
| 110 | 158 | 0.05979076 | 0.03298027 | -0.06659991 | -0.00366455 | 0.04481618 | -0.01362450 | -0.01453441 | 0.01496065 | 0.00412631 | -0.00363846 | 0.00037536 | -0.00008548 | -0.00027092 | 0.00070192 | 0.00028297 | -0.00091955 | -0.00019158 |
| 110 | 159 | 0.06009801 | 0.03225756 | -0.06719616 | -0.00300057 | 0.04596110 | -0.01346635 | -0.01399523 | 0.01530938 | 0.00405009 | -0.00378457 | 0.00045641 | 0.00005038 | -0.00041008 | 0.00067643 | 0.00018228 | -0.00088190 | -0.00029501 |
| 110 | 160 | 0.05968589 | 0.03247193 | -0.06659334 | -0.00320826 | 0.04450296 | -0.01418896 | -0.01412135 | 0.01505273 | 0.00387518 | -0.00370475 | 0.00045784 | -0.00008283 | -0.00028215 | 0.00069493 | 0.00026190 | -0.00093349 | -0.00019341 |
| 110 | 161 | 0.05996906 | 0.03173345 | -0.06720588 | -0.00253529 | 0.04566917 | -0.01398958 | -0.01357208 | 0.01544941 | 0.00382011 | -0.00386170 | 0.00052891 | 0.00004824 | -0.00042007 | 0.00067718 | 0.00016887 | -0.00090531 | -0.00029992 |
| 110 | 162 | 0.05958396 | 0.03201017 | -0.06654704 | -0.00277987 | 0.04419716 | -0.01469944 | -0.01371662 | 0.01513535 | 0.00364370 | -0.00376604 | 0.00053496 | -0.00008004 | -0.00029267 | 0.00068737 | 0.00024308 | -0.00094951 | -0.00019721 |
| 110 | 163 | 0.05984246 | 0.03126927 | -0.06716613 | -0.00210317 | 0.04537463 | -0.01446128 | -0.01316548 | 0.01557475 | 0.00361072 | -0.00393283 | 0.00059567 | 0.00004626 | -0.00042895 | 0.00067671 | 0.00015766 | -0.00093001 | -0.00030619 |
| 110 | 164 | 0.05948503 | 0.03159666 | -0.06646277 | -0.00237948 | 0.04389774 | -0.01515964 | -0.01332591 | 0.01520820 | 0.00343252 | -0.00382225 | 0.00060621 | -0.00007717 | -0.00030229 | 0.00067942 | 0.00022669 | -0.00096765 | -0.00020295 |
| 110 | 165 | 0.05971910 | 0.03086502 | -0.06708124 | -0.00170471 | 0.04507775 | -0.01488592 | -0.01278125 | 0.01568524 | 0.00342250 | -0.00399767 | 0.00065632 | 0.00004453 | -0.00043679 | 0.00067523 | 0.00014867 | -0.00095581 | -0.00031383 |
| 110 | 166 | 0.05938921 | 0.03123255 | -0.06634265 | -0.00200756 | 0.04360309 | -0.01557332 | -0.01295402 | 0.01527089 | 0.00324184 | -0.00387328 | 0.00067126 | -0.00007424 | -0.00031098 | 0.00067123 | 0.00021281 | -0.00098783 | -0.00021058 |
| 110 | 167 | 0.05959985 | 0.03051970 | -0.06695581 | -0.00134061 | 0.04477792 | -0.01526819 | -0.01242374 | 0.01578085 | 0.00325545 | -0.00405599 | 0.00071070 | 0.00004315 | -0.00044370 | 0.00067297 | 0.00014185 | -0.00098250 | -0.00032279 |
| 110 | 168 | 0.05929692 | 0.03091834 | -0.06618919 | -0.00166510 | 0.04331088 | -0.01594444 | -0.01260482 | 0.01532300 | 0.00307158 | -0.00391898 | 0.00072994 | -0.00007123 | -0.00031877 | 0.00066295 | 0.00020144 | -0.00100987 | -0.00022006 |
| 110 | 169 | 0.05948542 | 0.03023129 | -0.06679473 | -0.00101166 | 0.04447395 | -0.01561278 | -0.01209589 | 0.01586163 | 0.00310909 | -0.00410763 | 0.00075884 | 0.00004220 | -0.00044990 | 0.00067013 | 0.00013707 | -0.00100983 | -0.00033305 |
| 110 | 170 | 0.05920851 | 0.03065386 | -0.06600539 | -0.00135344 | 0.04301831 | -0.01627705 | -0.01228138 | 0.01536411 | 0.00292136 | -0.00395923 | 0.00078219 | -0.00006806 | -0.00032583 | 0.00065469 | 0.00019248 | -0.00103354 | -0.00023131 |
| 110 | 171 | 0.05937647 | 0.02999698 | -0.06660277 | -0.00071843 | 0.04416432 | -0.01592430 | -0.01179935 | 0.01592783 | 0.00298248 | -0.00415253 | 0.00080091 | 0.00004179 | -0.00045561 | 0.00066691 | 0.00013417 | -0.00103751 | -0.00034451 |
| 111 | 161 | 0.06023755 | 0.03190208 | -0.06659293 | -0.00244186 | 0.04783716 | -0.01208984 | -0.01335792 | 0.01544457 | 0.00447813 | -0.00376468 | 0.00045540 | 0.00021611 | -0.00058785 | 0.00066333 | 0.00012810 | -0.00088424 | -0.00044817 |
| 111 | 162 | 0.06011442 | 0.03151918 | -0.06721767 | -0.00212378 | 0.04624093 | -0.01379653 | -0.01316595 | 0.01575430 | 0.00398472 | -0.00387783 | 0.00055105 | 0.00008005 | -0.00047161 | 0.00068724 | 0.00017848 | -0.00095996 | -0.00035157 |
| 111 | 163 | 0.06006917 | 0.03147422 | -0.06652346 | -0.00205176 | 0.04738892 | -0.01263223 | -0.01300139 | 0.01558269 | 0.00426813 | -0.00384768 | 0.00051290 | 0.00020599 | -0.00059037 | 0.00066714 | 0.00012379 | -0.00091316 | -0.00045104 |
| 111 | 164 | 0.05997334 | 0.03112361 | -0.06710306 | -0.00174170 | 0.04584768 | -0.01427306 | -0.01279342 | 0.01584876 | 0.00377793 | -0.00394528 | 0.00061257 | 0.00007725 | -0.00047836 | 0.00068562 | 0.00016916 | -0.00098634 | -0.00035896 |
| 111 | 165 | 0.05990754 | 0.03110735 | -0.06641614 | -0.00169669 | 0.04694460 | -0.01313269 | -0.01267304 | 0.01570434 | 0.00407976 | -0.00392302 | 0.00056483 | 0.00019639 | -0.00059180 | 0.00066955 | 0.00012122 | -0.00094186 | -0.00045442 |
| 111 | 166 | 0.05983785 | 0.03078791 | -0.06695020 | -0.00139585 | 0.04545423 | -0.01470597 | -0.01244956 | 0.01592867 | 0.00359384 | -0.00400568 | 0.00066748 | 0.00007492 | -0.00048438 | 0.00068339 | 0.00016206 | -0.00101344 | -0.00036771 |
| 111 | 167 | 0.05975316 | 0.03079714 | -0.06627651 | -0.00137531 | 0.04650636 | -0.01359428 | -0.01237393 | 0.01581042 | 0.00391240 | -0.00399064 | 0.00061130 | 0.00018742 | -0.00059236 | 0.00067082 | 0.00012024 | -0.00097016 | -0.00045833 |
| 111 | 168 | 0.05970870 | 0.03050962 | -0.06676414 | -0.00108623 | 0.04506058 | -0.01509974 | -0.01213687 | 0.01599436 | 0.00343191 | -0.00405893 | 0.00071587 | 0.00007315 | -0.00048984 | 0.00068074 | 0.00015705 | -0.00104100 | -0.00037775 |
| 111 | 169 | 0.05960632 | 0.03053882 | -0.06610948 | -0.00108588 | 0.04607566 | -0.01401988 | -0.01210370 | 0.01590203 | 0.00376494 | -0.00405063 | 0.00065256 | 0.00017914 | -0.00059225 | 0.00067117 | 0.00012062 | -0.00099784 | -0.00046274 |
| 111 | 170 | 0.05958637 | 0.03028552 | -0.06654944 | -0.00081234 | 0.04466638 | -0.01545847 | -0.01185621 | 0.01604633 | 0.00329115 | -0.00410502 | 0.00075797 | 0.00007200 | -0.00049489 | 0.00067782 | 0.00015394 | -0.00106872 | -0.00038893 |
| 111 | 171 | 0.05946701 | 0.03032750 | -0.06591924 | -0.00082633 | 0.04565347 | -0.01441202 | -0.01186088 | 0.01598029 | 0.00363590 | -0.00410323 | 0.00068898 | 0.00017155 | -0.00059163 | 0.00067080 | 0.00012216 | -0.00102473 | -0.00046761 |
| 111 | 172 | 0.05947106 | 0.03011179 | -0.06631003 | -0.00057302 | 0.04427159 | -0.01578555 | -0.01160702 | 0.01608535 | 0.00317021 | -0.00414408 | 0.00079420 | 0.00007147 | -0.00049970 | 0.00067474 | 0.00015250 | -0.00109632 | -0.00040112 |
| 112 | 164 | 0.05990432 | 0.03177266 | -0.06648649 | -0.00226321 | 0.04513574 | -0.01438742 | -0.01310297 | 0.01549477 | 0.00395406 | -0.00376432 | 0.00059417 | -0.00001888 | -0.00037460 | 0.00067256 | 0.00025630 | -0.00104360 | -0.00028794 |
| 112 | 165 | 0.06008520 | 0.03112928 | -0.06689020 | -0.00152755 | 0.04605561 | -0.01414877 | -0.01254539 | 0.01603721 | 0.00395439 | -0.00394345 | 0.00061620 | 0.00010926 | -0.00052926 | 0.00069523 | 0.00018597 | -0.00104689 | -0.00041698 |
| 112 | 166 | 0.05978071 | 0.03146252 | -0.06628749 | -0.00193810 | 0.04468563 | -0.01484451 | -0.01276386 | 0.01553092 | 0.00375320 | -0.00381495 | 0.00065337 | -0.00001747 | -0.00038161 | 0.00066584 | 0.00024449 | -0.00106718 | -0.00029786 |
| 112 | 167 | 0.05994065 | 0.03085896 | -0.06667957 | -0.00123162 | 0.04557757 | -0.01459235 | -0.01224478 | 0.01609031 | 0.00377631 | -0.00399865 | 0.00066510 | 0.00010623 | -0.00053324 | 0.00069268 | 0.00018063 | -0.00107438 | -0.00042615 |
| 112 | 168 | 0.05966356 | 0.03120583 | -0.06606003 | -0.00164797 | 0.04423884 | -0.01526026 | -0.01245451 | 0.01555599 | 0.00357551 | -0.00385902 | 0.00070549 | -0.00001566 | -0.00038820 | 0.00065934 | 0.00023521 | -0.00109202 | -0.00030954 |
| 112 | 169 | 0.05980376 | 0.03064286 | -0.06644060 | -0.00096898 | 0.04510644 | -0.01499695 | -0.01197514 | 0.01613106 | 0.00362072 | -0.00404643 | 0.00070764 | 0.00010384 | -0.00053681 | 0.00068988 | 0.00017719 | -0.00110182 | -0.00043635 |
| 112 | 170 | 0.05955324 | 0.03100090 | -0.06580744 | -0.00139281 | 0.04379450 | -0.01563831 | -0.01217618 | 0.01557008 | 0.00342019 | -0.00389652 | 0.00075078 | -0.00001340 | -0.00039450 | 0.00065311 | 0.00022829 | -0.00111776 | -0.00032281 |

TABLE 3. Nuclear charge density distribution FB coefficients.

| $Z$ | $N$ | $a_1$ | $a_2$ | $a_3$ | $a_4$ | $a_5$ | $a_6$ | $a_7$ | $a_8$ | $a_9$ | $a_{10}$ | $a_{11}$ | $a_{12}$ | $a_{13}$ | $a_{14}$ | $a_{15}$ | $a_{16}$ | $a_{17}$ |
|---|---|---|---|---|---|---|---|---|---|---|---|---|---|---|---|---|---|---|
| 112 | 171 | 0.05967450 | 0.03047698 | -0.06617655 | -0.00073713 | 0.04464338 | -0.01536516 | -0.01173496 | 0.01616052 | 0.00348619 | -0.00408700 | 0.00074428 | 0.00010204 | -0.00054005 | 0.00068690 | 0.00017540 | -0.00112891 | -0.00044736 |
| 112 | 172 | 0.05944985 | 0.03084545 | -0.06553233 | -0.00117179 | 0.04335212 | -0.01598157 | -0.01192862 | 0.01557362 | 0.00328616 | -0.00392754 | 0.00078960 | -0.00001066 | -0.00040066 | 0.00064715 | 0.00022350 | -0.00114406 | -0.00033746 |
| 112 | 173 | 0.05955246 | 0.03035708 | -0.06588985 | -0.00053288 | 0.04418990 | -0.01569888 | -0.01152149 | 0.01617995 | 0.00337108 | -0.00412066 | 0.00077558 | 0.00010076 | -0.00054298 | 0.00068376 | 0.00017504 | -0.00115538 | -0.00045893 |
| 113 | 165 | 0.06027258 | 0.03167461 | -0.06593218 | -0.00164480 | 0.04742543 | -0.01258212 | -0.01263786 | 0.01605431 | 0.00465438 | -0.00385113 | 0.00048043 | 0.00026399 | -0.00069918 | 0.00071924 | 0.00017713 | -0.00105085 | -0.00057341 |
| 113 | 166 | 0.06017742 | 0.03128752 | -0.06651649 | -0.00142592 | 0.04604212 | -0.01408006 | -0.01240832 | 0.01615455 | 0.00414417 | -0.00392717 | 0.00060680 | 0.00014001 | -0.00057780 | 0.00070522 | 0.00020843 | -0.00111021 | -0.00047687 |
| 113 | 167 | 0.06009029 | 0.03142322 | -0.06571148 | -0.00135121 | 0.04687634 | -0.01308512 | -0.01237294 | 0.01612349 | 0.00447139 | -0.00391592 | 0.00052580 | 0.00025168 | -0.00069273 | 0.00071720 | 0.00017523 | -0.00107536 | -0.00057263 |
| 113 | 168 | 0.06002379 | 0.03107801 | -0.06624715 | -0.00116059 | 0.04550824 | -0.01452387 | -0.01214086 | 0.01618866 | 0.00397361 | -0.00397729 | 0.00065027 | 0.00013569 | -0.00057874 | 0.00070218 | 0.00020450 | -0.00113693 | -0.00048513 |
| 113 | 169 | 0.05991672 | 0.03121551 | -0.06547037 | -0.00108106 | 0.04635045 | -0.01354694 | -0.01213106 | 0.01618245 | 0.00430794 | -0.00397333 | 0.00056641 | 0.00024012 | -0.00068589 | 0.00071455 | 0.00017445 | -0.00109893 | -0.00057225 |
| 113 | 170 | 0.05987798 | 0.03091837 | -0.06595147 | -0.00092112 | 0.04499211 | -0.01492550 | -0.01189986 | 0.01621328 | 0.00382521 | -0.00402009 | 0.00068790 | 0.00013192 | -0.00057925 | 0.00069887 | 0.00020220 | -0.00116306 | -0.00049399 |
| 113 | 171 | 0.05975172 | 0.03104674 | -0.06521241 | -0.00083208 | 0.04584847 | -0.01396966 | -0.01190999 | 0.01623235 | 0.00416247 | -0.00402373 | 0.00060270 | 0.00022926 | -0.00067879 | 0.00071142 | 0.00017458 | -0.00112142 | -0.00057221 |
| 113 | 172 | 0.05973957 | 0.03080412 | -0.06563189 | -0.00070394 | 0.04449558 | -0.01528661 | -0.01168224 | 0.01622969 | 0.00369718 | -0.00405596 | 0.00072030 | 0.00012859 | -0.00057931 | 0.00069529 | 0.00020127 | -0.00118834 | -0.00050322 |
| 113 | 173 | 0.05959498 | 0.03091236 | -0.06494074 | -0.00060216 | 0.04537061 | -0.01435529 | -0.01170730 | 0.01627426 | 0.00403327 | -0.00406758 | 0.00063512 | 0.00021908 | -0.00067154 | 0.00070791 | 0.00017543 | -0.00114274 | -0.00057242 |
| 113 | 174 | 0.05960786 | 0.03073072 | -0.06529032 | -0.00050531 | 0.04402027 | -0.01560847 | -0.01148437 | 0.01623921 | 0.00358762 | -0.00408535 | 0.00074808 | 0.00012561 | -0.00057890 | 0.00069144 | 0.00020146 | -0.00121253 | -0.00051253 |
| 114 | 170 | 0.05987715 | 0.03166996 | -0.06525406 | -0.00162979 | 0.04419313 | -0.01504794 | -0.01228633 | 0.01565774 | 0.00398428 | -0.00379309 | 0.00068492 | 0.00004224 | -0.00046743 | 0.00065861 | 0.00027738 | -0.00120732 | -0.00042222 |
| 114 | 171 | 0.05991405 | 0.03133773 | -0.06522758 | -0.00083168 | 0.04483728 | -0.01481413 | -0.01184105 | 0.01627275 | 0.00404923 | -0.00398310 | 0.00065998 | 0.00015262 | -0.00060924 | 0.00070451 | 0.00023170 | -0.00122217 | -0.00054438 |
| 114 | 172 | 0.05974844 | 0.03157286 | -0.06489020 | -0.00141536 | 0.04366536 | -0.01540800 | -0.01205597 | 0.01565159 | 0.00384859 | -0.00382190 | 0.00071865 | 0.00004205 | -0.00046938 | 0.00065331 | 0.00027382 | -0.00123231 | -0.00043457 |
| 114 | 173 | 0.05976402 | 0.03126549 | -0.06485048 | -0.00060474 | 0.04434005 | -0.01515204 | -0.01162818 | 0.01628433 | 0.00392659 | -0.00401540 | 0.00068956 | 0.00014708 | -0.00060508 | 0.00069981 | 0.00023096 | -0.00124451 | -0.00055045 |
| 114 | 174 | 0.05962551 | 0.03152054 | -0.06449956 | -0.00121801 | 0.04316157 | -0.01572322 | -0.01184459 | 0.01564014 | 0.00373361 | -0.00384452 | 0.00074720 | 0.00004197 | -0.00047073 | 0.00064783 | 0.00027179 | -0.00125652 | -0.00044706 |
| 114 | 175 | 0.05962043 | 0.03122789 | -0.06445518 | -0.00039097 | 0.04387166 | -0.01544843 | -0.01142989 | 0.01629091 | 0.00382080 | -0.00404167 | 0.00071525 | 0.00014182 | -0.00060047 | 0.00069478 | 0.00023103 | -0.00126543 | -0.00055632 |
| 115 | 172 | 0.05989546 | 0.03189214 | -0.06422397 | -0.00061287 | 0.04474171 | -0.01460400 | -0.01173944 | 0.01633390 | 0.00429230 | -0.00393840 | 0.00062739 | 0.00016219 | -0.00062036 | 0.00070407 | 0.00026462 | -0.00127656 | -0.00058228 |
| 115 | 173 | 0.05970293 | 0.03202960 | -0.06339783 | -0.00030044 | 0.04562306 | -0.01371264 | -0.01169558 | 0.01645161 | 0.00458297 | -0.00396380 | 0.00054202 | 0.00024397 | -0.00070080 | 0.00072559 | 0.00023825 | -0.00122876 | -0.00064102 |
| 115 | 174 | 0.05973424 | 0.03184724 | -0.06380537 | -0.00037082 | 0.04425816 | -0.01491568 | -0.01152436 | 0.01634359 | 0.00417182 | -0.00396804 | 0.00065540 | 0.00015455 | -0.00061225 | 0.00069823 | 0.00026330 | -0.00129513 | -0.00058478 |
| 115 | 175 | 0.05953639 | 0.03191759 | -0.06310923 | -0.00007573 | 0.04514784 | -0.01407696 | -0.01150394 | 0.01647643 | 0.00444600 | -0.00400303 | 0.00057371 | 0.00023196 | -0.00068922 | 0.00072006 | 0.00023615 | -0.00124248 | -0.00063624 |
| 115 | 176 | 0.05957984 | 0.03183017 | -0.06337623 | -0.00014180 | 0.04380522 | -0.01518746 | -0.01132186 | 0.01634864 | 0.00406616 | -0.00399204 | 0.00068014 | 0.00014729 | -0.00060399 | 0.00069214 | 0.00026249 | -0.00131213 | -0.00058708 |
| 116 | 173 | 0.05981663 | 0.03255368 | -0.06296498 | -0.00024951 | 0.04473729 | -0.01428716 | -0.01157832 | 0.01640202 | 0.00454669 | -0.00388748 | 0.00059285 | 0.00015987 | -0.00061100 | 0.00069691 | 0.00029931 | -0.00132431 | -0.00060575 |
| 116 | 174 | 0.05977692 | 0.03267087 | -0.06300656 | -0.00110891 | 0.04358518 | -0.01486487 | -0.01186030 | 0.01573456 | 0.00437391 | -0.00371273 | 0.00065435 | 0.00006694 | -0.00049561 | 0.00064308 | 0.00033524 | -0.00133545 | -0.00051540 |
| 116 | 175 | 0.05964798 | 0.03251823 | -0.06253668 | -0.00000491 | 0.04426368 | -0.01458251 | -0.01136492 | 0.01640823 | 0.00442388 | -0.00391515 | 0.00062022 | 0.00015100 | -0.00060050 | 0.00069047 | 0.00029660 | -0.00133868 | -0.00060525 |
| 116 | 176 | 0.05962350 | 0.03269096 | -0.06249322 | -0.00086367 | 0.04311907 | -0.01510554 | -0.01163226 | 0.01572806 | 0.00427127 | -0.00372998 | 0.00067820 | 0.00006176 | -0.00048820 | 0.00063564 | 0.00033333 | -0.00135299 | -0.00052049 |
| 116 | 177 | 0.05948688 | 0.03250444 | -0.06210616 | 0.00022447 | 0.04381930 | -0.01484140 | -0.01116377 | 0.01640942 | 0.00431393 | -0.00393755 | 0.00064481 | 0.00014269 | -0.00059026 | 0.00068389 | 0.00029419 | -0.00135151 | -0.00060469 |
| 117 | 174 | 0.05968395 | 0.03328037 | -0.06153385 | 0.00021144 | 0.04479631 | -0.01390040 | -0.01137835 | 0.01646738 | 0.00479982 | -0.00383095 | 0.00055816 | 0.00014763 | -0.00058421 | 0.00068436 | 0.00033379 | -0.00136435 | -0.00061604 |
| 117 | 175 | 0.05945970 | 0.03327264 | -0.06092012 | 0.00058506 | 0.04560904 | -0.01320288 | -0.01134955 | 0.01662371 | 0.00498770 | -0.00388411 | 0.00048470 | 0.00022188 | -0.00065706 | 0.00071444 | 0.00030119 | -0.00130919 | -0.00066180 |
| 117 | 176 | 0.05951326 | 0.03323749 | -0.06113019 | 0.00044100 | 0.04432241 | -0.01419300 | -0.01117395 | 0.01646743 | 0.00467057 | -0.00385738 | 0.00058553 | 0.00013868 | -0.00057341 | 0.00067801 | 0.00032907 | -0.00137445 | -0.00061355 |
| 117 | 177 | 0.05930115 | 0.03313295 | -0.06071087 | 0.00076780 | 0.04512414 | -0.01357332 | -0.01117948 | 0.01663271 | 0.00483438 | -0.00392188 | 0.00051717 | 0.00021139 | -0.00064768 | 0.00070936 | 0.00029459 | -0.00131544 | -0.00065527 |
| 118 | 175 | 0.05951043 | 0.03402866 | -0.06003079 | 0.00070381 | 0.04487056 | -0.01349400 | -0.01117203 | 0.01651699 | 0.00503921 | -0.00376968 | 0.00052445 | 0.00012867 | -0.00054535 | 0.00066850 | 0.00036633 | -0.00139630 | -0.00061593 |
| 118 | 176 | 0.05957268 | 0.03414582 | -0.06020043 | -0.00028553 | 0.04383841 | -0.01391280 | -0.01143948 | 0.01585791 | 0.00494385 | -0.00357923 | 0.00057975 | 0.00005052 | -0.00045102 | 0.00061294 | 0.00040179 | -0.00141204 | -0.00054468 |
| 118 | 177 | 0.05934292 | 0.03396571 | -0.05967806 | 0.00090594 | 0.04438905 | -0.01379372 | -0.01098149 | 0.01650962 | 0.00490088 | -0.00379572 | 0.00055227 | 0.00012063 | -0.00053612 | 0.00066282 | 0.00035922 | -0.00140233 | -0.00061248 |

TABLE 4. Charge radii, 2nd moment, 4th moment and 6th moment obtained by DNN.

| $Z$ | $N$ | $R_c$(fm) | $R_2$(fm$^2$) | $R_4$(fm$^4$) | $R_6$(fm$^6$) | $Z$ | $N$ | $R_c$(fm) | $R_2$(fm$^2$) | $R_4$(fm$^4$) | $R_6$(fm$^6$) |
|---|---|---|---|---|---|---|---|---|---|---|---|
| 8 | 3 | 2.8822 | 8.3069 | 137.8741 | 4078.4644 | 13 | 15 | 3.0843 | 9.5130 | 138.0176 | 2659.4802 |
| 8 | 4 | 2.9361 | 8.6207 | 134.5239 | 3463.3629 | 13 | 16 | 3.0939 | 9.5724 | 142.3748 | 2958.9548 |
| 8 | 5 | 2.9123 | 8.4812 | 141.7719 | 4143.1752 | 13 | 17 | 3.1377 | 9.8452 | 146.3000 | 2874.1497 |
| 8 | 6 | 2.8764 | 8.2738 | 122.9965 | 3073.9470 | 13 | 18 | 3.1602 | 9.9869 | 152.1456 | 3184.6388 |
| 8 | 7 | 2.7411 | 7.5136 | 110.7293 | 3136.4024 | 13 | 19 | 3.2118 | 10.3158 | 157.3889 | 3127.3962 |
| 8 | 8 | 2.6867 | 7.2185 | 88.3467 | 1928.6845 | 13 | 20 | 3.2393 | 10.4929 | 163.0949 | 3388.3301 |
| 8 | 9 | 2.6774 | 7.1684 | 97.9343 | 2666.4488 | 13 | 21 | 3.2616 | 10.6378 | 167.0149 | 3446.5520 |
| 8 | 10 | 2.7559 | 7.5950 | 99.7584 | 2281.3328 | 13 | 22 | 3.2368 | 10.4766 | 166.4584 | 3659.5917 |
| 8 | 11 | 2.7352 | 7.4811 | 106.0622 | 2872.8639 | 13 | 23 | 3.2745 | 10.7222 | 173.6590 | 3824.0659 |
| 8 | 12 | 2.7551 | 7.5904 | 97.0066 | 2111.1122 | 13 | 24 | 3.2364 | 10.4743 | 170.8966 | 3992.7230 |
| 8 | 13 | 2.7248 | 7.4245 | 100.9361 | 2603.9270 | 13 | 25 | 3.3277 | 11.0737 | 184.3296 | 4196.9376 |
| 8 | 14 | 2.7118 | 7.3539 | 86.5119 | 1689.4031 | 13 | 26 | 3.3340 | 11.1158 | 185.4015 | 4318.7459 |
| 8 | 15 | 2.6972 | 7.2747 | 94.6152 | 2365.7149 | 13 | 27 | 3.4135 | 11.6517 | 197.4423 | 4495.9583 |
| 8 | 16 | 2.6874 | 7.2221 | 82.8879 | 1618.9969 | 13 | 28 | 3.4195 | 11.6933 | 198.2081 | 4590.1215 |
| 8 | 17 | 2.7021 | 7.3015 | 94.2430 | 2344.4062 | 13 | 29 | 3.4528 | 11.9215 | 205.8824 | 4808.6648 |
| 8 | 18 | 2.6681 | 7.1189 | 77.4546 | 1404.5331 | 13 | 30 | 3.4346 | 11.7963 | 206.7056 | 5099.6657 |
| 8 | 19 | 2.7291 | 7.4478 | 93.8561 | 2174.8448 | 14 | 8 | 3.1146 | 9.7008 | 142.2849 | 2802.4808 |
| 8 | 20 | 2.6625 | 7.0888 | 70.0373 | 955.7785 | 14 | 9 | 3.0681 | 9.4134 | 132.3126 | 2433.1244 |
| 9 | 4 | 3.0301 | 9.1818 | 151.4622 | 3984.3391 | 14 | 10 | 3.1123 | 9.6863 | 143.7931 | 2914.6206 |
| 9 | 5 | 2.8903 | 8.3538 | 121.5635 | 2888.3162 | 14 | 11 | 3.0813 | 9.4945 | 137.4660 | 2674.7439 |
| 9 | 6 | 2.9831 | 8.8987 | 141.3465 | 3623.0820 | 14 | 12 | 3.1133 | 9.6928 | 145.9611 | 3048.9008 |
| 9 | 7 | 2.7934 | 7.8030 | 101.2415 | 2123.2275 | 14 | 13 | 3.0948 | 9.5780 | 142.8485 | 2933.2486 |
| 9 | 8 | 2.8884 | 8.3428 | 120.0285 | 2807.7397 | 14 | 14 | 3.1140 | 9.6972 | 147.9496 | 3175.5283 |
| 9 | 9 | 2.7696 | 7.6705 | 94.6439 | 1827.3232 | 14 | 15 | 3.1113 | 9.6803 | 146.5516 | 3082.2001 |
| 9 | 10 | 2.8997 | 8.4084 | 122.2541 | 2900.5055 | 14 | 16 | 3.1379 | 9.8464 | 151.4121 | 3275.4719 |
| 9 | 11 | 2.8018 | 7.8502 | 100.2895 | 2023.2615 | 14 | 17 | 3.1609 | 9.9910 | 153.7973 | 3257.7425 |
| 9 | 12 | 2.8688 | 8.2301 | 115.4453 | 2657.6814 | 14 | 18 | 3.1859 | 10.1501 | 157.5920 | 3379.8656 |
| 9 | 13 | 2.8023 | 7.8529 | 99.8834 | 2006.8898 | 14 | 19 | 3.2331 | 10.4530 | 164.1203 | 3470.4121 |
| 9 | 14 | 2.8195 | 7.9496 | 105.5400 | 2325.6877 | 14 | 20 | 3.2432 | 10.5183 | 163.9054 | 3421.2154 |
| 9 | 15 | 2.8077 | 7.8834 | 100.0190 | 1994.8359 | 14 | 21 | 3.2753 | 10.7274 | 172.0328 | 3725.3649 |
| 9 | 16 | 2.8484 | 8.1136 | 109.3915 | 2423.5626 | 14 | 22 | 3.2421 | 10.5114 | 166.7004 | 3642.7435 |
| 9 | 17 | 2.8476 | 8.1089 | 104.7005 | 2070.7781 | 14 | 23 | 3.2801 | 10.7591 | 177.2371 | 4062.9250 |
| 9 | 18 | 2.9030 | 8.4272 | 116.5232 | 2580.1871 | 14 | 24 | 3.2424 | 10.5132 | 170.0239 | 3893.8109 |
| 9 | 19 | 2.9088 | 8.4609 | 110.9048 | 2100.2593 | 14 | 25 | 3.3304 | 11.0913 | 187.2632 | 4416.1674 |
| 9 | 20 | 2.9732 | 8.8399 | 123.7231 | 2618.1702 | 14 | 26 | 3.3280 | 11.0758 | 182.6748 | 4178.9386 |
| 9 | 21 | 2.9629 | 8.7786 | 118.4169 | 2273.1230 | 14 | 27 | 3.4242 | 11.7251 | 201.6938 | 4747.3225 |
| 9 | 22 | 2.9565 | 8.7408 | 121.7213 | 2632.7056 | 14 | 28 | 3.4073 | 11.6095 | 194.5518 | 4434.8656 |
| 10 | 5 | 3.0644 | 9.3908 | 148.6384 | 3553.0777 | 14 | 29 | 3.4697 | 12.0389 | 211.5138 | 5108.3533 |
| 10 | 6 | 3.0637 | 9.3861 | 142.7269 | 3154.9229 | 14 | 30 | 3.4264 | 11.7403 | 203.1356 | 4914.9345 |
| 10 | 7 | 3.0255 | 9.1535 | 138.6776 | 3149.8043 | 14 | 31 | 3.4874 | 12.1622 | 220.8320 | 5643.1489 |
| 10 | 8 | 3.0253 | 9.1525 | 133.3923 | 2793.1504 | 15 | 9 | 3.0693 | 9.4205 | 128.5264 | 2146.4485 |
| 10 | 9 | 2.9973 | 8.9841 | 131.3992 | 2857.0869 | 15 | 10 | 3.1109 | 9.6778 | 141.4613 | 2755.4085 |
| 10 | 10 | 2.9918 | 8.9511 | 127.0437 | 2604.7204 | 15 | 11 | 3.1007 | 9.6142 | 137.1402 | 2491.7692 |
| 10 | 11 | 2.9632 | 8.7805 | 125.1684 | 2680.2163 | 15 | 12 | 3.1253 | 9.7677 | 146.5431 | 2987.4383 |
| 10 | 12 | 2.9441 | 8.6676 | 117.8293 | 2318.2091 | 15 | 13 | 3.1244 | 9.7619 | 144.0999 | 2784.7031 |
| 10 | 13 | 2.9201 | 8.5270 | 117.3938 | 2457.0277 | 15 | 14 | 3.1341 | 9.8226 | 150.3579 | 3180.3969 |
| 10 | 14 | 2.8989 | 8.4034 | 109.6697 | 2085.0066 | 15 | 15 | 3.1475 | 9.9071 | 149.0124 | 2964.3157 |
| 10 | 15 | 2.9135 | 8.4882 | 115.6354 | 2401.6358 | 15 | 16 | 3.1645 | 10.0140 | 154.7784 | 3294.2642 |
| 10 | 16 | 2.9207 | 8.5303 | 112.4304 | 2151.9196 | 15 | 17 | 3.1973 | 10.2228 | 156.1102 | 3116.9914 |
| 10 | 17 | 2.9711 | 8.8273 | 124.1551 | 2622.6292 | 15 | 18 | 3.2201 | 10.3691 | 161.7488 | 3394.1120 |
| 10 | 18 | 2.9654 | 8.7939 | 117.8408 | 2246.3974 | 15 | 19 | 3.2674 | 10.6757 | 165.9367 | 3308.6814 |
| 10 | 19 | 3.0459 | 9.2773 | 134.3320 | 2826.7432 | 15 | 20 | 3.2881 | 10.8117 | 169.8669 | 3479.8190 |
| 10 | 20 | 3.0135 | 9.0809 | 121.6565 | 2201.7335 | 15 | 21 | 3.3159 | 10.9952 | 174.9537 | 3589.6798 |
| 10 | 21 | 3.0739 | 9.4490 | 137.8670 | 2900.9966 | 15 | 22 | 3.2942 | 10.8517 | 174.2214 | 3753.2565 |
| 10 | 22 | 2.9837 | 8.9026 | 117.2474 | 2138.0907 | 15 | 23 | 3.3293 | 11.0844 | 180.8383 | 3907.1536 |
| 10 | 23 | 3.0470 | 9.2842 | 135.2258 | 2948.6208 | 15 | 24 | 3.2966 | 10.8676 | 178.0777 | 4025.6922 |
| 10 | 24 | 2.9600 | 8.7616 | 115.0642 | 2191.6104 | 15 | 25 | 3.3765 | 11.4007 | 190.1139 | 4220.6689 |
| 11 | 6 | 3.0691 | 9.4195 | 142.7622 | 3123.8095 | 15 | 26 | 3.3854 | 11.4610 | 191.5114 | 4326.6418 |
| 11 | 7 | 2.9765 | 8.8594 | 118.4371 | 2080.6454 | 15 | 27 | 3.4437 | 11.8593 | 200.0403 | 4431.1572 |
| 11 | 8 | 3.0496 | 9.2998 | 134.8621 | 2730.9716 | 15 | 28 | 3.4480 | 11.8886 | 200.1590 | 4477.3776 |
| 11 | 9 | 2.9697 | 8.8192 | 115.6522 | 1941.0119 | 15 | 29 | 3.4717 | 12.0529 | 205.9560 | 4652.2042 |
| 11 | 10 | 3.0210 | 9.1262 | 130.5213 | 2643.4371 | 15 | 30 | 3.4605 | 11.9753 | 207.2968 | 4905.8273 |
| 11 | 11 | 2.9701 | 8.8212 | 117.5532 | 2073.9678 | 15 | 31 | 3.4894 | 12.1759 | 214.2951 | 5101.7276 |
| 11 | 12 | 2.9892 | 8.9352 | 125.6166 | 2536.7397 | 15 | 32 | 3.4802 | 12.1120 | 216.5665 | 5407.8661 |
| 11 | 13 | 2.9674 | 8.8055 | 118.9392 | 2191.7120 | 16 | 10 | 3.2184 | 10.3580 | 162.4035 | 3405.4623 |
| 11 | 14 | 2.9633 | 8.7811 | 122.3787 | 2503.2150 | 16 | 11 | 3.1742 | 10.0754 | 154.8259 | 3206.0512 |
| 11 | 15 | 2.9826 | 8.8959 | 121.9617 | 2307.8636 | 16 | 12 | 3.2267 | 10.4114 | 165.7953 | 3575.1831 |
| 11 | 16 | 3.0011 | 9.0068 | 127.3677 | 2613.3579 | 16 | 13 | 3.1866 | 10.1547 | 159.1609 | 3407.3142 |
| 11 | 17 | 3.0374 | 9.2256 | 130.0021 | 2500.6173 | 16 | 14 | 3.2283 | 10.4217 | 167.4070 | 3678.1525 |
| 11 | 18 | 3.0727 | 9.4413 | 137.9932 | 2869.1912 | 16 | 15 | 3.1999 | 10.2394 | 162.7762 | 3572.2274 |
| 11 | 19 | 3.1109 | 9.6778 | 140.1710 | 2699.3818 | 16 | 16 | 3.2476 | 10.5467 | 172.0325 | 3874.5730 |
| 11 | 20 | 3.1546 | 9.9518 | 149.0694 | 3067.5167 | 16 | 17 | 3.2370 | 10.4783 | 167.5949 | 3653.8238 |
| 11 | 21 | 3.1558 | 9.9590 | 147.9174 | 2935.5749 | 16 | 18 | 3.2744 | 10.7220 | 174.2401 | 3848.1554 |
| 11 | 22 | 3.1326 | 9.8132 | 147.6933 | 3166.2174 | 16 | 19 | 3.2880 | 10.8107 | 172.0705 | 3610.6041 |
| 11 | 23 | 3.1587 | 9.9775 | 151.9765 | 3223.9831 | 16 | 20 | 3.3077 | 10.9406 | 174.6436 | 3655.1638 |
| 11 | 24 | 3.1171 | 9.7165 | 148.6270 | 3380.0944 | 16 | 21 | 3.3224 | 11.0386 | 176.9939 | 3716.6472 |
| 11 | 25 | 3.2106 | 10.3077 | 161.5512 | 3544.4863 | 16 | 22 | 3.3165 | 10.9990 | 178.2338 | 3860.5819 |
| 11 | 26 | 3.2187 | 10.3602 | 162.9177 | 3676.0283 | 16 | 23 | 3.3295 | 11.0855 | 180.9672 | 3961.7938 |
| 11 | 27 | 3.3063 | 10.9316 | 175.7521 | 3852.0570 | 16 | 24 | 3.3217 | 11.0334 | 180.9870 | 4040.3487 |
| 11 | 28 | 3.3189 | 11.0149 | 178.1127 | 4012.5828 | 16 | 25 | 3.3680 | 11.3432 | 187.5762 | 4159.2677 |
| 12 | 7 | 3.0653 | 9.3960 | 136.2766 | 2718.1155 | 16 | 26 | 3.3772 | 11.4057 | 187.0504 | 4070.6153 |
| 12 | 8 | 3.0937 | 9.5709 | 139.8511 | 2774.3970 | 16 | 27 | 3.4243 | 11.7256 | 194.4812 | 4237.4686 |
| 12 | 9 | 3.0517 | 9.3126 | 132.2422 | 2546.4968 | 16 | 28 | 3.4135 | 11.6522 | 189.8168 | 4010.3210 |
| 12 | 10 | 3.0635 | 9.3851 | 135.1507 | 2672.8025 | 16 | 29 | 3.4462 | 11.8764 | 198.7801 | 4394.6552 |
| 12 | 11 | 3.0334 | 9.2016 | 130.5000 | 2555.0212 | 16 | 30 | 3.4320 | 11.7788 | 197.1050 | 4391.9356 |
| 12 | 12 | 3.0389 | 9.2348 | 131.7535 | 2620.0326 | 16 | 31 | 3.4606 | 11.9757 | 205.9734 | 4800.6842 |
| 12 | 13 | 3.0204 | 9.1230 | 130.0974 | 2619.1853 | 16 | 32 | 3.4940 | 12.2082 | 214.9700 | 5060.8297 |
| 12 | 14 | 3.0209 | 9.1257 | 129.8721 | 2624.5570 | 16 | 33 | 3.5232 | 12.4133 | 226.2326 | 5594.7673 |
| 12 | 15 | 3.0277 | 9.1667 | 131.6116 | 2689.8536 | 17 | 11 | 3.2087 | 10.2955 | 160.1821 | 3299.6652 |
| 12 | 16 | 3.0452 | 9.2733 | 132.9525 | 2695.6367 | 17 | 12 | 3.2444 | 10.5264 | 169.1675 | 3663.3061 |
| 12 | 17 | 3.0831 | 9.5058 | 139.7978 | 2892.6413 | 17 | 13 | 3.2298 | 10.4313 | 165.8753 | 3527.9584 |
| 12 | 18 | 3.0961 | 9.5858 | 139.4648 | 2808.8679 | 17 | 14 | 3.2530 | 10.5822 | 172.2319 | 3813.2876 |
| 12 | 19 | 3.1602 | 9.9869 | 150.8276 | 3124.0548 | 17 | 15 | 3.2494 | 10.5584 | 170.4618 | 3708.5944 |
| 12 | 20 | 3.1527 | 9.9397 | 145.3618 | 2832.2415 | 17 | 16 | 3.2781 | 10.7461 | 177.7791 | 4026.4784 |
| 12 | 21 | 3.1946 | 10.2054 | 156.7105 | 3305.6509 | 17 | 17 | 3.2894 | 10.8204 | 175.8049 | 3799.8752 |
| 12 | 22 | 3.1364 | 9.8371 | 144.8575 | 2946.2881 | 17 | 18 | 3.3132 | 10.9775 | 180.9464 | 4004.4377 |
| 12 | 23 | 3.1853 | 10.1459 | 158.8266 | 3549.7696 | 17 | 19 | 3.3410 | 11.1625 | 180.7761 | 3782.4059 |
| 12 | 24 | 3.1273 | 9.7801 | 146.4652 | 3157.2841 | 17 | 20 | 3.3550 | 11.2558 | 182.6404 | 3835.7446 |
| 12 | 25 | 3.2305 | 10.4363 | 167.3795 | 3850.7767 | 17 | 21 | 3.3755 | 11.3937 | 185.7044 | 3880.8391 |
| 12 | 26 | 3.2084 | 10.2940 | 157.2275 | 3359.7736 | 17 | 22 | 3.3688 | 11.3490 | 187.5224 | 4086.4076 |
| 12 | 27 | 3.3297 | 11.0867 | 182.0512 | 4168.8522 | 17 | 23 | 3.3859 | 11.4641 | 189.6818 | 4090.5290 |
| 12 | 28 | 3.2944 | 10.8528 | 169.7503 | 3624.5961 | 17 | 24 | 3.3739 | 11.3830 | 190.5000 | 4279.5948 |
| 12 | 29 | 3.3769 | 11.4035 | 191.9223 | 4530.8233 | 17 | 25 | 3.4104 | 11.6307 | 193.2695 | 4173.7155 |
| 13 | 8 | 3.0625 | 9.3791 | 131.1769 | 2402.2791 | 17 | 26 | 3.4188 | 11.6880 | 194.2759 | 4224.5828 |
| 13 | 9 | 3.0223 | 9.1344 | 119.6611 | 1851.6990 | 17 | 27 | 3.4383 | 11.8216 | 194.4979 | 4072.2187 |
| 13 | 10 | 3.0583 | 9.3531 | 132.5443 | 2516.6109 | 17 | 28 | 3.4389 | 11.8260 | 193.5126 | 4042.8546 |
| 13 | 11 | 3.0428 | 9.2589 | 126.4583 | 2160.1493 | 17 | 29 | 3.4514 | 11.9120 | 196.7078 | 4143.8718 |
| 13 | 12 | 3.0586 | 9.3551 | 135.0048 | 2674.1597 | 17 | 30 | 3.4517 | 11.9145 | 199.7130 | 4391.9955 |
| 13 | 13 | 3.0610 | 9.3700 | 132.8482 | 2459.4293 | 17 | 31 | 3.4903 | 12.1825 | 210.2908 | 4731.4728 |
| 13 | 14 | 3.0611 | 9.3702 | 138.1133 | 2864.7889 | 17 | 32 | 3.4960 | 12.2223 | 214.4629 | 4993.3162 |

TABLE 4. Charge radii, 2nd moment, 4th moment and 6th moment obtained by DNN.

| $Z$ | $N$ | $R_c$(fm) | $R_2$(fm$^2$) | $R_4$(fm$^4$) | $R_6$(fm$^6$) | $Z$ | $N$ | $R_c$(fm) | $R_2$(fm$^2$) | $R_4$(fm$^4$) | $R_6$(fm$^6$) |
|---|---|---|---|---|---|---|---|---|---|---|---|
| 17 | 33 | 3.6136 | 13.0580 | 268.0376 | 7894.8600 | 21 | 37 | 3.6414 | 13.2595 | 241.3678 | 5514.2545 |
| 17 | 34 | 3.5989 | 12.9519 | 265.5955 | 7955.0157 | 21 | 38 | 3.6397 | 13.2474 | 242.8279 | 5769.2399 |
| 17 | 35 | 3.7207 | 13.8432 | 285.6198 | 8051.5511 | 21 | 39 | 3.6858 | 13.5854 | 254.9497 | 6051.2712 |
| 18 | 11 | 3.2931 | 10.8445 | 177.5239 | 3868.7380 | 21 | 40 | 3.6638 | 13.4232 | 248.4724 | 5875.3767 |
| 18 | 12 | 3.3366 | 11.1329 | 186.1441 | 4133.3380 | 21 | 41 | 3.7013 | 13.6999 | 257.9602 | 6058.6565 |
| 18 | 13 | 3.3096 | 10.9534 | 182.1907 | 4065.2447 | 21 | 42 | 3.6982 | 13.6770 | 257.0040 | 6104.9896 |
| 18 | 14 | 3.3447 | 11.1868 | 188.5419 | 4243.4463 | 22 | 15 | 3.4861 | 12.1526 | 221.1984 | 5354.2842 |
| 18 | 15 | 3.3267 | 11.0667 | 186.9979 | 4275.9640 | 22 | 16 | 3.5191 | 12.3843 | 227.6076 | 5548.3633 |
| 18 | 16 | 3.3677 | 11.3411 | 194.2743 | 4468.6557 | 22 | 17 | 3.5321 | 12.4759 | 230.4513 | 5638.6687 |
| 18 | 17 | 3.3552 | 11.2573 | 191.8670 | 4409.9770 | 22 | 18 | 3.5560 | 12.6450 | 234.0314 | 5706.7533 |
| 18 | 18 | 3.3866 | 11.4690 | 196.3112 | 4479.7517 | 22 | 19 | 3.5650 | 12.7092 | 234.5851 | 5680.7830 |
| 18 | 19 | 3.3841 | 11.4522 | 192.8674 | 4283.7496 | 22 | 20 | 3.5748 | 12.7792 | 234.7162 | 5617.5204 |
| 18 | 20 | 3.4004 | 11.5629 | 194.1250 | 4248.0117 | 22 | 21 | 3.5806 | 12.8210 | 236.4721 | 5695.9165 |
| 18 | 21 | 3.4064 | 11.6032 | 194.9263 | 4276.8342 | 22 | 22 | 3.5931 | 12.9101 | 239.9326 | 5830.6047 |
| 18 | 22 | 3.4192 | 11.6907 | 199.5870 | 4495.7021 | 22 | 23 | 3.5911 | 12.8963 | 239.9630 | 5855.3164 |
| 18 | 23 | 3.4182 | 11.6843 | 199.0789 | 4490.8410 | 22 | 24 | 3.6063 | 13.0052 | 243.6773 | 5984.0217 |
| 18 | 24 | 3.4293 | 11.7599 | 202.8406 | 4662.2813 | 22 | 25 | 3.5994 | 12.9559 | 240.3694 | 5827.6227 |
| 18 | 25 | 3.4326 | 11.7824 | 199.8195 | 4454.9358 | 22 | 26 | 3.5886 | 12.8782 | 238.7861 | 5841.4895 |
| 18 | 26 | 3.4394 | 11.8292 | 199.2950 | 4361.1976 | 22 | 27 | 3.5744 | 12.7761 | 233.0149 | 5580.7085 |
| 18 | 27 | 3.4414 | 11.8430 | 195.8329 | 4135.5149 | 22 | 28 | 3.5641 | 12.7029 | 239.1549 | 6438.4577 |
| 18 | 28 | 3.4365 | 11.8092 | 193.0137 | 3968.2965 | 22 | 29 | 3.5848 | 12.8510 | 243.3528 | 6486.3123 |
| 18 | 29 | 3.4481 | 11.8894 | 196.1240 | 4121.8166 | 22 | 30 | 3.6001 | 12.9606 | 231.6725 | 5181.9706 |
| 18 | 30 | 3.4814 | 12.1198 | 206.2988 | 4499.3753 | 22 | 31 | 3.6110 | 13.0395 | 235.3795 | 5372.2141 |
| 18 | 31 | 3.4849 | 12.1448 | 208.5181 | 4646.6331 | 22 | 32 | 3.6129 | 13.0531 | 235.3316 | 5378.9131 |
| 18 | 32 | 3.4969 | 12.2282 | 224.5337 | 5855.4825 | 22 | 33 | 3.6186 | 13.0942 | 238.2416 | 5584.1004 |
| 18 | 33 | 3.5440 | 12.5600 | 237.5968 | 6305.7798 | 22 | 34 | 3.6352 | 13.2146 | 241.5199 | 5643.0125 |
| 18 | 34 | 3.5443 | 12.5618 | 219.2886 | 4918.9080 | 22 | 35 | 3.6402 | 13.2514 | 244.6658 | 5878.5686 |
| 18 | 35 | 3.5959 | 12.9307 | 239.8094 | 6025.5484 | 22 | 36 | 3.6593 | 13.3908 | 247.4990 | 5846.6348 |
| 18 | 36 | 3.5344 | 12.4920 | 219.5696 | 5243.3417 | 22 | 37 | 3.6674 | 13.4495 | 251.9575 | 6148.9449 |
| 19 | 12 | 3.3493 | 11.2178 | 189.3472 | 4254.0347 | 22 | 38 | 3.6864 | 13.5896 | 253.6962 | 6008.0334 |
| 19 | 13 | 3.3377 | 11.1405 | 187.1384 | 4182.1892 | 22 | 39 | 3.7012 | 13.6990 | 260.7071 | 6421.5982 |
| 19 | 14 | 3.3627 | 11.3079 | 192.8653 | 4400.4821 | 22 | 40 | 3.7075 | 13.7453 | 257.5726 | 6001.3396 |
| 19 | 15 | 3.3632 | 11.3112 | 193.0720 | 4402.8225 | 22 | 41 | 3.7221 | 13.8542 | 264.7919 | 6435.4023 |
| 19 | 16 | 3.3950 | 11.5259 | 200.0975 | 4653.1940 | 22 | 42 | 3.7378 | 13.9710 | 264.5233 | 6154.2738 |
| 19 | 17 | 3.4002 | 11.5613 | 199.1909 | 4553.5806 | 22 | 43 | 3.7668 | 14.1890 | 276.9789 | 6899.2132 |
| 19 | 18 | 3.4220 | 11.7103 | 203.1976 | 4672.5111 | 23 | 16 | 3.5309 | 12.4671 | 231.6527 | 5712.8934 |
| 19 | 19 | 3.4268 | 11.7428 | 200.1934 | 4438.8456 | 23 | 17 | 3.5555 | 12.6418 | 235.2151 | 5757.8741 |
| 19 | 20 | 3.4359 | 11.8054 | 200.6323 | 4412.2373 | 23 | 18 | 3.5786 | 12.8066 | 240.4133 | 5955.2863 |
| 19 | 21 | 3.4411 | 11.8412 | 200.8235 | 4386.0223 | 23 | 19 | 3.5910 | 12.8953 | 240.1962 | 5844.6469 |
| 19 | 22 | 3.4545 | 11.9339 | 206.2835 | 4667.3624 | 23 | 20 | 3.6039 | 12.9884 | 242.1841 | 5897.4448 |
| 19 | 23 | 3.4500 | 11.9026 | 203.8300 | 4533.7967 | 23 | 21 | 3.6058 | 13.0017 | 241.8864 | 5852.4479 |
| 19 | 24 | 3.4620 | 11.9857 | 209.2201 | 4828.1754 | 23 | 22 | 3.6201 | 13.1054 | 247.1037 | 6102.4534 |
| 19 | 25 | 3.4528 | 11.9217 | 202.1995 | 4415.7014 | 23 | 23 | 3.6177 | 13.0880 | 246.0158 | 6042.3703 |
| 19 | 26 | 3.4604 | 11.9742 | 203.0318 | 4433.8162 | 23 | 24 | 3.6319 | 13.1904 | 251.1230 | 6290.3985 |
| 19 | 27 | 3.4472 | 11.8833 | 195.6443 | 4025.7569 | 23 | 25 | 3.5439 | 12.5590 | 233.8915 | 5925.1722 |
| 19 | 28 | 3.4458 | 11.8735 | 194.0885 | 3950.6272 | 23 | 26 | 3.5759 | 12.7871 | 239.4402 | 6046.9564 |
| 19 | 29 | 3.4629 | 11.9918 | 199.1185 | 4140.1612 | 23 | 27 | 3.6017 | 12.9725 | 251.8600 | 7060.0868 |
| 19 | 30 | 3.4820 | 12.1240 | 206.0272 | 4466.6045 | 23 | 28 | 3.6012 | 12.9688 | 250.5639 | 6960.1780 |
| 19 | 31 | 3.5083 | 12.3085 | 221.5542 | 5414.9020 | 23 | 29 | 3.6209 | 13.1106 | 237.5520 | 5368.5987 |
| 19 | 32 | 3.5252 | 12.4274 | 226.7436 | 5645.3790 | 23 | 30 | 3.6346 | 13.2104 | 245.0320 | 5865.1837 |
| 19 | 33 | 3.5764 | 12.7904 | 222.2317 | 4621.7662 | 23 | 31 | 3.6405 | 13.2533 | 243.0544 | 5595.5148 |
| 19 | 34 | 3.5609 | 12.6800 | 221.5137 | 4899.0984 | 23 | 32 | 3.6513 | 13.3323 | 250.1432 | 6125.0597 |
| 19 | 35 | 3.5718 | 12.7580 | 224.7992 | 5037.2490 | 23 | 33 | 3.6673 | 13.4489 | 251.0801 | 5937.8686 |
| 19 | 36 | 3.5555 | 12.6415 | 223.4018 | 5273.3518 | 23 | 34 | 3.6740 | 13.4985 | 256.6658 | 6412.0265 |
| 19 | 37 | 3.6142 | 13.0628 | 240.1207 | 5803.8471 | 23 | 35 | 3.6954 | 13.6558 | 259.1108 | 6249.6346 |
| 19 | 38 | 3.5946 | 12.9213 | 236.5384 | 5905.8951 | 23 | 36 | 3.6988 | 13.6808 | 263.0412 | 6638.9040 |
| 19 | 39 | 3.6695 | 13.4649 | 258.8760 | 6653.1928 | 23 | 37 | 3.7279 | 13.8969 | 268.1100 | 6563.0426 |
| 19 | 40 | 3.6284 | 13.1651 | 246.1225 | 6219.5096 | 23 | 38 | 3.7280 | 13.8983 | 270.1538 | 6838.4068 |
| 20 | 13 | 3.3914 | 11.5017 | 198.4817 | 4569.4028 | 23 | 39 | 3.7673 | 14.1924 | 279.1086 | 6924.5797 |
| 20 | 14 | 3.4143 | 11.6575 | 202.6003 | 4686.2095 | 23 | 40 | 3.7525 | 14.0813 | 275.2754 | 6900.4532 |
| 20 | 15 | 3.4172 | 11.6771 | 204.2851 | 4770.8010 | 23 | 41 | 3.7833 | 14.3130 | 282.0394 | 6948.3964 |
| 20 | 16 | 3.4438 | 11.8599 | 208.4759 | 4853.9176 | 23 | 42 | 3.7869 | 14.3405 | 283.9611 | 7180.9366 |
| 20 | 17 | 3.4496 | 11.9000 | 209.7233 | 4892.5623 | 23 | 43 | 3.8185 | 14.5811 | 291.8799 | 7370.9352 |
| 20 | 18 | 3.4633 | 11.9947 | 210.2071 | 4821.7138 | 23 | 44 | 3.8333 | 14.6943 | 297.4182 | 7772.1997 |
| 20 | 19 | 3.4656 | 12.0103 | 209.2653 | 4747.3816 | 24 | 17 | 3.5734 | 12.7691 | 241.0750 | 6026.0135 |
| 20 | 20 | 3.4646 | 12.0033 | 206.2043 | 4544.9390 | 24 | 18 | 3.6083 | 13.0200 | 247.7475 | 6223.8680 |
| 20 | 21 | 3.4760 | 12.0822 | 209.7477 | 4709.4670 | 24 | 19 | 3.6168 | 13.0813 | 248.0545 | 6182.4356 |
| 20 | 22 | 3.4893 | 12.1751 | 213.0926 | 4827.1625 | 24 | 20 | 3.6382 | 13.2367 | 251.5109 | 6257.2594 |
| 20 | 23 | 3.4911 | 12.1879 | 214.5143 | 4926.1154 | 24 | 21 | 3.6376 | 13.2322 | 251.3870 | 6252.5625 |
| 20 | 24 | 3.5037 | 12.2759 | 217.2752 | 5007.1639 | 24 | 22 | 3.6616 | 13.4071 | 258.3494 | 6543.3827 |
| 20 | 25 | 3.4895 | 12.1764 | 212.3008 | 4797.8935 | 24 | 23 | 3.6557 | 13.3641 | 257.3693 | 6524.9025 |
| 20 | 26 | 3.4908 | 12.1855 | 210.1411 | 4620.2965 | 24 | 24 | 3.5702 | 12.7465 | 245.1743 | 6541.8697 |
| 20 | 27 | 3.4673 | 12.0224 | 202.0219 | 4277.5035 | 24 | 25 | 3.5427 | 12.5509 | 238.8003 | 6339.9830 |
| 20 | 28 | 3.4839 | 12.1374 | 205.1635 | 4336.6155 | 24 | 26 | 3.6602 | 13.3968 | 273.2337 | 8176.4773 |
| 20 | 29 | 3.4811 | 12.1181 | 204.9946 | 4370.6281 | 24 | 27 | 3.6423 | 13.2660 | 265.2345 | 7670.8523 |
| 20 | 30 | 3.5159 | 12.3613 | 222.9390 | 5503.2152 | 24 | 28 | 3.6398 | 13.2483 | 249.4701 | 6154.0259 |
| 20 | 31 | 3.5270 | 12.4394 | 225.1029 | 5512.9277 | 24 | 29 | 3.6510 | 13.3295 | 252.8478 | 6299.1317 |
| 20 | 32 | 3.5487 | 12.5934 | 211.9416 | 4178.6238 | 24 | 30 | 3.6702 | 13.4701 | 258.0933 | 6504.2744 |
| 20 | 33 | 3.5601 | 12.6743 | 217.1903 | 4515.4463 | 24 | 31 | 3.6790 | 13.5348 | 261.0070 | 6646.1963 |
| 20 | 34 | 3.5440 | 12.5602 | 211.0090 | 4269.0538 | 24 | 32 | 3.6995 | 13.6866 | 266.4087 | 6846.8341 |
| 20 | 35 | 3.5521 | 12.6176 | 216.7189 | 4699.6004 | 24 | 33 | 3.7064 | 13.7376 | 268.8364 | 6974.2868 |
| 20 | 36 | 3.5679 | 12.7296 | 217.7339 | 4560.0261 | 24 | 34 | 3.7258 | 13.8813 | 273.0022 | 7074.7513 |
| 20 | 37 | 3.5797 | 12.8143 | 225.3577 | 5102.9111 | 24 | 35 | 3.7319 | 13.9273 | 275.3146 | 7199.5004 |
| 20 | 38 | 3.5979 | 12.9446 | 225.5431 | 4836.1239 | 24 | 36 | 3.7506 | 14.0669 | 278.4350 | 7205.9659 |
| 20 | 39 | 3.6173 | 13.0848 | 236.2539 | 5532.7361 | 24 | 37 | 3.7587 | 14.1277 | 281.4922 | 7357.3654 |
| 20 | 40 | 3.6175 | 13.0861 | 229.6205 | 4856.7187 | 24 | 38 | 3.7759 | 14.2575 | 283.4832 | 7275.5637 |
| 20 | 41 | 3.6345 | 13.2094 | 239.6375 | 5500.7448 | 24 | 39 | 3.7894 | 14.3596 | 288.4898 | 7499.7501 |
| 21 | 14 | 3.4368 | 11.8113 | 209.3251 | 4942.2938 | 24 | 40 | 3.7976 | 14.4214 | 287.2159 | 7259.1222 |
| 21 | 15 | 3.4491 | 11.8963 | 211.6363 | 4993.2217 | 24 | 41 | 3.8118 | 14.5301 | 292.9912 | 7563.8050 |
| 21 | 16 | 3.4787 | 12.1012 | 218.4567 | 5243.8928 | 24 | 42 | 3.8268 | 14.6444 | 294.0727 | 7450.8067 |
| 21 | 17 | 3.4930 | 12.2008 | 219.8277 | 5224.3276 | 24 | 43 | 3.8511 | 14.8309 | 303.9260 | 8015.0939 |
| 21 | 18 | 3.5135 | 12.3444 | 223.7361 | 5344.2455 | 24 | 44 | 3.8641 | 14.9316 | 304.5130 | 7888.2844 |
| 21 | 19 | 3.5180 | 12.3760 | 221.7057 | 5175.2516 | 24 | 45 | 3.8933 | 15.1581 | 316.5110 | 8599.4534 |
| 21 | 20 | 3.5264 | 12.4354 | 222.1113 | 5149.0228 | 24 | 46 | 3.9005 | 15.2135 | 315.0580 | 8361.0380 |
| 21 | 21 | 3.5276 | 12.4440 | 221.8889 | 5122.2355 | 25 | 18 | 3.6336 | 13.2032 | 252.8229 | 6351.3966 |
| 21 | 22 | 3.5435 | 12.5564 | 227.3456 | 5376.3625 | 25 | 19 | 3.6439 | 13.2783 | 252.8889 | 6269.8459 |
| 21 | 23 | 3.5344 | 12.4921 | 224.2455 | 5233.5549 | 25 | 20 | 3.6547 | 13.3567 | 254.4199 | 6309.4064 |
| 21 | 24 | 3.5496 | 12.5997 | 229.7436 | 5501.5686 | 25 | 21 | 3.6607 | 13.4006 | 256.5255 | 6406.3011 |
| 21 | 25 | 3.5372 | 12.5117 | 223.0991 | 5142.0843 | 25 | 22 | 3.6766 | 13.5177 | 261.4565 | 6618.7786 |
| 21 | 26 | 3.5462 | 12.5755 | 224.3054 | 5167.8409 | 25 | 23 | 3.5329 | 12.4814 | 238.1663 | 6387.4335 |
| 21 | 27 | 3.5250 | 12.4259 | 217.6013 | 4893.7522 | 25 | 24 | 3.5255 | 12.4293 | 237.7659 | 6445.3870 |
| 21 | 28 | 3.5350 | 12.4962 | 218.5539 | 4890.3891 | 25 | 25 | 3.7278 | 13.8966 | 289.4167 | 8574.2882 |
| 21 | 29 | 3.5267 | 12.4374 | 226.0891 | 5732.0409 | 25 | 26 | 3.7066 | 13.7390 | 282.6218 | 8279.8117 |
| 21 | 30 | 3.5587 | 12.6641 | 234.3416 | 5978.7492 | 25 | 27 | 3.6710 | 13.4763 | 258.7799 | 6571.4204 |
| 21 | 31 | 3.5753 | 12.7830 | 219.0855 | 4390.6627 | 25 | 28 | 3.6668 | 13.4455 | 255.6811 | 6301.9031 |
| 21 | 32 | 3.5845 | 12.8486 | 225.5128 | 4896.0375 | 25 | 29 | 3.6762 | 13.5148 | 255.8174 | 6182.2286 |
| 21 | 33 | 3.5769 | 12.7945 | 220.7434 | 4590.3883 | 25 | 30 | 3.6992 | 13.6842 | 265.4709 | 6719.1016 |
| 21 | 34 | 3.5843 | 12.8473 | 226.6652 | 5088.4213 | 25 | 31 | 3.7093 | 13.7592 | 265.6039 | 6577.9757 |
| 21 | 35 | 3.6051 | 12.9969 | 230.0493 | 5032.2786 | 25 | 32 | 3.7282 | 13.8995 | 274.1492 | 7096.1709 |
| 21 | 36 | 3.6080 | 13.0175 | 233.9223 | 5426.8447 | 25 | 33 | 3.7362 | 13.9593 | 272.9820 | 6849.2267 |

TABLE 4. Charge radii, 2nd moment, 4th moment and 6th moment obtained by DNN.

| $Z$ | $N$ | $R_c$ (fm) | $R_2$ (fm$^2$) | $R_4$ (fm$^4$) | $R_6$ (fm$^6$) | $Z$ | $N$ | $R_c$ (fm) | $R_2$ (fm$^2$) | $R_4$ (fm$^4$) | $R_6$ (fm$^6$) |
|---|---|---|---|---|---|---|---|---|---|---|---|
| 25 | 34 | 3.7516 | 14.0743 | 280.2271 | 7319.1710 | 28 | 51 | 3.9633 | 15.7077 | 327.9399 | 8444.1002 |
| 25 | 35 | 3.7605 | 14.1416 | 278.9363 | 7021.5938 | 28 | 52 | 3.9822 | 15.8582 | 333.6292 | 8576.4448 |
| 25 | 36 | 3.7727 | 14.2333 | 284.6841 | 7415.7784 | 28 | 53 | 3.9927 | 15.9413 | 341.5996 | 9185.7665 |
| 25 | 37 | 3.7865 | 14.3377 | 284.8874 | 7150.3973 | 28 | 54 | 4.0135 | 16.1085 | 347.5326 | 9304.5036 |
| 25 | 38 | 3.7952 | 14.4034 | 288.8272 | 7440.6809 | 29 | 23 | 3.8065 | 14.4895 | 313.8440 | 9713.0412 |
| 25 | 39 | 3.8174 | 14.5727 | 292.2173 | 7304.9351 | 29 | 24 | 3.8266 | 14.6430 | 321.8935 | 10159.7422 |
| 25 | 40 | 3.8136 | 14.5435 | 291.7599 | 7397.6300 | 29 | 25 | 3.8034 | 14.4661 | 307.9850 | 9161.6493 |
| 25 | 41 | 3.8284 | 14.6564 | 293.5916 | 7269.9558 | 29 | 26 | 3.7900 | 14.3638 | 301.3130 | 8701.6795 |
| 25 | 42 | 3.8380 | 14.7305 | 297.4299 | 7557.7391 | 29 | 27 | 3.7756 | 14.2548 | 291.2187 | 7946.6396 |
| 25 | 43 | 3.8540 | 14.8530 | 300.3206 | 7552.7197 | 29 | 28 | 3.7757 | 14.2563 | 287.7855 | 7544.8885 |
| 25 | 44 | 3.8709 | 14.9841 | 306.5201 | 7948.3485 | 29 | 29 | 3.7923 | 14.3812 | 292.3381 | 7728.4998 |
| 25 | 45 | 3.8837 | 15.0829 | 308.8124 | 7967.8462 | 29 | 30 | 3.8211 | 14.6005 | 302.2663 | 8160.6787 |
| 25 | 46 | 3.9027 | 15.2314 | 315.4947 | 8359.7193 | 29 | 31 | 3.8289 | 14.6609 | 303.3784 | 8154.4752 |
| 25 | 47 | 3.9100 | 15.2882 | 316.1098 | 8329.1600 | 29 | 32 | 3.8552 | 14.8629 | 312.5641 | 8570.0545 |
| 25 | 48 | 3.9293 | 15.4395 | 322.4454 | 8656.4280 | 29 | 33 | 3.8569 | 14.8757 | 311.0813 | 8412.5694 |
| 26 | 19 | 3.6674 | 13.4498 | 258.7288 | 6482.8918 | 29 | 34 | 3.8805 | 15.0580 | 319.1842 | 8782.3939 |
| 26 | 20 | 3.6774 | 13.5231 | 260.3507 | 6542.5462 | 29 | 35 | 3.8798 | 15.0526 | 316.5443 | 8540.6958 |
| 26 | 21 | 3.6856 | 13.5833 | 262.7720 | 6629.9692 | 29 | 36 | 3.9001 | 15.2105 | 323.1955 | 8836.5894 |
| 26 | 22 | 3.5877 | 12.8713 | 249.1184 | 6706.4869 | 29 | 37 | 3.9013 | 15.2201 | 321.1050 | 8596.5588 |
| 26 | 23 | 3.5469 | 12.5807 | 242.1891 | 6585.7341 | 29 | 38 | 3.9178 | 15.3494 | 326.0403 | 8799.1631 |
| 26 | 24 | 3.7677 | 14.1958 | 305.1241 | 9558.8634 | 29 | 39 | 3.9247 | 15.4033 | 326.1465 | 8654.0620 |
| 26 | 25 | 3.7599 | 14.1365 | 298.5726 | 8971.5575 | 29 | 40 | 3.9359 | 15.4910 | 329.2878 | 8767.2768 |
| 26 | 26 | 3.7142 | 13.7956 | 275.6524 | 7555.1715 | 29 | 41 | 3.9377 | 15.5052 | 328.3264 | 8624.9754 |
| 26 | 27 | 3.7034 | 13.7153 | 269.0449 | 7024.9988 | 29 | 42 | 3.9555 | 15.6462 | 333.7073 | 8840.9439 |
| 26 | 28 | 3.6958 | 13.6588 | 261.5254 | 6396.5106 | 29 | 43 | 3.9536 | 15.6311 | 331.6534 | 8700.6852 |
| 26 | 29 | 3.7137 | 13.7915 | 268.1538 | 6729.8426 | 29 | 44 | 3.9722 | 15.7783 | 337.2881 | 8934.2476 |
| 26 | 30 | 3.7356 | 13.9547 | 273.8981 | 6928.7212 | 29 | 45 | 3.9665 | 15.7334 | 334.0831 | 8778.6975 |
| 26 | 31 | 3.7480 | 14.0473 | 278.9303 | 7206.6785 | 29 | 46 | 3.9825 | 15.8600 | 338.6725 | 8956.1962 |
| 26 | 32 | 3.7674 | 14.1932 | 283.1707 | 7302.4986 | 29 | 47 | 3.9739 | 15.7917 | 334.5403 | 8780.5582 |
| 26 | 33 | 3.7739 | 14.2423 | 286.3181 | 7503.9366 | 29 | 48 | 3.9870 | 15.8964 | 337.8687 | 8875.8509 |
| 26 | 34 | 3.7928 | 14.3852 | 289.6759 | 7519.6664 | 29 | 49 | 3.9763 | 15.8112 | 333.0958 | 8684.6727 |
| 26 | 35 | 3.7945 | 14.3979 | 291.0492 | 7635.2784 | 29 | 50 | 3.9882 | 15.9057 | 335.5696 | 8707.2809 |
| 26 | 36 | 3.8141 | 14.5476 | 294.1560 | 7606.3566 | 29 | 51 | 3.9901 | 15.9211 | 338.2679 | 8954.9509 |
| 26 | 37 | 3.8133 | 14.5415 | 294.3354 | 7645.2486 | 29 | 52 | 4.0114 | 16.0909 | 346.8910 | 9342.7694 |
| 26 | 38 | 3.8339 | 14.6991 | 297.5519 | 7604.5650 | 29 | 53 | 4.0239 | 16.1920 | 353.5241 | 9754.4007 |
| 26 | 39 | 3.8339 | 14.6989 | 297.5671 | 7601.4909 | 29 | 54 | 4.0437 | 16.3517 | 361.7161 | 10119.4137 |
| 26 | 40 | 3.8513 | 14.8325 | 300.2839 | 7547.3161 | 29 | 55 | 4.0644 | 16.5193 | 371.3245 | 10649.3690 |
| 26 | 41 | 3.8482 | 14.8085 | 299.9358 | 7571.6068 | 30 | 24 | 3.8404 | 14.7485 | 328.8444 | 10552.4790 |
| 26 | 42 | 3.8711 | 14.9851 | 304.4128 | 7606.5479 | 30 | 25 | 3.8297 | 14.6670 | 321.4845 | 9944.7520 |
| 26 | 43 | 3.8714 | 14.9878 | 305.5339 | 7757.1648 | 30 | 26 | 3.8165 | 14.5657 | 311.7200 | 9204.2454 |
| 26 | 44 | 3.8925 | 15.1513 | 309.4475 | 7775.2362 | 30 | 27 | 3.8105 | 14.5199 | 305.8330 | 8669.2917 |
| 26 | 45 | 3.8959 | 15.1780 | 311.8570 | 8027.1031 | 30 | 28 | 3.8134 | 14.5424 | 301.1106 | 8134.5347 |
| 26 | 46 | 3.9113 | 15.2986 | 313.8060 | 7944.5383 | 30 | 29 | 3.8360 | 14.7146 | 309.2600 | 8515.6695 |
| 26 | 47 | 3.9168 | 15.3416 | 316.9248 | 8244.2138 | 30 | 30 | 3.8600 | 14.8995 | 315.5455 | 8718.6803 |
| 26 | 48 | 3.9264 | 15.4168 | 316.7392 | 8041.8598 | 30 | 31 | 3.8783 | 15.0410 | 322.4472 | 9053.5615 |
| 26 | 49 | 3.9332 | 15.4703 | 320.1016 | 8347.4734 | 30 | 32 | 3.8964 | 15.1818 | 326.1810 | 9116.0967 |
| 26 | 50 | 3.9381 | 15.5089 | 318.1859 | 8044.9593 | 30 | 33 | 3.9095 | 15.2845 | 331.3516 | 9375.1929 |
| 27 | 20 | 3.6832 | 13.5662 | 261.6880 | 6578.5558 | 30 | 34 | 3.9238 | 15.3959 | 333.3180 | 9336.0068 |
| 27 | 21 | 3.5685 | 12.7339 | 245.6283 | 6634.9096 | 30 | 35 | 3.9323 | 15.4628 | 336.7064 | 9503.4569 |
| 27 | 22 | 3.5558 | 12.6435 | 244.2458 | 6675.1763 | 30 | 36 | 3.9446 | 15.5601 | 337.8290 | 9414.0612 |
| 27 | 23 | 3.7845 | 14.3222 | 307.3484 | 9449.7507 | 30 | 37 | 3.9504 | 15.6054 | 339.8950 | 9496.9183 |
| 27 | 24 | 3.8038 | 14.4685 | 313.7483 | 9762.2821 | 30 | 38 | 3.9623 | 15.7001 | 340.9849 | 9409.0184 |
| 27 | 25 | 3.7539 | 14.0918 | 289.5829 | 8270.5697 | 30 | 39 | 3.9680 | 15.7448 | 342.6388 | 9442.9872 |
| 27 | 26 | 3.7404 | 13.9906 | 283.4264 | 7861.3047 | 30 | 40 | 3.9809 | 15.8472 | 344.4932 | 9396.3079 |
| 27 | 27 | 3.7164 | 13.8116 | 270.1830 | 6964.1649 | 30 | 41 | 3.9858 | 15.8869 | 346.2522 | 9452.9866 |
| 27 | 28 | 3.7172 | 13.8174 | 267.6966 | 6642.6141 | 30 | 42 | 4.0000 | 15.9997 | 348.6139 | 9447.0369 |
| 27 | 29 | 3.7286 | 13.9024 | 270.1691 | 6719.1182 | 30 | 43 | 4.0021 | 16.0170 | 349.5509 | 9508.7965 |
| 27 | 30 | 3.7587 | 14.1276 | 281.0300 | 7229.1400 | 30 | 44 | 4.0142 | 16.1137 | 351.3155 | 9484.9333 |
| 27 | 31 | 3.7629 | 14.1595 | 280.5349 | 7134.9476 | 30 | 45 | 4.0130 | 16.1040 | 351.1091 | 9526.1716 |
| 27 | 32 | 3.7899 | 14.3635 | 290.5424 | 7627.9618 | 30 | 46 | 4.0220 | 16.1765 | 351.8045 | 9445.4093 |
| 27 | 33 | 3.7891 | 14.3575 | 287.7881 | 7391.7701 | 30 | 47 | 4.0186 | 16.1489 | 350.7479 | 9465.3362 |
| 27 | 34 | 3.8133 | 14.5416 | 296.7151 | 7841.0406 | 30 | 48 | 4.0249 | 16.1999 | 350.4093 | 9319.7281 |
| 27 | 35 | 3.8112 | 14.5256 | 293.0570 | 7524.5927 | 30 | 49 | 4.0209 | 16.1675 | 348.9981 | 9328.1647 |
| 27 | 36 | 3.8325 | 14.6878 | 300.5945 | 7902.1948 | 30 | 50 | 4.0252 | 16.2021 | 347.8400 | 9126.7813 |
| 27 | 37 | 3.8332 | 14.6931 | 297.6595 | 7586.5076 | 30 | 51 | 4.0318 | 16.2555 | 353.0045 | 9532.4457 |
| 27 | 38 | 3.8508 | 14.8289 | 303.4941 | 7868.9789 | 30 | 52 | 4.0470 | 16.3779 | 358.2743 | 9704.4217 |
| 27 | 39 | 3.8580 | 14.8844 | 302.9377 | 7646.6750 | 30 | 53 | 4.0589 | 16.4750 | 365.8752 | 10224.6132 |
| 27 | 40 | 3.8676 | 14.9583 | 306.2091 | 7813.9103 | 30 | 54 | 4.0761 | 16.6147 | 371.6458 | 10406.2736 |
| 27 | 41 | 3.8681 | 14.9620 | 304.2583 | 7594.8353 | 30 | 55 | 4.0938 | 16.7592 | 381.6984 | 11030.3294 |
| 27 | 42 | 3.8868 | 15.1074 | 310.4554 | 7889.5836 | 30 | 56 | 4.1113 | 16.9030 | 387.4283 | 11205.1611 |
| 27 | 43 | 3.8855 | 15.0972 | 308.1709 | 7715.8839 | 31 | 25 | 3.8560 | 14.8688 | 330.3145 | 10257.3423 |
| 27 | 44 | 3.9069 | 15.2642 | 315.1814 | 8049.6220 | 31 | 26 | 3.8406 | 14.7504 | 322.6629 | 9756.1323 |
| 27 | 45 | 3.9030 | 15.2337 | 312.2596 | 7890.6811 | 31 | 27 | 3.8400 | 14.7457 | 316.6612 | 9154.5436 |
| 27 | 46 | 3.9229 | 15.3893 | 318.5422 | 8178.3259 | 31 | 28 | 3.8417 | 14.7587 | 312.9499 | 8721.4135 |
| 27 | 47 | 3.9161 | 15.3357 | 314.7585 | 8003.8166 | 31 | 29 | 3.8660 | 14.9461 | 320.4027 | 9029.4639 |
| 27 | 48 | 3.9337 | 15.4744 | 319.9154 | 8208.9923 | 31 | 30 | 3.8931 | 15.1562 | 329.3325 | 9397.9165 |
| 27 | 49 | 3.9241 | 15.3985 | 315.2535 | 8006.7888 | 31 | 31 | 3.9082 | 15.2742 | 333.2079 | 9513.1940 |
| 27 | 50 | 3.9408 | 15.5296 | 319.6099 | 8135.4708 | 31 | 32 | 3.9326 | 15.4651 | 341.2732 | 9858.1284 |
| 27 | 51 | 3.9408 | 15.5297 | 321.5811 | 8343.1381 | 31 | 33 | 3.9403 | 15.5263 | 342.2273 | 9812.1140 |
| 28 | 20 | 3.5658 | 12.7149 | 245.4406 | 6706.2645 | 31 | 34 | 3.9609 | 15.6887 | 348.8784 | 10099.1210 |
| 28 | 21 | 3.5523 | 12.6186 | 243.8284 | 6705.7983 | 31 | 35 | 3.9654 | 15.7242 | 348.4076 | 9962.2566 |
| 28 | 22 | 3.7614 | 14.1484 | 304.2636 | 9660.7977 | 31 | 36 | 3.9815 | 15.8523 | 353.2817 | 10164.9298 |
| 28 | 23 | 3.8010 | 14.4477 | 313.7926 | 9852.4743 | 31 | 37 | 3.9873 | 15.8987 | 353.2390 | 10032.1760 |
| 28 | 24 | 3.7875 | 14.3449 | 305.4873 | 9355.3569 | 31 | 38 | 3.9989 | 15.9914 | 356.2263 | 10137.1540 |
| 28 | 25 | 3.7766 | 14.2628 | 298.4115 | 8776.2774 | 31 | 39 | 4.0098 | 16.0789 | 358.3300 | 10109.6336 |
| 28 | 26 | 3.7487 | 14.0530 | 284.4293 | 7873.9523 | 31 | 40 | 4.0163 | 16.1309 | 359.3307 | 10108.2721 |
| 28 | 27 | 3.7424 | 14.0056 | 278.9853 | 7382.4365 | 31 | 41 | 4.0225 | 16.1809 | 360.0912 | 10050.2760 |
| 28 | 28 | 3.7307 | 13.9180 | 270.2354 | 6699.5099 | 31 | 42 | 4.0320 | 16.2568 | 362.1789 | 10113.5223 |
| 28 | 29 | 3.7583 | 14.1248 | 279.9678 | 7155.0438 | 31 | 43 | 4.0343 | 16.2754 | 361.7320 | 10040.3943 |
| 28 | 30 | 3.7764 | 14.2611 | 284.3732 | 7279.7909 | 31 | 44 | 4.0428 | 16.3440 | 363.5534 | 10101.0415 |
| 28 | 31 | 3.7965 | 14.4132 | 291.9606 | 7660.4926 | 31 | 45 | 4.0421 | 16.3385 | 362.1989 | 10018.1053 |
| 28 | 32 | 3.8120 | 14.5311 | 294.8186 | 7680.3217 | 31 | 46 | 4.0485 | 16.3905 | 363.2016 | 10032.0856 |
| 28 | 33 | 3.8249 | 14.6299 | 300.1600 | 7971.5876 | 31 | 47 | 4.0458 | 16.3682 | 361.2297 | 9943.3907 |
| 28 | 34 | 3.8396 | 14.7424 | 302.1155 | 7916.3586 | 31 | 48 | 4.0510 | 16.4108 | 361.5691 | 9901.5247 |
| 28 | 35 | 3.8468 | 14.7979 | 305.4420 | 8113.3579 | 31 | 49 | 4.0467 | 16.3754 | 359.1475 | 9809.1217 |
| 28 | 36 | 3.8618 | 14.9139 | 307.1159 | 8021.2790 | 31 | 50 | 4.0522 | 16.4202 | 359.2401 | 9721.9723 |
| 28 | 37 | 3.8657 | 14.9438 | 309.0212 | 8134.6261 | 31 | 51 | 4.0597 | 16.4815 | 363.9804 | 10049.8685 |
| 28 | 38 | 3.8816 | 15.0669 | 310.9040 | 8045.5590 | 31 | 52 | 4.0756 | 16.6106 | 370.6976 | 10353.5342 |
| 28 | 39 | 3.8851 | 15.0940 | 312.3065 | 8106.0780 | 31 | 53 | 4.0909 | 16.7356 | 378.2016 | 10782.8921 |
| 28 | 40 | 3.9012 | 15.2196 | 314.7235 | 8033.8404 | 31 | 54 | 4.1063 | 16.8618 | 385.0104 | 11098.4421 |
| 28 | 41 | 3.9039 | 15.2404 | 316.4657 | 8132.9267 | 31 | 55 | 4.1270 | 17.0320 | 394.3852 | 11589.7931 |
| 28 | 42 | 3.9239 | 15.3972 | 320.0844 | 8120.5576 | 31 | 56 | 4.1424 | 17.1594 | 401.3673 | 11914.7257 |
| 28 | 43 | 3.9252 | 15.4075 | 321.5976 | 8264.1682 | 31 | 57 | 4.1651 | 17.3479 | 411.3896 | 12420.6337 |
| 28 | 44 | 3.9438 | 15.5539 | 324.7927 | 8243.1005 | 32 | 26 | 3.8681 | 14.9621 | 332.9434 | 10239.1162 |
| 28 | 45 | 3.9416 | 15.5360 | 325.0855 | 8369.7854 | 32 | 27 | 3.8639 | 14.9300 | 326.8125 | 9652.2154 |
| 28 | 46 | 3.9560 | 15.6500 | 326.8640 | 8281.5021 | 32 | 28 | 3.8803 | 15.0570 | 326.3796 | 9312.1904 |
| 28 | 47 | 3.9502 | 15.6041 | 325.7885 | 8364.1203 | 32 | 29 | 3.9002 | 15.2114 | 333.2531 | 9602.3292 |
| 28 | 48 | 3.9603 | 15.6839 | 325.9926 | 8190.1063 | 32 | 30 | 3.9307 | 15.4502 | 342.0250 | 9929.7259 |
| 28 | 49 | 3.9526 | 15.6228 | 324.0161 | 8233.3087 | 32 | 31 | 3.9487 | 15.5919 | 348.5015 | 10209.7346 |
| 28 | 50 | 3.9592 | 15.6750 | 322.9157 | 7983.2612 | 32 | 32 | 3.9698 | 15.7596 | 353.6392 | 10357.6690 |

TABLE 4. Charge radii, 2nd moment, 4th moment and 6th moment obtained by DNN.

| $Z$ | $N$ | $R_c$ (fm) | $R_2$ (fm$^2$) | $R_4$ (fm$^4$) | $R_6$ (fm$^6$) | $Z$ | $N$ | $R_c$ (fm) | $R_2$ (fm$^2$) | $R_4$ (fm$^4$) | $R_6$ (fm$^6$) |
|---|---|---|---|---|---|---|---|---|---|---|---|
| 32 | 33 | 3.9845 | 15.8764 | 359.0635 | 10591.5069 | 35 | 48 | 4.1606 | 17.3110 | 403.7398 | 11653.0761 |
| 32 | 34 | 3.9984 | 15.9875 | 361.3674 | 10600.0603 | 35 | 49 | 4.1612 | 17.3153 | 403.0804 | 11626.3048 |
| 32 | 35 | 4.0098 | 16.0787 | 365.5371 | 10767.5704 | 35 | 50 | 4.1573 | 17.2831 | 399.2412 | 11346.4730 |
| 32 | 36 | 4.0195 | 16.1561 | 366.2323 | 10698.6501 | 35 | 51 | 4.1747 | 17.4285 | 407.9873 | 11847.4184 |
| 32 | 37 | 4.0290 | 16.2330 | 369.5040 | 10806.8201 | 35 | 52 | 4.1876 | 17.5358 | 413.6436 | 12101.6053 |
| 32 | 38 | 4.0369 | 16.2969 | 369.7420 | 10722.2716 | 35 | 53 | 4.2080 | 17.7073 | 423.4654 | 12636.9121 |
| 32 | 39 | 4.0469 | 16.3775 | 372.9368 | 10809.1796 | 35 | 54 | 4.2197 | 17.8059 | 428.9578 | 12893.5007 |
| 32 | 40 | 4.0544 | 16.4385 | 373.1773 | 10723.9833 | 35 | 55 | 4.2406 | 17.9827 | 438.9109 | 13428.7922 |
| 32 | 41 | 4.0621 | 16.5006 | 375.3883 | 10764.6610 | 35 | 56 | 4.2515 | 18.0754 | 444.2514 | 13678.8758 |
| 32 | 42 | 4.0690 | 16.5568 | 375.6278 | 10704.4925 | 35 | 57 | 4.2708 | 18.2400 | 453.6156 | 14193.9896 |
| 32 | 43 | 4.0726 | 16.5860 | 376.3905 | 10705.9016 | 35 | 58 | 4.2820 | 18.3359 | 459.1762 | 14447.8498 |
| 32 | 44 | 4.0779 | 16.6291 | 376.2357 | 10637.6814 | 35 | 59 | 4.3015 | 18.5025 | 468.8640 | 14996.9658 |
| 32 | 45 | 4.0780 | 16.6299 | 375.7281 | 10605.0926 | 35 | 60 | 4.3205 | 18.6665 | 477.9024 | 15402.4941 |
| 32 | 46 | 4.0819 | 16.6615 | 375.1267 | 10509.4318 | 35 | 61 | 4.3642 | 19.0464 | 498.1436 | 16433.0537 |
| 32 | 47 | 4.0800 | 16.6463 | 373.8543 | 10458.1087 | 35 | 62 | 4.4220 | 19.5542 | 523.0774 | 17498.4659 |
| 32 | 48 | 4.0830 | 16.6706 | 372.9121 | 10331.1672 | 35 | 63 | 4.4486 | 19.7903 | 534.4964 | 18054.5337 |
| 32 | 49 | 4.0806 | 16.6511 | 371.3327 | 10272.0811 | 36 | 31 | 4.0737 | 16.5950 | 393.5857 | 12083.1233 |
| 32 | 50 | 4.0831 | 16.6714 | 370.1765 | 10118.7421 | 36 | 32 | 4.0976 | 16.7904 | 401.6255 | 12454.9972 |
| 32 | 51 | 4.0923 | 16.7470 | 375.6330 | 10480.6028 | 36 | 33 | 4.1138 | 16.9234 | 407.0889 | 12616.2549 |
| 32 | 52 | 4.1061 | 16.8602 | 381.2902 | 10726.7139 | 36 | 34 | 4.1282 | 17.0420 | 411.3049 | 12805.1067 |
| 32 | 53 | 4.1201 | 16.9753 | 388.8737 | 11183.6836 | 36 | 35 | 4.1451 | 17.1819 | 417.4727 | 13015.4590 |
| 32 | 54 | 4.1348 | 17.0966 | 394.7307 | 11431.7234 | 36 | 36 | 4.1526 | 17.2445 | 418.8960 | 13067.7622 |
| 32 | 55 | 4.1531 | 17.2484 | 404.1740 | 11962.3498 | 36 | 37 | 4.1709 | 17.3961 | 425.9028 | 13336.9912 |
| 32 | 56 | 4.1676 | 17.3686 | 409.8601 | 12201.2074 | 36 | 38 | 4.1745 | 17.4264 | 425.7424 | 13307.2868 |
| 32 | 57 | 4.1888 | 17.5460 | 420.4843 | 12768.7332 | 36 | 39 | 4.1943 | 17.5922 | 433.6865 | 13647.5214 |
| 32 | 58 | 4.2025 | 17.6611 | 425.9635 | 13004.7439 | 36 | 40 | 4.1904 | 17.5597 | 429.4912 | 13359.0545 |
| 33 | 27 | 3.8928 | 15.1539 | 336.6162 | 10038.1413 | 36 | 41 | 4.2001 | 17.6408 | 432.0708 | 13377.5036 |
| 33 | 28 | 3.9016 | 15.2225 | 334.8566 | 9693.6528 | 36 | 42 | 4.1951 | 17.5988 | 427.8165 | 13121.1357 |
| 33 | 29 | 3.9273 | 15.4236 | 342.9631 | 10017.2650 | 36 | 43 | 4.1988 | 17.6301 | 427.7296 | 13012.4350 |
| 33 | 30 | 3.9580 | 15.6657 | 353.0119 | 10428.8633 | 36 | 44 | 4.1947 | 17.5952 | 424.2305 | 12816.7076 |
| 33 | 31 | 3.9753 | 15.8029 | 358.0025 | 10589.4324 | 36 | 45 | 4.1939 | 17.5889 | 422.1432 | 12619.5119 |
| 33 | 32 | 4.0021 | 16.0170 | 366.7943 | 10959.5537 | 36 | 46 | 4.1917 | 17.5703 | 419.7834 | 12490.9688 |
| 33 | 33 | 4.0128 | 16.1026 | 369.2749 | 10987.7374 | 36 | 47 | 4.1880 | 17.5393 | 416.3361 | 12237.0404 |
| 33 | 34 | 4.0339 | 16.2723 | 376.0003 | 11272.5355 | 36 | 48 | 4.1880 | 17.5397 | 415.1623 | 12167.9705 |
| 33 | 35 | 4.0420 | 16.3381 | 377.4770 | 11241.8217 | 36 | 49 | 4.1826 | 17.4937 | 410.8096 | 11874.7791 |
| 33 | 36 | 4.0568 | 16.4577 | 381.7929 | 11415.4791 | 36 | 50 | 4.1848 | 17.5124 | 410.7088 | 11853.7162 |
| 33 | 37 | 4.0668 | 16.5390 | 384.0453 | 11420.6709 | 36 | 51 | 4.1962 | 17.6080 | 415.7625 | 12092.5376 |
| 33 | 38 | 4.0758 | 16.6122 | 386.0484 | 11477.4143 | 36 | 52 | 4.2128 | 17.7474 | 424.1105 | 12555.1614 |
| 33 | 39 | 4.0907 | 16.7338 | 390.5264 | 11608.6245 | 36 | 53 | 4.2290 | 17.8844 | 431.4041 | 12902.5376 |
| 33 | 40 | 4.0925 | 16.7485 | 389.2272 | 11469.5099 | 36 | 54 | 4.2422 | 17.9963 | 438.3504 | 13294.7235 |
| 33 | 41 | 4.1018 | 16.8251 | 391.3876 | 11486.6666 | 36 | 55 | 4.2611 | 18.1568 | 446.9465 | 13707.6831 |
| 33 | 42 | 4.1028 | 16.8334 | 389.8798 | 11360.8610 | 36 | 56 | 4.2720 | 18.2497 | 452.9628 | 14050.1920 |
| 33 | 43 | 4.1087 | 16.8812 | 390.8654 | 11352.2470 | 36 | 57 | 4.2909 | 18.4120 | 461.7305 | 14479.7434 |
| 33 | 44 | 4.1077 | 16.8730 | 388.7196 | 11210.4158 | 36 | 58 | 4.3058 | 18.5403 | 469.6914 | 14905.9004 |
| 33 | 45 | 4.1110 | 16.9006 | 388.9440 | 11193.1477 | 36 | 59 | 4.3228 | 18.6868 | 477.7330 | 15315.6366 |
| 33 | 46 | 4.1088 | 16.8821 | 386.3246 | 11025.3650 | 36 | 60 | 4.3771 | 19.1589 | 502.8795 | 16508.6612 |
| 33 | 47 | 4.1103 | 16.8948 | 386.0005 | 11004.2457 | 36 | 61 | 4.3960 | 19.3249 | 511.6526 | 16953.2585 |
| 33 | 48 | 4.1083 | 16.8780 | 383.3189 | 10813.9932 | 36 | 62 | 4.4480 | 19.7849 | 534.5631 | 17970.2228 |
| 33 | 49 | 4.1082 | 16.8774 | 382.5154 | 10788.1457 | 36 | 63 | 4.4529 | 19.8283 | 536.9019 | 18119.9932 |
| 33 | 50 | 4.1077 | 16.8728 | 380.1548 | 10583.6052 | 36 | 64 | 4.5651 | 20.8405 | 615.2590 | 23888.9838 |
| 33 | 51 | 4.1213 | 16.9852 | 387.2881 | 11014.1836 | 36 | 65 | 4.5783 | 20.9604 | 622.7976 | 24412.3320 |
| 33 | 52 | 4.1340 | 17.0898 | 392.8324 | 11265.1374 | 37 | 34 | 4.1591 | 17.2985 | 423.8932 | 13313.5706 |
| 33 | 53 | 4.1527 | 17.2448 | 401.7181 | 11750.5646 | 37 | 35 | 4.1704 | 17.3918 | 427.8584 | 13455.8614 |
| 33 | 54 | 4.1649 | 17.3468 | 407.4306 | 12021.3070 | 37 | 36 | 4.1891 | 17.5485 | 434.3995 | 13744.9380 |
| 33 | 55 | 4.1864 | 17.5256 | 417.1987 | 12527.4271 | 37 | 37 | 4.2038 | 17.6718 | 440.2164 | 14014.0563 |
| 33 | 56 | 4.1985 | 17.6277 | 423.0815 | 12809.5701 | 37 | 38 | 4.2154 | 17.7692 | 443.7308 | 14143.7003 |
| 33 | 57 | 4.2201 | 17.8091 | 432.8202 | 13307.9847 | 37 | 39 | 4.2332 | 17.9198 | 451.4136 | 14551.5503 |
| 33 | 58 | 4.2324 | 17.9136 | 438.8945 | 13595.5510 | 37 | 40 | 4.2306 | 17.8976 | 447.3528 | 14198.0630 |
| 33 | 59 | 4.2530 | 18.0879 | 448.3570 | 14093.1519 | 37 | 41 | 4.2368 | 17.9506 | 448.7962 | 14217.0085 |
| 34 | 29 | 3.9621 | 15.6981 | 355.3106 | 10538.7379 | 37 | 42 | 4.2298 | 17.8913 | 442.9931 | 13801.6750 |
| 34 | 30 | 3.9963 | 15.9700 | 366.2016 | 10998.9094 | 37 | 43 | 4.2327 | 17.9159 | 443.0785 | 13768.3253 |
| 34 | 31 | 4.0136 | 16.1090 | 371.9739 | 11200.7493 | 37 | 44 | 4.2239 | 17.8411 | 436.6234 | 13339.2738 |
| 34 | 32 | 4.0367 | 16.2952 | 378.6199 | 11458.9402 | 37 | 45 | 4.2243 | 17.8450 | 435.7566 | 13275.3214 |
| 34 | 33 | 4.0526 | 16.4234 | 384.1825 | 11658.7441 | 37 | 46 | 4.2161 | 17.7753 | 429.6863 | 12876.7796 |
| 34 | 34 | 4.0664 | 16.5356 | 387.2302 | 11746.7194 | 37 | 47 | 4.2145 | 17.7621 | 428.0295 | 12788.1304 |
| 34 | 35 | 4.0808 | 16.6533 | 392.4246 | 11930.7463 | 37 | 48 | 4.2084 | 17.7107 | 422.9728 | 12442.2418 |
| 34 | 36 | 4.0886 | 16.7168 | 393.1578 | 11909.1232 | 37 | 49 | 4.2051 | 17.6827 | 420.6350 | 12331.1641 |
| 34 | 37 | 4.1027 | 16.8323 | 398.2496 | 12086.9821 | 37 | 50 | 4.2019 | 17.6559 | 416.8245 | 12040.0106 |
| 34 | 38 | 4.1076 | 16.8723 | 397.9741 | 12019.0349 | 37 | 51 | 4.2191 | 17.8007 | 425.6483 | 12545.5912 |
| 34 | 39 | 4.1228 | 16.9972 | 403.5109 | 12221.6440 | 37 | 52 | 4.2359 | 17.9432 | 432.9646 | 12870.3976 |
| 34 | 40 | 4.1242 | 17.0087 | 401.4914 | 12042.9621 | 37 | 53 | 4.2543 | 18.0988 | 442.3424 | 13404.2457 |
| 34 | 41 | 4.1336 | 17.0865 | 404.0933 | 12076.1338 | 37 | 54 | 4.2691 | 18.2252 | 449.0332 | 13703.5865 |
| 34 | 42 | 4.1337 | 17.0871 | 401.9038 | 11920.5499 | 37 | 55 | 4.2858 | 18.3679 | 457.8569 | 14222.8199 |
| 34 | 43 | 4.1380 | 17.1230 | 402.4345 | 11869.3235 | 37 | 56 | 4.2993 | 18.4838 | 464.0897 | 14495.4525 |
| 34 | 44 | 4.1376 | 17.1197 | 400.3762 | 11736.4593 | 37 | 57 | 4.3129 | 18.6011 | 471.8399 | 14982.8795 |
| 34 | 45 | 4.1381 | 17.1235 | 399.3363 | 11627.9479 | 37 | 58 | 4.3280 | 18.7317 | 478.8306 | 15279.5257 |
| 34 | 46 | 4.1380 | 17.1230 | 397.6220 | 11512.6865 | 37 | 59 | 4.3462 | 18.8892 | 488.9040 | 15891.9135 |
| 34 | 47 | 4.1361 | 17.1077 | 395.6042 | 11371.8083 | 37 | 60 | 4.3791 | 19.1761 | 503.5993 | 16524.7541 |
| 34 | 48 | 4.1369 | 17.1138 | 394.3192 | 11268.6031 | 37 | 61 | 4.4243 | 19.5743 | 524.6124 | 17583.5440 |
| 34 | 49 | 4.1340 | 17.0897 | 391.7556 | 11108.1512 | 37 | 62 | 4.4562 | 19.8578 | 538.2119 | 18133.9637 |
| 34 | 50 | 4.1357 | 17.1037 | 390.9091 | 11014.1618 | 37 | 63 | 4.4968 | 20.2215 | 580.6926 | 22280.9442 |
| 34 | 51 | 4.1467 | 17.1953 | 396.4192 | 11323.8511 | 37 | 64 | 4.5651 | 20.8404 | 615.8454 | 23981.6339 |
| 34 | 52 | 4.1611 | 17.3150 | 403.1337 | 11669.1854 | 37 | 65 | 4.4929 | 20.1863 | 577.7265 | 22057.4202 |
| 34 | 53 | 4.1768 | 17.4458 | 410.7348 | 12073.5086 | 37 | 66 | 4.5732 | 20.9145 | 619.3255 | 24112.9957 |
| 34 | 54 | 4.1901 | 17.5572 | 416.9353 | 12389.9165 | 37 | 67 | 4.5170 | 20.4037 | 588.2436 | 22472.6525 |
| 34 | 55 | 4.2092 | 17.7171 | 426.0350 | 12856.8260 | 38 | 35 | 4.2028 | 17.6636 | 440.8538 | 14047.4535 |
| 34 | 56 | 4.2211 | 17.8177 | 431.6643 | 13144.6326 | 38 | 36 | 4.2115 | 17.7366 | 443.4551 | 14187.7372 |
| 34 | 57 | 4.2416 | 17.9913 | 441.4273 | 13635.4697 | 38 | 37 | 4.2324 | 17.9136 | 451.9183 | 14538.5019 |
| 34 | 58 | 4.2540 | 18.0961 | 447.3871 | 13941.0022 | 38 | 38 | 4.2365 | 17.9482 | 452.4889 | 14566.5545 |
| 34 | 59 | 4.2740 | 18.2673 | 457.0006 | 14424.5605 | 38 | 39 | 4.2592 | 18.1404 | 462.1560 | 15018.1774 |
| 34 | 60 | 4.3035 | 18.5203 | 470.7245 | 15093.4340 | 38 | 40 | 4.2519 | 18.0790 | 456.4925 | 14644.7316 |
| 34 | 61 | 4.3244 | 18.7005 | 480.7240 | 15600.5177 | 38 | 41 | 4.2602 | 18.1495 | 458.4615 | 14630.5733 |
| 35 | 30 | 4.0200 | 16.1605 | 374.9882 | 11335.0438 | 38 | 42 | 4.2524 | 18.0832 | 452.9529 | 14299.6893 |
| 35 | 31 | 4.0349 | 16.2805 | 379.4133 | 11462.6965 | 38 | 43 | 4.2541 | 18.0971 | 451.7479 | 14122.4984 |
| 35 | 32 | 4.0656 | 16.5291 | 389.9433 | 11923.0808 | 38 | 44 | 4.2481 | 18.0467 | 447.4738 | 13880.8122 |
| 35 | 33 | 4.0764 | 16.6170 | 392.9610 | 11983.1871 | 38 | 45 | 4.2447 | 18.0175 | 443.8089 | 13582.0816 |
| 35 | 34 | 4.1000 | 16.8097 | 400.9063 | 12331.8971 | 38 | 46 | 4.2419 | 17.9937 | 441.2976 | 13447.9790 |
| 35 | 35 | 4.1103 | 16.8947 | 403.9087 | 12397.4862 | 38 | 47 | 4.2348 | 17.9338 | 435.8379 | 13063.5681 |
| 35 | 36 | 4.1262 | 17.0253 | 408.8802 | 12606.0668 | 38 | 48 | 4.2354 | 17.9389 | 435.1544 | 13032.0243 |
| 35 | 37 | 4.1394 | 17.1350 | 413.2723 | 12756.8696 | 38 | 49 | 4.2259 | 17.8583 | 428.4096 | 12585.7874 |
| 35 | 38 | 4.1484 | 17.2094 | 415.4653 | 12823.5929 | 38 | 50 | 4.2296 | 17.8899 | 429.3608 | 12640.0738 |
| 35 | 39 | 4.1661 | 17.3562 | 422.0593 | 13116.2565 | 38 | 51 | 4.2398 | 17.9761 | 433.3830 | 12791.2427 |
| 35 | 40 | 4.1644 | 17.3419 | 418.8473 | 12846.1635 | 38 | 52 | 4.2594 | 18.1425 | 443.6979 | 13376.4823 |
| 35 | 41 | 4.1741 | 17.4228 | 421.3968 | 12896.9740 | 38 | 53 | 4.2745 | 18.2711 | 450.0951 | 13652.7246 |
| 35 | 42 | 4.1689 | 17.3801 | 416.9454 | 12593.3069 | 38 | 54 | 4.2887 | 18.3926 | 458.2030 | 14127.9850 |
| 35 | 43 | 4.1754 | 17.4342 | 418.3100 | 12607.7424 | 38 | 55 | 4.3057 | 18.5391 | 465.6980 | 14473.2095 |
| 35 | 44 | 4.1681 | 17.3729 | 413.1174 | 12286.3908 | 38 | 56 | 4.3175 | 18.6409 | 472.8836 | 14893.4340 |
| 35 | 45 | 4.1723 | 17.4079 | 413.7000 | 12284.9061 | 38 | 57 | 4.3334 | 18.7780 | 480.0700 | 15245.0995 |
| 35 | 46 | 4.1646 | 17.3442 | 408.4832 | 11967.3367 | 38 | 58 | 4.3569 | 18.9823 | 492.5715 | 15897.2256 |
| 35 | 47 | 4.1669 | 17.3633 | 408.4235 | 11953.8117 | 38 | 59 | 4.3705 | 19.1011 | 498.9474 | 16230.4252 |

TABLE 4. Charge radii, 2nd moment, 4th moment and 6th moment obtained by DNN.

| $Z$ | $N$ | $R_c$ (fm) | $R_2$ (fm$^2$) | $R_4$ (fm$^4$) | $R_6$ (fm$^6$) | $Z$ | $N$ | $R_c$ (fm) | $R_2$ (fm$^2$) | $R_4$ (fm$^4$) | $R_6$ (fm$^6$) |
|---|---|---|---|---|---|---|---|---|---|---|---|
| 38 | 60 | 4.4376 | 19.6926 | 529.8959 | 17673.0536 | 41 | 66 | 4.5688 | 20.8737 | 606.9105 | 22759.0335 |
| 38 | 61 | 4.4499 | 19.8019 | 535.4039 | 17959.4489 | 41 | 67 | 4.5395 | 20.6074 | 589.3471 | 21763.4695 |
| 38 | 62 | 4.5346 | 20.5622 | 599.6716 | 22943.6824 | 41 | 68 | 4.5935 | 21.1006 | 618.5510 | 23277.0079 |
| 38 | 63 | 4.5497 | 20.6999 | 607.8437 | 23479.5945 | 41 | 69 | 4.5677 | 20.8635 | 601.6463 | 22258.1132 |
| 38 | 64 | 4.5589 | 20.7836 | 610.9010 | 23463.2929 | 41 | 70 | 4.6173 | 21.3194 | 629.3149 | 23750.5980 |
| 38 | 65 | 4.5719 | 20.9021 | 618.3075 | 23977.3484 | 41 | 71 | 4.5939 | 21.1042 | 612.8797 | 22718.1964 |
| 38 | 66 | 4.5850 | 21.0218 | 622.2756 | 23934.7677 | 41 | 72 | 4.6381 | 21.5120 | 638.0741 | 24109.6902 |
| 38 | 67 | 4.5953 | 21.1166 | 628.8362 | 24422.2846 | 41 | 73 | 4.6167 | 21.3136 | 622.0272 | 23074.3293 |
| 38 | 68 | 4.6069 | 21.2233 | 633.0009 | 24462.6595 | 41 | 74 | 4.6548 | 21.6671 | 644.0946 | 24303.6539 |
| 38 | 69 | 4.6151 | 21.2990 | 637.2988 | 24759.8402 | 41 | 75 | 4.6349 | 21.4823 | 628.4367 | 23274.5263 |
| 39 | 36 | 4.2454 | 18.0233 | 458.1896 | 14833.8460 | 42 | 39 | 4.3576 | 18.9886 | 503.5785 | 16959.4544 |
| 39 | 37 | 4.2586 | 18.1359 | 464.1600 | 15158.6561 | 42 | 40 | 4.3454 | 18.8828 | 495.8818 | 16459.0455 |
| 39 | 38 | 4.2748 | 18.2737 | 469.8238 | 15387.4112 | 42 | 41 | 4.3651 | 19.0537 | 503.0407 | 16724.7141 |
| 39 | 39 | 4.2888 | 18.3941 | 476.8724 | 15821.1727 | 42 | 42 | 4.3495 | 18.9180 | 494.2138 | 16204.1108 |
| 39 | 40 | 4.2889 | 18.3949 | 473.5070 | 15461.9514 | 42 | 43 | 4.3633 | 19.0380 | 498.3797 | 16306.9441 |
| 39 | 41 | 4.2878 | 18.3850 | 472.2227 | 15375.3817 | 42 | 44 | 4.3472 | 18.8979 | 489.7488 | 15828.0707 |
| 39 | 42 | 4.2837 | 18.3501 | 467.0930 | 14946.4702 | 42 | 45 | 4.3552 | 18.9681 | 490.9957 | 15777.1842 |
| 39 | 43 | 4.2790 | 18.3094 | 464.2245 | 14790.3518 | 42 | 46 | 4.3407 | 18.8414 | 483.4877 | 15381.8058 |
| 39 | 44 | 4.2735 | 18.2626 | 458.5389 | 14349.0005 | 42 | 47 | 4.3431 | 18.8623 | 481.9157 | 15188.2085 |
| 39 | 45 | 4.2662 | 18.2004 | 454.5581 | 14147.6123 | 42 | 48 | 4.3313 | 18.7605 | 476.0800 | 14896.7766 |
| 39 | 46 | 4.2617 | 18.1620 | 449.4211 | 13747.7737 | 42 | 49 | 4.3282 | 18.7337 | 471.8403 | 14571.0849 |
| 39 | 47 | 4.2525 | 18.0834 | 444.5758 | 13511.4412 | 42 | 50 | 4.3202 | 18.6644 | 467.9625 | 14389.1376 |
| 39 | 48 | 4.2504 | 18.0656 | 440.6586 | 13184.1211 | 42 | 51 | 4.3359 | 18.8001 | 474.2659 | 14646.5375 |
| 39 | 49 | 4.2396 | 17.9746 | 435.1472 | 12919.7007 | 42 | 52 | 4.3478 | 18.9033 | 481.8883 | 15098.4801 |
| 39 | 50 | 4.2405 | 17.9819 | 432.6660 | 12670.6264 | 42 | 53 | 4.3626 | 19.0325 | 487.9810 | 15361.5829 |
| 39 | 51 | 4.2531 | 18.0891 | 439.8629 | 13107.8228 | 42 | 54 | 4.3746 | 19.1370 | 495.8875 | 15825.5640 |
| 39 | 52 | 4.2766 | 18.2895 | 449.9412 | 13551.7889 | 42 | 55 | 4.3859 | 19.2364 | 500.7013 | 16048.2832 |
| 39 | 53 | 4.2882 | 18.3887 | 457.0008 | 14002.8280 | 42 | 56 | 4.4107 | 19.4545 | 514.5073 | 16787.2381 |
| 39 | 54 | 4.3093 | 18.5697 | 466.1802 | 14400.5653 | 42 | 57 | 4.4204 | 19.5401 | 518.7752 | 16994.8525 |
| 39 | 55 | 4.3169 | 18.6360 | 471.8856 | 14811.9544 | 42 | 58 | 4.4572 | 19.8662 | 552.8277 | 20017.3665 |
| 39 | 56 | 4.3370 | 18.8099 | 480.6915 | 15180.7844 | 42 | 59 | 4.4702 | 19.9824 | 559.4755 | 20399.5335 |
| 39 | 57 | 4.3415 | 18.8488 | 485.3367 | 15564.5948 | 42 | 60 | 4.4955 | 20.2092 | 569.7928 | 20787.9451 |
| 39 | 58 | 4.3667 | 19.0681 | 496.3731 | 16022.6732 | 42 | 61 | 4.5056 | 20.3002 | 575.3142 | 21126.6268 |
| 39 | 59 | 4.3857 | 19.2341 | 507.5237 | 16709.7561 | 42 | 62 | 4.5243 | 20.4692 | 582.5611 | 21354.2875 |
| 39 | 60 | 4.4321 | 19.6435 | 527.4836 | 17547.3494 | 42 | 63 | 4.5291 | 20.5126 | 585.5142 | 21570.1873 |
| 39 | 61 | 4.4740 | 20.0166 | 570.0628 | 21573.9088 | 42 | 64 | 4.5477 | 20.6812 | 592.2147 | 21720.9901 |
| 39 | 62 | 4.5319 | 20.5380 | 598.7856 | 22907.8681 | 42 | 65 | 4.5495 | 20.6978 | 593.6605 | 21866.2229 |
| 39 | 63 | 4.4942 | 20.1978 | 578.6321 | 21956.4103 | 42 | 66 | 4.5682 | 20.8687 | 600.2627 | 21977.8850 |
| 39 | 64 | 4.5702 | 20.8869 | 617.1374 | 23815.3419 | 42 | 67 | 4.5709 | 20.8934 | 603.0702 | 22249.9775 |
| 39 | 65 | 4.5330 | 20.5477 | 596.3844 | 22770.8831 | 42 | 68 | 4.5897 | 21.0658 | 611.0966 | 22527.0452 |
| 39 | 66 | 4.5996 | 21.1560 | 630.6693 | 24448.6885 | 42 | 69 | 4.5953 | 21.1164 | 614.6628 | 22792.4892 |
| 39 | 67 | 4.5626 | 20.8174 | 608.6999 | 23215.3251 | 42 | 70 | 4.6116 | 21.2666 | 621.6684 | 23053.2560 |
| 39 | 68 | 4.6227 | 21.3689 | 640.2760 | 24772.5949 | 42 | 71 | 4.6183 | 21.3283 | 625.1338 | 23267.6573 |
| 39 | 69 | 4.5895 | 21.0633 | 619.0418 | 23513.8042 | 42 | 72 | 4.6314 | 21.4502 | 630.6044 | 23469.9697 |
| 39 | 70 | 4.6427 | 21.5544 | 648.1497 | 25029.6729 | 42 | 73 | 4.6382 | 21.5129 | 633.4453 | 23608.5673 |
| 40 | 37 | 4.2831 | 18.3446 | 473.0172 | 15519.3264 | 42 | 74 | 4.6482 | 21.6061 | 637.2238 | 23730.7770 |
| 40 | 38 | 4.2850 | 18.3616 | 472.9456 | 15517.4847 | 42 | 75 | 4.6544 | 21.6639 | 639.1769 | 23783.6928 |
| 40 | 39 | 4.3124 | 18.5970 | 485.0374 | 16109.5468 | 42 | 76 | 4.6619 | 21.7337 | 641.4815 | 23827.8897 |
| 40 | 40 | 4.3025 | 18.5112 | 478.2153 | 15664.9386 | 42 | 77 | 4.6670 | 21.7808 | 642.3262 | 23788.8897 |
| 40 | 41 | 4.3141 | 18.6116 | 481.7176 | 15734.8676 | 43 | 40 | 4.3835 | 19.2147 | 512.5751 | 17258.7790 |
| 40 | 42 | 4.3031 | 18.5165 | 474.8010 | 15319.4100 | 43 | 41 | 4.3869 | 19.2446 | 514.2512 | 17373.2920 |
| 40 | 43 | 4.3074 | 18.5539 | 474.7167 | 15200.7642 | 43 | 42 | 4.3878 | 19.2530 | 510.7982 | 16977.5564 |
| 40 | 44 | 4.2980 | 18.4729 | 468.9301 | 14870.8289 | 43 | 43 | 4.3889 | 19.2627 | 511.4831 | 17058.4024 |
| 40 | 45 | 4.2961 | 18.4567 | 465.7684 | 14594.8755 | 43 | 44 | 4.3850 | 19.2278 | 505.7880 | 16555.9197 |
| 40 | 46 | 4.2899 | 18.4033 | 461.8279 | 14380.8187 | 43 | 45 | 4.3843 | 19.2225 | 505.7370 | 16613.3823 |
| 40 | 47 | 4.2829 | 18.3433 | 456.1498 | 13979.9467 | 43 | 46 | 4.3769 | 19.1575 | 498.5602 | 16047.3277 |
| 40 | 48 | 4.2805 | 18.3231 | 454.2932 | 13886.7783 | 43 | 47 | 4.3745 | 19.1366 | 497.7321 | 16075.5219 |
| 40 | 49 | 4.2696 | 18.2291 | 446.6356 | 13388.0467 | 43 | 48 | 4.3651 | 19.0538 | 489.7790 | 15485.1779 |
| 40 | 50 | 4.2711 | 18.2420 | 446.7880 | 13404.7999 | 43 | 49 | 4.3606 | 19.0153 | 488.0534 | 15473.2135 |
| 40 | 51 | 4.2807 | 18.3246 | 450.4497 | 13531.6583 | 43 | 50 | 4.3504 | 18.9259 | 479.9048 | 14886.9758 |
| 40 | 52 | 4.3002 | 18.4914 | 461.1208 | 14137.5651 | 43 | 51 | 4.3652 | 19.0554 | 489.1025 | 15480.7718 |
| 40 | 53 | 4.3131 | 18.6033 | 466.5106 | 14365.6064 | 43 | 52 | 4.3770 | 19.1580 | 493.3339 | 15572.9357 |
| 40 | 54 | 4.3279 | 18.7307 | 475.3595 | 14878.5479 | 43 | 53 | 4.3845 | 19.2239 | 499.3638 | 16027.4667 |
| 40 | 55 | 4.3409 | 18.8430 | 481.0106 | 15142.8851 | 43 | 54 | 4.3980 | 19.3427 | 504.7331 | 16180.1140 |
| 40 | 56 | 4.3586 | 18.9970 | 491.3287 | 15712.4451 | 43 | 55 | 4.4041 | 19.3959 | 510.1275 | 16605.3956 |
| 40 | 57 | 4.3689 | 19.0877 | 496.1001 | 15958.4289 | 43 | 56 | 4.4233 | 19.5657 | 518.4911 | 16916.2396 |
| 40 | 58 | 4.4135 | 19.4788 | 518.2124 | 17053.2042 | 43 | 57 | 4.5051 | 20.2956 | 575.1641 | 21189.3093 |
| 40 | 59 | 4.4245 | 19.5759 | 523.2485 | 17316.2943 | 43 | 58 | 4.4637 | 19.9243 | 552.7102 | 19845.9682 |
| 40 | 60 | 4.4924 | 20.1820 | 576.3388 | 21535.5076 | 43 | 59 | 4.4468 | 19.7743 | 542.9299 | 19397.6500 |
| 40 | 61 | 4.5057 | 20.3013 | 583.3698 | 21986.6386 | 43 | 60 | 4.4964 | 20.2176 | 566.7963 | 20456.8731 |
| 40 | 62 | 4.5297 | 20.5180 | 593.4388 | 22349.2374 | 43 | 61 | 4.4733 | 20.0106 | 553.8132 | 19811.4198 |
| 40 | 63 | 4.5440 | 20.6478 | 601.2851 | 22851.9838 | 43 | 62 | 4.5180 | 20.4121 | 576.0104 | 20847.3772 |
| 40 | 64 | 4.5608 | 20.8007 | 607.4651 | 22987.0180 | 43 | 63 | 4.4932 | 20.1889 | 561.5063 | 20067.0046 |
| 40 | 65 | 4.5710 | 20.8941 | 613.3777 | 23398.3728 | 43 | 64 | 4.5370 | 20.5845 | 583.5965 | 21119.6042 |
| 40 | 66 | 4.5856 | 21.0273 | 617.8212 | 23396.4199 | 43 | 65 | 4.5132 | 20.3686 | 569.1574 | 20313.5892 |
| 40 | 67 | 4.5951 | 21.1145 | 623.8701 | 23822.9410 | 43 | 66 | 4.5547 | 20.7451 | 590.2705 | 21319.9943 |
| 40 | 68 | 4.6091 | 21.2438 | 629.2027 | 23933.7211 | 43 | 67 | 4.5347 | 20.5634 | 578.1527 | 20644.1872 |
| 40 | 69 | 4.6188 | 21.3334 | 634.3443 | 24246.5836 | 43 | 68 | 4.5769 | 20.9479 | 601.0696 | 21832.5243 |
| 40 | 70 | 4.6312 | 21.4481 | 639.4451 | 24408.5613 | 43 | 69 | 4.5604 | 20.7971 | 589.8055 | 21144.8462 |
| 40 | 71 | 4.6397 | 21.5267 | 643.0086 | 24582.3628 | 43 | 70 | 4.6001 | 21.1607 | 612.0977 | 22356.5011 |
| 40 | 72 | 4.6501 | 21.6237 | 647.4844 | 24752.6269 | 43 | 71 | 4.5860 | 21.0314 | 601.3561 | 21656.0233 |
| 40 | 73 | 4.6560 | 21.6784 | 648.8608 | 24765.9475 | 43 | 72 | 4.6216 | 21.3593 | 621.7846 | 22794.9272 |
| 41 | 38 | 4.3211 | 18.6718 | 488.4456 | 16246.8461 | 43 | 73 | 4.6094 | 21.2461 | 611.4606 | 22088.5803 |
| 41 | 39 | 4.3347 | 18.7897 | 495.9660 | 16736.0244 | 43 | 74 | 4.6401 | 21.5305 | 629.2934 | 23092.7382 |
| 41 | 40 | 4.3395 | 18.8317 | 494.2726 | 16426.6910 | 43 | 75 | 4.6294 | 21.4318 | 619.4622 | 22393.1462 |
| 41 | 41 | 4.3404 | 18.8395 | 494.5492 | 16455.6744 | 43 | 76 | 4.6554 | 21.6726 | 634.4975 | 23237.0108 |
| 41 | 42 | 4.3393 | 18.8291 | 490.1729 | 16024.6743 | 43 | 77 | 4.6463 | 21.5882 | 625.3204 | 22559.2682 |
| 41 | 43 | 4.3364 | 18.8045 | 488.8097 | 15985.2353 | 43 | 78 | 4.6679 | 21.7896 | 637.6734 | 23241.0868 |
| 41 | 44 | 4.3317 | 18.7633 | 482.8577 | 15486.0318 | 43 | 79 | 4.6603 | 21.7186 | 629.2607 | 22597.2930 |
| 41 | 45 | 4.3257 | 18.7117 | 480.1456 | 15392.8042 | 44 | 41 | 4.4055 | 19.4085 | 520.0126 | 17486.0668 |
| 41 | 46 | 4.3197 | 18.6599 | 473.7306 | 14882.7313 | 44 | 42 | 4.3885 | 19.2591 | 510.6911 | 16940.5646 |
| 41 | 47 | 4.3109 | 18.5842 | 469.8143 | 14739.9268 | 44 | 43 | 4.4070 | 19.4217 | 517.0966 | 17164.8341 |
| 41 | 48 | 4.3054 | 18.5367 | 463.7643 | 14261.0618 | 44 | 44 | 4.3869 | 19.2448 | 506.8035 | 16605.8994 |
| 41 | 49 | 4.2942 | 18.4404 | 458.7921 | 14070.8771 | 44 | 45 | 4.4034 | 19.3898 | 511.9127 | 16749.5460 |
| 41 | 50 | 4.2904 | 18.4074 | 453.6283 | 13646.2713 | 44 | 46 | 4.3824 | 19.2057 | 501.6828 | 16223.8090 |
| 41 | 51 | 4.3022 | 18.5087 | 461.2605 | 14152.1662 | 44 | 47 | 4.3957 | 19.3224 | 505.0388 | 16272.4232 |
| 41 | 52 | 4.3226 | 18.6849 | 469.4093 | 14454.6446 | 44 | 48 | 4.3757 | 19.1467 | 495.6151 | 15809.8326 |
| 41 | 53 | 4.3301 | 18.7497 | 475.5101 | 14910.9043 | 44 | 49 | 4.3847 | 19.2257 | 496.8149 | 15749.0849 |
| 41 | 54 | 4.3498 | 18.9210 | 483.5758 | 15206.2948 | 44 | 50 | 4.3670 | 19.0706 | 488.7355 | 15367.2277 |
| 41 | 55 | 4.3523 | 18.9425 | 487.7439 | 15589.9697 | 44 | 51 | 4.3904 | 19.2757 | 498.4819 | 15785.7903 |
| 41 | 56 | 4.3757 | 19.1469 | 497.6466 | 15969.0517 | 44 | 52 | 4.3914 | 19.2841 | 501.2357 | 16003.6759 |
| 41 | 57 | 4.3841 | 19.2203 | 504.4959 | 16477.3173 | 44 | 53 | 4.4099 | 19.4472 | 508.8550 | 16327.0346 |
| 41 | 58 | 4.4206 | 19.5421 | 520.3472 | 17130.4621 | 44 | 54 | 4.4140 | 19.4836 | 513.3816 | 16645.8674 |
| 41 | 59 | 4.4694 | 19.9754 | 563.1421 | 20933.7845 | 44 | 55 | 4.4280 | 19.6075 | 519.0727 | 16882.6896 |
| 41 | 60 | 4.4846 | 20.1117 | 569.9940 | 21134.7548 | 44 | 56 | 4.4477 | 19.7820 | 541.8739 | 19136.5545 |
| 41 | 61 | 4.4553 | 19.8497 | 553.9182 | 20369.0919 | 44 | 57 | 4.4643 | 19.9298 | 549.5987 | 19490.2198 |
| 41 | 62 | 4.5220 | 20.4481 | 587.1243 | 21944.8828 | 44 | 58 | 4.4845 | 20.1107 | 557.0670 | 19744.6505 |
| 41 | 63 | 4.4885 | 20.1466 | 568.3419 | 20981.6775 | 44 | 59 | 4.4946 | 20.2016 | 561.9357 | 19972.8691 |
| 41 | 64 | 4.5472 | 20.6770 | 598.2052 | 22434.1300 | 44 | 60 | 4.5117 | 20.3554 | 568.7006 | 20223.2182 |
| 41 | 65 | 4.5134 | 20.3706 | 578.3879 | 21355.3480 | 44 | 61 | 4.5145 | 20.3809 | 570.1904 | 20303.2059 |

TABLE 4. Charge radii, 2nd moment, 4th moment and 6th moment obtained by DNN.

| $Z$ | $N$ | $R_c$ (fm) | $R_2$ (fm$^2$) | $R_4$ (fm$^4$) | $R_6$ (fm$^6$) | $Z$ | $N$ | $R_c$ (fm) | $R_2$ (fm$^2$) | $R_4$ (fm$^4$) | $R_6$ (fm$^6$) |
|---|---|---|---|---|---|---|---|---|---|---|---|
| 44 | 62 | 4.5335 | 20.5528 | 577.6722 | 20550.6934 | 47 | 52 | 4.4375 | 19.6911 | 519.5136 | 16774.4800 |
| 44 | 63 | 4.5321 | 20.5399 | 577.1658 | 20548.3336 | 47 | 53 | 4.6077 | 21.2311 | 618.0989 | 22847.2002 |
| 44 | 64 | 4.5528 | 20.7279 | 585.3451 | 20796.0288 | 47 | 54 | 4.4760 | 20.0346 | 546.7311 | 18884.3761 |
| 44 | 65 | 4.5486 | 20.6899 | 583.4747 | 20739.9326 | 47 | 55 | 4.4864 | 20.1282 | 550.4701 | 19075.3624 |
| 44 | 66 | 4.5693 | 20.8783 | 591.7330 | 20975.4180 | 47 | 56 | 4.5045 | 20.2903 | 556.3532 | 19085.5121 |
| 44 | 67 | 4.5654 | 20.8428 | 590.7143 | 21019.9818 | 47 | 57 | 4.5110 | 20.3495 | 559.1455 | 19266.8959 |
| 44 | 68 | 4.5856 | 21.0275 | 599.9122 | 21389.0179 | 47 | 58 | 4.5266 | 20.4898 | 564.7970 | 19341.3389 |
| 44 | 69 | 4.5863 | 21.0343 | 600.7755 | 21497.3693 | 47 | 59 | 4.5277 | 20.5001 | 565.1122 | 19405.9045 |
| 44 | 70 | 4.6040 | 21.1966 | 608.8623 | 21838.8526 | 47 | 60 | 4.5451 | 20.6582 | 572.0909 | 19575.2544 |
| 44 | 71 | 4.6077 | 21.2310 | 610.7908 | 21968.4759 | 47 | 61 | 4.5441 | 20.6492 | 571.6232 | 19618.2630 |
| 44 | 72 | 4.6218 | 21.3614 | 617.0289 | 22228.9872 | 47 | 62 | 4.5624 | 20.8152 | 579.1408 | 19819.8532 |
| 44 | 73 | 4.6272 | 21.4110 | 619.3554 | 22346.0061 | 47 | 63 | 4.5594 | 20.7880 | 578.1109 | 19869.1040 |
| 44 | 74 | 4.6377 | 21.5079 | 623.5019 | 22501.3541 | 47 | 64 | 4.5773 | 20.9515 | 585.4621 | 20053.1408 |
| 44 | 75 | 4.6439 | 21.5654 | 625.8788 | 22590.0729 | 47 | 65 | 4.5723 | 20.9055 | 583.9117 | 20121.4547 |
| 44 | 76 | 4.6512 | 21.6334 | 628.1143 | 22642.0184 | 47 | 66 | 4.5896 | 21.0646 | 590.8947 | 20269.5639 |
| 44 | 77 | 4.6579 | 21.6959 | 630.4608 | 22701.9076 | 47 | 67 | 4.5827 | 21.0007 | 587.7659 | 20214.3852 |
| 44 | 78 | 4.6630 | 21.7433 | 631.2288 | 22669.4691 | 47 | 68 | 4.6013 | 21.1720 | 595.8036 | 20435.1983 |
| 44 | 79 | 4.6700 | 21.8091 | 633.5355 | 22704.5580 | 47 | 69 | 4.5959 | 21.1222 | 592.4683 | 20302.6805 |
| 44 | 80 | 4.6739 | 21.8457 | 633.3987 | 22615.5451 | 47 | 70 | 4.6153 | 21.3009 | 602.0352 | 20690.9870 |
| 44 | 81 | 4.6808 | 21.9095 | 635.4453 | 22618.9761 | 47 | 71 | 4.6120 | 21.2705 | 599.0900 | 20529.6259 |
| 45 | 43 | 4.4281 | 19.6082 | 527.7924 | 17776.6808 | 47 | 72 | 4.6300 | 21.4366 | 608.5804 | 20971.7046 |
| 45 | 44 | 4.4222 | 19.5557 | 521.6850 | 17276.1958 | 47 | 73 | 4.6287 | 21.4247 | 606.2465 | 20807.0606 |
| 45 | 45 | 4.4267 | 19.5952 | 523.7884 | 17433.1679 | 47 | 74 | 4.6438 | 21.5645 | 614.4658 | 21215.4790 |
| 45 | 46 | 4.4180 | 19.5189 | 516.4550 | 16874.6475 | 47 | 75 | 4.6442 | 21.5689 | 612.8260 | 21061.5982 |
| 45 | 47 | 4.4220 | 19.5539 | 518.3921 | 17033.1996 | 47 | 76 | 4.6558 | 21.6769 | 619.1603 | 21386.6763 |
| 45 | 48 | 4.4110 | 19.4571 | 510.0827 | 16435.9549 | 47 | 77 | 4.6579 | 21.6964 | 618.3099 | 21254.9645 |
| 45 | 49 | 4.4142 | 19.4848 | 511.6827 | 16582.3200 | 47 | 78 | 4.6663 | 21.7747 | 622.6906 | 21482.2886 |
| 45 | 50 | 4.4013 | 19.3716 | 502.6323 | 15960.5414 | 47 | 79 | 4.6700 | 21.8089 | 622.7312 | 21382.6831 |
| 45 | 51 | 4.4170 | 19.5096 | 511.8744 | 16540.1309 | 47 | 80 | 4.6760 | 21.8651 | 625.4696 | 21522.5999 |
| 45 | 52 | 4.4197 | 19.5338 | 512.2097 | 16448.2391 | 47 | 81 | 4.6811 | 21.9128 | 626.4294 | 21460.4495 |
| 45 | 53 | 4.4294 | 19.6194 | 518.5824 | 16885.9596 | 47 | 82 | 4.6857 | 21.9560 | 627.9919 | 21534.0236 |
| 45 | 54 | 4.4327 | 19.6491 | 519.7683 | 16862.6008 | 47 | 83 | 4.7045 | 22.1326 | 638.7323 | 22091.1215 |
| 45 | 55 | 4.5622 | 20.8137 | 598.9305 | 22098.5334 | 47 | 84 | 4.7199 | 22.2776 | 648.5589 | 22712.9791 |
| 45 | 56 | 4.4687 | 19.9694 | 548.6406 | 19266.0043 | 47 | 85 | 4.7428 | 22.4937 | 661.3097 | 23370.6284 |
| 45 | 57 | 4.4668 | 19.9522 | 546.2104 | 19172.6282 | 47 | 86 | 4.7584 | 22.6421 | 671.7382 | 24032.9504 |
| 45 | 58 | 4.4978 | 20.2302 | 559.7452 | 19639.1797 | 48 | 46 | 4.1207 | 16.9802 | 399.2329 | 12042.5895 |
| 45 | 59 | 4.4885 | 20.1465 | 554.2063 | 19395.1868 | 48 | 47 | 4.4360 | 19.6778 | 521.8142 | 17030.3737 |
| 45 | 60 | 4.5172 | 20.4051 | 567.5064 | 19922.8611 | 48 | 48 | 4.3677 | 19.0764 | 492.8909 | 15785.1300 |
| 45 | 61 | 4.5036 | 20.2826 | 559.6481 | 19534.5916 | 48 | 49 | 4.4390 | 19.7046 | 520.1958 | 16827.3165 |
| 45 | 62 | 4.5340 | 20.5568 | 574.1027 | 20146.6913 | 48 | 50 | 4.4119 | 19.4647 | 508.1784 | 16298.6542 |
| 45 | 63 | 4.5195 | 20.4255 | 565.6719 | 19717.3649 | 48 | 51 | 4.4369 | 19.6863 | 518.3771 | 16699.9986 |
| 45 | 64 | 4.5497 | 20.6995 | 580.2130 | 20338.2583 | 48 | 52 | 4.4465 | 19.7710 | 530.6130 | 18090.2526 |
| 45 | 65 | 4.5347 | 20.5634 | 571.6420 | 19913.8620 | 48 | 53 | 4.4765 | 20.0391 | 544.6821 | 18674.5041 |
| 45 | 66 | 4.5634 | 20.8245 | 585.4589 | 20489.6417 | 48 | 54 | 4.4773 | 20.0462 | 540.7420 | 18265.9467 |
| 45 | 67 | 4.5500 | 20.7027 | 577.7194 | 20105.3145 | 48 | 55 | 4.5040 | 20.2857 | 553.2346 | 18773.4425 |
| 45 | 68 | 4.5796 | 20.9730 | 593.0400 | 20823.3825 | 48 | 56 | 4.5101 | 20.3413 | 553.7831 | 18679.5348 |
| 45 | 69 | 4.5700 | 20.8854 | 586.3260 | 20432.5413 | 48 | 57 | 4.5279 | 20.5018 | 562.1994 | 19017.5098 |
| 45 | 70 | 4.5987 | 21.1483 | 602.0356 | 21241.2908 | 48 | 58 | 4.5360 | 20.5753 | 564.1521 | 19004.2482 |
| 45 | 71 | 4.5920 | 21.0861 | 596.1105 | 20848.3802 | 48 | 59 | 4.5480 | 20.6846 | 570.0908 | 19260.7245 |
| 45 | 72 | 4.6178 | 21.3240 | 610.7385 | 21638.9782 | 48 | 60 | 4.5592 | 20.7865 | 573.8205 | 19333.0157 |
| 45 | 73 | 4.6131 | 21.2803 | 605.4170 | 21246.2785 | 48 | 61 | 4.5671 | 20.8580 | 578.0088 | 19543.8593 |
| 45 | 74 | 4.6349 | 21.4823 | 618.0068 | 21945.0783 | 48 | 62 | 4.5793 | 20.9703 | 582.5845 | 19659.7147 |
| 45 | 75 | 4.6318 | 21.4535 | 613.2722 | 21560.1045 | 48 | 63 | 4.5839 | 21.0120 | 585.4154 | 19839.6975 |
| 45 | 76 | 4.6495 | 21.6180 | 623.4855 | 22134.0887 | 48 | 64 | 4.5963 | 21.1264 | 590.3655 | 19977.1681 |
| 45 | 77 | 4.6480 | 21.6035 | 619.4564 | 21769.5860 | 48 | 65 | 4.5981 | 21.1422 | 592.0133 | 20131.3302 |
| 45 | 78 | 4.6621 | 21.7355 | 627.4321 | 22216.8828 | 48 | 66 | 4.6107 | 21.2589 | 597.2830 | 20286.8245 |
| 45 | 79 | 4.6621 | 21.7350 | 624.2321 | 21885.3745 | 48 | 67 | 4.6096 | 21.2486 | 597.1101 | 20326.9493 |
| 45 | 80 | 4.6736 | 21.8426 | 630.3453 | 22220.3298 | 48 | 68 | 4.6230 | 21.3725 | 603.0745 | 20537.6333 |
| 45 | 81 | 4.6748 | 21.8542 | 627.9931 | 21928.9783 | 48 | 69 | 4.6218 | 21.3607 | 602.2170 | 20501.2041 |
| 45 | 82 | 4.6845 | 21.9445 | 632.6004 | 22166.5114 | 48 | 70 | 4.6355 | 21.4880 | 608.9731 | 20809.5147 |
| 45 | 83 | 4.7021 | 22.1094 | 641.9421 | 22632.5470 | 48 | 71 | 4.6348 | 21.4811 | 607.9242 | 20728.3321 |
| 46 | 44 | 4.4101 | 19.4487 | 516.4766 | 17042.2158 | 48 | 72 | 4.6472 | 21.5967 | 614.2999 | 21053.1447 |
| 46 | 45 | 4.4333 | 19.6537 | 524.4782 | 17309.6846 | 48 | 73 | 4.6475 | 21.5989 | 613.4514 | 20955.8409 |
| 46 | 46 | 4.4104 | 19.4519 | 513.5560 | 16767.5112 | 48 | 74 | 4.6576 | 21.6928 | 618.5972 | 21234.0635 |
| 46 | 47 | 4.4288 | 19.6143 | 519.3352 | 16928.2358 | 48 | 75 | 4.6589 | 21.7050 | 618.1597 | 21140.2281 |
| 46 | 48 | 4.4058 | 19.4113 | 508.6436 | 16417.4491 | 48 | 76 | 4.6663 | 21.7748 | 621.7141 | 21337.9441 |
| 46 | 49 | 4.4226 | 19.5597 | 513.5540 | 16531.1188 | 48 | 77 | 4.6687 | 21.7964 | 621.7946 | 21261.1271 |
| 46 | 50 | 4.4003 | 19.3629 | 503.3946 | 16059.3877 | 48 | 78 | 4.6740 | 21.8462 | 623.8573 | 21373.2820 |
| 46 | 51 | 4.4266 | 19.5952 | 514.4149 | 16525.1169 | 48 | 79 | 4.6773 | 21.8770 | 624.5361 | 21324.1366 |
| 46 | 52 | 4.4154 | 19.4960 | 511.6059 | 16494.6863 | 48 | 80 | 4.6813 | 21.9147 | 625.5019 | 21365.8196 |
| 46 | 53 | 4.4361 | 19.6786 | 519.9815 | 16829.2112 | 48 | 81 | 4.6856 | 21.9545 | 626.8417 | 21352.1131 |
| 46 | 54 | 4.4480 | 19.7847 | 536.2203 | 18575.4818 | 48 | 82 | 4.6892 | 21.9884 | 627.1957 | 21346.9200 |
| 46 | 55 | 4.4733 | 20.0101 | 548.0051 | 19080.1671 | 48 | 83 | 4.7050 | 22.1369 | 637.0634 | 21876.8426 |
| 46 | 56 | 4.4840 | 20.1061 | 549.9612 | 19023.5981 | 48 | 84 | 4.7185 | 22.2643 | 644.7141 | 22312.2130 |
| 46 | 57 | 4.5021 | 20.2685 | 558.3587 | 19366.8083 | 48 | 85 | 4.7386 | 22.4543 | 657.4082 | 23042.0328 |
| 46 | 58 | 4.5134 | 20.3711 | 562.1163 | 19470.3278 | 48 | 86 | 4.7526 | 22.5876 | 665.3116 | 23464.1266 |
| 46 | 59 | 4.5224 | 20.4521 | 566.2827 | 19629.9165 | 48 | 87 | 4.7758 | 22.8087 | 679.9554 | 24328.4255 |
| 46 | 60 | 4.5365 | 20.5802 | 571.5394 | 19794.2514 | 49 | 47 | 4.1368 | 17.1132 | 404.2570 | 12244.2567 |
| 46 | 61 | 4.5396 | 20.6079 | 573.0243 | 19851.5535 | 49 | 48 | 4.3803 | 19.1869 | 497.0855 | 15913.7020 |
| 46 | 62 | 4.5567 | 20.7637 | 579.8067 | 20069.8317 | 49 | 49 | 4.3842 | 19.2211 | 498.5780 | 16007.3208 |
| 46 | 63 | 4.5555 | 20.7527 | 579.3740 | 20060.8908 | 49 | 50 | 4.4343 | 19.6635 | 516.5697 | 16605.1802 |
| 46 | 64 | 4.5737 | 20.9190 | 586.8447 | 20301.0288 | 49 | 51 | 4.6313 | 21.4489 | 626.9253 | 23118.4121 |
| 46 | 65 | 4.5693 | 20.8781 | 584.9080 | 20242.4129 | 49 | 52 | 4.4749 | 20.0244 | 541.9537 | 18455.8955 |
| 46 | 66 | 4.5879 | 21.0486 | 592.7891 | 20499.3335 | 49 | 53 | 4.4873 | 20.1361 | 546.4545 | 18628.2986 |
| 46 | 67 | 4.5814 | 20.9895 | 590.0409 | 20424.0007 | 49 | 54 | 4.5003 | 20.2529 | 548.9594 | 18430.9424 |
| 46 | 68 | 4.5997 | 21.1569 | 598.5261 | 20771.5497 | 49 | 55 | 4.5172 | 20.4048 | 556.9718 | 18846.9020 |
| 46 | 69 | 4.5966 | 21.1285 | 596.9370 | 20717.9888 | 49 | 56 | 4.5260 | 20.4844 | 558.3424 | 18654.2564 |
| 46 | 70 | 4.6134 | 21.2831 | 605.0603 | 21088.4078 | 49 | 57 | 4.5381 | 20.5944 | 564.6423 | 19030.2367 |
| 46 | 71 | 4.6135 | 21.2844 | 604.6961 | 21065.6792 | 49 | 58 | 4.5475 | 20.6796 | 566.6616 | 18893.0333 |
| 46 | 72 | 4.6274 | 21.4125 | 611.3867 | 21384.5302 | 49 | 59 | 4.5589 | 20.7832 | 573.1819 | 19328.8084 |
| 46 | 73 | 4.6299 | 21.4363 | 611.9645 | 21383.5559 | 49 | 60 | 4.5682 | 20.8684 | 575.3370 | 19204.1231 |
| 46 | 74 | 4.6403 | 21.5324 | 616.6844 | 21607.6231 | 49 | 61 | 4.5788 | 20.9650 | 582.1548 | 19712.8632 |
| 46 | 75 | 4.6445 | 21.5714 | 617.9044 | 21618.3400 | 49 | 62 | 4.5871 | 21.0411 | 583.8241 | 19558.2484 |
| 46 | 76 | 4.6517 | 21.6381 | 620.6502 | 21736.7185 | 49 | 63 | 4.5963 | 21.1260 | 590.7039 | 20130.4983 |
| 46 | 77 | 4.6570 | 21.6880 | 622.3651 | 21756.6970 | 49 | 64 | 4.6035 | 21.1918 | 591.7200 | 19929.2226 |
| 46 | 78 | 4.6619 | 21.7329 | 623.5044 | 21782.2410 | 49 | 65 | 4.6107 | 21.2583 | 598.2437 | 20541.8858 |
| 46 | 79 | 4.6682 | 21.7920 | 625.7025 | 21815.5179 | 49 | 66 | 4.6173 | 21.3196 | 598.8696 | 20302.9293 |
| 46 | 80 | 4.6717 | 21.8245 | 625.7556 | 21772.5684 | 49 | 67 | 4.6204 | 21.3477 | 602.0548 | 20653.3713 |
| 46 | 81 | 4.6788 | 21.8913 | 628.4226 | 21821.9073 | 49 | 68 | 4.6290 | 21.4279 | 603.4260 | 20415.7855 |
| 46 | 82 | 4.6819 | 21.9202 | 627.9199 | 21738.1248 | 49 | 69 | 4.6301 | 21.4377 | 604.6498 | 20602.9368 |
| 46 | 83 | 4.7022 | 22.1107 | 640.5047 | 22439.1958 | 49 | 70 | 4.6416 | 21.5445 | 608.7689 | 20602.9448 |
| 46 | 84 | 4.7170 | 22.2497 | 648.8010 | 22906.4200 | 49 | 71 | 4.6418 | 21.5467 | 608.8746 | 20683.3511 |
| 46 | 85 | 4.7413 | 22.4804 | 664.0226 | 23787.8653 | 49 | 72 | 4.6539 | 21.6592 | 614.1647 | 20815.2425 |
| 47 | 45 | 4.4404 | 19.7175 | 528.8387 | 17622.7202 | 49 | 73 | 4.6543 | 21.6623 | 613.8980 | 20840.6907 |
| 47 | 46 | 4.4396 | 19.7105 | 525.2518 | 17257.8707 | 49 | 74 | 4.6651 | 21.7633 | 618.9828 | 21008.5979 |
| 47 | 47 | 4.4462 | 19.7686 | 528.0818 | 17436.3021 | 49 | 75 | 4.6662 | 21.7731 | 618.8776 | 21016.6657 |
| 47 | 48 | 4.4359 | 19.6771 | 520.6135 | 16913.6402 | 49 | 76 | 4.6746 | 21.8522 | 622.8302 | 21152.7722 |
| 47 | 49 | 4.4427 | 19.7377 | 523.6280 | 17104.9990 | 49 | 77 | 4.6768 | 21.8722 | 623.2564 | 21168.7033 |
| 47 | 50 | 4.4307 | 19.6315 | 515.4986 | 16559.5510 | 49 | 78 | 4.6826 | 21.9265 | 625.6740 | 21239.3330 |
| 47 | 51 | 4.4438 | 19.7470 | 522.9407 | 17014.5291 | 49 | 79 | 4.6861 | 21.9596 | 626.8838 | 21279.1323 |

TABLE 4. Charge radii, 2nd moment, 4th moment and 6th moment obtained by DNN.

| $Z$ | $N$ | $R_c$ (fm) | $R_2$ (fm$^2$) | $R_4$ (fm$^4$) | $R_6$ (fm$^6$) | $Z$ | $N$ | $R_c$ (fm) | $R_2$ (fm$^2$) | $R_4$ (fm$^4$) | $R_6$ (fm$^6$) |
|---|---|---|---|---|---|---|---|---|---|---|---|
| 49 | 80 | 4.6896 | 21.9926 | 627.8260 | 21281.0119 | 52 | 70 | 4.7126 | 22.2082 | 645.9000 | 22549.2283 |
| 49 | 81 | 4.6947 | 22.0398 | 629.9447 | 21352.1493 | 52 | 71 | 4.7148 | 22.2293 | 645.5687 | 22449.8436 |
| 49 | 82 | 4.6967 | 22.0591 | 629.7838 | 21302.8242 | 52 | 72 | 4.7221 | 22.2986 | 649.5432 | 22648.2809 |
| 49 | 83 | 4.7141 | 22.2227 | 640.4101 | 21898.0212 | 52 | 73 | 4.7229 | 22.3054 | 647.9819 | 22450.3537 |
| 49 | 84 | 4.7256 | 22.3310 | 647.1771 | 22293.7665 | 52 | 74 | 4.7303 | 22.3756 | 652.2985 | 22701.5297 |
| 49 | 85 | 4.7466 | 22.5302 | 659.8134 | 23016.4861 | 52 | 75 | 4.7300 | 22.3727 | 649.9532 | 22438.6820 |
| 49 | 86 | 4.7584 | 22.6425 | 667.2066 | 23452.1640 | 52 | 76 | 4.7369 | 22.4383 | 654.0853 | 22700.2964 |
| 49 | 87 | 4.7806 | 22.8541 | 680.0251 | 24172.6429 | 52 | 77 | 4.7361 | 22.4302 | 651.3724 | 22404.4625 |
| 49 | 88 | 4.7937 | 22.9791 | 688.6572 | 24689.5855 | 52 | 78 | 4.7424 | 22.4902 | 655.1074 | 22653.2514 |
| 50 | 49 | 4.1651 | 17.3479 | 409.4247 | 12150.9994 | 52 | 79 | 4.7414 | 22.4808 | 652.3699 | 22350.6890 |
| 50 | 50 | 4.4445 | 19.7537 | 526.4713 | 17773.4278 | 52 | 80 | 4.7474 | 22.5380 | 655.7662 | 22581.4127 |
| 50 | 51 | 4.4711 | 19.9903 | 538.5408 | 18223.9798 | 52 | 81 | 4.7467 | 22.5307 | 653.2798 | 22292.5405 |
| 50 | 52 | 4.4608 | 19.8990 | 528.3629 | 17454.0251 | 52 | 82 | 4.7528 | 22.5892 | 656.5284 | 22510.7420 |
| 50 | 53 | 4.4933 | 20.1900 | 543.4498 | 18053.1239 | 52 | 83 | 4.7618 | 22.6751 | 661.4924 | 22719.7277 |
| 50 | 54 | 4.4967 | 20.2203 | 541.8545 | 17797.1842 | 52 | 84 | 4.7772 | 22.8218 | 671.6287 | 23379.3678 |
| 50 | 55 | 4.5208 | 20.4373 | 553.2388 | 18253.5346 | 52 | 85 | 4.7895 | 22.9394 | 678.7276 | 23737.7601 |
| 50 | 56 | 4.5243 | 20.4689 | 552.3552 | 18060.9517 | 52 | 86 | 4.8054 | 23.0919 | 689.1664 | 24396.3334 |
| 50 | 57 | 4.5434 | 20.6428 | 561.8252 | 18475.7162 | 52 | 87 | 4.8199 | 23.2312 | 697.6166 | 24848.0552 |
| 50 | 58 | 4.5498 | 20.7006 | 562.7748 | 18389.8063 | 52 | 88 | 4.8355 | 23.3816 | 707.9050 | 25485.3520 |
| 50 | 59 | 4.5656 | 20.8446 | 571.0641 | 18799.9651 | 52 | 89 | 4.8502 | 23.5246 | 716.4671 | 25952.6276 |
| 50 | 60 | 4.5729 | 20.9110 | 572.9406 | 18780.0144 | 52 | 90 | 4.8644 | 23.6620 | 726.0585 | 26548.2268 |
| 50 | 61 | 4.5863 | 21.0340 | 580.4664 | 19199.5976 | 52 | 91 | 4.8771 | 23.7859 | 733.2893 | 26948.7839 |
| 50 | 62 | 4.5932 | 21.0976 | 582.6183 | 19211.6616 | 52 | 92 | 4.8893 | 23.9051 | 741.9503 | 27497.0344 |
| 50 | 63 | 4.6048 | 21.2043 | 589.5751 | 19642.7884 | 52 | 93 | 4.8978 | 23.9884 | 746.5605 | 27760.3549 |
| 50 | 64 | 4.6111 | 21.2620 | 591.7091 | 19666.0106 | 53 | 53 | 4.6252 | 21.3922 | 610.4944 | 21514.1937 |
| 50 | 65 | 4.6210 | 21.3536 | 598.1135 | 20104.3973 | 53 | 54 | 4.6276 | 21.4143 | 607.6883 | 21052.2461 |
| 50 | 66 | 4.6268 | 21.4070 | 600.1982 | 20130.9336 | 53 | 55 | 4.6469 | 21.5941 | 619.5089 | 21826.0555 |
| 50 | 67 | 4.6341 | 21.4745 | 604.2644 | 20373.1731 | 53 | 56 | 4.6482 | 21.6060 | 616.5364 | 21378.6714 |
| 50 | 68 | 4.6409 | 21.5379 | 606.6978 | 20391.6071 | 53 | 57 | 4.6670 | 21.7807 | 628.8626 | 22240.8564 |
| 50 | 69 | 4.6468 | 21.5926 | 609.4119 | 20521.5584 | 53 | 58 | 4.6675 | 21.7857 | 625.7300 | 21795.9516 |
| 50 | 70 | 4.6553 | 21.6719 | 613.5739 | 20699.5955 | 53 | 59 | 4.6860 | 21.9584 | 638.6806 | 22747.0890 |
| 50 | 71 | 4.6597 | 21.7129 | 615.0147 | 20726.9462 | 53 | 60 | 4.6854 | 21.9533 | 635.0733 | 22277.5727 |
| 50 | 72 | 4.6685 | 21.7951 | 619.8356 | 20987.4925 | 53 | 61 | 4.7032 | 22.1200 | 648.3129 | 23293.6844 |
| 50 | 73 | 4.6716 | 21.8243 | 620.3047 | 20938.1946 | 53 | 62 | 4.7015 | 22.1045 | 644.0682 | 22780.5281 |
| 50 | 74 | 4.6796 | 21.8987 | 624.8407 | 21209.2710 | 53 | 63 | 4.7177 | 22.2567 | 657.0390 | 23828.6667 |
| 50 | 75 | 4.6818 | 21.9196 | 624.7306 | 21115.5134 | 53 | 64 | 4.7153 | 22.2342 | 652.2409 | 23268.2721 |
| 50 | 76 | 4.6883 | 21.9803 | 628.3709 | 21345.2495 | 53 | 65 | 4.7288 | 22.3617 | 664.3304 | 24317.3658 |
| 50 | 77 | 4.6901 | 21.9966 | 628.0596 | 21237.5187 | 53 | 66 | 4.7264 | 22.3393 | 659.3168 | 23723.2763 |
| 50 | 78 | 4.6951 | 22.0438 | 630.6030 | 21400.5546 | 53 | 67 | 4.7336 | 22.4067 | 665.8162 | 24320.8101 |
| 50 | 79 | 4.6967 | 22.0590 | 630.4157 | 21304.3576 | 53 | 68 | 4.7327 | 22.3985 | 660.2027 | 23582.3637 |
| 50 | 80 | 4.7008 | 22.0972 | 632.0170 | 21399.8216 | 53 | 69 | 4.7357 | 22.4268 | 663.3014 | 23922.4613 |
| 50 | 81 | 4.7026 | 22.1146 | 632.2074 | 21334.0501 | 53 | 70 | 4.7392 | 22.4597 | 661.2590 | 23464.6373 |
| 50 | 82 | 4.7065 | 22.1507 | 633.2257 | 21376.9801 | 53 | 71 | 4.7393 | 22.4614 | 661.7758 | 23594.2365 |
| 50 | 83 | 4.7187 | 22.2660 | 640.8595 | 21785.1524 | 53 | 72 | 4.7455 | 22.5198 | 662.2927 | 23357.5672 |
| 50 | 84 | 4.7325 | 22.3970 | 648.7824 | 22240.3821 | 53 | 73 | 4.7443 | 22.5085 | 661.2227 | 23339.6073 |
| 50 | 85 | 4.7476 | 22.5398 | 658.4557 | 22807.2033 | 53 | 74 | 4.7514 | 22.5757 | 663.1337 | 23250.7110 |
| 50 | 86 | 4.7624 | 22.6808 | 666.9925 | 23286.1998 | 53 | 75 | 4.7501 | 22.5631 | 661.3795 | 23145.1713 |
| 50 | 87 | 4.7793 | 22.8420 | 677.9235 | 23952.2662 | 53 | 76 | 4.7567 | 22.6260 | 663.6859 | 23137.2882 |
| 50 | 88 | 4.7945 | 22.9874 | 686.6872 | 24431.7829 | 53 | 77 | 4.7561 | 22.6202 | 661.9412 | 22991.6543 |
| 50 | 89 | 4.8118 | 23.1530 | 697.7857 | 25118.7849 | 53 | 78 | 4.7615 | 22.6717 | 664.0120 | 23018.3241 |
| 50 | 90 | 4.8263 | 23.2932 | 706.2300 | 25572.9819 | 53 | 79 | 4.7621 | 22.6781 | 662.7450 | 22865.4221 |
| 51 | 51 | 4.5019 | 20.2669 | 553.7404 | 19027.6704 | 53 | 80 | 4.7663 | 22.7173 | 664.3454 | 22903.7251 |
| 51 | 52 | 4.5147 | 20.3823 | 555.3685 | 18752.1553 | 53 | 81 | 4.7685 | 22.7387 | 663.8341 | 22764.2574 |
| 51 | 53 | 4.5426 | 20.6353 | 568.6837 | 19403.7508 | 53 | 82 | 4.7716 | 22.7686 | 665.0106 | 22809.1556 |
| 51 | 54 | 4.5493 | 20.6961 | 568.2119 | 19075.0590 | 53 | 83 | 4.7851 | 22.8972 | 672.6061 | 23198.9848 |
| 51 | 55 | 4.5671 | 20.8580 | 577.6874 | 19619.3021 | 53 | 84 | 4.7976 | 23.0169 | 681.2240 | 23770.3841 |
| 51 | 56 | 4.5722 | 20.9046 | 576.8870 | 19300.5045 | 53 | 85 | 4.8133 | 23.1676 | 689.7670 | 24197.5914 |
| 51 | 57 | 4.5890 | 21.0587 | 586.7398 | 19932.0111 | 53 | 86 | 4.8272 | 23.3017 | 699.6550 | 24852.4184 |
| 51 | 58 | 4.5935 | 21.1004 | 585.8778 | 19619.5905 | 53 | 87 | 4.8417 | 23.4419 | 707.0386 | 25194.1866 |
| 51 | 59 | 4.6103 | 21.2547 | 596.5676 | 20363.1465 | 53 | 88 | 4.8579 | 23.5990 | 718.7698 | 25967.4055 |
| 51 | 60 | 4.6135 | 21.2842 | 595.1257 | 20021.2178 | 53 | 89 | 4.8669 | 23.6869 | 722.5134 | 26094.5559 |
| 51 | 61 | 4.6300 | 21.4368 | 606.4959 | 20865.5610 | 53 | 90 | 4.8860 | 23.8726 | 736.4428 | 27007.8603 |
| 51 | 62 | 4.6315 | 21.4509 | 604.1864 | 20470.5453 | 53 | 91 | 4.8859 | 23.8719 | 734.4543 | 26814.7550 |
| 51 | 63 | 4.6471 | 21.5959 | 615.8234 | 21387.9373 | 53 | 92 | 4.9081 | 24.0896 | 750.7847 | 27880.9045 |
| 51 | 64 | 4.6473 | 21.5976 | 612.7124 | 20936.5260 | 53 | 93 | 4.8973 | 23.9832 | 742.0080 | 27306.1359 |
| 51 | 65 | 4.6611 | 21.7260 | 624.0131 | 21890.1209 | 53 | 94 | 4.9230 | 24.2362 | 760.9235 | 28537.0899 |
| 51 | 66 | 4.6608 | 21.7230 | 620.4878 | 21399.2035 | 54 | 54 | 4.6528 | 21.6489 | 620.7675 | 21749.2671 |
| 51 | 67 | 4.6692 | 21.8019 | 627.1932 | 21978.7516 | 54 | 55 | 4.6726 | 21.8335 | 630.9664 | 22230.4148 |
| 51 | 68 | 4.6708 | 21.8163 | 623.7284 | 21408.3072 | 54 | 56 | 4.6745 | 21.8506 | 631.0312 | 22195.1080 |
| 51 | 69 | 4.6758 | 21.8633 | 627.6469 | 21777.3884 | 54 | 57 | 4.6918 | 22.0130 | 640.2189 | 22659.1722 |
| 51 | 70 | 4.6815 | 21.9168 | 627.6836 | 21486.4856 | 54 | 58 | 4.6947 | 22.0399 | 641.4166 | 22705.9069 |
| 51 | 71 | 4.6841 | 21.9412 | 629.5310 | 21691.0332 | 54 | 59 | 4.7099 | 22.1836 | 649.7899 | 23160.8528 |
| 51 | 72 | 4.6920 | 22.0149 | 631.7223 | 21593.0214 | 54 | 60 | 4.7133 | 22.2153 | 651.6139 | 23248.4732 |
| 51 | 73 | 4.6933 | 22.0272 | 632.3280 | 21688.5171 | 54 | 61 | 4.7266 | 22.3405 | 659.1928 | 23692.6995 |
| 51 | 74 | 4.7014 | 22.1033 | 635.3137 | 21691.9450 | 54 | 62 | 4.7300 | 22.3726 | 661.1647 | 23783.5930 |
| 51 | 75 | 4.7024 | 22.1125 | 635.4258 | 21728.7283 | 54 | 63 | 4.7410 | 22.4772 | 667.8350 | 24210.0991 |
| 51 | 76 | 4.7093 | 22.1778 | 638.1370 | 21758.7874 | 54 | 64 | 4.7442 | 22.5076 | 669.6872 | 24283.4076 |
| 51 | 77 | 4.7107 | 22.1911 | 638.3429 | 21775.4428 | 54 | 65 | 4.7527 | 22.5880 | 675.2739 | 24684.9602 |
| 51 | 78 | 4.7158 | 22.2392 | 640.1581 | 21785.5545 | 54 | 66 | 4.7559 | 22.6187 | 677.0557 | 24742.2955 |
| 51 | 79 | 4.7182 | 22.2618 | 640.8797 | 21808.4248 | 54 | 67 | 4.7598 | 22.6553 | 678.4612 | 24803.3840 |
| 51 | 80 | 4.7215 | 22.2927 | 641.6349 | 21781.7456 | 54 | 68 | 4.7630 | 22.6864 | 678.9195 | 24691.2772 |
| 51 | 81 | 4.7253 | 22.3282 | 643.1247 | 21825.6168 | 54 | 69 | 4.7644 | 22.6994 | 678.2641 | 24596.7393 |
| 51 | 82 | 4.7272 | 22.3465 | 643.0027 | 21768.3332 | 54 | 70 | 4.7698 | 22.7505 | 680.4579 | 24627.3639 |
| 51 | 83 | 4.7423 | 22.4896 | 652.3940 | 22309.9811 | 54 | 71 | 4.7689 | 22.7425 | 677.9230 | 24386.4296 |
| 51 | 84 | 4.7533 | 22.5940 | 659.0368 | 22703.3475 | 54 | 72 | 4.7758 | 22.8083 | 681.4680 | 24536.3184 |
| 51 | 85 | 4.7711 | 22.7638 | 669.8433 | 23332.9181 | 54 | 73 | 4.7733 | 22.7843 | 677.4736 | 24176.2579 |
| 51 | 86 | 4.7829 | 22.8761 | 677.3595 | 23783.4290 | 54 | 74 | 4.7810 | 22.8583 | 681.8978 | 24414.2564 |
| 51 | 87 | 4.8009 | 23.0486 | 687.7765 | 24376.6525 | 54 | 75 | 4.7775 | 22.8243 | 676.9428 | 23969.9601 |
| 51 | 88 | 4.8142 | 23.1766 | 696.7300 | 24923.1103 | 54 | 76 | 4.7856 | 22.9016 | 681.8437 | 24266.1375 |
| 51 | 89 | 4.8290 | 23.3195 | 704.6305 | 25350.3055 | 54 | 77 | 4.7815 | 22.8630 | 676.4013 | 23771.9310 |
| 51 | 90 | 4.8445 | 23.4688 | 715.3901 | 26018.7751 | 54 | 78 | 4.7897 | 22.9417 | 681.5467 | 24104.7940 |
| 51 | 91 | 4.8526 | 23.5480 | 718.7461 | 26168.9990 | 54 | 79 | 4.7857 | 22.9031 | 676.0211 | 23590.0981 |
| 52 | 52 | 4.5494 | 20.6974 | 570.1029 | 19376.8021 | 54 | 80 | 4.7941 | 22.9836 | 681.3293 | 23947.2793 |
| 52 | 53 | 4.5766 | 20.9452 | 583.2148 | 19924.8633 | 54 | 81 | 4.7905 | 22.9491 | 676.0526 | 23435.9936 |
| 52 | 54 | 4.5788 | 20.9658 | 581.6279 | 19700.0382 | 54 | 82 | 4.7992 | 23.0323 | 681.5059 | 23810.7221 |
| 52 | 55 | 4.6007 | 21.1667 | 592.4822 | 20177.9350 | 54 | 83 | 4.8061 | 23.0983 | 684.2757 | 23847.0990 |
| 52 | 56 | 4.6030 | 21.1881 | 591.7783 | 20039.2727 | 54 | 84 | 4.8243 | 23.2737 | 697.1080 | 24714.3164 |
| 52 | 57 | 4.6216 | 21.3589 | 601.3237 | 20496.7138 | 54 | 85 | 4.8348 | 23.3757 | 702.1808 | 24898.2467 |
| 52 | 58 | 4.6251 | 21.3914 | 601.8791 | 20451.3988 | 54 | 86 | 4.8530 | 23.5512 | 714.9462 | 25740.1137 |
| 52 | 59 | 4.6411 | 21.5398 | 610.5127 | 20904.8373 | 54 | 87 | 4.8659 | 23.6772 | 721.5132 | 26020.7921 |
| 52 | 60 | 4.6450 | 21.5761 | 611.7732 | 20915.2029 | 54 | 88 | 4.8826 | 23.8396 | 733.5168 | 26810.8117 |
| 52 | 61 | 4.6589 | 21.7058 | 619.6655 | 21367.7772 | 54 | 89 | 4.8955 | 23.9658 | 740.1231 | 27107.9796 |
| 52 | 62 | 4.6628 | 21.7415 | 621.1859 | 21399.2542 | 54 | 90 | 4.9097 | 24.1047 | 750.8537 | 27829.6657 |
| 52 | 63 | 4.6747 | 21.8531 | 628.3470 | 21847.5674 | 54 | 91 | 4.9197 | 24.2031 | 755.8162 | 28054.1794 |
| 52 | 64 | 4.6783 | 21.8868 | 629.8728 | 21877.4289 | 54 | 92 | 4.9317 | 24.3220 | 765.4536 | 28717.3828 |
| 52 | 65 | 4.6882 | 21.9789 | 636.2204 | 22316.6909 | 54 | 93 | 4.9365 | 24.3687 | 767.3298 | 28791.1204 |
| 52 | 66 | 4.6917 | 22.0123 | 637.7336 | 22338.8984 | 54 | 94 | 4.9418 | 24.4212 | 773.3140 | 29273.1147 |
| 52 | 67 | 4.6979 | 22.0700 | 640.7208 | 22500.6793 | 54 | 95 | 4.9372 | 24.3756 | 769.9704 | 29102.1352 |
| 52 | 68 | 4.7021 | 22.1098 | 641.7532 | 22431.8160 | 54 | 96 | 4.7104 | 22.1879 | 642.1135 | 27841.6558 |
| 52 | 69 | 4.7063 | 22.1489 | 643.0152 | 22457.9388 | 55 | 56 | 4.7135 | 22.2175 | 653.6493 | 23472.1633 |

TABLE 4. Charge radii, 2nd moment, 4th moment and 6th moment obtained by DNN.

| $Z$ | $N$ | $R_c$(fm) | $R_2$(fm$^2$) | $R_4$(fm$^4$) | $R_6$(fm$^6$) | $Z$ | $N$ | $R_c$(fm) | $R_2$(fm$^2$) | $R_4$(fm$^4$) | $R_6$(fm$^6$) |
|---|---|---|---|---|---|---|---|---|---|---|---|
| 55 | 57 | 4.7327 | 22.3986 | 667.3721 | 24485.5555 | 57 | 86 | 4.9177 | 24.1840 | 748.9968 | 27360.8942 |
| 55 | 58 | 4.7319 | 22.3912 | 663.4629 | 23994.0801 | 57 | 87 | 4.9266 | 24.2717 | 751.8425 | 27343.9148 |
| 55 | 59 | 4.7505 | 22.5670 | 677.4417 | 25061.5502 | 57 | 88 | 4.9477 | 24.4795 | 768.2093 | 28485.6011 |
| 55 | 60 | 4.7492 | 22.5548 | 673.3081 | 24556.7378 | 57 | 89 | 4.9476 | 24.4790 | 765.8111 | 28194.1945 |
| 55 | 61 | 4.7663 | 22.7177 | 687.0347 | 25646.4705 | 57 | 90 | 4.9719 | 24.7195 | 784.5331 | 29485.3997 |
| 55 | 62 | 4.7645 | 22.7003 | 682.5286 | 25111.2764 | 57 | 91 | 4.9109 | 24.1168 | 750.1878 | 27756.8401 |
| 55 | 63 | 4.7791 | 22.8396 | 695.3529 | 26188.9113 | 57 | 92 | 4.9849 | 24.8492 | 794.3358 | 30143.8146 |
| 55 | 64 | 4.7770 | 22.8198 | 690.5257 | 25619.1539 | 57 | 93 | 4.7770 | 22.8193 | 698.4626 | 31853.1437 |
| 55 | 65 | 4.7881 | 22.9255 | 701.8825 | 26660.3063 | 57 | 94 | 4.9200 | 24.2062 | 812.2554 | 40319.4729 |
| 55 | 66 | 4.7863 | 22.9091 | 697.0193 | 26067.8323 | 57 | 95 | 5.0903 | 25.9107 | 882.5448 | 37541.2685 |
| 55 | 67 | 4.7907 | 22.9508 | 702.2822 | 26610.2412 | 57 | 96 | 5.1215 | 26.2299 | 914.1667 | 40433.8747 |
| 55 | 68 | 4.7897 | 22.9416 | 696.0541 | 25806.9274 | 57 | 97 | 5.1248 | 26.2632 | 906.3638 | 38734.4754 |
| 55 | 69 | 4.7903 | 22.9470 | 697.8864 | 26078.9453 | 57 | 98 | 5.1552 | 26.5759 | 938.8712 | 41874.8005 |
| 55 | 70 | 4.7927 | 22.9703 | 694.5924 | 25512.9779 | 57 | 99 | 5.1274 | 26.2897 | 907.5693 | 38785.7490 |
| 55 | 71 | 4.7909 | 22.9530 | 693.8360 | 25559.9520 | 57 | 100 | 5.1645 | 26.6718 | 945.5165 | 42329.4933 |
| 55 | 72 | 4.7956 | 22.9979 | 692.8824 | 25203.5522 | 58 | 61 | 4.8353 | 23.3799 | 729.5971 | 28076.2508 |
| 55 | 73 | 4.7929 | 22.9718 | 690.5334 | 25087.2496 | 58 | 62 | 4.8426 | 23.4505 | 734.4499 | 28363.5777 |
| 55 | 74 | 4.7985 | 23.0259 | 691.1507 | 24897.4273 | 58 | 63 | 4.8484 | 23.5070 | 738.4630 | 28664.4148 |
| 55 | 75 | 4.7961 | 23.0023 | 688.0985 | 24677.1611 | 58 | 64 | 4.8549 | 23.5700 | 742.7692 | 28909.1215 |
| 55 | 76 | 4.8016 | 23.0554 | 689.5750 | 24609.0202 | 58 | 65 | 4.8578 | 23.5979 | 745.3568 | 29162.7261 |
| 55 | 77 | 4.8002 | 23.0419 | 686.4795 | 24331.4474 | 58 | 66 | 4.8637 | 23.6558 | 749.3050 | 29381.7676 |
| 55 | 78 | 4.8051 | 23.0885 | 688.3312 | 24349.1665 | 58 | 67 | 4.8623 | 23.6415 | 747.0893 | 29201.3606 |
| 55 | 79 | 4.8051 | 23.0891 | 685.6161 | 24046.8547 | 58 | 68 | 4.8676 | 23.6936 | 748.5565 | 29120.1019 |
| 55 | 80 | 4.8092 | 23.1283 | 687.6142 | 24127.1323 | 58 | 69 | 4.8639 | 23.6574 | 744.0864 | 28759.5176 |
| 55 | 81 | 4.8109 | 23.1444 | 685.5146 | 23821.5222 | 58 | 70 | 4.8703 | 23.7194 | 746.4736 | 28761.6801 |
| 55 | 82 | 4.8144 | 23.1780 | 687.6019 | 23950.8342 | 58 | 71 | 4.8647 | 23.6653 | 740.0384 | 28237.5853 |
| 55 | 83 | 4.8280 | 23.3091 | 694.1805 | 24217.7880 | 58 | 72 | 4.8720 | 23.7365 | 743.4415 | 28338.5667 |
| 55 | 84 | 4.8422 | 23.4465 | 704.8453 | 24964.1467 | 58 | 73 | 4.8652 | 23.6703 | 735.5531 | 27686.7088 |
| 55 | 85 | 4.8571 | 23.5913 | 711.8887 | 25230.8594 | 58 | 74 | 4.8733 | 23.7491 | 739.9206 | 27888.1468 |
| 55 | 86 | 4.8734 | 23.7498 | 724.1961 | 26083.7424 | 58 | 75 | 4.8659 | 23.6773 | 731.2009 | 27153.0805 |
| 55 | 87 | 4.8851 | 23.8642 | 729.1202 | 26219.8075 | 58 | 76 | 4.8745 | 23.7610 | 736.3373 | 27442.3051 |
| 55 | 88 | 4.9043 | 24.0524 | 743.6205 | 27210.1652 | 58 | 77 | 4.8672 | 23.6901 | 727.4115 | 26668.7233 |
| 55 | 89 | 4.9081 | 24.0893 | 743.7382 | 27086.3025 | 58 | 78 | 4.8760 | 23.7756 | 733.0161 | 27021.9106 |
| 55 | 90 | 4.9307 | 24.3121 | 760.7286 | 28230.7268 | 58 | 79 | 4.8694 | 23.7112 | 724.4257 | 26248.9661 |
| 55 | 91 | 4.9237 | 24.2433 | 754.3131 | 27754.9712 | 58 | 80 | 4.8780 | 23.7951 | 730.1393 | 26635.5727 |
| 55 | 92 | 4.9500 | 24.5025 | 773.9030 | 29061.0837 | 58 | 81 | 4.8725 | 23.7411 | 722.2921 | 25894.3781 |
| 55 | 93 | 4.8756 | 23.7718 | 732.8348 | 27071.1565 | 58 | 82 | 4.8806 | 23.8201 | 727.7412 | 26281.4879 |
| 55 | 94 | 4.9545 | 24.5471 | 778.8013 | 29484.6873 | 58 | 83 | 4.8898 | 23.9099 | 730.9385 | 26277.0353 |
| 55 | 95 | 4.5596 | 20.7903 | 543.3824 | 22204.1588 | 58 | 84 | 4.9102 | 24.1100 | 745.9791 | 27313.4290 |
| 55 | 96 | 4.7211 | 22.2891 | 665.8132 | 31194.2703 | 58 | 85 | 4.9222 | 24.2279 | 750.9397 | 27416.5727 |
| 55 | 97 | 4.9810 | 24.8108 | 779.1639 | 28996.4689 | 58 | 86 | 4.9403 | 24.4065 | 764.9151 | 28393.8972 |
| 56 | 57 | 4.7512 | 22.5741 | 675.8999 | 24763.1608 | 58 | 87 | 4.9531 | 24.5330 | 770.6365 | 28560.4727 |
| 56 | 58 | 4.7558 | 22.6176 | 678.7640 | 24941.7007 | 58 | 88 | 4.9681 | 24.6818 | 782.9888 | 29453.3891 |
| 56 | 59 | 4.7691 | 22.7443 | 686.1964 | 25356.7932 | 58 | 89 | 4.9791 | 24.7915 | 788.0585 | 29614.9468 |
| 56 | 60 | 4.7741 | 22.7924 | 689.5076 | 25555.5207 | 58 | 90 | 4.9913 | 24.9131 | 797.8805 | 30300.8315 |
| 56 | 61 | 4.7854 | 22.8996 | 696.0550 | 25952.5976 | 58 | 91 | 4.9961 | 24.9610 | 800.0188 | 30366.8517 |
| 56 | 62 | 4.7902 | 22.9461 | 699.2733 | 26133.6156 | 58 | 92 | 5.0208 | 25.2081 | 871.8209 | 42102.3933 |
| 56 | 63 | 4.7990 | 23.0302 | 704.7582 | 26503.8525 | 58 | 93 | 4.9958 | 24.9580 | 866.5168 | 43271.1925 |
| 56 | 64 | 4.8034 | 23.0727 | 707.6465 | 26652.3548 | 58 | 94 | 5.1403 | 26.4225 | 931.4357 | 41730.9426 |
| 56 | 65 | 4.8093 | 23.1292 | 711.8461 | 26986.1401 | 58 | 95 | 5.1593 | 26.6188 | 951.2413 | 43536.3792 |
| 56 | 66 | 4.8136 | 23.1703 | 714.5489 | 27113.1881 | 58 | 96 | 5.1805 | 26.8372 | 966.6577 | 44449.6408 |
| 56 | 67 | 4.8145 | 23.1794 | 714.0614 | 27055.6276 | 58 | 97 | 5.1920 | 26.9572 | 977.6505 | 45355.5172 |
| 56 | 68 | 4.8183 | 23.2162 | 714.7429 | 26941.4479 | 58 | 98 | 5.2019 | 27.0602 | 986.5860 | 46166.6337 |
| 56 | 69 | 4.8166 | 23.1997 | 711.9726 | 26711.7314 | 58 | 99 | 5.2033 | 27.0747 | 987.3669 | 46170.6903 |
| 56 | 70 | 4.8222 | 23.2540 | 714.0878 | 26713.4303 | 58 | 100 | 5.2148 | 27.1937 | 997.1187 | 47013.7581 |
| 56 | 71 | 4.8183 | 23.2161 | 709.2762 | 26322.1242 | 58 | 101 | 5.2074 | 27.1173 | 989.3567 | 46281.6439 |
| 56 | 72 | 4.8255 | 23.2857 | 712.7753 | 26443.2337 | 59 | 62 | 4.8646 | 23.6641 | 750.0955 | 29407.6447 |
| 56 | 73 | 4.8200 | 23.2326 | 706.4002 | 25920.2449 | 59 | 63 | 4.8732 | 23.7478 | 760.0906 | 30346.4238 |
| 56 | 74 | 4.8284 | 23.3135 | 711.0692 | 26150.8409 | 59 | 64 | 4.8758 | 23.7732 | 758.1041 | 29968.7005 |
| 56 | 75 | 4.8220 | 23.2520 | 703.7042 | 25535.3942 | 59 | 65 | 4.8800 | 23.8139 | 765.5962 | 30786.9320 |
| 56 | 76 | 4.8312 | 23.3403 | 709.2504 | 25855.9592 | 59 | 66 | 4.8829 | 23.8428 | 763.8984 | 30428.6629 |
| 56 | 77 | 4.8246 | 23.2766 | 701.4532 | 25187.2745 | 59 | 67 | 4.8826 | 23.8394 | 765.5795 | 30690.7450 |
| 56 | 78 | 4.8342 | 23.3693 | 707.5902 | 25574.9203 | 59 | 68 | 4.8858 | 23.8709 | 761.6695 | 30020.0448 |
| 56 | 79 | 4.8279 | 23.3088 | 699.8454 | 24887.3745 | 59 | 69 | 4.8842 | 23.8557 | 761.1671 | 30069.1163 |
| 56 | 80 | 4.8377 | 23.4037 | 706.3195 | 25319.5864 | 59 | 70 | 4.8874 | 23.8867 | 757.9526 | 29496.6631 |
| 56 | 81 | 4.8323 | 23.3509 | 699.0166 | 24641.4610 | 59 | 71 | 4.8854 | 23.8670 | 755.8276 | 29369.6956 |
| 56 | 82 | 4.8421 | 23.4459 | 705.5836 | 25096.2327 | 59 | 72 | 4.8881 | 23.8935 | 753.2604 | 28906.2156 |
| 56 | 83 | 4.8489 | 23.5117 | 707.5448 | 25043.8888 | 59 | 73 | 4.8862 | 23.8753 | 750.0359 | 28639.4409 |
| 56 | 84 | 4.8693 | 23.7106 | 722.3759 | 26056.1053 | 59 | 74 | 4.8884 | 23.8961 | 748.1998 | 28300.6106 |
| 56 | 85 | 4.8799 | 23.8131 | 726.6508 | 26146.0544 | 59 | 75 | 4.8872 | 23.8844 | 744.3495 | 27928.0084 |
| 56 | 86 | 4.8992 | 24.0025 | 740.9086 | 27109.3777 | 59 | 76 | 4.8887 | 23.8994 | 743.3309 | 27723.3743 |
| 56 | 87 | 4.9118 | 24.1257 | 746.5907 | 27294.8696 | 59 | 77 | 4.8885 | 23.8973 | 739.2386 | 27274.8824 |
| 56 | 88 | 4.9284 | 24.2891 | 759.4045 | 28176.4074 | 59 | 78 | 4.8895 | 23.9073 | 739.0518 | 27202.0737 |
| 56 | 89 | 4.9402 | 24.4058 | 764.9282 | 28377.9648 | 59 | 79 | 4.8904 | 23.9157 | 734.9683 | 26701.4231 |
| 56 | 90 | 4.9538 | 24.5406 | 776.1158 | 29171.4133 | 59 | 80 | 4.8910 | 23.9218 | 735.5308 | 26745.4940 |
| 56 | 91 | 4.9621 | 24.6223 | 779.7878 | 29300.4576 | 59 | 81 | 4.8928 | 23.9396 | 731.6059 | 26212.0992 |
| 56 | 92 | 4.9708 | 24.7088 | 787.7980 | 29897.7619 | 59 | 82 | 4.8931 | 23.9425 | 732.7468 | 26348.5333 |
| 56 | 93 | 4.9708 | 24.7092 | 787.1634 | 29845.0677 | 59 | 83 | 4.9093 | 24.1010 | 739.3676 | 26507.4586 |
| 56 | 94 | 4.8966 | 23.9770 | 781.8453 | 37051.4570 | 59 | 84 | 4.9253 | 24.2582 | 752.5226 | 27471.6026 |
| 56 | 95 | 4.8265 | 23.2954 | 743.7316 | 36190.4009 | 59 | 85 | 4.9379 | 24.3833 | 757.2232 | 27523.1206 |
| 56 | 96 | 5.0819 | 25.8258 | 868.4872 | 35863.3434 | 59 | 86 | 4.9566 | 24.5677 | 772.3901 | 28618.2507 |
| 56 | 97 | 5.0717 | 25.7226 | 864.9172 | 36194.3163 | 59 | 87 | 4.9627 | 24.6287 | 773.3604 | 28474.1415 |
| 56 | 98 | 5.1255 | 26.2705 | 901.8730 | 37974.1837 | 59 | 88 | 4.9838 | 24.8387 | 790.5089 | 29703.9606 |
| 57 | 59 | 4.8033 | 23.0715 | 711.6527 | 27188.0267 | 59 | 89 | 4.9363 | 24.3667 | 762.7029 | 28213.1133 |
| 57 | 60 | 4.8035 | 23.0738 | 708.2858 | 26723.2556 | 59 | 90 | 5.0043 | 25.0430 | 804.2256 | 30521.4042 |
| 57 | 61 | 4.8184 | 23.2165 | 721.3606 | 27807.8811 | 59 | 91 | 4.9315 | 24.3193 | 806.4985 | 37729.7801 |
| 57 | 62 | 4.8184 | 23.2170 | 717.8076 | 27325.5221 | 59 | 92 | 5.0498 | 25.5009 | 902.8300 | 44836.7230 |
| 57 | 63 | 4.8300 | 23.3289 | 729.4167 | 28357.8174 | 59 | 93 | 5.1601 | 26.6268 | 947.1846 | 42562.8145 |
| 57 | 64 | 4.8300 | 23.3287 | 725.6951 | 27855.1105 | 59 | 94 | 5.1804 | 26.8363 | 971.7323 | 45189.1387 |
| 57 | 65 | 4.8375 | 23.4012 | 735.3324 | 28813.4187 | 59 | 95 | 5.1937 | 26.9741 | 974.7543 | 44475.7467 |
| 57 | 66 | 4.8379 | 23.4052 | 731.7205 | 28304.9109 | 59 | 96 | 5.2124 | 27.1688 | 999.3345 | 47288.8733 |
| 57 | 67 | 4.8394 | 23.4195 | 735.2278 | 28731.5317 | 59 | 97 | 5.2031 | 27.0728 | 984.0144 | 45332.5236 |
| 57 | 68 | 4.8400 | 23.4254 | 729.6350 | 27952.3423 | 59 | 98 | 5.2253 | 27.3042 | 1011.5342 | 48399.6345 |
| 57 | 69 | 4.8388 | 23.4141 | 730.1740 | 28124.4888 | 59 | 99 | 5.2045 | 27.0866 | 985.3237 | 45539.2471 |
| 57 | 70 | 4.8412 | 23.4372 | 726.4632 | 27515.9420 | 59 | 100 | 5.2305 | 27.3577 | 1015.4228 | 48760.9607 |
| 57 | 71 | 4.8386 | 23.4125 | 724.8090 | 27477.4345 | 59 | 101 | 5.2011 | 27.0513 | 981.7711 | 45327.3954 |
| 57 | 72 | 4.8421 | 23.4460 | 722.7382 | 27037.3078 | 59 | 102 | 5.2301 | 27.3544 | 1013.3026 | 48554.5626 |
| 57 | 73 | 4.8393 | 23.4188 | 719.7268 | 26841.4982 | 60 | 64 | 4.9003 | 24.0134 | 775.5437 | 31062.1565 |
| 57 | 74 | 4.8432 | 23.4562 | 718.9955 | 26560.0471 | 60 | 65 | 4.9012 | 24.0220 | 776.9094 | 31239.1020 |
| 57 | 75 | 4.8410 | 23.4354 | 715.3606 | 26256.8537 | 60 | 66 | 4.9084 | 24.0920 | 781.9210 | 31554.4967 |
| 57 | 76 | 4.8447 | 23.4715 | 715.6564 | 26117.3045 | 60 | 67 | 4.9064 | 24.0727 | 778.7455 | 31264.8949 |
| 57 | 77 | 4.8438 | 23.4625 | 711.9366 | 25746.2171 | 60 | 68 | 4.9131 | 24.1382 | 781.0895 | 31242.0803 |
| 57 | 78 | 4.8471 | 23.4946 | 713.0072 | 25728.9213 | 60 | 69 | 4.9096 | 24.1046 | 775.9694 | 30779.5947 |
| 57 | 79 | 4.8476 | 23.4997 | 709.5274 | 25317.6548 | 60 | 70 | 4.9161 | 24.1684 | 778.5465 | 30804.4094 |
| 57 | 80 | 4.8505 | 23.5272 | 711.1918 | 25402.2577 | 60 | 71 | 4.9114 | 24.1217 | 771.7263 | 30192.2605 |
| 57 | 81 | 4.8524 | 23.5461 | 708.1154 | 24970.1151 | 60 | 72 | 4.9177 | 24.1837 | 774.6048 | 30275.5055 |
| 57 | 82 | 4.8549 | 23.5699 | 710.2286 | 25135.8946 | 60 | 73 | 4.9119 | 24.1265 | 766.4956 | 29549.7109 |
| 57 | 83 | 4.8699 | 23.7164 | 716.6707 | 25329.1606 | 60 | 74 | 4.9180 | 24.1868 | 769.7065 | 29695.8121 |
| 57 | 84 | 4.8853 | 23.8666 | 728.8916 | 26209.3977 | 60 | 75 | 4.9116 | 24.1238 | 760.8540 | 28899.5998 |
| 57 | 85 | 4.8998 | 24.0079 | 734.8908 | 26360.6832 | 60 | 76 | 4.9175 | 24.1818 | 764.3318 | 29102.6449 |

TABLE 4. Charge radii, 2nd moment, 4th moment and 6th moment obtained by DNN.

| $Z$ | $N$ | $R_c\,(\mathrm{fm})$ | $R_2\,(\mathrm{fm}^2)$ | $R_4\,(\mathrm{fm}^4)$ | $R_6\,(\mathrm{fm}^6)$ | $Z$ | $N$ | $R_c\,(\mathrm{fm})$ | $R_2\,(\mathrm{fm}^2)$ | $R_4\,(\mathrm{fm}^4)$ | $R_6\,(\mathrm{fm}^6)$ |
|---|---|---|---|---|---|---|---|---|---|---|---|
| 60 | 77 | 4.9111 | 24.1190 | 755.3367 | 28280.5810 | 63 | 69 | 4.9747 | 24.7480 | 822.1770 | 33727.1486 |
| 60 | 78 | 4.9166 | 24.1731 | 758.9091 | 28523.8909 | 63 | 70 | 4.9819 | 24.8195 | 823.1856 | 33518.9348 |
| 60 | 79 | 4.9108 | 24.1159 | 750.3034 | 27714.6937 | 63 | 71 | 4.9788 | 24.7882 | 818.1820 | 33063.7658 |
| 60 | 80 | 4.9157 | 24.1642 | 753.7290 | 27974.1221 | 63 | 72 | 4.9833 | 24.8328 | 818.2618 | 32866.4838 |
| 60 | 81 | 4.9108 | 24.1163 | 745.8888 | 27206.1852 | 63 | 73 | 4.9789 | 24.7895 | 811.7792 | 32279.0717 |
| 60 | 82 | 4.9150 | 24.1572 | 748.9252 | 27456.0131 | 63 | 74 | 4.9810 | 24.8108 | 811.2631 | 32118.3460 |
| 60 | 83 | 4.9271 | 24.2759 | 753.8397 | 27535.8004 | 63 | 75 | 4.9757 | 24.7571 | 803.4749 | 31412.4713 |
| 60 | 84 | 4.9453 | 24.4557 | 768.1444 | 28563.6822 | 63 | 76 | 4.9761 | 24.7614 | 802.8067 | 31314.6889 |
| 60 | 85 | 4.9584 | 24.5853 | 773.6704 | 28677.4164 | 63 | 77 | 4.9701 | 24.7020 | 794.1146 | 30517.6227 |
| 60 | 86 | 4.9740 | 24.7407 | 786.9142 | 29665.0286 | 63 | 78 | 4.9695 | 24.6959 | 793.6889 | 30500.3457 |
| 60 | 87 | 4.9862 | 24.8626 | 792.1166 | 29775.0694 | 63 | 79 | 4.9635 | 24.6367 | 784.5825 | 29644.9442 |
| 60 | 88 | 5.0018 | 25.0184 | 803.7879 | 30550.4968 | 63 | 80 | 4.9625 | 24.6259 | 784.6530 | 29712.8398 |
| 60 | 89 | 5.0095 | 25.0951 | 807.0912 | 30624.7202 | 63 | 81 | 4.9570 | 24.5716 | 775.5593 | 28829.2666 |
| 60 | 90 | 5.0537 | 25.5398 | 885.2248 | 41274.6410 | 63 | 82 | 4.9559 | 24.5605 | 776.2218 | 28974.9972 |
| 60 | 91 | 5.0784 | 25.7904 | 917.7075 | 44802.1879 | 63 | 83 | 4.9661 | 24.6618 | 778.9512 | 28872.7991 |
| 60 | 92 | 5.1255 | 26.2710 | 922.0088 | 41251.6762 | 63 | 84 | 4.9819 | 24.8192 | 793.3196 | 29982.2345 |
| 60 | 93 | 5.1801 | 26.8332 | 971.4832 | 45097.4599 | 63 | 85 | 4.9444 | 24.4472 | 767.5883 | 28307.2144 |
| 60 | 94 | 5.1619 | 26.6456 | 955.6122 | 44091.9860 | 63 | 86 | 5.0094 | 25.0943 | 808.4310 | 30661.2383 |
| 60 | 95 | 5.2106 | 27.1505 | 998.7414 | 47295.4667 | 63 | 87 | 4.9748 | 24.7482 | 806.6530 | 33673.8213 |
| 60 | 96 | 5.1837 | 26.8706 | 976.7631 | 46050.7230 | 63 | 88 | 5.0476 | 25.4780 | 859.7776 | 36927.4913 |
| 60 | 97 | 5.2241 | 27.2915 | 1011.9112 | 48549.3752 | 63 | 89 | 5.0905 | 25.9130 | 878.2191 | 36349.2934 |
| 60 | 98 | 5.1972 | 27.0106 | 989.1073 | 47194.0794 | 63 | 90 | 5.0891 | 25.8986 | 879.3106 | 36814.6458 |
| 60 | 99 | 5.2295 | 27.3475 | 1016.6312 | 49050.7266 | 63 | 91 | 5.1209 | 26.2231 | 903.4881 | 38288.3598 |
| 60 | 100 | 5.2041 | 27.0822 | 994.1056 | 47624.4533 | 63 | 92 | 5.1162 | 26.1759 | 902.7373 | 38787.5504 |
| 60 | 101 | 5.2290 | 27.3429 | 1015.0603 | 48963.2180 | 63 | 93 | 5.1373 | 26.3920 | 918.0560 | 39561.8573 |
| 60 | 102 | 5.2058 | 27.1003 | 993.2393 | 47466.5364 | 63 | 94 | 5.1329 | 26.3471 | 917.6832 | 40177.2165 |
| 60 | 103 | 5.2246 | 27.2966 | 1009.0314 | 48442.0442 | 63 | 95 | 5.1459 | 26.4807 | 925.7468 | 40316.1067 |
| 61 | 65 | 4.9201 | 24.2078 | 794.3806 | 32622.9083 | 63 | 96 | 5.1438 | 26.4582 | 927.0203 | 41083.3157 |
| 61 | 66 | 4.9254 | 24.2592 | 794.9669 | 32468.0779 | 63 | 97 | 5.1491 | 26.5130 | 928.5107 | 40690.5428 |
| 61 | 67 | 4.9244 | 24.2495 | 794.8884 | 32529.4197 | 63 | 98 | 5.1507 | 26.5294 | 932.4344 | 41622.8196 |
| 61 | 68 | 4.9308 | 24.3131 | 793.5678 | 32050.1727 | 63 | 99 | 5.1490 | 26.5121 | 928.2979 | 40827.7006 |
| 61 | 69 | 4.9296 | 24.3005 | 792.0360 | 31949.7653 | 63 | 100 | 5.1554 | 26.5781 | 935.4559 | 41913.6643 |
| 61 | 70 | 4.9344 | 24.3483 | 790.3729 | 31506.2087 | 63 | 101 | 5.1476 | 26.4980 | 926.8355 | 40856.1500 |
| 61 | 71 | 4.9329 | 24.3339 | 787.5907 | 31263.4511 | 63 | 102 | 5.1592 | 26.6173 | 937.3221 | 42062.3564 |
| 61 | 72 | 4.9359 | 24.3634 | 785.6185 | 30870.3587 | 63 | 103 | 5.1465 | 26.4863 | 925.4882 | 40881.5345 |
| 61 | 73 | 4.9343 | 24.3474 | 781.7213 | 30498.8618 | 63 | 104 | 5.1629 | 26.6553 | 938.9306 | 42156.8963 |
| 61 | 74 | 4.9357 | 24.3609 | 779.7368 | 30183.7698 | 63 | 105 | 5.1585 | 26.6101 | 927.6273 | 40263.9260 |
| 61 | 75 | 4.9340 | 24.3446 | 774.9352 | 29701.4395 | 63 | 106 | 5.2035 | 27.0768 | 961.4724 | 42558.0378 |
| 61 | 76 | 4.9342 | 24.3461 | 773.2898 | 29489.0899 | 63 | 107 | 5.1870 | 26.9046 | 937.7648 | 39535.8241 |
| 61 | 77 | 4.9327 | 24.3317 | 767.8833 | 28920.0138 | 64 | 69 | 5.0016 | 25.0165 | 839.6719 | 34712.5530 |
| 61 | 78 | 4.9321 | 24.3251 | 766.8121 | 28820.5991 | 64 | 70 | 5.0077 | 25.0770 | 843.4312 | 34914.4172 |
| 61 | 79 | 4.9310 | 24.3147 | 761.1182 | 28191.1408 | 64 | 71 | 5.0038 | 25.0378 | 835.3309 | 34092.1354 |
| 61 | 80 | 4.9298 | 24.3030 | 760.6753 | 28197.3148 | 64 | 72 | 5.0079 | 25.0787 | 838.1807 | 34272.5348 |
| 61 | 81 | 4.9293 | 24.2979 | 754.9706 | 27532.4833 | 64 | 73 | 5.0016 | 25.0162 | 828.4460 | 33347.9411 |
| 61 | 82 | 4.9277 | 24.2825 | 755.0469 | 27622.6619 | 64 | 74 | 5.0045 | 25.0455 | 830.8155 | 33525.2296 |
| 61 | 83 | 4.9425 | 24.4282 | 760.8622 | 27721.4126 | 64 | 75 | 4.9960 | 24.9596 | 819.6066 | 32517.7814 |
| 61 | 84 | 4.9586 | 24.5874 | 774.6611 | 28754.6654 | 64 | 76 | 4.9986 | 24.9863 | 822.0056 | 32715.4649 |
| 61 | 85 | 4.9685 | 24.6858 | 777.4289 | 28672.6943 | 64 | 77 | 4.9881 | 24.8807 | 809.6664 | 31650.1219 |
| 61 | 86 | 4.9864 | 24.8644 | 793.0577 | 29845.9211 | 64 | 78 | 4.9913 | 24.9134 | 812.5663 | 31890.2644 |
| 61 | 87 | 4.9476 | 24.4790 | 768.6466 | 28398.5191 | 64 | 79 | 4.9793 | 24.7938 | 799.5247 | 30792.5967 |
| 61 | 88 | 5.0117 | 25.1169 | 808.6744 | 30686.4192 | 64 | 80 | 4.9839 | 24.8390 | 803.2798 | 31091.2128 |
| 61 | 89 | 4.9962 | 24.9619 | 842.7752 | 38360.2037 | 64 | 81 | 4.9710 | 24.7107 | 789.9056 | 29980.8072 |
| 61 | 90 | 5.0808 | 25.8150 | 909.8134 | 43057.0810 | 64 | 82 | 4.9772 | 24.7728 | 794.7284 | 30346.0449 |
| 61 | 91 | 5.1612 | 26.6379 | 946.6716 | 42224.8082 | 64 | 83 | 4.9794 | 24.7948 | 793.3766 | 30058.4711 |
| 61 | 92 | 5.1575 | 26.5998 | 949.3187 | 43158.0631 | 64 | 84 | 5.0118 | 25.1179 | 811.2928 | 30890.5528 |
| 61 | 93 | 5.1949 | 26.9866 | 975.8868 | 44455.3635 | 64 | 85 | 5.0085 | 25.0848 | 808.0604 | 30618.6851 |
| 61 | 94 | 5.1859 | 26.8932 | 974.7548 | 45276.6977 | 64 | 86 | 5.0480 | 25.4827 | 846.4683 | 34362.9004 |
| 61 | 95 | 5.2097 | 27.1412 | 990.4019 | 45785.6540 | 64 | 87 | 5.0404 | 25.4052 | 844.1606 | 34691.2350 |
| 61 | 96 | 5.1992 | 27.0319 | 987.6943 | 46536.2562 | 64 | 88 | 5.0746 | 25.7518 | 863.0984 | 35061.6187 |
| 61 | 97 | 5.2158 | 27.2044 | 996.6909 | 46469.0835 | 64 | 89 | 5.0717 | 25.7219 | 859.0410 | 34699.9366 |
| 61 | 98 | 5.2048 | 27.0896 | 992.7061 | 47095.5596 | 64 | 90 | 5.1111 | 26.1228 | 896.2946 | 37957.2400 |
| 61 | 99 | 5.2159 | 27.2058 | 997.2289 | 46684.1668 | 64 | 91 | 5.1002 | 26.0125 | 883.2833 | 36723.6570 |
| 61 | 100 | 5.2049 | 27.0905 | 991.8910 | 47108.9962 | 64 | 92 | 5.1396 | 26.4152 | 922.9968 | 40402.9169 |
| 61 | 101 | 5.2124 | 27.1687 | 994.0688 | 46587.1184 | 64 | 93 | 5.1200 | 26.2143 | 900.5125 | 38273.7730 |
| 61 | 102 | 5.2014 | 27.0548 | 987.1326 | 46730.2974 | 64 | 94 | 5.1624 | 26.6507 | 944.2699 | 42379.8076 |
| 61 | 103 | 5.2069 | 27.1115 | 988.8593 | 46306.0982 | 64 | 95 | 5.1344 | 26.3623 | 912.8911 | 39419.4377 |
| 61 | 104 | 5.1960 | 26.9988 | 980.0424 | 46102.1169 | 64 | 96 | 5.1806 | 26.8386 | 960.7495 | 43908.3084 |
| 62 | 66 | 4.9514 | 24.5164 | 814.0862 | 33694.6815 | 64 | 97 | 5.1453 | 26.4736 | 921.8003 | 40249.8947 |
| 62 | 67 | 4.9506 | 24.5080 | 810.3348 | 33277.9272 | 64 | 98 | 5.1950 | 26.9877 | 973.1724 | 45031.9097 |
| 62 | 68 | 4.9584 | 24.5855 | 814.0602 | 33384.6510 | 64 | 99 | 5.1540 | 26.5632 | 928.5266 | 40858.8698 |
| 62 | 69 | 4.9564 | 24.5663 | 808.5800 | 32805.7000 | 64 | 100 | 5.2063 | 27.1056 | 982.2554 | 45807.8309 |
| 62 | 70 | 4.9630 | 24.6311 | 811.9346 | 32931.9260 | 64 | 101 | 5.1616 | 26.6423 | 934.1076 | 41332.5853 |
| 62 | 71 | 4.9596 | 24.5980 | 804.8298 | 32214.4286 | 64 | 102 | 5.2152 | 27.1979 | 988.6359 | 46298.5233 |
| 62 | 72 | 4.9648 | 24.6492 | 807.7577 | 32358.3602 | 64 | 103 | 5.1689 | 26.7180 | 939.2872 | 41743.4590 |
| 62 | 73 | 4.9600 | 24.6020 | 799.2747 | 31532.2242 | 64 | 104 | 5.2220 | 27.2691 | 992.8861 | 46567.9765 |
| 62 | 74 | 4.9639 | 24.6402 | 801.8175 | 31695.2257 | 64 | 105 | 5.1923 | 26.9599 | 952.1974 | 42007.9550 |
| 62 | 75 | 4.9580 | 24.5818 | 792.3623 | 30796.1193 | 64 | 106 | 5.2679 | 27.7503 | 1022.6899 | 47747.5679 |
| 62 | 76 | 4.9608 | 24.6094 | 794.6399 | 30980.8240 | 64 | 107 | 5.2323 | 27.3773 | 975.0676 | 42481.7113 |
| 62 | 77 | 4.9543 | 24.5448 | 784.7120 | 30046.5299 | 64 | 108 | 5.3104 | 28.2006 | 1051.2634 | 49003.8157 |
| 62 | 78 | 4.9564 | 24.5656 | 786.8959 | 30255.3410 | 65 | 70 | 5.0215 | 25.2151 | 849.9013 | 35146.7005 |
| 62 | 79 | 4.9497 | 24.4996 | 776.9387 | 29317.4259 | 65 | 71 | 5.0222 | 25.2226 | 843.6637 | 34363.6542 |
| 62 | 80 | 4.9516 | 24.5181 | 779.2179 | 29551.6667 | 65 | 72 | 5.0196 | 25.1961 | 843.2960 | 34418.7136 |
| 62 | 81 | 4.9450 | 24.4534 | 769.5002 | 28629.2242 | 65 | 73 | 5.0173 | 25.1734 | 835.0913 | 33516.9696 |
| 62 | 82 | 4.9472 | 24.4745 | 772.0607 | 28889.7937 | 65 | 74 | 5.0139 | 25.1392 | 834.5499 | 33591.4517 |
| 62 | 83 | 4.9575 | 24.5770 | 775.4743 | 28855.4799 | 65 | 75 | 5.0091 | 25.0907 | 824.6409 | 32584.5844 |
| 62 | 84 | 4.9753 | 24.7540 | 790.5588 | 29979.7972 | 65 | 76 | 5.0056 | 25.0562 | 824.4310 | 32709.0825 |
| 62 | 85 | 4.9847 | 24.8469 | 793.3936 | 29916.6775 | 65 | 77 | 4.9989 | 24.9895 | 813.2723 | 31621.2465 |
| 62 | 86 | 5.0061 | 25.0613 | 807.6234 | 30754.8908 | 65 | 78 | 4.9961 | 24.9611 | 813.8106 | 31819.1032 |
| 62 | 87 | 5.0112 | 25.1119 | 808.9700 | 30697.3844 | 65 | 79 | 4.9884 | 24.8844 | 801.9470 | 30679.6911 |
| 62 | 88 | 5.0387 | 25.3889 | 854.3750 | 36837.3226 | 65 | 80 | 4.9866 | 24.8666 | 803.4834 | 30962.6123 |
| 62 | 89 | 5.0652 | 25.6558 | 885.6173 | 40050.9980 | 65 | 81 | 4.9787 | 24.7874 | 791.4106 | 29799.2101 |
| 62 | 90 | 5.0803 | 25.8091 | 876.2110 | 37061.2168 | 65 | 82 | 4.9782 | 24.7825 | 794.0203 | 30166.8251 |
| 62 | 91 | 5.1215 | 26.2302 | 913.2485 | 39966.9210 | 65 | 83 | 4.9381 | 24.3846 | 764.5277 | 28115.2772 |
| 62 | 92 | 5.1157 | 26.1700 | 908.7970 | 39906.4374 | 65 | 84 | 5.0096 | 25.0959 | 808.5862 | 30608.1798 |
| 62 | 93 | 5.1484 | 26.5064 | 936.9574 | 41966.6600 | 65 | 85 | 4.9548 | 24.5502 | 773.3390 | 29306.3460 |
| 62 | 94 | 5.1407 | 26.4268 | 932.6743 | 42125.9317 | 65 | 86 | 5.0451 | 25.4533 | 840.8850 | 33582.0875 |
| 62 | 95 | 5.1626 | 26.6528 | 950.1822 | 43239.5360 | 65 | 87 | 5.0348 | 25.3495 | 822.0265 | 31174.2488 |
| 62 | 96 | 5.1593 | 26.6179 | 949.9021 | 43734.1923 | 65 | 88 | 5.0699 | 25.7037 | 853.4981 | 33783.1160 |
| 62 | 97 | 5.1700 | 26.7286 | 956.6297 | 43918.1947 | 65 | 89 | 5.0633 | 25.6367 | 844.2833 | 32843.7304 |
| 62 | 98 | 5.1726 | 26.7554 | 961.4579 | 44788.2120 | 65 | 90 | 5.1001 | 26.0106 | 878.8369 | 35869.0226 |
| 62 | 99 | 5.1727 | 26.7569 | 958.2596 | 44143.2789 | 65 | 91 | 5.0826 | 25.8326 | 859.7583 | 34103.0653 |
| 62 | 100 | 5.1818 | 26.8508 | 968.4047 | 45369.9623 | 65 | 92 | 5.1224 | 26.2394 | 898.0444 | 37550.5839 |
| 62 | 101 | 5.1728 | 26.7574 | 956.8679 | 44058.1545 | 65 | 93 | 5.0961 | 25.9702 | 870.5143 | 35028.6769 |
| 62 | 102 | 5.1879 | 26.9139 | 971.7516 | 45570.9772 | 65 | 94 | 5.1397 | 26.4166 | 912.7554 | 38874.8745 |
| 62 | 103 | 5.1716 | 26.7454 | 953.9320 | 43791.9530 | 65 | 95 | 5.1054 | 26.0651 | 877.8420 | 35712.2142 |
| 62 | 104 | 5.1916 | 26.9525 | 972.4130 | 45485.0085 | 65 | 96 | 5.1533 | 26.5569 | 924.1192 | 39908.3545 |
| 62 | 105 | 5.1903 | 26.9395 | 962.9050 | 43814.4211 | 65 | 97 | 5.1123 | 26.1355 | 883.2584 | 36260.4916 |
| 62 | 106 | 5.2426 | 27.4844 | 1007.5453 | 47232.7155 | 65 | 98 | 5.1647 | 26.6740 | 933.2662 | 40727.0880 |
| 63 | 67 | 4.9672 | 24.6726 | 823.7538 | 34253.8030 | 65 | 99 | 5.1185 | 26.1986 | 888.1771 | 36774.4403 |
| 63 | 68 | 4.9769 | 24.7693 | 825.7406 | 34047.7934 | 65 | 100 | 5.1748 | 26.7781 | 941.1148 | 41403.0548 |

TABLE 4. Charge radii, 2nd moment, 4th moment and 6th moment obtained by DNN.

| $Z$ | $N$ | $R_c\,(\mathrm{fm})$ | $R_2\,(\mathrm{fm}^2)$ | $R_4\,(\mathrm{fm}^4)$ | $R_6\,(\mathrm{fm}^6)$ | $Z$ | $N$ | $R_c\,(\mathrm{fm})$ | $R_2\,(\mathrm{fm}^2)$ | $R_4\,(\mathrm{fm}^4)$ | $R_6\,(\mathrm{fm}^6)$ |
|---|---|---|---|---|---|---|---|---|---|---|---|
| 65 | 101 | 5.1252 | 26.2674 | 893.7146 | 37337.9886 | 68 | 97 | 5.2187 | 27.2348 | 972.3014 | 42895.1895 |
| 65 | 102 | 5.1842 | 26.8757 | 948.3296 | 41998.5660 | 68 | 98 | 5.2553 | 27.6183 | 1010.1192 | 46301.3208 |
| 65 | 103 | 5.1332 | 26.3501 | 900.6257 | 38012.5966 | 68 | 99 | 5.2336 | 27.3909 | 984.8566 | 43941.3598 |
| 65 | 104 | 5.1933 | 26.9704 | 955.3590 | 42564.7783 | 68 | 100 | 5.2673 | 27.7444 | 1020.3282 | 47184.4024 |
| 65 | 105 | 5.1508 | 26.5312 | 907.4082 | 37682.0814 | 68 | 101 | 5.2471 | 27.5319 | 996.3113 | 44905.3422 |
| 65 | 106 | 5.2304 | 27.3571 | 974.5080 | 42596.8372 | 68 | 102 | 5.2769 | 27.8459 | 1028.3639 | 47871.6590 |
| 65 | 107 | 5.1793 | 26.8255 | 916.7700 | 36735.5739 | 68 | 103 | 5.2594 | 27.6615 | 1007.0710 | 45828.4697 |
| 65 | 108 | 5.2667 | 27.7376 | 995.1847 | 42985.9391 | 68 | 104 | 5.2847 | 27.9280 | 1034.7432 | 48413.6572 |
| 65 | 109 | 5.2092 | 27.1353 | 929.8899 | 36363.4338 | 68 | 105 | 5.2759 | 27.8353 | 1014.5073 | 45676.8591 |
| 66 | 72 | 5.0783 | 25.7889 | 870.9491 | 35304.0500 | 68 | 106 | 5.3133 | 28.2314 | 1049.2329 | 48369.1090 |
| 66 | 73 | 5.0351 | 25.3522 | 849.3452 | 34520.3142 | 68 | 107 | 5.2972 | 28.0604 | 1020.6156 | 44776.9170 |
| 66 | 74 | 5.0700 | 25.7044 | 861.4716 | 34498.8491 | 68 | 108 | 5.3397 | 28.5119 | 1063.5522 | 48519.6037 |
| 66 | 75 | 5.0269 | 25.2700 | 839.4926 | 33657.6027 | 68 | 109 | 5.3174 | 28.2749 | 1028.1509 | 44256.4064 |
| 66 | 76 | 5.0592 | 25.5958 | 850.6298 | 33631.5397 | 68 | 110 | 5.3626 | 28.7572 | 1076.6469 | 48787.0649 |
| 66 | 77 | 5.0172 | 25.1720 | 828.8762 | 32770.4431 | 68 | 111 | 5.3356 | 28.4687 | 1036.2821 | 44055.5394 |
| 66 | 78 | 5.0469 | 25.4717 | 839.0005 | 32736.3826 | 68 | 112 | 5.3815 | 28.9604 | 1087.8470 | 49108.7589 |
| 66 | 79 | 5.0070 | 25.0704 | 818.2383 | 31898.6601 | 69 | 75 | 4.2884 | 18.3907 | 458.9588 | 14805.6281 |
| 66 | 80 | 5.0340 | 25.3410 | 827.1599 | 31845.0716 | 69 | 76 | 4.3530 | 18.9490 | 487.2232 | 16013.4405 |
| 66 | 81 | 4.9976 | 24.9756 | 808.2111 | 31074.0909 | 69 | 77 | 4.2451 | 18.0210 | 433.5561 | 13261.0835 |
| 66 | 82 | 5.0211 | 25.2110 | 815.5584 | 30979.7402 | 69 | 78 | 4.2974 | 18.4678 | 456.7586 | 14290.8413 |
| 66 | 83 | 5.0128 | 25.1285 | 810.1095 | 30638.5432 | 69 | 79 | 4.2065 | 17.6945 | 410.8953 | 11871.4210 |
| 66 | 84 | 5.0751 | 25.7562 | 859.6092 | 34131.8080 | 69 | 80 | 4.2455 | 18.0246 | 428.8534 | 12726.6528 |
| 66 | 85 | 5.0576 | 25.5798 | 846.3110 | 33342.6718 | 69 | 81 | 5.0099 | 25.0991 | 812.6829 | 31849.1949 |
| 66 | 86 | 5.0970 | 25.9797 | 876.4897 | 35274.8759 | 69 | 82 | 5.0407 | 25.4082 | 839.2744 | 33932.2518 |
| 66 | 87 | 5.0787 | 25.7928 | 858.2986 | 33756.2534 | 69 | 83 | 5.0446 | 25.4479 | 828.9983 | 31844.4885 |
| 66 | 88 | 5.1349 | 26.3672 | 910.4347 | 38144.4327 | 69 | 84 | 5.0568 | 25.5708 | 836.8792 | 32060.3745 |
| 66 | 89 | 5.1104 | 26.1160 | 884.7295 | 35890.3213 | 69 | 85 | 5.0846 | 25.8536 | 860.2518 | 34074.7314 |
| 66 | 90 | 5.1658 | 26.6850 | 938.7193 | 40647.0690 | 69 | 86 | 5.0985 | 25.9948 | 869.9071 | 34462.9544 |
| 66 | 91 | 5.1349 | 26.3673 | 905.5139 | 37661.9436 | 69 | 87 | 5.1165 | 26.1782 | 886.5560 | 36113.1983 |
| 66 | 92 | 5.1909 | 26.9453 | 961.8850 | 42756.4841 | 69 | 88 | 5.1330 | 26.3475 | 898.4474 | 36690.0974 |
| 66 | 93 | 5.1543 | 26.5665 | 921.8894 | 39102.4174 | 69 | 89 | 5.1459 | 26.4802 | 910.9893 | 38002.6029 |
| 66 | 94 | 5.2112 | 27.1563 | 980.4554 | 44477.6503 | 69 | 90 | 5.1639 | 26.6661 | 924.2198 | 38718.0660 |
| 66 | 95 | 5.1698 | 26.7269 | 934.8559 | 40262.5445 | 69 | 91 | 5.1719 | 26.7486 | 932.7705 | 39704.3457 |
| 66 | 96 | 5.2275 | 27.3272 | 995.1058 | 45837.3810 | 69 | 92 | 5.1908 | 26.9440 | 946.7120 | 40509.1756 |
| 66 | 97 | 5.1828 | 26.8613 | 945.4410 | 41204.7225 | 69 | 93 | 5.1943 | 26.9813 | 951.8200 | 41228.6104 |
| 66 | 98 | 5.2408 | 27.4657 | 1006.4835 | 46874.6286 | 69 | 94 | 5.2137 | 27.1826 | 966.0563 | 42070.6051 |
| 66 | 99 | 5.1942 | 26.9797 | 954.5055 | 41991.0473 | 69 | 95 | 5.2138 | 27.1838 | 968.6589 | 42620.2008 |
| 66 | 100 | 5.2514 | 27.5776 | 1015.1538 | 47635.2326 | 69 | 96 | 5.2333 | 27.3878 | 982.7587 | 43438.1064 |
| 66 | 101 | 5.2047 | 27.0884 | 962.6898 | 42678.0716 | 69 | 97 | 5.2312 | 27.3650 | 984.0317 | 43934.6816 |
| 66 | 102 | 5.2599 | 27.6670 | 1021.6060 | 48168.1175 | 69 | 98 | 5.2503 | 27.5660 | 997.4001 | 44656.6019 |
| 66 | 103 | 5.2145 | 27.1913 | 970.4329 | 43313.2716 | 69 | 99 | 5.2472 | 27.5332 | 998.6249 | 45221.7517 |
| 66 | 104 | 5.2667 | 27.7377 | 1026.2951 | 48524.0733 | 69 | 100 | 5.2653 | 27.7233 | 1010.5252 | 45772.3927 |
| 66 | 105 | 5.2362 | 27.4174 | 982.1079 | 43472.0472 | 69 | 101 | 5.2626 | 27.6945 | 1012.9630 | 46519.7649 |
| 66 | 106 | 5.3046 | 28.1387 | 1048.7662 | 49053.9772 | 69 | 102 | 5.2787 | 27.8646 | 1022.6082 | 46828.1768 |
| 66 | 107 | 5.2694 | 27.7665 | 999.0639 | 43431.7701 | 69 | 103 | 5.2776 | 27.8530 | 1027.3865 | 47853.9191 |
| 66 | 108 | 5.3393 | 28.5083 | 1070.1238 | 49682.2562 | 69 | 104 | 5.2910 | 27.9943 | 1034.0682 | 47862.5234 |
| 66 | 109 | 5.3005 | 28.0948 | 1016.3327 | 43657.0694 | 69 | 105 | 5.2875 | 27.9581 | 1028.1786 | 47097.5463 |
| 66 | 110 | 5.3693 | 28.8297 | 1088.9935 | 50310.8537 | 69 | 106 | 5.3058 | 28.1520 | 1033.4388 | 46332.9566 |
| 67 | 73 | 4.5583 | 20.7779 | 595.6691 | 21879.2350 | 69 | 107 | 5.2939 | 28.0256 | 1018.0288 | 44617.5033 |
| 67 | 74 | 5.0392 | 25.3938 | 845.2065 | 33858.2790 | 69 | 108 | 5.3207 | 28.3103 | 1035.1397 | 45245.1567 |
| 67 | 75 | 4.5282 | 20.5043 | 574.7283 | 20453.2386 | 69 | 109 | 5.3017 | 28.1081 | 1011.6011 | 42708.9406 |
| 67 | 76 | 5.0292 | 25.2927 | 834.3633 | 32955.2121 | 69 | 110 | 5.3348 | 28.4606 | 1038.5580 | 44567.6078 |
| 67 | 77 | 4.4957 | 20.2115 | 553.1766 | 19035.8798 | 69 | 111 | 5.3103 | 28.1991 | 1008.5835 | 41378.4476 |
| 67 | 78 | 5.0181 | 25.1810 | 823.0475 | 32040.3472 | 69 | 112 | 5.3476 | 28.5969 | 1043.1426 | 44254.6134 |
| 67 | 79 | 4.4628 | 19.9167 | 532.0356 | 17677.2926 | 69 | 113 | 5.3191 | 28.2923 | 1008.4860 | 40593.4107 |
| 67 | 80 | 5.0069 | 25.0694 | 811.9301 | 31149.3025 | 70 | 78 | 4.2493 | 18.0563 | 427.1127 | 12200.2585 |
| 67 | 81 | 4.4312 | 19.6353 | 512.1635 | 16414.2521 | 70 | 79 | 4.2029 | 17.6646 | 409.1645 | 11745.7944 |
| 67 | 82 | 4.9966 | 24.9662 | 801.5031 | 30306.1267 | 70 | 80 | 5.0692 | 25.6970 | 852.3860 | 33330.7878 |
| 67 | 83 | 4.9441 | 24.4444 | 765.1132 | 28783.2842 | 70 | 81 | 5.0104 | 25.1036 | 813.1646 | 31818.8368 |
| 67 | 84 | 5.0473 | 25.4754 | 839.8090 | 33194.8404 | 70 | 82 | 5.0484 | 25.4861 | 826.8454 | 30783.5654 |
| 67 | 85 | 5.0332 | 25.3330 | 819.3297 | 30853.1418 | 70 | 83 | 5.0559 | 25.5620 | 835.4637 | 31939.0901 |
| 67 | 86 | 5.0685 | 25.6902 | 848.9480 | 33037.7156 | 70 | 84 | 5.0918 | 25.9261 | 859.6198 | 32985.7086 |
| 67 | 87 | 5.0659 | 25.6637 | 844.8669 | 32723.4611 | 70 | 85 | 5.1003 | 26.0132 | 869.3813 | 34272.0906 |
| 67 | 88 | 5.1040 | 26.0510 | 878.0923 | 35308.3218 | 70 | 86 | 5.1274 | 26.2907 | 889.0549 | 35268.3676 |
| 67 | 89 | 5.0914 | 25.9221 | 865.2093 | 34300.0016 | 70 | 87 | 5.1361 | 26.3797 | 898.6815 | 36516.4754 |
| 67 | 90 | 5.1324 | 26.3413 | 901.9059 | 37258.3896 | 70 | 88 | 5.1621 | 26.6471 | 918.3112 | 37594.2380 |
| 67 | 91 | 5.1122 | 26.1346 | 881.8709 | 35620.2673 | 70 | 89 | 5.1699 | 26.7275 | 926.6073 | 38669.7695 |
| 67 | 92 | 5.1557 | 26.5816 | 921.5227 | 38900.4388 | 70 | 90 | 5.1937 | 26.9743 | 945.5333 | 39811.4118 |
| 67 | 93 | 5.1290 | 26.3064 | 895.3487 | 36727.4770 | 70 | 91 | 5.2001 | 27.0410 | 951.9182 | 40644.2486 |
| 67 | 94 | 5.1750 | 26.7807 | 937.6237 | 40268.5144 | 70 | 92 | 5.2210 | 27.2592 | 969.5275 | 41811.8690 |
| 67 | 95 | 5.1429 | 26.4490 | 906.6218 | 37692.4786 | 70 | 93 | 5.2262 | 27.3137 | 974.1063 | 42404.5765 |
| 67 | 96 | 5.1913 | 26.9493 | 951.0839 | 41415.3919 | 70 | 94 | 5.2439 | 27.4986 | 989.8705 | 43544.3967 |
| 67 | 97 | 5.1552 | 26.5758 | 916.8063 | 38593.2970 | 70 | 95 | 5.2485 | 27.5471 | 993.3008 | 43959.8009 |
| 67 | 98 | 5.2055 | 27.0969 | 962.7054 | 42397.2777 | 70 | 96 | 5.2626 | 27.6951 | 1006.6894 | 45005.0400 |
| 67 | 99 | 5.1670 | 26.6983 | 926.8656 | 39498.7809 | 70 | 97 | 5.2675 | 27.7467 | 1009.9718 | 45345.6985 |
| 67 | 100 | 5.2182 | 27.2300 | 973.1195 | 43267.1112 | 70 | 98 | 5.2777 | 27.8542 | 1020.3925 | 46217.7531 |
| 67 | 101 | 5.1793 | 26.8247 | 937.4976 | 40462.3289 | 70 | 99 | 5.2838 | 27.9190 | 1024.6734 | 46605.9849 |
| 67 | 102 | 5.2300 | 27.3530 | 982.7856 | 44071.0704 | 70 | 100 | 5.2898 | 27.9825 | 1031.5282 | 47223.0435 |
| 67 | 103 | 5.1922 | 26.9594 | 949.1244 | 41519.7758 | 70 | 101 | 5.2981 | 28.0702 | 1037.9399 | 47783.7478 |
| 67 | 104 | 5.2410 | 27.4685 | 992.0602 | 44850.3005 | 70 | 102 | 5.2997 | 28.0870 | 1040.7012 | 48069.7083 |
| 67 | 105 | 5.2081 | 27.1242 | 954.5628 | 41073.7300 | 70 | 103 | 5.3109 | 28.2059 | 1050.2626 | 48918.9263 |
| 67 | 106 | 5.2688 | 27.7602 | 1002.8244 | 44189.2014 | 70 | 104 | 5.3080 | 28.1748 | 1048.5320 | 48809.8408 |
| 67 | 107 | 5.2278 | 27.3302 | 956.1034 | 39469.0726 | 70 | 105 | 5.3212 | 28.3153 | 1052.9591 | 48480.0085 |
| 67 | 108 | 5.2956 | 28.0437 | 1015.0721 | 43892.7202 | 70 | 106 | 5.3265 | 28.3711 | 1054.7033 | 48216.4914 |
| 67 | 109 | 5.2485 | 27.5464 | 960.9313 | 38399.8541 | 70 | 107 | 5.3303 | 28.4119 | 1048.5274 | 46822.4545 |
| 67 | 110 | 5.3201 | 28.3037 | 1027.5839 | 43879.3343 | 70 | 108 | 5.3440 | 28.5588 | 1061.9383 | 47917.9099 |
| 67 | 111 | 5.2685 | 27.7574 | 967.9233 | 37802.5660 | 70 | 109 | 5.3396 | 28.5111 | 1046.6456 | 45630.3864 |
| 68 | 74 | 4.4244 | 19.5752 | 520.3700 | 17344.5603 | 70 | 110 | 5.3600 | 28.7291 | 1069.4949 | 47857.6400 |
| 68 | 75 | 4.9447 | 24.4504 | 784.6819 | 30262.3736 | 70 | 111 | 5.3487 | 28.6084 | 1046.9881 | 44886.7066 |
| 68 | 76 | 4.3678 | 19.0776 | 489.2141 | 15598.1275 | 70 | 112 | 5.3737 | 28.8764 | 1076.7976 | 47979.7109 |
| 68 | 77 | 4.9169 | 24.1762 | 764.3739 | 28907.2994 | 70 | 113 | 5.3572 | 28.7001 | 1049.1451 | 44550.3502 |
| 68 | 78 | 4.3123 | 18.5962 | 459.3115 | 13952.2040 | 70 | 114 | 5.3850 | 28.9982 | 1083.4660 | 48233.7967 |
| 68 | 79 | 4.8859 | 23.8725 | 742.6631 | 27500.7437 | 70 | 115 | 5.3650 | 28.7829 | 1052.6532 | 44560.8671 |
| 68 | 80 | 4.2611 | 18.1570 | 432.0554 | 12465.8531 | 71 | 79 | 5.1386 | 26.4050 | 917.6308 | 39453.6373 |
| 68 | 81 | 4.8526 | 23.5481 | 720.1211 | 26073.6430 | 71 | 80 | 5.0823 | 25.8296 | 866.3021 | 35023.0534 |
| 68 | 82 | 5.0607 | 25.6110 | 847.5219 | 33395.2717 | 71 | 81 | 5.1060 | 26.0710 | 880.8333 | 35931.4184 |
| 68 | 83 | 5.0406 | 25.4078 | 836.7453 | 33356.3788 | 71 | 82 | 5.0587 | 25.5909 | 838.1708 | 32203.1491 |
| 68 | 84 | 5.0635 | 25.6389 | 842.1537 | 32168.6792 | 71 | 83 | 5.1118 | 26.1310 | 880.2937 | 35457.1833 |
| 68 | 85 | 5.0611 | 25.6149 | 841.6284 | 32456.3742 | 71 | 84 | 5.1063 | 26.0741 | 872.6445 | 34367.8867 |
| 68 | 86 | 5.1057 | 26.0685 | 877.8267 | 34963.2013 | 71 | 85 | 5.1453 | 26.4737 | 907.9427 | 37586.4083 |
| 68 | 87 | 5.1000 | 26.0095 | 873.0510 | 34828.1919 | 71 | 86 | 5.1417 | 26.4375 | 901.0365 | 36488.0036 |
| 68 | 88 | 5.1418 | 26.4384 | 909.5244 | 37594.6489 | 71 | 87 | 5.1789 | 26.8213 | 936.2291 | 39761.5479 |
| 68 | 89 | 5.1318 | 26.3351 | 899.5926 | 36941.4936 | 71 | 88 | 5.1769 | 26.8003 | 929.7558 | 38656.0070 |
| 68 | 90 | 5.1734 | 26.7640 | 937.5390 | 39971.7604 | 71 | 89 | 5.2112 | 27.1566 | 963.7444 | 41884.2120 |
| 68 | 91 | 5.1591 | 26.6163 | 922.4360 | 38785.9661 | 71 | 90 | 5.2096 | 27.1404 | 957.0143 | 40743.4440 |
| 68 | 92 | 5.2001 | 27.0411 | 961.3779 | 42032.8377 | 71 | 91 | 5.2405 | 27.4624 | 989.1410 | 43873.1210 |
| 68 | 93 | 5.1822 | 26.8549 | 941.7524 | 40366.7367 | 71 | 92 | 5.2386 | 27.4425 | 981.5774 | 42663.7731 |
| 68 | 94 | 5.2223 | 27.2722 | 981.1442 | 43762.6283 | 71 | 93 | 5.2660 | 27.7310 | 1011.8263 | 45697.1672 |
| 68 | 95 | 5.2018 | 27.0584 | 958.1401 | 41720.1925 | 71 | 94 | 5.2633 | 27.7021 | 1003.0448 | 44387.5462 |
| 68 | 96 | 5.2405 | 27.4627 | 997.2269 | 45174.7373 | 71 | 95 | 5.2880 | 27.9634 | 1031.8818 | 47368.0973 |

TABLE 4. Charge radii, 2nd moment, 4th moment and 6th moment obtained by DNN.

| $Z$ | $N$ | $R_c$(fm) | $R_2$(fm$^2$) | $R_4$(fm$^4$) | $R_6$(fm$^6$) | $Z$ | $N$ | $R_c$(fm) | $R_2$(fm$^2$) | $R_4$(fm$^4$) | $R_6$(fm$^6$) |
|---|---|---|---|---|---|---|---|---|---|---|---|
| 71 | 96 | 5.2841 | 27.9219 | 1021.6153 | 45928.5930 | 74 | 93 | 5.3212 | 28.3153 | 1043.5119 | 46466.4436 |
| 71 | 97 | 5.3071 | 28.1654 | 1049.7670 | 48920.3938 | 74 | 94 | 5.2981 | 28.0699 | 1017.7883 | 43908.1437 |
| 71 | 98 | 5.3017 | 28.1080 | 1037.7748 | 47322.1521 | 74 | 95 | 5.3403 | 28.5190 | 1061.4535 | 48010.6778 |
| 71 | 99 | 5.3239 | 28.3438 | 1066.0298 | 50391.3877 | 74 | 96 | 5.3133 | 28.2310 | 1032.4968 | 45262.8726 |
| 71 | 100 | 5.3167 | 28.2674 | 1052.0834 | 48608.9766 | 74 | 97 | 5.3560 | 28.6870 | 1077.0425 | 49438.0246 |
| 71 | 101 | 5.3390 | 28.5050 | 1081.1636 | 51812.7240 | 74 | 98 | 5.3260 | 28.3661 | 1045.4810 | 46523.5261 |
| 71 | 102 | 5.3298 | 28.4071 | 1065.0993 | 49830.0348 | 74 | 99 | 5.3690 | 28.8262 | 1090.6929 | 50766.1446 |
| 71 | 103 | 5.3530 | 28.6542 | 1095.5718 | 53209.1232 | 74 | 100 | 5.3368 | 28.4819 | 1057.1329 | 47704.8349 |
| 71 | 104 | 5.3417 | 28.5334 | 1077.3189 | 51021.0134 | 74 | 101 | 5.3800 | 28.9446 | 1102.9456 | 52023.8871 |
| 71 | 105 | 5.3573 | 28.7002 | 1092.2524 | 52241.9643 | 74 | 102 | 5.3466 | 28.5866 | 1068.0075 | 48839.0227 |
| 71 | 106 | 5.3451 | 28.5702 | 1067.1669 | 48856.8880 | 74 | 103 | 5.3898 | 29.0502 | 1114.3710 | 53244.7545 |
| 71 | 107 | 5.3535 | 28.6601 | 1073.9577 | 49272.1409 | 74 | 104 | 5.3561 | 28.6881 | 1078.7128 | 49966.7142 |
| 71 | 108 | 5.3494 | 28.6161 | 1059.9530 | 47168.0768 | 74 | 105 | 5.3912 | 29.0655 | 1111.0226 | 52570.6741 |
| 71 | 109 | 5.3515 | 28.6383 | 1059.5051 | 46863.2870 | 74 | 106 | 5.3599 | 28.7285 | 1073.8099 | 48791.0057 |
| 71 | 110 | 5.3544 | 28.6699 | 1055.7213 | 45976.4916 | 74 | 107 | 5.3859 | 29.0080 | 1095.6410 | 50354.7362 |
| 71 | 111 | 5.3514 | 28.6371 | 1049.3416 | 45080.6616 | 74 | 108 | 5.3654 | 28.7873 | 1072.1112 | 48054.8606 |
| 71 | 112 | 5.3600 | 28.7294 | 1054.2866 | 45269.1202 | 74 | 109 | 5.3827 | 28.9740 | 1084.2025 | 48667.5531 |
| 71 | 113 | 5.3531 | 28.6558 | 1043.4466 | 43929.2469 | 74 | 110 | 5.3723 | 28.8617 | 1073.3388 | 47733.7850 |
| 71 | 114 | 5.3658 | 28.7915 | 1055.2642 | 45001.2278 | 74 | 111 | 5.3822 | 28.9678 | 1077.1567 | 47552.3151 |
| 71 | 115 | 5.3563 | 28.6896 | 1041.2989 | 43349.8115 | 74 | 112 | 5.3803 | 28.9473 | 1077.0348 | 47780.9270 |
| 71 | 116 | 5.3715 | 28.8527 | 1058.1215 | 45101.2917 | 74 | 113 | 5.3842 | 28.9894 | 1074.4619 | 47000.0201 |
| 71 | 117 | 5.3602 | 28.7314 | 1041.9554 | 43226.1656 | 74 | 114 | 5.3888 | 29.0390 | 1082.6234 | 48131.6227 |
| 72 | 81 | 5.1085 | 26.0970 | 879.8921 | 35303.8776 | 74 | 115 | 5.3884 | 29.0343 | 1075.5920 | 46948.8803 |
| 72 | 82 | 5.1058 | 26.0696 | 868.5973 | 33317.8800 | 74 | 116 | 5.3974 | 29.1319 | 1089.5061 | 48710.9087 |
| 72 | 83 | 5.1175 | 26.1890 | 881.0333 | 34882.3422 | 74 | 117 | 5.3940 | 29.0949 | 1079.6461 | 47293.2922 |
| 72 | 84 | 5.1335 | 26.3528 | 889.0744 | 34725.8266 | 74 | 118 | 5.4058 | 29.2231 | 1097.1778 | 49444.3576 |
| 72 | 85 | 5.1509 | 26.5318 | 906.8197 | 36730.2305 | 74 | 119 | 5.4003 | 29.1628 | 1085.5649 | 47903.7407 |
| 72 | 86 | 5.1636 | 26.6629 | 912.2470 | 36389.4517 | 74 | 120 | 5.4140 | 29.3119 | 1105.3003 | 50266.5333 |
| 72 | 87 | 5.1856 | 26.8904 | 934.4793 | 38756.0006 | 74 | 121 | 5.4066 | 29.2314 | 1092.4053 | 48654.0072 |
| 72 | 88 | 5.1944 | 26.9822 | 937.0098 | 38249.5351 | 74 | 122 | 5.4222 | 29.4000 | 1113.7279 | 51126.7158 |
| 72 | 89 | 5.2194 | 27.2417 | 962.1918 | 40830.9903 | 74 | 123 | 5.4128 | 29.2983 | 1099.5308 | 49441.7819 |
| 72 | 90 | 5.2235 | 27.2853 | 961.3739 | 40162.8105 | 75 | 84 | 5.2335 | 27.3895 | 969.7482 | 40699.1202 |
| 72 | 91 | 5.2501 | 27.5631 | 988.1217 | 42825.8382 | 75 | 85 | 5.2766 | 27.8429 | 1009.4410 | 44203.8852 |
| 72 | 92 | 5.2493 | 27.5547 | 983.7523 | 41994.8895 | 75 | 86 | 5.2593 | 27.6607 | 989.0305 | 41972.1152 |
| 72 | 93 | 5.2766 | 27.8430 | 1011.2591 | 44665.3244 | 75 | 87 | 5.3020 | 28.1117 | 1030.5254 | 45738.9362 |
| 72 | 94 | 5.2710 | 27.7830 | 1003.2996 | 43660.6499 | 75 | 88 | 5.2849 | 27.9297 | 1009.0088 | 43368.7149 |
| 72 | 95 | 5.2990 | 28.0796 | 1031.3749 | 46328.5595 | 75 | 89 | 5.3257 | 28.3627 | 1050.9829 | 47303.5099 |
| 72 | 96 | 5.2887 | 27.9706 | 1019.8547 | 45128.9759 | 75 | 90 | 5.3086 | 28.1812 | 1028.6564 | 44834.8241 |
| 72 | 97 | 5.3176 | 28.2774 | 1048.7559 | 47831.9173 | 75 | 91 | 5.3466 | 28.5866 | 1070.0219 | 48840.2380 |
| 72 | 98 | 5.3031 | 28.1227 | 1033.7056 | 46407.7794 | 75 | 92 | 5.3295 | 28.4040 | 1047.0148 | 46300.1140 |
| 72 | 99 | 5.3332 | 28.4434 | 1063.9222 | 49208.9807 | 75 | 93 | 5.3646 | 28.7791 | 1087.1657 | 50307.4620 |
| 72 | 100 | 5.3148 | 28.2470 | 1045.3886 | 47529.2659 | 75 | 94 | 5.3473 | 28.5933 | 1063.5215 | 47711.7545 |
| 72 | 101 | 5.3465 | 28.5855 | 1077.4540 | 50497.9901 | 75 | 95 | 5.3797 | 28.9409 | 1102.3319 | 51689.8400 |
| 72 | 102 | 5.3246 | 28.3518 | 1055.5379 | 48537.1966 | 75 | 96 | 5.3619 | 28.7505 | 1078.0798 | 49047.4731 |
| 72 | 103 | 5.3583 | 28.7109 | 1089.9003 | 51735.0036 | 75 | 97 | 5.3923 | 29.0764 | 1115.7570 | 52994.6544 |
| 72 | 104 | 5.3334 | 28.4452 | 1064.8181 | 49481.1941 | 75 | 98 | 5.3741 | 28.8808 | 1090.9458 | 50311.0586 |
| 72 | 105 | 5.3632 | 28.7635 | 1088.7715 | 51115.8288 | 75 | 99 | 5.4029 | 29.1918 | 1127.8539 | 54242.0126 |
| 72 | 106 | 5.3431 | 28.5488 | 1064.1568 | 48489.5897 | 75 | 100 | 5.3844 | 28.9915 | 1102.5782 | 51522.9972 |
| 72 | 107 | 5.3629 | 28.7602 | 1076.8031 | 49002.7246 | 75 | 101 | 5.4124 | 29.2941 | 1139.1027 | 55457.4872 |
| 72 | 108 | 5.3533 | 28.6582 | 1065.7301 | 47874.9905 | 75 | 102 | 5.3936 | 29.0908 | 1113.5262 | 52713.2343 |
| 72 | 109 | 5.3638 | 28.7709 | 1068.1724 | 47395.6422 | 75 | 103 | 5.4212 | 29.3897 | 1149.9759 | 56667.3312 |
| 72 | 110 | 5.3636 | 28.7683 | 1069.0744 | 47601.1094 | 75 | 104 | 5.4024 | 29.1861 | 1124.3455 | 53915.0173 |
| 72 | 111 | 5.3662 | 28.7963 | 1063.0390 | 46317.6417 | 75 | 105 | 5.4193 | 29.3683 | 1142.7722 | 55634.2948 |
| 72 | 112 | 5.3735 | 28.8741 | 1073.6894 | 47618.8028 | 75 | 106 | 5.3949 | 29.1053 | 1106.6635 | 51499.0150 |
| 72 | 113 | 5.3698 | 28.8350 | 1061.2304 | 45750.2496 | 75 | 107 | 5.4080 | 29.2470 | 1120.1700 | 52698.3347 |
| 72 | 114 | 5.3825 | 28.9717 | 1079.1017 | 47871.5918 | 75 | 108 | 5.3897 | 29.0486 | 1092.8396 | 49586.4183 |
| 72 | 115 | 5.3743 | 28.8833 | 1062.2691 | 45635.0771 | 75 | 109 | 5.3998 | 29.1580 | 1102.2064 | 50334.9890 |
| 72 | 116 | 5.3907 | 29.0592 | 1084.9364 | 48302.4833 | 75 | 110 | 5.3873 | 29.0231 | 1083.5750 | 48243.0859 |
| 72 | 117 | 5.3793 | 28.9364 | 1065.4769 | 45883.0105 | 75 | 111 | 5.3954 | 29.1104 | 1089.7749 | 48625.8646 |
| 72 | 118 | 5.3979 | 29.1370 | 1090.9910 | 48861.7158 | 75 | 112 | 5.3881 | 29.0316 | 1079.1012 | 47486.2374 |
| 73 | 82 | 5.1348 | 26.3665 | 895.2070 | 35866.4583 | 75 | 113 | 5.3950 | 29.1065 | 1083.0375 | 47577.9192 |
| 73 | 83 | 5.1799 | 26.8311 | 933.9607 | 39231.7559 | 75 | 114 | 5.3918 | 29.0716 | 1079.1044 | 47277.0155 |
| 73 | 84 | 5.1649 | 26.6762 | 917.4244 | 37369.0291 | 75 | 115 | 5.3982 | 29.1404 | 1081.3578 | 47118.4539 |
| 73 | 85 | 5.2115 | 27.1599 | 959.7623 | 41135.9867 | 75 | 116 | 5.3977 | 29.1356 | 1082.7668 | 47523.0909 |
| 73 | 86 | 5.1971 | 27.0103 | 942.1913 | 39101.9195 | 75 | 117 | 5.4037 | 29.2000 | 1083.4430 | 47108.7298 |
| 73 | 87 | 5.2437 | 27.4968 | 986.7830 | 43161.6542 | 75 | 118 | 5.4050 | 29.2140 | 1088.9755 | 48097.7787 |
| 73 | 88 | 5.2299 | 27.3522 | 968.4046 | 41000.2146 | 75 | 119 | 5.4102 | 29.2706 | 1087.6989 | 47376.6766 |
| 73 | 89 | 5.2746 | 27.8212 | 1013.4066 | 45203.2709 | 75 | 120 | 5.4127 | 29.2972 | 1096.5630 | 48861.6741 |
| 73 | 90 | 5.2608 | 27.6762 | 994.0862 | 42931.9315 | 75 | 121 | 5.4166 | 29.3400 | 1092.6815 | 47758.4386 |
| 73 | 91 | 5.3025 | 28.1162 | 1038.2278 | 47163.9549 | 75 | 122 | 5.4202 | 29.3788 | 1104.5865 | 49690.0727 |
| 73 | 92 | 5.2882 | 27.9649 | 1017.7355 | 44785.9153 | 75 | 123 | 5.4223 | 29.4015 | 1097.4158 | 48128.1442 |
| 73 | 93 | 5.3267 | 28.3733 | 1060.4740 | 48989.0605 | 75 | 124 | 5.4274 | 29.4572 | 1112.5096 | 50493.1474 |
| 73 | 94 | 5.3114 | 28.2110 | 1038.6331 | 46501.5935 | 76 | 85 | 5.2787 | 27.8650 | 1002.2373 | 42549.6985 |
| 73 | 95 | 5.3471 | 28.5917 | 1079.9832 | 50664.3674 | 76 | 86 | 5.2655 | 27.7257 | 982.2628 | 40062.7813 |
| 73 | 96 | 5.3307 | 28.4160 | 1056.7438 | 48067.1889 | 76 | 87 | 5.3004 | 28.0941 | 1018.8101 | 43671.9670 |
| 73 | 97 | 5.3643 | 28.7755 | 1097.0362 | 52205.5413 | 76 | 88 | 5.2821 | 27.9004 | 994.8838 | 40973.0622 |
| 73 | 98 | 5.3466 | 28.5858 | 1072.4296 | 49499.7211 | 76 | 89 | 5.3204 | 28.3067 | 1034.9577 | 44846.0272 |
| 73 | 99 | 5.3788 | 28.9316 | 1112.1121 | 53640.5816 | 76 | 90 | 5.2974 | 28.0622 | 1007.3042 | 41943.8740 |
| 73 | 100 | 5.3599 | 28.7281 | 1086.2221 | 50829.3465 | 76 | 91 | 5.3381 | 28.4956 | 1050.1872 | 46044.0694 |
| 73 | 101 | 5.3914 | 29.0671 | 1125.7299 | 54999.3278 | 76 | 92 | 5.3112 | 28.2090 | 1019.3970 | 42971.2112 |
| 73 | 102 | 5.3713 | 28.8510 | 1098.7060 | 52091.7288 | 76 | 93 | 5.3533 | 28.6575 | 1064.1272 | 47233.4786 |
| 73 | 103 | 5.4027 | 29.1887 | 1138.3784 | 56309.8268 | 76 | 94 | 5.3235 | 28.3398 | 1031.0209 | 44038.2264 |
| 73 | 104 | 5.3816 | 28.9619 | 1110.4204 | 53320.1089 | 76 | 95 | 5.3659 | 28.7933 | 1076.6803 | 48395.0209 |
| 73 | 105 | 5.4029 | 29.1910 | 1132.3536 | 55258.6452 | 76 | 96 | 5.3345 | 28.4566 | 1042.1601 | 45130.4809 |
| 73 | 106 | 5.3776 | 28.9187 | 1094.6790 | 50884.8602 | 76 | 97 | 5.3765 | 28.9072 | 1088.0287 | 49528.0828 |
| 73 | 107 | 5.3936 | 29.0910 | 1110.3641 | 52196.0336 | 76 | 98 | 5.3445 | 28.5635 | 1052.9969 | 46246.3932 |
| 73 | 108 | 5.3751 | 28.8914 | 1082.2696 | 48932.8087 | 76 | 99 | 5.3857 | 29.0060 | 1098.5677 | 50648.5392 |
| 73 | 109 | 5.3865 | 29.0143 | 1092.4304 | 49685.5360 | 76 | 100 | 5.3541 | 28.6669 | 1063.8899 | 47399.3170 |
| 73 | 110 | 5.3745 | 28.8848 | 1073.7213 | 47521.8175 | 76 | 101 | 5.3942 | 29.0971 | 1108.8065 | 51782.7823 |
| 73 | 111 | 5.3823 | 28.9689 | 1079.4291 | 47817.7869 | 76 | 102 | 5.3641 | 28.7736 | 1075.2931 | 48612.6266 |
| 73 | 112 | 5.3759 | 28.9001 | 1069.1661 | 46665.1423 | 76 | 103 | 5.4026 | 29.1881 | 1119.2860 | 52962.1755 |
| 73 | 113 | 5.3812 | 28.9573 | 1071.5928 | 46612.9081 | 76 | 104 | 5.3749 | 28.8900 | 1087.6683 | 49913.4514 |
| 73 | 114 | 5.3791 | 28.9348 | 1068.3074 | 46327.0854 | 76 | 105 | 5.4026 | 29.1881 | 1115.4073 | 52344.5246 |
| 73 | 115 | 5.3828 | 28.9751 | 1068.4099 | 46010.4845 | 76 | 106 | 5.3751 | 28.8917 | 1080.5570 | 48700.9344 |
| 73 | 116 | 5.3836 | 28.9837 | 1070.4855 | 46427.8549 | 76 | 107 | 5.3955 | 29.1116 | 1099.5433 | 50241.0578 |
| 73 | 117 | 5.3863 | 29.0120 | 1068.7326 | 45879.6238 | 76 | 108 | 5.3779 | 28.9214 | 1077.3649 | 47968.7935 |
| 73 | 118 | 5.3889 | 29.0400 | 1074.8255 | 46856.7881 | 76 | 109 | 5.3914 | 29.0676 | 1088.1993 | 48687.1968 |
| 73 | 119 | 5.3904 | 29.0560 | 1071.1104 | 46051.1492 | 76 | 110 | 5.3831 | 28.9779 | 1077.9233 | 47697.6945 |
| 73 | 120 | 5.3943 | 29.0983 | 1080.4625 | 47494.6078 | 76 | 111 | 5.3908 | 29.0610 | 1081.8495 | 47722.7110 |
| 73 | 121 | 5.3941 | 29.0968 | 1074.2103 | 46359.0374 | 76 | 112 | 5.3905 | 29.0571 | 1081.7794 | 47838.7055 |
| 74 | 83 | 5.1845 | 26.8794 | 933.1303 | 38395.1251 | 76 | 113 | 5.3937 | 29.0920 | 1080.4305 | 47333.3651 |
| 74 | 84 | 5.1895 | 26.9305 | 929.8201 | 37179.5320 | 76 | 114 | 5.3993 | 29.1527 | 1088.2580 | 48318.0132 |
| 74 | 85 | 5.2130 | 27.1753 | 954.1350 | 39787.0016 | 76 | 115 | 5.3995 | 29.1550 | 1083.3295 | 47449.6937 |
| 74 | 86 | 5.2125 | 27.1701 | 946.4815 | 38288.6502 | 76 | 116 | 5.4091 | 29.2580 | 1096.5841 | 49047.0657 |
| 74 | 87 | 5.2428 | 27.4871 | 977.1155 | 41375.3029 | 76 | 117 | 5.4074 | 29.2403 | 1089.4972 | 47956.7198 |
| 74 | 88 | 5.2362 | 27.4182 | 964.6092 | 39577.1529 | 76 | 118 | 5.4191 | 29.3668 | 1106.0096 | 49933.5773 |
| 74 | 89 | 5.2719 | 27.7925 | 1000.6215 | 43085.6240 | 76 | 119 | 5.4164 | 29.3369 | 1097.6813 | 48714.6943 |
| 74 | 90 | 5.2592 | 27.6591 | 983.2504 | 41000.7528 | 76 | 120 | 5.4291 | 29.4751 | 1115.9185 | 50891.6747 |
| 74 | 91 | 5.2983 | 28.0723 | 1023.1032 | 44810.8248 | 76 | 121 | 5.4254 | 29.4351 | 1106.7138 | 49585.3860 |
| 74 | 92 | 5.2800 | 27.8789 | 1001.2586 | 42471.5603 | 76 | 122 | 5.4389 | 29.5815 | 1125.9016 | 51851.1832 |

TABLE 4. Charge radii, 2nd moment, 4th moment and 6th moment obtained by DNN.

| $Z$ | $N$ | $R_c$(fm) | $R_2$(fm$^2$) | $R_4$(fm$^4$) | $R_6$(fm$^6$) | $Z$ | $N$ | $R_c$(fm) | $R_2$(fm$^2$) | $R_4$(fm$^4$) | $R_6$(fm$^6$) |
|---|---|---|---|---|---|---|---|---|---|---|---|
| 76 | 123 | 5.4341 | 29.5295 | 1115.7361 | 50453.9622 | 79 | 107 | 5.4134 | 29.3046 | 1107.4789 | 50150.9281 |
| 76 | 124 | 5.4486 | 29.6868 | 1135.7338 | 52758.5805 | 79 | 108 | 5.4124 | 29.2936 | 1097.2533 | 48456.6810 |
| 76 | 125 | 5.4424 | 29.6194 | 1124.3005 | 51241.4286 | 79 | 109 | 5.4101 | 29.2693 | 1095.0566 | 48351.7536 |
| 76 | 126 | 5.4583 | 29.7930 | 1145.2865 | 53571.9470 | 79 | 110 | 5.4124 | 29.2939 | 1090.9746 | 47439.2793 |
| 76 | 127 | 5.4614 | 29.8267 | 1142.1551 | 52681.5992 | 79 | 111 | 5.4102 | 29.2703 | 1087.2693 | 47064.6857 |
| 77 | 86 | 5.3155 | 28.2544 | 1026.5189 | 43663.5807 | 79 | 112 | 5.4150 | 29.3223 | 1088.6431 | 46887.2751 |
| 77 | 87 | 5.3459 | 28.5782 | 1056.2658 | 46396.1331 | 79 | 113 | 5.4135 | 29.3060 | 1083.9996 | 46278.7231 |
| 77 | 88 | 5.3323 | 28.4333 | 1039.8162 | 44609.8640 | 79 | 114 | 5.4202 | 29.3782 | 1090.1616 | 46783.1004 |
| 77 | 89 | 5.3601 | 28.7303 | 1068.8852 | 47406.4343 | 79 | 115 | 5.4195 | 29.3707 | 1084.7320 | 45942.2169 |
| 77 | 90 | 5.3468 | 28.5883 | 1052.0677 | 45564.8934 | 79 | 116 | 5.4275 | 29.4579 | 1095.0671 | 47070.8800 |
| 77 | 91 | 5.3721 | 28.8597 | 1080.4590 | 48424.1694 | 79 | 117 | 5.4274 | 29.4566 | 1088.6357 | 45969.4300 |
| 77 | 92 | 5.3591 | 28.7194 | 1063.2509 | 46525.3503 | 79 | 118 | 5.4364 | 29.5548 | 1102.5913 | 47661.8791 |
| 77 | 93 | 5.3823 | 28.9690 | 1091.1111 | 49449.0020 | 79 | 119 | 5.4364 | 29.5545 | 1094.6850 | 46249.7311 |
| 77 | 94 | 5.3693 | 28.8294 | 1073.4626 | 47489.8951 | 79 | 120 | 5.4462 | 29.6614 | 1111.7917 | 48445.9371 |
| 77 | 95 | 5.3910 | 29.0628 | 1101.0706 | 50486.3596 | 79 | 121 | 5.4458 | 29.6564 | 1101.8458 | 46663.8927 |
| 77 | 96 | 5.3780 | 28.9233 | 1082.9762 | 48467.6969 | 79 | 122 | 5.4563 | 29.7708 | 1121.7026 | 49304.6267 |
| 77 | 97 | 5.3987 | 29.1464 | 1110.6629 | 51549.0154 | 79 | 123 | 5.4549 | 29.7563 | 1109.1961 | 47094.6816 |
| 77 | 98 | 5.3859 | 29.0077 | 1092.2148 | 49478.8835 | 79 | 124 | 5.4661 | 29.8781 | 1131.4957 | 50126.8220 |
| 77 | 99 | 5.4061 | 29.2258 | 1120.2606 | 52653.8927 | 79 | 125 | 5.4637 | 29.8517 | 1116.0510 | 47441.0079 |
| 77 | 100 | 5.3935 | 29.0900 | 1101.6856 | 50550.7960 | 79 | 126 | 5.4755 | 29.9813 | 1140.5547 | 50818.2530 |
| 77 | 101 | 5.4135 | 29.3065 | 1130.2358 | 53818.8860 | 79 | 127 | 5.4792 | 30.0221 | 1129.5982 | 48370.0509 |
| 77 | 102 | 5.4016 | 29.1777 | 1111.9048 | 51712.8909 | 79 | 128 | 5.5015 | 30.2665 | 1164.1157 | 52605.8121 |
| 77 | 103 | 5.4216 | 29.3934 | 1140.9225 | 55060.3508 | 79 | 129 | 5.5017 | 30.2686 | 1150.1398 | 49937.7989 |
| 77 | 104 | 5.4108 | 29.2769 | 1123.3330 | 52992.2217 | 79 | 130 | 5.5266 | 30.5437 | 1186.6064 | 54241.8980 |
| 77 | 105 | 5.4195 | 29.3709 | 1134.4134 | 54175.8613 | 79 | 131 | 5.5234 | 30.5082 | 1169.8794 | 51403.5485 |
| 77 | 106 | 5.4025 | 29.1866 | 1105.9597 | 50741.5149 | 80 | 90 | 5.3128 | 28.2259 | 996.5292 | 38813.8290 |
| 77 | 107 | 5.4087 | 29.2539 | 1113.0663 | 51464.7509 | 80 | 91 | 5.3434 | 28.5522 | 1029.2009 | 41945.2026 |
| 77 | 108 | 5.3968 | 29.1258 | 1092.7435 | 48995.7095 | 80 | 92 | 5.3223 | 28.3273 | 1005.8722 | 39714.3635 |
| 77 | 109 | 5.4015 | 29.1757 | 1096.7219 | 49325.2973 | 80 | 93 | 5.3505 | 28.6282 | 1037.2022 | 42806.9415 |
| 77 | 110 | 5.3946 | 29.1018 | 1084.3691 | 47813.9240 | 80 | 94 | 5.3329 | 28.4399 | 1016.9220 | 40817.5299 |
| 77 | 111 | 5.3985 | 29.1433 | 1086.0466 | 47815.2579 | 80 | 95 | 5.3588 | 28.7165 | 1046.9418 | 43870.9138 |
| 77 | 112 | 5.3961 | 29.1177 | 1081.0621 | 47209.0697 | 80 | 96 | 5.3450 | 28.5692 | 1030.0527 | 42144.0021 |
| 77 | 113 | 5.3997 | 29.1571 | 1081.0371 | 46926.0699 | 80 | 97 | 5.3687 | 28.8225 | 1058.7557 | 45152.1584 |
| 77 | 114 | 5.4010 | 29.1708 | 1082.4954 | 47139.6526 | 80 | 98 | 5.3591 | 28.7197 | 1045.5476 | 43707.5753 |
| 77 | 115 | 5.4047 | 29.2105 | 1081.0271 | 46585.2267 | 80 | 99 | 5.3805 | 28.9501 | 1072.8721 | 46658.3645 |
| 77 | 116 | 5.4087 | 29.2536 | 1087.8440 | 47514.9602 | 80 | 100 | 5.3753 | 28.8939 | 1063.5157 | 45508.0303 |
| 77 | 117 | 5.4122 | 29.2922 | 1084.8147 | 46668.0177 | 80 | 101 | 5.3946 | 29.1013 | 1089.3702 | 48386.1067 |
| 77 | 118 | 5.4181 | 29.3553 | 1095.9420 | 48208.8553 | 80 | 102 | 5.3937 | 29.0915 | 1083.8520 | 47528.3452 |
| 77 | 119 | 5.4211 | 29.3885 | 1090.9484 | 47022.7568 | 80 | 103 | 5.4108 | 29.2765 | 1108.1795 | 50320.8013 |
| 77 | 120 | 5.4282 | 29.4650 | 1105.5416 | 49082.3609 | 80 | 104 | 5.4139 | 29.3099 | 1106.2394 | 49735.0079 |
| 77 | 121 | 5.4302 | 29.4875 | 1098.0683 | 47500.8072 | 80 | 105 | 5.4190 | 29.3656 | 1113.1649 | 50531.2012 |
| 77 | 122 | 5.4382 | 29.5739 | 1115.5454 | 50004.3879 | 80 | 106 | 5.4171 | 29.3446 | 1102.2543 | 48815.4347 |
| 77 | 123 | 5.4389 | 29.5816 | 1105.1279 | 47975.6036 | 80 | 107 | 5.4189 | 29.3641 | 1104.6597 | 49120.6435 |
| 77 | 124 | 5.4477 | 29.6778 | 1125.1955 | 50870.5564 | 80 | 108 | 5.4213 | 29.3908 | 1100.4884 | 48201.9024 |
| 77 | 125 | 5.4469 | 29.6685 | 1111.5235 | 48356.4512 | 80 | 109 | 5.4204 | 29.3810 | 1098.9856 | 48061.6292 |
| 77 | 126 | 5.4568 | 29.7764 | 1134.1039 | 51609.2376 | 80 | 110 | 5.4267 | 29.4490 | 1101.0175 | 47901.2174 |
| 77 | 127 | 5.4632 | 29.8469 | 1126.1105 | 49429.6416 | 80 | 111 | 5.4238 | 29.4175 | 1096.3388 | 47374.8843 |
| 77 | 128 | 5.4863 | 30.0994 | 1161.1043 | 53699.3068 | 80 | 112 | 5.4331 | 29.5190 | 1103.8230 | 47906.5178 |
| 78 | 87 | 5.3389 | 28.5039 | 1038.8066 | 43824.1502 | 80 | 113 | 5.4290 | 29.4736 | 1096.7543 | 47062.2841 |
| 78 | 88 | 5.3127 | 28.2249 | 1008.9559 | 40845.4939 | 80 | 114 | 5.4406 | 29.6000 | 1108.7585 | 48194.4345 |
| 78 | 89 | 5.3511 | 28.6343 | 1048.8760 | 44597.5924 | 80 | 115 | 5.4358 | 29.5477 | 1100.0632 | 47102.3106 |
| 78 | 90 | 5.3239 | 28.3442 | 1018.3725 | 41625.7191 | 80 | 116 | 5.4489 | 29.6900 | 1115.5346 | 48723.1952 |
| 78 | 91 | 5.3612 | 28.7429 | 1058.0135 | 45387.2299 | 80 | 117 | 5.4439 | 29.6365 | 1105.8645 | 47446.6402 |
| 78 | 92 | 5.3338 | 28.4496 | 1027.4164 | 42447.7496 | 80 | 118 | 5.4577 | 29.7863 | 1123.7208 | 49432.6347 |
| 78 | 93 | 5.3697 | 28.8341 | 1066.5103 | 46207.3334 | 80 | 119 | 5.4530 | 29.7350 | 1113.5431 | 48020.4744 |
| 78 | 94 | 5.3429 | 28.5463 | 1036.4251 | 43331.7779 | 80 | 120 | 5.4668 | 29.8862 | 1132.8056 | 50249.8050 |
| 78 | 95 | 5.3772 | 28.9141 | 1074.7539 | 47077.3308 | 80 | 121 | 5.4624 | 29.8383 | 1122.3564 | 48730.0080 |
| 78 | 96 | 5.3517 | 28.6407 | 1045.8443 | 44304.2184 | 80 | 122 | 5.4760 | 29.9869 | 1142.2454 | 51094.3592 |
| 78 | 97 | 5.3842 | 28.9900 | 1083.2264 | 48023.6504 | 80 | 123 | 5.4719 | 29.9418 | 1131.5330 | 49471.5205 |
| 78 | 98 | 5.3610 | 28.7401 | 1056.1928 | 45396.4079 | 80 | 124 | 5.4852 | 30.0869 | 1151.5298 | 51886.3332 |
| 78 | 99 | 5.3916 | 29.0694 | 1092.4498 | 49076.2101 | 80 | 125 | 5.4811 | 30.0428 | 1140.3766 | 50142.7273 |
| 78 | 100 | 5.3714 | 28.8517 | 1067.9801 | 46638.6883 | 80 | 126 | 5.4941 | 30.1853 | 1160.1954 | 52550.1182 |
| 78 | 101 | 5.3999 | 29.1592 | 1102.9226 | 50263.8988 | 80 | 127 | 5.4976 | 30.2231 | 1155.2023 | 51227.9179 |
| 78 | 102 | 5.3834 | 28.9815 | 1081.6160 | 48054.1521 | 80 | 128 | 5.5199 | 30.4693 | 1183.7369 | 54368.3684 |
| 78 | 103 | 5.4097 | 29.2651 | 1115.0579 | 51610.1078 | 80 | 129 | 5.5208 | 30.4797 | 1175.9525 | 52751.4302 |
| 78 | 104 | 5.3975 | 29.1331 | 1097.3426 | 49653.8065 | 80 | 130 | 5.5450 | 30.7469 | 1206.6137 | 56093.8322 |
| 78 | 105 | 5.4115 | 29.2843 | 1113.4467 | 51244.1257 | 80 | 131 | 5.5431 | 30.7256 | 1195.6153 | 54153.2991 |
| 78 | 106 | 5.3968 | 29.1250 | 1089.9130 | 48482.2852 | 80 | 132 | 5.5689 | 31.0125 | 1228.4049 | 57703.8421 |
| 78 | 107 | 5.4055 | 29.2196 | 1099.3222 | 49369.9346 | 80 | 133 | 5.5640 | 30.9583 | 1214.0592 | 55432.8853 |
| 78 | 108 | 5.3982 | 29.1406 | 1086.0269 | 47753.1485 | 80 | 134 | 5.5912 | 31.2619 | 1248.8009 | 59183.2806 |
| 78 | 109 | 5.4023 | 29.1848 | 1089.3752 | 47994.8211 | 80 | 135 | 5.5836 | 31.1771 | 1231.2538 | 56596.6861 |
| 78 | 110 | 5.4020 | 29.1817 | 1085.7627 | 47464.6203 | 80 | 136 | 5.6118 | 31.4925 | 1267.5989 | 60524.1340 |
| 78 | 111 | 5.4023 | 29.1853 | 1084.0710 | 47155.5010 | 81 | 95 | 5.3437 | 28.5549 | 1036.5250 | 43515.8500 |
| 78 | 112 | 5.4080 | 29.2470 | 1088.8933 | 47583.4456 | 81 | 96 | 5.3444 | 28.5622 | 1031.8549 | 42551.5262 |
| 78 | 113 | 5.4057 | 29.2220 | 1083.4317 | 46843.9443 | 81 | 97 | 5.3561 | 28.6875 | 1051.5751 | 45124.0496 |
| 78 | 114 | 5.4160 | 29.3327 | 1094.9193 | 48048.6836 | 81 | 98 | 5.3589 | 28.7176 | 1048.7700 | 44312.1744 |
| 78 | 115 | 5.4121 | 29.2911 | 1086.9951 | 47004.1335 | 81 | 99 | 5.3706 | 28.8434 | 1068.8764 | 46932.5329 |
| 78 | 116 | 5.4252 | 29.4330 | 1103.1541 | 48779.2140 | 81 | 100 | 5.3761 | 28.9021 | 1068.5549 | 46334.1622 |
| 78 | 117 | 5.4208 | 29.3849 | 1093.8931 | 47539.8655 | 81 | 101 | 5.3871 | 29.0205 | 1088.1541 | 48911.0668 |
| 78 | 118 | 5.4352 | 29.5417 | 1112.8295 | 49682.5368 | 81 | 102 | 5.3956 | 29.1129 | 1090.8549 | 48578.6367 |
| 78 | 119 | 5.4308 | 29.4934 | 1103.0077 | 48327.7701 | 81 | 103 | 5.4052 | 29.2160 | 1109.0904 | 51027.4388 |
| 78 | 120 | 5.4455 | 29.6535 | 1123.2206 | 50666.1051 | 81 | 104 | 5.4172 | 29.3456 | 1115.1806 | 50996.8541 |
| 78 | 121 | 5.4412 | 29.6071 | 1113.1884 | 49236.1565 | 81 | 105 | 5.4138 | 29.3087 | 1113.9140 | 51173.6451 |
| 78 | 122 | 5.4557 | 29.7649 | 1133.7328 | 51646.0200 | 81 | 106 | 5.4190 | 29.3655 | 1108.4996 | 49714.9813 |
| 78 | 123 | 5.4515 | 29.7191 | 1123.4682 | 50144.8754 | 81 | 107 | 5.4137 | 29.3080 | 1104.1369 | 49534.0050 |
| 78 | 124 | 5.4657 | 29.8744 | 1143.9283 | 52551.2961 | 81 | 108 | 5.4221 | 29.3996 | 1104.0718 | 48713.7975 |
| 78 | 125 | 5.4613 | 29.8260 | 1133.1653 | 50955.9138 | 81 | 109 | 5.4162 | 29.3351 | 1097.7779 | 48259.9213 |
| 78 | 126 | 5.4756 | 29.9820 | 1153.4952 | 53323.9628 | 81 | 110 | 5.4266 | 29.4480 | 1101.9843 | 48006.7788 |
| 78 | 127 | 5.4801 | 30.0316 | 1150.5839 | 52306.7065 | 81 | 111 | 5.4210 | 29.3869 | 1094.6892 | 47346.6291 |
| 78 | 128 | 5.5065 | 30.3210 | 1182.1796 | 55582.2970 | 81 | 112 | 5.4324 | 29.5105 | 1102.2716 | 47599.4619 |
| 78 | 129 | 5.5079 | 30.3373 | 1175.7354 | 54187.6516 | 81 | 113 | 5.4276 | 29.4589 | 1094.5450 | 46771.2793 |
| 78 | 130 | 5.5369 | 30.6568 | 1210.0654 | 57684.4825 | 81 | 114 | 5.4393 | 29.5862 | 1104.8812 | 47485.4573 |
| 79 | 89 | 5.3596 | 28.7250 | 1053.2621 | 44777.1576 | 81 | 115 | 5.4356 | 29.5455 | 1096.8803 | 46494.0959 |
| 79 | 90 | 5.3544 | 28.6693 | 1044.7123 | 43638.5218 | 81 | 116 | 5.4473 | 29.6734 | 1109.6412 | 47642.7786 |
| 79 | 91 | 5.3667 | 28.8019 | 1061.0547 | 45577.1721 | 81 | 117 | 5.4444 | 29.6410 | 1101.1412 | 46460.3451 |
| 79 | 92 | 5.3619 | 28.7499 | 1052.2405 | 44378.5575 | 81 | 118 | 5.4562 | 29.7697 | 1116.2084 | 48027.6662 |
| 79 | 93 | 5.3737 | 28.8762 | 1069.3583 | 46488.3621 | 81 | 119 | 5.4535 | 29.7402 | 1106.7087 | 46600.0648 |
| 79 | 94 | 5.3690 | 28.8262 | 1060.1465 | 45220.9057 | 81 | 120 | 5.4655 | 29.8716 | 1124.0844 | 48575.1335 |
| 79 | 95 | 5.3808 | 28.9530 | 1078.4600 | 47519.7428 | 81 | 121 | 5.4625 | 29.8392 | 1112.9540 | 46832.1959 |
| 79 | 96 | 5.3764 | 28.9052 | 1068.9018 | 46191.4583 | 81 | 122 | 5.4750 | 29.9756 | 1132.6564 | 49202.1568 |
| 79 | 97 | 5.3885 | 29.0363 | 1088.5795 | 48676.5475 | 81 | 123 | 5.4713 | 29.9356 | 1119.2813 | 47069.3308 |
| 79 | 98 | 5.3846 | 28.9934 | 1078.9637 | 47315.5991 | 81 | 124 | 5.4844 | 30.0788 | 1141.2643 | 49814.1712 |
| 79 | 99 | 5.3972 | 29.1293 | 1099.8822 | 49961.1751 | 81 | 125 | 5.4798 | 30.0286 | 1125.1519 | 47222.5829 |
| 79 | 100 | 5.3942 | 29.0969 | 1090.7417 | 48615.9839 | 81 | 126 | 5.4936 | 30.1794 | 1149.2864 | 50315.6134 |
| 79 | 101 | 5.4069 | 29.2348 | 1112.4993 | 51374.9844 | 81 | 127 | 5.4941 | 30.1849 | 1136.8973 | 47934.8219 |
| 79 | 102 | 5.4056 | 29.2200 | 1104.5335 | 50107.0475 | 81 | 128 | 5.5153 | 30.4186 | 1168.4815 | 51727.8522 |
| 79 | 103 | 5.4180 | 29.3546 | 1126.5108 | 52916.2008 | 81 | 129 | 5.5139 | 30.4031 | 1154.7354 | 49277.0783 |
| 79 | 104 | 5.4190 | 29.3655 | 1120.4823 | 51791.7034 | 81 | 130 | 5.5361 | 30.6479 | 1186.6964 | 53036.7435 |
| 79 | 105 | 5.4198 | 29.3742 | 1124.2374 | 52431.6294 | 81 | 131 | 5.5330 | 30.6141 | 1171.8583 | 50546.6965 |
| 79 | 106 | 5.4147 | 29.3190 | 1107.2504 | 49922.5720 | 81 | 132 | 5.5557 | 30.8656 | 1203.8425 | 54241.7144 |

TABLE 4. Charge radii, 2nd moment, 4th moment and 6th moment obtained by DNN.

| $Z$ | $N$ | $R_c$ (fm) | $R_2$ (fm$^2$) | $R_4$ (fm$^4$) | $R_6$ (fm$^6$) | $Z$ | $N$ | $R_c$ (fm) | $R_2$ (fm$^2$) | $R_4$ (fm$^4$) | $R_6$ (fm$^6$) |
|---|---|---|---|---|---|---|---|---|---|---|---|
| 81 | 133 | 5.5513 | 30.8165 | 1188.1464 | 51732.6615 | 84 | 123 | 5.5430 | 30.7247 | 1199.3100 | 54593.5789 |
| 81 | 134 | 5.5741 | 31.0709 | 1219.9071 | 55346.9903 | 84 | 124 | 5.5525 | 30.8306 | 1212.3231 | 56110.8460 |
| 81 | 135 | 5.5686 | 31.0098 | 1203.5459 | 52829.5937 | 84 | 125 | 5.5524 | 30.8297 | 1206.4784 | 54927.7142 |
| 81 | 136 | 5.5914 | 31.2639 | 1234.9036 | 56357.6938 | 84 | 126 | 5.5616 | 30.9309 | 1219.4396 | 56481.3304 |
| 81 | 137 | 5.5852 | 31.1940 | 1218.0627 | 53838.0168 | 84 | 127 | 5.5674 | 30.9962 | 1218.7686 | 55684.0091 |
| 82 | 96 | 5.3124 | 28.2211 | 995.1368 | 38881.0286 | 84 | 128 | 5.5834 | 31.1744 | 1238.3545 | 57840.9720 |
| 82 | 97 | 5.3317 | 28.4273 | 1020.1171 | 41651.3990 | 84 | 129 | 5.5873 | 31.2177 | 1235.9086 | 56881.7282 |
| 82 | 98 | 5.3344 | 28.4560 | 1019.3349 | 41259.1626 | 84 | 130 | 5.6045 | 31.4101 | 1256.7855 | 59156.5670 |
| 82 | 99 | 5.3521 | 28.6449 | 1043.2141 | 43974.4024 | 84 | 131 | 5.6063 | 31.4303 | 1252.3519 | 58017.2262 |
| 82 | 100 | 5.3591 | 28.7198 | 1046.4356 | 43905.1829 | 84 | 132 | 5.6246 | 31.6362 | 1274.6334 | 60429.7109 |
| 82 | 101 | 5.3751 | 28.8918 | 1069.0912 | 46540.6447 | 84 | 133 | 5.6244 | 31.6335 | 1268.1100 | 59097.0382 |
| 82 | 102 | 5.3855 | 29.0032 | 1075.4808 | 46728.7581 | 84 | 134 | 5.6436 | 31.8508 | 1291.7683 | 61655.6653 |
| 82 | 103 | 5.4001 | 29.1606 | 1096.9787 | 49276.9033 | 84 | 135 | 5.6415 | 31.8270 | 1283.1612 | 60122.5688 |
| 82 | 104 | 5.4126 | 29.2965 | 1105.4707 | 49637.2300 | 84 | 136 | 5.6615 | 32.0522 | 1308.0394 | 62825.0013 |
| 82 | 105 | 5.4146 | 29.3176 | 1108.9084 | 50124.3604 | 84 | 137 | 5.6578 | 32.0102 | 1297.4507 | 61090.8217 |
| 82 | 106 | 5.4176 | 29.3505 | 1103.8947 | 48952.3737 | 84 | 138 | 5.6779 | 32.2391 | 1323.2892 | 63925.1661 |
| 82 | 107 | 5.4185 | 29.3597 | 1105.2645 | 49175.9555 | 84 | 139 | 5.6730 | 32.1826 | 1310.9117 | 61996.6873 |
| 82 | 108 | 5.4233 | 29.4116 | 1103.9151 | 48493.4788 | 84 | 140 | 5.6930 | 32.4106 | 1337.3867 | 64944.3710 |
| 82 | 109 | 5.4233 | 29.4123 | 1103.4732 | 48464.6539 | 84 | 141 | 5.6872 | 32.3438 | 1323.4950 | 62836.1591 |
| 82 | 110 | 5.4296 | 29.4803 | 1105.5952 | 48267.6528 | 84 | 142 | 5.7067 | 32.5662 | 1350.2489 | 65873.8890 |
| 82 | 111 | 5.4291 | 29.4749 | 1103.5718 | 47998.2222 | 84 | 143 | 5.7003 | 32.4938 | 1335.1789 | 63607.4700 |
| 82 | 112 | 5.4366 | 29.5571 | 1108.9934 | 48276.9121 | 85 | 106 | 5.5473 | 30.7728 | 1246.7562 | 62259.9779 |
| 82 | 113 | 5.4358 | 29.5477 | 1105.5953 | 47780.5295 | 85 | 107 | 5.5270 | 30.5475 | 1227.7638 | 60856.2070 |
| 82 | 114 | 5.4445 | 29.6425 | 1114.1242 | 48515.0093 | 85 | 108 | 5.5427 | 30.7213 | 1234.0567 | 60500.1343 |
| 82 | 115 | 5.4434 | 29.6306 | 1109.5306 | 47806.3720 | 85 | 109 | 5.5230 | 30.5034 | 1215.0171 | 59035.0342 |
| 82 | 116 | 5.4531 | 29.7366 | 1120.8821 | 48959.9001 | 85 | 110 | 5.5391 | 30.6811 | 1223.0417 | 58951.2015 |
| 82 | 117 | 5.4519 | 29.7228 | 1115.2504 | 48054.6332 | 85 | 111 | 5.5214 | 30.4862 | 1205.3346 | 57529.4107 |
| 82 | 118 | 5.4624 | 29.8383 | 1129.0210 | 49572.3010 | 85 | 112 | 5.5370 | 30.6581 | 1214.3223 | 57668.9008 |
| 82 | 119 | 5.4611 | 29.8231 | 1122.4703 | 48483.6728 | 85 | 113 | 5.5225 | 30.4982 | 1198.9864 | 56365.7947 |
| 82 | 120 | 5.4723 | 29.9456 | 1138.1463 | 50294.7582 | 85 | 114 | 5.5369 | 30.6575 | 1208.4095 | 56697.0379 |
| 82 | 121 | 5.4707 | 29.9290 | 1130.7335 | 49029.4575 | 85 | 115 | 5.5262 | 30.5391 | 1196.0022 | 55546.4974 |
| 82 | 122 | 5.4823 | 30.0561 | 1147.7557 | 51055.6361 | 85 | 116 | 5.5392 | 30.6823 | 1205.6016 | 56056.7599 |
| 82 | 123 | 5.4807 | 30.0378 | 1139.4574 | 49609.9007 | 85 | 117 | 5.5323 | 30.6060 | 1196.1381 | 55045.5949 |
| 82 | 124 | 5.4925 | 30.1673 | 1157.2888 | 51774.4038 | 85 | 118 | 5.5437 | 30.7328 | 1205.8951 | 55737.9125 |
| 82 | 125 | 5.4906 | 30.1468 | 1147.9957 | 50131.4765 | 85 | 119 | 5.5402 | 30.6939 | 1198.8744 | 54806.8936 |
| 82 | 126 | 5.5025 | 30.2772 | 1166.1808 | 52368.2656 | 85 | 120 | 5.5503 | 30.8062 | 1208.9388 | 55693.6708 |
| 82 | 127 | 5.5058 | 30.3135 | 1161.0765 | 51016.9421 | 85 | 121 | 5.5494 | 30.7962 | 1203.4685 | 54747.5623 |
| 82 | 128 | 5.5246 | 30.5215 | 1185.8480 | 53851.1626 | 85 | 122 | 5.5586 | 30.8975 | 1214.0798 | 55844.4155 |
| 82 | 129 | 5.5255 | 30.5314 | 1178.4556 | 52289.7395 | 85 | 123 | 5.5593 | 30.9059 | 1209.0414 | 54765.4872 |
| 82 | 130 | 5.5458 | 30.7556 | 1204.5450 | 55235.0667 | 85 | 124 | 5.5678 | 31.0001 | 1220.4537 | 56086.1163 |
| 82 | 131 | 5.5442 | 30.7385 | 1194.7481 | 53454.6518 | 85 | 125 | 5.5693 | 31.0166 | 1214.6789 | 54750.2946 |
| 82 | 132 | 5.5657 | 30.9768 | 1222.1235 | 56514.7188 | 85 | 126 | 5.5774 | 31.1073 | 1227.1175 | 56304.5263 |
| 82 | 133 | 5.5619 | 30.9342 | 1209.9454 | 54513.3334 | 85 | 127 | 5.5835 | 31.1760 | 1225.6174 | 55307.2457 |
| 82 | 134 | 5.5842 | 31.1837 | 1238.4974 | 57688.6811 | 85 | 128 | 5.5981 | 31.3389 | 1244.8206 | 57527.5556 |
| 82 | 135 | 5.5784 | 31.1187 | 1224.0849 | 55471.2685 | 85 | 129 | 5.6021 | 31.3833 | 1242.0870 | 56503.7512 |
| 82 | 136 | 5.6014 | 31.3752 | 1253.6118 | 58756.5312 | 85 | 130 | 5.6182 | 31.5639 | 1262.1016 | 58712.1525 |
| 82 | 137 | 5.5939 | 31.2922 | 1237.2255 | 56336.0317 | 85 | 131 | 5.6201 | 31.5859 | 1258.2076 | 57668.9725 |
| 82 | 138 | 5.6170 | 31.5512 | 1267.4616 | 59721.1503 | 85 | 132 | 5.6375 | 31.7817 | 1278.9889 | 59870.3781 |
| 83 | 101 | 5.4182 | 29.3572 | 1122.8187 | 52146.7078 | 85 | 133 | 5.6377 | 31.7832 | 1273.9411 | 58799.6173 |
| 83 | 102 | 5.4376 | 29.5680 | 1136.7026 | 52788.2713 | 85 | 134 | 5.6561 | 31.9915 | 1295.4439 | 61004.6235 |
| 83 | 103 | 5.4398 | 29.5915 | 1147.3496 | 54579.5393 | 85 | 135 | 5.6546 | 31.9748 | 1289.2300 | 59890.0128 |
| 83 | 104 | 5.4621 | 29.8344 | 1164.3572 | 55520.3816 | 85 | 136 | 5.6738 | 32.1924 | 1311.3685 | 62109.4726 |
| 83 | 105 | 5.4501 | 29.7035 | 1154.3178 | 54937.6072 | 85 | 137 | 5.6710 | 32.1600 | 1303.9878 | 60931.8855 |
| 83 | 106 | 5.4632 | 29.8465 | 1157.2978 | 54220.7072 | 85 | 138 | 5.6906 | 32.3831 | 1326.6109 | 63172.3474 |
| 83 | 107 | 5.4499 | 29.7019 | 1145.0576 | 53380.2999 | 85 | 139 | 5.6866 | 32.3379 | 1318.1214 | 61917.0233 |
| 83 | 108 | 5.4653 | 29.8691 | 1151.9574 | 53139.1197 | 85 | 140 | 5.7063 | 32.5623 | 1341.0261 | 64180.1333 |
| 83 | 109 | 5.4522 | 29.7260 | 1138.8357 | 52138.7941 | 85 | 141 | 5.7016 | 32.5077 | 1331.5564 | 62839.7057 |
| 83 | 110 | 5.4684 | 29.9030 | 1148.4939 | 52294.9243 | 85 | 142 | 5.7209 | 32.7291 | 1354.4901 | 65120.6379 |
| 83 | 111 | 5.4565 | 29.7737 | 1135.5652 | 51212.7940 | 85 | 143 | 5.7157 | 32.6690 | 1344.2427 | 63697.0549 |
| 83 | 112 | 5.4726 | 29.9495 | 1147.0714 | 51705.2105 | 85 | 144 | 5.7344 | 32.8832 | 1366.9294 | 65985.9762 |
| 83 | 113 | 5.4627 | 29.8414 | 1135.0070 | 50586.5941 | 86 | 107 | 5.5781 | 31.1149 | 1275.4670 | 64525.2837 |
| 83 | 114 | 5.4781 | 30.0094 | 1147.8030 | 51378.8426 | 86 | 108 | 5.5759 | 31.0902 | 1265.4834 | 63019.4038 |
| 83 | 115 | 5.4704 | 29.9248 | 1136.8163 | 50231.4766 | 86 | 109 | 5.5701 | 31.0261 | 1259.4112 | 62487.1522 |
| 83 | 116 | 5.4848 | 30.0829 | 1150.6874 | 51309.9495 | 86 | 110 | 5.5699 | 31.0241 | 1252.6311 | 61369.2041 |
| 83 | 117 | 5.4790 | 30.0191 | 1140.5358 | 50102.2322 | 86 | 111 | 5.5634 | 30.9515 | 1245.2614 | 60673.4704 |
| 83 | 118 | 5.4926 | 30.1687 | 1155.5422 | 51470.7894 | 86 | 112 | 5.5654 | 30.9734 | 1241.8480 | 59962.8308 |
| 83 | 119 | 5.4882 | 30.1199 | 1145.6406 | 50139.5112 | 86 | 113 | 5.5588 | 30.8998 | 1233.8446 | 59157.2736 |
| 83 | 120 | 5.5013 | 30.2646 | 1161.9941 | 51810.2330 | 86 | 114 | 5.5628 | 30.9450 | 1233.7723 | 58852.9684 |
| 83 | 121 | 5.4976 | 30.2231 | 1151.5468 | 50269.2720 | 86 | 115 | 5.5567 | 30.8774 | 1225.8024 | 57992.0978 |
| 83 | 122 | 5.5106 | 30.3672 | 1169.4845 | 52253.2796 | 86 | 116 | 5.5627 | 30.9431 | 1228.8013 | 58066.9326 |
| 83 | 123 | 5.5069 | 30.3259 | 1157.6511 | 50407.0584 | 86 | 117 | 5.5577 | 30.8881 | 1221.4617 | 57199.8027 |
| 83 | 124 | 5.5202 | 30.4727 | 1177.3334 | 52707.3831 | 86 | 118 | 5.5650 | 30.9693 | 1227.0213 | 57600.1176 |
| 83 | 125 | 5.5160 | 30.4263 | 1163.3642 | 50463.7092 | 86 | 119 | 5.5616 | 30.9314 | 1220.7511 | 56762.5177 |
| 83 | 126 | 5.5297 | 30.5780 | 1184.8315 | 53072.7704 | 86 | 120 | 5.5697 | 31.0220 | 1228.1849 | 57414.0431 |
| 83 | 127 | 5.5299 | 30.5801 | 1174.4362 | 51076.4937 | 86 | 121 | 5.5680 | 31.0032 | 1223.1808 | 56620.0250 |
| 83 | 128 | 5.5499 | 30.8009 | 1202.4028 | 54340.7880 | 86 | 122 | 5.5765 | 31.0972 | 1231.7544 | 57441.2870 |
| 83 | 129 | 5.5485 | 30.7854 | 1191.0570 | 52315.8371 | 86 | 123 | 5.5764 | 31.0964 | 1227.9194 | 56676.8441 |
| 83 | 130 | 5.5690 | 31.0139 | 1219.0480 | 55518.0218 | 86 | 124 | 5.5847 | 31.1888 | 1236.9777 | 57593.6279 |
| 83 | 131 | 5.5663 | 30.9842 | 1206.9907 | 53486.0565 | 86 | 125 | 5.5859 | 31.2028 | 1233.9473 | 56817.5298 |
| 83 | 132 | 5.5872 | 31.2165 | 1234.7710 | 56609.7249 | 86 | 126 | 5.5937 | 31.2900 | 1243.0000 | 57773.8942 |
| 83 | 133 | 5.5835 | 31.1756 | 1222.1619 | 54578.3659 | 86 | 127 | 5.6016 | 31.3783 | 1246.1736 | 57499.0954 |
| 83 | 134 | 5.6043 | 31.4084 | 1249.5837 | 57620.0465 | 86 | 128 | 5.6159 | 31.5387 | 1261.8165 | 59071.2976 |
| 83 | 135 | 5.5999 | 31.3592 | 1236.5347 | 55588.0118 | 86 | 129 | 5.6228 | 31.6160 | 1264.2513 | 58727.7760 |
| 83 | 136 | 5.6205 | 31.5896 | 1263.4959 | 58552.0684 | 86 | 130 | 5.6379 | 31.7854 | 1280.8615 | 60397.0496 |
| 83 | 137 | 5.6156 | 31.5348 | 1250.1045 | 56514.3090 | 86 | 131 | 5.6436 | 31.8497 | 1282.2826 | 59960.6000 |
| 83 | 138 | 5.6356 | 31.7600 | 1276.5175 | 59408.5152 | 86 | 132 | 5.6593 | 32.0278 | 1300.0191 | 61753.8744 |
| 83 | 139 | 5.6305 | 31.7026 | 1262.8900 | 57360.0299 | 86 | 133 | 5.6638 | 32.0783 | 1300.2371 | 61204.7017 |
| 83 | 140 | 5.6498 | 31.9199 | 1288.6665 | 60192.6219 | 86 | 134 | 5.6801 | 32.2635 | 1319.0952 | 63132.9341 |
| 83 | 141 | 5.6447 | 31.8626 | 1274.9341 | 58131.5518 | 86 | 135 | 5.6833 | 32.3003 | 1317.9879 | 62454.7365 |
| 84 | 102 | 5.4868 | 30.1055 | 1183.5500 | 56509.5187 | 86 | 136 | 5.7000 | 32.4899 | 1337.8258 | 64515.2114 |
| 84 | 103 | 5.4973 | 30.2205 | 1200.9968 | 58696.2605 | 86 | 137 | 5.7021 | 32.5138 | 1335.3245 | 63694.4866 |
| 84 | 104 | 5.5114 | 30.3758 | 1211.4724 | 59279.0487 | 86 | 138 | 5.7188 | 32.7041 | 1355.9023 | 65874.1953 |
| 84 | 105 | 5.5081 | 30.3395 | 1209.2540 | 59228.4496 | 86 | 139 | 5.7198 | 32.7165 | 1351.9991 | 64901.8908 |
| 84 | 106 | 5.5114 | 30.3756 | 1204.1101 | 58027.8132 | 86 | 140 | 5.7362 | 32.9037 | 1373.0244 | 67181.7191 |
| 84 | 107 | 5.5067 | 30.3233 | 1199.7038 | 57727.2099 | 86 | 141 | 5.7364 | 32.9065 | 1367.7725 | 66053.8899 |
| 84 | 108 | 5.5117 | 30.3790 | 1197.8027 | 56943.9373 | 86 | 142 | 5.7521 | 33.0868 | 1388.9382 | 68412.0247 |
| 84 | 109 | 5.5061 | 30.3172 | 1191.7893 | 56435.4820 | 86 | 143 | 5.7517 | 33.0823 | 1382.4537 | 67130.7606 |
| 84 | 110 | 5.5127 | 30.3895 | 1192.9285 | 56063.3507 | 86 | 144 | 5.7665 | 33.2522 | 1403.4608 | 69544.6443 |
| 84 | 111 | 5.5068 | 30.3246 | 1185.8820 | 55389.7017 | 86 | 145 | 5.7657 | 33.2432 | 1395.9265 | 68119.2996 |
| 84 | 112 | 5.5146 | 30.4107 | 1189.8580 | 55417.6029 | 87 | 110 | 5.5890 | 31.2373 | 1275.9617 | 63711.4395 |
| 84 | 113 | 5.5089 | 30.3485 | 1182.3187 | 54620.2802 | 87 | 111 | 5.5738 | 31.0671 | 1260.9278 | 62495.2546 |
| 84 | 114 | 5.5178 | 30.4457 | 1188.8804 | 55026.9806 | 87 | 112 | 5.5798 | 31.1344 | 1259.1587 | 61663.2979 |
| 84 | 115 | 5.5128 | 30.3910 | 1181.3097 | 54142.6515 | 87 | 113 | 5.5677 | 30.9989 | 1246.6174 | 60598.3850 |
| 84 | 116 | 5.5223 | 30.4963 | 1190.1233 | 54893.0793 | 87 | 114 | 5.5732 | 31.0608 | 1245.5815 | 59945.1979 |
| 84 | 117 | 5.5184 | 30.4525 | 1182.8701 | 53950.4446 | 87 | 115 | 5.5650 | 30.9693 | 1236.3356 | 59089.8080 |
| 84 | 118 | 5.5283 | 30.5625 | 1193.4992 | 54993.2977 | 87 | 116 | 5.5699 | 31.0235 | 1235.9172 | 58613.8030 |
| 84 | 119 | 5.5255 | 30.5313 | 1186.7567 | 54008.6467 | 87 | 117 | 5.5660 | 30.9805 | 1230.3267 | 57986.4795 |
| 84 | 120 | 5.5356 | 30.6425 | 1198.6853 | 55278.7913 | 87 | 118 | 5.5701 | 31.0258 | 1230.4594 | 57687.4894 |
| 84 | 121 | 5.5339 | 30.6235 | 1192.4689 | 54252.6511 | 87 | 119 | 5.5705 | 31.0299 | 1228.3435 | 57258.1644 |
| 84 | 122 | 5.5437 | 30.7331 | 1205.1742 | 55679.7611 | 87 | 120 | 5.5737 | 31.0664 | 1229.0349 | 57139.0049 |

TABLE 4. Charge radii, 2nd moment, 4th moment and 6th moment obtained by DNN.

| $Z$ | $N$ | $R_c$(fm) | $R_2$(fm$^2$) | $R_4$(fm$^4$) | $R_6$(fm$^6$) | $Z$ | $N$ | $R_c$(fm) | $R_2$(fm$^2$) | $R_4$(fm$^4$) | $R_6$(fm$^6$) |
|---|---|---|---|---|---|---|---|---|---|---|---|
| 87 | 121 | 5.5777 | 31.1108 | 1229.6886 | 56830.8520 | 90 | 134 | 5.7082 | 32.5841 | 1322.6057 | 61048.7437 |
| 87 | 122 | 5.5803 | 31.1396 | 1231.0119 | 56896.3541 | 90 | 135 | 5.7186 | 32.7023 | 1333.3684 | 61874.3387 |
| 87 | 123 | 5.5869 | 31.2131 | 1233.3065 | 56594.1047 | 90 | 136 | 5.7329 | 32.8662 | 1346.2659 | 62850.4726 |
| 87 | 124 | 5.5890 | 31.2365 | 1235.4166 | 56853.7590 | 90 | 137 | 5.7448 | 33.0023 | 1358.9133 | 63856.7671 |
| 87 | 125 | 5.5970 | 31.3261 | 1238.0042 | 56422.4606 | 90 | 138 | 5.7570 | 33.1435 | 1370.2653 | 64744.9476 |
| 87 | 126 | 5.5989 | 31.3472 | 1241.1102 | 56889.5556 | 90 | 139 | 5.7702 | 33.2949 | 1384.4403 | 65893.8393 |
| 87 | 127 | 5.6120 | 31.4944 | 1249.0686 | 56946.2254 | 90 | 140 | 5.7803 | 33.4117 | 1394.1190 | 66686.3746 |
| 87 | 128 | 5.6210 | 31.5953 | 1259.7284 | 58141.7461 | 90 | 141 | 5.7943 | 33.5744 | 1409.3494 | 67928.0696 |
| 87 | 129 | 5.6317 | 31.7163 | 1266.5753 | 58207.6883 | 90 | 142 | 5.8023 | 33.6665 | 1417.3446 | 68626.5063 |
| 87 | 130 | 5.6430 | 31.8436 | 1278.7109 | 59430.5771 | 90 | 143 | 5.8169 | 33.8362 | 1433.0963 | 69904.3545 |
| 87 | 131 | 5.6514 | 31.9379 | 1284.3045 | 59493.5988 | 90 | 144 | 5.8227 | 33.9044 | 1439.5040 | 70519.3512 |
| 87 | 132 | 5.6649 | 32.0909 | 1298.0424 | 60769.0227 | 90 | 145 | 5.8375 | 34.0768 | 1455.2310 | 71774.5464 |
| 87 | 133 | 5.6709 | 32.1588 | 1302.2254 | 60805.1734 | 90 | 146 | 5.8415 | 34.1225 | 1460.2267 | 72323.7919 |
| 87 | 134 | 5.6864 | 32.3351 | 1317.5905 | 62155.0653 | 90 | 147 | 5.8561 | 34.2937 | 1475.4338 | 73501.8813 |
| 87 | 135 | 5.6901 | 32.3776 | 1320.2093 | 62132.9038 | 90 | 148 | 5.8582 | 34.3191 | 1479.2287 | 74006.2220 |
| 87 | 136 | 5.7073 | 32.5736 | 1337.1024 | 63571.3985 | 90 | 149 | 5.8725 | 34.4859 | 1493.5076 | 75060.7945 |
| 87 | 137 | 5.7090 | 32.5924 | 1338.0409 | 63458.1527 | 91 | 120 | 5.5936 | 31.2886 | 1237.3518 | 56637.1638 |
| 87 | 138 | 5.7274 | 32.8033 | 1356.2502 | 64990.2944 | 91 | 121 | 5.5972 | 31.3284 | 1240.0399 | 56768.7161 |
| 87 | 139 | 5.7272 | 32.8008 | 1355.4634 | 64757.9810 | 91 | 122 | 5.5935 | 31.2876 | 1229.8280 | 55313.1478 |
| 87 | 140 | 5.7464 | 33.0211 | 1374.6780 | 66378.5306 | 91 | 123 | 5.5995 | 31.3544 | 1234.5705 | 55575.5589 |
| 87 | 141 | 5.7446 | 33.0006 | 1372.2221 | 66009.2148 | 91 | 124 | 5.5969 | 31.3250 | 1226.1850 | 54343.5463 |
| 87 | 142 | 5.7641 | 33.2243 | 1392.0666 | 67704.2758 | 91 | 125 | 5.6049 | 31.4148 | 1232.6114 | 54696.4757 |
| 87 | 143 | 5.7610 | 33.1897 | 1388.0979 | 67191.7741 | 91 | 126 | 5.6030 | 31.3941 | 1225.8143 | 53671.2406 |
| 87 | 144 | 5.7802 | 33.4111 | 1408.1699 | 68941.3253 | 91 | 127 | 5.6195 | 31.5791 | 1242.2420 | 55035.2519 |
| 87 | 145 | 5.7764 | 33.3667 | 1402.9408 | 68292.0647 | 91 | 128 | 5.6276 | 31.6701 | 1245.9204 | 54974.0023 |
| 87 | 146 | 5.7949 | 33.5804 | 1422.8315 | 70071.2518 | 91 | 129 | 5.6428 | 31.8413 | 1262.9194 | 56546.2150 |
| 88 | 113 | 5.5901 | 31.2497 | 1266.2794 | 62016.3778 | 91 | 130 | 5.6535 | 31.9620 | 1268.0335 | 56474.9476 |
| 88 | 114 | 5.5889 | 31.2354 | 1258.9305 | 60923.5428 | 91 | 131 | 5.6674 | 32.1192 | 1285.4685 | 58235.0877 |
| 88 | 115 | 5.5824 | 31.1636 | 1251.2874 | 60142.7797 | 91 | 132 | 5.6804 | 32.2667 | 1292.0043 | 58179.5418 |
| 88 | 116 | 5.5847 | 31.1892 | 1248.4375 | 59515.1084 | 91 | 133 | 5.6930 | 32.4103 | 1309.7419 | 60099.7018 |
| 88 | 117 | 5.5784 | 31.1190 | 1240.5788 | 58677.4987 | 91 | 134 | 5.7079 | 32.5797 | 1317.5174 | 60073.2983 |
| 88 | 118 | 5.5838 | 31.1783 | 1241.6544 | 58461.3532 | 91 | 135 | 5.7193 | 32.7106 | 1335.3815 | 62116.0173 |
| 88 | 119 | 5.5784 | 31.1183 | 1234.3750 | 57633.1661 | 91 | 136 | 5.7355 | 32.8956 | 1344.1030 | 62122.1297 |
| 88 | 120 | 5.5859 | 31.2024 | 1238.5170 | 57746.8039 | 91 | 137 | 5.7458 | 33.0144 | 1361.8447 | 64239.8865 |
| 88 | 121 | 5.5820 | 31.1591 | 1232.4122 | 56975.9366 | 91 | 138 | 5.7627 | 33.2087 | 1371.1910 | 64277.6651 |
| 88 | 122 | 5.5909 | 31.2582 | 1238.6306 | 57324.7245 | 91 | 139 | 5.7719 | 33.3151 | 1388.4814 | 66414.0487 |
| 88 | 123 | 5.5888 | 31.2344 | 1233.9645 | 56627.8204 | 91 | 140 | 5.7890 | 33.5129 | 1398.1618 | 66482.6370 |
| 88 | 124 | 5.5981 | 31.3393 | 1241.3224 | 57123.0049 | 91 | 141 | 5.7971 | 33.6063 | 1414.6218 | 68576.7271 |
| 88 | 125 | 5.5977 | 31.3343 | 1237.9772 | 56479.8318 | 91 | 142 | 5.8140 | 33.8027 | 1424.4058 | 68677.1650 |
| 88 | 126 | 5.6069 | 31.4378 | 1245.7424 | 57054.1552 | 91 | 143 | 5.8209 | 33.8824 | 1439.6581 | 70669.6931 |
| 88 | 127 | 5.6141 | 31.5177 | 1249.9991 | 57071.2508 | 91 | 144 | 5.8373 | 34.0736 | 1449.3781 | 70805.0085 |
| 88 | 128 | 5.6298 | 31.6942 | 1264.6625 | 58302.2711 | 91 | 145 | 5.8429 | 34.1391 | 1463.1029 | 72644.5352 |
| 88 | 129 | 5.6371 | 31.7770 | 1269.5429 | 58383.0608 | 91 | 146 | 5.8585 | 34.3219 | 1472.6266 | 72816.9778 |
| 88 | 130 | 5.6529 | 31.9554 | 1284.5552 | 59654.1329 | 91 | 147 | 5.8629 | 34.3734 | 1484.6194 | 74466.1811 |
| 88 | 131 | 5.6604 | 32.0401 | 1289.9055 | 59783.6984 | 91 | 148 | 5.8775 | 34.5453 | 1493.8313 | 74675.6442 |
| 88 | 132 | 5.6761 | 32.2187 | 1305.2900 | 61113.7920 | 91 | 149 | 5.8808 | 34.5840 | 1504.0179 | 76113.0830 |
| 88 | 133 | 5.6837 | 32.3047 | 1310.9726 | 61276.9417 | 91 | 150 | 5.8943 | 34.7428 | 1512.7912 | 76354.5924 |
| 88 | 134 | 5.6992 | 32.4809 | 1326.6272 | 62670.2589 | 92 | 123 | 5.5974 | 31.3314 | 1225.9658 | 54174.1030 |
| 88 | 135 | 5.7068 | 32.5677 | 1332.4774 | 62847.8114 | 92 | 124 | 5.6103 | 31.4760 | 1233.6095 | 54404.1047 |
| 88 | 136 | 5.7218 | 32.7385 | 1348.2154 | 64297.4705 | 92 | 125 | 5.6004 | 31.3643 | 1221.6661 | 53115.5061 |
| 88 | 137 | 5.7293 | 32.8251 | 1354.0365 | 64466.1106 | 92 | 126 | 5.6149 | 31.5276 | 1231.4448 | 53572.4421 |
| 88 | 138 | 5.7435 | 32.9875 | 1369.6371 | 65959.3388 | 92 | 127 | 5.6151 | 31.5289 | 1230.4922 | 53317.3054 |
| 88 | 139 | 5.7509 | 33.0727 | 1375.2007 | 66091.5870 | 92 | 128 | 5.6381 | 31.7887 | 1249.6730 | 54657.7750 |
| 88 | 140 | 5.7640 | 33.2239 | 1390.4539 | 67614.6122 | 92 | 129 | 5.6405 | 31.8157 | 1251.6666 | 54712.1202 |
| 88 | 141 | 5.7712 | 33.3067 | 1395.5225 | 67681.0882 | 92 | 130 | 5.6623 | 32.0617 | 1269.6441 | 55933.8796 |
| 88 | 142 | 5.7831 | 33.4444 | 1410.2559 | 69222.2340 | 92 | 131 | 5.6673 | 32.1188 | 1274.9665 | 56329.3117 |
| 88 | 143 | 5.7900 | 33.5238 | 1414.6169 | 69195.2207 | 92 | 132 | 5.6871 | 32.3437 | 1291.2316 | 57406.0208 |
| 88 | 144 | 5.8005 | 33.6463 | 1428.7004 | 70745.7056 | 92 | 133 | 5.6951 | 32.4341 | 1300.1445 | 58162.6769 |
| 88 | 145 | 5.8071 | 33.7219 | 1432.1954 | 70602.3676 | 92 | 134 | 5.7124 | 32.6314 | 1314.2156 | 59065.7778 |
| 88 | 146 | 5.8162 | 33.8281 | 1445.5384 | 72156.1956 | 92 | 135 | 5.7233 | 32.7567 | 1326.8095 | 60187.8789 |
| 88 | 147 | 5.8224 | 33.8998 | 1448.0822 | 71881.0350 | 92 | 136 | 5.7377 | 32.9214 | 1338.2911 | 60891.9684 |
| 89 | 116 | 5.5886 | 31.2325 | 1252.7962 | 59862.4757 | 92 | 137 | 5.7516 | 33.0809 | 1354.4598 | 62365.4604 |
| 89 | 117 | 5.5870 | 31.2148 | 1250.3949 | 59572.0422 | 92 | 138 | 5.7628 | 33.2099 | 1363.0876 | 62853.0421 |
| 89 | 118 | 5.5845 | 31.1863 | 1241.5183 | 58294.3488 | 92 | 139 | 5.7794 | 33.4010 | 1382.5172 | 64644.2110 |
| 89 | 119 | 5.5864 | 31.2083 | 1242.1120 | 58198.3713 | 92 | 140 | 5.7873 | 33.4929 | 1388.1878 | 64909.6481 |
| 89 | 120 | 5.5844 | 31.1854 | 1234.8018 | 57145.5648 | 92 | 141 | 5.8062 | 33.7114 | 1410.3841 | 66967.4705 |
| 89 | 121 | 5.5899 | 31.2470 | 1238.3515 | 57229.0083 | 92 | 142 | 5.8109 | 33.7668 | 1413.1558 | 67017.9861 |
| 89 | 122 | 5.5881 | 31.2266 | 1232.3351 | 56380.3877 | 92 | 143 | 5.8315 | 34.0068 | 1437.4730 | 69276.3227 |
| 89 | 123 | 5.5966 | 31.3216 | 1238.2202 | 56576.0322 | 92 | 144 | 5.8333 | 34.0278 | 1437.5610 | 69132.7574 |
| 89 | 124 | 5.5949 | 31.3026 | 1233.3725 | 55922.6055 | 92 | 145 | 5.8552 | 34.2829 | 1463.2584 | 71515.6050 |
| 89 | 125 | 5.6053 | 31.4198 | 1240.4826 | 56119.2118 | 92 | 146 | 5.8543 | 34.2730 | 1461.0038 | 71210.1580 |
| 89 | 126 | 5.6038 | 31.4028 | 1236.8569 | 55667.2784 | 92 | 147 | 5.8767 | 34.5361 | 1487.3050 | 73637.3275 |
| 89 | 127 | 5.6207 | 31.5919 | 1251.2826 | 56583.1866 | 92 | 148 | 5.8736 | 34.4996 | 1483.1274 | 73209.4840 |
| 89 | 128 | 5.6274 | 31.6676 | 1256.4441 | 56959.5933 | 92 | 149 | 5.8961 | 34.7643 | 1509.2886 | 75603.3516 |
| 89 | 129 | 5.6422 | 31.8343 | 1270.4065 | 57967.7183 | 92 | 150 | 5.8912 | 34.7057 | 1503.6478 | 75096.6937 |
| 89 | 130 | 5.6517 | 31.9415 | 1277.3186 | 58377.3214 | 92 | 151 | 5.9132 | 34.9661 | 1528.9991 | 77386.2308 |
| 89 | 131 | 5.6643 | 32.0842 | 1290.5663 | 59451.7411 | 93 | 126 | 5.6035 | 31.3998 | 1216.9168 | 51864.8039 |
| 89 | 132 | 5.6764 | 32.2220 | 1299.3999 | 59931.2023 | 93 | 127 | 5.6171 | 31.5517 | 1231.5754 | 53220.7503 |
| 89 | 133 | 5.6868 | 32.3402 | 1311.6854 | 61036.5177 | 93 | 128 | 5.6281 | 31.6759 | 1236.7060 | 53118.9908 |
| 89 | 134 | 5.7014 | 32.5057 | 1322.4380 | 61611.6094 | 93 | 129 | 5.6413 | 31.8248 | 1253.0083 | 54794.3161 |
| 89 | 135 | 5.7096 | 32.5996 | 1333.5277 | 62706.5224 | 93 | 130 | 5.6542 | 31.9704 | 1258.7329 | 54598.8547 |
| 89 | 136 | 5.7261 | 32.7882 | 1346.0351 | 63390.2262 | 93 | 131 | 5.6673 | 32.1186 | 1276.8085 | 56595.4647 |
| 89 | 137 | 5.7323 | 32.8587 | 1355.7060 | 64429.8916 | 93 | 132 | 5.6815 | 32.2797 | 1282.8222 | 56306.0012 |
| 89 | 138 | 5.7502 | 33.0644 | 1369.6857 | 65223.9764 | 93 | 133 | 5.6947 | 32.4299 | 1302.7715 | 58617.4902 |
| 89 | 139 | 5.7544 | 33.1129 | 1377.7507 | 66165.1134 | 93 | 134 | 5.7096 | 32.5995 | 1308.6749 | 58226.4866 |
| 89 | 140 | 5.7732 | 33.3294 | 1392.8456 | 67062.2525 | 93 | 135 | 5.7231 | 32.7538 | 1330.4920 | 60832.6336 |
| 89 | 141 | 5.7756 | 33.3577 | 1399.1797 | 67867.8136 | 93 | 136 | 5.7380 | 32.9250 | 1335.8839 | 60331.6701 |
| 89 | 142 | 5.7947 | 33.5786 | 1415.0009 | 68854.1668 | 93 | 137 | 5.7519 | 33.0842 | 1359.4013 | 63194.8705 |
| 89 | 143 | 5.7956 | 33.5891 | 1419.5639 | 69497.1562 | 93 | 138 | 5.7664 | 33.2510 | 1363.9657 | 62581.5295 |
| 89 | 144 | 5.8145 | 33.8086 | 1435.7237 | 70554.8655 | 93 | 139 | 5.7805 | 33.4141 | 1388.8236 | 65644.8932 |
| 89 | 145 | 5.8142 | 33.8044 | 1438.5830 | 71021.7929 | 93 | 140 | 5.7941 | 33.5721 | 1392.3879 | 64927.5827 |
| 89 | 146 | 5.8324 | 34.0170 | 1454.7028 | 72129.3896 | 93 | 141 | 5.8083 | 33.7367 | 1418.0553 | 68118.0189 |
| 89 | 147 | 5.8311 | 34.0016 | 1456.0274 | 72420.2624 | 93 | 142 | 5.8209 | 33.8834 | 1420.5984 | 67316.2269 |
| 89 | 148 | 5.8483 | 34.2027 | 1471.7474 | 73553.6124 | 93 | 143 | 5.8349 | 34.0457 | 1446.4276 | 70550.2109 |
| 90 | 118 | 5.5972 | 31.3284 | 1249.5068 | 58553.4949 | 93 | 144 | 5.8464 | 34.1799 | 1448.0588 | 69692.5882 |
| 90 | 119 | 5.5893 | 31.2404 | 1240.1168 | 57577.7391 | 93 | 145 | 5.8597 | 34.3357 | 1473.3662 | 72884.6027 |
| 90 | 120 | 5.5967 | 31.3227 | 1242.1390 | 57320.1336 | 93 | 146 | 5.8701 | 34.4578 | 1474.2753 | 72004.1622 |
| 90 | 121 | 5.5893 | 31.2398 | 1232.9850 | 56335.8694 | 93 | 147 | 5.8824 | 34.6031 | 1498.4216 | 75074.8902 |
| 90 | 122 | 5.5991 | 31.3505 | 1238.1801 | 56400.1020 | 93 | 148 | 5.8918 | 34.7135 | 1498.8268 | 74204.0306 |
| 90 | 123 | 5.5929 | 31.2801 | 1229.9642 | 55465.4876 | 93 | 149 | 5.9030 | 34.8452 | 1521.2873 | 77087.5797 |
| 90 | 124 | 5.6043 | 31.4079 | 1237.2837 | 55758.9907 | 93 | 150 | 5.9114 | 34.9449 | 1521.3880 | 76254.0645 |
| 90 | 125 | 5.5995 | 31.3539 | 1230.3512 | 54897.8171 | 93 | 151 | 5.9212 | 35.0610 | 1541.7971 | 78902.4738 |
| 90 | 126 | 5.6115 | 31.4890 | 1238.8637 | 55340.8419 | 93 | 152 | 5.9288 | 35.1507 | 1541.7351 | 78125.5703 |
| 90 | 127 | 5.6155 | 31.5343 | 1241.2740 | 55330.7354 | 94 | 127 | 5.6170 | 31.5511 | 1223.9953 | 51782.1264 |
| 90 | 128 | 5.6348 | 31.7509 | 1257.6629 | 56520.9713 | 94 | 128 | 5.6431 | 31.8441 | 1245.4066 | 53278.8656 |
| 90 | 129 | 5.6403 | 31.8125 | 1262.0548 | 56714.6264 | 94 | 129 | 5.6420 | 31.8318 | 1244.3533 | 53094.2870 |
| 90 | 130 | 5.6589 | 32.0227 | 1277.9944 | 57864.8813 | 94 | 130 | 5.6663 | 32.1070 | 1264.1036 | 54418.1845 |
| 90 | 131 | 5.6659 | 32.1024 | 1284.4960 | 58271.4284 | 94 | 131 | 5.6682 | 32.1287 | 1266.8155 | 54630.7258 |
| 90 | 132 | 5.6834 | 32.3016 | 1299.7257 | 59377.8180 | 94 | 132 | 5.6902 | 32.3781 | 1284.3202 | 55747.1669 |
| 90 | 133 | 5.6921 | 32.4006 | 1308.3923 | 59998.9897 | 94 | 133 | 5.6955 | 32.4382 | 1291.1624 | 56385.8087 |

TABLE 4. Charge radii, 2nd moment, 4th moment and 6th moment obtained by DNN.

| $Z$ | $N$ | $R_c$(fm) | $R_2$(fm$^2$) | $R_4$(fm$^4$) | $R_6$(fm$^6$) | $Z$ | $N$ | $R_c$(fm) | $R_2$(fm$^2$) | $R_4$(fm$^4$) | $R_6$(fm$^6$) |
|---|---|---|---|---|---|---|---|---|---|---|---|
| 94 | 134 | 5.7144 | 32.6549 | 1305.9068 | 57262.9409 | 98 | 158 | 5.9474 | 35.3714 | 1541.3770 | 76151.5896 |
| 94 | 135 | 5.7233 | 32.7561 | 1317.0960 | 58342.6432 | 99 | 140 | 5.7702 | 33.2951 | 1346.4879 | 58738.0470 |
| 94 | 136 | 5.7389 | 32.9349 | 1328.6674 | 58954.2361 | 99 | 141 | 5.7826 | 33.4383 | 1371.6955 | 62097.6575 |
| 94 | 137 | 5.7514 | 33.0782 | 1344.2385 | 60472.8500 | 99 | 142 | 5.7923 | 33.5512 | 1368.3099 | 60502.2542 |
| 94 | 138 | 5.7633 | 33.2155 | 1352.3545 | 60801.1699 | 99 | 143 | 5.8070 | 33.7209 | 1396.9392 | 64266.6056 |
| 94 | 139 | 5.7793 | 33.4000 | 1372.1692 | 62740.5504 | 99 | 144 | 5.8146 | 33.8092 | 1390.9691 | 62398.3655 |
| 94 | 140 | 5.7874 | 33.4939 | 1376.6852 | 62777.5014 | 99 | 145 | 5.8313 | 34.0039 | 1422.7099 | 66521.0548 |
| 94 | 141 | 5.8066 | 33.7169 | 1400.4160 | 65101.6428 | 99 | 146 | 5.8367 | 34.0675 | 1414.3023 | 64409.9430 |
| 94 | 142 | 5.8110 | 33.7674 | 1401.3458 | 64851.6230 | 99 | 147 | 5.8553 | 34.2847 | 1448.7168 | 68832.3510 |
| 94 | 143 | 5.8331 | 34.0245 | 1428.4978 | 67508.6078 | 99 | 148 | 5.8587 | 34.3245 | 1438.0979 | 66514.5762 |
| 94 | 144 | 5.8338 | 34.0332 | 1425.9955 | 66987.3011 | 99 | 149 | 5.8788 | 34.5603 | 1474.6315 | 71167.4116 |
| 94 | 145 | 5.8582 | 34.3186 | 1455.9263 | 69910.8801 | 99 | 150 | 5.8803 | 34.5780 | 1462.1036 | 68684.9747 |
| 94 | 146 | 5.8557 | 34.2887 | 1450.2902 | 69146.6356 | 99 | 151 | 5.9015 | 34.8274 | 1500.1052 | 73490.5385 |
| 94 | 147 | 5.8818 | 34.5954 | 1482.2456 | 72259.4957 | 99 | 152 | 5.9013 | 34.8257 | 1486.0331 | 70889.6370 |
| 94 | 148 | 5.8763 | 34.5313 | 1473.8916 | 71291.2452 | 99 | 153 | 5.9231 | 35.0831 | 1524.7882 | 75765.3503 |
| 94 | 149 | 5.9035 | 34.8517 | 1507.0535 | 74509.5709 | 99 | 154 | 5.9216 | 35.0651 | 1509.5829 | 73094.8088 |
| 94 | 150 | 5.8956 | 34.7585 | 1496.4790 | 73383.8896 | 99 | 155 | 5.9434 | 35.3245 | 1548.3493 | 77956.7760 |
| 94 | 151 | 5.9233 | 35.0852 | 1530.0176 | 76622.3246 | 99 | 156 | 5.9409 | 35.2937 | 1532.4475 | 75266.1695 |
| 94 | 152 | 5.9134 | 34.9686 | 1517.7728 | 75391.0666 | 99 | 157 | 5.9623 | 35.5492 | 1570.5041 | 80034.2476 |
| 94 | 153 | 5.9409 | 35.2944 | 1550.8983 | 78567.8978 | 99 | 158 | 5.9590 | 35.5093 | 1554.3370 | 77370.8352 |
| 95 | 128 | 5.6312 | 31.7100 | 1232.4383 | 51874.2279 | 99 | 159 | 5.9796 | 35.7554 | 1591.0217 | 81972.1984 |
| 95 | 129 | 5.6411 | 31.8225 | 1245.8523 | 53341.4023 | 100 | 141 | 5.7756 | 33.3576 | 1349.6256 | 58720.6239 |
| 95 | 130 | 5.6560 | 31.9903 | 1252.9628 | 53216.9044 | 100 | 142 | 5.7912 | 33.5375 | 1358.4782 | 58853.6036 |
| 95 | 131 | 5.6667 | 32.1119 | 1269.0377 | 55092.0978 | 100 | 143 | 5.7964 | 33.5983 | 1370.1112 | 60378.2108 |
| 95 | 132 | 5.6819 | 32.2840 | 1275.3671 | 54771.0339 | 100 | 144 | 5.8091 | 33.7452 | 1376.2980 | 60288.0627 |
| 95 | 133 | 5.6938 | 32.4189 | 1294.3858 | 57066.8634 | 100 | 145 | 5.8174 | 33.8416 | 1391.4777 | 62169.5098 |
| 95 | 134 | 5.7086 | 32.5878 | 1299.4419 | 56528.8224 | 100 | 146 | 5.8270 | 33.9542 | 1394.8992 | 61852.9958 |
| 95 | 135 | 5.7218 | 32.7394 | 1321.5665 | 59243.9103 | 100 | 147 | 5.8384 | 34.0864 | 1413.5904 | 64080.6095 |
| 95 | 136 | 5.7357 | 32.8983 | 1324.9198 | 58473.5483 | 100 | 148 | 5.8450 | 34.1636 | 1414.1851 | 63536.4844 |
| 95 | 137 | 5.7505 | 33.0683 | 1350.1297 | 61587.4240 | 100 | 149 | 5.8593 | 34.3313 | 1436.2755 | 66092.9433 |
| 95 | 138 | 5.7630 | 33.2119 | 1351.4914 | 60581.3981 | 100 | 150 | 5.8628 | 34.3728 | 1434.0221 | 65322.1231 |
| 95 | 139 | 5.7793 | 33.4001 | 1379.5538 | 64052.4141 | 100 | 151 | 5.8800 | 34.5743 | 1459.3147 | 68182.6617 |
| 95 | 140 | 5.7901 | 33.5249 | 1378.8012 | 62821.1271 | 100 | 152 | 5.8805 | 34.5806 | 1454.2512 | 67190.4264 |
| 95 | 141 | 5.8077 | 33.7291 | 1409.2667 | 66586.6149 | 100 | 153 | 5.9003 | 34.8136 | 1482.4533 | 70321.5915 |
| 95 | 142 | 5.8167 | 33.8336 | 1406.4594 | 65155.7229 | 100 | 154 | 5.8979 | 34.7857 | 1474.6955 | 69119.8959 |
| 95 | 143 | 5.8352 | 34.0497 | 1438.6905 | 69135.1246 | 100 | 155 | 5.9200 | 35.0467 | 1505.4142 | 72478.8540 |
| 95 | 144 | 5.8425 | 34.1342 | 1434.0466 | 67543.0113 | 100 | 156 | 5.9150 | 34.9868 | 1495.1545 | 71086.3196 |
| 95 | 145 | 5.8615 | 34.3568 | 1467.2715 | 71643.5592 | 100 | 157 | 5.9390 | 35.2714 | 1527.9087 | 74622.0901 |
| 95 | 146 | 5.8671 | 34.4231 | 1461.1325 | 69937.9433 | 100 | 158 | 5.9315 | 35.1825 | 1515.4193 | 73064.7051 |
| 95 | 147 | 5.8861 | 34.6460 | 1494.5208 | 74062.4605 | 100 | 159 | 5.9570 | 35.4855 | 1549.6533 | 76719.3024 |
| 95 | 148 | 5.8904 | 34.6967 | 1487.2971 | 72295.0821 | 100 | 160 | 5.9474 | 35.3710 | 1535.2749 | 75029.4660 |
| 95 | 149 | 5.9088 | 34.9138 | 1520.0308 | 76349.0750 | 101 | 144 | 5.7996 | 33.6355 | 1371.3005 | 60230.9184 |
| 95 | 150 | 5.9120 | 34.9520 | 1512.1508 | 74570.9140 | 101 | 145 | 5.8107 | 33.7644 | 1396.9600 | 63807.1916 |
| 95 | 151 | 5.9294 | 35.1578 | 1543.4998 | 78470.3373 | 101 | 146 | 5.8193 | 33.8646 | 1391.4043 | 61918.0184 |
| 95 | 152 | 5.9318 | 35.1866 | 1535.3631 | 76727.2546 | 101 | 147 | 5.8326 | 34.0188 | 1420.1293 | 65847.5810 |
| 95 | 153 | 5.9478 | 35.3768 | 1564.7337 | 80403.2745 | 101 | 148 | 5.8392 | 34.0962 | 1412.3154 | 63728.5069 |
| 95 | 154 | 5.9497 | 35.3986 | 1556.6709 | 78732.2574 | 101 | 149 | 5.8544 | 34.2744 | 1443.8228 | 67968.6993 |
| 96 | 135 | 5.7205 | 32.7242 | 1307.2147 | 56695.6799 | 101 | 150 | 5.8591 | 34.3290 | 1433.8921 | 65647.0288 |
| 96 | 136 | 5.7384 | 32.9296 | 1320.0235 | 57331.8974 | 101 | 151 | 5.8761 | 34.5289 | 1467.8014 | 70146.2224 |
| 96 | 137 | 5.7466 | 33.0231 | 1331.8313 | 58590.0128 | 101 | 152 | 5.8789 | 34.5615 | 1455.9435 | 67652.4818 |
| 96 | 138 | 5.7610 | 33.1888 | 1341.2887 | 58939.1227 | 101 | 153 | 5.8974 | 34.7799 | 1491.7858 | 72351.3740 |
| 96 | 139 | 5.7727 | 33.3246 | 1357.4620 | 60639.0310 | 101 | 154 | 5.8985 | 34.7919 | 1478.2456 | 69719.9362 |
| 96 | 140 | 5.7834 | 33.4481 | 1363.3941 | 60692.6555 | 101 | 155 | 5.9182 | 35.0246 | 1515.4665 | 74551.9299 |
| 96 | 141 | 5.7988 | 33.6257 | 1383.8279 | 62817.8781 | 101 | 156 | 5.9176 | 35.0182 | 1500.5488 | 71821.4809 |
| 96 | 142 | 5.8057 | 33.7059 | 1386.1540 | 62573.8956 | 101 | 157 | 5.9380 | 35.2603 | 1538.5334 | 76715.3192 |
| 96 | 143 | 5.8244 | 33.9236 | 1410.6073 | 65095.4066 | 101 | 158 | 5.9362 | 35.2381 | 1522.5836 | 73926.9003 |
| 96 | 144 | 5.8276 | 33.9605 | 1409.3522 | 64559.5752 | 101 | 159 | 5.9569 | 35.4844 | 1560.6878 | 78809.9330 |
| 96 | 145 | 5.8494 | 34.2149 | 1437.4429 | 67435.1904 | 101 | 160 | 5.9540 | 35.4496 | 1544.0802 | 76005.8043 |
| 96 | 146 | 5.8489 | 34.2100 | 1432.7396 | 66621.7910 | 101 | 161 | 5.9745 | 35.6946 | 1581.6588 | 80806.7446 |
| 96 | 147 | 5.8734 | 34.4966 | 1463.9547 | 69797.0758 | 102 | 146 | 5.8169 | 33.8364 | 1379.6684 | 60037.6913 |
| 96 | 148 | 5.8696 | 34.4524 | 1456.0522 | 68730.2863 | 102 | 147 | 5.8207 | 33.8805 | 1390.9086 | 61653.4100 |
| 96 | 149 | 5.8962 | 34.7653 | 1489.7597 | 72139.3810 | 102 | 148 | 5.8329 | 34.0228 | 1396.2879 | 61441.6276 |
| 96 | 150 | 5.8895 | 34.6857 | 1479.0064 | 70852.0833 | 102 | 149 | 5.8395 | 34.0993 | 1410.6585 | 63365.5404 |
| 96 | 151 | 5.9176 | 35.0183 | 1514.4865 | 74420.7230 | 102 | 150 | 5.8490 | 34.2109 | 1413.6576 | 62966.0779 |
| 96 | 152 | 5.9083 | 34.9078 | 1501.3264 | 72954.6984 | 102 | 151 | 5.8584 | 34.3203 | 1431.1433 | 65190.3299 |
| 96 | 153 | 5.9374 | 35.2529 | 1537.8030 | 76603.1003 | 102 | 152 | 5.8652 | 34.4004 | 1431.6844 | 64598.9381 |
| 96 | 154 | 5.9259 | 35.1168 | 1522.7444 | 75006.2466 | 102 | 153 | 5.8773 | 34.5423 | 1452.2043 | 67110.0116 |
| 96 | 155 | 5.9554 | 35.4672 | 1559.4320 | 78653.7026 | 102 | 154 | 5.8814 | 34.5903 | 1450.2459 | 66324.6900 |
| 96 | 156 | 5.9423 | 35.3111 | 1543.0131 | 76976.8974 | 102 | 155 | 5.8961 | 34.7636 | 1473.6424 | 69102.2621 |
| 97 | 136 | 5.7311 | 32.8453 | 1314.2185 | 56858.4669 | 102 | 156 | 5.8974 | 34.7798 | 1469.2023 | 68125.8856 |
| 97 | 137 | 5.7435 | 32.9882 | 1337.0352 | 59779.0156 | 102 | 157 | 5.9146 | 34.9826 | 1495.2363 | 71142.1546 |
| 97 | 138 | 5.7559 | 33.1300 | 1337.7057 | 58675.6255 | 102 | 158 | 5.9134 | 34.9677 | 1488.3934 | 69982.8515 |
| 97 | 139 | 5.7704 | 33.2980 | 1364.1095 | 62036.3334 | 102 | 159 | 5.9327 | 35.1971 | 1516.7372 | 73201.7327 |
| 97 | 140 | 5.7808 | 33.4171 | 1362.1638 | 60641.5875 | 102 | 160 | 5.9290 | 35.1529 | 1507.6448 | 71874.4293 |
| 97 | 141 | 5.7974 | 33.6098 | 1391.9201 | 64402.6154 | 102 | 161 | 5.9502 | 35.4051 | 1537.8957 | 75252.9724 |
| 97 | 142 | 5.8056 | 33.7045 | 1387.3804 | 62737.1906 | 102 | 162 | 5.9442 | 35.3339 | 1526.7696 | 73778.2464 |
| 97 | 143 | 5.8241 | 33.9197 | 1420.0722 | 66840.7763 | 103 | 148 | 5.8216 | 33.8911 | 1390.1206 | 61383.4257 |
| 97 | 144 | 5.8301 | 33.9899 | 1413.0963 | 64937.0090 | 103 | 149 | 5.8305 | 33.9952 | 1414.7398 | 64995.6317 |
| 97 | 145 | 5.8501 | 34.2236 | 1448.1415 | 69309.5884 | 103 | 150 | 5.8393 | 34.0977 | 1408.7575 | 63000.3871 |
| 97 | 146 | 5.8541 | 34.2705 | 1439.0066 | 67209.4900 | 103 | 151 | 5.8503 | 34.2257 | 1436.1085 | 66918.8057 |
| 97 | 147 | 5.8752 | 34.5176 | 1475.6980 | 71766.1563 | 103 | 152 | 5.8572 | 34.3073 | 1428.1837 | 64733.4204 |
| 97 | 148 | 5.8774 | 34.5436 | 1464.7836 | 69519.7823 | 103 | 153 | 5.8701 | 34.4582 | 1458.0447 | 68922.9707 |
| 97 | 149 | 5.8990 | 34.7980 | 1502.3247 | 74167.9336 | 103 | 154 | 5.8753 | 34.5189 | 1448.2632 | 66567.0920 |
| 97 | 150 | 5.8997 | 34.8062 | 1490.0739 | 71829.2540 | 103 | 155 | 5.8899 | 34.6911 | 1480.3349 | 70985.7538 |
| 97 | 151 | 5.9213 | 35.0614 | 1527.6410 | 76475.0690 | 103 | 156 | 5.8933 | 34.7311 | 1468.8254 | 68482.0483 |
| 97 | 152 | 5.9208 | 35.0556 | 1514.5259 | 74098.7866 | 103 | 157 | 5.9095 | 34.9220 | 1502.7294 | 73081.0147 |
| 97 | 153 | 5.9418 | 35.3053 | 1551.3260 | 78653.0766 | 103 | 158 | 5.9112 | 34.9423 | 1489.6722 | 70455.8542 |
| 97 | 154 | 5.9405 | 35.2893 | 1537.8141 | 76291.4491 | 103 | 159 | 5.9286 | 35.1486 | 1524.9505 | 75179.4488 |
| 97 | 155 | 5.9605 | 35.5277 | 1573.1357 | 80674.6755 | 103 | 160 | 5.9288 | 35.1507 | 1510.5834 | 72463.6157 |
| 97 | 156 | 5.9586 | 35.5050 | 1559.6471 | 78373.3961 | 103 | 161 | 5.9471 | 35.3683 | 1546.7182 | 77251.5624 |
| 97 | 157 | 5.9772 | 35.7274 | 1592.9046 | 82520.1565 | 103 | 162 | 5.9460 | 35.3546 | 1531.3186 | 74478.2683 |
| 98 | 139 | 5.7637 | 33.2200 | 1342.5576 | 58724.4743 | 103 | 163 | 5.9648 | 35.5787 | 1567.7523 | 79267.5769 |
| 98 | 140 | 5.7781 | 33.3870 | 1351.1552 | 58915.0351 | 104 | 149 | 5.8222 | 33.8979 | 1389.1440 | 61114.2548 |
| 98 | 141 | 5.7872 | 33.4917 | 1365.7214 | 60593.1894 | 104 | 150 | 5.8370 | 34.0711 | 1396.7677 | 61047.6465 |
| 98 | 142 | 5.7983 | 33.6205 | 1371.1540 | 60517.1289 | 104 | 151 | 5.8389 | 34.0929 | 1406.7223 | 62640.0141 |
| 98 | 143 | 5.8107 | 33.7645 | 1389.6807 | 62592.4655 | 104 | 152 | 5.8515 | 34.2400 | 1412.3901 | 62424.2609 |
| 98 | 144 | 5.8184 | 33.8538 | 1391.8688 | 62248.2448 | 104 | 153 | 5.8559 | 34.2915 | 1425.1270 | 64283.7242 |
| 98 | 145 | 5.8341 | 34.0366 | 1414.2273 | 64701.5325 | 104 | 154 | 5.8661 | 34.4112 | 1428.7230 | 63910.8757 |
| 98 | 146 | 5.8383 | 34.0859 | 1413.1551 | 64091.7168 | 104 | 155 | 5.8731 | 34.4929 | 1444.2468 | 66032.5586 |
| 98 | 147 | 5.8571 | 34.3056 | 1439.1099 | 66894.1559 | 104 | 156 | 5.8808 | 34.5841 | 1445.6750 | 65495.4215 |
| 98 | 148 | 5.8580 | 34.3156 | 1434.8347 | 66026.6311 | 104 | 157 | 5.8903 | 34.6960 | 1463.9326 | 67869.3939 |
| 98 | 149 | 5.8796 | 34.5692 | 1464.0369 | 69139.0260 | 104 | 158 | 5.8956 | 34.7580 | 1463.1402 | 67164.2034 |
| 98 | 150 | 5.8772 | 34.5413 | 1456.6992 | 68028.3066 | 104 | 159 | 5.9076 | 34.8993 | 1484.0099 | 69774.2368 |
| 98 | 151 | 5.9012 | 34.8247 | 1488.6929 | 71401.5871 | 104 | 160 | 5.9103 | 34.9322 | 1480.9889 | 68900.9264 |
| 98 | 152 | 5.8959 | 34.7614 | 1478.5265 | 70070.2845 | 104 | 161 | 5.9246 | 35.1012 | 1504.2788 | 71724.4052 |
| 98 | 153 | 5.9220 | 35.0695 | 1512.7458 | 73644.9004 | 104 | 162 | 5.9250 | 35.1057 | 1499.0828 | 70688.5115 |
| 98 | 154 | 5.9139 | 34.9743 | 1500.0755 | 72123.9195 | 104 | 163 | 5.9414 | 35.3001 | 1524.5315 | 73696.3933 |
| 98 | 155 | 5.9415 | 35.3012 | 1535.8797 | 75833.3454 | 104 | 164 | 5.9395 | 35.2774 | 1517.2692 | 72508.4123 |
| 98 | 156 | 5.9311 | 35.1783 | 1521.1062 | 74160.6008 | 105 | 150 | 5.8528 | 34.2551 | 1425.3797 | 64701.2681 |
| 98 | 157 | 5.9596 | 35.5174 | 1557.8014 | 77933.3535 | 105 | 151 | 5.8650 | 34.3984 | 1454.9270 | 68882.5292 |